%% file: thesis.tex
\pdfoutput=1
\documentclass{outhesis}

\usepackage[top=1in, bottom=1in, left=1.6in, right=1.2in]{geometry}

\newcommand*{\ATLASLATEXPATH}{latex/}
\usepackage{\ATLASLATEXPATH atlasmisc}
\usepackage{\ATLASLATEXPATH atlasbsm}
\usepackage{\ATLASLATEXPATH atlasphysics}
\usepackage{\ATLASLATEXPATH myadditions}

\begin{document}
\author{Othmane Rifki}
\university{UNIVERSITY OF OKLAHOMA}
\college{GRADUATE COLLEGE}
\department{HOMER L. DODGE DEPARTMENT OF PHYSICS AND ASTRONOMY}
\mytitle{SEARCH FOR SUPERSYMMETRY IN FINAL STATES WITH TWO SAME SIGN
LEPTONS OR THREE LEPTONS AND JETS WITH THE ATLAS DETECTOR AT THE LHC
}
\address{Norman, Oklahoma}
\yr{2017}
\dgname{DOCTOR OF PHILOSOPHY}

\committee{{Dr. Brad Abbott, Chair}, {Dr. Peter Barker}, Dr. Michael Strauss, Dr. Chung Kao, Dr. Eric Abraham}


\begin{acknowledgements}
\input{texfiles/acknow.tex}

\end{acknowledgements}

\frontmatter
\maketitle


\chapter*{Abstract}
\addcontentsline{toc}{chapter}{Abstract}%
\input{texfiles/abstract.tex}


\chapter*{Preface}\label{chap:preface}
\addcontentsline{toc}{chapter}{Preface}%
\input{texfiles/preface.tex}


\mainmatter

\chapter{Introduction}\label{chap:intro}
\graphicspath{{figures/intro/}}
\input{texfiles/intro.tex}

\chapter{Theoretical Background}\label{chap:theory}
\graphicspath{{figures/theory/}}
\section{The Standard Model of Particle Physics}\label{sec:theory.sm}
\input{texfiles/sec.theory.sm.tex}
\section{Supersymmetry}\label{sec:theory.susy}
\input{texfiles/sec.theory.susy.tex}
\section{Discovery at the LHC}\label{sec:theory.disc}
\input{texfiles/sec.theory.disc.tex}

\chapter{Experimental Apparatus}\label{chap:exp}
\graphicspath{{figures/exp/}}
\section{The Large Hadron Collider}\label{sec:exp.lhc}
\input{texfiles/sec.exp.lhc.tex}

\section{A Toroidal LHC Apparatus (ATLAS)}\label{sec:exp.atlas}
\input{texfiles/sec.exp.atlas.tex}
\section{ATLAS Trigger and Data Acquisition System}\label{sec:exp.tdaq}
\input{texfiles/sec.exp.tdaq.tex}

\section{ATLAS Operations}\label{sec:exp.op}
\input{texfiles/sec.exp.op.tex}

\section{Event Simulation}\label{sec:exp.sim}
\input{texfiles/sec.exp.sim.tex}
\section{Reconstruction and Identification Techniques}\label{sec:exp.reco}
\input{texfiles/sec.exp.reco.tex}

\chapter[The Region of Interest Builder]{The Region of Interest Builder\raisebox{.3\baselineskip}{\normalsize\footnotemark}}\label{chap:roib}
\footnotetext{This chapter is largely based on the author's published work referenced in \cite{pcroib_orifki,Rifki:2016ufa}.}
\graphicspath{{figures/roib/}}
\section{Overview}\label{sec:roib.overview}
\input{texfiles/sec.roib.overview.tex}

\section{VMEbus based RoIB}\label{sec:roib.overview}
\input{texfiles/sec.roib.vme.tex}
\section{PC based RoIB}\label{sec:roib.pc}
\input{texfiles/sec.roib.pc.tex}
\section{Prototype Tests}\label{sec:roib.proto}
\input{texfiles/sec.roib.proto.tex}
\section{Online Performance in Run-2}\label{sec:roib.perf}
\input{texfiles/sec.roib.perf.tex}

\chapter{Analysis Strategy}\label{chap:strategy}
\graphicspath{{figures/strategy/}}
\section{Overview}\label{sec:strategy.overview}
\input{texfiles/sec.strategy.overview.tex}
\section{Benchmarking Models}\label{sec:strategy.pheno}
\input{texfiles/sec.strategy.pheno.tex}
\section{Dataset and Simulated Event Samples}\label{sec:strategy.samples}
\input{texfiles/sec.strategy.samples.tex}

\section{Event Selection}\label{sec:strategy.sel}
\input{texfiles/sec.strategy.sel.tex}
\section{Signal Regions}\label{sec:strategy.sr}
\input{texfiles/sec.strategy.sr.tex}
\section{Analysis Acceptance and Efficiency}\label{sec:strategy.sr}
\input{texfiles/sec.strategy.acc.tex}

\chapter{Data-driven Background Estimation Techniques}\label{chap:fake}
\graphicspath{{figures/fake/}}
\section{The problem of fakes}\label{sec:fake.prob}
\input{texfiles/sec.fake.prob.tex}
\section{Common processes for faking leptons}\label{sec:fake.proc}
\input{texfiles/sec.fake.proc.tex}
\section{The Monte Carlo Template Method}\label{sec:fake.mct}
\input{texfiles/sec.fake.mct.tex}
\section{Matrix Method}\label{sec:fake.mxm}
\input{texfiles/sec.fake.mxm.tex}

\chapter{Background Estimation}\label{chap:bkg}
\graphicspath{{figures/bkg/}}
\section{Overview}\label{sec:bkg.overview}
\input{texfiles/sec.bkg.overview.tex}

\section{Irreducible Backgrounds}\label{sec:bkg.irred}
\input{texfiles/sec.bkg.irred.tex}

\section{Reducible Backgrounds}\label{sec:bkg.red}
\input{texfiles/sec.bkg.red.tex}

\section{Systematic uncertainties}\label{chap:syst}
\input{texfiles/sec.syst.overview.tex}
\subsection{Theoretical Uncertainties}\label{sec:syst.theory}
\input{texfiles/sec.syst.theory.tex}
\subsection{Experimental Uncertainties}\label{sec:syst.exp}
\input{texfiles/sec.syst.exp.tex}

\chapter{Statistical Treatment}\label{chap:stat}
\graphicspath{{figures/stat/}}
\input{texfiles/sec.stat.overview.tex}

\section{Likelihood Function}
\input{texfiles/sec.stat.like.tex}
\section{Limit Setting Procedure}
\input{texfiles/sec.stat.limit.tex}
\section{Statistical Implementation}
\input{texfiles/sec.stat.impl.tex}

\chapter{Results and Interpretation}\label{chap:res}
\graphicspath{{figures/res/}}
\input{texfiles/chap.res.tex}

\chapter{Conclusions}\label{chap:concl}
\graphicspath{{figures/concl/}}
\input{texfiles/concl.tex}


\bibliography{thesis,bibtex/bib/ATLAS,bibtex/bib/PubNotes,bibtex/bib/CMS,bibtex/bib/ConfNotes}

\clearpage

\appendix

\chapter{Auxiliary material}\label{chap:aux}
\graphicspath{{figures/app/}}
\input{texfiles/app.aux.tex}

\backmatter

\end{document}

%% file: texfiles/acknow.tex
While I have not found evidence for supersymmetry in the work presented in this dissertation, 
my search for it led me to meet some talented, passionate, and knowledgeable physicists, 
many of whom became close friends and mentors to me during this exciting journey. 
I am deeply grateful for their guidance and encouragement to think critically and independently,
to take initiatives, and to appreciate the elegance in the world of elementary particles 
and their interactions.

First and foremost, thanks to my advisor, Brad Abbott, for his continuous support and encouragement over these past five years. 
Brad gave me enough freedom and space to pursue the projects I felt most passionate about and at the same time be always
there to guide me through all the decisions I had to make during my PhD. I could always rely on his perspective to set realistic 
expectations, and for that I am very grateful.

Thanks to the faculty of the High Energy Physics (HEP) group at OU, Michael Strauss, Patrick Skubic, Phillip Gutierrez, 
and John Stupak,
for their useful feedback on my many presentations and for financially supporting my research.
I would like to thank my doctoral committee for evaluating my progress and reviewing my work especially towards the end 
of my PhD. 
Thanks to the OU HEP students and postdocs, Ben Peasron, Callie Bradley, David Bertsche, David Shope, Muhammad Alhroob, 
Muhammad Saleem, Scarlet Norberg, and Yu-Ting Shen, who made my experience 
either at CERN or OU richer and more enjoyable.

Thanks to the whole Argonne National Laboratory HEP group. 
I would like to particularly thank Bob Blair, Jeremy Love, and Jinlong Zhang in working together to make the evolution of the ATLAS 
Region of Interest Builder (RoIB) a success. 
We replaced the two-decade-old RoIB that ``discovered the Higgs'', as Jeremy likes to say, with a modern solution 
that worked flawlessly.
I would like to thank Sergei Chekanov with whom I made my first contribution to an ATLAS paper.
I am particularly grateful for having worked with Sasha Paramonov who showed me how to think like an experimental particle physicist.
I would like to also thank the theorists in the group, in particular Ahmed Ismail, Carlos Wagner, and Ian Low for the many 
discussions we had about new signatures of supersymmetry.
I am thankful to James Proudfoot who continued offering his advice and guidance throughout my stay at Argonne and afterwards.

Thanks to William Vazquez and Wainer Vandelli in their invaluable help while developing the RoIB software.
I still remember my excitement the Friday afternoon at CERN when we finally figured out how to get the RoIB to operate at 
over the targeted 100 kHz rate. It was one of the highlights of the project.

Thanks to my collaborators in the ``SS/3L'' supersymmetry seach group. In particular, I am thankful to 
Otilia Ducu, Julien Maurer, Ximo Poveda, Peter Tornambe, and Fabio Cardillo for the countless hours of debugging, 
checking and cross-checking results, and the long Skype conversations. 
I have not only learned from them the intricate details of searching for rare signals and estimating 
the difficult detector backgrounds but also crisis management skills with the three conference rushes we had to endure. 
We had a very pleasant and efficient work environment which translated into being among the first groups to 
publish their results.
It was a real pleasure working with this team.

Thanks to all my teachers and professors who had an impact on my trajectory.
Thanks to my graduate school professors, 
in particular Ronald Kantowski, Chung Kao, and Howard Baer, who made the subjects of 
electrodynamics, particle physics, and quantum field theory much more understandable.
Thanks to my undergraduate professors at Drexel University,
Chuck Lane, David Goldberg, Michael Vogeley, and Robert Gilmore, who gave me a solid foundation
in physics and played an important role in my orientation.
Thanks to my high school physics teacher, Mohammed El Baki, who 
decided in of our classes not to follow 
the curriculum and instead took us on a journey to the fundamental 
constituents of matter down to the level of quarks. 

Thanks to the many friends at CERN and the Geneva area for the good times in the mountains, on the slopes, aboard boats, 
and under water. The hiking, skiing, sailing, and diving experiences made going through graduate school much more enjoyable and fun.

Thanks to my family and friends in Morocco, Philadelphia, Norman, Chicago, and around the world for their continuous support and care. 
The list of people that I would like to thank is very long.
I would like to particularly mention Abdelfattah, Debbie and Mohamed, 
Heidar and his family, Mak, Salah, Sameed, and many who should be in this list but aren't. 

Finally, I want to thank my parents, Latifa and Chouaib, and my sister, Hafsa, for their  
endless support and love during this adventure. This dissertation is dedicated to them.


%% file: texfiles/abstract.tex
The Standard Model of particle physics is the culmination of decades of experimental and theoretical advancements
to successfully describe the elementary particles and their interactions at low energies, up to 100 \GeV. 
Beyond this scale lies the realm of new physics needed to remedy problems that arise at higher energies, the \TeV~scale and above. 
Supersymmetry (SUSY) is the most favored extension of the Standard Model that solves many of its limitations, if predicted 
SUSY particles exist at the \TeV~scale.
The Large Hadron Collider (LHC) at CERN has opened a new phase of exploration into new physics at the \TeV~scale 
after increasing the center-of-mass energy of the proton-proton collisions to 13 \TeV. 
The ATLAS experiment has collected this collision data with over 90\% efficiency due to the excellent performance 
of many of its systems, in particular the data acquisition system.
The work realized and described in this dissertation ensures the efficient collection of ATLAS data as well as the analysis 
of this data to search for SUSY.
The first part is devoted to the migration of the functionality of the multi-card custom electronics 
Region of Interest Builder (RoIB), a central part of the data acquisition system which processes every event recorded by ATLAS, to a single 
PCI-Express card hosted in a commodity computing node. This evolution was undertaken to 
increase the system flexibility and reduce the operational overload associated with custom electronics.
The second part deals with the search for strongly produced supersymmetric particles decaying into final 
states with multiple energetic jets and either two leptons (electrons or muons) with the same electric 
charge or at least three leptons using the whole proton-proton collision dataset of 36 \ifb~at 
$\sqrt{s}$ = 13 \TeV~recorded with the ATLAS detector in 2015 and 2016.
The analysis pioneers the search for supersymmetry with a novel experimental signature of three leptons of the same 
electric charge.
Due to the low Standard Model background, these final states are particularly adapted to searches for gluinos or third generation
squarks in several SUSY production topologies determined from a variety of simplified and phenomenological models.
The main aspects of the analysis are described, in particular the methods used to estimate the various backgrounds that come 
from known Standard Model production processes with a final state similar to the SUSY models being targeted, 
as well as detector measurement effects.
The absence of excess over the Standard Model prediction  is interpreted in 
terms of limits on the masses of superpartners derived at 95\% confidence 
level. In the studied decay modes and depending on the decay topology, the existence of gluinos with masses below
1.9 \TeV, sbottoms with masses below 700 \GeV, and neutralinos with masses 
below 1.2 \TeV~are excluded.

%% file: texfiles/preface.tex
Over the past five years, 
I have always been asked by family, friends, 
and people I meet, what do you do.
I have written the introduction of Chapter 1 with a general audience in mind to answer this
question.

I conducted  my research in order to  better understand the fundamental interactions of elementary particles 
and to search for unknown physics phenomena by studying proton-proton collisions recorded by the ATLAS detector at the world's
most powerful particle accelerator, the Large Hadron Collider (LHC) at CERN.
I have performed this work over five years while being part of the  experimental particle physics group at the University of Oklahoma.
I spent one year at Argonne National Laboratory as an ATLAS Support Center fellow and I spent the last two years of my PhD at CERN. 

During my PhD, I contributed to two standard model measurements with the LHC Run-1 data of proton-proton collisions at 
a center-of-mass energy of 8 \TeV.
I performed the higher order calculations of the photon production rate which was found to agree with data over ten orders of 
magnitude. The result helped reduce the uncertainties associated with the dynamic structure of the proton impacting
every physics result at the LHC \cite{paper-2015-Photon}.
I also participated in the analysis that made the first experimental observation of the  associated production of a top quark pair
and a vector boson \cite{paper-2015-ttV}, an important background for the search of unknown physics that I present in this dissertation.

The search for supersymmetry is one of the highest priorities of the 
LHC program. The LHC experiments has carried out a 
vigorous search program to analyze the fast incoming data at the higher 
center of mass energy of 13 \TeV. 
The main part of my research is in analyzing the LHC proton-proton collision data at 13 \TeV~
in final states with two same sign leptons or three leptons and jets to search for supersymmetry
which has been the subject of several publications \cite{Aaboud:2017dmy,conf-2016-SS3L,paper-2015-SS3L,conf-2015-SS3L}. 
In this dissertation, I describe in detail the work I performed in searching for supersymmetry in this final state.
I present my own work, except where explicit reference is made to the work of others.
It is worth noting that the regulations of the ATLAS collaboration require me to include 
the ``ATLAS'' label in the plots that have used ATLAS data or ATLAS simulation.

The conclusions of this search provide relevant results to help guide the particle physics 
community in setting constraints on a large variety of supersymmetric models. 
This search, along with other complementary searches performed by ATLAS, has
put stringent constraints on the masses of the strongly produced 
supersymmetric particles.

In addition to analyzing the physics data, I have carried out other projects as part of the ATLAS collaboration.
I participated in the evolution of the Region of Interest Builder (RoIB), a system that processes 
every event recorded by ATLAS \cite{pcroib_orifki}. 
I increased the ATLAS data acquisition system flexibility and reduced the operational overload associated 
with custom electronics 
by migrating the functionality of the RoIB from a custom multi-card crate of VME-based electronics to a single custom PCI-Express 
(PCIe) card in a commodity-computer. I have helped install the new system in ATLAS at the start of 2016.
This new system has been used to collect all the data analyzed in this dissertation.

I have also insured the good operation of the ATLAS detector by actively participating in data-taking and fulfilling various 
supporting roles for the ATLAS data acquisition system. I have taken shift leader shifts at the ATLAS control room 
where I was responsible for coordinating the activities of the different detectors of ATLAS to have ATLAS ready for collisions
and for communicating to the LHC operators the readiness of ATLAS for 
proton collisions.
I have also provided 
operational support as an on-call expert for two critical elements of the ATLAS data acquisition system; the RoIB and 
the readout system that buffers all the ATLAS data.

%% file: texfiles/intro.tex
\subsection*{Historical Background}

Of what is the universe made? This question has intrigued human curiosity since the dawn of time. 
Today, we are confident that we do not know the complete answer to this question.
However, a lot of progress has been made with the aim of reducing the diversity of the physical phenomena 
observed around us to a limited number of constituents following fundamental principles.
Over two thousand years ago, the ancient Greeks  postulated that all is made of Earth, Air, Fire and Water.
Fast-forward to the end of the 19$^\text{th}$ century, Mendeleev and others made the astonishing remark that by organizing the 
relative atomic masses of chemical elements, elements with similar chemical properties followed a pattern.
The periodic table of elements was born. 
The predictive power of the periodic table led to the anticipation of new elements that were later discovered.
However, the table lacked compactness and needed a more fundamental underlying structure that could 
connect the different elements together. 

At the turn of the 20$^\text{th}$ century, several important discoveries established 
the existence of the atom and its constituents. 
The atom is formed by electrons bound via the electromagnetic force to a nucleus, where nearly all the mass resides.
The nucleus itself is formed from 
protons and neutrons that are ``glued'' together by the strong nuclear force (or strong force).
These elements formed the underlying substructure that explained qualitatively the systematic organization of the periodic table.
After 1913, quantum ideas were applied to the atom offering a quantitative description of the origin of structure in atoms and molecules, 
including the chemical elements and their properties.
The decades that followed refined our understanding of the composition of matter through a series of experimental results.
By studying the collisions of protons, neutrons, and electrons in the 1950's and 1960's, a plethora of new particles were discovered 
which belonged to the same family as the proton and neutron, called \textit{hadrons}. 
These particles could not all be elementary
\footnote{Elementary particles refer to  particles that cannot be decomposed into further constituents.}.
By invoking a similar argument that atoms were composite based on Mendeleev's table,
a new layer of structure was unfolded to reveal the existence of \textit{quarks} as basic constituents of all hadrons.
Six types of quarks were discovered over the years with the top quark discovered at Fermilab in Chicago, Illinois 
in 1995, being the most massive elementary particle
\cite{PhysRevLett.74.2626,Abachi:1995iq}.

The observation of the continuous energy spectra in the radioactive $\beta$ decay of nuclei 
led to the discover of neutrinos to remedy  the energy conservation law in the decay. 
Neutrinos are very light neutral particles that interact via the weak nuclear force 
responsible for radioactivity and nuclear fusion, the process that powers the stars.
Electrons and neutrinos had other relatives collectively called \textit{leptons}.
The quarks and leptons are referred to as \textit{fermions} and
have a half integer spin, an intrinsic property of elementary particles.
The strong, weak, and electromagnetic interactions are mediated by 
gluons, $W$ and $Z$ particles, and photons, respectively. These particles 
have an integer spin and are called \textit{bosons}. 
%
The latest addition to the known elementary particles happened in 2012 
 with the discovery of 
a new boson, the Higgs boson, 
that allows the quarks and leptons and the $W$ and $Z$ bosons to 
acquire mass\cite{Aad:2012tfa,Chatrchyan:2012xdj}.

\subsection*{The Standard Model}
The physics of elementary particles became the most ambitious and organized attempt to answer the question of what the universe is 
made out of. 
Through a mixture of both theoretical insight and experimental input, we now 
know that everything we see in our daily life is formed from quarks and leptons
that interact via the strong, weak, and electromagnetic forces.\footnote{ 
The fourth fundamental force of gravity is extremely weak and
only acts at the macroscopic scale.}
The forms of these forces are determined from basic principles of 
symmetry and invariance.
As a result, a theoretical framework was constructed to synthesize all these 
developments in a quantitative 
calculational tool that became known as the \textit{Standard Model of particle physics} (SM). 
The only inputs needed by the SM are the interaction strengths of the forces and quark and lepton masses to make 
very accurate predictions about the behavior of elementary particles.
Over the past 30 years, the SM has been vigorously tested by many 
experiments and has been shown to accurately describe particle 
interactions at the highest energies produced in the laboratory.
 Yet, we know it is not the complete story. 

\subsection*{Limitations}
In 1933, an observation of the Coma Cluster by Fritz Zwicky suggested that the galaxies in the cluster were moving too fast to be explained 
by the luminous matter present\cite{1933AcHPh}. 
The same observation was repeated when looking at the rotation speeds of individual galaxies which 
suggested an invisible component of mass, dark matter. 
The experimental evidence established that dark matter is not made out of baryons
and is more abundant than ordinary matter.
For example, anisotropies in the cosmic microwave background, a radiation 
left over from the Big Bang, that were consistent with 
quantum fluctuations from an inflationary epoch \cite{Hu:2001bc,2009AIPC}, 
encoded details about the density of matter 
in the form of 
cosmological parameters as they traveled through space and time to reach 
our experiments.
The astonishing conclusion was that the universe has nearly five times 
as much dark matter as ordinary matter \cite{Bertone:2004pz}.

Supernovae surveys gave direct evidence for an accelerating universe
 \cite{Perlmutter:1998np},
a view that was cemented by the measurement of cosmological parameters
\cite{Adam:2015rua,Ade:2015xua}
which led to the startling discovery that most of the energy density of 
the universe is in the 
form of an unknown negative-pressure, called dark energy \cite{Scranton:2003in}.
There is an extensive program of experiments 
which will probe the dark energy. 

Astrophysics and cosmology told us about 
the existence of dark matter and measured its density to a remarkable 
precision. Particle physics holds the hope to uncover what dark matter is.
In short, all experimental evidence is consistent with a universe 
constructed of 
\begin{itemize}
\item baryons (everyday matter): $\sim 5\%$ 
\item dark matter: $\sim 20\%$ 
\item dark energy: $\sim 75\%$ 
\item neutrinos, photons: a tiny fraction
\end{itemize}

Today, we are confronted by many puzzles related to our view of the universe.
Everything we know of, namely all the particles of the Standard Model, 
constitute only 5\% of the energy budget of the universe. 
The universe is also predominately composed of matter as opposed to 
anti-matter even though at the start of the universe, they were in equal 
amounts
. The Standard Model describes the content of 
everyday matter 
and how it interacts but without telling us why it is that way.
Moreover, the Standard Model only describes these phenomena 
up to an energy scale of $\mathcal{O}\left(100\right)$ \GeV, called the weak scale.
Beyond this scale lies the realm of phenomena not described by the standard 
model that extend all the way to the Planck scale of 
$\mathcal{O}\left(10^{19}\right)$ \GeV. There is no mechanism to generate mass for 
neutrinos in the Standard Model. Last but not least, the Standard Model 
does not incorporate gravity, the fourth fundamental force.
The SM is unable to account for these observed 
features in the universe. 
Thus, there is a need for a theory beyond the Standard Model.

\subsection*{Supersymmetry}

One of the most prominent extensions of the Standard Model, 
that addresses many of the shortcomings mentioned above, is a theory based on 
a new symmetry, called supersymmetry.
This symmetry is between the matter particles, fermions, and particles whose
exchange mediates the forces, bosons. Our current description of the world
treats fermions and bosons differently. Supersymmetry puts forward the idea
that fermions and bosons can be treated in a fully symmetric way. 
In other words,
if we exchange fermions and bosons in the equations of the theory, the 
equations will still look the same. An immediate consequence of the theory
is that every Standard Model particle will have a ``superpartner,'' 
none of which have yet been discovered.
As a result, we can design experiments to search for these 
supersymmetric particles. The work presented in this dissertation is about the search for supersymmetric particles with a specific signature.
The many benefits of supersymmetry will be discussed later but here it is worth 
mentioning two important features of the theory: 
it unifies the three interactions, electromagnetic, strong, and weak forces,
at very high energies 
and it provides a dark matter candidate particle. Now that we understand 
what we are trying to do, it is time to address the question of how to do it.

\subsection*{Experimental techniques}

The human eye can resolve pieces of dust up to $10^{-5}$ m.
The subatomic distances we are interested in probing range
from  $10^{-15}$ m, the size of a proton, 
down to $10^{-18}$ m, the experimental limit to the maximum size of a quark.
Instruments are needed to extend our senses to 
probe these very small scales.
For instance, light microscopes can reveal the structure of things down to 
$10^{-6}$ m, the scale of bacteria and molecules. 
A special type of microscope is needed to probe smaller distances, 
 a particle accelerator.
The basic idea is that in order to see an object, a wave must scatter off 
this object and must have a wavelength smaller than the object being 
probed.
Since particles have a wavelike character, they can be used to 
probe ever shorter distances according to
\[
E = \frac{hc}{\lambda}
\]
where $E$ is the energy of the particle, $\lambda$ is its wavelength, 
and $hc \sim 10^{-6}$ eVm. As a result, 
the higher the speed of the particles, the greater their 
energy and momentum and the shorter their associated wavelength.
Modern accelerators can generate energy in the \TeV~scale and thus probe 
a distance of $10^{-18}$ m.
All the development that we have made 
describes phenomena happening at distances larger than about $10^{-18}$ m.
Thus, it is possible that electrons and quarks have some structure which is 
smaller than what we can resolve in experiment. For this reason, we 
currently consider them as not having any deeper structure, i.e. they are 
called pointlike objects. 

Over the last century, beams of particles were used to study the 
composition of matter.
Initially, beams originated from phenomena that were 
already naturally occurring, such as alpha and beta particles coming from radioactive 
decays and cosmic rays.
Some cosmic rays are much more energetic than what we can produce 
in the laboratory today, however, they occur at random and 
with a low intensity. Instead, 
high energy particle accelerators were used to deliver high intensity beams of 
electrons, protons, and other particles under controlled conditions.
For this reason, particle physics is also known as high energy physics.
By colliding two sufficiently energetic particles,  new particles will be created 
according to Einstein's equation $E = mc^2$ (or more generally $E = \sqrt{\left(mc^2\right)^2+\left(pc\right)^2}$),
where energy can be exchanged for mass, and vice versa, the exchange rate being $c^2$, the square of the 
speed of light. For example, an electron has a mass of 0.5 \MeV~ 
and can only be created in an electron-positron pair, thus 
1 \MeV~of energy is needed for an electron--positron pair to be produced at rest.
Energies in the \TeV~ range were present about a billionth of a second after the Big Bang.
In other words, by colliding high energy particles, it is possible to 
recreate momentarily conditions similar to those of the universe when it 
was newly born.
At such energies, particles and antiparticles were created, including 
exotic forms no longer common today.
Most of the particles generated in these collisions are extremely short lived 
with lifetimes less than $10^{-20}$ seconds, 
producing radiation and decaying to stable particles, such as electrons and quarks, that make up most of what we see today.
One of the exotic forms of matter that may exist is supersymmetry.
The search for evidence for supersymmetric particles using data collected at a
high energy particle accelerator is the subject of this dissertation.

The Large Hadron Collider (LHC) 
is the world's most energetic 
particle accelerator and the pinnacle of colliding beam technology.
Is it located at CERN, 
the European Laboratory for Particle Physics
\footnote{
The acronym comes from French ``Conseil Européen pour la Recherche Nucléaire''
which was established to do fundamental physics research.
In 1952, this research concentrated on understanding the atom and its nucleus, hence the word ``nuclear''.
Today, our knowledge goes deeper than the nucleus which motivates the modified name.
}, near Geneva, Switzerland.
The LHC accelerates counter rotating beams of protons 
to 99.9999991\% the speed of light in a 27 km ring reaching an energy of 
6.5 \TeV~per beam. 
Magnets, cooled 
by the largest cryogenic system in the world to 1.9 K (-271.3 $^{\circ}{\textrm C}$), that keep the 
protons on track and bring the counter-rotating needle-like
beams 
to meet head on 40 million times per second. 
The debris of each collision fly off in all directions, 
briefly producing less common exotic forms of matter
captured by large particle detectors in the form of ``snapshots'' of these collisions, called events.
The teams of scientists analyze these events to identify the different particles that were produced 
and reconstruct the full collision process.
With this information, it is possible to make precision measurements of rare Standard Model processes, like 
the production of the Higgs boson, or search for physics beyond the Standard Model, like evidence 
for supersymmetry.
ATLAS is one of the general-purpose particle detectors at the LHC that supplied the events 
analyzed in this dissertation to search for supersymmetry. 
The ATLAS detector is the largest-volume particle detector ever built --
the size of a seven-story building 46 meters high and 26 meters in diameter, 
weight 7000 tonnes, 
and able to measure particle trajectories down to 0.01 meters.
Bunches of protons pass through each other at the heart of 
the ATLAS detector 40 million times per second.
Each time they cross there are on average 25 proton-proton collisions, leading to 
about a billion proton collisions per second. 
The data generated in these collisions amounts to about 60 terabytes per second, 
an amount far beyond what is technologically possible to store.
In fact, the processes of interest are extremely rare.
For example, the Higgs boson is produced once in 20 million million collisions.
In more practical terms, a Higgs boson might appear once a day during the LHC operations.
ATLAS has a big computational challenge to recognize this one Higgs event and record it to tape 
out of 35 million million other collisions each day.
The topic of this dissertation is to search for supersymmetric particles that 
are even rarer and thus more challenging to look for. 

This dissertation will give a detailed explanation on how we searched for supersymmetric 
particles using the ATLAS detector.
First, the motivation behind the work will begin with an overview of the 
Standard Model of particle physics and supersymmetry in 
Chapter~\ref{chap:theory} followed by the design of the ATLAS detector 
at the LHC in Chapter~\ref{chap:exp}.
The Region of Interest Builder that processes every event recorded by ATLAS 
is covered in Chapter~\ref{chap:roib}.
The detailed description of the search starts in Chapter~\ref{chap:strategy}
covering the basic analysis strategy and the supersymmetric models considered.
The most challenging part of the analysis is the estimation of Standard Model and 
detector backgrounds with novel techniques developed by the author and covered in 
Chapters~\ref{chap:fake} and ~\ref{chap:bkg}. The statistical 
methodology and interpretation of the results is presented in 
Chapters~\ref{chap:stat} and ~\ref{chap:res}.
This analysis represents an important search for supersymmetric particles 
with the early data-set collected by ATLAS at a new center of mass energy of 13 \TeV.
The strength of the search lies in exploring regions of the parameter space 
with a small mass difference between the supersymmetric particles, regions
that are difficult to probe with other searches for unknown physics.

%% file: texfiles/sec.theory.sm.tex
The Standard Model (SM) of particle physics is a description of the physical world around us in terms of fundamental particles 
and their interactions. 
The development of the SM has been guided by both theoretical predictions and experimental discoveries.
The SM includes three of the four fundamental forces: electromagnetism, the strong interaction, and the weak interaction.
The mathematical formalism used relies on quantum field theory. 

The fundamental particles are represented by the states of quantized fields.
Quarks and leptons constitute matter and are associated with fields of half integer spin, called fermion fields.
The dynamics of the system is defined by the Lagrangian, $\mathcal{L}$, a quantity that describes the motion and excitations 
in the fields.
The Lagrangian of the SM is invariant under spacetime dependent continuous internal transformations of the group 
$ SU\left(3\right) \times SU\left(2\right) \times U\left(1\right) $.
This invariance is called gauge invariance and is necessary to ensure that the theory is renormalizable.
The renormalizability condition guarantees the predictive power of the theory.
To preserve gauge invariance, additional quantum fields with spin one are required, called gauge bosons.
As a result, twelve gauge fields are required to write a gauge invariant Lagrangian, 
 eight for the generators of $ SU\left(3\right) $, three for the generators of $SU\left(2\right)$, 
and one for the $U\left(1\right)$ generator.
The elements described are enough to write down the Lagrangian of the SM.

\subsection{Quantum Chromodynamics}

The $ SU\left(3\right) $ gauge symmetry coupled to the quarks describes Quantum Chromodynamics (QCD), the theory of strong interactions.
The eight $ SU\left(3\right) $ gauge fields are associated to the different colored states of the gluon.
The QCD Lagrangian is given by
\begin{equation}
\mathcal{L}_{QCD} = -\frac{1}{4} G\indices{_{A\mu\nu}}G\indices{_{A}^{\mu\nu}} + \sum_{i=\text{flavors}} \overline{q}_i \left(i\slashed{D} - m_i\right) q_i,
\end{equation}
where $G$'s are the gauge fields of QCD given by 
\begin{equation}
G\indices{_{A\mu\nu}} = \partial_{\mu} G\indices{_{A\nu}} - \partial_{\nu} G\indices{_{A\mu}} - g_S f_{ABC} G\indices{_{B\mu}} G\indices{_{C\nu}},
\end{equation}
and the covariant derivative of QCD, $D\mu$,  defined as
\begin{equation}
D_\mu = \partial_{\mu} + i g_S \frac{\lambda_A}{2}G\indices{_{A\mu}},
\end{equation}
where $g_S$ is the strong coupling constant, and $\lambda_A$ are the eight Gell-Mann matrices.
The indices of the quarks, $i=1,2,3$, run over the colors: red, blue, green, and their anticolors.
While the indices of the gluons, $A,B,C = 1, \cdots, 8$, correspond to the combinations of colors and anticolors.
Color must be conserved in all QCD interactions, similar to the  electric charge.
Gluons have been inferred experimentally  and interact with quarks as predicted by the SM \cite{BRANDELIK1979243}.

\subsection{The Electroweak Theory}

The $ SU\left(2\right) \times U\left(1\right) $ gauge symmetry describes the
electroweak theory that unifies the electromagnetic and weak interactions.
There are two problems with this part of the SM.
The four gauge fields of $ SU\left(2\right) $ and $ U\left(1\right) $ 
must be added without mass to preserve gauge invariance.
However, the gauge bosons of the weak force have a large mass according to observation,
and thus in direct contradiction with the prediction.
In addition, the weak interaction violates parity where it couples differently 
to the left and right-handed quark and lepton helicity states.
The solution is to treat the two helicity states of the leptons as different fields
with different couplings. A fermion mass term in the Lagrangian would couple to 
these different fields but will break gauge invariance.
Again to maintain gauge invariance, the fermion fields should be massless in 
direct contradiction with observation.

Both of the problems described can be resolved by introducing 
spontaneous symmetry breaking. The principle is to introduce new scalar fields with 
zero spin that couple to the electroweak  $ SU\left(2\right) \times U\left(1\right) $
gauge fields while preserving the gauge invariance of the Lagrangian.
The form of the potential describing this new interaction is chosen in such a way that 
zero values of the fields do not correspond to the lowest energy state.
As a consequence, the ground state of the field will ``break'' the  
$ SU\left(2\right) \times U\left(1\right) $ symmetry 
even though the Lagrangian preserves it.
The scalar fields will take a non-zero value, called the vacuum expectation value 
(vev), to allow the fermions and weak gauge bosons to appear as massive particles.
A consequence of this mechanism is that one additional scalar field obtains mass
and thus the theory predicts a neutral massive spin zero particle, the Higgs boson. 

The complete Lagrangian of the electroweak theory, including the mechanism 
of electroweak symmetry breaking, can then be expressed as
\begin{equation}
\mathcal{L}_{EW} = \mathcal{L}_\text{gauge} + \mathcal{L}_\text{matter} + \mathcal{L}_\text{Higgs} +   \mathcal{L}_\text{Yukawa}  .
\end{equation}
The kinetic portion of the Lagrangian introduces the gauge isotriplet, $W\indices{_{\mu}^{i=1,2,3}}$, for 
 $ SU\left(2\right) $ and the gauge single, $B_\mu$, of $ U\left(1\right) $, in
\begin{equation}
\mathcal{L}_\text{gauge} =  -\frac{1}{4} \vect{W}\indices{_{\mu\nu}}\vect{W}\indices{^{\mu\nu}} - B\indices{_{\mu\nu}}B\indices{^{\mu\nu}},
\end{equation}
where 
\begin{equation}
\vect{W}^{\mu\nu} = \partial^{\mu}\vect{W}^{\nu} - \partial^{\nu} \vect{W}^{\mu} - g \vect{W}^{\mu} \times \vect{W}^{\nu},
\end{equation}
\begin{equation}
B\indices{^{\mu\nu}} = \partial^{\mu} B^{\nu} - \partial^{\nu} B^{\mu}
\end{equation}
and $g$ is the $ SU\left(2\right) $ gauge coupling constant. A linear superposition of the 
fields  $W\indices{_{\mu}^{i=1,2,3}}$ and $B_\mu$ lead to the SM $W^{\pm}$, $Z$, and photon.
The matter Lagrangian is
\begin{equation}
\mathcal{L}_\text{matter} = i\overline{\psi}\slashed{D}\psi
\end{equation}
where the covariant derivative of the electroweak theory is defined as
\begin{equation}
\label{eq:DmuEWK}
D_{\mu} = \partial_\mu + i g \vect{W}_\mu \cdot \vect{T} + \frac{1}{2}i g' B_{\mu}Y.
\end{equation}
where $g'$ is  $ U\left(1\right) $ gauge coupling constant, $\vect{T}$ is the weak isospin, and $Y$ is the weak hypercharge.
The Higgs potential introduces a doublet of complex scalar fields, $\Phi$, expressed as
\begin{equation}
 \mathcal{L}_\text{Higgs} = \left(D_\mu\Phi\right)^{\dagger} \left(D^\mu\Phi\right) + \mu^2 \Phi^{\dagger} \Phi - \lambda  \left(\Phi^{\dagger} \Phi \right)^2
\end{equation}
where $D_\mu$ is given in Eq. \ref{eq:DmuEWK} and 
\begin{equation}
\Phi = \left( \begin{matrix} \phi^+ \\ \phi^0\end{matrix} \right)
\end{equation}
The shape of the Higgs potential is determined by the parameters $\mu$ and $\lambda$. 
If $\mu^2 < 0$, the Higgs field will acquire a set of identical minima with a vev of 
$v=-\frac{\mu^2}{2\lambda} \equiv \frac{v^2}{2}$. The physical mass of the Higgs particle in the SM is
\begin{equation}
m_h = \sqrt{-2\mu^2},
\label{eq:theory.sm.mh}
\end{equation}
observed by ATLAS and CMS in 2012 with a mass of \\
$m_h = 125.09 \pm 0.24$ \GeV~\cite{atlas_higgs,cms_higgs}.
The Yukawa interactions are introduced to the Lagrangian manually to describe the interaction between the fermions and the Higgs field, 
expressed as 
\begin{equation}
 \mathcal{L}_\text{Yukawa} = \sum_\text{generations} \left[-\lambda_e \overline{L} \cdot \phi e_R - \lambda_d \overline{Q} \cdot \phi d_R 
 - \lambda_u \epsilon^{ab} \overline{Q}_a \phi_b^{\dagger} u_R + h.c. \right]
\end{equation}
where $\lambda$ is the Yukawa coupling of the particular fermion, $L$ and $e_R$ are the lepton fields, 
$Q$, $u_R$, and $d_R$ are the quark fields, $\epsilon^{ab} $ is the completely antisymmetric  $ SU\left(2\right) $ tensor 
with $\epsilon^{12} = 1$ ($\epsilon^{11} = \epsilon^{22} = 0$, $\epsilon^{21} = -1$).

\subsection{Limitations of the Standard Model}

\begin{figure}[htb!]
\centering
\includegraphics[width=0.75\textwidth]{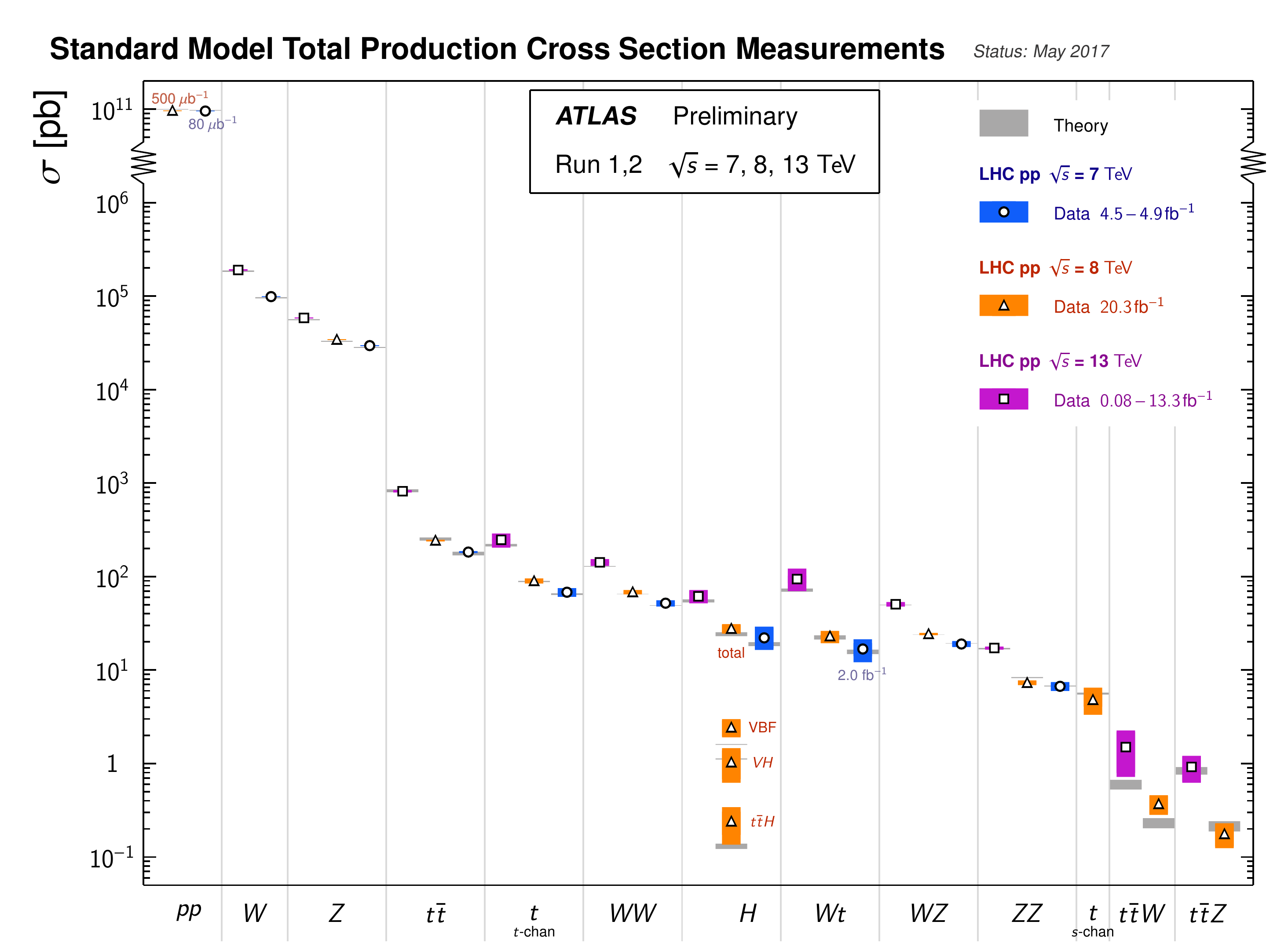}
\caption{Summary of several Standard Model total production cross section measurements, corrected for leptonic branching fractions, 
compared to the corresponding theoretical expectations. All theoretical expectations were calculated at next-to-leading order or higher.}
\label{fig:theory.sm.summary}
\end{figure}

\begin{figure}[htb!]
\centering
\includegraphics[width=0.75\textwidth]{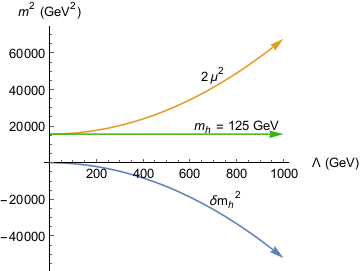}
\caption{Illustration of how the Higgs mass parameter $m_h^2 = -2\mu^2$ needs to be adjusted to compensate for the quantum corrections $\delta m_h^2$ to ensure 
that $m_\text{h,physical} \sim 125$ \GeV~\cite{Bae:2015jea}.}
\label{fig:theory.sm.deltamh2}
\end{figure}

The SM has now been tested successfully over the past decades which validated its dynamics in the gauge sector and in the flavor structure.
As an illustration of this remarkable achievement, Figure~\ref{fig:theory.sm.summary} shows the agreement between the SM total production cross section
of several processes that span twelve orders of magnitude, 
measured by ATLAS compared to theoretical expectations 
at 7 \TeV, 8 \TeV, and 13 \TeV.
The most obvious shortcoming is that the SM makes no attempt to include the fourth fundamental force of gravity. 
The reason is that the addition of the gravitational terms results in a theory that is not renormalizable, hence 
it looses its predictive power.
In the energy regime explored by modern accelerators, the impact of gravity is negligible. 
However, these effects become important at the Planck scale that corresponds to energies of 
$E_\text{Planck} = m_\text{Planck} c^2 = \sqrt{\hbar c^5/G_\text{Newton}} \sim 1.2 \times 10^{16}$ \TeV~(the LHC reaches $\sqrt{s}=13$ \TeV), 
which is well beyond our reach \cite{Ade:2013zuv}. 
Putting this problem aside, there are several problems in the energy range accessible to our accelerators:
\begin{itemize}
\item Dark matter: There is now overwhelming evidence for its existence; rotation curves, Cosmic Microwave Background, primordial abundance of the light elements, etc. 
Yet, the SM does not have a dark matter candidate\cite{Bertone:2004pz}.
\item Baryon asymmetry: The ratio of matter to antimatter is asymmetric with complete absence of antimatter except 
when temporarily formed, as in cosmic rays.
The asymmetry can be explained with the presence of $CP$-violating
\footnote{$CP$ refers to invariance under conjugation of (C) charge and (P) parity symmetries. 
The charge conjugation transforms a particle to its antiparticle while parity transforms the coordinate system to its mirror image.}
interactions. While the SM contains such $CP$ violating terms in the form of the
Cabibbo-Kobayashi-Maskawa matrix, describing  quark mixing, and the Pontecorvo-Maki-Nakagawa-Sakata matrix, describing neutrino 
mixing, the size of this effect is too small to account for the observed asymmetry \cite{Canetti:2012zc}.
\item Anomalous magnetic moment of the muon: The measurement of the magnetic moment anomaly $a_\mu = \frac{g-2}{2}$, where $g$ is the gyromagnetic ratio of the muon, 
deviates from the SM prediction by 3.3$\sigma$ \cite{PhysRevD.73.072003,Hagiwara:2011af}.
\item Neutrino masses: The direct consequence of the observation of solar and atmospheric neutrino oscillations is that neutrinos are massive. The SM does not have a mechanism to include 
mass terms in its Lagrangian\cite{pdg}.
\end{itemize}
These arguments are not considered flaws of the SM but rather limitations that need to be overcome by adding new elements to the theory, i.e. new interactions and new particles.
None of the arguments mentioned address the question of the energy scale at which the new physics should appear.
For this, we turn to the two known scales in physics: the scale of electroweak physics of $\mathcal{O}\left(10^2\right)$ \GeV~and the scale of gravity of $\mathcal{O}\left(10^{19}\right)$ \GeV,
also known as the Planck scale.
The difference between the two scales is in the order of $\mathcal{O}\left(10^{16}\right)$. 
Since the SM is a renormalizable theory, it can be effectively valid up to the Planck scale and 
used to evaluate radiative corrections to any precision.
This causes a problem that can be best illustrated by calculating the mass of the Higgs boson in the SM.
The physical mass of the Higgs boson ( $m_\text{h,physical} \sim 125$ \GeV) can be written as $ m_\text{h,physical}^2 \simeq m_h^2 + \delta m_h^2$, 
where  $m_h$ is the Higgs mass parameter in the Lagrangian given in Eq. \ref{eq:theory.sm.mh}, and $\delta m_h$ is the one-loop radiative corrections 
obtained  by evaluating the diagrams of Figure~\ref{fig:theory.sm.oneloopH}.
\begin{figure}[htb!]
\centering
\begin{subfigure}[htb!]{0.32\textwidth}
\centering
\begin{fmffile}{sunset}
\begin{fmfgraph*}(75,75)
\fmfleft{i}
\fmfright{o}
\fmf{dashes,tension=5,label=$h$}{i,v1}
\fmf{dashes,tension=5,label=$h$}{v1,o}
\fmf{dashes,left,tension=0.75,label=$h$}{v1,v1}
\end{fmfgraph*}
\end{fmffile} 
\subcaption{}
\label{fig:}
\end{subfigure}
\begin{subfigure}[htb!]{0.32\textwidth}
\centering
\begin{fmffile}{sunset}
\begin{fmfgraph*}(75,75)
\fmfleft{i}
\fmfright{o}
\fmf{dashes,tension=5,label=$h$}{i,v1}
\fmf{dashes,tension=5,label=$h$}{v1,o}
\fmf{boson,left,tension=.75,label=$W,,Z$}{v1,v1}
\end{fmfgraph*}
\end{fmffile} 
\subcaption{}
\label{fig:}
\end{subfigure}
\begin{subfigure}[htb!]{0.32\textwidth}
\centering
\begin{fmffile}{sunset}
\begin{fmfgraph*}(75,75)
\fmfleft{i}
\fmfright{o}
\fmf{dashes,tension=5,label=$h$}{i,v1}
\fmf{dashes,tension=5,label=$h$}{v2,o}
\fmf{fermion,left,tension=1.5,label=$\ell,,q$}{v1,v2,v1}
\end{fmfgraph*}
\end{fmffile} 
\subcaption{}
\label{fig:}
\end{subfigure}
\vspace{-0.25cm}
\caption{The one-loop contribution of (a) quarks and leptons, (b) Higgs bosons, and (c) $W$,$Z$ bosons to the mass of the Higgs bosons.}
\label{fig:theory.sm.oneloopH}
\end{figure} 
The Higgs mass can then be expressed as
\begin{equation}
 m_\text{h,physical}^2 \simeq m_h^2 + \frac{C}{16\pi^2}\Lambda^2,
\end{equation}
where the coefficient $C$ embodies the various coupling constants of the SM\footnote{Expression can be found in equation (3) of \cite{Baer:2015fsa}.}.
The diagrams contributing to the Higgs mass diverge quadratically with $\Lambda$, the cut-off scale at which the SM is no longer valid.
If the diagrams do not mutually compensate for one another, the cut-off of quadratic divergences, $\Lambda$, can be as high as the Planck scale ($\mathcal{O}\left(10^{19}\right)$ \GeV).
In other words, the mass scale of the Higgs boson should be of the order of the Planck scale, while the observation is sixteen orders of magnitude below.
The mass parameter $m_h$ must be ``tuned'' to cancel out this huge correction.
This fine-tuning is regarded as \textit{unnatural} and a sign of undiscovered principles that would explain this hierarchy paradox, known as the ``hierarchy problem''.
By requiring that the quantum corrections, encoded in the cut-off scale $\Lambda$, are not too far off from the mass parameter $m_h$, we can make an educated 
guess that the SM can only be valid up to an energy scale of $\Lambda \sim \mathcal{O}\left(1\right)$ \TeV~as illustrated in Figure~\ref{fig:theory.sm.deltamh2}.
This scale can be experimentally probed with the LHC to verify if new physics exists. Hence, the work presented in this dissertation is to search for new phenomena 
at this energy scale.
It is time to address the model studied in this dissertation that would resolve the hierarchy problem and some of the other limitations of the SM described in this section.

%% file: texfiles/sec.theory.susy.tex
The hierarchy problem stated in the last section can be solved if the 
diagrams of Figure~\ref{fig:theory.sm.oneloopH} cancelled out. 
It is possible since the radiative contributions to the Higgs mass 
coming from the fermion loop has a minus sign while 
the boson loop contributes with a positive sign. 
By introducing a new symmetry between 
fermions and bosons, directly linking matter and gauge fields, 
the diagrams can mutually compensate one another.
This new symmetry is called \textit{supersymmetry} and commonly referred to as SUSY.
In this section, we describe briefly the principles of supersymmetry by focusing 
on the concepts rather than the technical implementation of the theory.
We also cover the motivations for examining supersymmetry and how it mitigates
many of the problems of the SM. Last but not least,
we will cover the phenomenology of the theory and its implications in a hadron 
collider like the LHC.

\subsection{Principles of Supersymmetry}



The symmetries encountered in the SM are a direct product of the Poincaré group that encodes the 
symmetries of space-time (translations, rotations, and boosts) and the internal symmetries 
($SU\left(3\right) \times SU\left(2\right) \times U\left(1\right)$). These symmetries
 do not affect the space-time geometric properties of the transformed states.
For instance, an isotropic rotation can transform a neutron into a proton, preserving the same spin, but cannot
transform a neutron into a pion, a particle with a different spin. 
Transformations of supersymmetry are very different in the sense that they directly associate 
fields of integer and half-integer spins, allowing fermions and bosons to be transformed into one another.

The development of the formalism of supersymmetry started with the famous no-go theorem from Coleman and 
Mandula \cite{Coleman:1967ad} who showed that there is no non-trivial way to mix the space-time symmetry 
group with the internal symmetry group in four dimensions and maintain non-zero scattering amplitudes. 
In this theorem, 
only commuting symmetry generators were considered which describe bosonic generators with integer spin.
Haag, Lopuszanski, and Sohnius generalized the theorem by extending the symmetry group to 
anticommutating generators that describe fermions \cite{Haag:1974qh}.
The super-Poincaré group, which includes supersymmetry transformations linking bosons and 
fermions in addition to the other space-time symmetries, was constructed. 
The conclusion is that the most general framework for the symmetries of physics is a direct product of the 
super-Poincaré group with the internal symmetry group.
 This group is represented by four supersymmetry generators $Q_\alpha$ and $\overline{Q}_{\dot{\alpha}}$, 
 where  $\alpha$ and  $\dot{\alpha}$ represent a left-handed and right-handed Weyl spinor index, respectively.
 They can act on a scalar state $\phi$ to obtain a spinor particle
 \begin{equation}
 Q_\alpha \left|\phi\right>  = \left|\psi_\alpha\right>,
 \end{equation}
 where the state $\psi$ represents a fermion. 
The momentum operator and gauge transformation generators of internal symmetries commute with the operators 
$Q$ and $\bar{Q}$. As a result, the supersymmetric states contain bosonic and fermionic fields, 
commonly referred to as \textit{supermultiplets}. The supermultiplets can be either chiral, containing a boson
and a left-handed fermion, or anti-chiral, with a right-handed fermion.
As a consequence of the commutation properties, the particles within a supermultiplet have identical charges, 
such as the electric charge and color charge, under all gauge symmetries.
The implication of this statement is that in a supersymmetric extension of the SM, there will be two 
superpartners, a boson and a fermion, with the same quantum numbers except spin. 

\subsection*{SUSY breaking} 

The momentum operator $P^{\mu}$ also commutes with $Q$, $\left[Q,P^\mu\right]=0$, which implies that if 
supersymmetry is exact, then that every bosonic state must have a corresponding fermionic partner with 
an identical mass. In other words, the supersymmetric partners must come in mass-degenerate pairs.
However, this possibility has been ruled out experimentally since we know that there are no superpartners 
with similar masses as the SM particles. Supersymmetry must then be a broken symmetry. 
However, supersymmetry breaking is not well understood. There is an appealing scheme that preserves 
most of the attractive features of supersymmetry which is known as \textit{soft supersymmetry breaking}.
In this scheme, the superpartner masses can be increased to an acceptable range within the current 
experimental bounds. Also, the scale of the mass splitting should be in the range of 
~$\mathcal{O}\left(100\right)$ \GeV~to $\mathcal{O}\left(1\right)$ \TeV, since it can be linked to electroweak 
symmetry breaking\cite{Chung:2003fi}. 

\subsection{Supersymmetric Phenomenology}

\subsection*{Minimal Supersymmetric Standard Model} 

The simplest implementation of a supersymmetric SM is known as the Minimal Supersymmetric Standard Model, or 
MSSM. It is minimal since it contains the smallest number of new particle states and new interactions 
necessary such that the SM particles still exist in their current forms and within a supersymmetric framework.

In this model, each SM fermion is placed within a supermultiplet containing an additional boson. 
These new bosonic particles are called the same name as their fermionic counterpart with a prepended `\textit{s-}'.
For instance, an electron ($e$) is partnered with a selectron ($\tilde{e}$), a quark ($q$) with a generic 
squark ($\tilde{q}$), etc. On the other hand, the SM bosons with spin 1, that is $B^0$, $W^{\pm}$, $W^0$ before 
electroweak symmetry breaking, are paired with fermionic superpartners with spin $\frac{1}{2}$ into gauge 
supermultiplets. These new fermionic particles are called the same as their bosonic counterpart but with the postfix 
`\textit{-ino}'. So we obtain gluinos (\gluino), winos ($\tilde{W}$) and binos ($\tilde{B}$).

The Higgs sector is chosen to consist of two left-chiral scalar superfields, $H_u$ and $H_d$, with different 
charges under $U\left(1\right)_Y$, $Y=1$ and $Y=-1$, respectively. The $H_u$ and $H_d$ supermultiplets are required 
since each gives mass to only the up or the down quarks. They are also introduced to ensure the cancellation 
of triangle anomalies in the SM, which would otherwise make the theory non-renormalizable.

\begin{table}
\scriptsize
\begin{center}
\vspace*{-0.035\textwidth}
\resizebox{1.\textwidth}{!}{
\begin{tabular}{ccccc}
\hline
\textbf{Particle group} & \textbf{Spin} & \textbf{$P_R$} & \textbf{Gauge eigenstates} & \textbf{Mass eigenstates} \\
\hline\hline
Higgs bosons & $0$ & $+1$ & \psm{H}{u}{0}, \psm{H}{d}{0}, \psm{H}{u}{+}, \psm{H}{d}{-} & \psm{h}{}{0}, \psm{H}{}{0}, \psm{A}{}{0}, \psm{H}{}{\pm}  \\
\hline
 & & & \psusy{u}{L}{}, \psusy{u}{R}{}, \psusy{d}{L}{}, \psusy{d}{R}{}  & (same) \\
squarks & $0$ & $-1$ &  \psusy{s}{L}{}, \psusy{s}{R}{}, \psusy{c}{L}{}, \psusy{c}{R}{}  & (same) \\
 & & & \stopL, \stopR, \sbottomL, \sbottomR & \stopone, \stoptwo, \sbottomone, \sbottomtwo\\
\hline
 & & & \psusy{e}{L}{}, \psusy{e}{R}{}, \psusy{\nu}{e}{} & (same) \\
sleptons & $0$ & $-1$ & \psusy{\mu}{L}{}, \psusy{\mu}{R}{}, \psusy{\nu}{\mu}{} &  (same) \\
 & & & \stauL, \stauR, \snustau & \stauone, \stautwo, \psusy{\nu}{\tau}{} \\
\hline
Neutralinos & $1/2$ & $-1$ & \psusy{B}{}{0}, \psusy{W}{}{0}, \psusy{H}{u}{0}, \psusy{H}{d}{0} & \ninoone, \ninotwo, \ninothree, \ninofour \\
\hline
Charginos & & & \psusy{W}{}{\pm}, \psusy{H}{u}{+}, \psusy{H}{d}{-} & \chinoonepm, \chinotwopm \\
\hline
gluino &  $1/2$ & $-1$ & \gluino & (same) \\
\hline
\end{tabular}}
\vspace*{-0.01\textheight}
\caption{Superpartners of the SM particles in the Minimal Supersymmetric Standard Model showing the mass eigenstates and which gauge eigenstates are mixed.
The two first generations of the squarks and sleptons are assumed to have negligible mixing.}
\label{tab:theory.susy.mssm}
\end{center}
\end{table}

The superpartners of the SM particles in the MSSM are shown in Table~\ref{tab:theory.susy.mssm}.
Since SUSY is a broken symmetry, the gaugino eigenstates mix with the Higgs multiplets to form a set of 
neutralinos ($\tilde{\chi}_{i}^{0}$, $i=1,2,3,4$), charginos ($\tilde{\chi}_{i}^{\pm}$, $i=1,2$), and Higgs bosons 
as a result of the SUSY breaking terms that are added. 
The neutralino and chargino states are ordered in terms of mass as
$m_{\ninoone} \leq m_{\ninotwo} \leq m_{\ninothree} \leq m_{\ninofour}$ and $m_{\chinoonepm} \leq m_{\chinotwopm}$.
It is worth noting that the MSSM expects to have five physical Higgs bosons: two CP-even Higgs bosons
$h^0$ and $H^0$, one CP-odd state $A^0$, and a pair of charged Higgs $H^\pm$. The observed Higgs boson of
125 \GeV~can be one of the two CP-even Higgs bosons. Unlike in the SM where the Higgs mass is one free parameter, 
the masses of the Higgs bosons at tree level and the mixing angle are expressed in terms of 
two parameters chosen to be the mass of $A^0$ ($m_A$) and the ratio of the two vacuum expectation values
($\tan \beta = v_u/v_d$). These vacuum expectation values, $v_u$ and $v_d$, correspond to the local minima 
of the scalar potential in which electroweak symmetry is spontaneously broken.

\subsection*{Simplified Models}

The details given so far have described the particle content of the MSSM. 
As far as the free parameters are concerned, in contrast to the SM which has 
nineteen free parameters, the MSSM has 124 free parameters. While a large portion of the 
parameter space is excluded, there are many degrees of freedom still remaining.
In principle, it is possible to reduce the number of parameters by making well-motivated assumptions
on the physics at higher energy scales. In fact, model builders attempt to formulate reasonable and 
economical models that are phenomenologically viable and falsifiable based on the current experimental results.

Another strategy is to completely decouple many of the particles in the SUSY spectrum, and assume a 100\% 
branching ratio for one specific decay mode, in what is known as \textit{simplified models}.
 In practice, the decoupling is achieved by arbitrarily 
tuning the SUSY breaking parameters in the Lagrangian to include the desired mass terms and couplings.
While such models are known to be not viable and may even 
break the renormalizability of the theory, they are considered as indicative  
of the reach of the analysis in probing the SUSY parameter space and can also be recast by theorists in terms 
of their own models. This is the strategy followed in most of the results shown in the analysis presented in this 
dissertation.

\subsection*{$R$-parity}

It is desirable to write down supersymmetric interactions that preserve baryon or lepton numbers since 
they are putatively good symmetries in the SM.
This can be achieved by requiring the conservation of a new quantity called \textit{$R$-parity}. 
For baryon number $B$, lepton number $L$, and particle spin $s$, the $R$-parity is defined as
\begin{equation}
P_R = \left(-1\right)^{3\left(B-L\right)+2s}.
\end{equation}
The MSSM is formulated as an $R$-parity conserving (RPC) theory. 
However, it can be extended to include a superpotential for the $R$-parity violating (RPV) interactions 
that can be written as
\begin{equation}
W_{\slashed{P}_R} = \frac{1}{2}\lambda^{ijk}L_iL_jE_k + \lambda'^{ijk}L_iQ_jD_k - \kappa^i L_i H_d  + \frac{1}{2}\lambda''^{ijk}U_iD_jD_k,
\end{equation}
in which chiral quark and lepton superfields are denoted by $Q$, $U$, $D$ and $L$, $E$, respectively,
where $i$, $j$, and $k$ are flavour indices. 
The terms show the only interactions that violate baryon or lepton number conservation where 
$\lambda$ and $\lambda'$ couplings break lepton number conservation, while  $\lambda''$ coupling breaks 
baryon number conservation. The work presented in this dissertation will not address RPV scenarios
\cite{Barbier:2004ez}.

\subsection{The Hierarchy Problem}

As described previously, there are scalar and fermion loops that contribute to the radiative corrections 
to the Higgs mass that diverges as $\Lambda^2$. By introducing supersymmetric partners, the 
large fermionic  contribution  to the Higgs mass will be compensated by the scalar particle loop of the 
same mass but with an opposite sign. In the case of unbroken supersymmetry, this cancellation is exact and 
will thus eliminate the fine-tuning problem. However, we know that supersymmetry must be broken.
Naturalness is introduced to place limits on the masses of certain superpartners in order not to replace the 
fine-tuning problem of the SM with another in a supersymmetric model. As a result, there is strong 
motivation for having supersymmetry in the weak scale which will inevitably stabilize the electroweak 
symmetry breaking of the SM which suffered from the fine-tuning problem.

There are other benefits of supersymmetry that are beyond the scope of this 
work. We refer the reader to the literature 
\cite{howiebook,Martin:1997ns,Ellis:2015cva}.

%% file: texfiles/sec.theory.disc.tex
Some notable discoveries in particles physics are those of the 
$W$ and $Z$ bosons \cite{Arnison:1984qu,Bagnaia:1983zx}, the top quark  
\cite{PhysRevLett.74.2626,Abachi:1995iq}, and the Higgs boson \cite{atlas_higgs, cms_higgs}.
The path towards the discovery of these particles was guided by 
theoretical insight which gave great confidence that these particles 
should exist. 
For instance, the features of the $W$ and $Z$ bosons, such as their mass 
and production rates, were known in advance. 
Their signals stood out from the backgrounds without ambiguity.
The top quark discovery was harder, but its production and decay 
properties were predicted.
For the Higgs, there was reasonable evidence for its existence.
The production and decay of the Higgs were all known as a function of 
mass, the only missing parameter in the theory. 
In fact, these properties were also known for alternative models 
to the SM implementation of the Higgs mechanism.

Today, we do not have such theoretical guidance, and thus our task 
is notably more difficult.
The strategy followed at the LHC is to aim at establishing 
significant deviations from the SM by carefully examining 
if the observed signal is not consistent with the Standard Model expectation.
The second step is to understand what this deviation corresponds to 
in the vast space of possible beyond the SM scenarios.

There are three possible scenarios to establish a deviation from  the 
SM expectation: invariant mass peaks, 
anomalous shapes of kinematic distributions, and excess in counting 
experiments.
By examining invariant mass distributions of dilepton, diphoton, or 
dijet final states, a peak that stands out from the background 
continuum that is not predicted by the SM is the clearest indication 
of a new physics signal.
The benefit of this type of signal is that the background can be directly 
taken from data by mere extrapolation of the sidebands of the invariant 
mass below and above the peak. As a result, the simulation will not 
be important in this scenario, which is desirable to avoid any 
mis-modelling or inaccuracies of the simulation.

The second strategy aims at establishing a clear difference in the shape 
of a given kinematic variable between the 
observed data from the expected SM background. 
Distributions, like the missing transverse momentum or the effective 
mass defined as the sum of all reconstructed objects 
and missing transverse momentum in the event, 
are chosen to be sensitive to new physics scenarios.
This approach relies 
heavily on a  precise knowledge of the SM background shapes.
For this reason, special care must be taken to validate the accuracy 
of the SM modeling. 
The claim that a new signal exists is far too important to only rely 
on a direct comparison with Monte Carlo simulation. For this reason, 
a combination of data-driven and correction techniques are employed. 
Often times, data is used internally to correct 
the shape and normalization of the SM backgrounds and to validate the estimate 
before extrapolating to the search region represented by a kinematic 
distribution of a given variable.

The last strategy aims at defining some selection criteria expected to 
increase the probability for observing a new signal, then counting the 
number of observed events passing the cuts and comparing it to 
the expected background.
In a sense, this strategy is similar to the shape discrepancy case 
except that an integral over the  full sample 
passing the cuts is taken since the statistics are typically low.
As a consequence, counting experiments require an even more careful
assessment of the background to achieve the most robust 
understanding of the expected prediction.

The work described in this dissertation follows the last strategy 
of designing several counting experiments. The essential part of the 
work is in establishing a reliable background estimate in these 
experiments to access the compatibility of the observed data with the 
predicted background.

%% file: texfiles/sec.exp.lhc.tex
The Large Hadron Collider (LHC) is the largest 
particle accelerator and collider in the world.
The LHC is built in a circular tunnel 27 km in circumference 
that is buried between 50 m to 175 m underground 
and straddles the Swiss and the French borders 
at CERN.
The LHC is a synchrotron that accelerates two counter-rotating beams of
protons  to 6.5 \TeV~then brings them into head-on collisions at the center of 
four large detectors: ALICE, ATLAS, CMS, and LHCb. The center of mass energy of the proton-proton (or $pp$)\
collision is $\sqrt{s}=$13 \TeV, the energy of the collision data analyzed in this dissertation.
The beam itself has a total energy of 336 MJ requiring an accurate and careful steering 
of the beam at all times.
This is achieved by a strong magnetic field, generated by 
superconducting magnets, that guides the protons around the accelerator.
There are 1232 dipoles magnets, each 15 meters long operating at 1.9 K and generating a magnetic field of 8.33 T.
The dipoles are comprised of 7600 km of superconducting cable which is formed from filaments of Niobium-titanium (NbTi).

\begin{figure}[!htb]
\centering
\includegraphics[width=1.\textwidth]{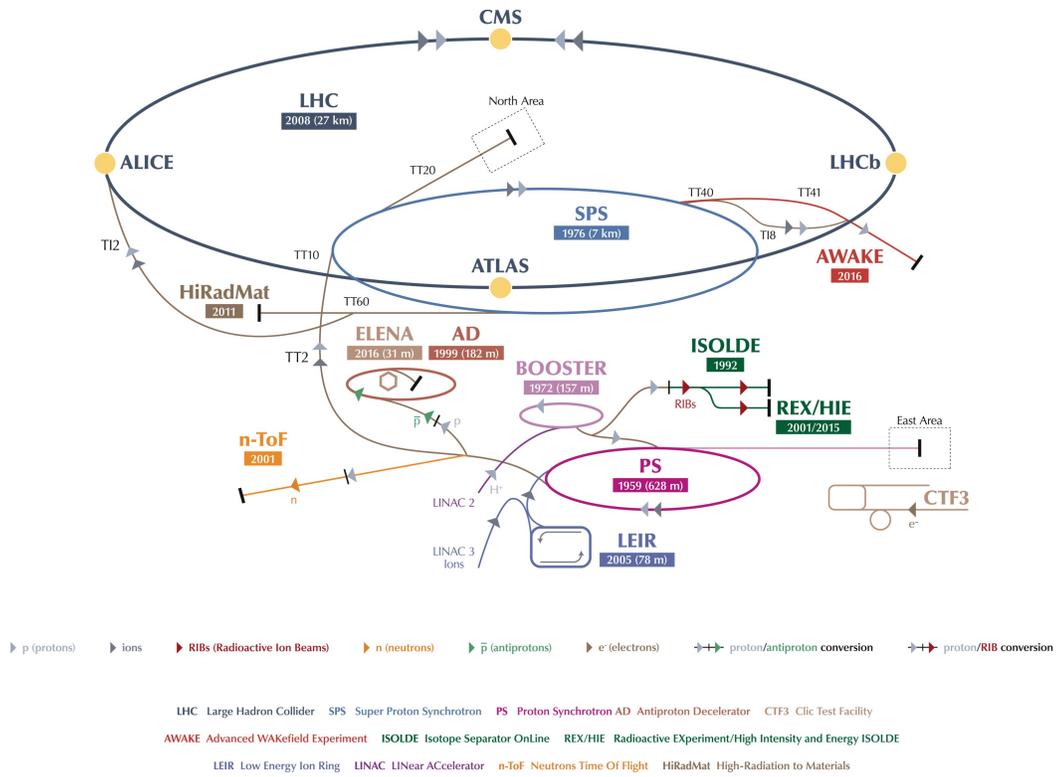}
\caption{The CERN accelerator complex 
composed of a chain of  particle accelerators with
the LHC as the last ring (dark blue line) \cite{DeMelis:2197559}.
}
\label{fig:exp.lhc.CCC}
\end{figure}

A complex of smaller accelerators boost the protons before
injecting them to the LHC, the last accelerator in the chain as shown in 
Figure~\ref{fig:exp.lhc.CCC}.
Protons, obtained from hydrogen atoms, start their journey in a linear 
accelerator
called the Linac2. The Linac2 accelerates the protons to 50 \MeV. Then, they are injected into
the PS Booster, which accelerates them to 1.4 \GeV. After the PS Booster, the protons are sent to
the Proton Synchrotron (PS) where they are accelerated to 25 GeV. They are then sent to the Super
Proton Synchrotron (SPS) where they are accelerated to 450 \GeV. 
At this stage, the protons are injected into the LHC and accelerated to the target 
energy of 6.5 \TeV~per proton.
 The beams are then focused at each of the interaction points to produce proton-proton collisions.
Under normal operating conditions, the colliding beams will circulate for $\mathcal{O}\left(10\right)$ hours at a time.

\begin{figure}[!htb]
\centering
\begin{subfigure}[t]{0.48\textwidth}
\includegraphics[width=0.95\textwidth]{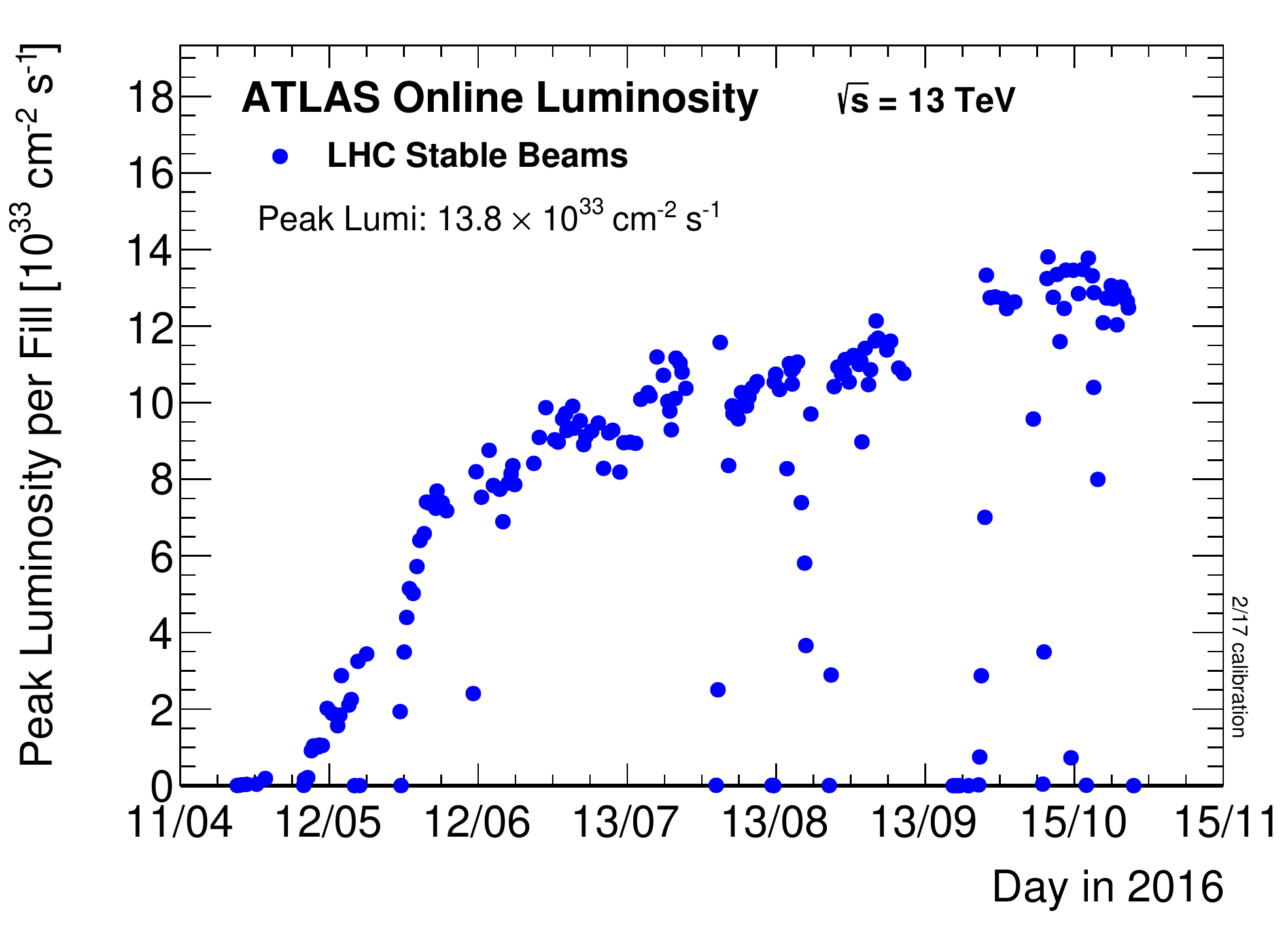}
\subcaption{}
\label{fig:exp.lhc.peakLumiByFill}
\end{subfigure}
\begin{subfigure}[t]{0.48\textwidth}
\includegraphics[width=0.95\textwidth]{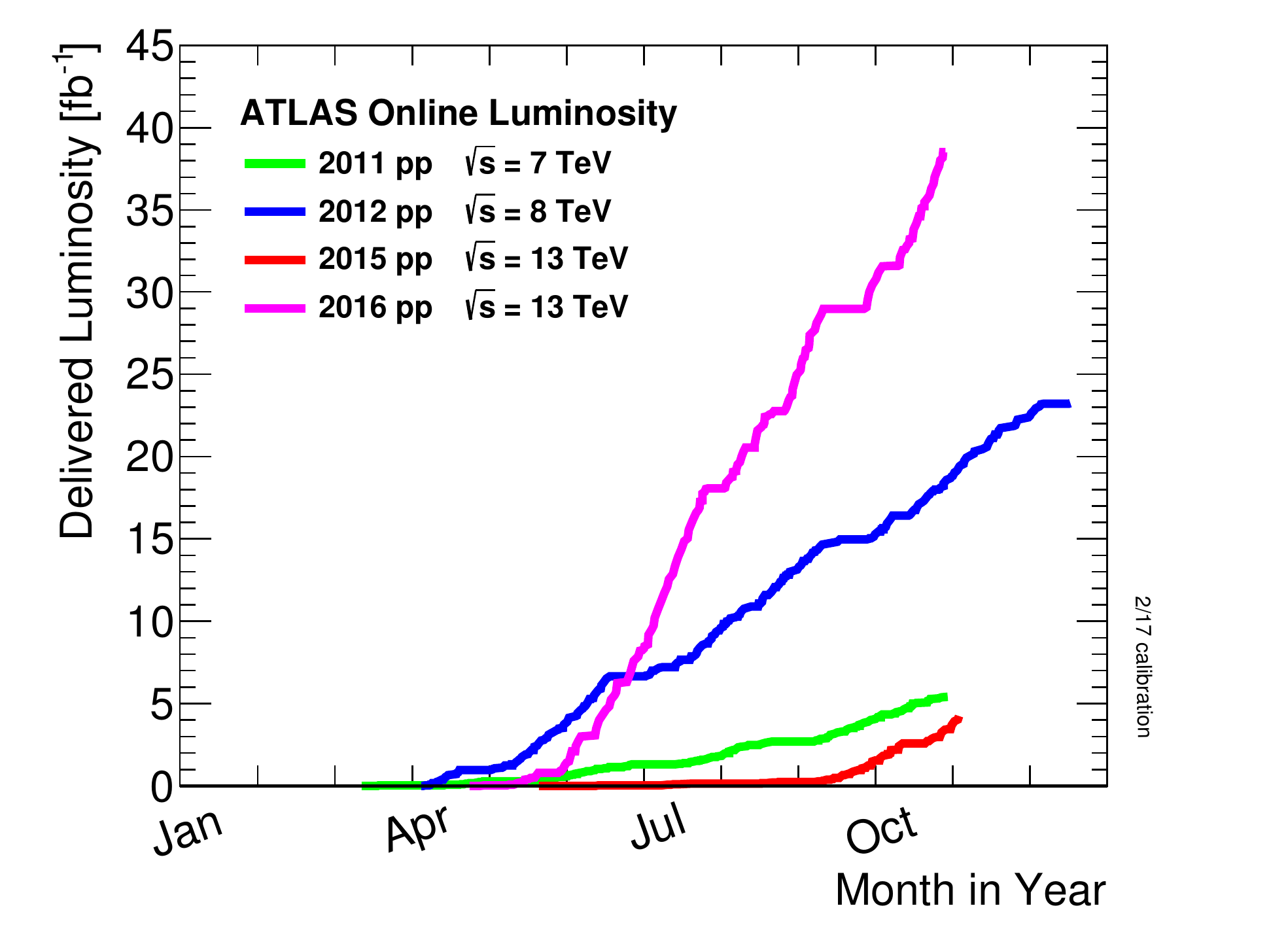}
\subcaption{}
\label{fig:exp.lhc.intlumivsyear}
\end{subfigure}
\vspace{-0.25cm}
\caption{(a) The peak instantaneous luminosity delivered to ATLAS in 2016
and (b) the cumulative luminosity delivered to ATLAS between 2011 and 2016, 
during stable beams for $pp$ collisions}
\label{fig:exp.lhc.peak}
\end{figure}

The protons are grouped in ``bunches'' when circulated in the LHC as a result of the acceleration 
scheme.
In normal operation of the LHC, each proton beam has 2808 bunches, with each bunch containing 
about 100 billion protons. 
These bunches are a few centimetres long and a few millimeters wide when they are far from a 
collision point but squeezed to about 16 micrometers when they collide.
The rate of their interaction is defined in terms of the luminosity, a 
measure of the number of collisions produced per second by the accelerator.
Generally, the event rate $\frac{dN}{dt}$ of a physics process with cross section $\sigma$ is
\begin{equation}
\frac{dN}{dt} = \sigma  \mathcal{L} 
\label{eq:exp.lhc.lumi}
\end{equation}
where the constant of proportionality, $\mathcal{L}$, is called the instantaneous luminosity,
and has units of $\textrm{cm}^{-2}\textrm{s}^{-1}$.
The LHC has exceeded its design luminosity of $10^{34}\textrm{cm}^{-2}\textrm{s}^{-1}$ or 10 $\textrm{nb}^{-1}\textrm{s}^{-1}$ 
(1 barn = $10^{-24} \textrm{cm}^2$) 
by almost 40\% as shown in Figure~\ref{fig:exp.lhc.peakLumiByFill}.
Given the total inelastic cross section of 60 mb, 
the collision rate of protons is then $ \sigma  \mathcal{L} \sim 10^9$ Hz: a billion proton
interactions per second.
The integral of the instantaneous luminosity, $L = \int  \mathcal{L} dt$, refers to the amount 
of data collected. 
The large integrated luminosities allow for the study of rare processes, such as the search for 
supersymmetric particles.
Figure~\ref{fig:exp.lhc.intlumivsyear} shows the data sets collected by the LHC,
 where the data collected in 2015 and 2016 at $\sqrt{s}=13$ \TeV~is the basis of the work 
presented in this dissertation. 

\begin{figure}[!htb]
\centering
\begin{subfigure}[t]{0.48\textwidth}
\includegraphics[width=0.95\textwidth]{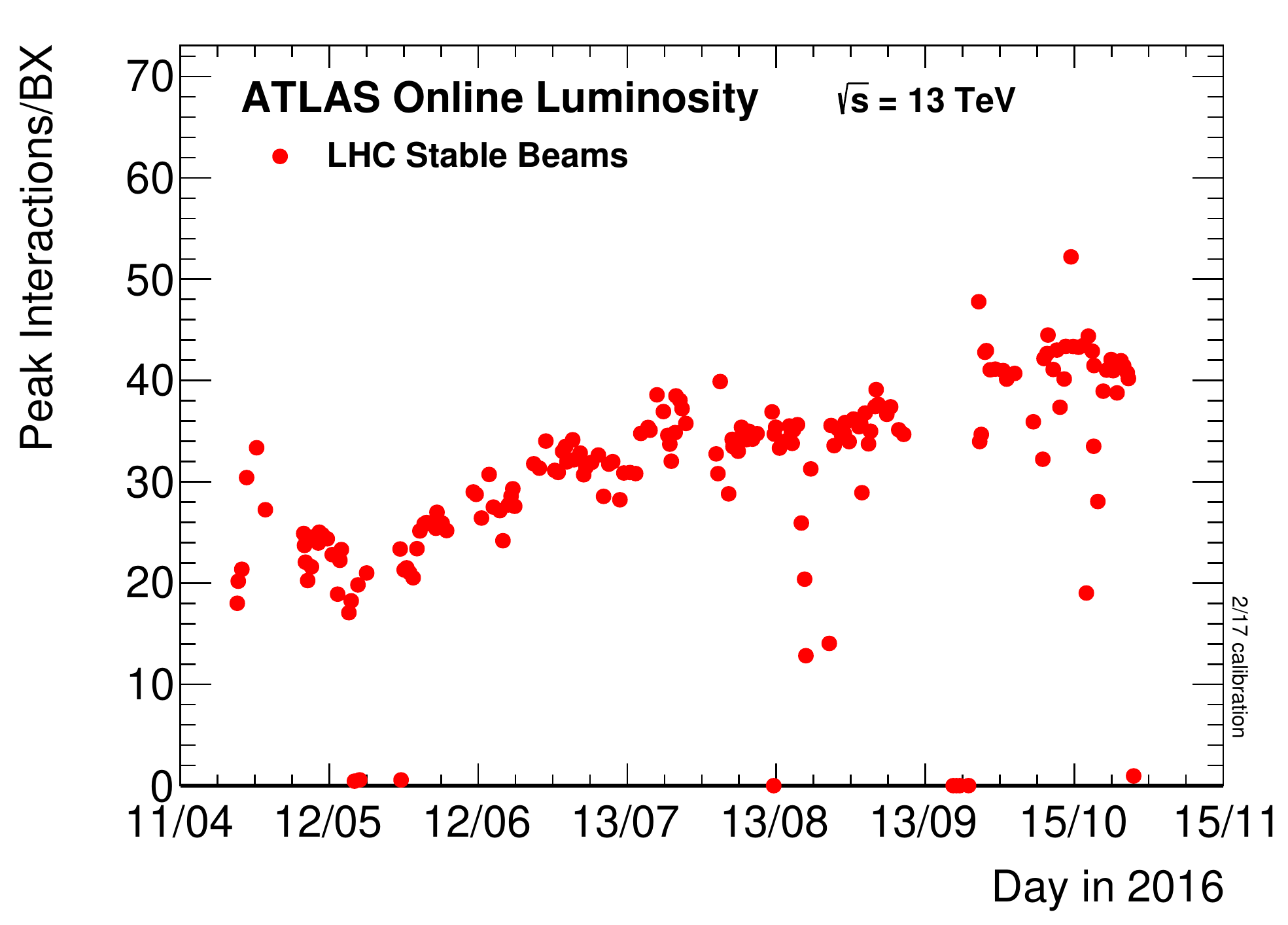}
\subcaption{}
\label{fig:exp.lhc.peakMuByFill}
\end{subfigure}
\begin{subfigure}[t]{0.48\textwidth}
\includegraphics[width=0.95\textwidth]{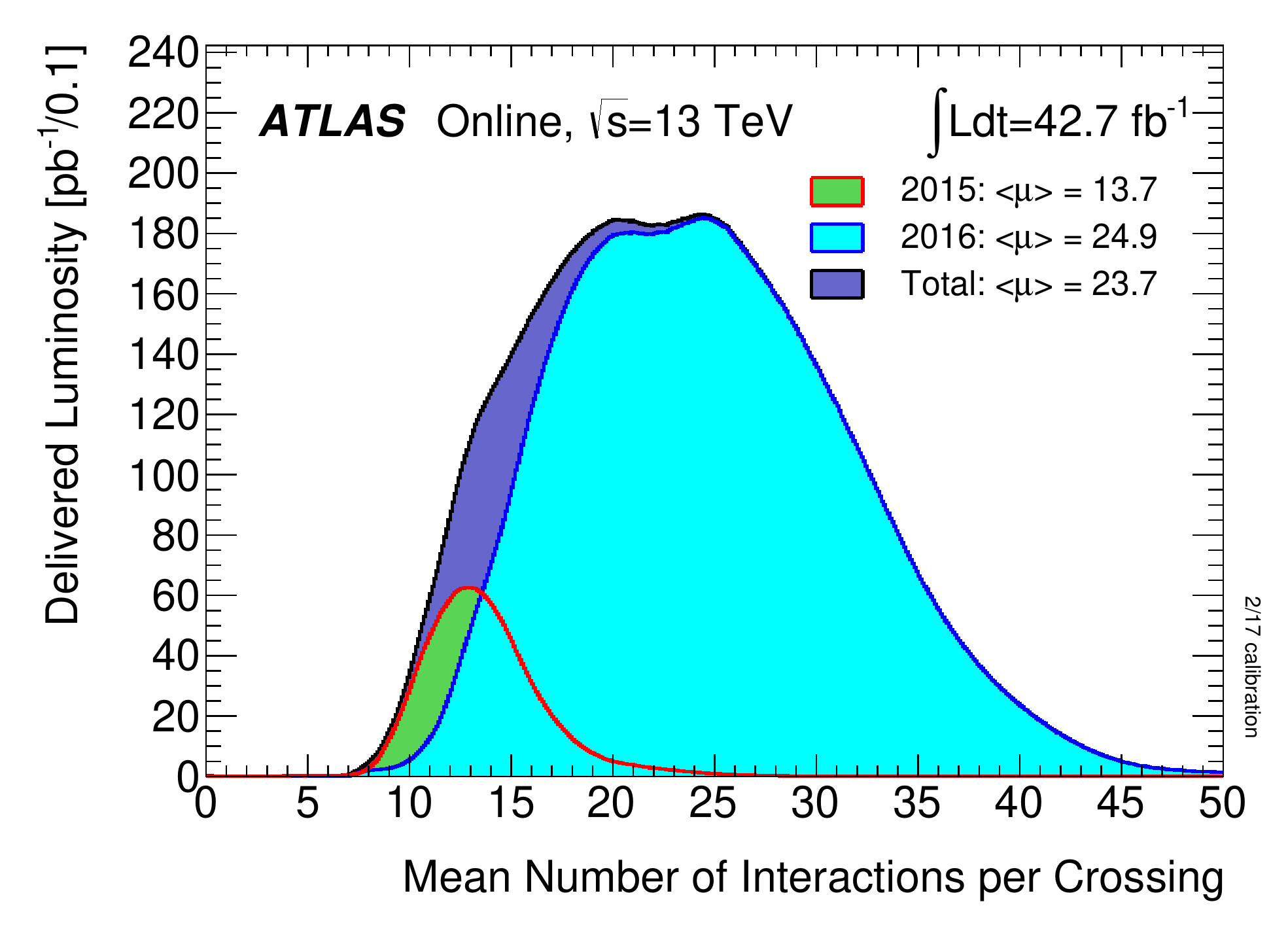}
\subcaption{}
\label{fig:exp.mu_2015_2016}
\end{subfigure}
\vspace{-0.25cm}
\caption{(a) the peak number of inelastic collisions per beam crossing during 2016 
  and (b) the mean number of these collisions per crossing for 2015 and 2016, during stable beams for $pp$ collisions}
\label{fig:exp.lhc.int}
\end{figure}

The other important characteristic of the LHC is that multiple $pp$ interactions occur at every 
bunch crossing.  This quantity is correlated with the instantaneous luminosity as can be seen 
by comparing Figure~\ref{fig:exp.lhc.peakLumiByFill} and Figure~\ref{fig:exp.lhc.peakMuByFill}. Figure~\ref{fig:exp.mu_2015_2016}
shows that the mean number of interactions per bunch crossing was 25 in 2016 with the peak number of interactions 
reaching up to 50. This causes a computational challenge in reconstructing the physics objects coming from 
one interesting interaction.
A typical hard scattering of two protons has a large impact parameter leading to low momentum particles in the final 
state. These types of collisions are known as ``minimum bias'' collisions and are considered as background to 
the more spectacular hard scattering that is typical of an interesting event. 
The minimum bias is generally not well understood since it comes from nonperturbative QCD. There are several Monte Carlo generators, such as {\sc PYTHIA} and
{\sc HERWIG}, that are used to estimate these processes by tuning them to 
data.

%% file: texfiles/sec.exp.atlas.tex
\begin{figure}[t!]
\centering
\includegraphics[width=0.95\textwidth]{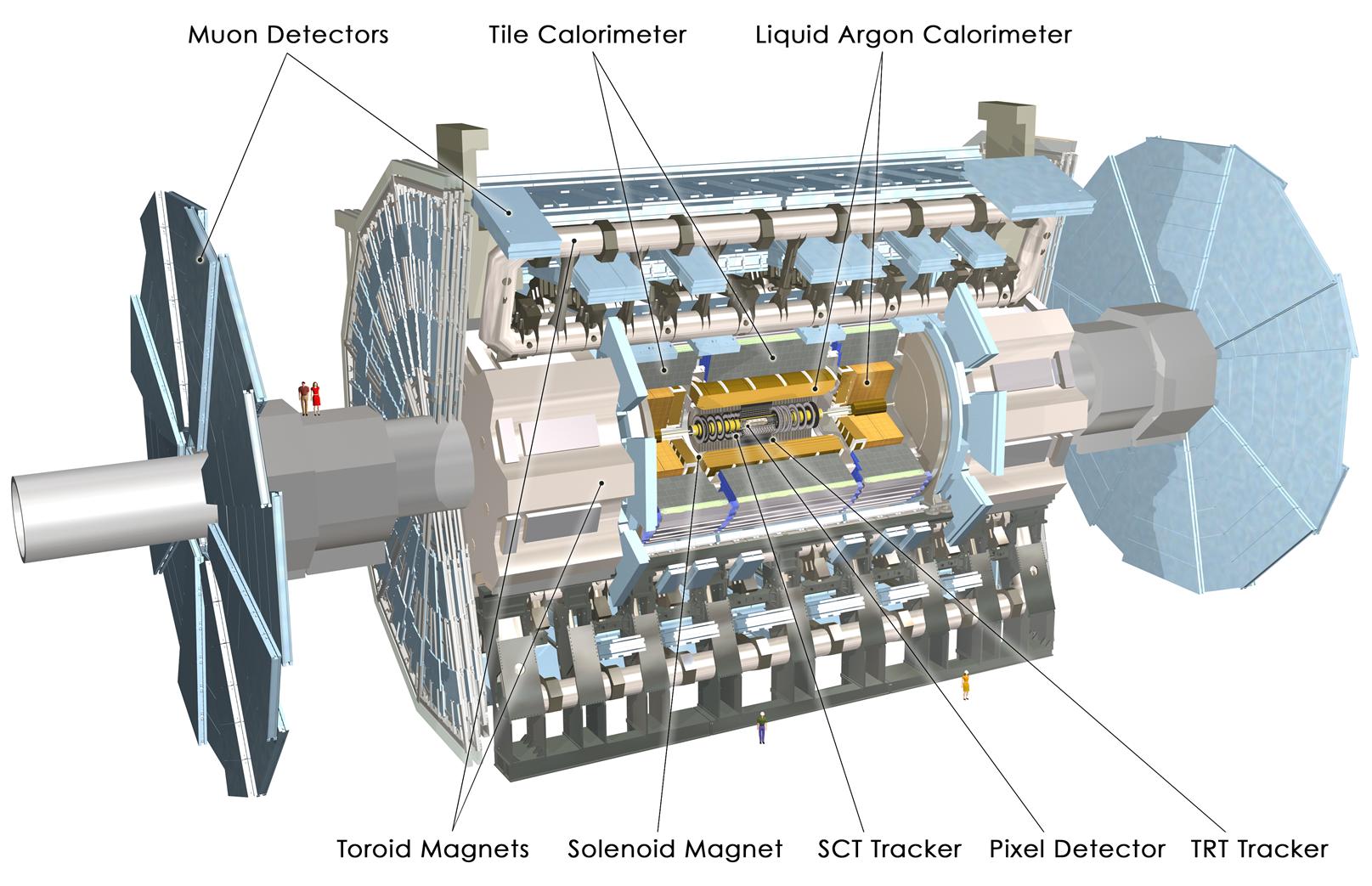}
\caption{Overview of all the subsystems of the ATLAS detector.}
\label{fig:exp.atlas.atlas}
\end{figure}

ATLAS (\textbf{A} \textbf{T}oroidal \textbf{L}HC \textbf{A}pparatu\textbf{S}) is a 
multi-purpose particle detector located at one of the 
LHC interaction points 100 meters underground. 
It is the largest particle detector ever built with a weight of about 7000 tonnes, a length of 44 m, 
and a diameter of 25 m as shown in Figure~\ref{fig:exp.atlas.atlas}.
It is designed to probe Higgs physics, QCD, and flavour physics, as well as a multitude of beyond the Standard Model (BSM) physics scenarios including 
supersymmetry.

ATLAS covers a solid angle of almost $4\pi$ to capture as much information from the collisions 
as possible. 
It is composed of multiple layers of detectors to ensure that all particles produced in the 
collision are identified and measured with high accuracy.
These subsystems are shown in Figure~\ref{fig:exp.atlas.atlas}.
The first detector resides in the part closest to the
LHC pipe and is composed of silicon tracking sensors designed to reconstruct the paths
of charged particles. 
Next, comes the electromagnetic and hadronic calorimeter cells that 
measure the energy of particles. Last comes the muon spectrometer in the outermost part of the 
detector to detect muons since they penetrate the calorimeters. 
An axial magnetic field of 2 T is applied across the inner detector while a toroidal magnetic field 
of approximately 0.5 T is applied across the muon detectors.
The remainder of this chapter will describe in more details these detectors.

\subsection{Co-ordinate System}

\begin{figure}[htb!]
\centering
\begin{tikzpicture}[scale=1.5]
    \draw [<->,thick] (0,3) node (yaxis) [above] {$\eta=0$}
        |- (3,0) node (xaxis) [right] {}; 
    \draw (0,0) coordinate (a_1) -- (2.65,1.41) coordinate (a_2);
    \draw (0,0) coordinate (b_1) -- (2.96,0.49) coordinate (b_2);
    \draw (0,0) coordinate (c_1) -- (1.39,2.66) coordinate (c_2);
    \draw (0,0) coordinate (d_1) -- (2.998,0.11) coordinate (d_2);
    \draw (3.5,0.375) node[above]      
      {$\eta = 2.5$};
    \draw (3.2,1.375) node[above]      
      {$\eta = 1.4$};
    \draw (1.5,2.7) node[above]      
      {$\eta = 0.5$};
    \draw (3.5,-0.1) node[above]      
      {$\eta = 4.0$};
    \draw (2.75,-0.45) node[above]      
      {$z$-axis};
    \draw (-0.1,2.85) node[left]      
      {$y$-axis};
\end{tikzpicture}
\caption{Illustration of some pseudorapidity values relevant for ATLAS.}
\label{fig:exp.atlas.pseudo}
\end{figure}
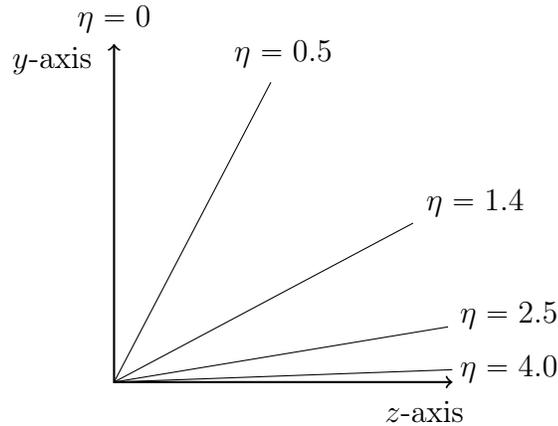

The common coordinate system of ATLAS is right-handed Cartesian, 
with its origin at the nominal interaction point. 
The axes are oriented such that the $x$-axis is pointing towards the center of the LHC ring, the $y$-axis is directed
vertically upward, and the $z$-axis defines one of the beam directions.
The $(x, y)$ plane defines the transverse plane, usually represented by polar coordinates $(r,\phi)$
with $\phi=0$ on the $x$-axis.
The polar angle $\theta$ is replaced by the pseudorapidity
\begin{equation}
\eta = - \ln\left(\tan\left(\frac{\theta}{2}\right)\right),
\end{equation}
shown in Figure~\ref{fig:exp.atlas.pseudo}.
It is named after the rapidity ($y$) since it yields the same quantity for massless particles
\begin{equation}
y = \frac{1}{2}\ln\left(\frac{E+p_Z}{E-p_Z}\right),
\end{equation}
which is invariant under boosts in the $z$-direction.
It is common to describe the separation between two physical objects in the detector by 
\begin{equation}
\Delta R = \sqrt{\Delta \eta^2 + \Delta \phi^2}
\end{equation}

\subsection{Inner detector}

\begin{figure}[htb!]
\centering
\includegraphics[width=0.95\textwidth]{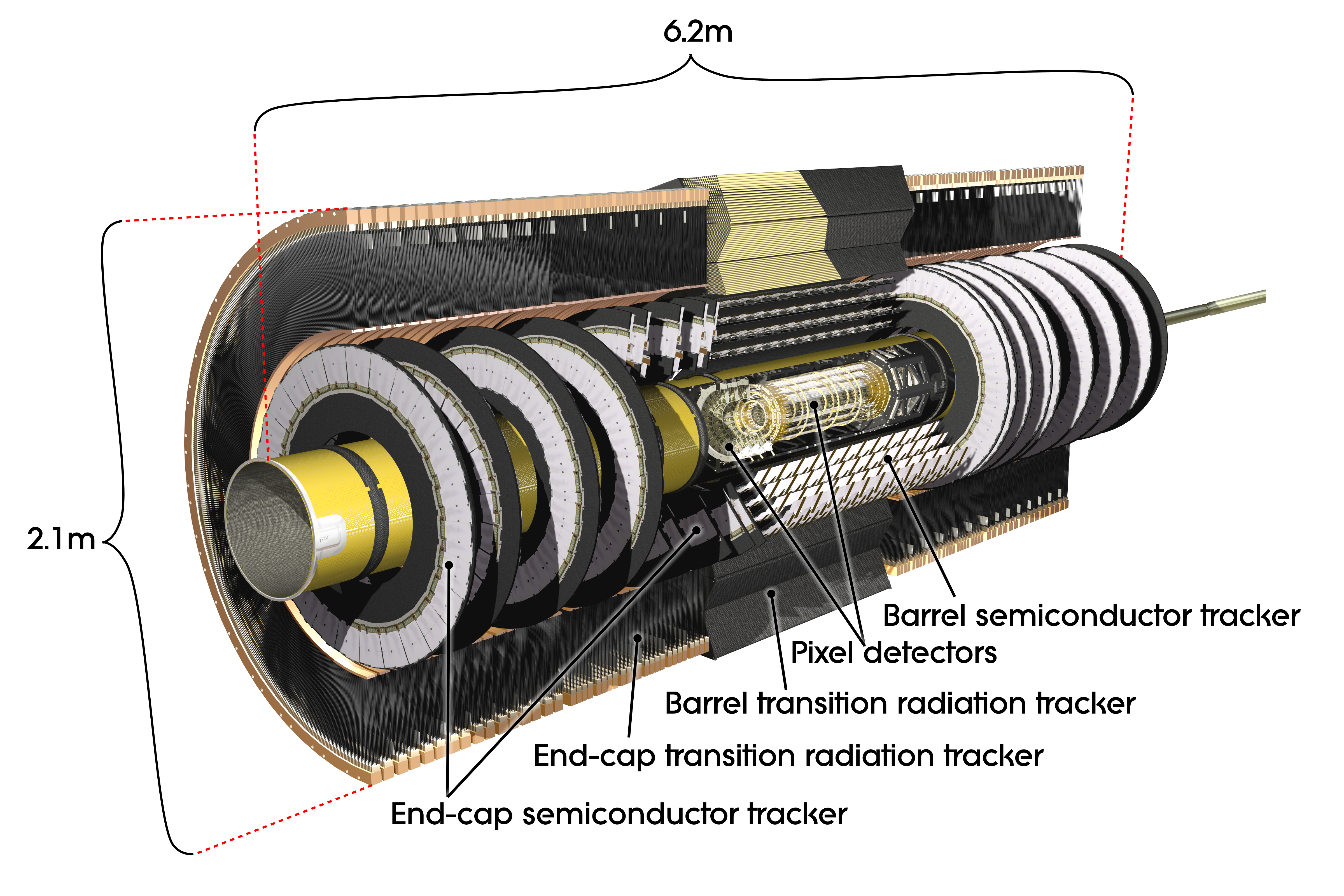}
\caption{Overview of the subsystems of the inner detector of ATLAS.}
\label{fig:exp.atlas.id.all}
\end{figure}

\begin{figure}[htb!]
\centering
\includegraphics[width=0.85\textwidth]{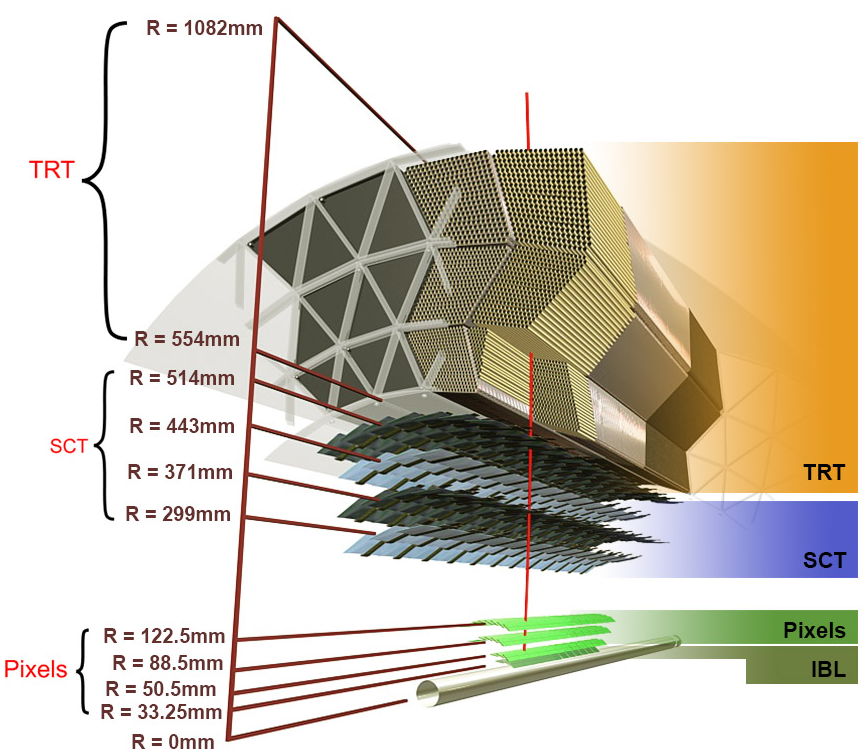}
\caption{Drawing showing the detector elements crossed by a charged particle with 10 \GeV~\pt
in the barrel of the Inner Detector. The particle emerges from the interaction point and traverses the beam-pipe, 
one IBL layer, three pixel layers, four double layers of SCT sensors, and around 35 TRT straws.}
\label{fig:exp.atlas.id.rad}
\end{figure}

In the innermost part of ATLAS is placed a tracking detector referred to as the Inner Detector (ID). 
It has finely segmented detectors to reconstruct the tracks of charged particles in the magnetic field of the solenoid.
The main subsystems of the ID are the pixel detector, the SemiConductor Tracker (SCT), and the Transition Radiation Tracker (TRT)
shown in Figure~\ref{fig:exp.atlas.id.all}.
Overall these give coverage of the solid angle defined
by $|\eta| < 2.5$, and occupy the volume with $33.25 < r < 1082$ mm. Using these systems,
its purpose is to detect the path taken by charged particles as they bend through the
magnetic field, and hence determine their momenta.
Particles from the main $pp$ interaction pass through several layers of silicon detectors each providing a 2-dimensional coordinate. 
To reduce correlations between individual points the layers are spread out evenly along the tracks.
Figure~\ref{fig:exp.atlas.id.rad} shows a charged particle with 10 \GeV~transverse momentum, denoted by \pt, that
emerges from the interaction point and traverses the beam-pipe, four pixel layers, 
four double layers of SCT sensors, and around 35 TRT straws.
These elements will be described next.

\subsection*{Pixel detector}
Since the pixel detector is the closest to the beam pipe ($33.25 < r <242$ mm), it is
the highest resolution detector. It contains 140 million semiconductor pixels each of just
50 $\times$ 400 $\mu$m, 
thus giving a 2-dimensional coordinate with just one layer. The best resolution is in the $\varphi$ coordinate. 
As a result, it is able to measure the charged particle track intersection 
position at the different layers  up
to a precision of 10 $\times$ 115 $\mu$m.
This resolution is important since the 
the area subtended by a given solid angle is at its smallest value near the interaction point.
 It is also designed to tolerate the very high radiation
doses that it is exposed to at such proximity to the interaction point. The detector is
formed of three barrel layers as well as two end-cap structures. Each of the end-caps
comprises four discs of sensors, arranged such that most tracks are more likely to hit pixels in at
least three distinct layers.
The pixel detectors provide the position of the main $pp$
interaction, called the primary vertex, and subsequent vertices from 
$B$-meson decays. These are important parameters for the identification of jets originating from $b$-hadrons,
essential for most of the physics program of ATLAS, including the search for unknown physics presented in this dissertation. 
 Further details can be found in the corresponding technical design report [8].

\subsection*{SCT}

The SCT surrounds the pixel layers.
The SCT is formed of four stereo layers in the barrel,
along with nine discs in each end-cap. 
Each SCT layer is composed of a double layer of silicon
strips, whose axes are tilted by 40 mrad with respect to one another. The pair of measurements at
each SCT layer locates charged particles in $r - \phi$ with an accuracy of 17 $\mu$m, 
and along $z$, with an accuracy of 580 $\mu$m.
The SCT provides between four and nine measurements per particle, with
coverage up to $|\eta|=2.5$.
Further details can be found in the technical design report of the inner
detector [9, 10].

\subsection*{TRT}

The TRT is the largest of the sub-detectors in the ID.
It is composed of  $\sim$300,000 straw drift tubes 
that provide position measurements with an accuracy of $\sim$130 $\mu$m in $\phi$.
A large number of hits, around 35 per particle, is provided, with coverage up to
$|\eta|$=2.0. 
It operates based on the ionization of the gas inside the tubes (70\% Xe, 27\% CO2 and 3\% O2) 
when traversed by charged particles; the ions then drift radially due to
the potential difference, and the excess charge is collected and detected. 
The tubes are arranged parallel to the beam axis in the barrel region, and radially in the end-caps. 
In addition to providing particle tracks, 
the TRT also provides particle identification through
the detection of transition radiation\footnote{Transition radiation is emitted whenever a charged particle 
crosses the boundary between two media.}. For example, 
electrons will emit more transition radiation photons than charged hadrons.

\subsection{Calorimeters}

\begin{figure}[htb!]
\centering
\includegraphics[width=0.95\textwidth]{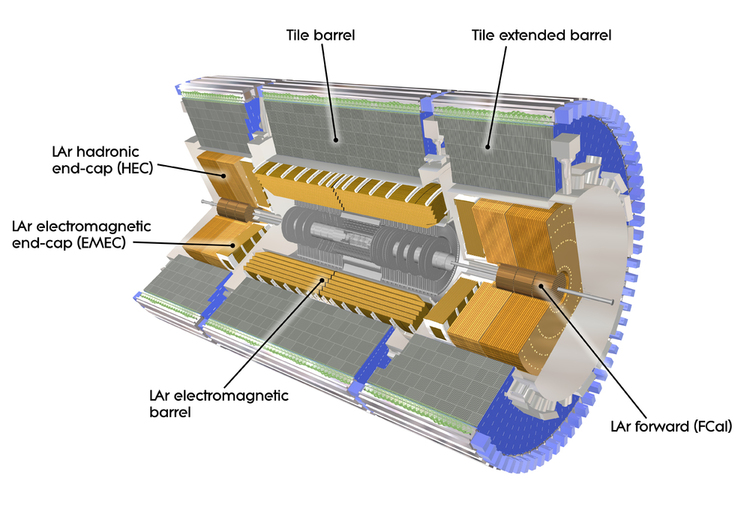}
\caption{Overview of the different calorimeters in ATLAS.}
\label{fig:exp.atlas.calo}
\end{figure}

The calorimeter system measures the energy of hadrons, electrons and photons.
The ATLAS calorimeter is divided into an electromagnetic calorimeter based on liquid argon (LAr)
and a hadronic calorimeter based on  iron-scintillator ``tiles'' (Tile).
The distinction is due to the
different interaction behaviour between the calorimeter and electrons/photons on one side and hadrons on the other side. 
 [12].
An overview of the calorimeter system can be seen in Figure~\ref{fig:exp.atlas.calo}. 
Overall they cover solid angles up to $|\eta| < 4.9$, with the electromagnetic calorimetry providing finer 
grained measurements to augment the inner detector for electron and photon measurements, while
the hadronic calorimeter is coarser but sufficient for jet reconstruction and measurements
of missing transverse momentum.

The ATLAS calorimeters are sampling calorimeters. 
Incident particles produce showers of energy in the calorimeter. 
Only a fraction of the energy produced by the particle is
measured by active detector sensors. 
The energy of the full shower can be inferred from the observed
energy. Thus a calibration must be used to estimate the true energy
of any observed shower in the calorimeter. Each calorimeter is also segmented in $\eta$ and
$\phi$  to provide some directional information, although it is coarser than that from
the inner detector. Finally, the calorimeter is designed to limit ``punch-through'' of high
energy jets into the muon chambers.

\subsection*{LAr Calorimeters}

\begin{figure}[htb!]
\centering
\includegraphics[width=0.75\textwidth]{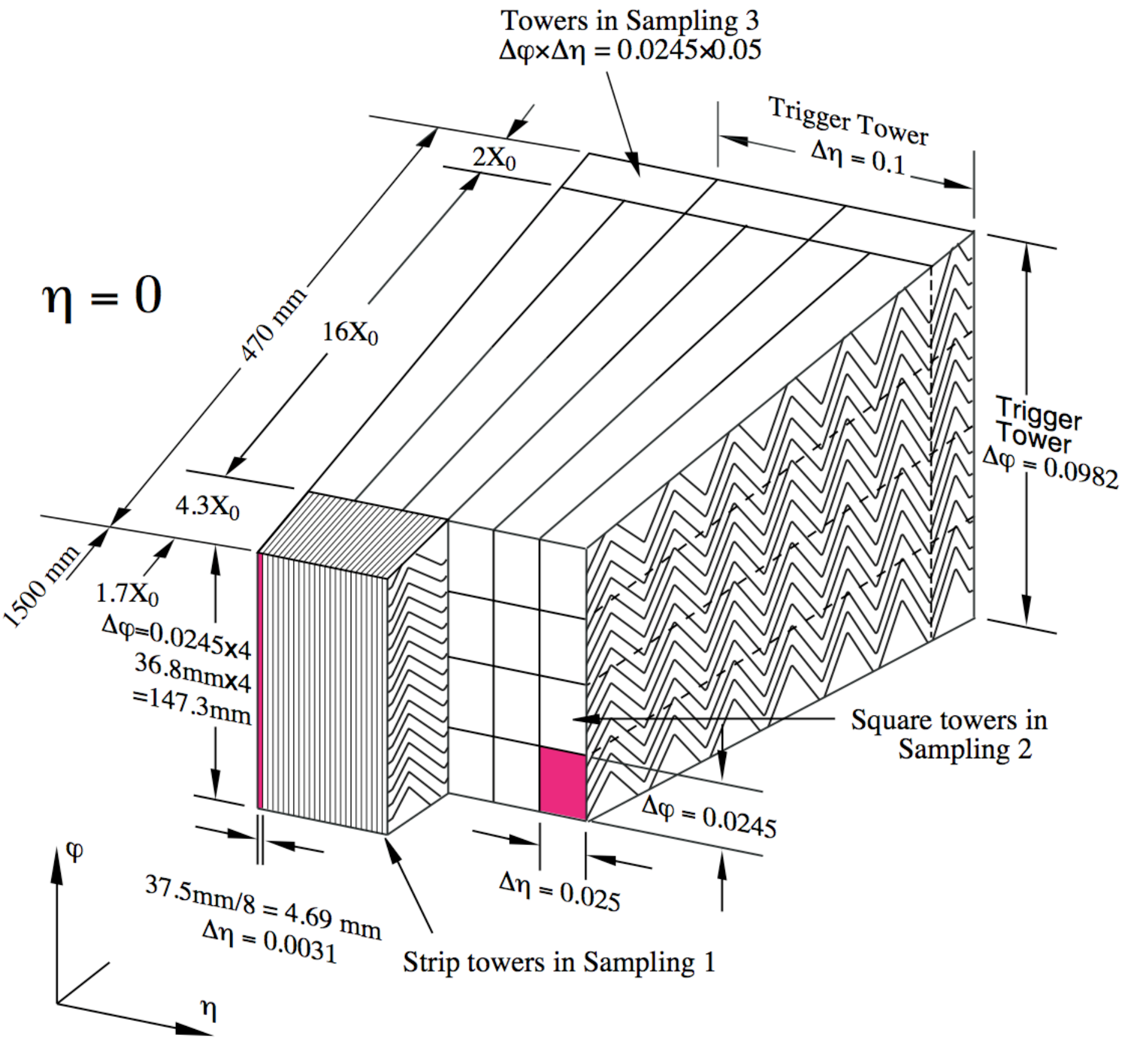}
\caption{Sketch of the accordion structure of the LAr EM calorimeter
where the different layers are clearly visible. 
The granularity in $\eta$ and $\phi$ of the cells of each of the three layers is shown.}
\label{fig:exp.atlas.accordion}
\end{figure}

The energies of electrons and photons are measured by the LAr electromagnetic calorimeters composed of 
the barrel section with $|\eta| <$ 1.475, and end-cap sections with $1.375 < |\eta| < 3.2$.
These detectors provide complete $\phi$ coverage and fast readout, in addition to 
high granularity measurements, critical for particle identification in the range $|\eta|<2.5$
There is a region of slightly degraded performance where the barrel and end-cap sections 
do overlap. Most ATLAS analyses, including the one presented in this dissertation, ignore
electron and photon candidates that fall into this ``crack'' region.
Figure~\ref{fig:exp.atlas.accordion}
shows a cut-away of the different layers in the electromagnetic barrel calorimeter. The first layer, referred to as
the ``strips'', provides very fine segmentation in $\eta$. The strips can separate between showers initiated
by electrons or photons and showers initiated by neutral pions. The second sampling provides most
of the energy measurement and has fine segmentation in both $\eta$ and $\phi$. The third sampling is coarser
and adds additional depth to the calorimeter.

\subsection*{Tile Calorimeters}

The tile calorimeter is the hadron calorimeter covering the range of $|\eta| < 1.7$.
The tile calorimeter uses steel tiles as an absorber and scintillating tiles as the detector.
The scintillator tile calorimeter is separated into a barrel and two extended barrel cylinders.
The light produced in the scintillators is read out with wavelength-shifting optical fibers to photomultipliers (PMTs) placed on the outside of the calorimeter. 

\subsection{Muon System}

The ATLAS muon system is used as a trigger to select events with high energy muons and to measure the position of muons
as they traverse the detector.

\begin{figure}[htb!]
\centering
\includegraphics[width=0.95\textwidth]{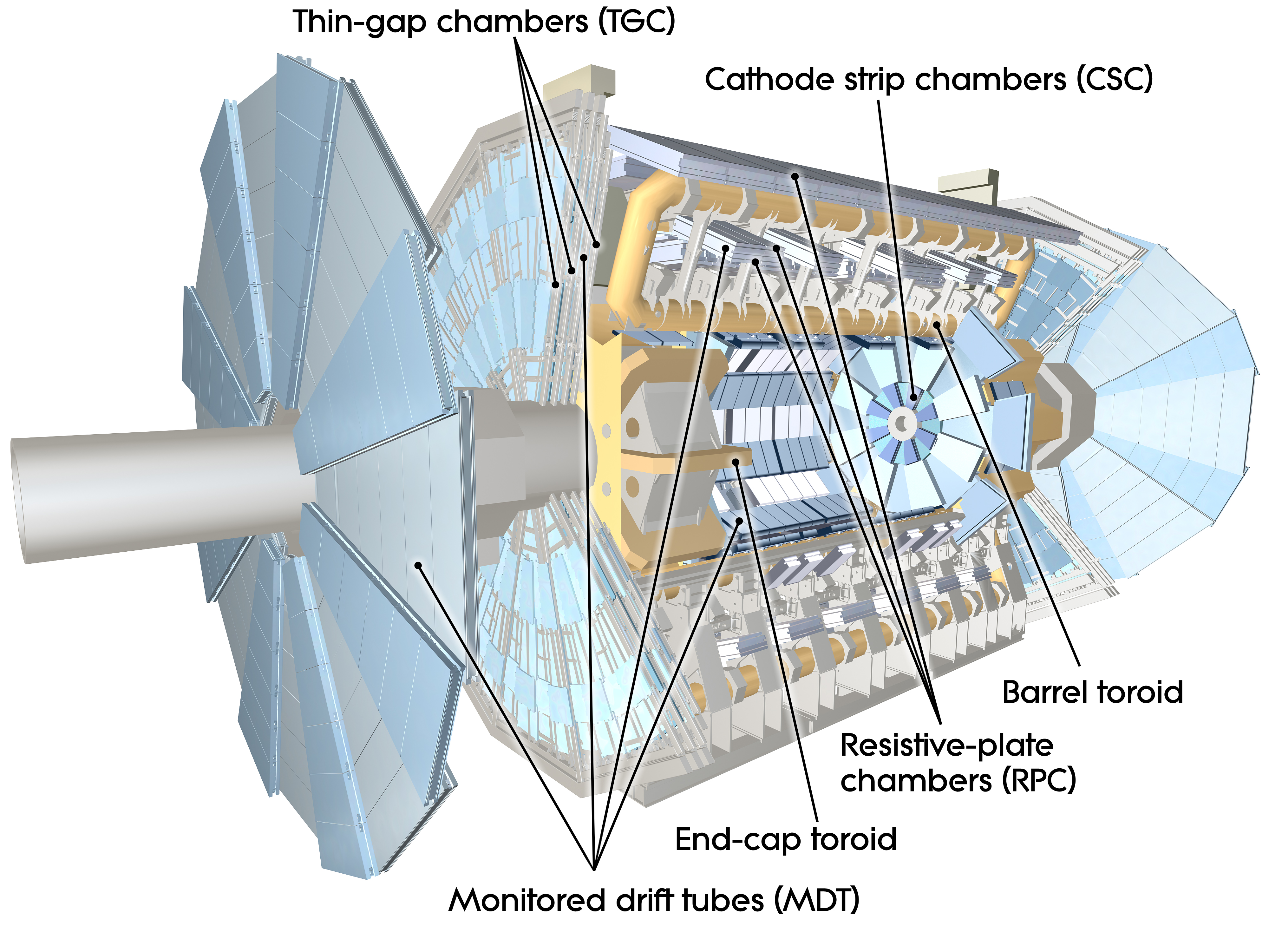}
\caption{Overview of the ATLAS muon system.}
\label{fig:exp.atlas.muon.all}
\end{figure}

The system covers the range of $|\eta| < 2.7$ and operates on the principle of measuring the deflection of tracks due to magnetic fields.
There is then a magnetic field in the barrel section, $|\eta| < 1.4$, induced by the main barrel coils, 
and a magnetic field in the end-cap region, $1.6 < |\eta| < 2.7$, induced by separate end-cap coils, as can be 
seen in Figure~\ref{fig:exp.atlas.muon.all}.
In the region $1.4 < |\eta| < 1.6$, the bending will occur by a combination of the barrel and end-cap fields.

\begin{figure}[htb!]
\centering
\includegraphics[width=0.95\textwidth]{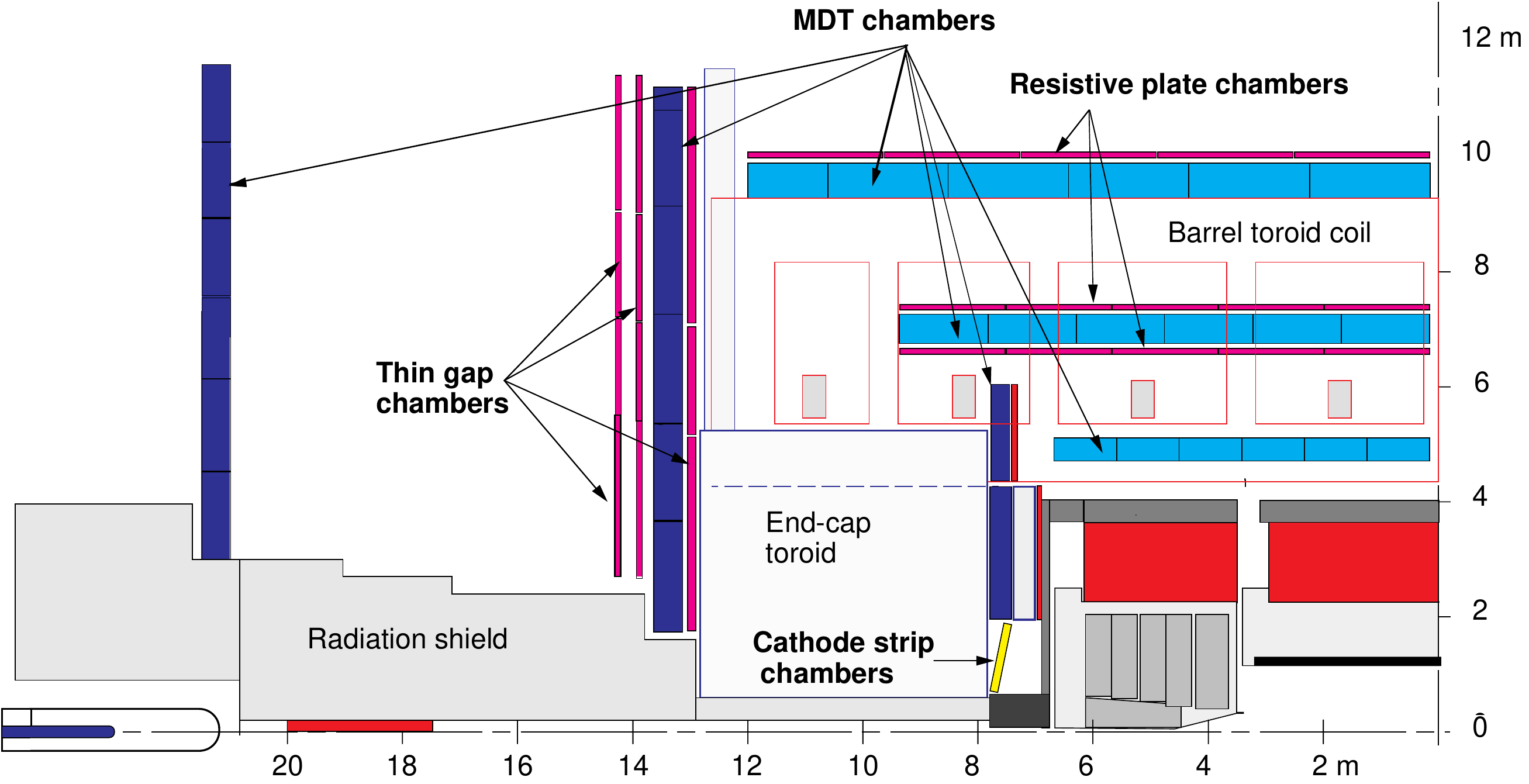}
\caption{Cross-sectional view of the muon detectors of ATLAS in the $\left(y-z\right)$-plane.}
\label{fig:exp.atlas.muon}
\end{figure}

Several technologies are used to select the events and make measurements.
The barrel region has resistive-plate chambers (RPC) for $|\eta| < 1.05$  that provide very fast timing information, $\sim 10$ ns,
used for triggering. The barrel also has monitored drift tubes (MDT) for $|\eta| < 2.0$ that give precise measurements, $\sim 35 \mu$m per chamber,
in the $\left(\eta,z\right)$-plane where the bending occurs.
The forward region of the detector, $2.0 < |\eta| < 2.7$,  has cathode strip detectors (CSCs)  nearest to the interaction point,
followed by thin-gap chambers (TGCs) and additional MDTs. The CSCs achieve a resolution of 40$\mu$m in the $\left(\eta,z\right)$-plane
and 5 mm in the transverse plane.
The layout of these components is more clearly shown in Figure~\ref{fig:exp.atlas.muon}.

%% file: texfiles/sec.exp.tdaq.tex
\begin{figure}[!t]
\centering
\includegraphics[width=1.\textwidth]{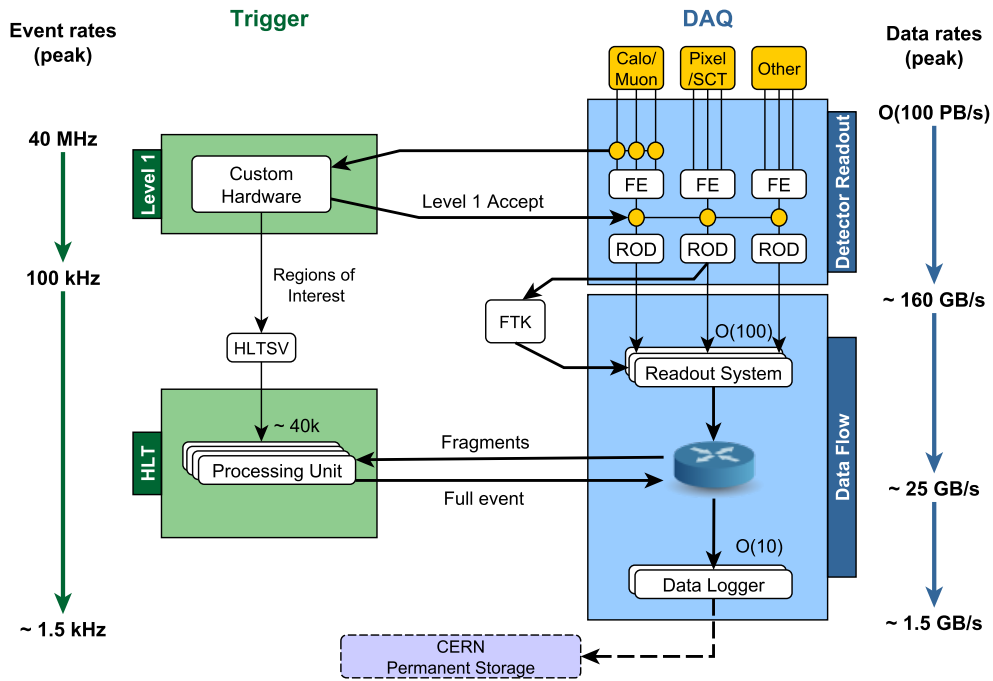}
\vspace{-0.5cm}
\caption{ATLAS TDAQ architecture.}
\label{fig:tdaq_diagram}
\end{figure}

The ATLAS  detector's data acquisition system, illustrated in Figure \ref{fig:tdaq_diagram}, makes use of a multi-tiered trigger to reduce the
bandwidth from the LHC proton bunch crossing rate of 40 MHz
to the 1 kHz written to disk \cite{evolution1,evolution2}. The first tier (Level-1 or L1) \cite{l1}, implemented in real time with custom electronics, 
makes an early event selection to determine if any objects of interest are present and reduces the data flow to 
100 kHz. The second tier, referred to as the High Level Trigger (HLT) \cite{hlt}, is implemented on a commodity computing cluster running custom triggering software. The HLT uses information from the
hardware based L1 system to guide the retrieval of information from the Readout System (ROS) \cite{ros}. 

\subsection{Hardware Trigger (L1)}
The L1 trigger has access to raw data from the calorimeters and the 
muon system. The L1 calorimeter trigger (L1Calo) uses reduced-granularity
information from 7200 trigger towers of the calorimeters. These trigger 
towers are divided in a $\Delta\eta\times\Delta\phi$ space by 
$0.1 \times 0.1$ over most of the calorimeter, and larger in the forward 
region. A decision is made based on the multiplicities and $E_{\rm{T}}$ 
thresholds of the objects 
identified by the L1Calo algorithms: electromagnetic (EM) clusters, 
$\tau$-leptons, jets, missing transverse energy, scalar sum $E_{\rm{T}}$, 
and total transverse energy of the L1 jets.
The L1 muon trigger (L1muon) uses measurements of the trajectories of muons 
in the RPC and TGC trigger chambers, located in the barrel and end-cap regions 
of the muon spectrometer. The multiplicity of the various muon \pt thresholds
is input to the trigger decision.

The central trigger processor (CTP) combines results from the L1Muon and L1Calo
triggers to issue an overall L1 accept or reject decision.
To facilitate this task, the CTP programs up to 256 configurations that 
consist of various combinations of $E_{\rm{T}}$ and \pt requirements, 
or thresholds.
The CTP has the capability of implementing different isolation criteria 
to the different objects such as the L1 EM clusters. 
A trigger menu is implemented as a collection of L1 items, each containing a 
logical combination of one or more configured L1 thresholds. 
For example, the item L1\_EM30i refers to an event requiring at least one 
isolated EM object with a transverse energy of $E_{\rm{T}} > 30$ \GeV. 
If the rates of a particular 
object is high, such as EM objects with low momentum, a prescale factor $\alpha$ is applied to the L1 item in the menu, where only 1 in $\alpha$
events is passed to the HLT. The L1 prescales are generally adjusted to 
maintain the optimal use of the allocated bandwidth for an L1 item during 
data-taking since the luminosity drops over the course of a run.

The L1 trigger has a 2.5 $\mu$s latency where the data fragments are 
held in pipeline buffers located within detector-specific front-end 
electronics. 
Once the CTP issues an accept, the data is pushed to detector-specific 
Readout Drivers (RODs), then transferred to the Readout System (ROS). 
The rest of the chain is described next. 

\begin{figure}[t!]
\centering
\includegraphics[width=0.95\textwidth]{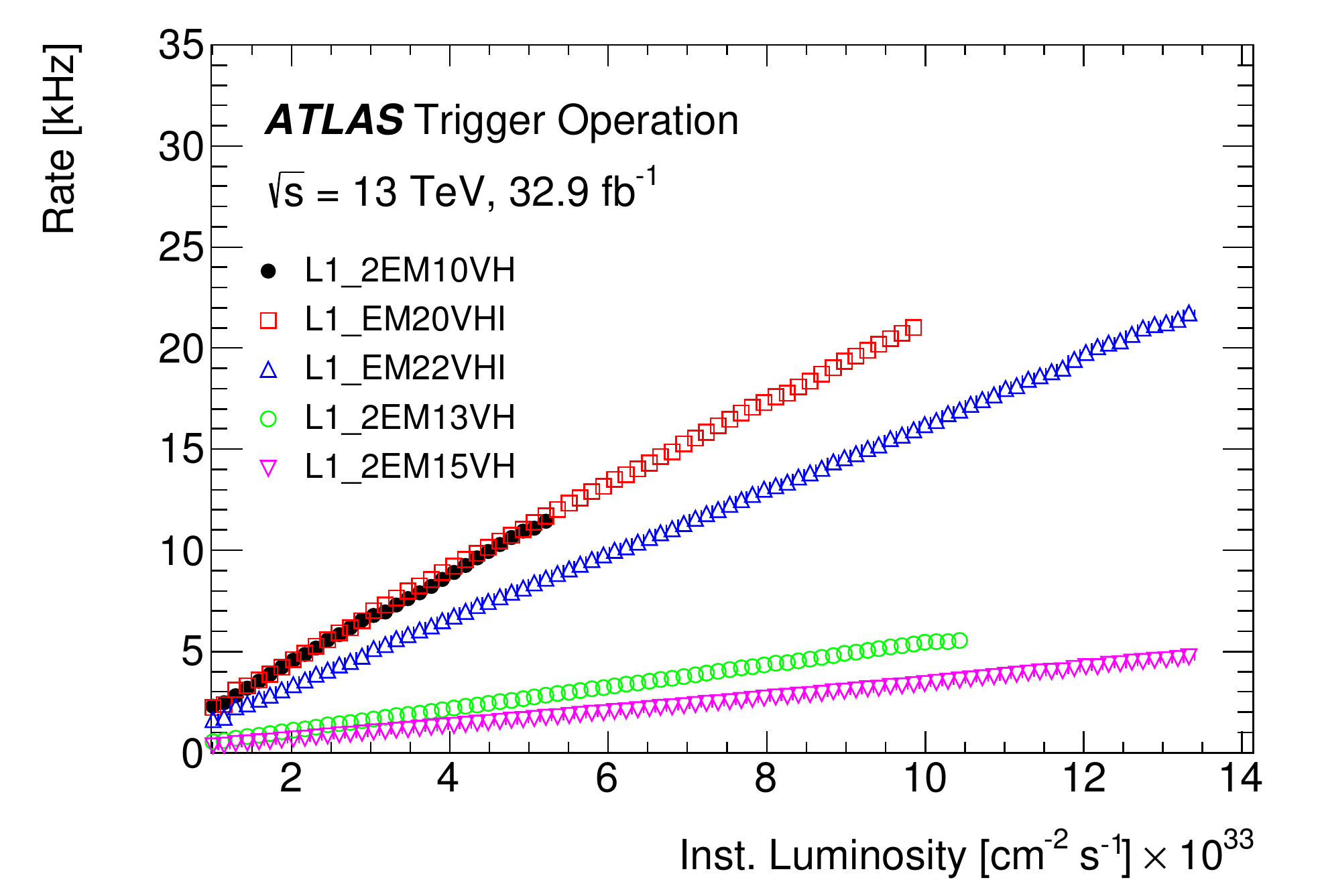}
\caption{Output rates of Level-1 EM triggers as a function of the uncalibrated instantaneous luminosity measured online during the 2016 proton-proton data taking at a center-of-mass energy of 13 TeV.
Rates are shown only for unprescaled triggers. All trigger rates show a linear dependency with instantaneous luminosity.}
\label{fig:}
\end{figure}

\subsection{Dataflow Challenges in Run-2}
The function of the DAQ system is to efficiently buffer, transport, and record the events that were selected by the trigger system. 
Its performance is affected by the instantaneous luminosity that leads to busy events with multiple proton-proton interactions occurring in each 
bunch crossing, referred to as pileup. The high pileup results in a higher data volume collected by the detector that needs to be 
processed at the required rate to avoid exerting back-pressure on the L1 system. 
In Run 2, the LHC has exceeded the designed instantaneous luminosity of \\$10^{34}$ cm$^{-2}$ s$^{-1}$ leading to pileup of~ $<\mu>=30$ or more as shown 
in Figures \ref{fig:exp.lhc.peakLumiByFill} and \ref{fig:exp.mu_2015_2016}.
The L1 accept 
rate has also increased from 75 kHz in Run 1 to 100 kHz in Run 2 and the average output rate of the data logger system has 
increased from 400-600 Hz in Run 1 to about 3 kHz with 1.5 kHz for physics data. 
Moreover, there were new detectors that were added in Run 2
(Insertable B-layer (IBL), L1 topological trigger, Fast Tracker (FTK))\cite{Aad:1602235} leading to 
an increase of 20\% in the number of readout channels. 
To cope with these changes, the ATLAS TDAQ system was upgraded during Run-2
simplifying its architecture and increasing its flexibility.
To be able to deliver more rate to the High Level Trigger (HLT), the upgrade also targeted the Readout System (ROS)\cite{PanduroVazquez2016939}. 
For the same reason the two levels of the HLT system were collapsed into a single level which made the system more flexible 
 allowing for incremental data retrieval and analysis. 
The dataflow network system was re-designed to increase its capacity and simplify its architecture\cite{1742-6596-396-1-012033}.

\subsection{ATLAS Dataflow Design}

In Run 1, the computing farm was subdivided into several slices, with 
each slice managed by a dedicated supervisor. This layout has been 
dropped in favor of global management by a single farm master 
operating at 100 kHz referred to as the HLT supervisor (HLTSV). 
The Region of Interest Builder (RoIB) that assembles the RoIs
previously implemented on a VMEbus 
system is now integrated with the HLTSV and the RoI building done in software.
Chapter~\ref{chap:roib} is dedicated to the work of the author in the RoIB evolution. 
The change in the HLT architecture from two to one level
required re-writing the HLT software and algorithms in such a way that 
each computing node in the farm can perform all processing steps. The handling of these
processing steps is done by a single Data Collection Manager (DCM) process 
running on each HLT node to manage the L1 RoIs, the dataflow 
between the ROS and the HLT processing units (HLTPU), 
the event building processes, and the data logging.
 In the new architecture, the computing resources are managed more efficiently
by balancing the utilization of all cluster nodes depending on the active HLT 
algorithms and by sharing the HLT code and services to reduce memory and 
resource usage. 

\begin{figure}[t!]
\centering
\includegraphics[width=1.2\textwidth]{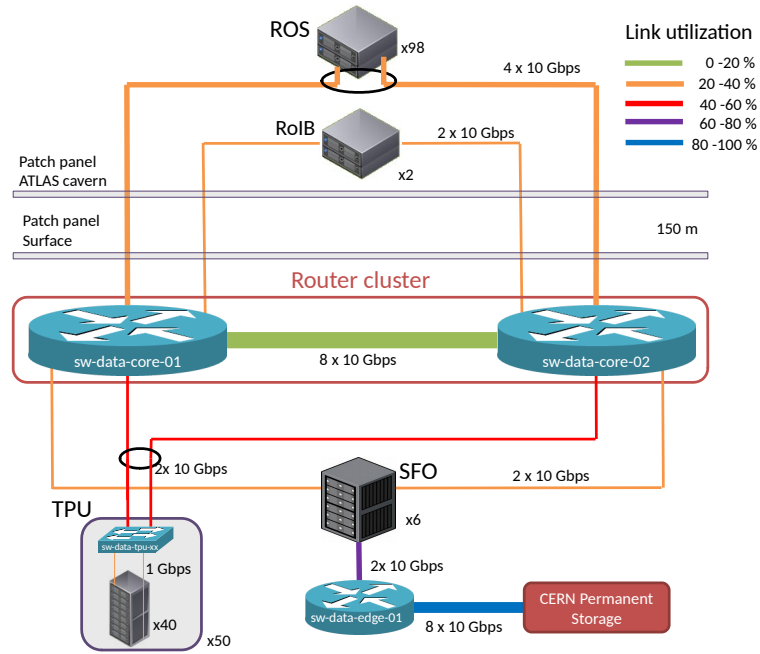} 
\caption{ATLAS Dataflow network}
\label{fig:net_diagram}
\end{figure}

The dataflow network shown in Figure~\ref{fig:net_diagram} was simplified and upgraded to handle a larger data volume.
A single network is used for 
RoI based access from the ROS, event 
building in the HLT processing nodes, and sending data for logging. 
A 10 GbE connectivity has been adopted throughout  the dataflow system
resulting in a factor of four increase in bandwidth between the data loggers and
the permanent storage, and a 4$\times$10 GbE output from each ROS PC to the core routers. 
The HLTSV and the HLT racks are all connected directly to each of the two core routers via 
 2$\times$10 GbE connections. Each HLT rack is hosting up to 40 nodes connected by 2$\times$1 GbE to the top-rack switches. 
The capacity of the routers can accommodate
an increase in the number of HLT server racks and ROS PCs by a factor of two, 
which will be needed when the system scales as run conditions 
change. The core routers also provide load balancing and traffic shaping protocols \cite{1742-6596-396-1-012033}
to distribute the data throughout the system more evenly. A duplication of core routers provide link redundancy at every level in 
case of link or switch failures.

To take advantage of multi-core architectures, the dataflow
 software is using multi-threaded software design for CPU consuming operations.
The Input/Output of the dataflow is based on asynchronous communication using industry standard libraries
such as the Boost::ASIO library. All the ATLAS software suite was switched to exclusively 64 bit operation in 2016.

In summary, the elements of the Run-2 ATLAS dataflow are:

\begin{itemize}
\item The Readout Sytem (ROS) buffers front-end data from the detectors and provides a standard interface to the DAQ system.
\item The Region of Interest Builder (RoIB) receives the L1 trigger information from the RoIs and combines the information for the HLT supervisor.
\item The HLT Supervisor (HLTSV) can handle the input from the RoIB and manage the HLT computing farm of about 2000 machines at 
over 100 kHz.
\item The Data Collection Manager (DCM) handles all Input/Output on the HLT nodes, including RoI requests from the HLT and full event building.
\item The HLT processing units (HLTPU) run the actual HLT algorithms which 
are forked from a single mother process to maximize memory sharing.
\item The Data loggers or SubFarm Output (SFO) are responsible for saving the 
accepted events to disk, and sending the files to CERN permanent storage infrastructure.
\end{itemize}

%% file: texfiles/sec.exp.op.tex
The reliable operation of the different ATLAS systems directly impacts the
efficiency of the ATLAS experiment
in recording the $p-p$ collisions delivered by the LHC.
As a result, high data-taking efficiency is crucial
for the ATLAS physics program.

All the ATLAS sub-detectors have operated with a very high efficiency
($93-95\%$) as shown in Table~\ref{tab:atlas.detstatus} for the 2016
data taking run.

\input{texfiles/dqphys.tex}

\begin{figure}[t!]
\centering
\includegraphics[width=0.75\textwidth]{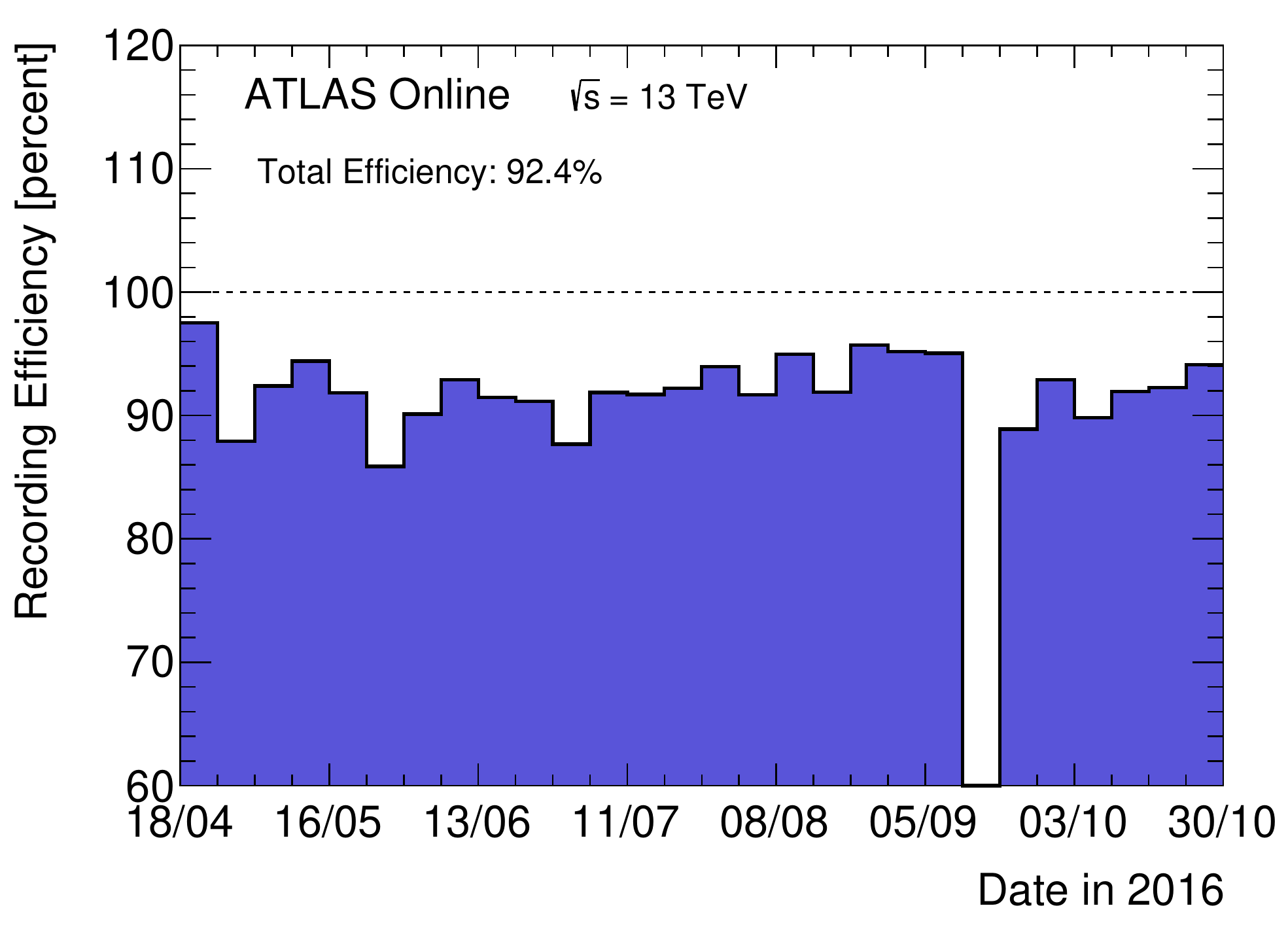}
\caption{ATLAS recorded efficiency \cite{atlasTwiki}.}
\label{fig:tdaq_diagram}
\end{figure}

\begin{figure}[t!]
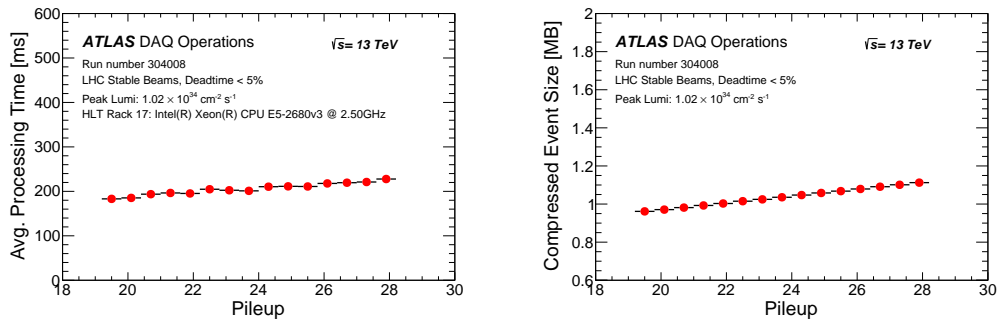

\centering
\begin{subfigure}[t]{0.48\textwidth}
\includegraphics[width=.95\textwidth]{hProf_run_304008_pileup_AvgProcessingTime.pdf}
\end{subfigure}
\begin{subfigure}[t]{0.48\textwidth}
\includegraphics[width=.95\textwidth]{hProf_run_304008_pileup_EventSize.pdf}
\end{subfigure}
\vspace{-0.3cm}
\caption{Performance in Run 2: Average processing time as a function of pileup (left), compressed event size as a function of pileup (right).}
\label{fig:run_pileup}
\end{figure}

The ATLAS recorded efficiency in 2016 is over 90\%, as shown in
Figure \ref{fig:tdaq_diagram} with a negligible fraction of data loss due to
the ATLAS DAQ system.
The ATLAS dataflow architecture is scaling well with the increased
instantaneous luminosity during 2016 data-taking and is capable of handling
larger pileup ($<\mu>$) and thus larger event sizes.
For illustration,  Figure \ref{fig:run_pileup} shows the evolution of the
average processing time per event and
the event size where there is relatively mild increase as a function of pileup
which is well within the system capacity.

As a result of this excellent performance of all the sub-detectors,
ATLAS has recorded  almost 92\% of the luminosity delivered by the LHC during
2015 and 2016 as illustrated in  Figure~\ref{fig:exp.op.intlumi}.
The total integrated luminosity used in this analysis after
applying a large number of checks amounts to
 36.1 \ifb~ divided between 3.2 \ifb~in 2015 and 32.9 \ifb~in 2016.

\begin{figure}[t!]
\centering
\begin{subfigure}[t]{0.48\textwidth}
\includegraphics[width=.95\textwidth]{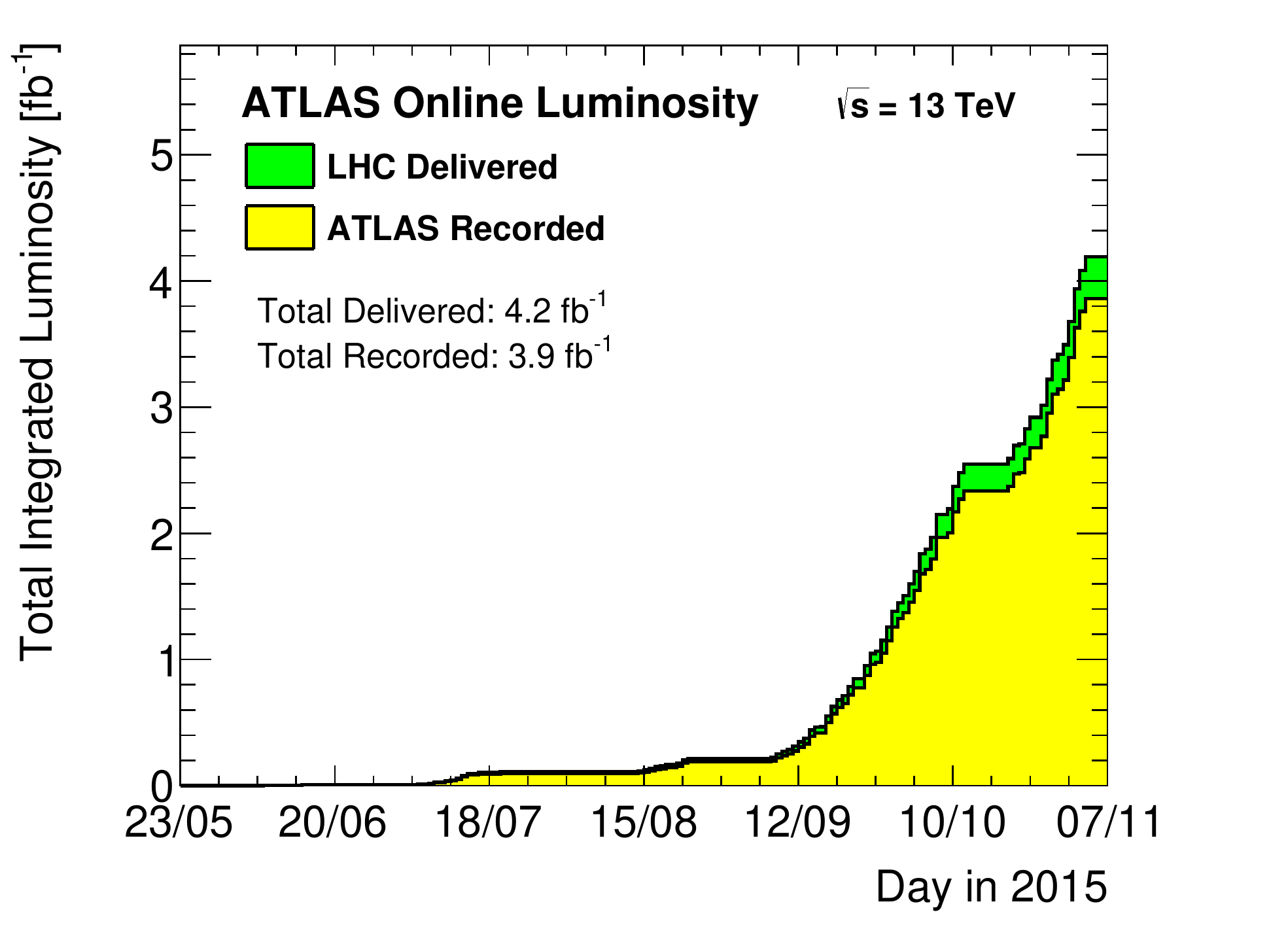}
\subcaption{2015}
\end{subfigure}
\begin{subfigure}[t]{0.48\textwidth}
\includegraphics[width=.95\textwidth]{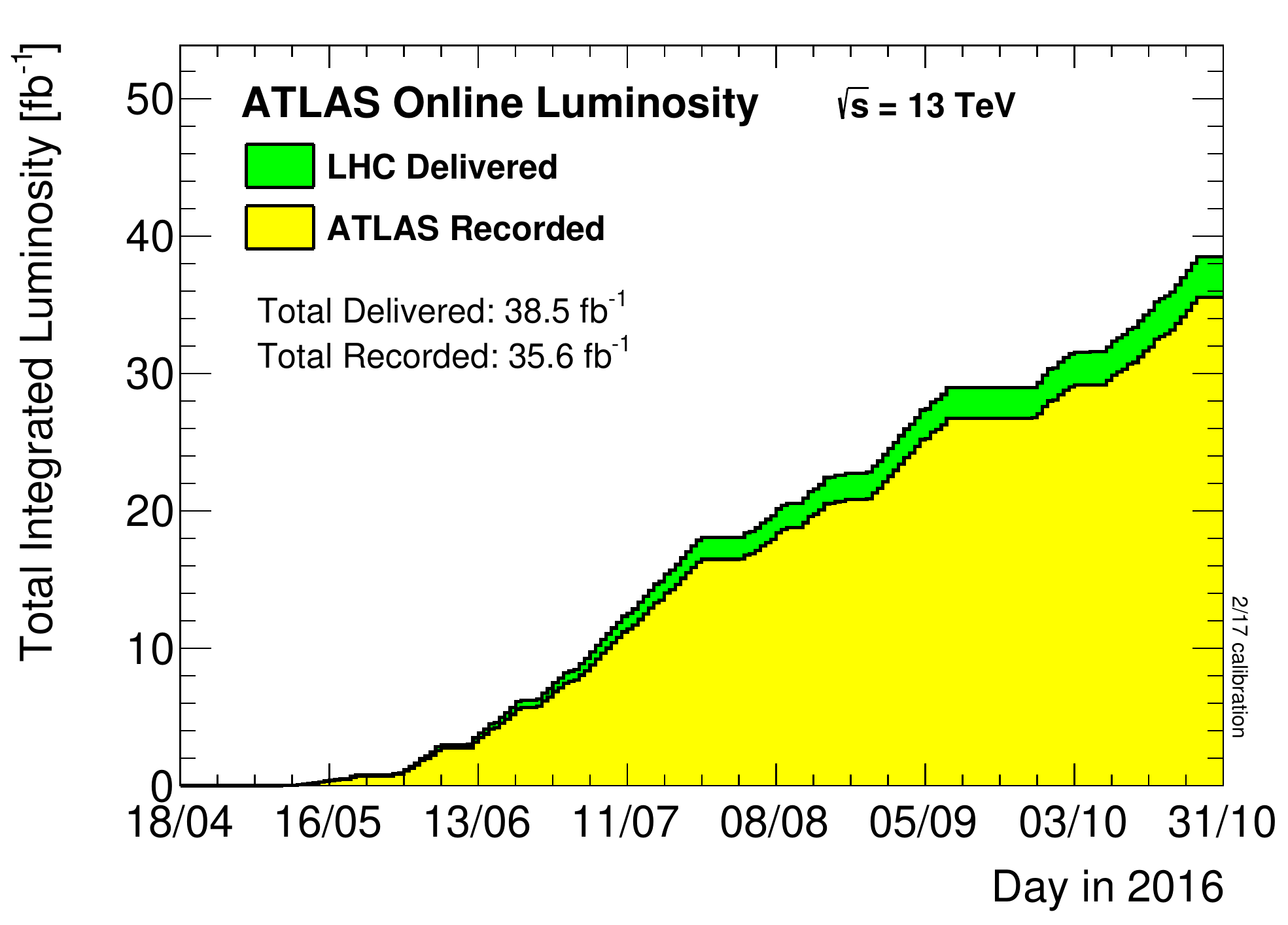}
\subcaption{2016}
\end{subfigure}
\vspace{-0.3cm}
\caption{
Cumulative luminosity versus time delivered to (green) and recorded by ATLAS
(yellow) during stable beams for pp collisions at 13 \TeV~centre-of-mass energy
in (a) 2015 and (b) 2016.
}
\label{fig:exp.op.intlumi}
\end{figure}

%% file: texfiles/dqphys.tex
\begin{table}[!htb]
  \caption{Luminosity weighted relative fraction of good quality data delivery efficiencies (\%) by the various components of the ATLAS 
detector and trigger subsystems during LHC stable beams in $pp$ collisions at $\sqrt{s}=13$ \TeV~with 25 ns bunch spacing between 
April-October 2016, corresponding to a recorded integrated luminosity of 35.9 \ifb. The toroid magnet was off for some runs, 
leading to a loss of 0.7 \ifb.}
\label{tab:atlas.detstatus}
\centering\def\arraystretch{1.2}
\resizebox{\textwidth}{!}{
\begin{tabular}{|c|c|c|c|c|c|c|c|c|c|c|c|}\hline
   \multicolumn{3}{|c|}{Inner Tracker} & 
   \multicolumn{2}{|c|}{Calorimeters} &
   \multicolumn{4}{|c|}{Muon Spectrometer} &
   \multicolumn{2}{|c|}{Magnets} &
   Trigger\\\hline
Pixels & SCT & TRT & LAr & Tile & MDT & RPC & CSC & TGC & Solenoid & Toroid & L1\\ \hline
98.9 & 99.9 & 99.7 & 99.3 & 98.9 & 99.8 & 99.8 & 99.9 & 99.9& 99.1 & 97.2 & 98.3 \\ \hline \hline
\multicolumn{12}{|c|}{Good for physics: 93-95\% (33.3-33.9 \ifb)} \\
\hline\hline\end{tabular}}
\end{table}


%% file: texfiles/sec.exp.sim.tex
In order to interpret the LHC data, it is essential to compare the observations to the expected outcomes from a physical model, typically the Standard Model and a SUSY scenario.
The event simulation starts from a proton-proton ($pp$) collision leading to the process of interest all the way to the expected detector response. 
These steps are the following:

\begin{itemize} 
\item Event generation: The process of interest $pp \to X$ is generated by relying on random sampling using Monte Carlo (MC) techniques, which repeatedly draw samples that represent 
  a possible outcome of a given process. The processes are generated using a software package, \textsc{Madgraph} ~\cite{Alwall:2014hca} for example, which calculates the matrix element 
  for each process to some order in QCD. Generators start from partons\footnote{Partons refer to all the particles that can exist inside 
the proton: quarks, anti-quarks, and gluons.} of a $pp$ collision, using parton distribution functions, and calculate the processes up to leading order or next-to-leading order. 
  For processes where the cross-section at a higher order is non-negligible a factor called the $k$-factor\footnote{The $k$-factor 
    describes the difference between the leading order cross section and higher order cross sections.} 
  is applied to the expected cross-section. 
  The partons from the hard interaction are colored and radiate gluons described by a parton showering software, as \textsc{Pythia} ~\cite{Sjostrand:2007gs}. The generators used in the analysis 
  are given in Section~\ref{sec:strategy.samples}. 
  The raw output of such generators 
  is an input to the next steps of simulation. They can also be used to perform generator level studies, also called ``truth level'', undergoing minimal processing to evaluate the 
  sensitivity and  acceptance of the analysis (for example Table~\ref{tab:strategy.cut}).
\item Detector simulation: The event generator gives particle momenta at the hadron level which are then processed by \textsc{Geant4}~\cite{Agostinelli:2002hh}, which simulates
  the propagation of particles through the different materials comprising the detector. The simulation includes  the best knowledge of the detector 
geometry, material budget and modeling of the particle interactions.
The full simulation of the detector is a slow process. For many applications, such as the generation 
  of SUSY signals, it is faster to use a parametrized response of the calorimeters~\cite{ATL-PHYS-PUB-2010-013}.
\item Digitisation: The detector simulation records the interaction of particles with the different components of the detector in the form of hits and energy deposits in the detector.
  The latter are used as inputs to emulate the response of the readout electronics of the detector. The output from this step is identical to the data recorded by the detector.
\item Reconstruction: At this stage, both the events recorded by the detector and simulated events are used to identify objects associated with fundamental particles, namely 
  electrons, muons, photons, and jets. The energy deposits not matched to physics objects are collected into a ``soft terms'' category used in the computation of the missing transverse momentum.  
\end{itemize} 

There are two other important elements of the simulation that are less understood. The simulated MC samples must handle the underlying event, which is the remainder of the non-hard scattered partons of 
the original interacting protons. Also, the MC generators must simulate the interactions between the other particles in the beam crossing, also referred to as pileup. In practice, 
the MC is generated with an expected pileup profile which is later corrected based on the observed pileup profile.

The final physics objects from the reconstruction step do not reflect all the knowledge we have about the detector. For instance, the energies of the objects must be calibrated
or certain parts of the detector may not always be working with the desired specification. 
 Once the reconstructed objects are defined and calibrated, the data is ready for analysis. Typically, the size of the dataset is reduced from petabytes to a size of few terabytes
  by only selecting the objects of interest in the analysis. For instance, this analysis requires at least two leptons applied to the samples used which significantly reduces the sample size.

%% file: texfiles/sec.exp.reco.tex
The identification and reconstruction step relies on the properties that particles display when they interact with the different components of the detector
described in Section~\ref{sec:exp.atlas}.
Once the particle is identified, it is desirable to determine its momentum and
 its origin among other properties.
The important part for the analysis is how well these objects are reconstructed which can be determined by measuring the reconstruction and 
identification efficiencies as a function of kinematic particles, 
usually \pt and $\eta$.
In the next sections we briefly describe how the main physics objects used in 
this analysis are identified and their efficiencies.

\subsection{Basic Principle}

\begin{figure}[htb!]
\centering
\includegraphics[width=0.95\textwidth]{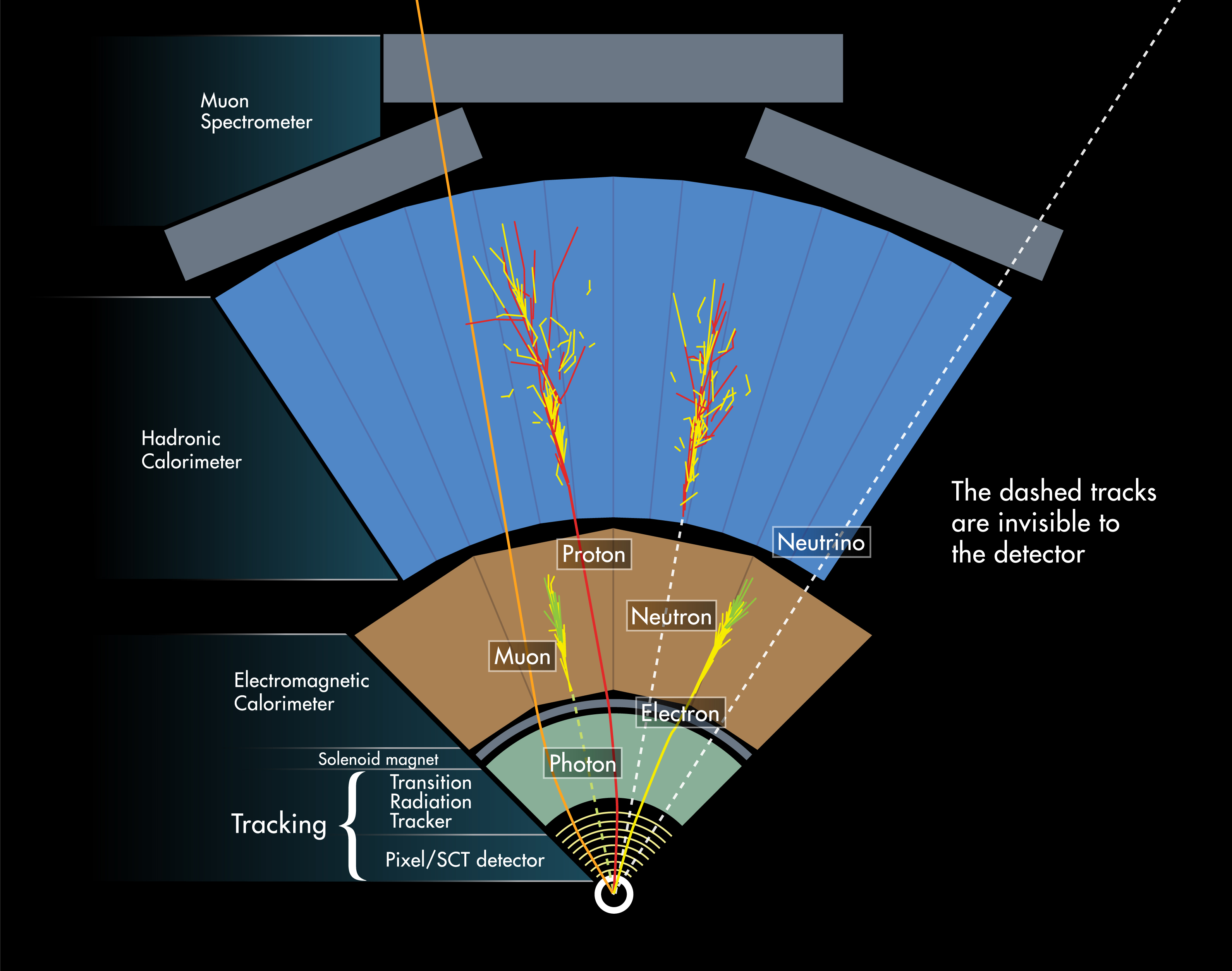}
\caption{ The different signatures of particles traversing 
the detector are shown in the transverse plane of the ATLAS detector.}
\label{fig:exp.atlas.reco.det}
\end{figure}

In this section we describe the higher-level reconstruction of particles
as they interact with the different components of the detector.
The schematic of Figure~\ref{fig:exp.atlas.reco.det} 
summarizes the signatures of the different particles in the ATLAS 
detector.

The most important reconstructed particles for the analysis presented 
in this dissertation are charged leptons. 
Charged leptons come from  electroweak processes and  
provide  clear signals which can be accurately reconstructed.
In the remainder of this dissertation, leptons will refer to 
only electrons or muons.

Muons are the simplest to identify since they traverse
the entire ATLAS detector: all other 
interacting particles are stopped before reaching the muon spectrometer.
They are reconstructed as tracks in the inner detector matched to tracks in the
muon spectrometer, 
and leave little energy in the electromagnetic and hadronic calorimeters.  
Muons are produced from decays of $W$ and $Z$ bosons with a relatively 
large momentum, above 15 \GeV, and are produced in isolation from surrounding
detector activity, qualities that we will use in this analysis.
The latter is called ``isolation'' which requires 
the energy of the reconstructed tracks and clusters near the reconstructed 
muon not exceed a certain value. 
This requirement is effective at suppressing muons produced
from background processes such as meson decay in flight and heavy-flavor decay
detailed in Chapter~\ref{chap:fake}.

Electrons are identified by a track in the inner detector that 
initiates  an electromagnetic
shower in the electromagnetic calorimeter. Most of the time,  
all of the energy of the electron is absorbed before reaching the hadronic 
calorimeter.
The electron is identified  by matching reconstructed EM clusters to tracks
reconstructed in the inner detector. 
This signature suffers from 
large backgrounds from other types of charged particles that can mimic this 
signature. For these reason, ATLAS uses several tools to effectively 
distinguish between the desired electrons and background.
Similar to muons, isolation is one of these tools.

Photons also produce an electromagnetic shower upon entering the calorimeter, 
except they leave no track in the inner detector since they are neutral.
In practice, photons might undergo a conversion to an $e^+e^-$ pair 
in the detector material before entering the calorimeter which will 
result in a track in the inner detector. Photons of the former case 
are called un-converted, while the latter are called converted photons.
ATLAS has dedicated algorithms to identify photon conversions from 
pairs of reconstructed tracks.

Tau leptons are charged particles that decay to the other leptons 
(40\% of the time) or to hadrons (60 \% of the time) before entering the detector.
If they decay to electrons or muons and neutrinos, they are indistinguishable from 
electrons or muons coming from $W$ or $Z$ bosons.
The experimental signatures of hadronically decaying taus are
multiple hadronic showers matched to tracks in the inner detector.
The latter suffers from large backgrounds 
from other types of particles that cannot be suppressed 
as efficiently as background from the leptonic tau decay.

Neutrinos only interact via the weak force
and are thus not directly detected by ATLAS.
As shown in Figure~\ref{fig:exp.atlas.reco.det}, 
they pass through all the sub-detectors.
However, their presence is inferred from an overall 
transverse momentum imbalance of the measured momenta in the event.
Thus, the transverse momentum of the neutrinos can be inferred.
This type of signature is similar to potential new particles 
that will not interact with our detector.
Nevertheless, neutrinos are very well understood
and any potential contribution from them can be accurately 
predicted.

The reconstruction of jets is an essential part of the analysis presented in this dissertation. 
Colored quarks and gluons from the primary interaction  undergo a process referred to as hadronization,
where they convert to sprays of colorless hadrons. The collection of this spray of particles is 
referred to as a jet. 
The reconstruction of a jet is based on regrouping of reconstructed clusters and tracks into larger collections using
various clustering algorithms as will be described next.
By measuring the energy and direction of a jet, we can infer information 
about the initial quarks or gluons that participated in the physics processes under study.
It can also be used to determine the energy of the initial parton in the hard interaction, 
a challenging aspect of jet reconstruction.
The energy of the jet must be calibrated by determining the 
Jet Energy Scale (JES) and the Jet Energy Resolution (JER).
The uncertainties associated with the JES and JER are one of the largest 
experimental uncertainties in this analysis.

The jet reconstruction algorithms cannot determine the type of parton that initiated a
given jet, except for $b$-quarks.
Since $b$-quark hadrons decay with suppressed weak interactions, they are 
relatively long-lived travelling a few millimeters before they decay.
Given the fine tracking resolution of ATLAS, 
a millimeter displacement from the interaction point is large enough to be resolved.
It is thus possible to identify jets containing $b$-hadrons  in a process called ``b-tagging''.
These types of jet, called $b$-jets, are used in this analysis.

More technical details about the reconstruction procedure and the efficiencies 
of the reconstructed objects will be given next.

\subsection{Electrons}
Electrons are reconstructed and identified for $\pt > 5$ \GeV~ and
 $|\eta| < 4.9$ \cite{ATLAS-CONF-2016-024}.
The electrons  are identified by a cluster in the electromagnetic calorimeter 
matched to a track in the inner detector.
The electron trajectory is determined using information from the inner detector. This involves the measurement of the track associated parameters : the
position in the transverse ($d_0$) and longitudinal ($z_0$) planes of the perigee, the particle direction ($\phi$, $\theta$)
and the parameter which provides the inverse track momentum multiplied by the charge ($q/p$). The track
parameters and the associated uncertainties are obtained from the track fitting procedure performed with
the ATLAS Global $\chi^2$ Track Fitter.

\begin{figure}[htb!]
\centering
\begin{subfigure}[t]{0.49\textwidth}
\includegraphics[width=1.\textwidth]{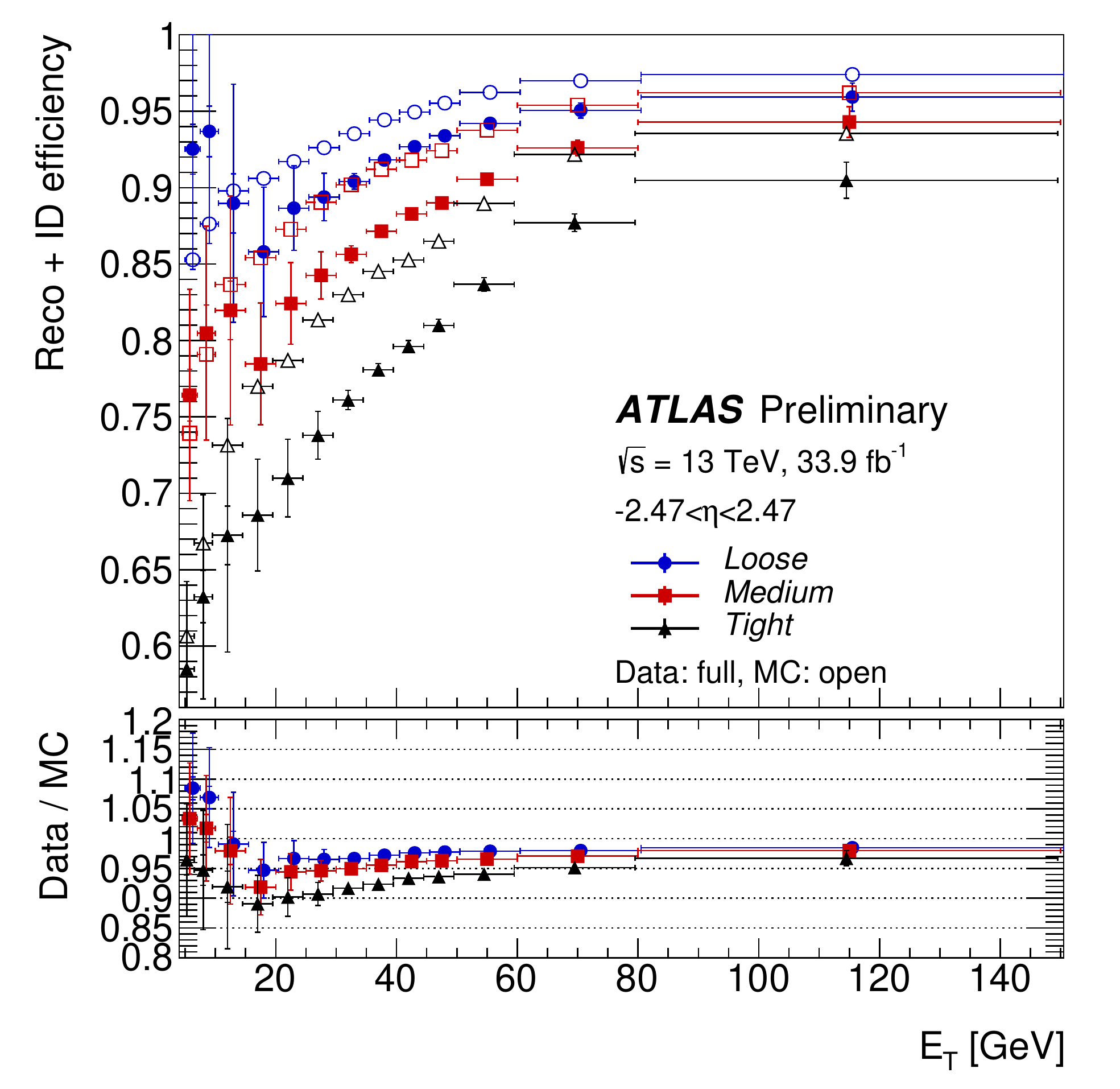}
\subcaption{}
\label{fig:exp.reco.eff_pt}
\end{subfigure}
\begin{subfigure}[t]{0.49\textwidth}
\includegraphics[width=1.\textwidth]{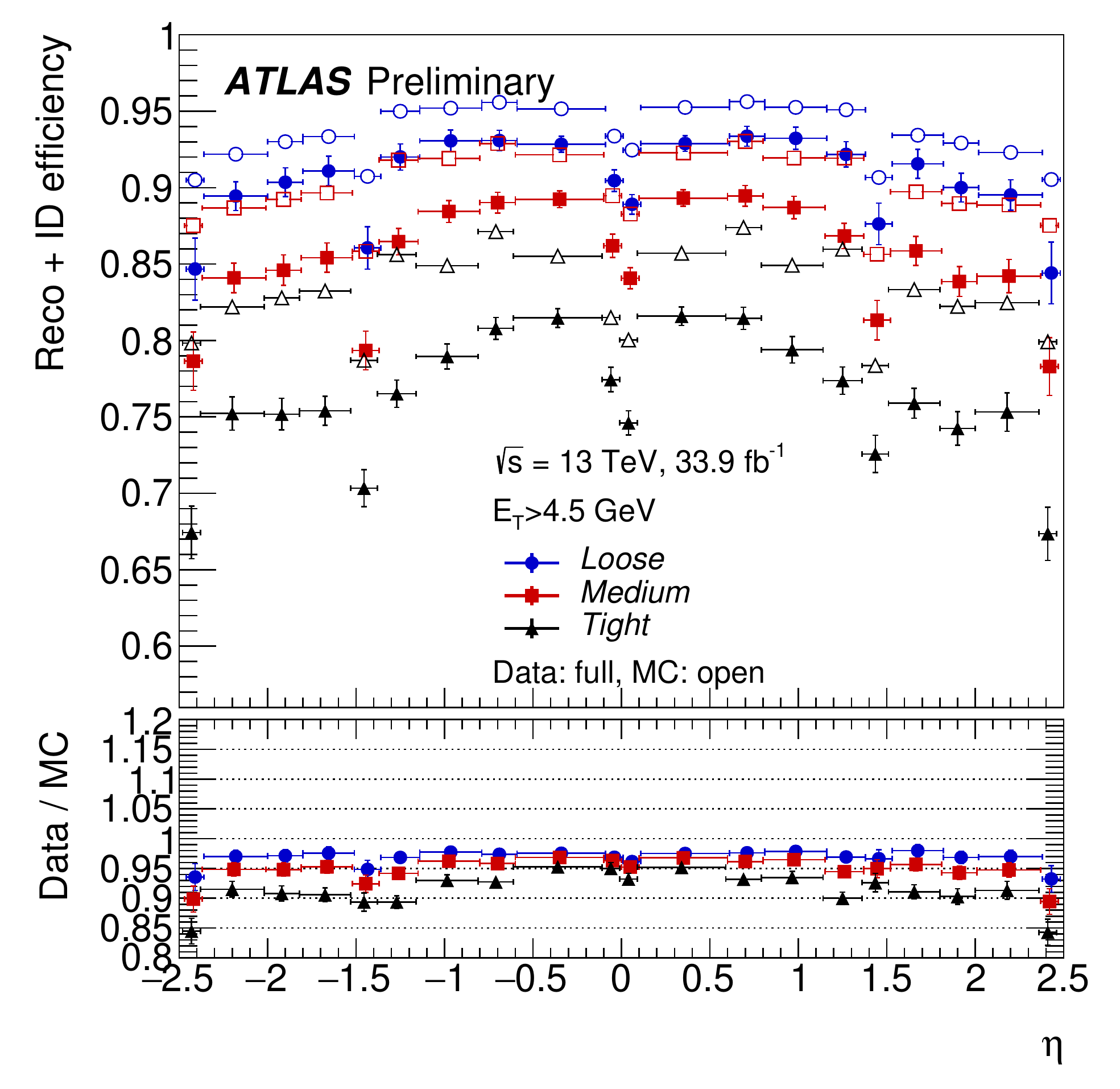}
\subcaption{}
\label{fig:exp.reco.eff_eta}
\end{subfigure}
\vspace{-0.25cm}
\caption{
Electron reconstruction and identification efficiencies in $Z\to\ee$ events as a function of (a) pseudo-rapidity $\eta$ (b) transverse energy \et.
The total statistical and systematic uncertainty is displayed. 
}
\label{fig:exp.reco.eff.el}
\end{figure}

The goal is to improve the efficiency of identifying electrons while rejecting background electrons arising from hadronic jets mistaken for electrons, 
electrons from photon conversion, Dalitz decays and from semileptonic heavy-flavour hadron decays.
To do so, three identification (ID) criteria, loose, medium, and tight are defined via likelihoods based on calorimetric cluster shower shapes, 
track and track-to-cluster matching variables.
The tighter the ID criteria, the higher the rejection of background electrons but the lower the identification efficiency.
Figure ~\ref{fig:exp.reco.eff.el} shows the  efficiencies in data and MC for three operating points that are based on a likelihood approach, Loose, Medium and Tight. 
The data efficiencies are obtained by applying data/MC efficiency ratios that were measured in  $J/\psi\to\ee$ and $Z\to\ee$ events to MC simulation. 
The lower efficiency in data than in MC arises from the fact that the MC does not properly represent the 2016 TRT conditions, 
in addition to the known mis-modelling of calorimeter shower shapes in the GEANT4 detector simulation.
The reconstruction efficiency of electrons is around 95\% for low \pt and goes up to 99.9\% for electron $\pt > 45$ \GeV.

To study and compute the corrections needed to account for the detector geometry and material distribution
two sets of MC samples are used. For the first set an ideal geometry (no mis-alignments) with the best
knowledge of the dead material is implemented. For the second scenario, the dead material between the
tracker and calorimeters is increased and the mis-alignments are included. The latter is used to assign the
systematic uncertainties.



\subsection{Muons}

\begin{figure}[htb!]
\centering
\begin{subfigure}[t]{0.56\textwidth}
\includegraphics[width=1.\textwidth]{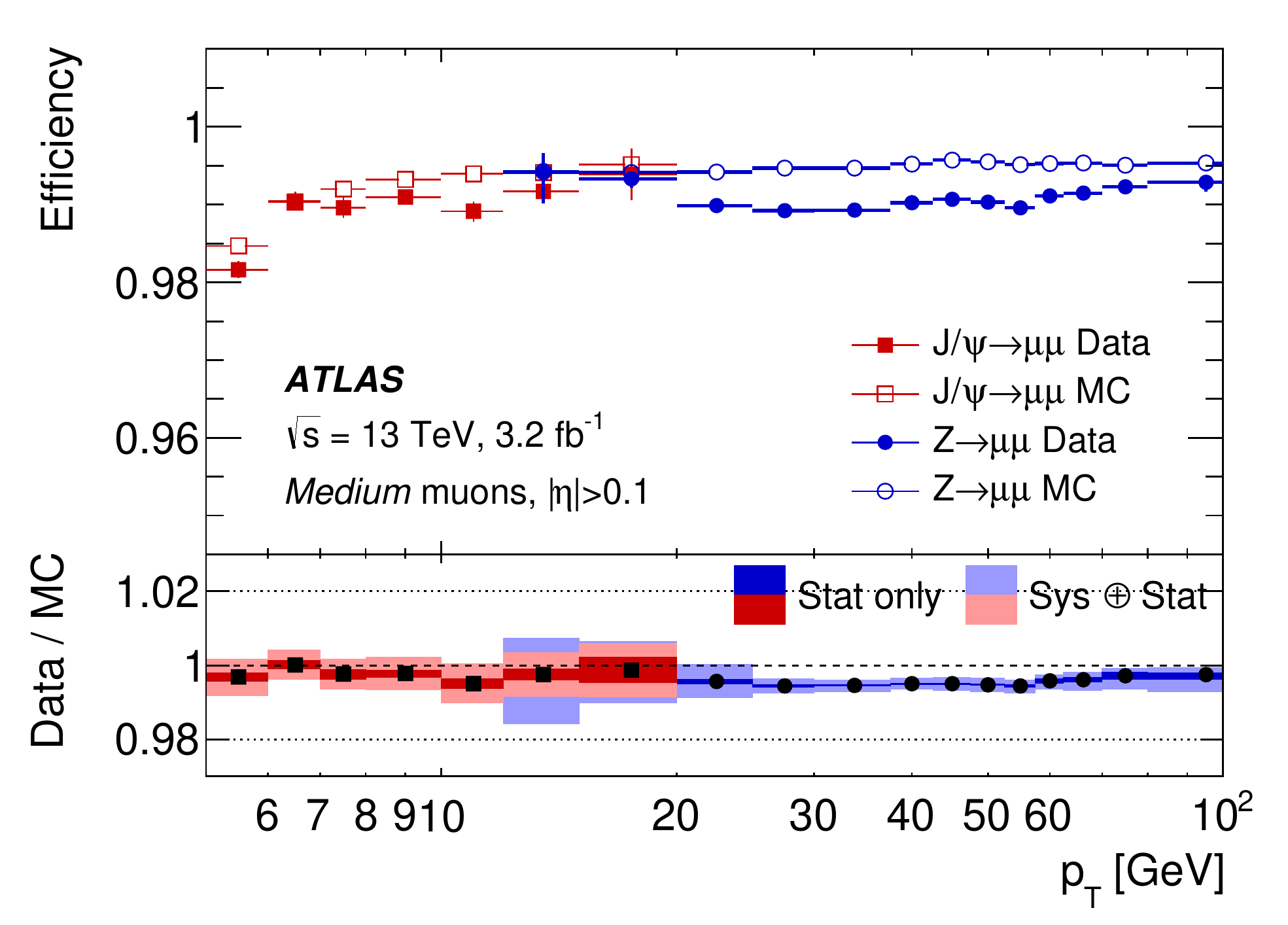}
\subcaption{}
\label{fig:}
\end{subfigure}
\begin{subfigure}[t]{0.43\textwidth}
\includegraphics[width=1.\textwidth]{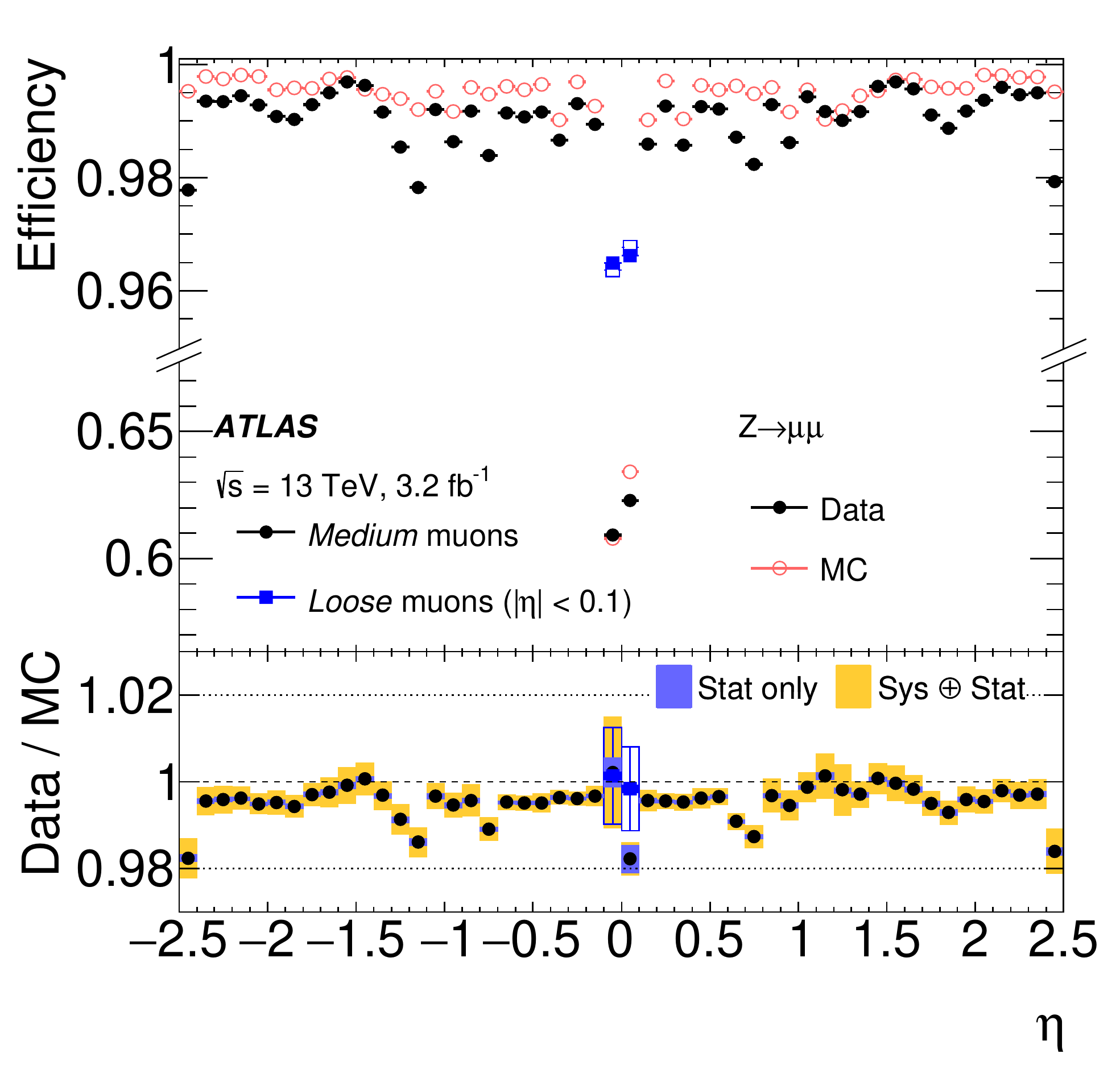}
\subcaption{}
\label{fig:}
\end{subfigure}
\vspace{-0.25cm}
\caption{(a) Reconstruction efficiency for the medium muon selection as a function of (a) \pt (b) $\eta$ of the muon.
 The error bars on the efficiencies indicate the statistical uncertainty. The panel at the bottom shows the ratio of the measured to predicted efficiencies, with statistical and systematic uncertainties.
}
\label{fig:exp.reco.muon}
\end{figure}

The ATLAS detector has been designed to provide clean and efficient 
muon identification and precise muon momentum measurements over 
a wide range of energy and solid angle. 
The muon  reconstruction starts in the inner 
detector where tracks are identified up to $|\eta|<2.5$ in a solenoidal field of 2 Tesla.
The muon spectrometer  measures  muons up to $|\eta|<2.7$  
providing momentum measurements with a design relative resolution of better than 3\% over a wide 
\pt range and to 10\% at 1 \TeV.
Similar to electrons, four muon identification selections are defined in order to meet 
the specific needs of different physics analyses:
\begin{itemize}
\item loose muons: maximize efficiency: ideal for multilepton final states analysis
\item medium muons: minimize systematics uncertainties
\item tight muons: optimize  purity
\item high-\pt muons: maximize momentum resolution for high \pt tracks ($>100$ \GeV)  
\end{itemize}
Figure~\ref{fig:exp.reco.muon} shows the reconstruction efficiency of the muon for different selections measured in  $Z\to\mu\mu$ and $J/\psi\to\mu\mu$ events.

The reconstruction efficiency is measured to be close to 99\% over most of the phase space relevant for the analysis. 
The isolation efficiency is between 93\% and 100\% based on the selection and muon momentum. The simulation reproduce both efficiencies very well.
The  momentum resolution is measured to be as low as 1.7\% and the momentum scale uncertainty is less than  0.05\%.

\subsection{Jets}
Jets are reconstructed using the anti-$k_{t}$ jet algorithm~\cite{Cacciari:2008gp} 
with the distance parameter $R$ set to $0.4$ and 
a three dimensional input of topological energy clusters in the 
calorimeter~\cite{PERF-2014-07}. 
The advantage of using the anti-$k_{t}$ algorithm is that it is infrared
 and collinear safe\footnote{These two problems arise when defining a seed used 
as a starting point of an iterative process of re-clustering energy 
depositions in the calorimeter cells.
If only particles above some momentum threshold are used as seeds then the 
procedure is collinearly unsafe. On the other hand, if the addition of an 
infinitely soft particle leads to a new stable energy cone being found then 
the procedure is infrared unsafe.
}.
It is also resilient to soft-QCD emissions, a process that is common in the hadron colliders.
The jets are constructed by defining two distances:
\begin{itemize}
\item $d_{ij} = \min\left((k_{kj}^{2p},k_{kj}^{2p}\right)\frac{\Delta_{ij}^2}{R^2}$: the distance between two particles i and j.
\item $d_{iB} = k_{ti}^{2p}$: the distance between a particle $i$ and the beam $B$.
\end{itemize}
where $\Delta_{ij}^2 = \left( y_i - y_j\right)^2 + \left(\phi_i - \phi_j\right)^2$ and $k_{ti}$, $y_i$, and $\phi_i$ are the transverse momentum, rapidity, and azimuth of the 
particle $i$. The radius parameter $R$ scales $d_{ij}$ with respect to $d_{iB}$ such that any pair of final jets $a$ and $b$ are separated by at least $R^2=\Delta_{ab}^2$.
The parameter $p$ governs the relative power of of energy with respect to the geometrical scales, and is set to $p=-1$ for the anti-$k_{t}$ algorithm.

\begin{figure}[htb!]
\centering
\includegraphics[width=0.65\textwidth]{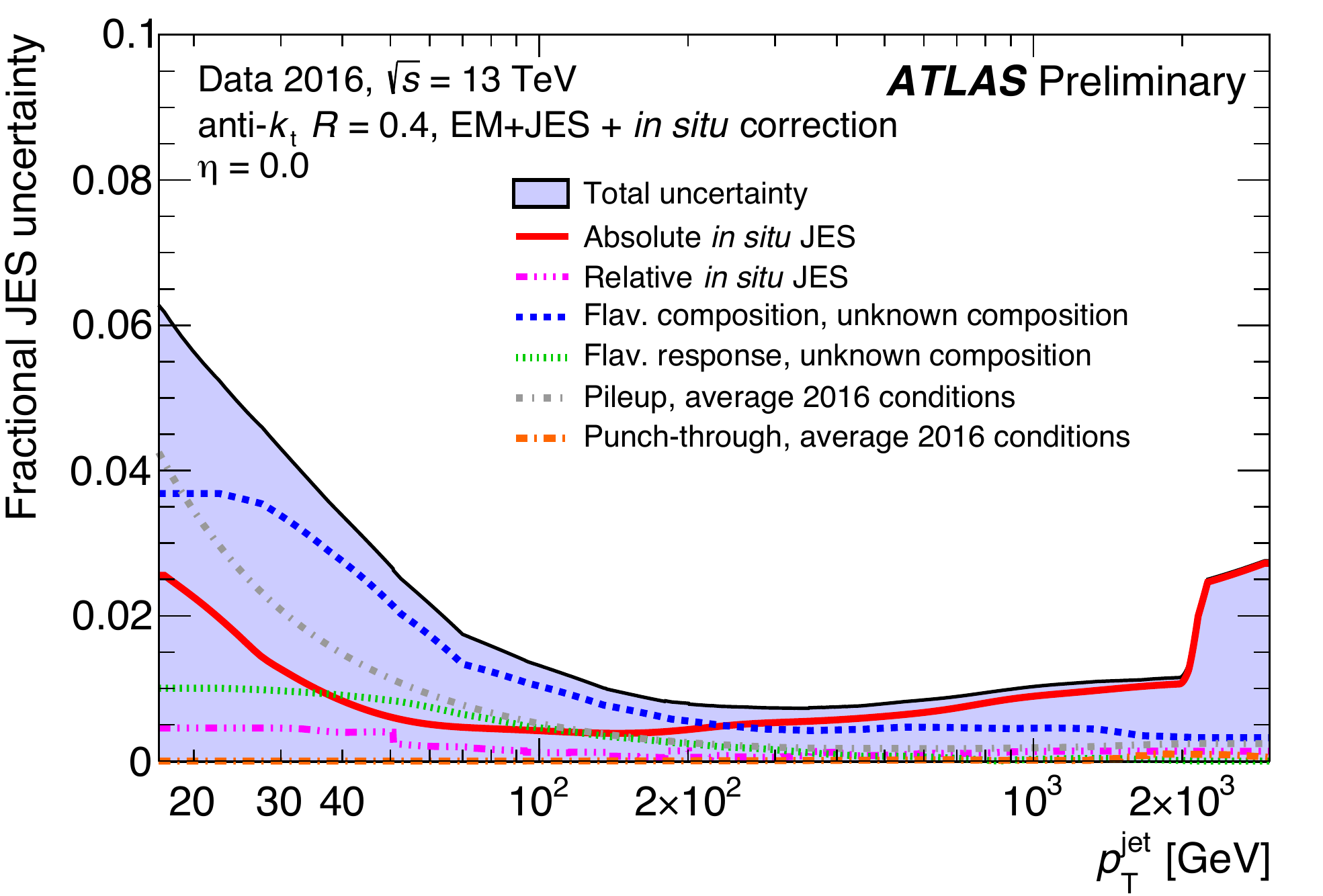}
\caption{Fractional jet energy scale systematic uncertainty components as a function of \pt for anti-$k_{t}$ jets at $\eta$ = 0.0 
with distance parameter of $R$ = 0.4 after calibration.
The total uncertainty (all components summed in quadrature) is shown as a filled blue region topped by a solid black line.}
\label{fig:exp.JES_pt_sys}
\end{figure}

The energies of the jets are calibrated to  account the necessary losses associated with sampling calorimeter,
the presence of dead material,  energy loss in non-instrumented regions, etc. This is performed using the local cluster weighting
(LCW) scheme \cite{Aad:2016upyew}, which uses calibrated topological clusters as input to the anti-$k_{t}$ jet algorithm,
and takes into account jet energy scale (JES) and jet energy resolution (JER) calibrations.
Both JES and JER can have an important contribution to the systematic uncertainties of this analysis.
The fractional JES uncertainty in Figure~\ref{fig:exp.JES_pt_sys} shows that jets with \pt below 50 \GeV~can have a larger uncertainty.
The choice was made to only require jets above 50 \GeV~in the analysis for this reason.
The fractional JER is around 17\% for jets with \pt of 30 \GeV~decreasing down to 5\% for more energetic jets.
An additional variable is used to suppress jets from pileup, called the jet vertex tagger (JVT).
JVT is a multivariate combination of the fraction of the total momentum of tracks in the jet which is associated with the primary vertex
and  track-based variables to suppress pileup jets \cite{ATL-PHYS-PUB-2015-034}. 

\subsection{Heavy flavor}

\begin{figure}[htb!]
\centering
\begin{subfigure}[t]{0.48\textwidth}
\includegraphics[width=0.95\textwidth]{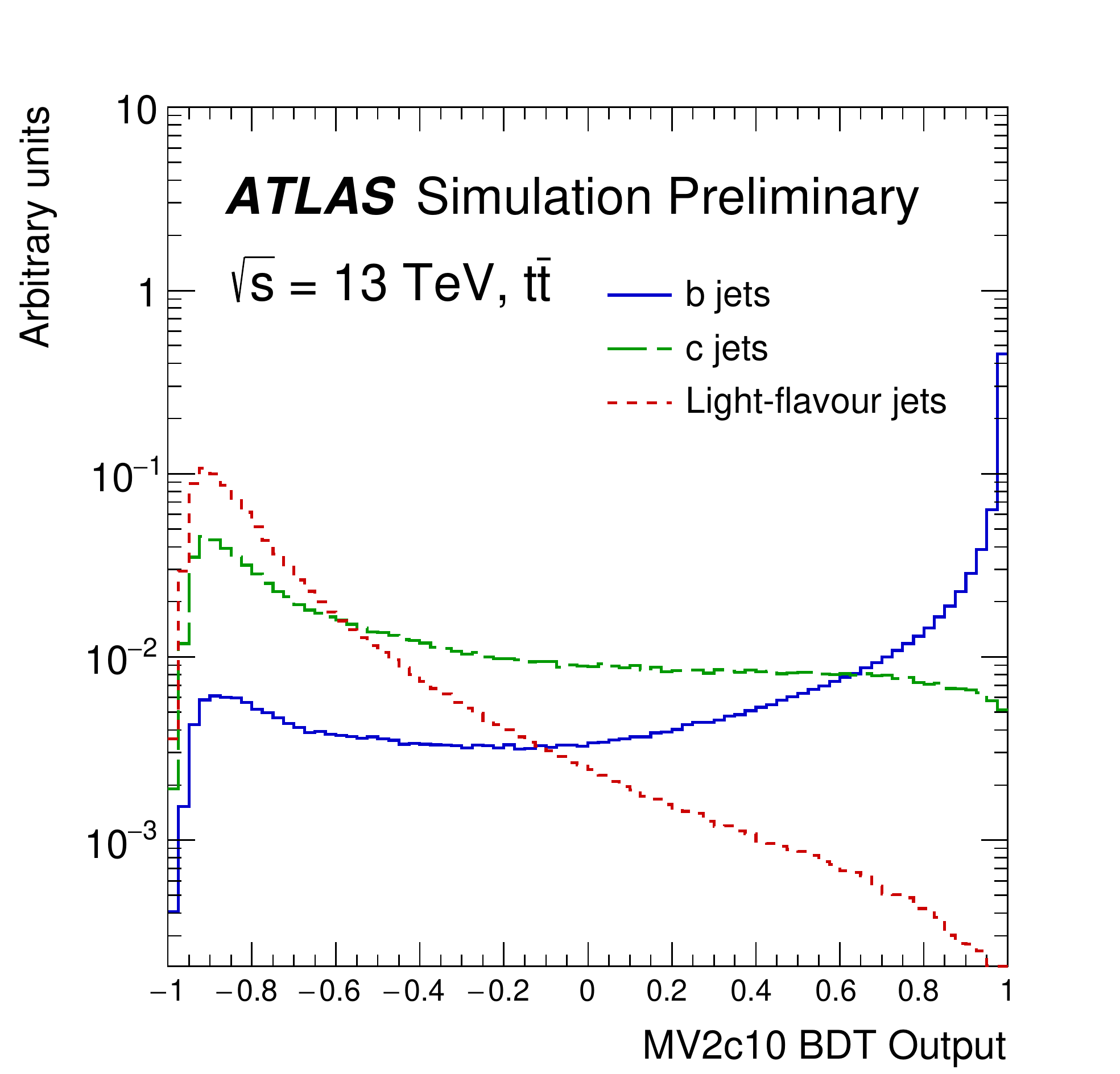}
\subcaption{}
\label{fig:exp.btag.bdt}
\end{subfigure}
\begin{subfigure}[t]{0.48\textwidth}
\includegraphics[width=0.95\textwidth]{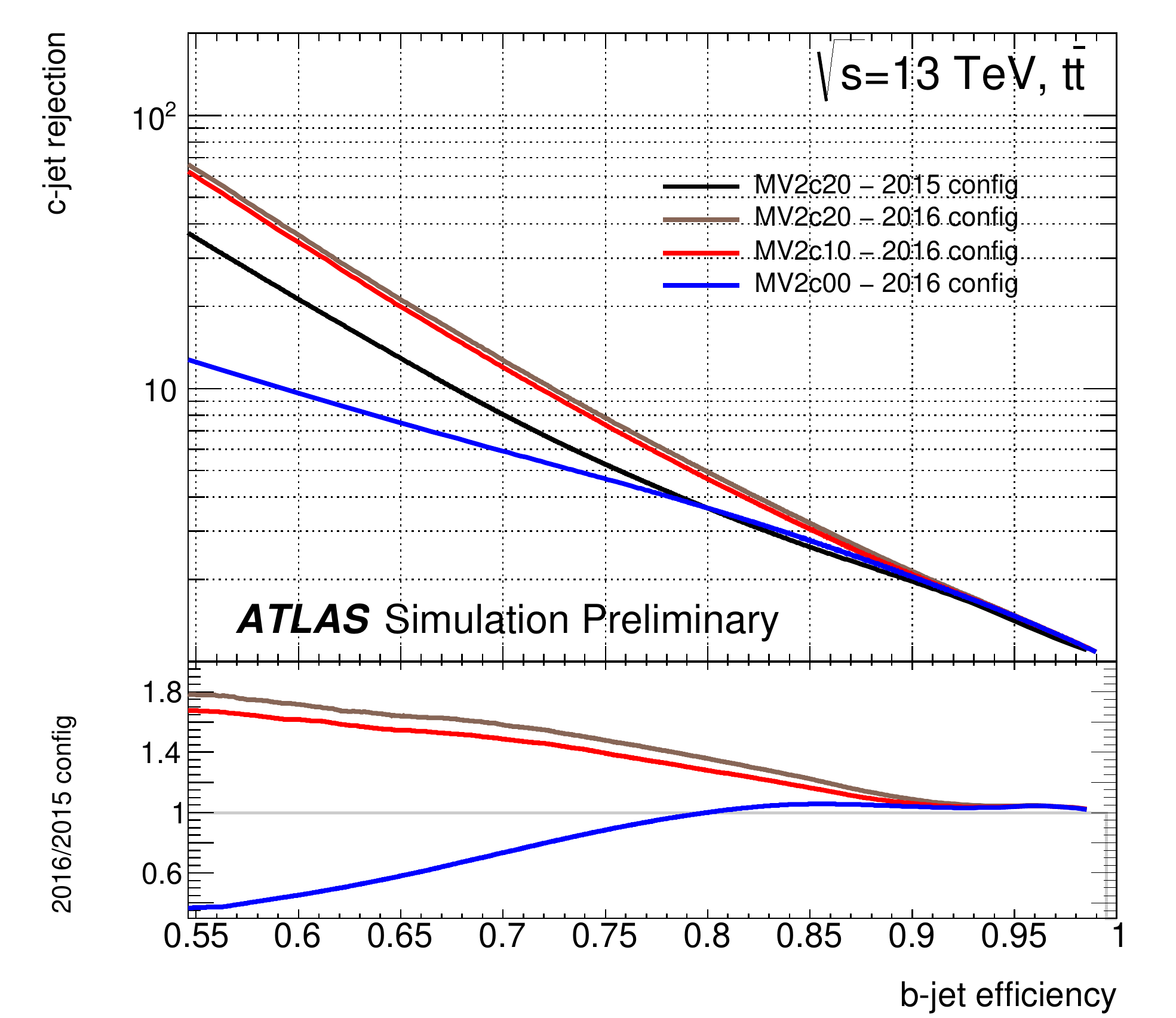}
\subcaption{}
\label{fig:exp.btag.eff}
\end{subfigure}
\vspace{-0.25cm}
\caption{
(a) MV2c10 BDT output for b- (solid blue), c- (dashed green) and light-flavour (dotted red) jets evaluated with tt events. 
(b) c-jet rejection versus b-jet efficiency for the 2015 and 2016 configurations of the algorithm.
 (for the 2016 configuration). 
}
\label{fig:exp.btag}
\end{figure}

The jet reconstruction algorithms cannot identify which type of parton initiated the jet except in the case of 
jets containing $b$-hadrons. 
Bottom-quark flavored hadrons live relatively longer which gives them specific characteristics that can be used 
to identify them. The procedure is commonly referred to as $b$-tagging and
is performed with the MV2 algorithm, a multivariate discriminant making 
use of track impact parameters 
and reconstructed secondary vertices~\cite{Aad:2015ydr,ATL-PHYS-PUB-2015-022}
to provide the best separation among the different jet flavour hypotheses.
An example of the output of the multivariate discriminant is shown in Figure~\ref{fig:exp.btag.bdt} 
where typically a cut must be applied on the score to identify an operating working point. 
Three MV2 algorithms were released that correspond to MV2c00, MV2c10, and 
MV2c20.
MV2c00 denotes the MV2 algorithm where no c-jet contribution was present in the training and MV2c10 (MV2c20) denote the MV2 outputs where a 7\% (15\%) c-jet fractions was present 
in the background sample.
The performance of the optimized  MV2c00, MV2c10 and MV2c20 b-tagging algorithms is shown in Figure~\ref{fig:exp.btag.eff}
for the c-jet rejection as a function of the b-jet efficiency.
The MV2c20 is the best performing algorithm leading to an optimal rejection of c-jets at a given b-tagging 
efficiency. As a result, it is the algorithm used in this analysis.

\subsection{Missing transverse momentum}
The missing transverse momentum is an important quantity in searching for new physics scenarios which expect a stable, 
non-electromagnetically, and non-hadronically interacting 
particle. Such a particle does not interact with the detector and can only be identified through an imbalance of the momentum 
distribution between the particles 
of the event. 
Since the momentum of the colliding protons is almost completely along the beam, i.e. longitudinally, 
the transverse component of the momenta of the 
scattered objects should add up to zero.
Based on the conservation of momentum, the sum of all visible four momenta projected in the transverse plane should be close to zero if 
no particles are missed.
However, this quantity will be large if a particle, potentially from new physics models, escaped detection. 
The missing transverse momentum is defined as the negative vector sum of the transverse momenta of the visible reconstructed objects in the
 event: 
\begin{equation}
\pt^{\rm{miss}} = - \sum_\text{visible} \pt
\end{equation}
where the visible objects include electrons, muons, jets, photons, taus, and a soft term.
In the rest of the dissertation, the magnitude of the missing transverse momentum vector is denoted by \met.
The soft term  is  a fundamental quantity in the reconstruction of \met and can be estimated by 
\begin{itemize}
\item Calorimeter based Soft Term (CST): accounts for both neutral and charged particle energies.
\item Track based Soft Term (TST): incorporates a natural pileup suppression by selecting only tracks from primary vertices.
\end{itemize}

\begin{figure}[htb!]
\centering
\begin{subfigure}[t]{0.48\textwidth}
\includegraphics[width=0.95\textwidth]{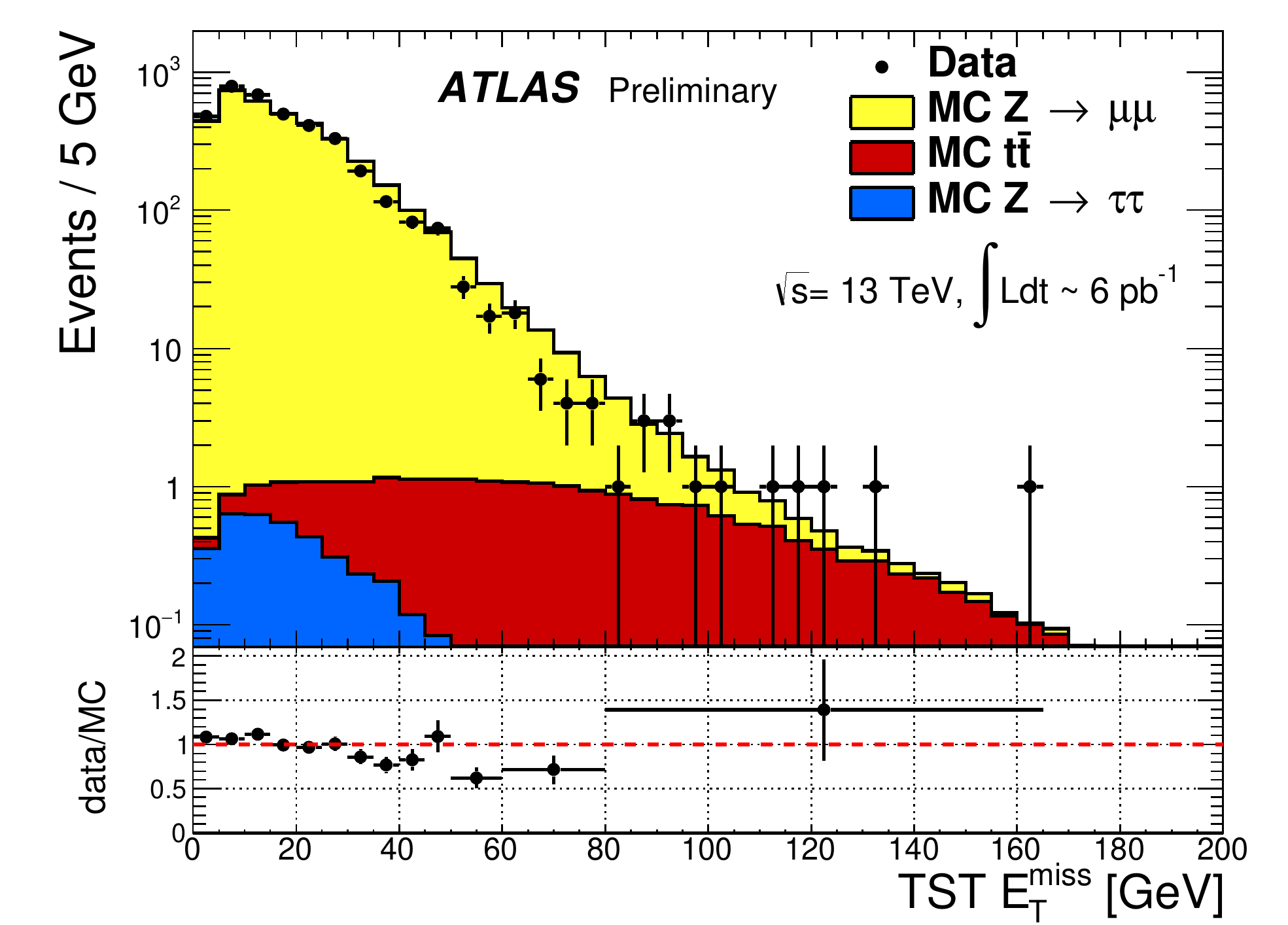}
\subcaption{}
\label{fig:exp.reco.met_dist}
\end{subfigure}
\begin{subfigure}[t]{0.48\textwidth}
\includegraphics[width=0.95\textwidth]{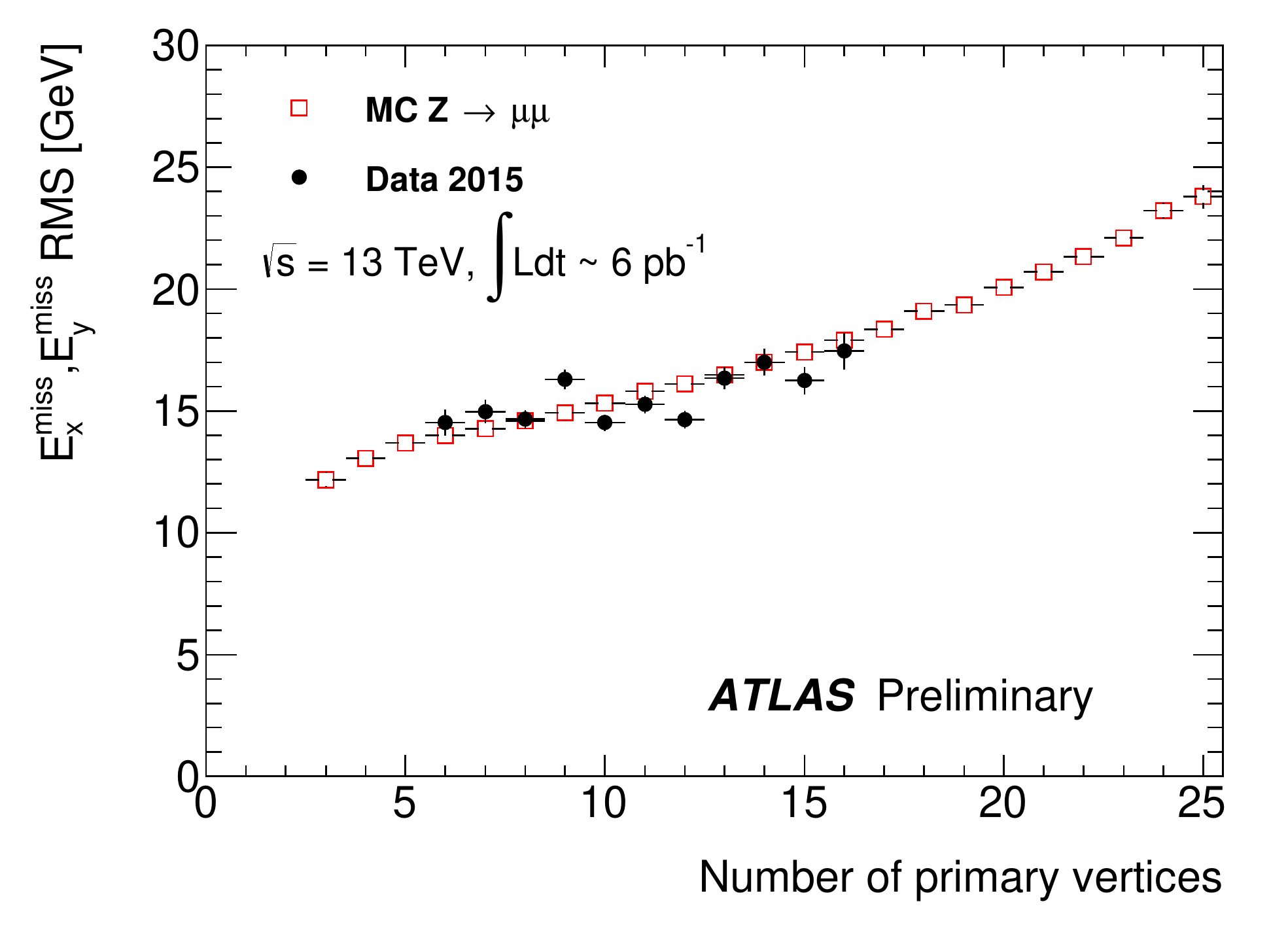}
\subcaption{}
\label{fig:exp.reco.met_res}
\end{subfigure}
\vspace{-0.25cm}
\caption{
(a) \met distribution as measured in data with $Z\to\mu\mu$ events without pile-up suppression.
(b) \met resolution as a function of the number of primary vertices in $Z\to\mu\mu$ events. The data (black circles) and MC simulation (red squares) are overlaid. 
}
\label{fig:exp.reco.met}
\end{figure}

The \met performance depends on the event topology affected by the presence of true \met, from neutrinos for example,
charged leptons, jet activity, and others.
The \met performance is generally studied with processes with and without genuine \met, such as 
$W \to e \nu$ and $Z\to\mu\mu$ events. 
The scale and resolution for the reconstructed \met in these processes are indicative of the reconstruction quality.
For illustration, results obtained with  $Z\to\mu\mu$ events are shown in Figure~\ref{fig:exp.reco.met}.
Generally, the \met has a resolution in the order of 10 to 20 \GeV.

%% file: texfiles/sec.roib.overview.tex
The Trigger and Data Acquisition (TDAQ) system reduces the 
proton interaction rate from 40 MHz to the ATLAS data storage capacity of about 1.5 kHz. 
A hardware First Level Trigger (L1) reduces the rate to 100 kHz and a software High Level Trigger (HLT) selects events for offline analysis. 
Jet, electromagnetic and tau clusters, missing transverse momentum ($E_{\mathrm{T}}^{\mathrm{miss}}$), $\sum E_{\mathrm{T}}$, 
jet $E_{\mathrm{T}}$, and muon candidate information from L1 determine detector Regions of Interest (RoIs) that seed HLT processing. These RoIs are provided to the HLT by a custom VMEbus based system, referred to as the Region of Interest Builder (RoIB) \cite{vme_roib}.
The RoIB collects data from L1 
trigger sources and assembles the data fragments into a complete record of L1 RoIs. These RoIs are made available to the HLT to initiate event processing. In order to improve maintainability and scalability, and to minimize the amount of custom hardware needing to be supported, 
the RoIB was implemented using commodity server hardware and an interface technology deployed 
within the ATLAS Trigger and Data Acquisition (TDAQ) system. The approach of implementing the RoIB functionality in software has been investigated in the past 
and the conclusion at that time was that a software based approach is possible but requires a card with a higher readout rate \cite{swroib_past}. 
Since data readout cards operating at high rates became available and the capabilities of computers have improved with the increase 
in CPU clock speed and number of cores, it became possible to implement the RoIB functionality using a PC based approach. 
The PC based RoIB must duplicate the functionality of the VMEbus based RoIB,
 which means that the PC based solution must receive and assemble the
individual L1 fragments and pass them as a single L1 result to the HLT. Modern computers have multicore CPU architectures 
with the possibility of running multi-threaded application, a feature which is being fully exploited in the RoIB software to achieve 
the desired performance of 100 kHz over 12 input links for fragment sizes of 400 bytes.  
This chapter describes the work of the author in evolving the RoIB from the VMEbus based system to the PC based system and gives details on the hardware, 
firmware, and software designs used to achieve the full RoIB functionality.


%% file: texfiles/sec.roib.vme.tex
\subsection{Hardware implementation}\label{sec:roib_current}

\begin{figure}[htb!] 
\centering
\includegraphics[width=.65\textwidth]{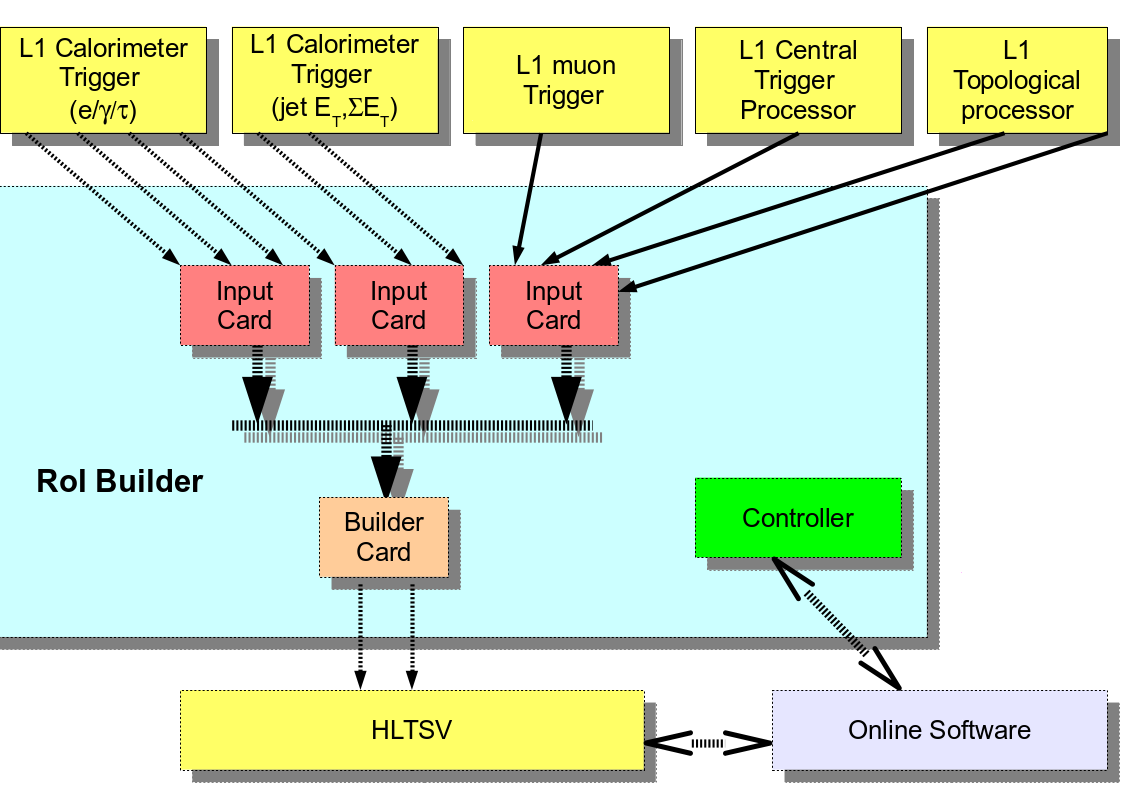}
\caption{Block scheme of the RoI Builder and overview of connections to external systems.  The
custom input and builder cards and the controller, a commercially available single board computer,
are installed in a single 9U VMEbus crate. The controller connects to the Control Network to interact with the rest of the 
data acquisition system.}
\label{roib_run1}
\end{figure}

The RoIB was implemented as a custom 9U VMEbus system that includes a controller which configures and monitors the system along with custom cards 
that receive and assemble the event fragments and send them to the HLT. Figure \ref{roib_run1} shows a block of the RoIB and 
its connection to external systems.

The RoIB contains four input cards and uses one builder card in the Run-2 configuration. Each input card accepts three inputs from 
L1 subsystems. 
The builder card assembles the input data of the events and passes the results via two optical links 
to another receiver card in a PC running the HLT supervisor (HLTSV) application. The receiver card in the HLTSV is a TILAR card \cite{tilar}
 that implements four PCI-express Generation 1 (PCIe Gen1) with 8 lanes
\footnote{PCI stands for Peripheral Component Interconnect which is 
a high-speed input/output (I/O) serial bus that can be installed on
motherboard of a computer.
It can transfer data at a speed of 250 megabytes per second per lane.
}
 to interface with the two optical links. The HLTSV manages the HLT processing farm by using L1 results provided by the RoIB, retrieves events from the ROS, assigns events to HLT farm nodes, and handles event bookkeeping including requesting removal of data from ROS storage when no longer required. 

The fragments received by the RoIB are identified by a 32 bit identifier, the extended L1 ID (L1ID). 
The RoIB input cards use the L1ID and the number of outputs enabled to assign keys to the various fragments and send them to the output channel in the builder card that was 
assigned that key value. The input data is transferred over a custom J3 back-plane. The back-plane operates at 20 MHz and transfers 16 data bits per 
clock cycle simultaneously for up to 12 inputs. The total maximum data throughput is therefore 480 MB/s, 40 MB/s per input.  
The maximum size of any single fragment is limited to 512 bytes imposed by resources available in the FPGA 
\footnote{FPGA stands for Field Programmable Gate Arrays 
 which are semiconductor devices composed of configurable logic blocks 
that can be reprogrammed for a desired application.}
firmware. The current RoIB input 
links are listed in Table \ref{tab:roib_links}.

\begin{table}[tbp]
\caption{L1 input sources to the RoIB.}
\label{tab:roib_links}
\smallskip
\centering
\begin{tabular}{|c|c|}
\hline
Source & Links\\
\hline
Central Trigger Processor (CTP)  & 1  \\
L1 calorimeters (e/$\gamma$, $\tau$, jet, $\sum E_\mathrm{T}$) & 6  \\
Muon Trigger to CTP Interface (MUCTPI) & 1  \\
Topological processor (L1Topo) & 2  \\
Spare & 2 \\
\hline
\end{tabular}
\end{table}

\subsection{System Performance and Evolution}\label{sec:roib_limit}

The custom VMEbus based RoIB operated reliably during the first run of the LHC, however, it is desirable to have a more flexible RoIB. 
In addition, the RoIB is getting close to its design limitation, as seen 
in Figure \ref{fig:roib_proto}. For fragments of 400 bytes and inputs from eight L1 systems, referred to as channels, the current RoIB rate limit is 60 kHz which is below the required 100 kHz at 
L1. While the current fragment size coming from L1 is around 160 bytes, the sizes are expected to grow due to the increase of instantaneous 
luminosity and the complexity of the L1 triggers. 
The current VMEbus system will be replaced by a PCI-express card hosted in the HLTSV PC with the 
possibility to upgrade the commodity hardware (e.g. ability to upgrade CPUs). 
The new configuration simplifies the readout architecture of ATLAS. The targeted rate for event building is 100 kHz over 12 input channels for 
fragment sizes on the order of 400 bytes.

%% file: texfiles/sec.roib.pc.tex
 A custom PCIe card developed by the ALICE collaboration, the Common ReadOut Receiver Card (C-RORC) \cite{alice}, was deployed as an 
upgraded detector readout interface within the ATLAS ROS with ATLAS specific firmware and software called the RobinNP \cite{crorc}. 
The new PC based RoIB uses the RobinNP firmware and a dedicated program 
interface to facilitate the implementation of the RoIB functionality 
on a commodity PC. In this section, we describe the C-RORC hardware as well as the RobinNP firmware, API, and the event building software. 
\subsection{The Common Readout Receiver Card}\label{sec:crorc}

\begin{figure}[tbp] 
\centering
\includegraphics[width=\textwidth]{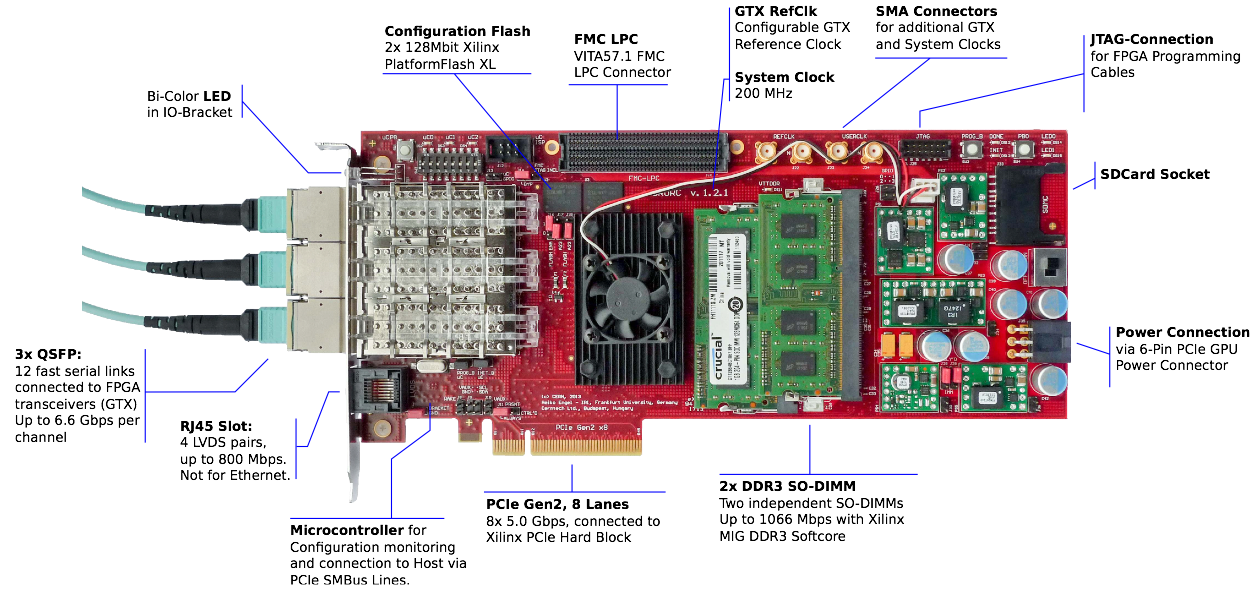}
\caption{Photo of the C-RORC board with the major components and features annotated \cite{crorc}.}
\label{fig:crorc}
\end{figure}

The C-RORC implements 8 PCIe Gen1 lanes with 1.4 GB/s bandwidth to the CPU fed via 12 optical links each running 200 MB/s on 3 QSFP 
\footnote{QSFP stands for a Quad Small Form-factor Pluggable which is 
a hot-pluggable transceiver used for data transfer.}
transceivers. It utilizes a single Xilinx Virtex-6 series FPGA that handles data input from the 12 links and buffers the data in two on-board DDR3 memories. It is also capable of processing and initiating DMA transfer of event data from the on-board memory to its host PC's memory. The major components of the C-RORC are annotated 
in the picture shown in Figure \ref{fig:crorc}.

\begin{figure}[tbp] 
\centering
\includegraphics[trim = 0cm 0cm 0cm 2.5cm, clip, width=.71\textwidth]{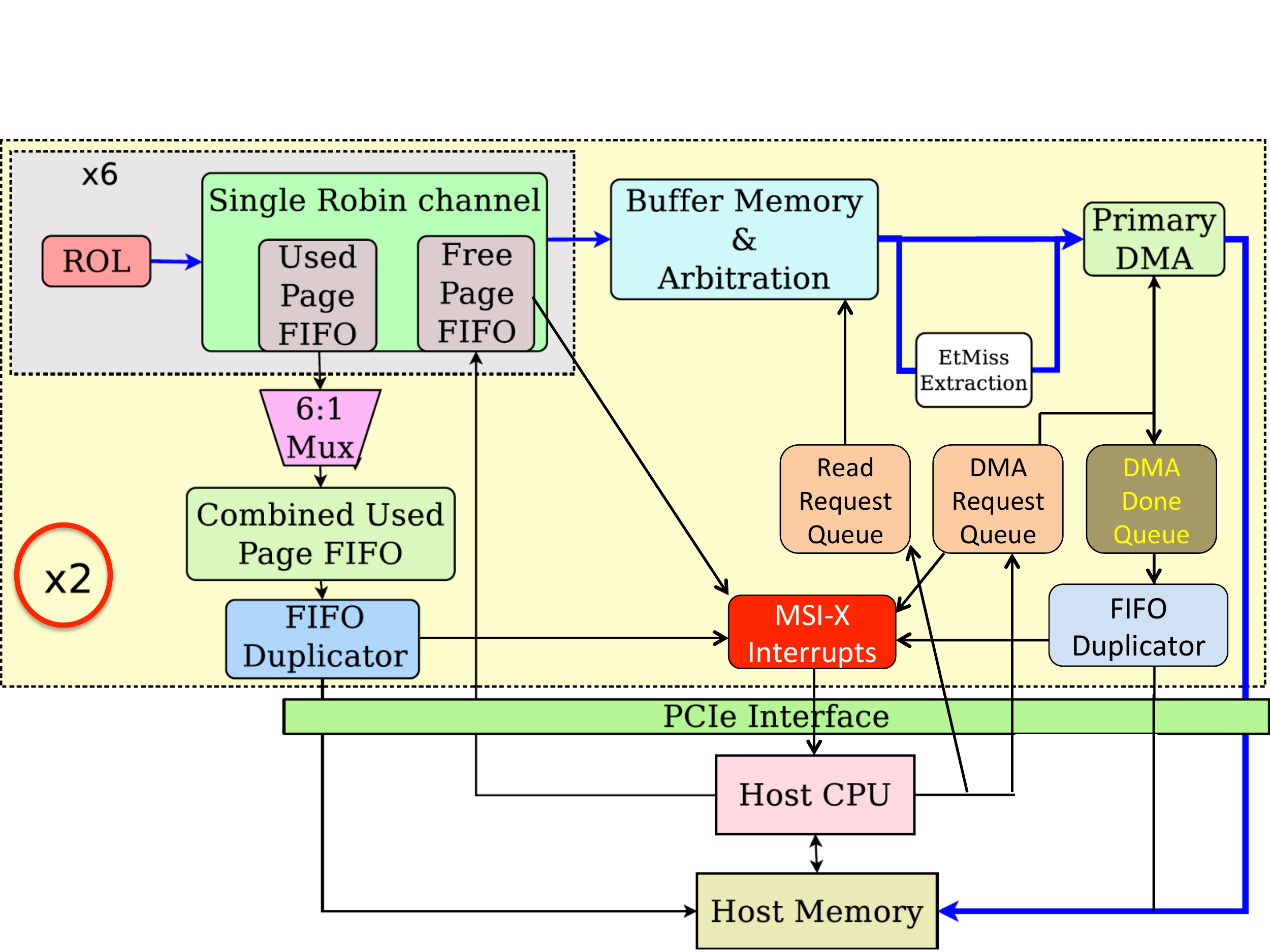}

\caption{RobinNP firmware organization
and  flow  of  data  from  host  CPU  to  the
firmware  (by  means  of  programmed  I/O)
and from the firmware to the host memory
(by means of DMA).}
\label{fig:robinnp_fw}
\end{figure}

\subsection{Readout System Firmware and Software}\label{sec:crorc_fw}

\begin{figure}[tbp] 
\centering
\includegraphics[trim = 0cm 0cm 0cm 5cm, width=.7\textwidth]{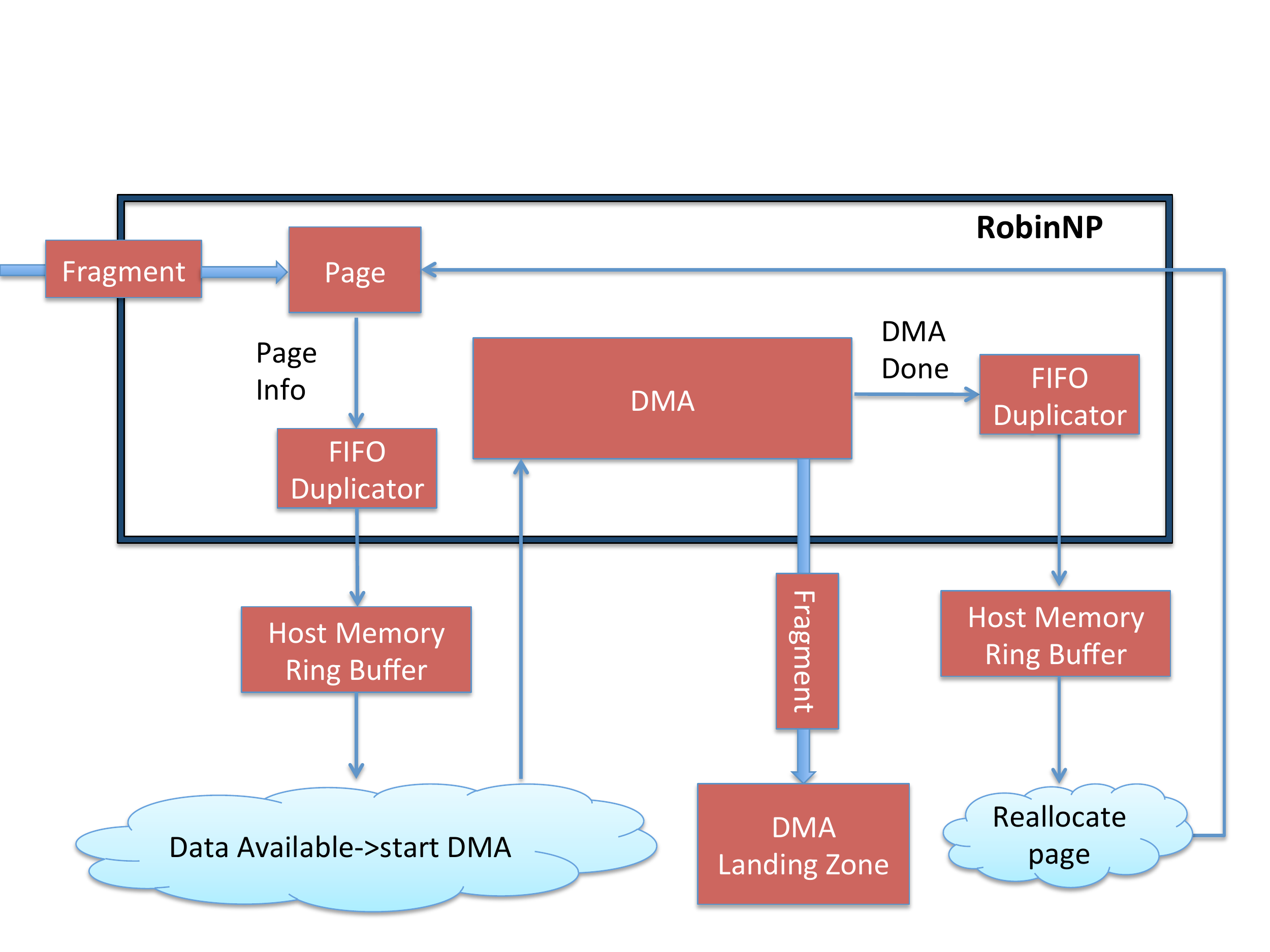}
\caption{Layout of the readout system firmware and software specific to the RoIB.}
\label{fig:roib_swfw}
\end{figure}

The RobinNP firmware used for the RoIB is identical to that used in the ATLAS ROS\cite{ros}. As shown in 
the schematic of Figure \ref{fig:robinnp_fw}, the logic is divided into two functional blocks, known as sub-ROBs, 
each servicing six input links and one DDR3 memory module. Event data fragments arriving via a link are subjected 
to a range of error checks before being stored in the memory module for the relevant sub-ROB. At the same time a token
representing the address of a region of the memory, referred to as a page, is passed to a listening software process via 
a `FIFO duplicator'. To avoid a costly read across the PCIe bus, data is continuously streamed from firmware to 
software via a chain of firmware and software FIFOs. Notification of new data arriving in the software FIFO is managed via coalesced 
interrupts to allow for efficient use of CPU resources.
For the RoIB application, the receipt of page information immediately triggers a DMA of fragment data from the RobinNP memory into 
the host PC memory. The fragments are then passed via a queue (one per sub-ROB) to the RoIB process along with any relevant fragment 
error information. A schematic of this shortened dataflow path is presented in Figure \ref{fig:roib_swfw}. 
The API for the RoIB process consists of these queues, return queues for processed pages now available for re-use and 
a configuration interface. The software is implemented with multiple threads each handling specific tasks such as supply of free pages, receipt
of used pages, DMA control and bulk receipt of fragment data.

\subsection{RoIB Software}\label{sec:crorc_sw}

The HLTSV is a multi-threaded application that obtains a L1 result from a variety of possible input sources and exchanges information with the 
rest of the HLT computing farm. 
For the RoIB, the L1 source is a RobinNP interface that performs fragment assembly and is used as a plug-in to the HLTSV application.
The RobinNP plug-in has two receive threads, each 
thread services six channels by pulling fragments from the RobinNP on-board memories to the host PC.
Fragments with the same L1ID are copied 
to a contiguous memory space and a queue of completed events is prepared. 
Upon request by the HLTSV, a pointer to the contiguous memory space is passed back to the 
HLTSV process for further handling. In order to optimize concurrent access to RoIB data structures, containers from the  Intel 
threading building block (TBB) library were used. These containers allow multiple threads to concurrently access and update items 
in the container while maintaining high performance.  

%% file: texfiles/sec.roib.proto.tex
In order to understand the requirements for the underlying server PC, a validation system based on Intel(R) Xeon(R) CPU E5-1650 v2 
@ 3.5 GHz with six cores was used to perform tests of the PC based RoIB. 
The goal was to perform software based fragment assembly at a rate of 100 kHz over 12 channels for a typical 
fragment size of 400 bytes. The current system offers flexibility in terms of the fragment size allowed which was not the case in the 
VMEbus based RoIB. The initial tests were performed with a standalone application that implements a minimal interface for event building. 
Once the system was validated, the relevant code modules were integrated into an HLTSV process running within the full ATLAS TDAQ 
software suite with appropriately scaled test hardware to represent the remaining elements of the system.

\subsection{Standalone Tests}\label{sec:perf_alone}

\begin{figure}[tbp] 
   \centering
   \includegraphics[width=.7\textwidth]{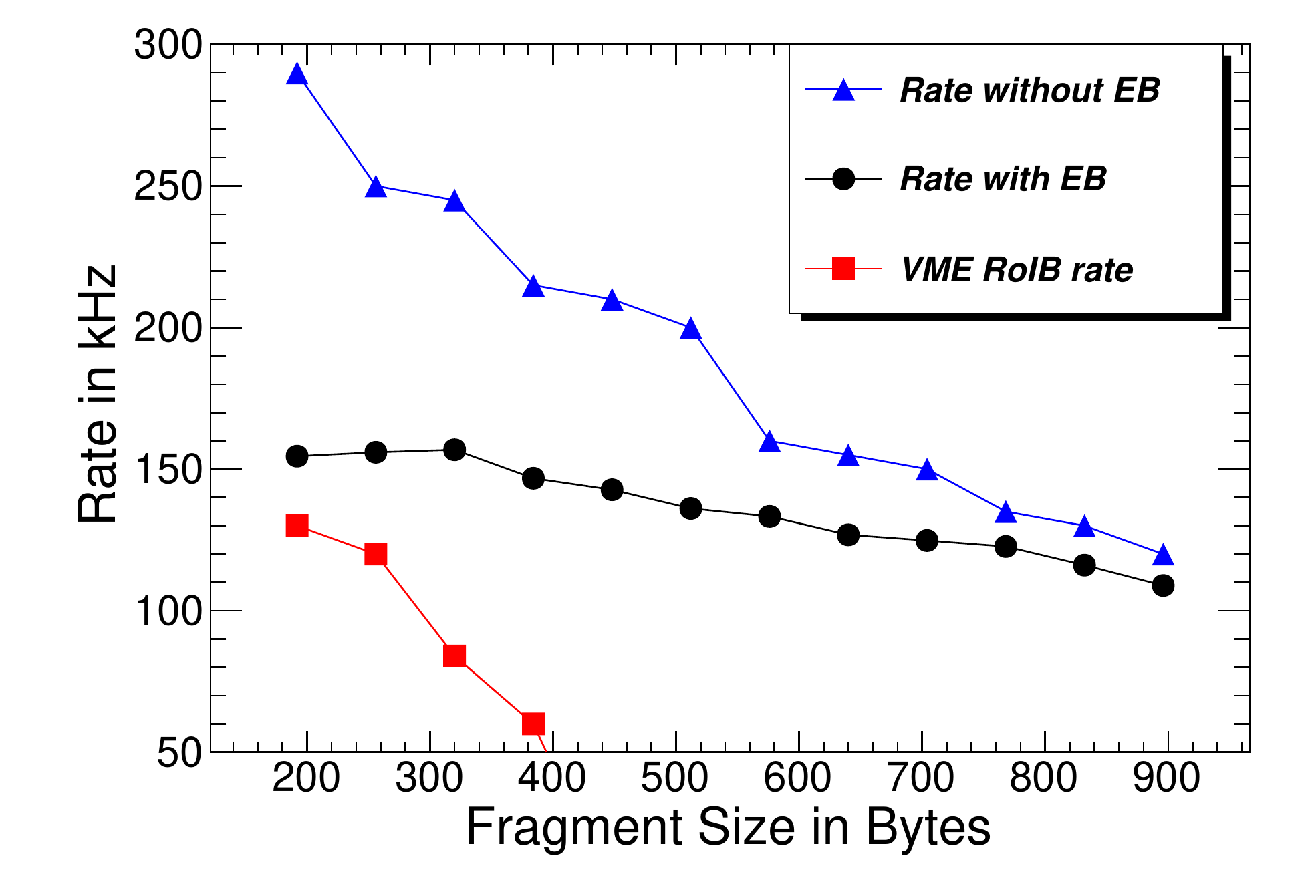}
   \caption{Rate as a function of the fragment size (in bytes) with an external source that emulates the L1 trigger input. 
     The rates shown are for the input rate to the RobinNP without event building (EB) (triangle), rate with EB (circle), and 
     for comparison, the current VMEbus RoIB rate is also shown (square).  }
   \label{fig:roib_proto}
 \end{figure}

The goal was to test input/output bandwidth limitations of the RobinNP and the rate of event building. Initial performance testing used 
a standalone RobinNP application and an external source that emulates the L1 trigger data 
in the form of 32-bit word fragments with 12 channels. In this test, the host PC was running the assembly routine with a single threaded application.  Figure \ref{fig:roib_proto} shows the input rate without 
event building as a function of fragment size. For 400 byte fragments the input rate to the RobinNP is 215 kHz. 
The same figure 
shows the event building rate which is 150 kHz. This performance shows that the event building 
at the required rate of 100 kHz with 12 channels is achievable in a standalone application.

\subsection{Full System Tests}\label{sec:perf_tdaq}

\begin{figure}[tbp] 
\centering
\includegraphics[width=.98\textwidth]{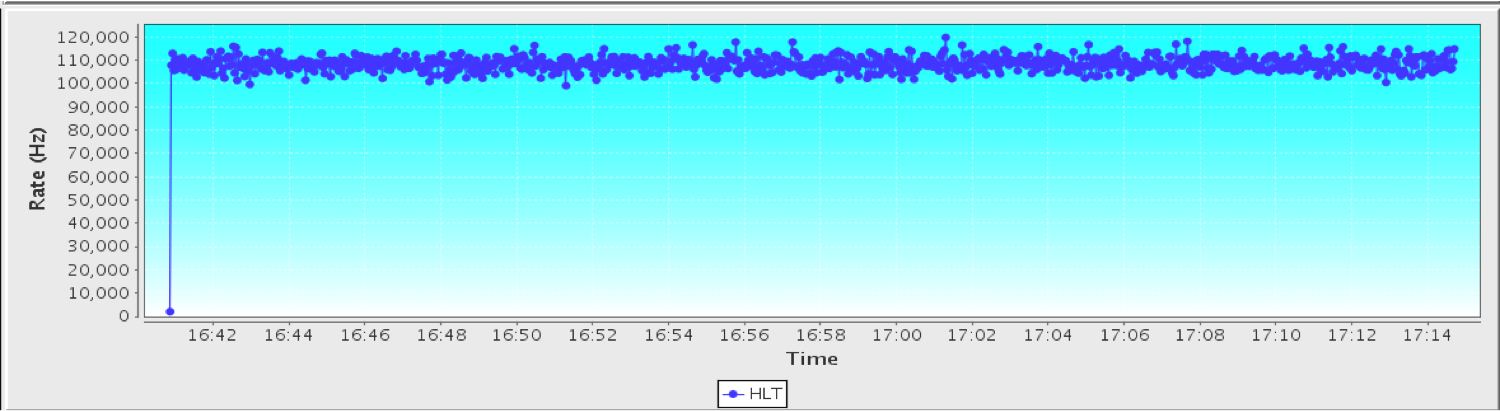}
\caption{Screenshot of a monitoring tool which shows the HLTSV processing rate using the ATLAS TDAQ software.}
\label{fig:partition}
\end{figure}

Since the HLTSV is performing tasks other than the event building, there is overhead associated with additional operations 
that reduces the performance. For this reason, we use the full ATLAS TDAQ software in a test environment that emulates the major components of the ATLAS data acquisition system, shown in Figure \ref{fig:tdaq_diagram}. The setup includes an emulated input from L1 trigger sources, 
the HLTSV and other PCs to simulate the HLT computing farm, and the ROS that buffers the full event data. 
 In this test setup, an external source sends data that emulates L1 RoIs via 12 links connected to the 
RobinNP hosted by the HLTSV. When the HLTSV requests a built RoI event, the software RoIB plug-in provides the RoI event which will be used 
to seed requests for the event data to be processed.
 Figure \ref{fig:partition} shows an event building rate of 110 kHz measured with 400 byte fragments with the HLTSV application in a setup close to the ATLAS TDAQ system. 

\begin{figure}[t!]
\centering
\includegraphics[width=0.7\textwidth]{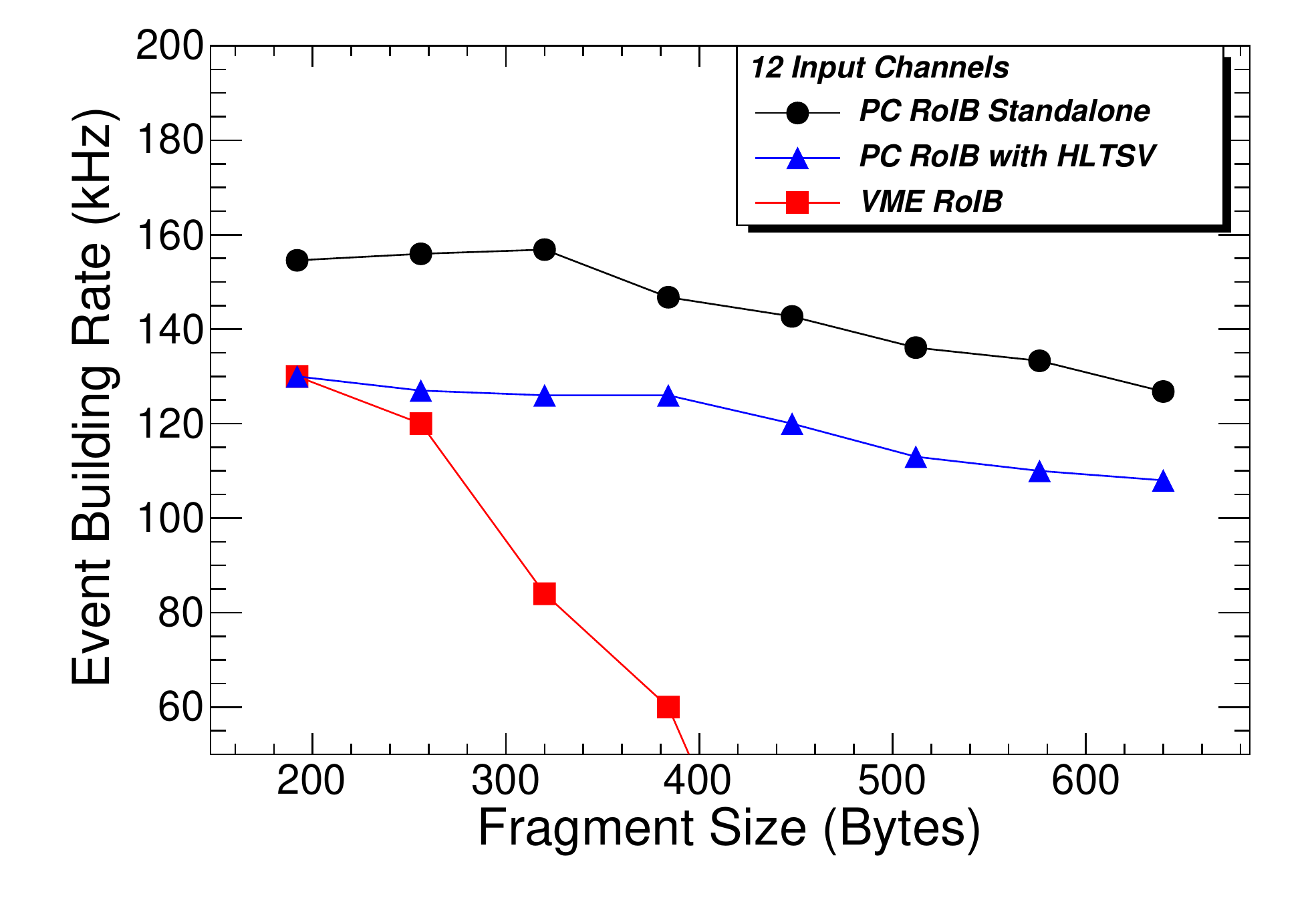} 
\caption{The event building rate as a function of the RoI record size in bytes. The rates are shown for a standalone application that implements 
  a minimal interface for event building, the integrated RoIB software into an HLTSV process running within the full ATLAS TDAQ software suite, and for comparison the VME-RoIB performance.}
\label{fig:roib_summary}
\end{figure}

As shown in Figure \ref{fig:roib_summary}, the performance of the PC-RoIB with realistic running ATLAS conditions is improved over the VME-RoIB particularly at high RoI sizes and maintains a rate of over 100 kHz with 12 channels. 

\begin{figure}[p!]
  \centering
  \begin{subfigure}{0.48\textwidth}
    \includegraphics[width=\textwidth]{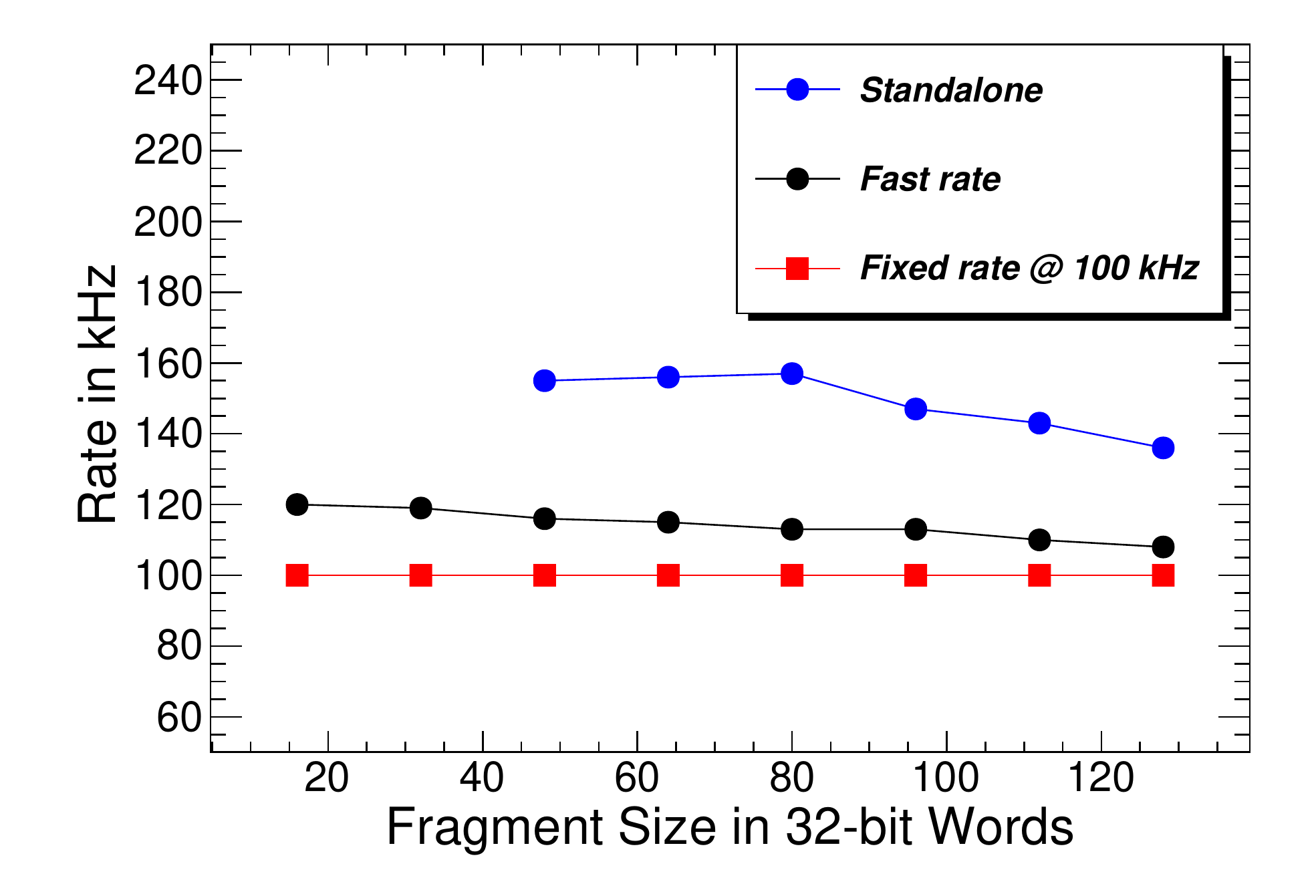}
    \subcaption{}
    \label{fig:roib.proto.rate_finala}
  \end{subfigure}
  \begin{subfigure}{0.48\textwidth}
    \includegraphics[width=\textwidth]{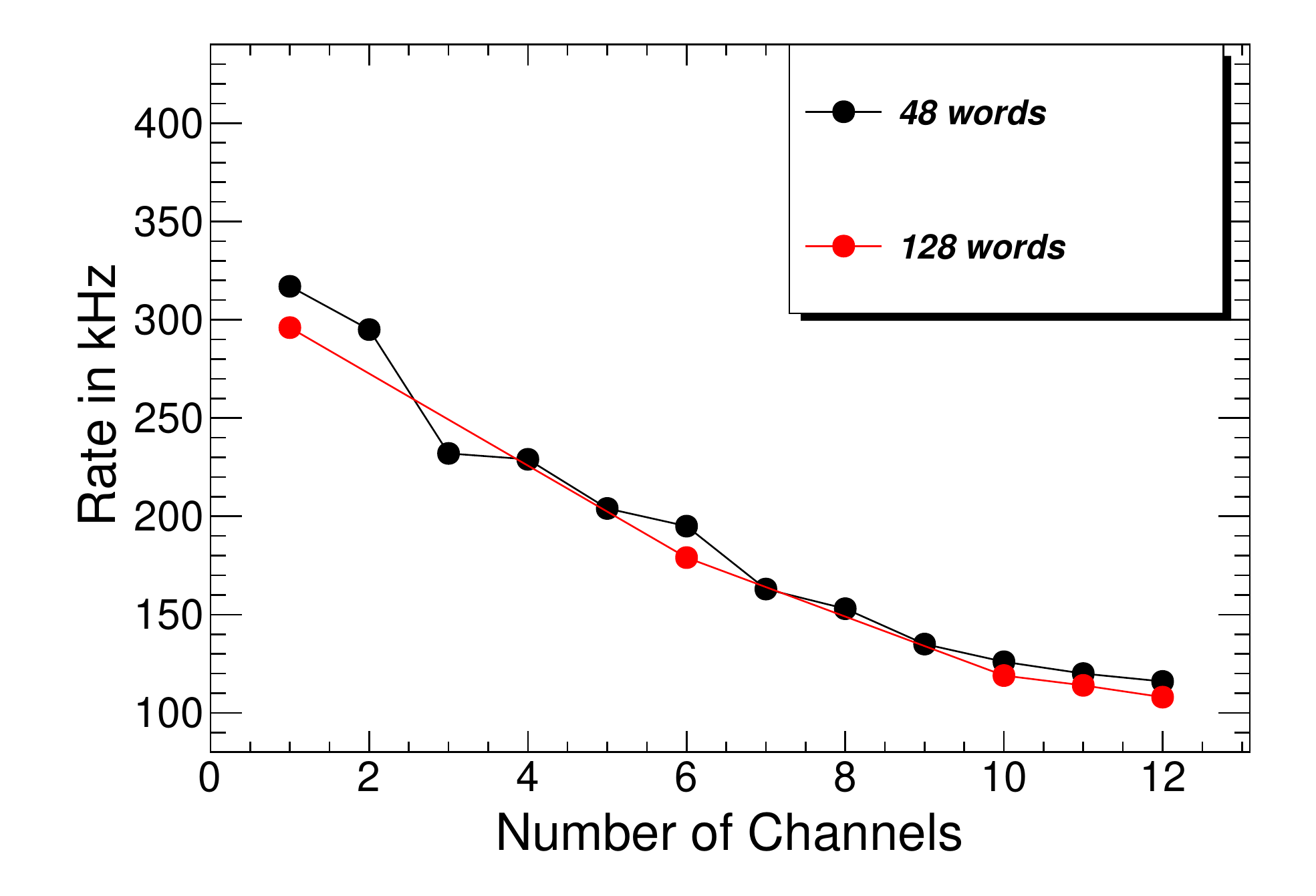}
    \subcaption{}
    \label{fig:roib.perf.chan}
  \end{subfigure}
  \caption{The event building rate as a function of (a) the fragment size
(b) the number of channels. The rates are shown for a standalone application that implements 
  a minimal interface for event building, the integrated RoIB software into an HLTSV process running within the full ATLAS TDAQ software suite.}
  \label{fig:roib.perf.other}
\end{figure}

The design specification of the ATLAS L1 trigger is to send data at 100 kHz.
While the tests above showed that the PC RoIB meets the desired rate 
requirement in the case that an external source is sending data as fast 
as possible (much more than 100 kHz), it is important to test that the 
PC RoIB will sustain the 100 kHz rate if the external source sends data 
at exactly 100 kHz. Figure~\ref{fig:roib.proto.rate_finala} demonstrates 
that in the event that the incoming data to the PC RoIB is fixed at 
100 kHz, the event building in the PC RoIB still operates at this rate.
The other important variable that can affect the rate is the number of 
channels. In particular, the ATLAS detector might decide to disable 
some of the channels which should not affect the rate of operation of the 
PC RoIB. Figure~\ref{fig:roib.perf.chan} shows that the PC RoIB will 
operate at even higher rates if the number of channels is reduced.

With these tests, the author validated the operation of the new PC RoIB
which deemed it ready to be deployed in the ATLAS system.

%% file: texfiles/sec.roib.perf.tex
The author deployed the new PC RoIB in the ATLAS trigger and data acquisition 
system during the LHC winter shutdown in January 2016. 
Initially, the PC RoIB was used as the main system with the VMEbus RoIB 
was used as a backup system. Later, the VMEbus RoIB was removed completely
and the PC RoIB became the only system running in the ATLAS trigger.
The PC RoIB operated reliably since its installation without any problems 
and without deadtime for the ATLAS data collection. 
Figure \ref{fig:roib.perf.buildtime} shows that
the RoIB event assembly does not depend on pileup conditions and
Figure \ref{fig:roib.perf.mem} shows that the 
memory usage of the HLTSV is at the level of 5\%.
It has now participated in collecting a dataset of over  35 \ifb exceeding 
the dataset collected by the VMEbus RoIB (22 \ifb).
The performance of the PC RoIB during the data taking of ATLAS has been 
very stable.

\begin{figure}[t!]
\centering
\begin{subfigure}[t]{0.48\textwidth}
\includegraphics[width=0.95\textwidth]{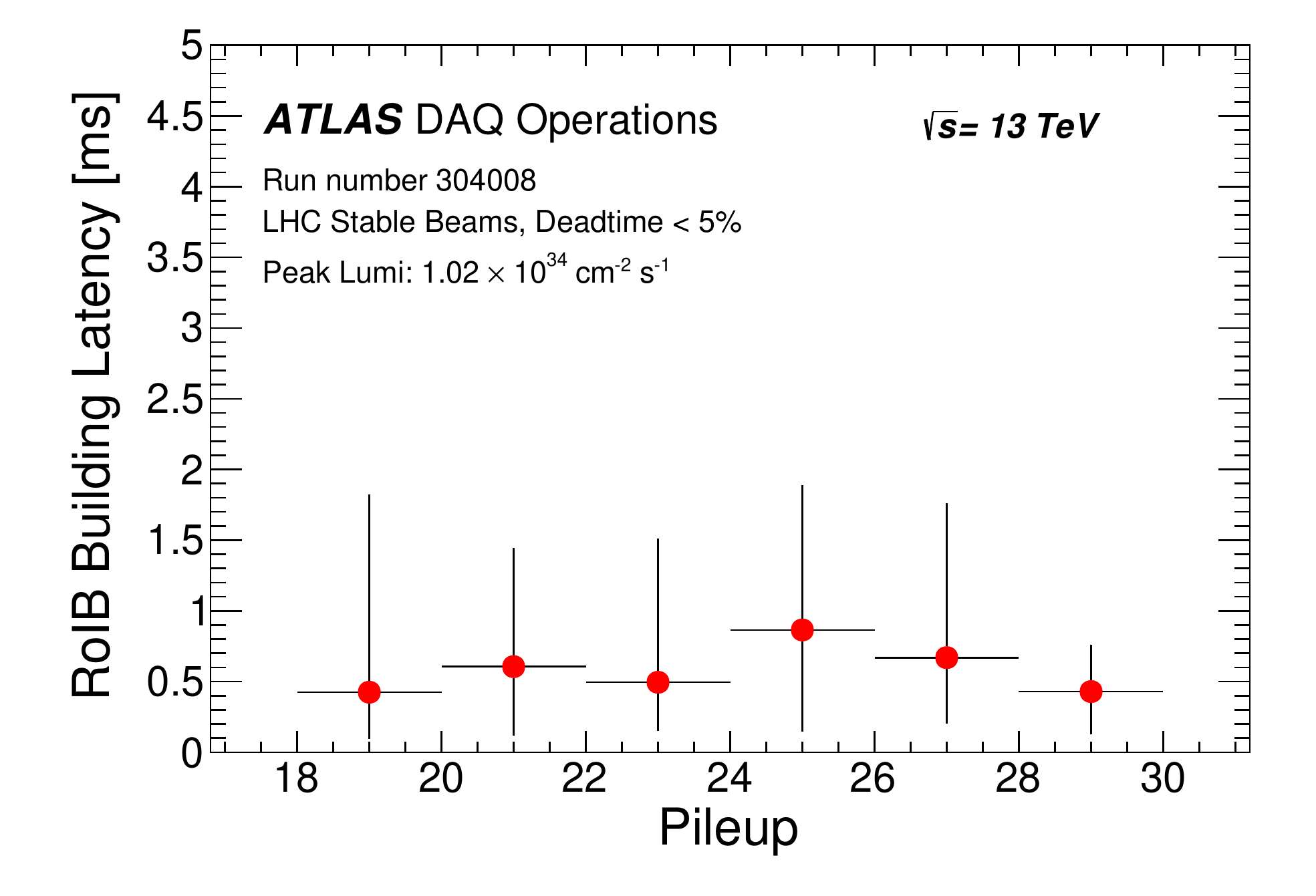}
\subcaption{}
\label{fig:roib.perf.buildtime}
\end{subfigure}
\begin{subfigure}[t]{0.48\textwidth}
\includegraphics[width=0.95\textwidth]{hProf_run_304008_l1rate_RoIBMemOccup}
\subcaption{}
\label{fig:roib.perf.mem}
\end{subfigure}
\vspace{-0.25cm}
\caption{RoIB performance: RoIB building latency as a function of pileup (left), RoIB memory occupancy as a function of L1 rate (right).}
\label{fig:roib_pileup_l1rate}
\end{figure}


%% file: texfiles/sec.strategy.overview.tex
The chapters thus far described supersymmetry as an extension of the standard 
model, motivated the need to search for it, and described the experimental 
apparatus used for this search, the LHC and ATLAS. 
This chapter begins the discussion of the search for supersymmetry that the 
author has performed.

The main task in designing a search for new physics signatures is to ensure 
that the search regions, expected to have an enhancement of the signal
 (also referred to as the signal regions), are sensitivity to 
 a wide 
range of new physics models that are well motivated.
In addition, it is important to design the search in a way that minimizes the
contamination from the known physics processes of the Standard Model.
In other words, a typical signal region should have a maximum expected 
signal with a minimum expected background. 

The search for supersymmetry with two leptons of the same-electric charge
or more than three leptons meets both criteria. 
The search targets the strong production of supersymmetric particles 
which mainly involves gluino pair production. Since gluinos are Majorana 
fermions, they can decay to either a positive or negative lepton in each branch
of the pair production. As a result, each branch is as likely to have two leptons 
of the same electric charge as it is to have two leptons of opposite electric
charge. While opposite-sign lepton production is a common signature in the 
Standard Model, the same-sign lepton signature is very rare. 
Also, processes that involve more than two leptons are rare in the 
Standard Model. On one hand the electroweak processes leading to 
$W$ and $Z$ bosons have a low cross sections. On the other hand the low 
branching ratio to leptons leads to an extreme background reduction.
The requirement of three or more leptons allows the analysis 
to target supersymmetric models with longer decay chains. The presence of 
a third softer lepton also increases the sensitivity to scenarios with 
small mass difference between the supersymmetric particles.

The next important step in a search for new physics is to estimate the 
backgrounds present in the signal regions.
The Standard Model backgrounds with a same-sign lepton pair or three or more 
leptons predominantly comes
from the associated production of a top quark pair and a vector boson (\ttbar + $W$, \ttbar + $Z$), and multi-boson production (di-boson and tri-boson). 
These backgrounds that lead to a signature with same-sign leptons or three 
or more leptons are referred to as \textit{irreducible backgrounds}.
However, the high cross section processes from the Standard Model such as \ttbar, might contribute to the signal regions via a mis-reconstruction of this 
process by the detector. As a result, there are two very important backgrounds
 that affect the analysis that are referred to as 
\textit{reducible backgrounds}.
The top quark pair production (\ttbar) process may decay fully leptonically 
($\ttbar \to \left(b\ell^+\bar{\nu}\right)\left(\bar{b}\ell^{-}\nu\right)$) 
and contribute to the signal regions if the lepton charge is mis-measured.
In the case of a semi-leptonic decay of \ttbar 
($\ttbar \to \left(b\ell^+\bar{\nu}\right)\left(\bar{b}q \bar{q}'\right)$)
where the hadronic decay is mis-identified as a leptonic decay, 
the process will contribute with a ``fake'' lepton in the signal regions.
The background estimation methodology will aim at estimating both 
reducible and irreducible backgrounds with Monte Carlo simulation and 
data-driven methods.

Finally, we assess the compatibility between the observed data and the 
predicted background in one counting experiment for each signal region 
by doing a hypothesis test
of the background-only or the background as well as the sought 
after signal hypotheses.
If an excess is found in data, we proceed to evaluate if this excess can lead
to a rejection of the background-only hypothesis and we check the plausibility that 
the new signal can describe the data.
Otherwise, we set exclusion limits for a certain region of the parameter 
space of a defined model or we set model-independent upper limits on the number 
of events from a beyond the Standard Model process. 

%% file: texfiles/sec.strategy.pheno.tex

Final states with two same-sign leptons or three leptons and multiple jets can probe a variety of supersymmetric models 
represented by decays of heavy superpartners involving massive gauge bosons, sleptons or top quarks. 
The decays of the superpartners can lead to many experimental signatures that may lead to different lepton, jet, and $b$-tagged jet multiplicities.
To exploit this wide range of possible signatures, the analysis uses six $R$-parity-conserving SUSY scenarios 
featuring gluino, bottom squark (sbottom) or top squark (stop) pair production. 
These scenarios were used as benchmarks to identify regions of the phase space 
where the analysis can bring particularly useful complementarity to other SUSY 
searches, 
and subsequently define signal regions with a particular focus on these 
regions. 
In this section, the scenarios considered are presented with details about 
the assumed superpartner masses and decay modes.  
To highlight the improvement in reach that this analysis brings, exclusion limits obtained prior to the work of the author will also
be shown.

\begin{figure}[t!]
\centering
\begin{tabular}{rrrr}
\begin{subfigure}[t]{0.24\textwidth}\includegraphics[width=\textwidth]{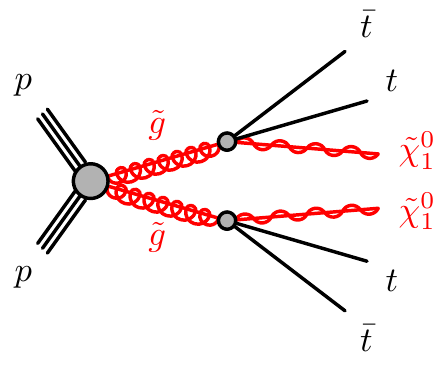}\caption{}\label{fig:strategy.pheno.feynman_gtt}\end{subfigure}&
\begin{subfigure}[t]{0.24\textwidth}\includegraphics[width=\textwidth]{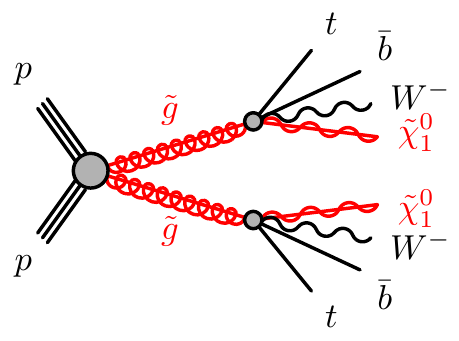}\caption{}\label{fig:strategy.pheno.feynman_gttOffshell}\end{subfigure}&
\begin{subfigure}[t]{0.24\textwidth}\includegraphics[width=\textwidth]{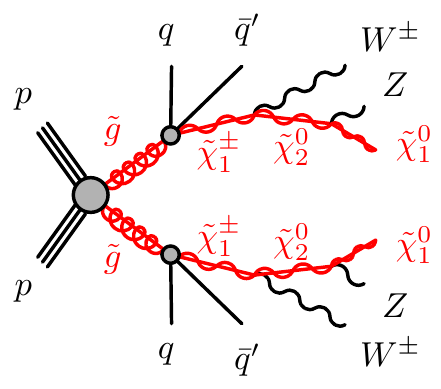}\caption{}\label{fig:strategy.pheno.feynman_gg2WZ}\end{subfigure}&
\begin{subfigure}[t]{0.24\textwidth}\includegraphics[width=\textwidth]{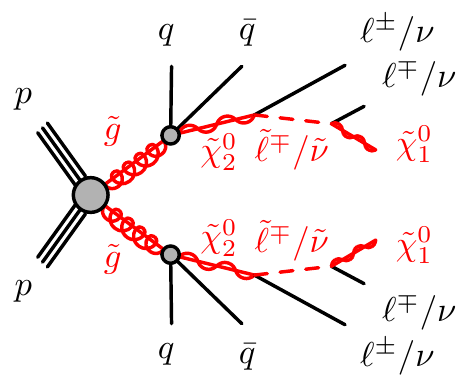}\caption{}\label{fig:strategy.pheno.feynman_gg2sl}\end{subfigure} \\
&
\begin{subfigure}[t]{0.24\textwidth}\includegraphics[width=\textwidth]{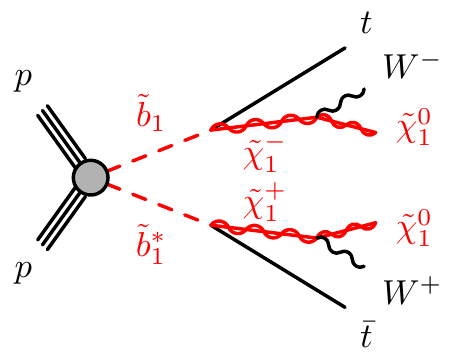}\caption{}\label{fig:strategy.pheno.feynman_b1b1}\end{subfigure} &
\begin{subfigure}[t]{0.24\textwidth}\includegraphics[width=\textwidth]{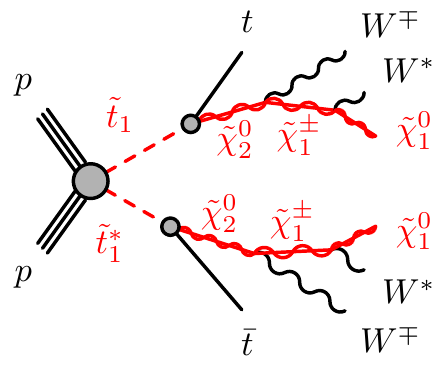}\caption{}\label{fig:strategy.pheno.feynman_t1t1}\end{subfigure} &
 \\
\end{tabular}
\caption{SUSY processes featuring gluino ((a), (b), (c), (d)) or third-generation squark ((e), (f)) pair production studied in this analysis. 
 In Figure~\ref{fig:strategy.pheno.feynman_gg2sl}, $\tilde{\ell} \equiv \tilde{e}, \tilde{\mu}, \tilde{\tau}$ and 
$\tilde{\nu} \equiv \tilde{\nu}_e, \tilde{\nu}_{\mu}, \tilde{\nu}_{\tau}$. In Figure~\ref{fig:strategy.pheno.feynman_t1t1}, the $W^*$ labels indicate 
largely off-shell $W$ bosons -- the mass difference between $\chinoonepm$ and $\ninoone$ is around 1~GeV.}
\label{fig:strategy.pheno.feynman}
\end{figure}

\begin{figure}[t]
\centering
\begin{subfigure}[t]{0.55\textwidth}\includegraphics[width=\textwidth]{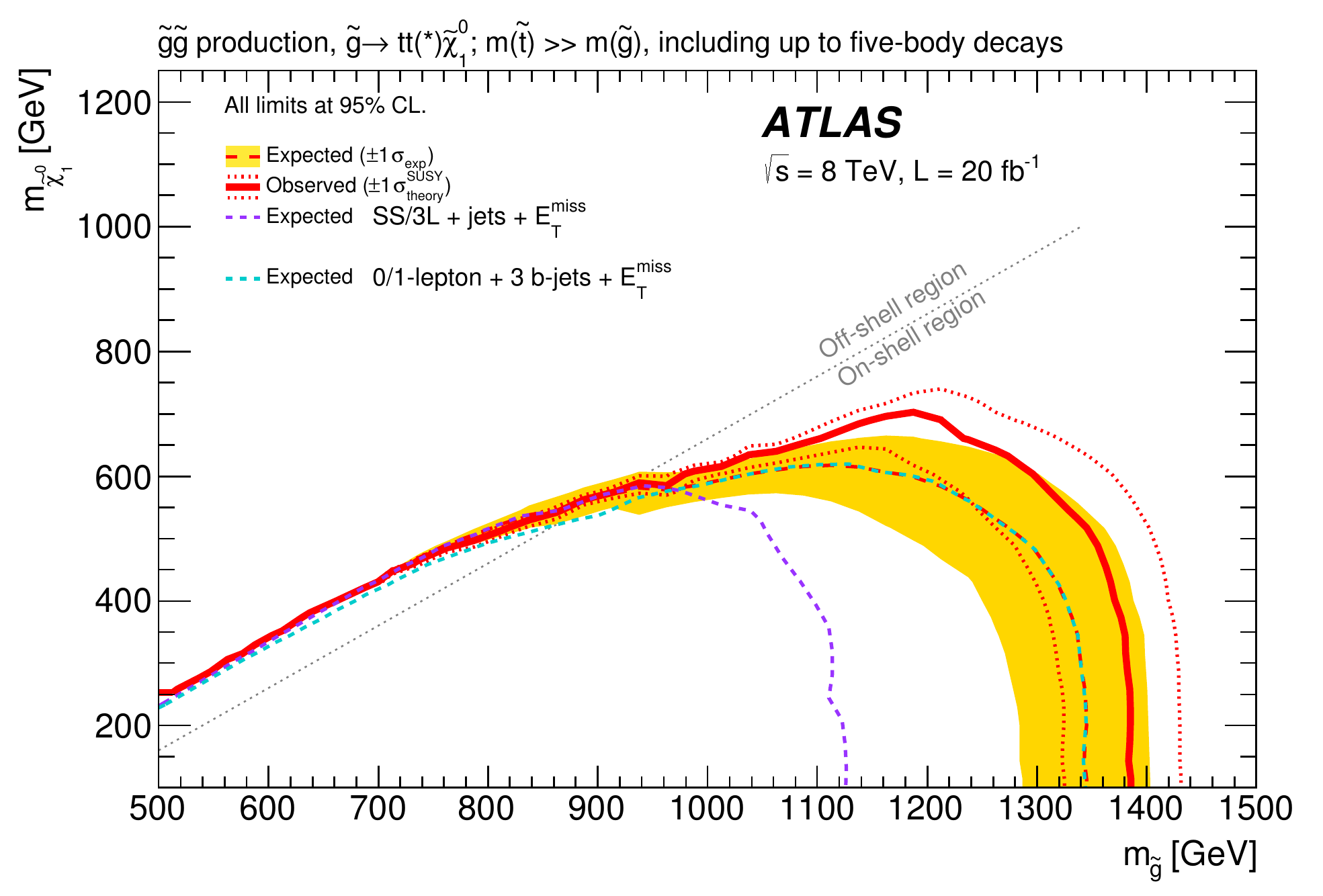}\caption{}\label{fig:strategy.pheno.run1_gluinoGtt}\end{subfigure}
\begin{subfigure}[t]{0.38\textwidth}\includegraphics[width=\textwidth]{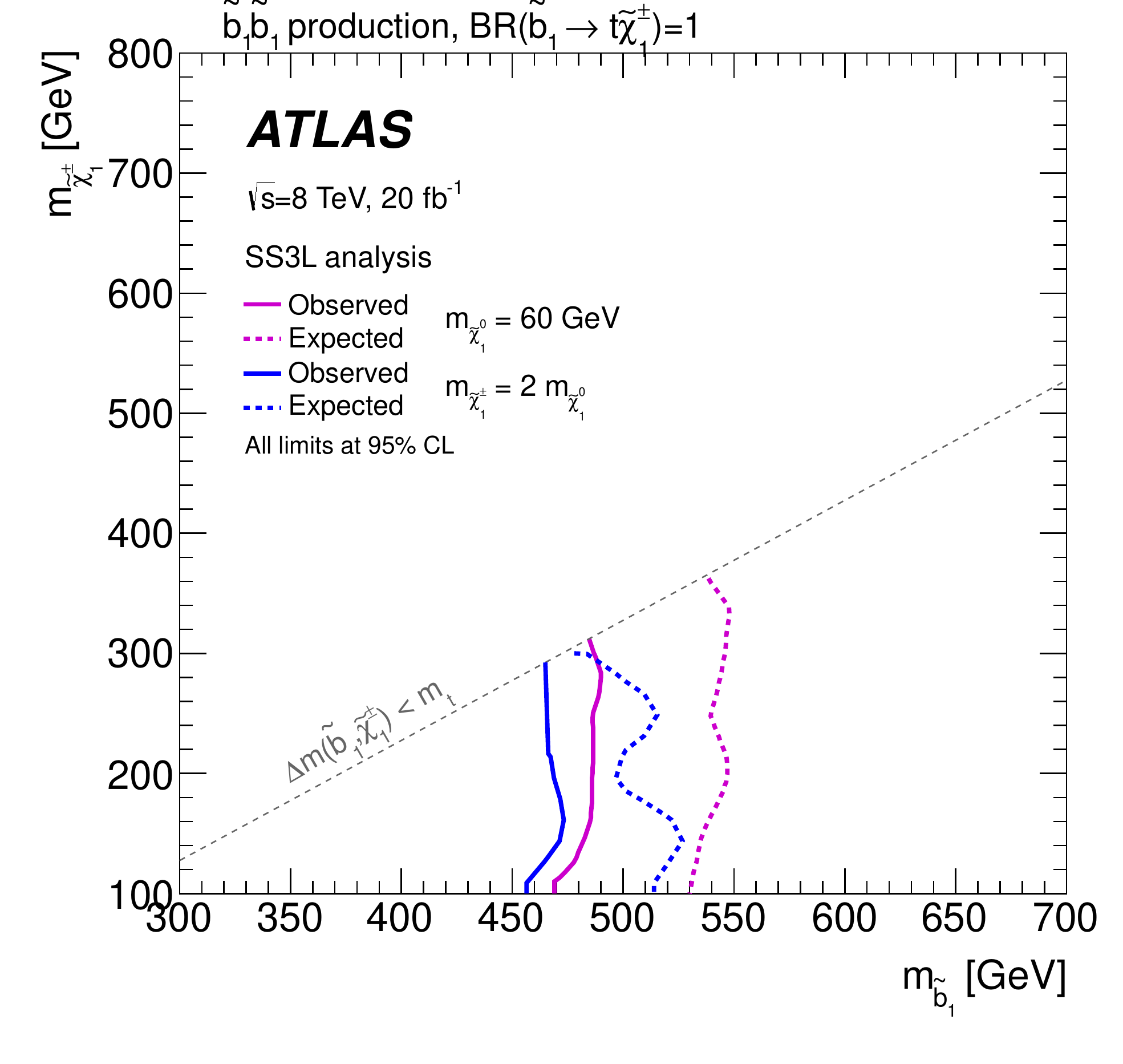}\caption{}\label{fig:strategy.pheno.sbottom_topC1}\end{subfigure}
\caption{Exclusion limits on the gluino-stop offshell~\cite{SUSY-2014-06} (left) and direct sbottom~\cite{SUSY-2014-07} (right) scenarios 
set by ATLAS with the 2012 dataset prior to the author's work.}
\label{fig:strategy.pheno.run1excl_3rdgen}
\end{figure}

\begin{figure}[t]
\centering
\begin{subfigure}[t]{0.49\textwidth}\includegraphics[width=\textwidth]{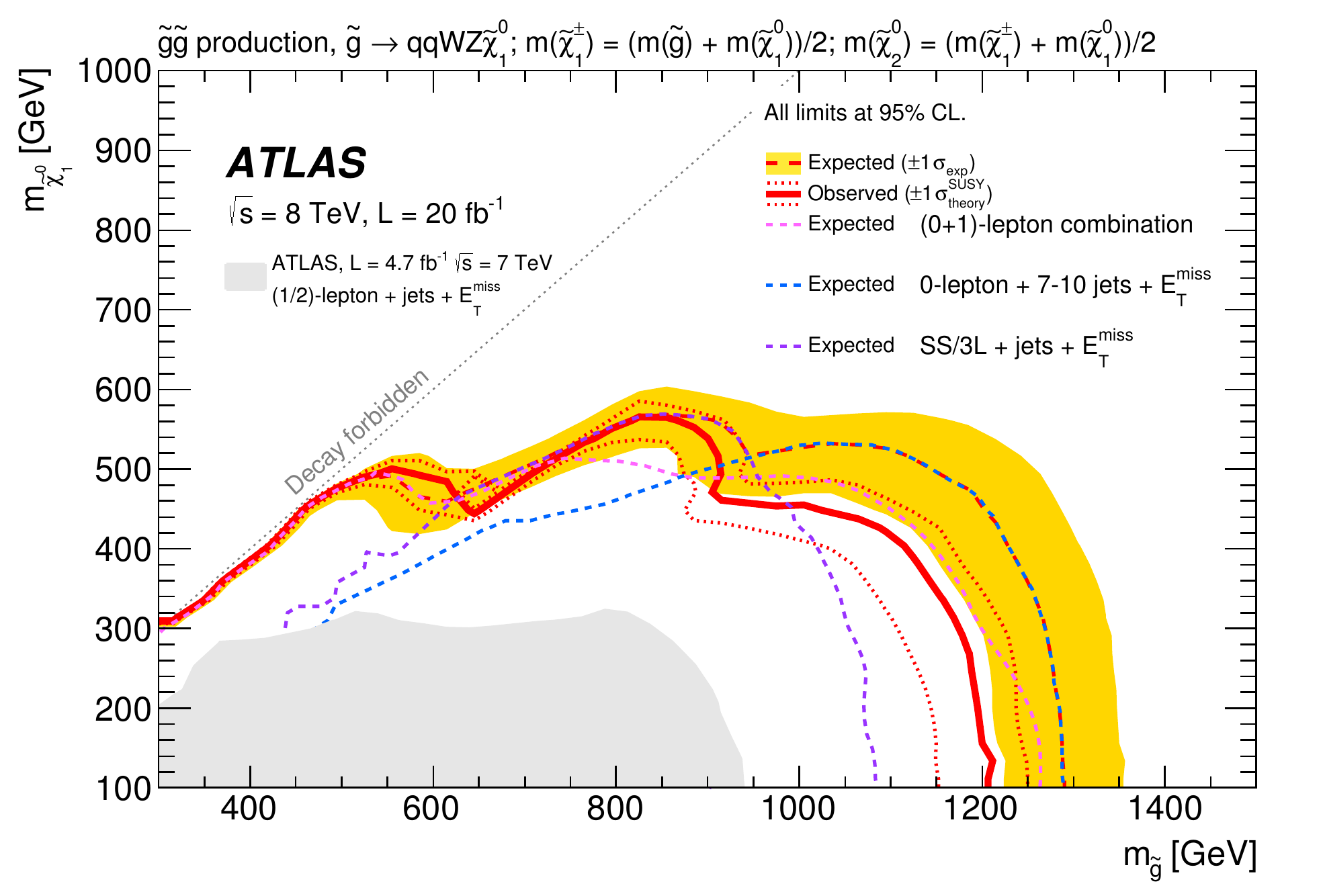}\caption{}\label{fig:strategy.pheno.run1excluded_gluino2stepWZ}\end{subfigure} 
\begin{subfigure}[t]{0.49\textwidth}\includegraphics[width=\textwidth]{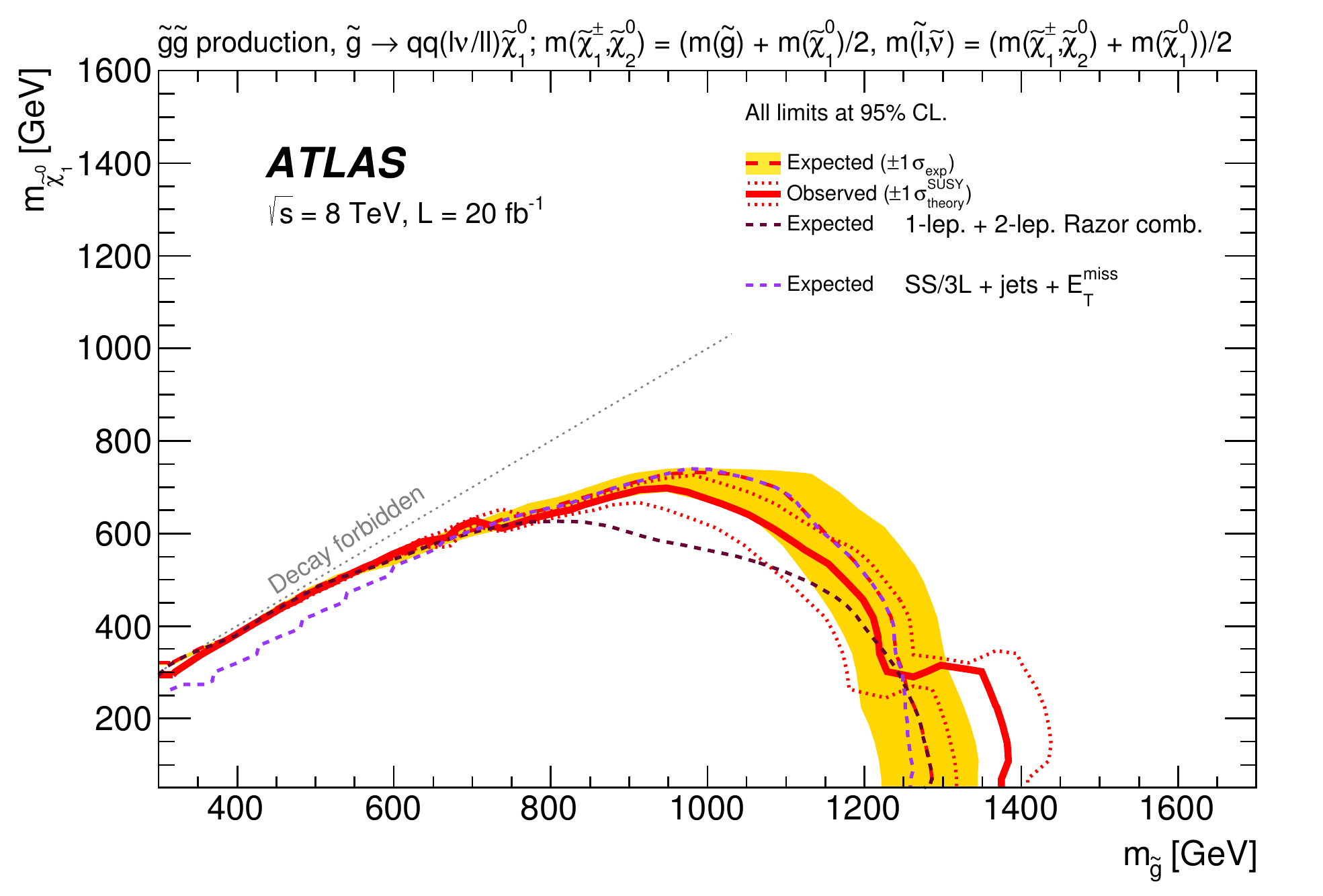}\caption{}\label{fig:strategy.pheno.run1excluded_gluino2stepSleptons}\end{subfigure}
\caption{Exclusion limits on scenarios featuring gluino pair production followed by two-step decays via heavy gauge bosons or sleptons 
set by ATLAS with the 2012 dataset prior to the author's work~\cite{SUSY-2014-06}.}
\label{fig:strategy.pheno.run1excluded_1stgen}
\end{figure}

\subsection*{Gluino pair production with slepton-mediated two-step decay $\gl\to q\bar q\ell\bar\ell\neut$}
\label{subsec:signals_g2slep}

This scenario (Figure~\ref{fig:strategy.pheno.feynman_gg2sl}) features gluino pair-production with two-step decays via neutralinos \neuttwo\ and sleptons, 
$\gl\to q\bar{q}'\neuttwo \to q\bar{q}'(\slep\ell/\snu\nu) \to q\bar{q}'(\ell\ell/\nu\nu)\neut$. 
The decays are mediated by generic heavy squarks, therefore the $b$-jet multiplicity in this scenario is low. 
The final state is made of charged leptons, four additional jets and invisible particles (neutrinos and neutralinos). 
The average jet multiplicity per event is the smallest among the four scenarios;  
another characteristic is the large fraction of events with several leptons, 
unlike the other scenarios that have a rather low acceptance due to the branching ratios of $W\to\ell\nu$ or $Z\to\ell\ell$. 
The exclusion limits obtained in Run 1 (Figure~\ref{fig:strategy.pheno.run1excluded_gluino2stepSleptons}) show again that the SS/3L+jets final state 
is very competitive to probe those models. 
This scenario is used as as benchmark to define the signal regions with $\ge 3$ leptons and no $b$-jet. 

The signal grid is built with variable gluino and \neut\ masses; the \neuttwo\ mass is chosen half-way between the gluino and LSP masses, 
and the sleptons masses are also set equal and half-way between the \neuttwo\ and LSP masses. 
The \neuttwo\ may decay to any of the six ``left-handed'' sleptons (\slep, \snu) with equal probability. 
``Right-handed'' sleptons are assumed heavy and do not participate to the decay. 

\subsection*{Gluino pair production with gaugino-mediated two-step decay $\gl\to q\bar q'WZ\neut$}
\label{subsec:signals_g2wz}

This scenario (Figure~\ref{fig:strategy.pheno.feynman_gg2WZ}) features gluino pair-production with two-step decays via gauginos and $W$ and $Z$ bosons, 
$\gl\to q\bar{q}'\chargino\to q\bar{q}'W\neuttwo\to q\bar{q}'WZ\neut$, 
mediated by generic heavy squarks of the first and second generations. 
The final state is made of two $W$ and two $Z$ bosons (possibly offshell), 
four additional jets and invisible particles (neutrinos and neutralinos). 
This generally leads to events with large jet multiplicities and a fair branching ratio for dileptonic final states. 
The exclusion limits obtained in Run 1 indeed illustrate the competitiveness of the SS/3L+jets search (Figure~\ref{fig:strategy.pheno.run1excluded_gluino2stepWZ})
particularly the heavy-\neut\ region of the phase space. 
This scenario is used as as benchmark to define the signal regions with many jets but none tagged as a $b$-jet. 

The signal grid is built with variable gluino and \neut\ masses, 
and the \chargino\ and \neuttwo\ masses are set such that the former lies half-way between the gluino and \neut\ masses, 
and the latter half-way between \chargino\ and \neut\ masses. 

\subsection*{Sbottom pair production with one-step decay $\sbot\to t\chargino$}
\label{subsec:signals_sbot}

In this scenario (Figure~\ref{fig:strategy.pheno.feynman_b1b1}), sbottoms are rather light and assumed to decay to a top quark and a chargino $\chargino$, 
with a subsequent $\chargino\to W^\pm\neut$ decay, 
providing complementarity to the mainstream search~\cite{ATLAS-CONF-2015-066} which focuses on the channel $\sbot\to b\neut$. 
The final state resulting from the production of a \sbsb\ pair contains two top quarks, two $W$ bosons and two neutralinos. 
While this final state may lead to various experimental signatures, 
the only model considered in Run-1~\cite{SUSY-2014-06} had
same-sign leptons and jets in the final state, leading to the exclusion limits presented in Figure~\ref{fig:strategy.pheno.run1excl_3rdgen}. 
Signal events typically contain one or two $b$-tagged jets. 
Therefore this scenario is used as benchmark to define the signal regions with one or more $b$-jets. 

The model adopts a fixed chargino-neutralino mass difference of 100 GeV, 
which always produces on-shell $W$ bosons in the $\chargino\to W\neut$ decay
\footnote{A different chargino mass assumption is adopted in the current 
work compared to the Run 1 paper~\cite{SUSY-2014-06}.
Figure~\ref{fig:strategy.pheno.run1excl_3rdgen} is shown for illustration only.
The reduced chargino-neutralino mass gap in the current analysis 
allows us to study signal scenarios with heavy neutralinos, which were not considered previously.}.
Only pair production of the lightest sbottom is considered, followed by an exclusive decay in the aforementioned channel.

\subsection*{Gluino pair production with stop-mediated decay $\gl\to t\bar t\neut$}
\label{subsec:signals_gtt}

In this scenario inspired by naturalness arguments, gluinos are coupling preferentially to stops which are lighter than the other squarks. 
Gluinos are however considered lighter than stops, and decay directly into a $t\bar t\neut$ triplet via a virtual stop (Figure~\ref{fig:strategy.pheno.feynman_gtt}). 
The pair production of gluinos leads to a final state containing four top quarks and two neutralinos. 
This characteristic final state is accessible through various experimental signatures, which is why this model 
is commonly used as a benchmark to compare analyses' sensitivities. 
The searches performed with Run-1 data~\cite{SUSY-2014-06}, 
summarized in Figure~\ref{fig:strategy.pheno.run1_gluinoGtt}, showed that the same-sign leptons final state is competitive only at large neutralino mass. 
This region of the phase space is consequently given particular attention in the choice of signal regions described further on. 
For instance, the region of phase-space with $\Delta m(\gl,\neut)<2m_t$, where gluinos decay via one or two offshell top quarks, is only accessible for this 
analysis.
In the signal samples referenced in this document, the mass of the lightest stop is fixed to 10 \TeV~and is mostly a $\widetilde{t}_R$ state. 
Only gluino pair production is considered, followed by an exclusive decay in the aforementioned channel. 
Signal events typically contain many $b$-tagged jets, 
therefore this scenario is used as benchmark to define the signal regions with $\ge 2$ $b$-jets. 

\subsection*{\stst\ with ``three-same-sign leptons'' signature}
\label{subsec:signals_3lss}

Inspired by Ref.~\cite{Huang:2015fba}, a simplified model featuring a stop pair-production with two-step 
decays via a neutralino \neuttwo\ and a chargino $\chargino$ is added in this version of the analysis, according to the decay illustrated on 
Figure~\ref{fig:strategy.pheno.feynman_t1t1}: \\
$\stop_1\to t \neuttwo \to t \chargino W^\mp \to t W^\pm W^\mp \neut$. 

This simplified model is a well-motivated representation of a MSSM model. 
The lightest stop ($\stop_1$) is right-handed and \neuttwo\ is bino-like 
which leads to a large branching ratio in the decay $\stop_1\to t \neuttwo$. 
Furthermore, the decay $\neuttwo \to \chargino W^\mp$ is also enhanced since $\chargino$ is wino-like, 
as long as $\chargino$ and \neut~ are nearly mass degenerate 
and $m_{\neuttwo} - m_{\neut} < m_{H} = 125$ \GeV~to suppress the decay $\neuttwo \to \neut + H$ 
(the decay $\neuttwo \to \neut + Z$ is suppressed).
By respecting these conditions and evading the bottom squark limit shown in Figure~\ref{fig:strategy.pheno.sbottom_topC1}, we consider
 a one-dimensional grid with a $\stop_1$ mass varying between 550 \GeV~and 800 \GeV~with a 50 \GeV~gap\footnote{Only the points at $\stop_1$ mass of 550~GeV~are available at the moment.}, 
a two body decay to an on-shell top quark and a \neuttwo~ which has a 100 \GeV~mass difference from \neut.
The mass difference between the $\chargino$ and \neut~ is taken to be 500 \MeV~which is not excluded by the disappearing track 
analysis. In fact, this mass gap could easily be increased by introducing a small amount of higgsino mixing~\cite{Aad:2013di}.

While the stop pair production is similar to the sbottom pair production in terms of kinematics, the stop pair production offers 
a unique topology that leads to three leptons of the same electric charge. This final state benefits from an extreme reduction of 
the SM background while maintaining a good signal acceptance which helps loosen the kinematic cuts to access a more compressed 
SUSY phase space. As a result, this scenario is complementary to the search for sbottoms.

\subsection*{Non-Universal Higgs Models}
\label{subsec:signals_nuhm2}

In references~\cite{Baer:2013xua,Baer:2013yha,Baer:2016usl}, 
theorists studied a complete two-extra-parameter non-universal Higgs model (NUHM2) 
that can have low fine tuning (natural) and
predicts final state signatures that allow large background rejection while retaining high 
signal efficiency. 
The NUHM2 model allows the soft SUSY breaking masses of the Higgs multiplets, $m_{H_{u}}$ and $m_{H_{d}}$, to be different from 
matter scalar masses ($m_{0}$) at the grand unification scale. The NUHM2 model is expected to form the effective theory for energies 
lower than $m_\textrm{GUT}$ resulting from SO(10) grand unified theories.
The scalar mass $m_{0}$, the soft SUSY breaking gaugino mass $m_{1/2}$, the pseudoscalar Higgs boson mass $m_{A}$, the trilinear SUSY breaking parameter $A_{0}$, the weak scale ratio of Higgs field vacuum expectation values $\tan\beta$, and the superpotential Higgs mass $\mu$ are the free parameters.
Both $m_{1/2}$ and $\mu$ are varied while the other parameters are fixed to $m_{0} = 5$ TeV, $A_{0} = -1.6m_{0}$, $\tan\beta = 15$, $m_{A} = 1$ TeV, and sign($\mu$)$>$0. 
These parameter choices lead directly to a Higgs mass of 125~GeV~in accord with experiment.  In this ``radiatively-driven natural'' SUSY approach, the higgsino is
required to have a mass below 200-300~\GeV, the stop to have a mass below
$\sim$3~\TeV, and the gluino below $\sim$4~TeV.
The model mainly involves gluino pair production with gluinos decaying 
predominantly to $\ttbar\ninoone$ and $tb\chinoonepm$, giving rise to final 
states with two same-sign leptons and \met.
Table~\ref{tab:NUHM2} shows the branching ratios of the dominant gluino decay modes for $m_{1/2} = 400$ \GeV.
Simulated NUHM2 signal samples with mass $(m_{1/2})$ values from 300-800 GeV~and $\mu = 150$ GeV~were generated where 
the gluino mass in this model is approximately $2.5\times m_{1/2}$.

\begin{table}[t!]
\begin{center}
\begin{tabular}{|c|c||c|c|}
\hline
\hline
Decay & BR & Decay & BR\\
\hline
$t\bar{t}\chi^{0}_{1}$ & 0.13 & $tb\chi^{\pm}_{1}$ & 0.45\\
$t\bar{t}\chi^{0}_{2}$ & 0.21 & $tb\chi^{\pm}_{2}$ & 0.04\\
$t\bar{t}\chi^{0}_{3}$ & 0.13 & - & - \\
$t\bar{t}\chi^{0}_{4}$ & 0.02 & - & - \\
\hline
$t\bar{t}\chi^{0}_{i}$ & 0.49 & $tb\chi^{\pm}_{i}$ & 0.49\\
\hline
\hline
\end{tabular}
\caption{The dominant gluino decay modes for $m_{1/2} = 400$ GeV~for the NUHM2 model.}
\label{tab:NUHM2}
\end{center}
\end{table}

%% file: texfiles/sec.strategy.samples.tex
\subsection{Collision Data}
The analysis uses $pp$--collisions data at $\sqrt s=13$ \TeV~ 
collected by the ATLAS detector during 2015 and 2016
 with a peak instantaneous luminosity of 
$L=1.4\times~10^{34}$~cm$^{-2}$s$^{-1}$.
The total integrated luminosity considered corresponds to 36.1 \ifb~ 
(3.2 \ifb~in 2015 and 32.9 \ifb~in 2016) recorded 
after applying beam, detector, and data-quality requirements.
The combined luminosity uncertainty for 2015 and 2016 is 3.2\%, 
assuming partially correlated uncertainties in 2015 and 2016.
The integrated luminosity was established following the same methodology as 
that detailed in Ref.~\cite{Aaboud:2016hhf},
from a preliminary calibration of the luminosity scale using a pair of $x$-$y$ 
beam separation scans.

\subsection{Simulated Event Samples}

Monte Carlo (MC) simulated event samples are used to model the SUSY signal 
and SM backgrounds. 
The irreducible SM backgrounds refer to processes that lead to two 
same-sign and/or three ``prompt'' leptons where the prompt leptons
are produced directly in the hard-scattering process, 
or in the subsequent decays 
of $W,Z,H$ bosons or prompt $\tau$ leptons. 
The reducible backgrounds, mainly 
arising from $\ttbar$ and $V$+jets production, are estimated either from data 
or from MC simulation as described in Section~\ref{chap:fake}. 

Table~\ref{tab:MC} presents the event generator, parton shower, cross-section 
normalization, PDF
set and the set of tuned parameters for the modelling of the parton shower, 
hadronization and underlying event. 
Apart from the MC samples produced by the \SHERPA generator, all MC samples
used the \textsc{EvtGen}~v1.2.0 program~\cite{EvtGen} 
to model the properties of bottom and charm hadron decays. 

\begin{table*}[!ht]
\begin{center}
\Large
\resizebox{\textwidth}{!}
{
\begin{tabular}{|l|l|c|c|c|c|c|}
\hline
Physics process    & Event generator & Parton shower & Cross-section & Cross-section & PDF set & Set of tuned \\
                   &          &               & order & value (fb)&         & parameters  \\
\hline
\hline
Signal                  & \AMCATNLO 2.2.3~\cite{Alwall:2014hca}         & \PYTHIA 8.186~\cite{Sjostrand:2007gs} & NLO+NLL
& See Table~\ref{tab:signal_xsections} & NNPDF2.3LO~\cite{Ball:2012cx} & A14~\cite{ATL-PHYS-PUB-2014-021} \\
\hline
$\ttbar +X$            &                                        &                                       & 
             &               &    &   \\
$\ttbar W$,$\ttbar Z/\gamma^{*}$ & \AMCATNLO 2.2.2            & \PYTHIA 8.186                         & NLO~\cite{YR4} & $600.8$, $123.7$                     & NNPDF2.3LO    & A14    \\
$\ttbar H$	   & \AMCATNLO 2.3.2        			& \PYTHIA 8.186  			& NLO~\cite{YR4}  & 	$507.1$		& NNPDF2.3LO	& A14  \\
4$t$    	& \AMCATNLO 2.2.2       			& \PYTHIA 8.186        			& NLO~\cite{Alwall:2014hca}	&  
$9.2$	& NNPDF2.3LO	& A14  \\
\hline
Diboson            &                   &   			&                      			&                               	&               &      \\
$ZZ$, $WZ$       & \SHERPA 2.2.1~\cite{gleisberg:2008ta}      & \SHERPA 2.2.1& NLO~\cite{ATL-PHYS-PUB-2016-002}& 
$1.3\cdot 10^3$,$4.5\cdot 10^3$
&NNPDF2.3LO & \SHERPA default \\
inc. $W^{\pm}W^{\pm}$   & \SHERPA 2.1.1 		& \SHERPA 2.1.1				& NLO~\cite{ATL-PHYS-PUB-2016-002}  &	
$86$
&CT10~\cite{Lai:2010vv} & \SHERPA default \\
\hline
Rare               &                  &    			&                      			&                               	&               &      \\
$\ttbar WW$, $\ttbar WZ$     & \AMCATNLO 2.2.2       & \PYTHIA 8.186      & NLO~\cite{Alwall:2014hca}  & 
$9.9$, $0.36$
& NNPDF2.3LO & A14  \\
$tZ$, $tWZ$, $t\ttbar$    & \AMCATNLO 2.2.2        & \PYTHIA 8.186       & LO                   & 
$240$, $16$, $1.6$
& NNPDF2.3LO     & A14  \\
$WH$, $ZH$   & \AMCATNLO 2.2.2        & \PYTHIA 8.186      & NLO~\cite{Dittmaier:2012vm}   & 
$1.4\cdot 10^3$, $868$  
& NNPDF2.3LO     & A14  \\
Triboson	   & \SHERPA 2.1.1         			& \SHERPA 2.1.1        			& NLO~\cite{ATL-PHYS-PUB-2016-002}
& $14.9$
& CT10	     	& \SHERPA default \\
\hline
Irreducible (Incl.)            &                      			&                      		&	&                               	&               &      \\
$W$+Jets      & \POWHEGBOX       		& \PYTHIA 8.186      			& NNLO	 & 
2.0 $\cdot 10^7$	& CT10      	 & AZNLO\cite{AZNLO:2014}\\
$Z$+Jets      & \POWHEGBOX       		& \PYTHIA 8.186      			& NNLO	 & 1.9 $\cdot 10^7$	& CT10      	 & AZNLO\cite{AZNLO:2014}\\
$\ttbar$    	   & \POWHEGBOX       		& \PYTHIA 6.428      			& NNLO+NNLL~\cite{Czakon:2011xx}	&  
8.3 $\cdot 10^5$ 	& CT10      	 & PERUGIA2012 (P2012) \cite{tt:perugia}\\
\hline
\end{tabular}
}
\caption{Simulated signal and background event samples: the corresponding event generator, parton shower, cross-section normalization, PDF set and 
set of tuned parameters are shown for each sample. Because of their very small contribution to the signal-region background estimate, 
$\ttbar WW$, $\ttbar WZ$, $tZ$, $tWZ$, $t\ttbar$, $WH$, $ZH$ and triboson are summed and labelled ``rare''.}
\label{tab:MC}
\end{center}
\end{table*}

The MC samples were processed through either a full ATLAS detector 
simulation~\cite{Aad:2010ah} based on 
\textsc{Geant4}~\cite{Agostinelli:2002hh} or a fast simulation using a 
parameterization of the calorimeter response 
and \textsc{Geant4} for the inner detector and muon spectrometer
~\cite{ATL-PHYS-PUB-2010-013},
and are reconstructed in the same manner as the data. 
All simulated samples are generated with a range of minimum-bias interactions 
using {\sc Pythia 8}~\cite{Sjostrand:2007gs} 
with the MSTW2008LO PDF set~\cite{Sherstnev:2007nd} and the A2 tune overlaid on the hard-scattering event 
to account for the multiple $pp$ interactions in the same bunch crossing 
(in-time pileup) 
and neighbouring bunch crossing (out-of-time pileup). 
The distribution of the average number of interactions per bunch crossing 
$\langle\mu\rangle$ ranges from 0.5 to 39.5, 
with a profile set as an estimate of the combined 2015+2016 data 
$\langle\mu\rangle$ profile. 
With larger luminosity collected during this year and the $\mu$ distribution in data being closer to that in the MC profile,
the simulated samples are re-weighted to reproduce the observed distribution 
of the average number of collisions per bunch crossing ($\mu$).

\subsection*{Background process simulation}

The two dominant irreducible background processes are $\ttbar V$ (with $V$ being a $W$ or $Z/\gamma^*$ boson) 
and diboson production with final states of four charged leptons $\ell$,\footnote{All lepton flavours are included here and $\tau$
leptons subsequently decay leptonically or hadronically.} three charged leptons and one neutrino, or 
two same-sign charged leptons and two neutrinos. 

The production of a $\ttbar V$ 
constitutes the main source of background with prompt same-sign leptons for event selections including $b$-jets. 
Simulated events for these processes were generated at NLO with \AMCATNLO v2.2.2~\cite{Alwall:2014hca} interfaced to \textsc{Pythia} 8,
with up to two ($ttW$) or one ($ttZ^{(*)}$) extra parton included in the matrix elements~\cite{ATL-PHYS-PUB-2016-005}. 
The samples are normalised to the inclusive process NLO cross-section using appropriate $k$-factors~\cite{Alwall:2014hca}.

The production of multiple $W,Z$ bosons decaying leptonically 
constitutes the main source of background with prompt same-sign leptons for event selections vetoing $b$-jets. 
Diboson processes with four charged leptons, three charged leptons and one neutrino, or two charged leptons and two neutrinos 
were simulated at NLO using the \textsc{Sherpa} 2.2.1 generator~\cite{gleisberg:2008ta}, as described in detail in Ref.~\cite{ATL-PHYS-PUB-2016-002}. 
The main samples simulate $qq \to VV\to\text{leptons}$ production including the doubly resonant $WZ$ and $ZZ$ processes, 
non-resonant contributions as well as Higgs-mediated contributions, and their interferences; 
up to three extra partons were included (at LO) in the matrix elements. 
Simulated events for the $W^\pm W^\pm jj$ process (including non-resonant contributions) were produced at LO with up to one extra parton, 
separately for QCD-induced $\left(\mathcal{O}(\alpha_\text{em}^4)\right)$ 
and VBS-induced $\left(\mathcal{O}(\alpha_\text{em}^6)\right)$ production -- the interferences being neglected. 
Additional samples for VBS-induced $qq\to 3\ell\nu jj$ and $qq\to 4\ell$ and loop-induced $gg\to WZ^{(*)}/ZZ^{(*)}$ processes
were also produced with the same configuration.
The samples generated at NLO are directly normalized to the cross-sections provided by the generator. 

Production of a Higgs boson in association with a $\ttbar$ pair is simulated using \AMCATNLO~\cite{Alwall:2014hca} 
(in \MADGRAPH v2.2.2) interfaced to \HERWIG 2.7.1~\cite{Corcella:2000bw}.  
The UEEE5 underlying-event tune is used together with the CTEQ6L1~\cite{Pumplin:2002vw} (matrix element) and CT10~\cite{Lai:2010vv} (parton shower) PDF sets.
Simulated samples of SM Higgs boson production in association with a $W$ or $Z$ boson are produced with \PYTHIA 8.186, using the \textsc{A14} tune and the \textsc{NNPDF23LO} PDF set. Events are normalised with cross-sections calculated at NLO~\cite{Dittmaier:2012vm}.

\MADGRAPH v2.2.2~\cite{Alwall:2011uj} is used to simulate the $t\bar{t}WW$, $tZ$, $\ttbar\ttbar$ and $\ttbar t$ processes, and the generator cross-section is used for $tZ$ and $\ttbar t$. \MADGRAPH interfaced to \textsc{Pythia} 8 is used to generate $t\bar{t} WZ$ processes, and appropriate $k$-factors are taken from~\cite{Alwall:2014hca}. \AMCATNLO interfaced to \PYTHIA 8 is used for the generation of the $tWZ$ process, with an alternative sample generated with \AMCATNLO interfaced to \HERWIG used to evaluate the parton shower uncertainty.  
Fully leptonic triboson processes ($WWW$, $WWZ$, $WZZ$ and $ZZZ$) with up to six charged leptons are simulated using \SHERPA~v2.1.1 
and described in Ref.~\cite{ATL-PHYS-PUB-2016-002}. 
The $4\ell$ and $2\ell+2\nu$ processes are calculated at next-to-leading order (NLO) for up to one additional parton; 
final states with two and three additional partons are calculated at leading order (LO). 
The $WWZ\to 4\ell+2\nu$ or $2\ell+4\nu$ processes are calculated at LO with up to two additional partons. 
The $WWW/WZZ\to 3\ell+3\nu$, $WZZ\to 5\ell+1\nu$, $ZZZ\to 6\ell+0\nu$, $4\ell+2\nu$ or $2\ell+4\nu$ processes 
are calculated at NLO with up to two extra partons at LO. 
The CT10~\cite{Lai:2010vv} parton distribution function (PDF) set is used for all \SHERPA samples in conjunction with 
a dedicated tuning of the parton shower parameters developed by the \SHERPA authors. 
The generator cross-sections (at NLO for most of the processes) are used when normalising these backgrounds.

Double parton scattering (DPS) occurs when two partons interact simultaneously in a proton-proton collision leading to two hard scattering 
processes overlapping in a detector event. 
Accordingly, two single $W$ production processes can lead to a $W^\pm$ + $W^\pm$ final state via DPS. 
This background is expected to have a negligible contribution to signal regions with high jet multiplicities.
To estimate a conservative upper bound on cross-section for $WW$ events which might arise from DPS, a standard ansatz is adopted: 
in this, for a collision in which a hard process (X) occurs, the probability that 
an additional (distinguishable) process (Y) occurs is parametrized as:
\begin{equation}
\sigma^{DPS}_{XY} = \sigma^{}_{X}\sigma^{}_{Y}/\sigma^{}_\text{eff}
\end{equation} 
where $\sigma^{}_{X}$ is the production cross section of the hard 
process X and $\sigma^{}_\text{eff}$ (effective area parameter) 
parametrizes the double-parton interaction part of the production 
cross section for the composite system (X+Y). 
A value of $\sigma^{}_\text{eff} = 10-20$ mb is assumed in this study (as obtained from 7~TeV~measurements, and with no observed dependence on $\sqrt{s}$), and it is independent on the processes involved. For the case of $W^\pm+W^\pm$ production:
\begin{equation}
\sigma^{DPS}_{W^\pm W^\pm} = \frac{ \sigma^{}_{W^+}\sigma^{}_{W^+} + \sigma^{}_{W^-}\sigma^{}_{W^-} + 2\sigma^{}_{W^+}\sigma^{}_{W^-}}{\sigma^{}_\text{eff} } \simeq 0.19-0.38\text{ pb.}
\end{equation} 

After the application of the SR criteria, only 4 raw MC events in the DPS $WW\to\ell\nu\ell\nu$ remain. 
Table~\ref{tab:DPS_SR} shows the expected contribution in the SRs where some MC event survives all the cuts. The ranges quoted in the tables reflect the range in the predicted $\sigma^{DPS}_{W^\pm W^\pm}$ cross-section above, as well as the combinatorics for scaling the jet multiplicity\footnote{For instance, a DPS event with 6 jets can be due to the overlap of two events with 6+0 jets, or 5+1, 4+2 or 3+3 jets. All possible combinations are considered and the range quoted in the table shows the combinations leading to the smallest and largest correction factors.}.
Due to the large uncertainties involved in these estimates, some of them difficult to quantify (such as the modelling of DPS by {\sc Pythia} at LO), the contribution from this background is not included in the final SR background estimates. 
Note that the estimated DPS contribution is typically much smaller than the uncertainty on the total background for the SRs.

\begin{table}[!htb]
\caption{Number of raw MC events and its equivalent for 36.1 \ifb with and without the correction as a function of the jet multiplicity. 
Only the SRs where at least one MC event passes all the cuts are shown.}
\label{tab:DPS_SR}
\centering
\begin{tabular}{l|c|c|c}
\hline
SR       & Raw MC events & Without $N_{\text{jet}}$ correction & With $N_{\text{jet}}$ correction \\\hline
Rpc2L0bS & 2 & 0.016-0.033 & 0.09-0.38 \\ 
Rpc2L0bH & 1 & 0.006-0.012 & 0.05-0.17 \\ 
\hline
\end{tabular}
\end{table}

\subsection*{Signal cross-sections and simulations}

\begin{table}[t!]
\centering
\caption{Signal cross-sections [pb] and related uncertainties [\%] for scenarios featuring \glgl\ (top table) or \sbsb\  (bottom table) production, 
as a function of the pair-produced superpartner mass, reproduced from Ref.~\cite{twiki-SusyCrossSections}.}
\label{tab:signal_xsections}
\resizebox{\textwidth}{!}{
\begin{tabular}{|c|c|c|c|c|c|c|c|c|c|c|}
\hline\hline
Gluino mass (GeV) & 500 & 550 & 600 & 650 & 700 \\\hline
Cross section (pb) & $27.4 \pm 14\%$ & $15.6 \pm 14\%$ & $9.20 \pm 14\%$ & $5.60 \pm 14\%$ & $3.53 \pm 14\%$\\\hline\hline
750 & 800 & 850 & 900 & 950 & 1000\\\hline
$2.27 \pm 14\%$ & $1.49 \pm 15\%$ & $0.996 \pm 15\%$ & $0.677 \pm 16\%$ & $0.466 \pm 16\%$ & $0.325 \pm 17\%$\\\hline\hline
1050 & 1100 & 1150 & 1200 & 1250 & 1300\\\hline
$0.229 \pm 17\%$ & $0.163 \pm 18\%$ & $0.118 \pm 18\%$ & $0.0856 \pm 18\%$ & $0.0627 \pm 19\%$ & $0.0461 \pm 20\%$\\\hline\hline
1350 & 1400 & 1450 & 1500 & 1550 & 1600\\\hline
$0.0340 \pm 20\%$ & $0.0253 \pm 21\%$ & $0.0189 \pm 22\%$ & $0.0142 \pm 23\%$ & $0.0107 \pm 23\%$ & $0.00810 \pm 24\%$\\\hline\hline
\end{tabular}}

\resizebox{0.8\textwidth}{!}{
\begin{tabular}{|c|c|c|c|c|}
\hline\hline
Sbottom mass (GeV) & 400 & 450 & 500 & 550 \\\hline
Cross section (pb) & $1.84 \pm 14\%$ & $0.948 \pm 13\%$ & $0.518 \pm 13\%$ & $0.296 \pm 13\%$\\\hline\hline
600 & 650 & 700 & 750 & 800\\\hline
$0.175 \pm 13\%$ & $0.107 \pm 13\%$ & $0.0670 \pm 13\%$ & $0.0431 \pm 14\%$ & $0.0283 \pm 14\%$\\\hline\hline
\end{tabular}}
\end{table}

The signal processes are generated from leading order (LO) matrix elements with up to two extra partons (only one for the grid featuring slepton-mediated gluino decays), 
using the \textsc{Madgraph v5.2.2.3} generator~\cite{Alwall:2014hca} interfaced to \textsc{Pythia} 8.186~\cite{Sjostrand:2007gs} 
with the \textit{ATLAS 14} tune~\cite{ATL-PHYS-PUB-2014-021} for the modelling of the SUSY decay chain, parton showering, 
hadronization and the description of the underlying event. 
Parton luminosities are provided by the \textsc{NNPDF23LO}~\cite{Carrazza:2013axa} set of parton distribution functions. 
Jet-parton matching is realized following the CKKW-L prescription~\cite{Lonnblad:2011xx}, 
with a matching scale set to one quarter of the pair-produced superpartner mass. 

The signal samples are normalised to the next-to-next-to-leading order cross-section from Ref.~\cite{twiki-SusyCrossSections} 
including the re-summation of soft gluon emission at next-to-next-to-leading-logarithmic accuracy (NLO+NLL), 
as detailed in Ref.~\cite{Borschensky:2014cia}; 
some of these cross-sections are shown for illustration in Table~\ref{tab:signal_xsections}. 

Cross-section uncertainties are also taken from Ref.~\cite{twiki-SusyCrossSections} as well, 
and include contributions from varied normalization and factorization scales, as well as PDF uncertainties. 
They typically vary between 15 and 25\%. 
Uncertainties on the signal acceptance are not considered since 
 these are generally smaller than the uncertainties on the inclusive production cross-section.

%% file: texfiles/sec.strategy.sel.tex
\subsection{Pre-selection and event cleaning}
\label{subsec:sec.strategy.selection_cleaning}

A sample of two same-sign or three lepton events are selected applying the following criteria:
\begin{itemize}
\item[$\bullet$] \textbf{Jet Cleaning}: 
Events are required to pass a set of cleaning requirements. 
An event is rejected if any pre-selected jets ($|\eta|<4.9$, after 
jet-electron overlap removal) fails the jet quality criteria. 
The cleaning requirements are intended to remove events where significant 
energy was deposited in the calorimeters 
due to instrumental effects such as cosmic rays, beam-induced (non-collision) 
particles, and noise. Around 0.5\% of data events are lost after applying 
this cut.

\item \textbf{Primary Vertex}:
Events are required to have a reconstructed vertex~\cite{ATL-PHYS-PUB-2015-026} 
with at least two associated tracks with $\pt >400$~MeV. The vertex with the largest $\Sigma \pt^2$ of the associated tracks 
is chosen as the primary vertex of the event.
This cut is found to be 100\% efficient.

\item \textbf{Bad Muon Veto}: 
Events containing at least one pre-selected muon satisfying $\sigma(q/p)/|q/p| >$ 0.2 before the overlap removal are rejected. 
Around 0.1\% of data events are removed by this cut.

\item \textbf{Cosmic Muon Veto}: 
Events containing a cosmic muon candidate are rejected. 
Cosmic muon candidates are looked for among pre-selected muons, 
if they fail the requirements $|z_0| <1.0$ mm and $|d_0|<0.2$ mm, 
where the longitudinal and transverse impact parameters $z_0$ and $d_0$ 
are calculated with respect to the primary vertex. 
Up to 6\% of data events are lost at this cleaning cut.

\item \textbf{At least two leptons}: 
Events are required to contain at least two signal leptons 
with $\pt>20$ \GeV for the two leading leptons. 
If the event contains a third signal lepton with $\pt>10$ \GeV the event is regarded as a three-lepton event, otherwise as a two-lepton event. 
The data sample obtained is then divided into three channels depending on the flavor of the two  leptons forming a same-sign pair ($ee$, $\mu\mu$, $e\mu$). 
If more than one same-sign pairs can be built, the one involving the leading 
lepton will be considered for the channel selection. 

\item \textbf{Same-sign}: 
if the event has exactly two leptons, then these two leptons
 have to be of identical electric charge (``same-sign'').
\end{itemize}

The following event variables are also used in the definition of the signal and validation regions in the analysis:
\begin{itemize}
\item The inclusive effective mass \meff~ defined as the scalar sum of
  all the signal leptons \pt , all signal jets \pt\ and \met. 
\end{itemize}

\subsection{Trigger strategy}
\label{subsec:sec.strategy.sel.selection_trigger}
  
Events are selected using a combination of dilepton and $\met$ triggers, the latter being used only for events with $\met>250$ \GeV. 
Since the trigger thresholds have been raised between 2015 and 2016 due to the 
continuous increase of the instantaneous luminosity, 
the dilepton triggers used for: 
\begin{itemize}
\item 2015 data:
logical \texttt{or} of a trigger with two electrons of 12 \GeV, 
with an electron of 17 \GeV~and a muon of 14 \GeV,
with two muons of 18 \GeV~and 8 \GeV. 
\item 2016 data:
logical \texttt{or} of a trigger with two electrons of 17 \GeV, 
with an electron of 17 \GeV~and a muon of 14 \GeV,
with two muons of 22 \GeV~and 8 \GeV. 
\end{itemize}
The $\met$ trigger was also raised from 70 \GeV~to a 100 \GeV~and 110 \GeV.
The trigger-level requirements on $\met$ and the leading and subleading lepton \pt are looser than those applied offline 
to ensure that trigger efficiencies are constant in the relevant phase space.


\par{\bfseries Trigger matching\\}
For events exclusively selected via one or several of the dilepton triggers, 
we require a matching between the online and offline leptons with $\pt>20$ \GeV.
with the exception of the di-muon trigger for which muons with $\pt>10$ \GeV 
are also considered.
In addition, for the di-muon trigger in the 2016 configuration, 
the \pt\ requirement of the leading matched muon is raised to 23 \GeV~
to remain on the trigger efficiency plateau. 

\par{\bfseries Trigger scale factors\\}
The simulated events are corrected for any potential differences in 
the trigger efficiency between data and MC simulation.
Assuming no correlation between the \met\ and dilepton triggers, 
trigger scale factors are applied to MC events which were not selected 
by the \met\ trigger.
These scale factors are computed for each event, considering the combination of fired triggers, the number and flavours of the leptons, 

\subsection{Object definition}
\label{subsec:strategy.sel.obj}

This section presents the definitions of the objects used in the analysis: 
jets, electrons, muons and $\met$ (the taus are not considered).

\subsection*{Jets}
\label{subsec:sec.strategy.sel.objects_jets}

The jets are kept only if they have $p_\mathrm{T}>20$~GeV~and lie 
within $|\eta|<2.8$. 
To mitigate the effects of pileup, the pile-up contribution is subtracted 
from the expected average energy contribution according to the jet area~\cite{Cacciari:2007fd,Aaboud:2017jcu}.
In order to reduce the effects of pile-up, 
a significant fraction of the tracks in jets with $\pt<60$ \GeV and $|\eta|<2.4$ must originate from the primary vertex, 
as defined by the jet vertex tagger (JVT)~\cite{ATLAS-CONF-2014-018}. 
The jet calibration follows the prescription in Ref.~\cite{Aaboud:2017jcu}.

The 70\% efficiency operating point of the MV2c10 algorithm  was chosen which 
corresponds to the
average efficiency for tagging $b$-jets in simulated $\ttbar$ events. 
This efficiency working point was favored by optimisation studies performed in 
simulated signal and background samples.
The rejection factors for light-quark/gluon jets, $c$-quark jets and hadronically decaying $\tau$ leptons in simulated $\ttbar$ events 
are approximately 380, 12 and 54, respectively~\cite{ATL-PHYS-PUB-2015-022,ATL-PHYS-PUB-2016-012}. 
Jets with $|\eta|<2.5$ which satisfy the $b$-tagging and JVT requirements are identified as $b$-jets. 
Correction factors and uncertainties determined from data for the $b$-tagging efficiencies and mis-tag rates
are applied to the simulated samples~\cite{ATL-PHYS-PUB-2015-022}. 

For the data-driven background estimations, two categories of electrons and muons are used: 
``candidate'' and ``signal'' with the latter being a subset of the ``candidate'' leptons satisfying tighter selection criteria. 

\subsection*{Electrons}
\label{subsec:sec.strategy.sel.objects_electrons}

Electron candidates are reconstructed from energy depositions in the 
electromagnetic calorimeter and required to be matched to an 
inner detector track, 
to have $\pT> 10$ \GeV and $|\eta|<2.47$, and to pass the 
``Loose'' likelihood-based electron identification 
requirement~\cite{ATLAS-CONF-2016-024}.
Electrons in the transition region between the barrel and endcap 
electromagnetic calorimeters ($1.37<|\eta|<1.52$) are rejected to reduce 
the contribution from fake/non-prompt electrons. 
The transverse impact parameter $d_0$ 
with respect to the reconstructed primary vertex 
must satisfy $|d_0/\sigma(d_0)|<5$.
This last requirement helps reduce the contribution from charge 
mis-identification. 

Signal electrons are additionally required to pass the ``Medium'' 
likelihood-based identification requirement~\cite{ATLAS-CONF-2016-024}.
Only signal electrons with $|\eta|<2.0$ are considered, 
to reduce the level of charge-flip background.
In addition, signal electrons that are likely to be reconstructed with an incorrect charge assignment are rejected using a few electron cluster and track properties: the track impact parameter, the track curvature significance, the cluster width and the quality of the matching between the cluster and its associated track, both in terms of energy and position. These variables, as well as the electron \pt and $\eta$, are combined into a single classifier using a boosted decision tree (BDT). A selection requirement on the BDT output is chosen to achieve a rejection factor between 7 and 8 for electrons with a wrong charge assignment while selecting properly measured electrons with an efficiency of 97\% (in $Z\rightarrow ee$ MC). 

A multiplicative event weight is applied for each signal electron in MC to the overall event weight 
in order to correct for differences in efficiency between data and MC.

\subsection*{Muons}
\label{subsec:sec.strategy.sel.objects_muons}

Muons candidates are reconstructed from muon spectrometer tracks matched to 
the inner detector tracks in the region $|\eta|<2.5$.
Muon candidates must pass the ``Medium'' identification 
requirements~\cite{Aad:2016jkr} and have  $p_\mathrm{T} > 10\GeV$ and 
$|\eta| < 2.4$. 
Signal muons are required to pass $\vert d_0\vert/\sigma(d_0) < 3$
and $|z_0 \cdot\sin(\theta)|<0.5$ mm.
 
A multiplicative event weight is applied for each selected muon in MC to the overall event weight 
in order to correct for differences in efficiency between data and MC.

\subsection*{Overlap removal}
\label{subsec:sec.strategy.sel.objects_overlap_removal}

According to the above definitions, one single final state object may fall in more than one category, being therefore effectively double-counted. 
For example, one isolated electron is typically reconstructed both as an electron and as a jet. 
A procedure to remove overlaps between final state objects was therefore put in place, and applied on pre-selected objects. 
Any jet within a distance $\Delta R_y \equiv \sqrt{(\Delta y)^2+(\Delta\phi)^2} =$ 0.2 of a lepton candidate is discarded, 
unless the jet is $b$-tagged,\footnote{In this case the $b$-tagging operating point corresponding to an efficiency of 85\% is used.} 
in which case the lepton is discarded since it probably originated from a semileptonic $b$-hadron decay. 
Any remaining lepton within $\Delta R_y \equiv \operatorname{min}\{0.4, 0.1 + 9.6 \GeV/\pt(\ell)\}$ of a jet is discarded. 
In the case of muons, the muon is retained and the jet is discarded if the jet has fewer than three associated tracks. This reduces 
inefficiencies for high-energy muons undergoing significant energy loss in the calorimeter. 

\subsection*{Missing transverse energy}
\label{subsec:sec.strategy.sel.objects_met}

The missing transverse energy (\met) is computed as a negative vector sum of 
the transverse momenta 
of all identified candidate objects (electrons, photons~\cite{Aaboud:2016yuq}, muons and jets) and an additional soft term. 
The soft term is constructed from all tracks associated with the primary vertex but not with any physics object. 
In this way, the $\met$ is adjusted for the best calibration of the jets and the other identified physics objects listed above, 
while maintaining approximate pile-up independence in the soft term~\cite{ATL-PHYS-PUB-2015-027, ATL-PHYS-PUB-2015-023}.

\subsection{Data-MC comparisons}
\label{subsec:sec.strategy.selection_DataMC}

\begin{figure}[htb!]
\centering
{\includegraphics[width=.75\textwidth]{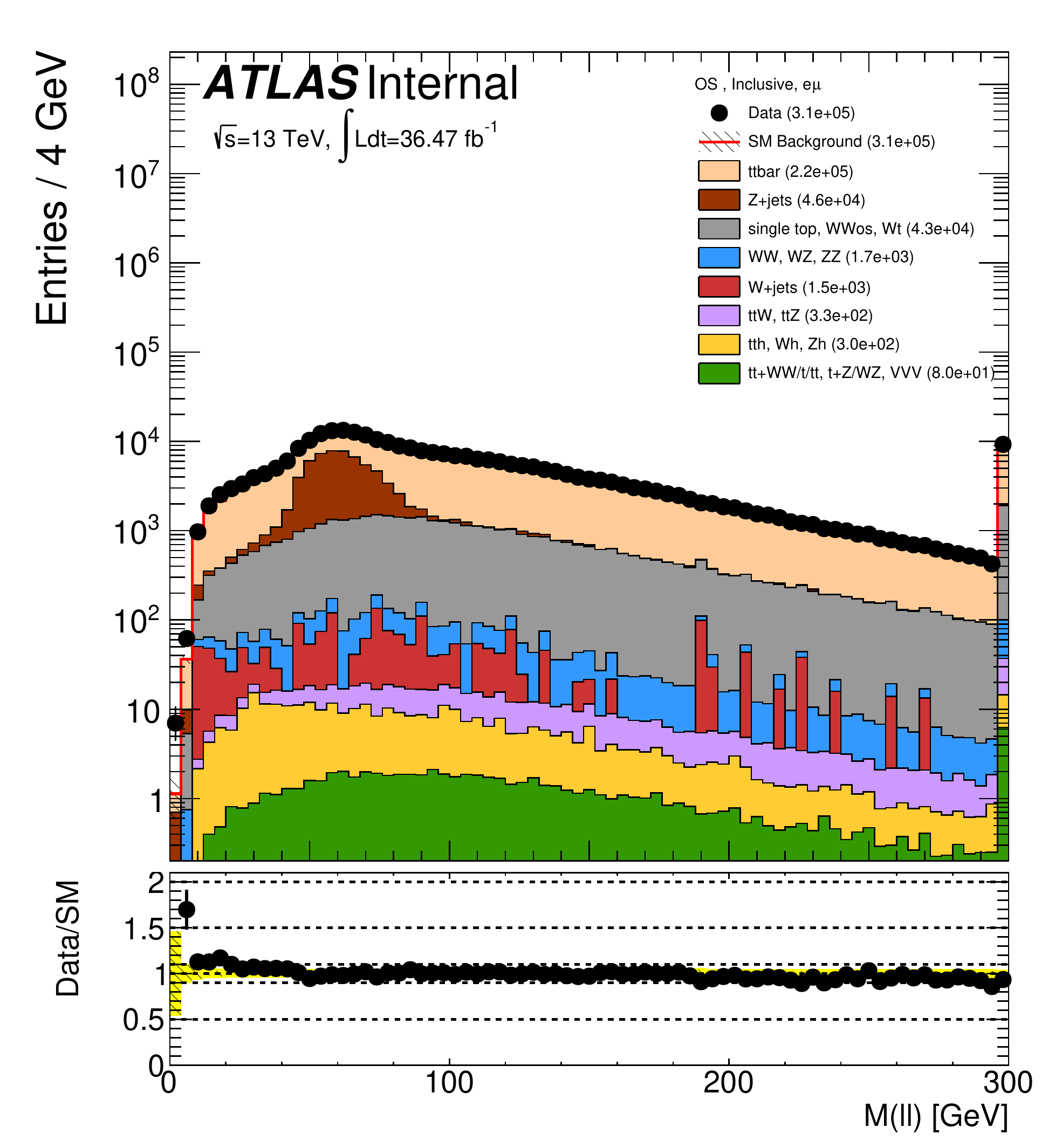}}
\caption{Dilepton invariant mass distributions for opposite-sign pairs for events selected in the $e\mu$ channel. 
No low-mass Drell-Yan sample is included. 
 The prediction is taken from MC only.
Only luminosity and MC statistical uncertainties are included.
}
\label{fig:dataMC_2em.os}
\end{figure}

\begin{figure}[htb!]
\centering
{\includegraphics[width=.49\textwidth]{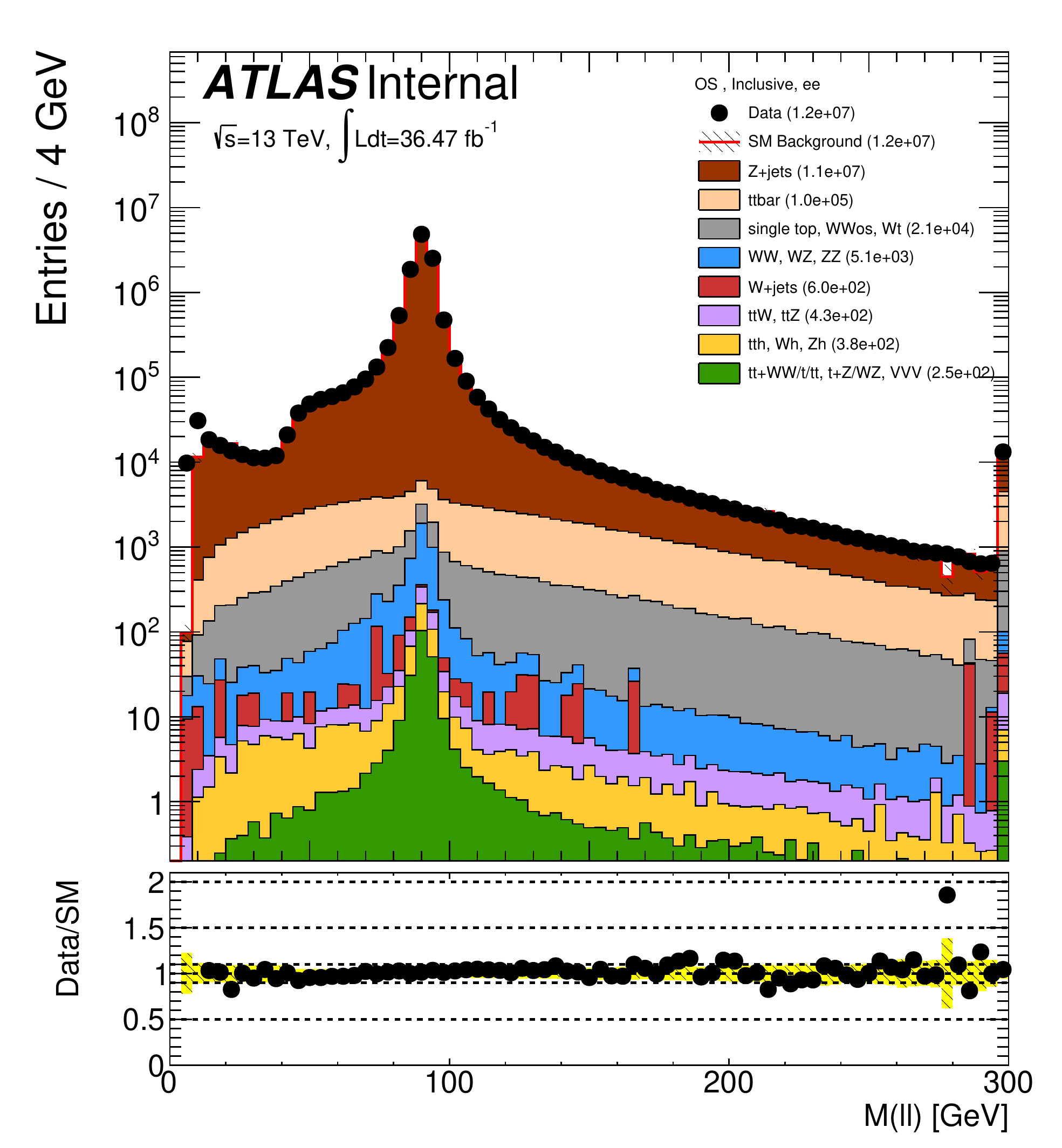}}
{\includegraphics[width=.49\textwidth]{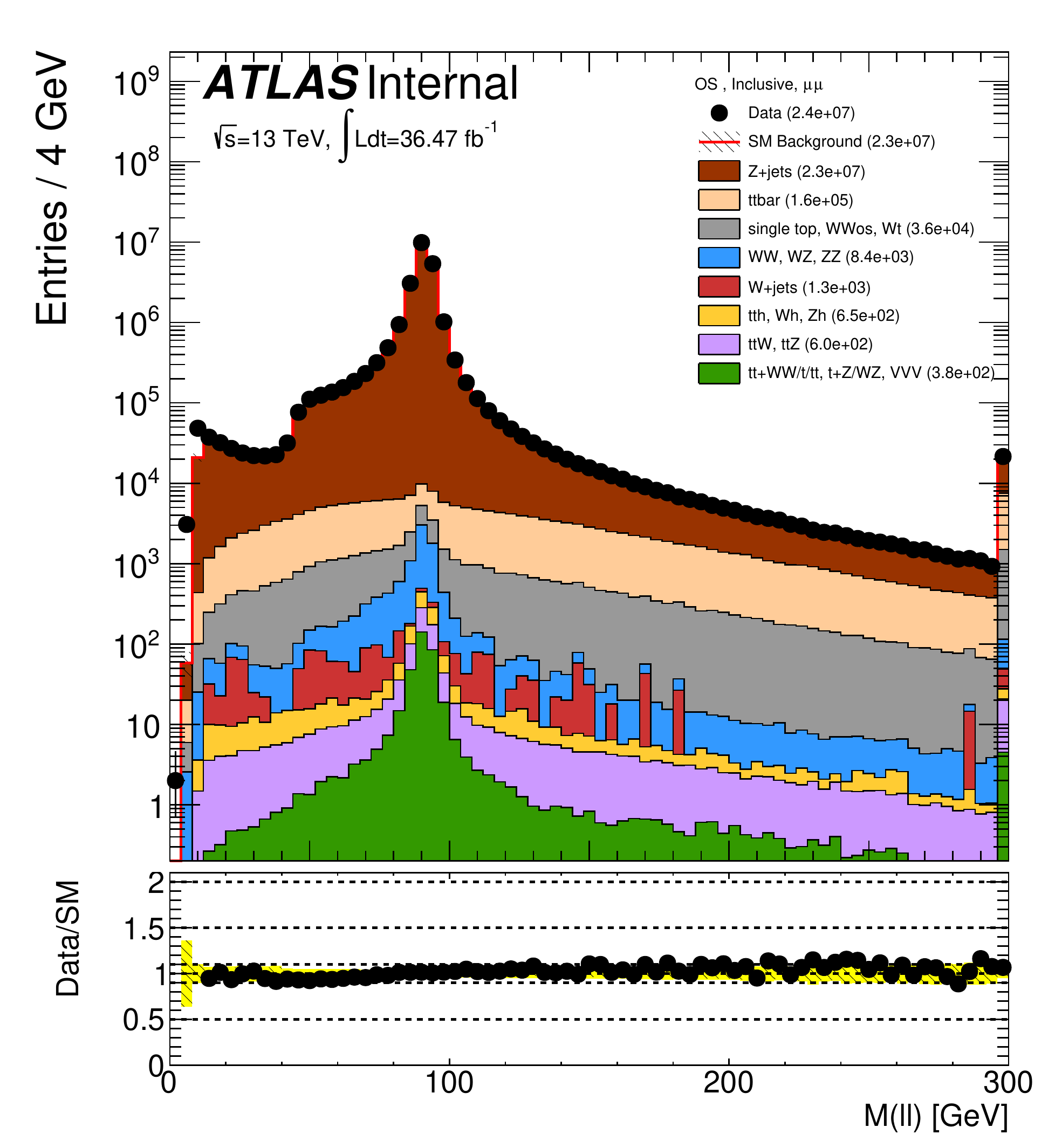}}
\caption{Dilepton invariant mass distributions for opposite-sign pairs for events selected in the $ee$ (left) and $e\mu$ (right) channels. 
No low-mass Drell-Yan sample is included. 
 The prediction is taken from MC only.
Only luminosity and MC statistical uncertainties are included.
}
\label{fig:dataMC_2ll.os}
\end{figure}

In order to validate the various choices made regarding the object definitions and event selection, 
check their sensible behavior and their reasonable modelling in the simulations, 
we looked at the distributions of several kinematic variables obtained with the full available data set.
A selection with two leptons of opposite-sign (OS) is used for this purpose since the modelling in the simulation 
is expected to be accurate.
For illustration, Figures~\ref{fig:dataMC_2em.os}-\ref{fig:dataMC_2ll.os} 
show the dilepton invariant mass distributions in data compared to MC for OS dilepton events.
A very good agreement with MC is observed in the OS selection with a clear $Z$-boson mass peak in the $ee$ and $\mu\mu$ channels. 

\begin{figure}[htb!]
\centering
{\includegraphics[width=.75\textwidth]{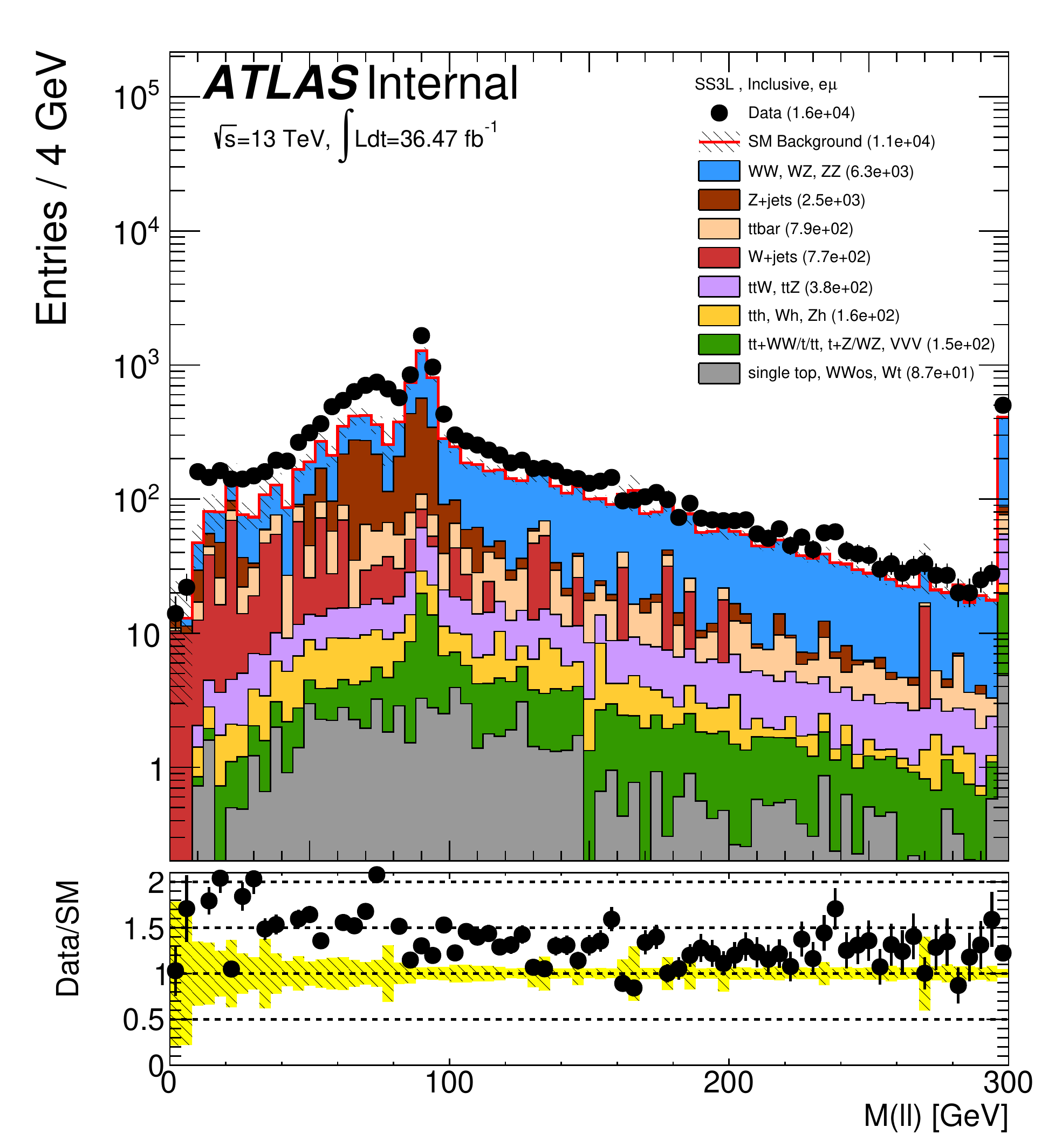}}
\caption{Dilepton invariant mass distributions for opposite-sign pairs for events selected in the $e\mu$ channel. 
No low-mass Drell-Yan sample is included. 
 The prediction is taken from MC only.
Only luminosity and MC statistical uncertainties are included.
}
\label{fig:dataMC_2em.ss}
\end{figure}

\begin{figure}[htb!]
\centering
{\includegraphics[width=.49\textwidth]{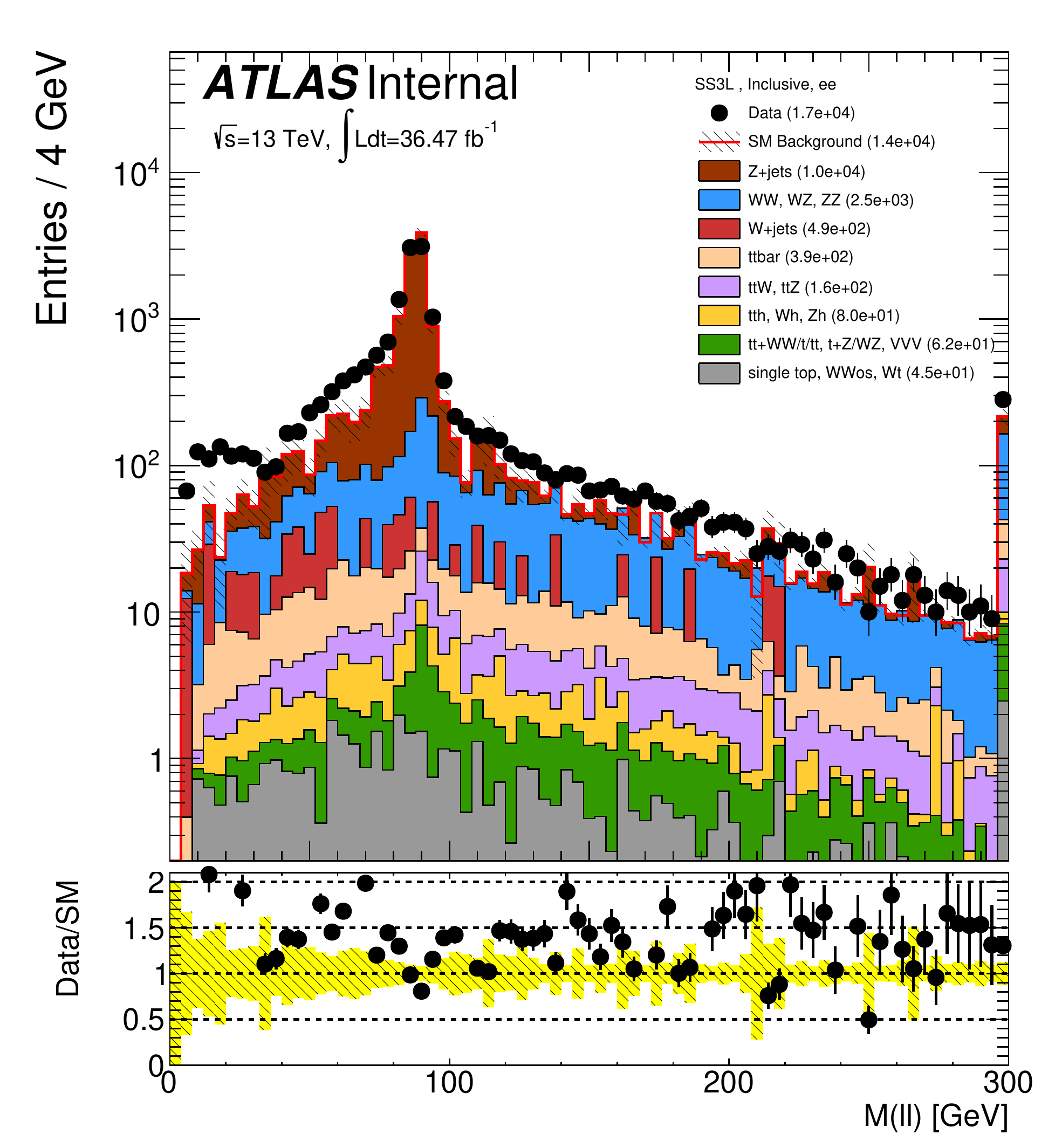}}
{\includegraphics[width=.49\textwidth]{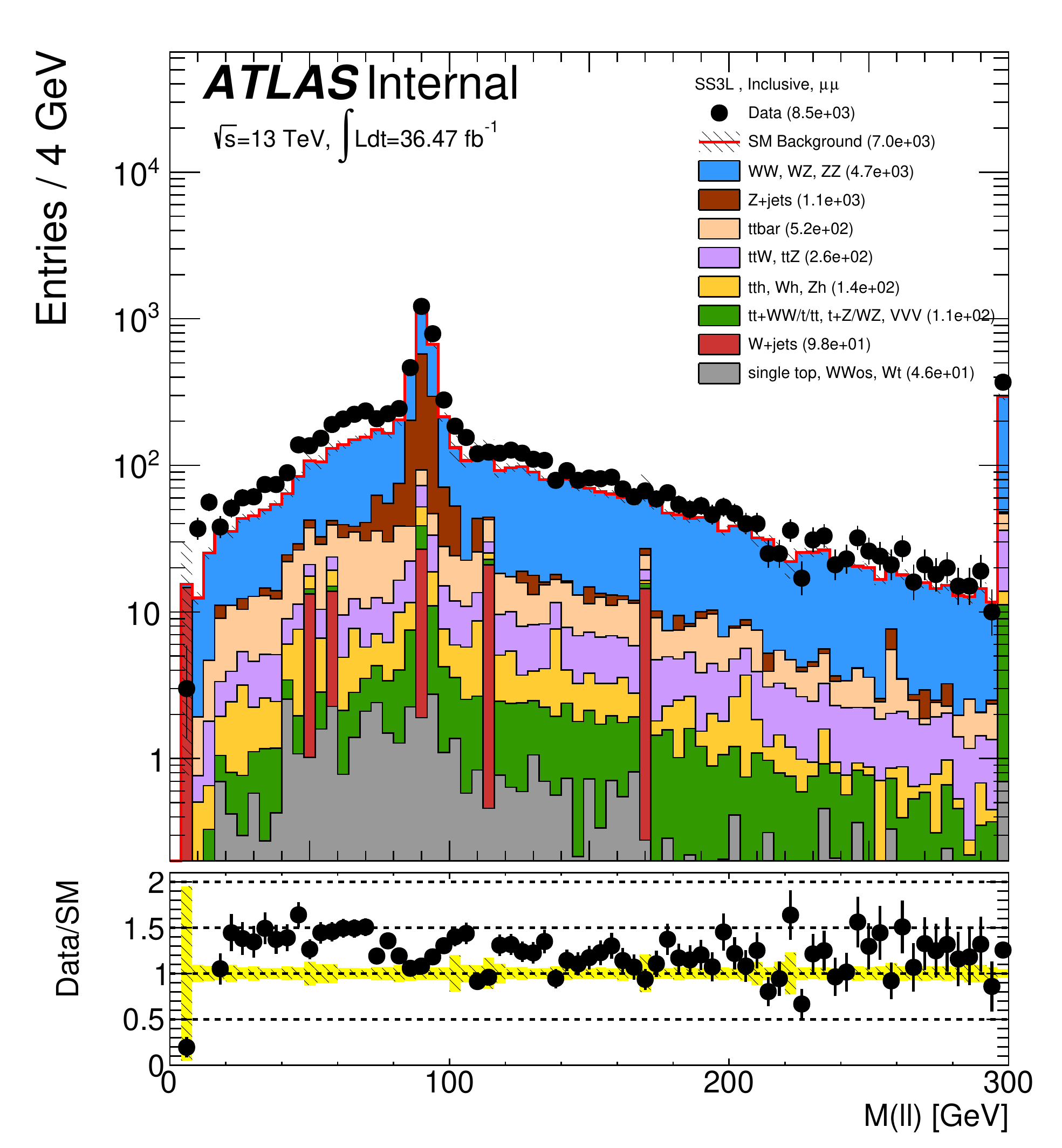}}
\caption{Dilepton invariant mass distributions for same-sign pairs for events selected in the $ee$ (left) and $e\mu$ (right) channels. 
No low-mass Drell-Yan sample is included. 
 The prediction is taken from MC only.
Only luminosity and MC statistical uncertainties are included.
}
\label{fig:dataMC_2ll.ss}
\end{figure}

Since the analysis uses a selection with two leptons of same-sign (SS), we must evaluate the level of agreement between data and simulation.
For illustration, Figures~\ref{fig:dataMC_2em.ss}-\ref{fig:dataMC_2ll.ss} show the dilepton invariant mass 
distributions in data compared to MC for SS dilepton events. 
The background distributions are taken directly from MC with no data-driven estimation of the charge flip or non-prompt lepton backgrounds.
In the SS channels, the $Z$-boson mass peak is also observed in the $ee$ channel due to electron charge mis-identification, with MC overestimating data. 
An accumulation of events at the $Z$-boson mass is also observed in the SS $e\mu$ and $\mu\mu$ channels due to three-lepton events 
from either $Z$+jets with a fake lepton or from $WZ$ production.  

\begin{figure}[htb!]
 \centering
 {\includegraphics[width=0.49\textwidth]{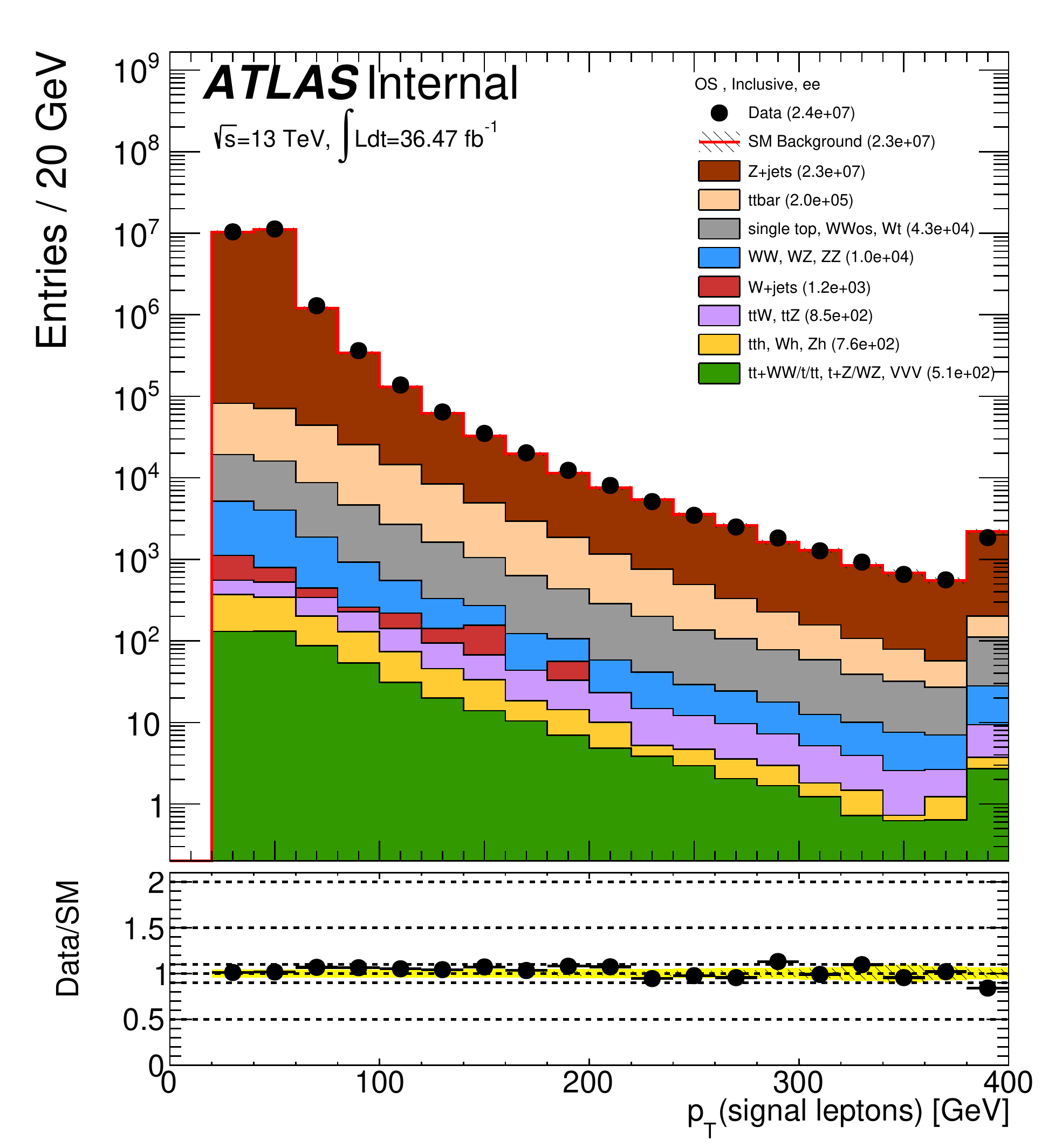}}
 {\includegraphics[width=0.49\textwidth]{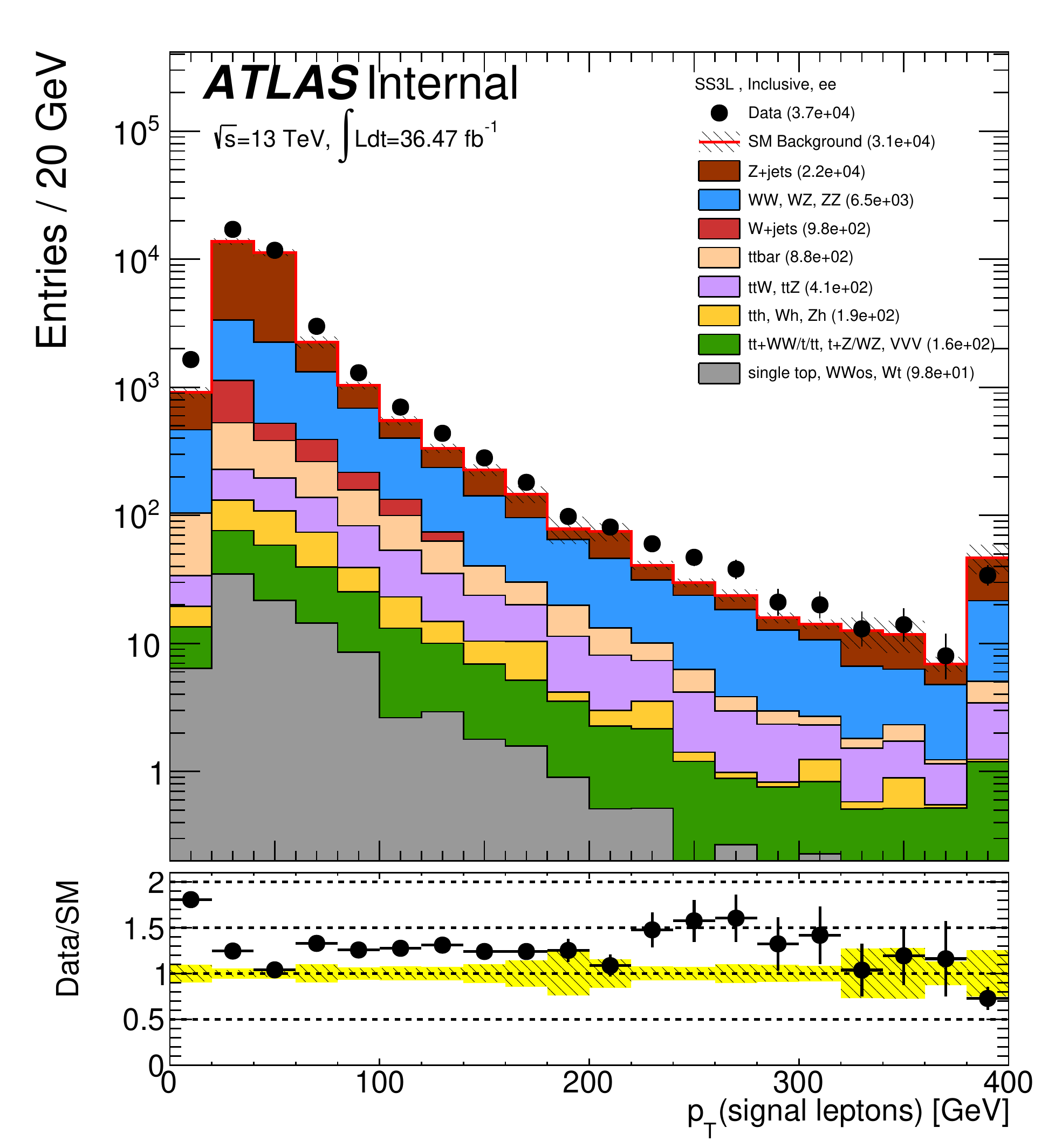}}
\caption{Signal lepton transverse momentum distributions for (left) OS and (right) SS  pairs for events selected in the $ee$ channel.  
The prediction is taken from MC only.
Only luminosity and MC statistical uncertainties are included.
}
\label{fig:dataMC_lep1.ee}
\end{figure}

\begin{figure}[htb!]
 \centering
{\includegraphics[width=0.49\textwidth]{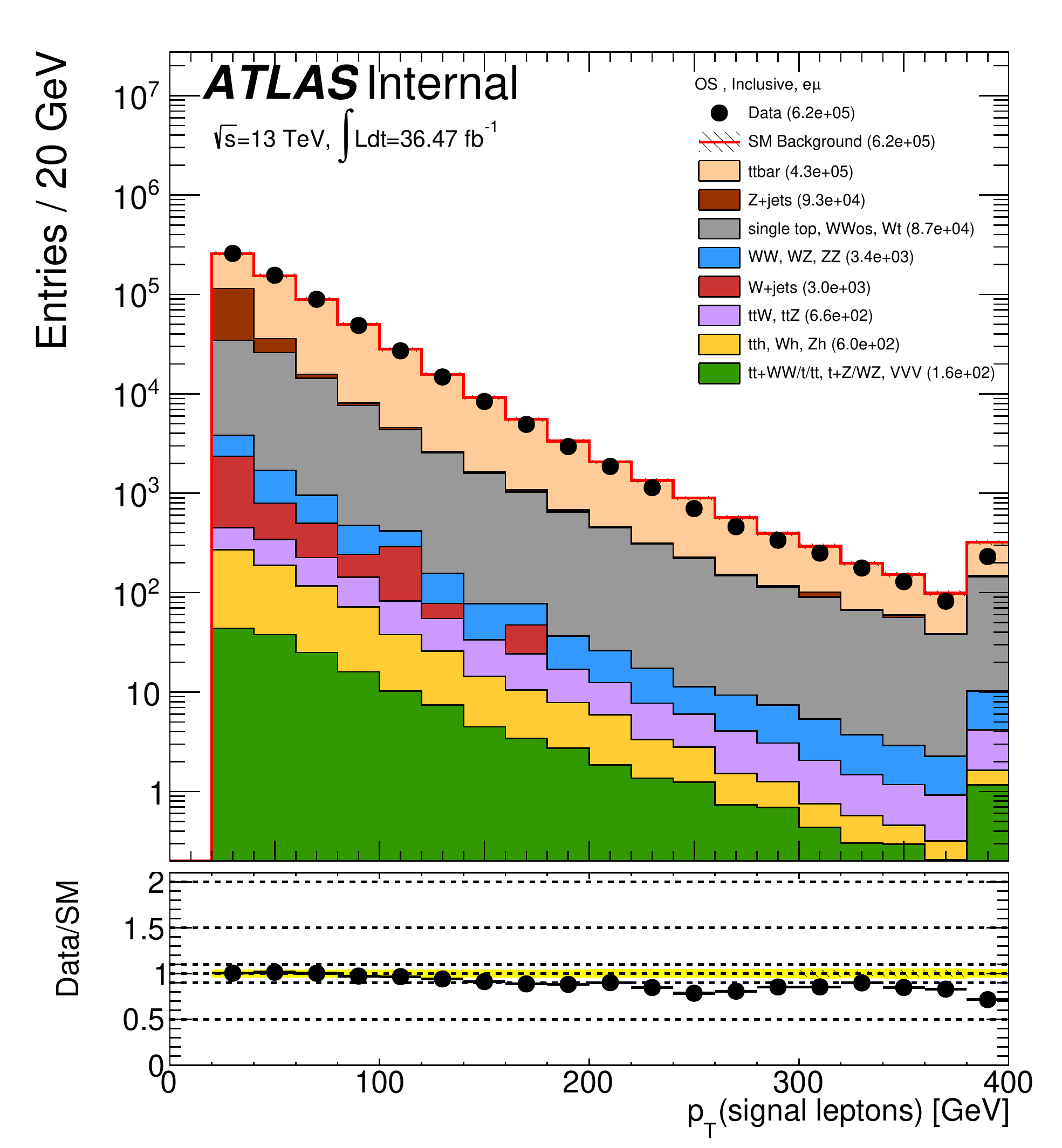}}
{\includegraphics[width=0.49\textwidth]{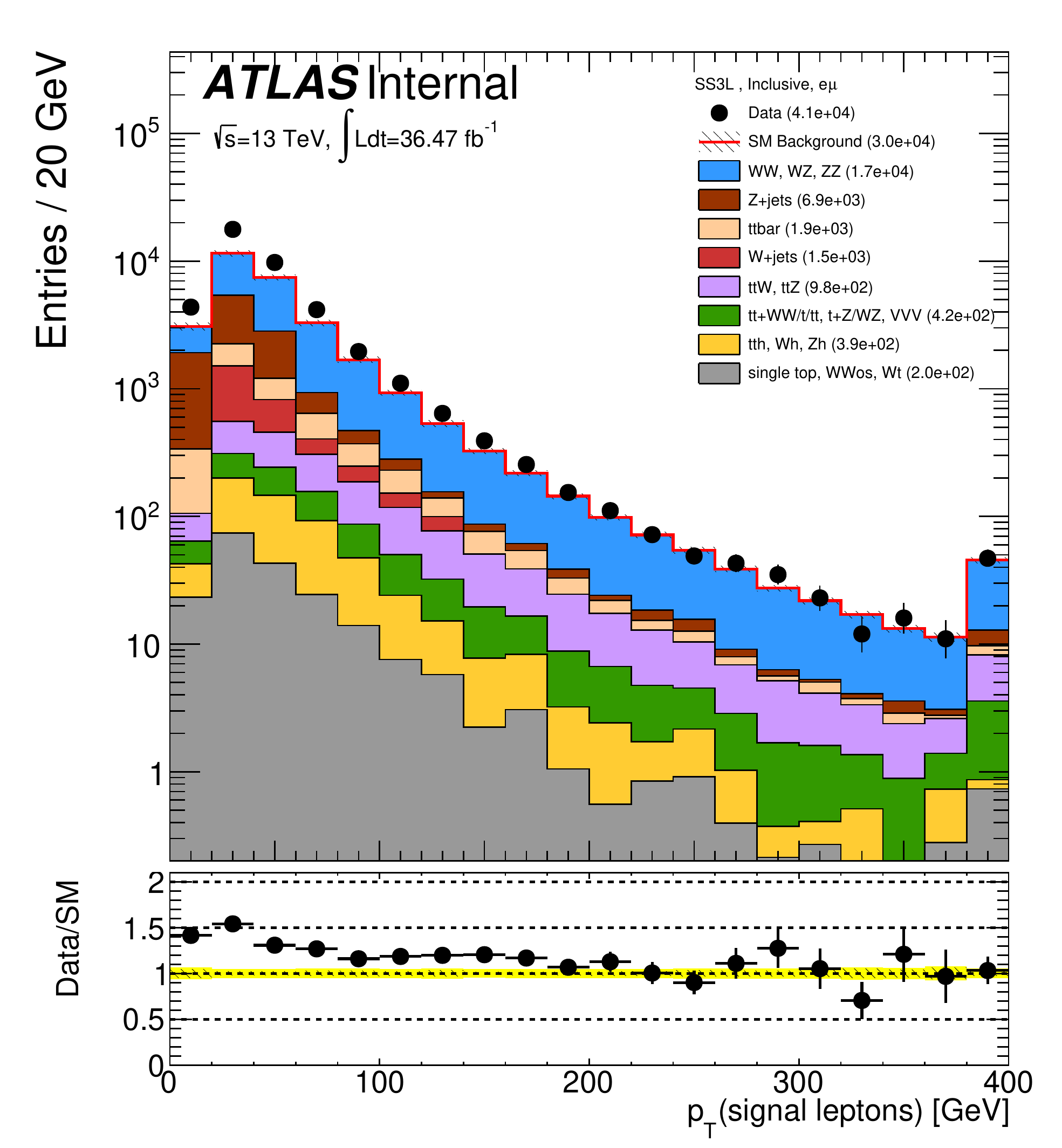}}
\caption{Signal lepton transverse momentum distributions for (left) OS and (right) SS  pairs for events selected in the $e\mu$ channel.
The prediction is taken from MC only.
Only luminosity and MC statistical uncertainties are included.
}
\label{fig:dataMC_lep1.em}
\end{figure}

\begin{figure}[htb!]
 \centering
{\includegraphics[width=0.49\textwidth]{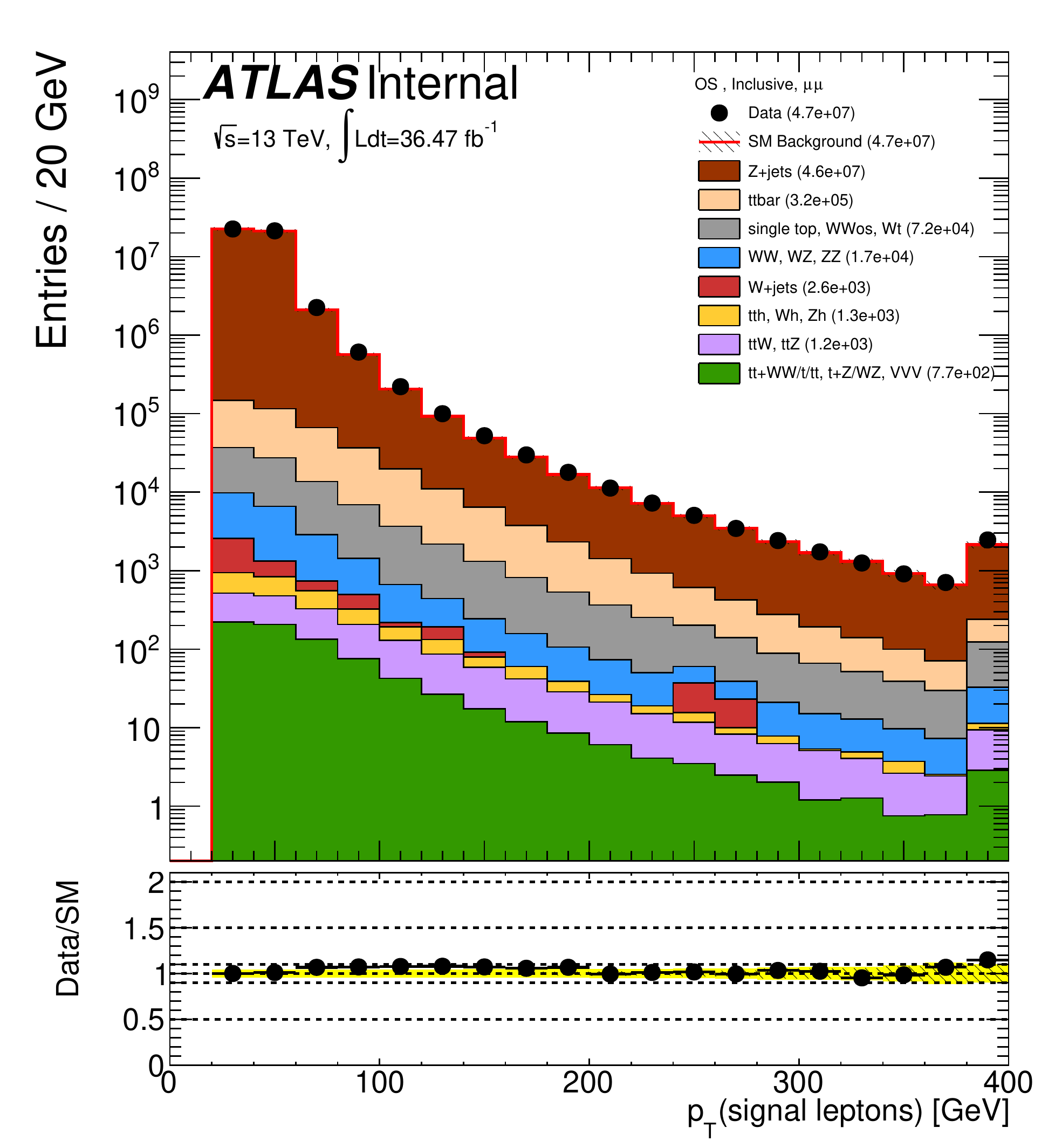}}
{\includegraphics[width=0.49\textwidth]{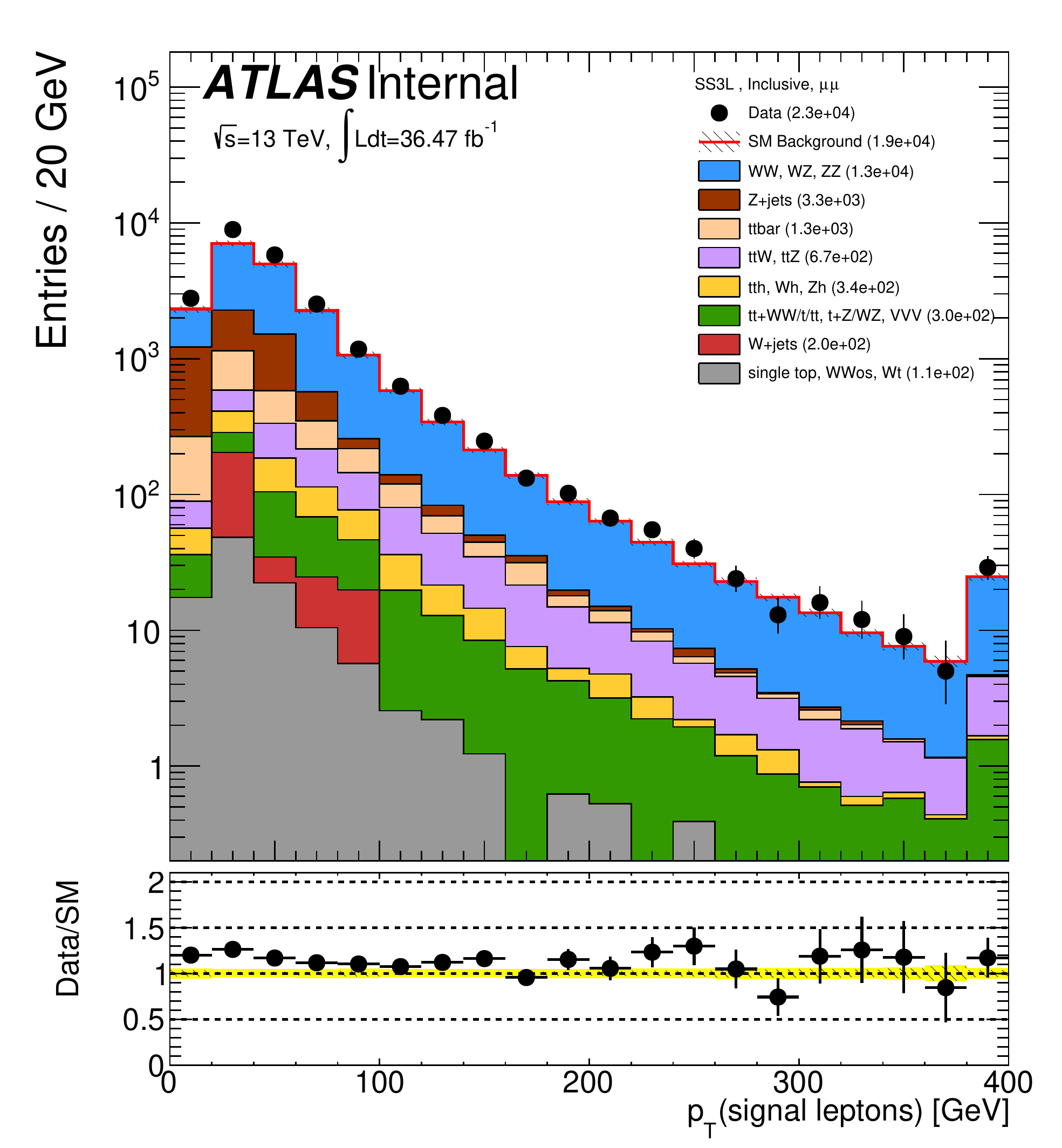}}
\caption{Signal lepton transverse momentum distributions for (left) OS and (right) SS pairs for events selected in the $\mu\mu$ channel.  
The prediction is taken from MC only.
Only luminosity and MC statistical uncertainties are included.
}
\label{fig:dataMC_lep1.mm}
\end{figure}

Similarly, the transverse momentum distributions of the signal leptons used in the analysis for OS (left) and SS (right) selections are shown in 
Figures~\ref{fig:dataMC_lep1.ee}-\ref{fig:dataMC_lep1.mm}. These distributions show that the OS selection has reasonable agreement between the observation 
and simulation while the SS selection has a mis-modelling particularly at low
lepton $\pt$ where some discrepancies and accumulation of events involving fake leptons ($Z$+jets, $W$+jets, $\ttbar$) are observed. 

\begin{figure}[htb!]
 \centering
 {\includegraphics[width=0.49\textwidth]{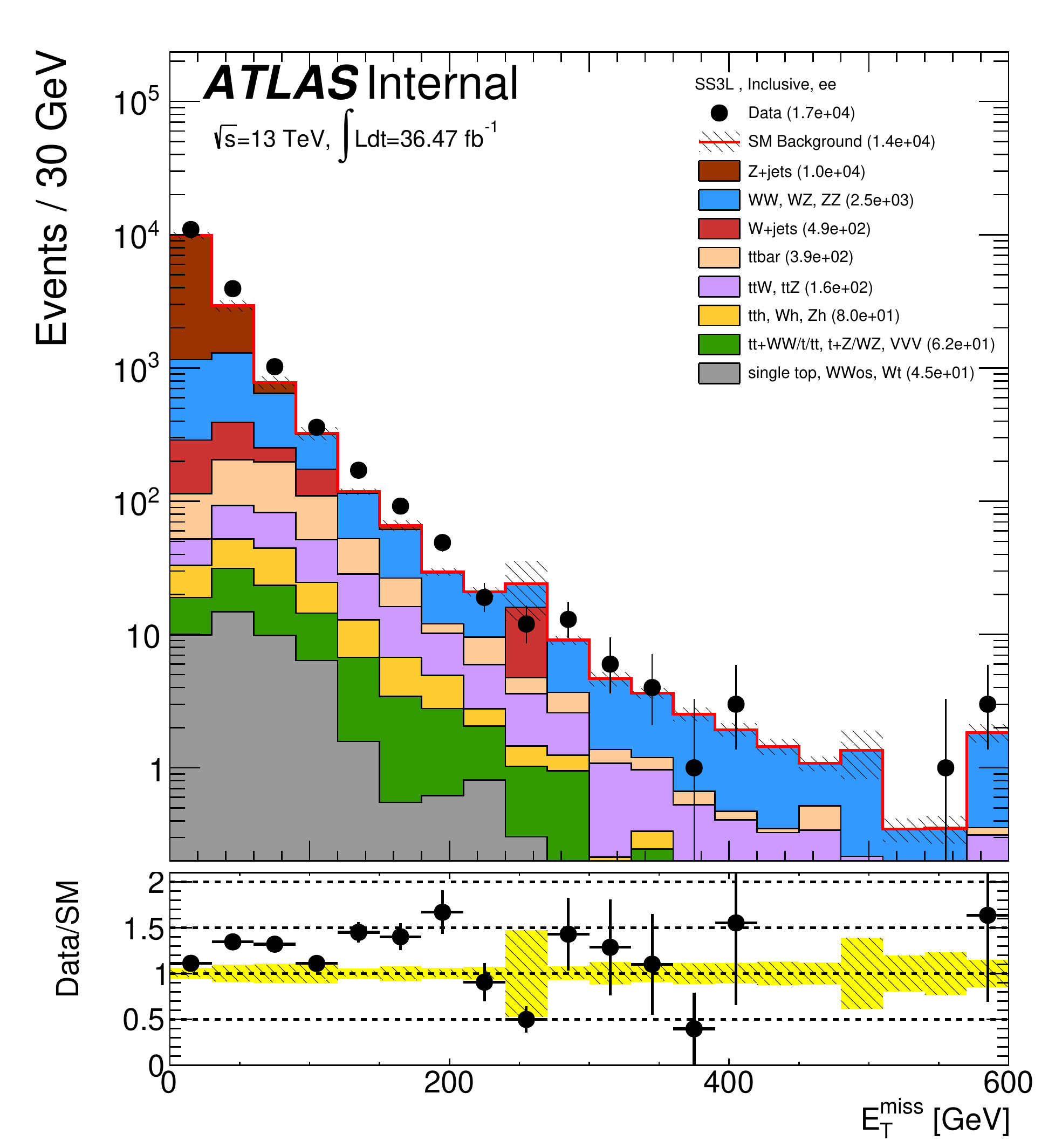}}
 {\includegraphics[width=0.49\textwidth]{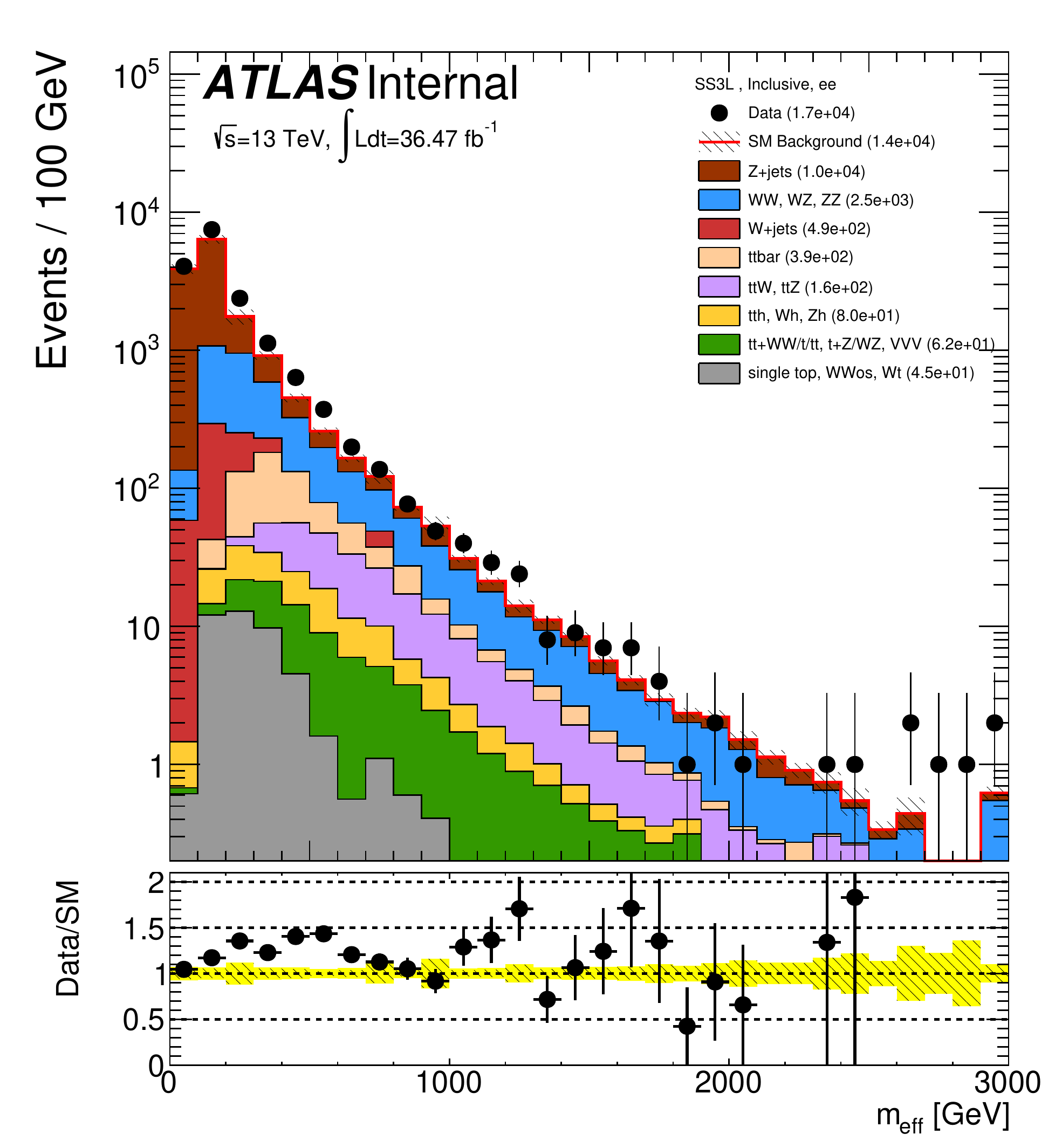}}
\caption{Distributions of the $\met$ (left) and effective mass (right) for events selected in the $ee$ channel.  The prediction is taken from MC only. 
Only luminosity and MC statistical uncertainties are included.}
\label{fig:dataMC_metmeff.ee}
\end{figure}

\begin{figure}[htb!]
 \centering
{\includegraphics[width=0.49\textwidth]{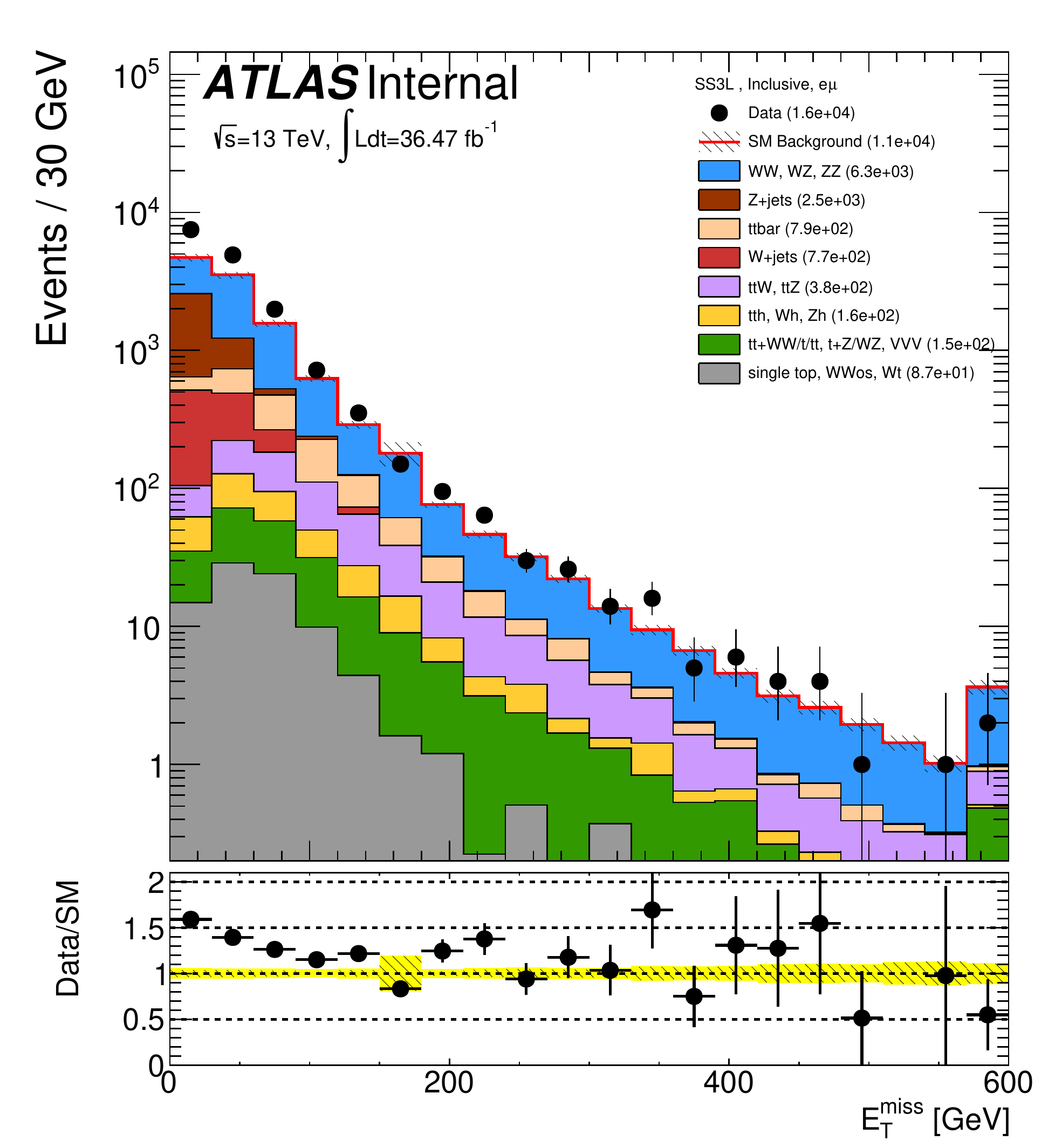}}
{\includegraphics[width=0.49\textwidth]{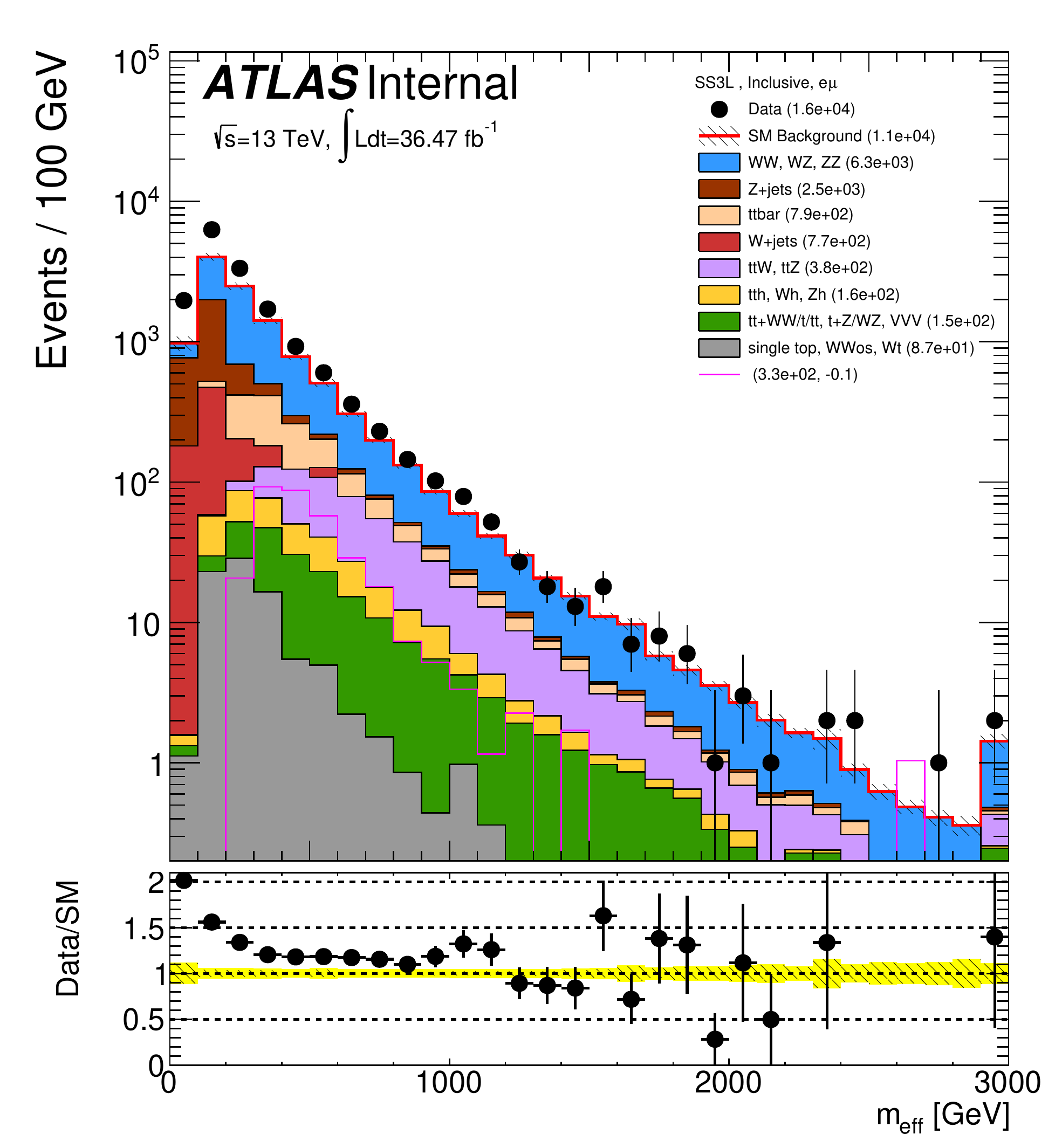}}
\caption{Distributions of the $\met$ (left) and effective mass (right) for events selected in the $e\mu$ channel.  The prediction is taken from MC only.  
Only luminosity and MC statistical uncertainties are included.}
\label{fig:dataMC_metmeff.em}
\end{figure}

\begin{figure}[htb!]
 \centering
{\includegraphics[width=0.49\textwidth]{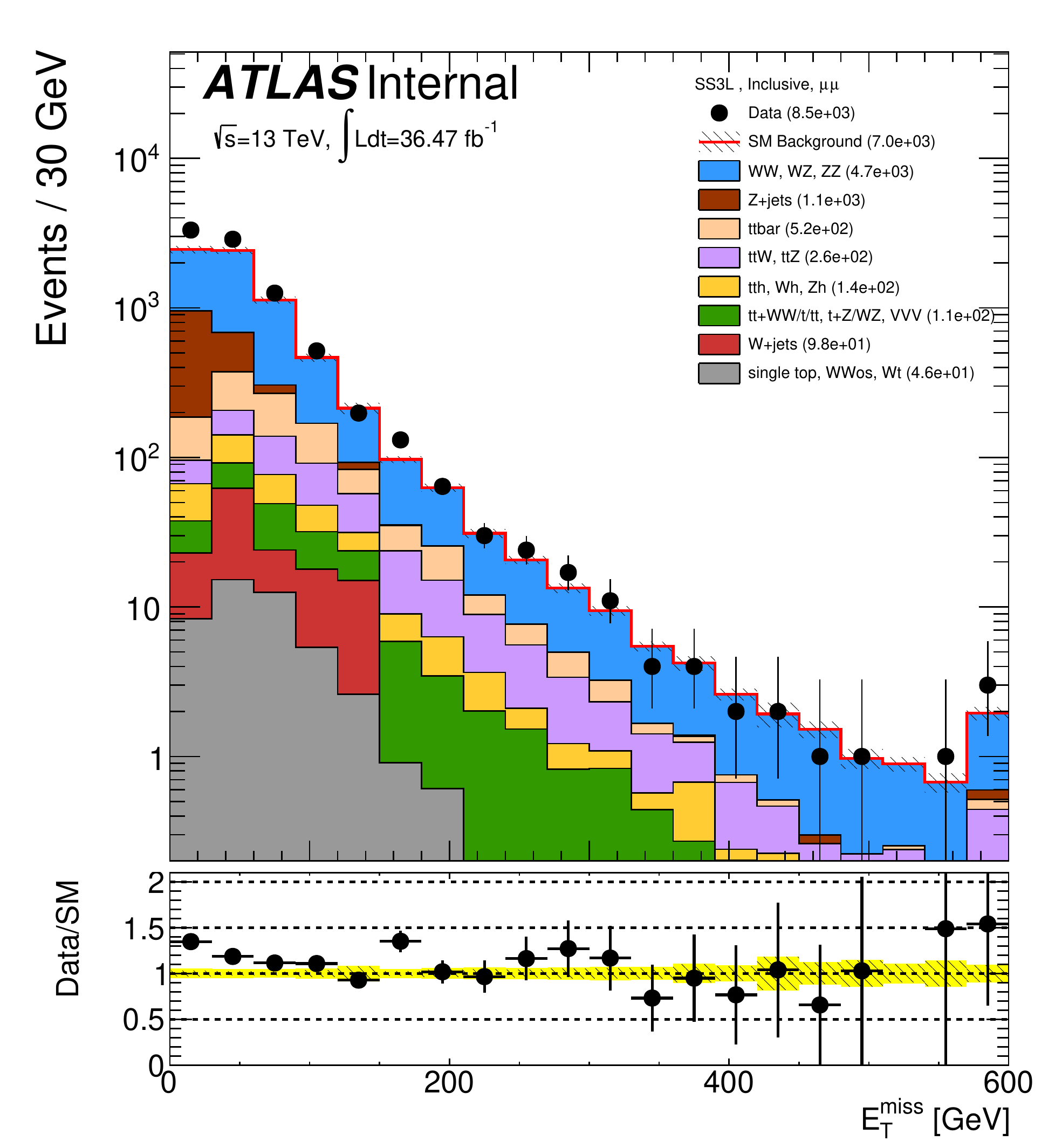}}
{\includegraphics[width=0.49\textwidth]{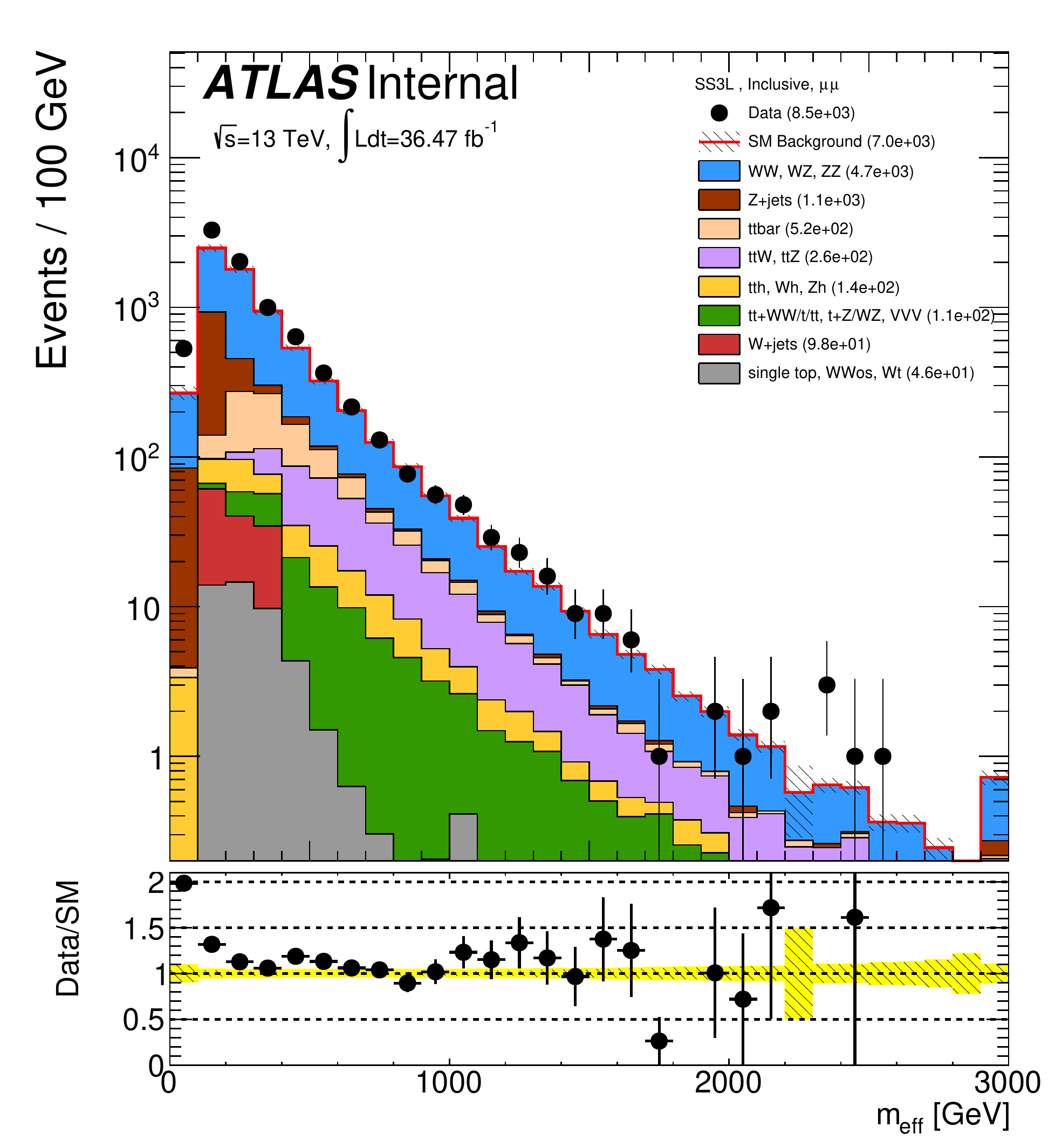}}
\caption{Distributions of the $\met$ (left) and effective mass (right) for events selected in the $\mu\mu$ channel.  The prediction is taken from MC only.  Only luminosity and MC statistical uncertainties are included.}
\label{fig:dataMC_metmeff.mm}
\end{figure}

Other variables that are used in the analysis to discriminate between signal and background processes are \met and \meff. 
Figures~\ref{fig:dataMC_metmeff.ee}-\ref{fig:dataMC_metmeff.mm} show the \met and \meff distributions for the $ee$, $e\mu$, $\mu\mu$ channels for a SS selection.

It is already clear from these distributions that simulation is not reliable in the SS selection. The disagreement can be at the level 
of 50\% suggesting that different techniques must be used to estimate the backgrounds. 
The background estimation, in Chapter ~\ref{chap:fake} and Chapter~\ref{chap:bkg}, will be dedicated to improving the estimates of 
the background prediction using data-driven methods and validating the estimates.

%% file: texfiles/sec.strategy.sr.tex
In order to maximize the sensitivity to the signal models of Figure~\ref{fig:strategy.pheno.feynman}, 13 non-exclusive signal regions are defined in Table~\ref{tab:SRdef3}. 
\begin{table}[tbh!]
\centering
\resizebox{\textwidth}{!}
{
\hspace{0.5cm}
\def\arraystretch{1.2}
\Large
\begin{tabular}{|l|c|c|c|c|c|r|c|c|l|}
\hline
Signal region  &  $N_{\textrm{leptons}}^{\textrm{signal}}$   & $N_{b\textrm{-jets}}$ & $N_{\textrm{jets}}$  & $\pt^{\textrm{jet}}$ & \met\ & \meff\ & \met/\meff  & Other & Targeted  \\
               &                                  &                   &                  &    [GeV]             & [GeV] & [GeV] &   &  & Signal  \\
\hline\hline

Rpc2L2bS         & $\ge 2$SS  & $\ge 2$ & $\ge 6$ & $>25$ & $>200$ & $>600$  & $>0.25$    & --			        & Figure~\ref{fig:strategy.pheno.feynman_gtt}\\ 
Rpc2L2bH         & $\ge 2$SS  & $\ge 2$ & $\ge 6$ & $>25$ & --     & $>1800$  & $>0.15$	  & -- 			        & Figure~\ref{fig:strategy.pheno.feynman_gtt}, NUHM2\\ 
\hline
Rpc2Lsoft1b    & $\ge 2$SS  & $\ge 1$ & $\ge 6$ & $>25$ & $>100$ &  --\hphantom{00}      & $>0.3\hphantom{0}$    & 20,10 $<$\ptlone,\ptltwo $<$ 100 GeV~& Figure~\ref{fig:strategy.pheno.feynman_gttOffshell}\\ 
Rpc2Lsoft2b      & $\ge 2$SS  & $\ge 2$ & $\ge 6$ & $>25$ & $>200$ & $>600$   & $>0.25$   & 20,10 $<$\ptlone,\ptltwo $<$ 100 GeV~& Figure~\ref{fig:strategy.pheno.feynman_gttOffshell} \\ 
\hline
Rpc2L0bS         & $\ge 2$SS  & $=0$    & $\ge 6$ & $>25$ & $>150$ & --\hphantom{00}      & $>0.25$   & -- 				& Figure~\ref{fig:strategy.pheno.feynman_gg2WZ}\\
Rpc2L0bH         & $\ge 2$SS  & $=0$    & $\ge 6$ & $>40$ & $>250$ & $>900$   & --	  & --				& Figure~\ref{fig:strategy.pheno.feynman_gg2WZ}\\
\hline
Rpc3L0bS       & $\ge 3$    & $=0$    & $\ge 4$ & $>40$ & $>200$ & $>600$   & --	  & --				& Figure~\ref{fig:strategy.pheno.feynman_gg2sl}\\ 
Rpc3L0bH       & $\ge 3$    & $=0$    & $\ge 4$ & $>40$ & $>200$ & $>1600$  & --  & --				& Figure~\ref{fig:strategy.pheno.feynman_gg2sl}\\
Rpc3L1bS       & $\ge 3$    & $\ge 1$ & $\ge 4$ & $>40$ & $>200$ & $>600$   & --  & --				& Other \\ 
Rpc3L1bH       & $\ge 3$    & $\ge 1$ & $\ge 4$ & $>40$ & $>200$ & $>1600$  & --  & --				& Other  \\
\hline
Rpc2L1bS         & $\ge 2$SS  & $\ge 1$ & $\ge 6$ & $>25$ & $>150$ & $>600$   & $>0.25$   & --				& Figure~\ref{fig:strategy.pheno.feynman_b1b1}\\
Rpc2L1bH         & $\ge 2$SS  & $\ge 1$ & $\ge 6$ & $>25$ & $>250$ & --\hphantom{00}      & $>0.2\hphantom{0}$    & --				& Figure~\ref{fig:strategy.pheno.feynman_b1b1}\\ 
\hline
Rpc3LSS1b    & $\ge \ell^\pm\ell^\pm\ell^\pm$ & $\ge 1$ & -- & --   & --  & --\hphantom{00}       & -- & veto 81$<$\mee$<$101 GeV~	& Figure~\ref{fig:strategy.pheno.feynman_t1t1}\\ 
\hline
\end{tabular}
}
\caption{
Summary of the signal region definitions. 
Requirements are placed on the number of signal leptons ($N_{\textrm{leptons}}^{\textrm{signal}}$) with $\pt>$20 \GeV~and a same sign (SS) pair (except for Rpc2Lsoft), the number of 
$b$-jets with $\pt>20 \GeV$ ($N_{b\textrm{-jets}}$), the number of jets ($N_{\textrm{jets}}$) above a certain \pt threshold ($\pt^{\textrm{jet}}$), 
\met, \meff\ and/or \met/\meff. 
The last column indicates the targeted signal model. 
}
\label{tab:SRdef3}
\end{table}

The SRs are named in the form Rpc\textit{N}L{\textit M}b{\textit X}, where {\textit N} indicates the number of leptons required, {\textit M} the number of $b$-jets required, and {\textit X} indicates the severity 
of the \met or \meff\ requirements (Soft, Medium or Hard). All signal regions allow any number of additional leptons in addition to a $e^\pm e^\pm$,
$e^\pm \mu^\pm$ or $\mu^\pm \mu^\pm$ pair. Signal regions with 3 leptons can be either any charge combination or all three with the same charge (Rpc3LSS1b).
For each lepton/$b$-jet multiplicity, two signal regions are defined targeting either compressed spectra or large mass splittings. 

The optimization of the definitions of signal regions relies on a brute-force
scan of several discriminating variables in a loose classification of events 
in terms of number of $b$-jets and/or leptons in the final state, each being 
associated to the signal scenario favouring this final state.
The other main discriminant variables (e.g number of jets above a certain \pt threshold, \meff, \met, $\met/\meff$ ratio) are then 
allowed to vary, 
to determine for each point of the parameter space the best configuration. 
The figure of merit used to rank configurations is the discovery significance ($Z_0$)
defined in Eq.~\ref{eq:Z0.FoM} which represents
a statistical test based on a ratio of two Poisson means~\cite{Cousins:2009}:

\begin{equation}
Z_0 = \sqrt{2\left(\left(s+b\right)\ln\left(1+\frac{s}{b}\right)-s\right)}
\label{eq:Z0.FoM}
\end{equation}
where $s$ and $b$ represent the expected number of signal and background events\footnote{Note that Eq.~\ref{eq:Z0.FoM} simplifies to the commonly used figure of merit 
$\frac{s}{\sqrt{b}} + \mathcal{O}\left(\frac{s}{b}\right)$ if $\frac{s}{b} \ll 1$.}.
A realistic systematic uncertainty of $\Delta b=30\%$ on the expected background yield 
was included in the statistical test by replacing $b$ with $b + \Delta b$ in Eq.~\ref{eq:Z0.FoM}.
To preserve the discovery potential, only configurations leading to at least two signal events were considered for a given signal point. The total number of background events should not be smaller than 1; to model in a more realistic way the effect of non-prompt and fake leptons and electron charge mis-identification backgrounds, 
which are determined from data in the analysis, 
the MC predictions for those processes in $\ttbar$ and $Z$+jets MC were scaled 
using the factors obtained from the MC template method (Section~\ref{sec:fake.mct}), as shown in Table~\ref{tab:mctemplateF}.
Note that different corrections are applied depending on the showering (Pythia or Sherpa) used for each sample, 
and for the fake and non-prompt leptons originated from heavy-flavour (HF) and light-flavour (LF).

\begin{table}[!htb]
\caption{Scaling factors applied to the electron charge-flip and non-prompt/fake lepton background in the SR optimization procedure.}
\label{tab:mctemplateF}
\def\arraystretch{1.1}
\centering
\begin{tabular}{|c||c|c|c|c|c|}
\hline 
& Charge mis-id & HF $e$ & HF $\mu$ & LF $e$ & LF $\mu$ \\ \hline\hline
Pythia & 0.96 $\pm$ 0.08 & 1.80 $\pm$ 0.45 & 2.10 $\pm$ 0.58 & 1.55 $\pm$ 0.14 & 0.74 $\pm$ 0.81\\\hline
Sherpa & 1.02 $\pm$ 0.09 & 2.72 $\pm$ 0.57 & 1.81 $\pm$ 0.75 & 1.16 $\pm$ 0.18 & 1.84 $\pm$ 1.16\\
\hline
\end{tabular}
\end{table}

Since the signal regions defined out of the scanning procedure may not be mutually exclusive,
the results expressed in terms of exclusion limits will be obtained for each signal point 
by using the signal region that leads to the best expected sensitivity.

\begin{figure}[htb!]  
\centering
{\includegraphics[width=0.49\textwidth,page=5]{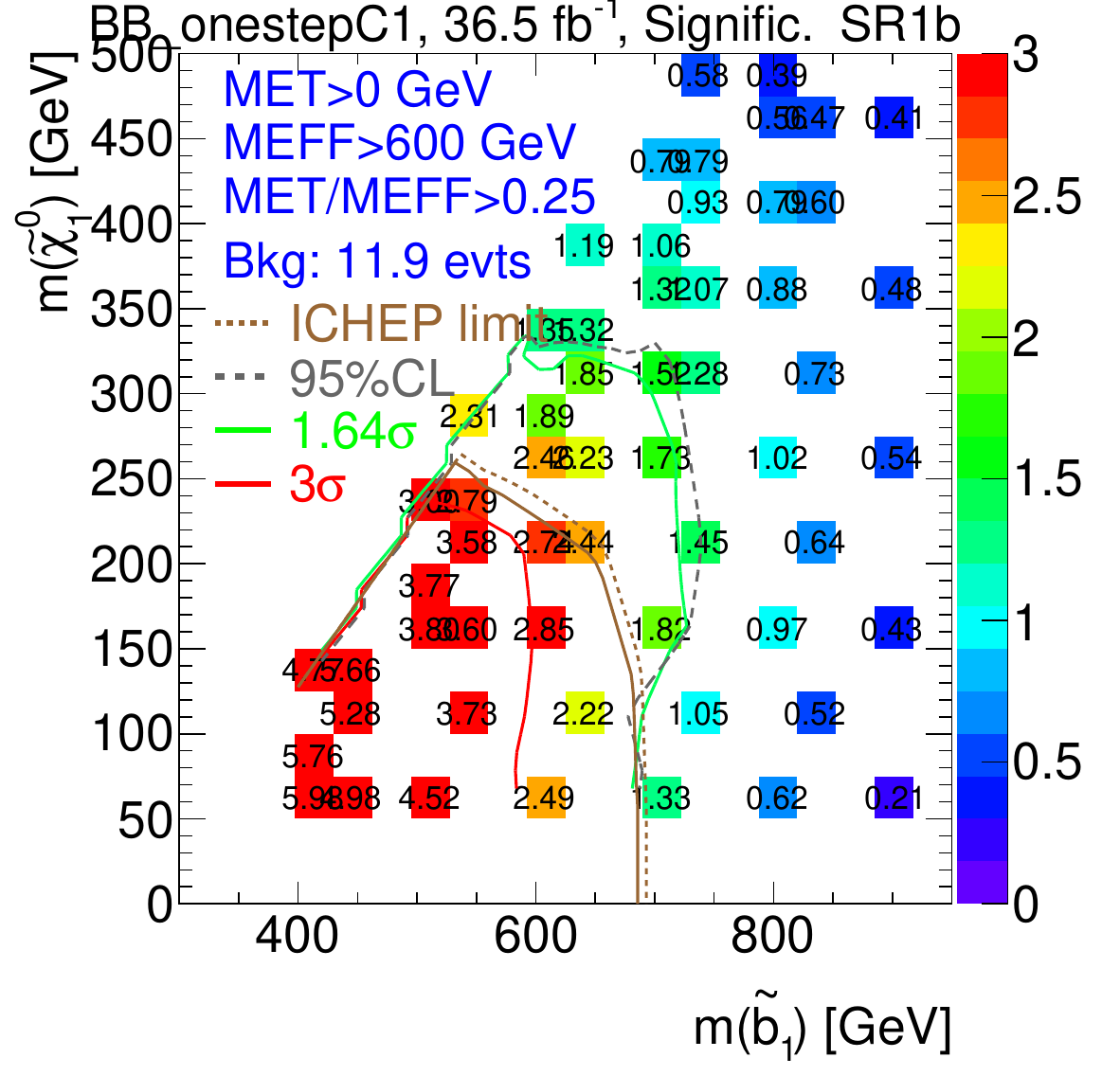} }
{\includegraphics[width=0.49\textwidth,page=6]{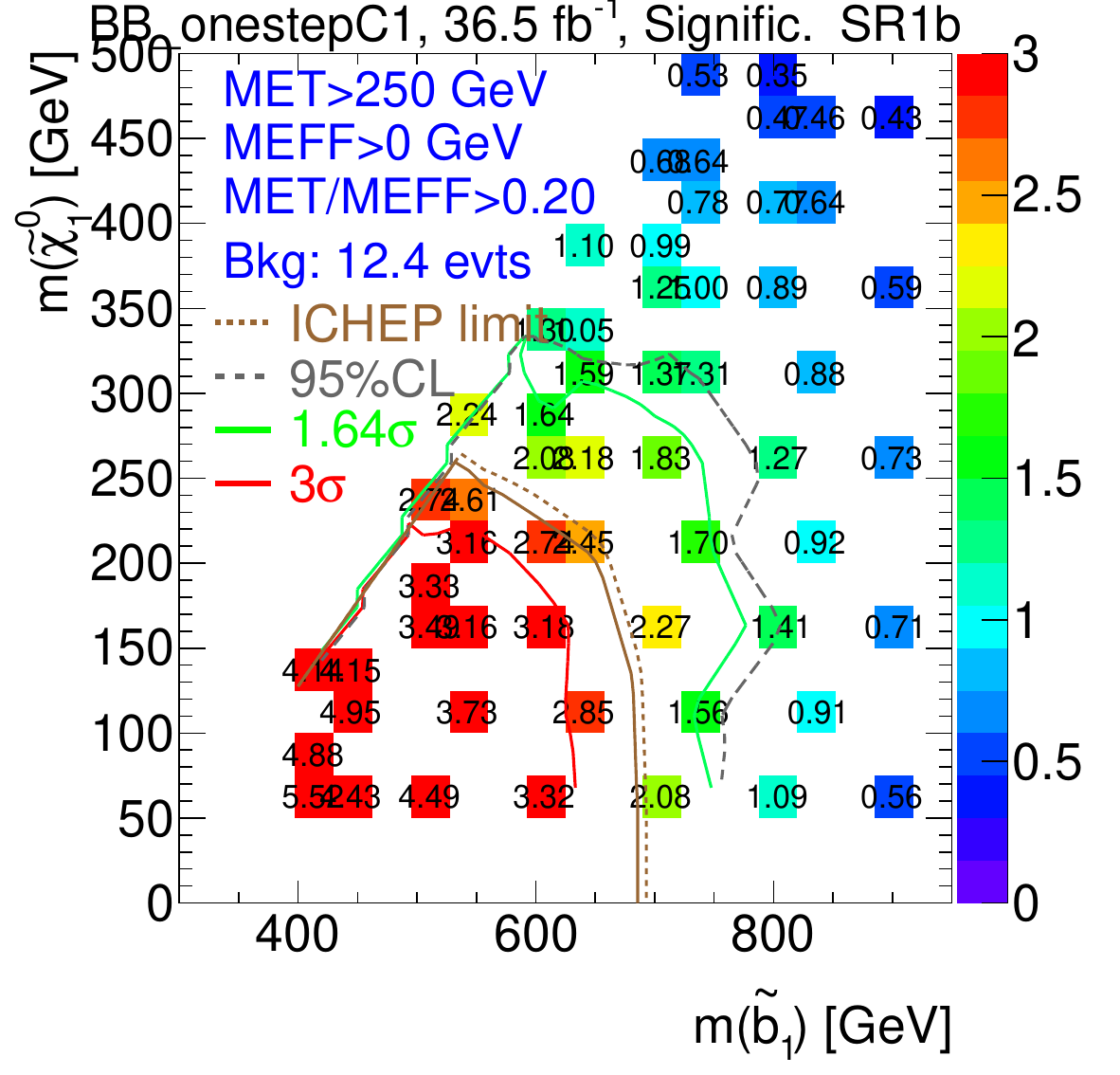} }
\vspace{-1cm}
\caption{Discovery significance for the (left) Rpc2L1bS and (right) Rpc2L1bH defined in 
Table~\ref{tab:SRdef3} for 36.5~\ifb.
The 95\% CL, 
1.64$\sigma$, and 3$\sigma$ discovery contours from the proposed signal 
regions are shown in grey, green, and red, respectively. 
 }
 \label{fig:SR_withB}
 \end{figure}

To illustrate the procedure, we show the performance of the optimization procedure for the sbottom pair production only, 
$\sbot\sbot^*\to t\bar t\tilde\chi_1^+\tilde\chi_1^-$, in Figure~\ref{fig:SR_withB}.
The discovery significance for each signal point is shown, together with the contours corresponding to a 
3$\sigma$ discovery sensitivity, 1.64$\sigma$ discovery sensitivity and 95\% confidence level limits. 

\begin{figure}[htb!]
\centering
\includegraphics[width=0.75\textwidth,page=3]{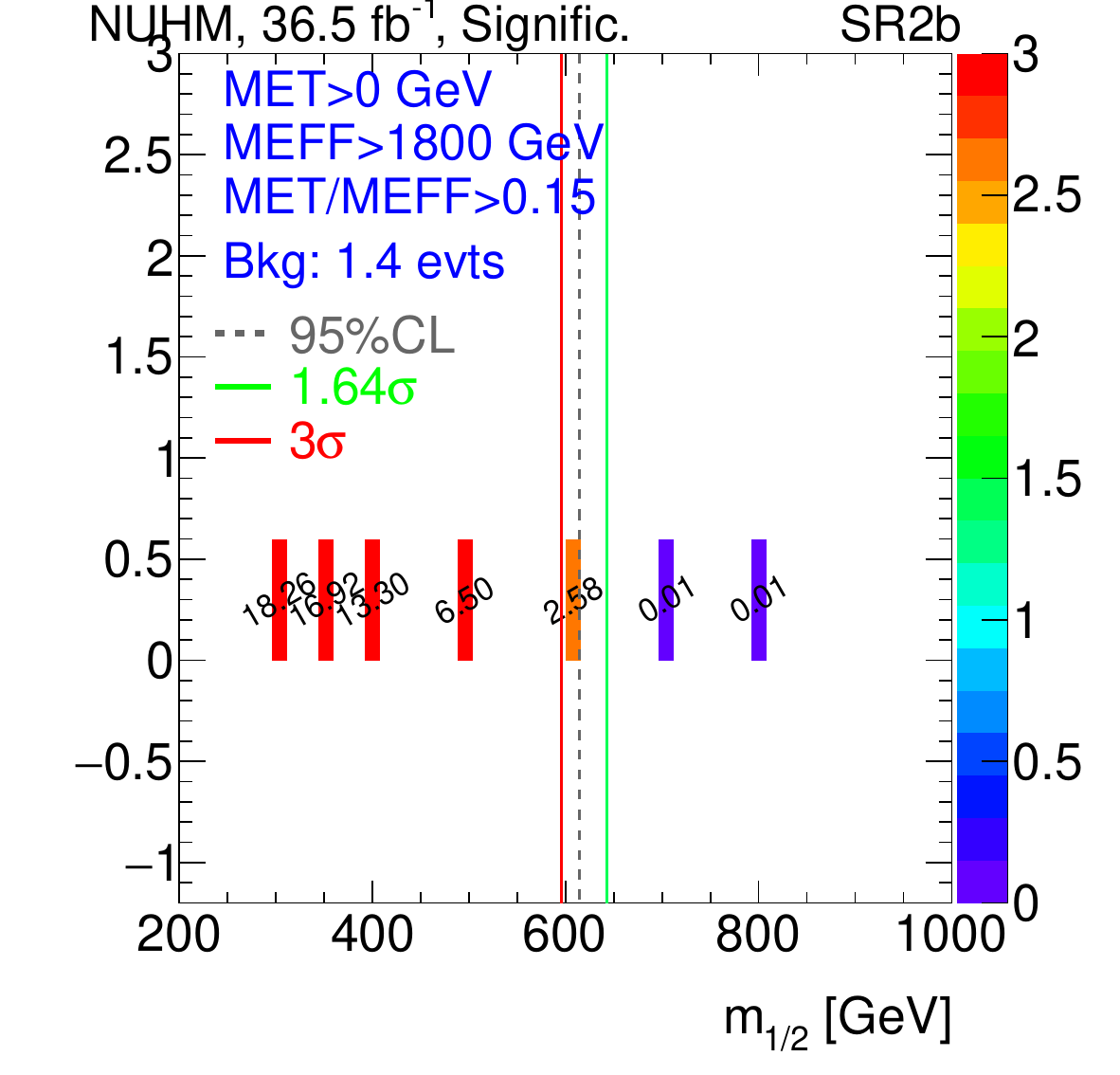}
\caption{Discovery significance for Rpc2L2bH signal region for 36.5~\ifb, NUHM2 model. The 95\% CL, 1.64$\sigma$, and 3$\sigma$ discovery contours from the proposed signal regions are shown in grey, green, and red, respectively.
}
\label{fig:SR_nuhm}
\end{figure}

Dedicated new SRs have been optimized for the gluino pair production with stop-mediated decay $\gl\to t\bar t\neut$ with off-shell tops.
The \glgl\ production with $\gl\to t\bar t\neut$ scenario at low LSP masses (where the multi-$b$ analysis has a much better sensitivity
\cite{ATLAS-CONF-2017-021}) is not the only motivation for Rpc2L2bH signal region, but also the NUHM2 model, which features large branching ratios for the $\gluino\to\ttbar\tilde{\chi}_{1,2,3}^0$ and $\gluino\to t\bar{b}\tilde{\chi}_{1,2}^\pm$ decays. 
As shown in Figure~\ref{fig:SR_nuhm}, with the Rpc2L2bH signal region,  $m_{1/2}$ values of 600~\GeV~can be excluded at 95\% CL or observed with a 3$\sigma$ significance.
This SR will be then used for the first interpretation in this model.


\begin{figure}
\centering
\begin{subfigure}[t]{0.48\textwidth}
\caption{$m_{\gl}$-$m_{\neut}$ mass plane}
\includegraphics[width=\textwidth]{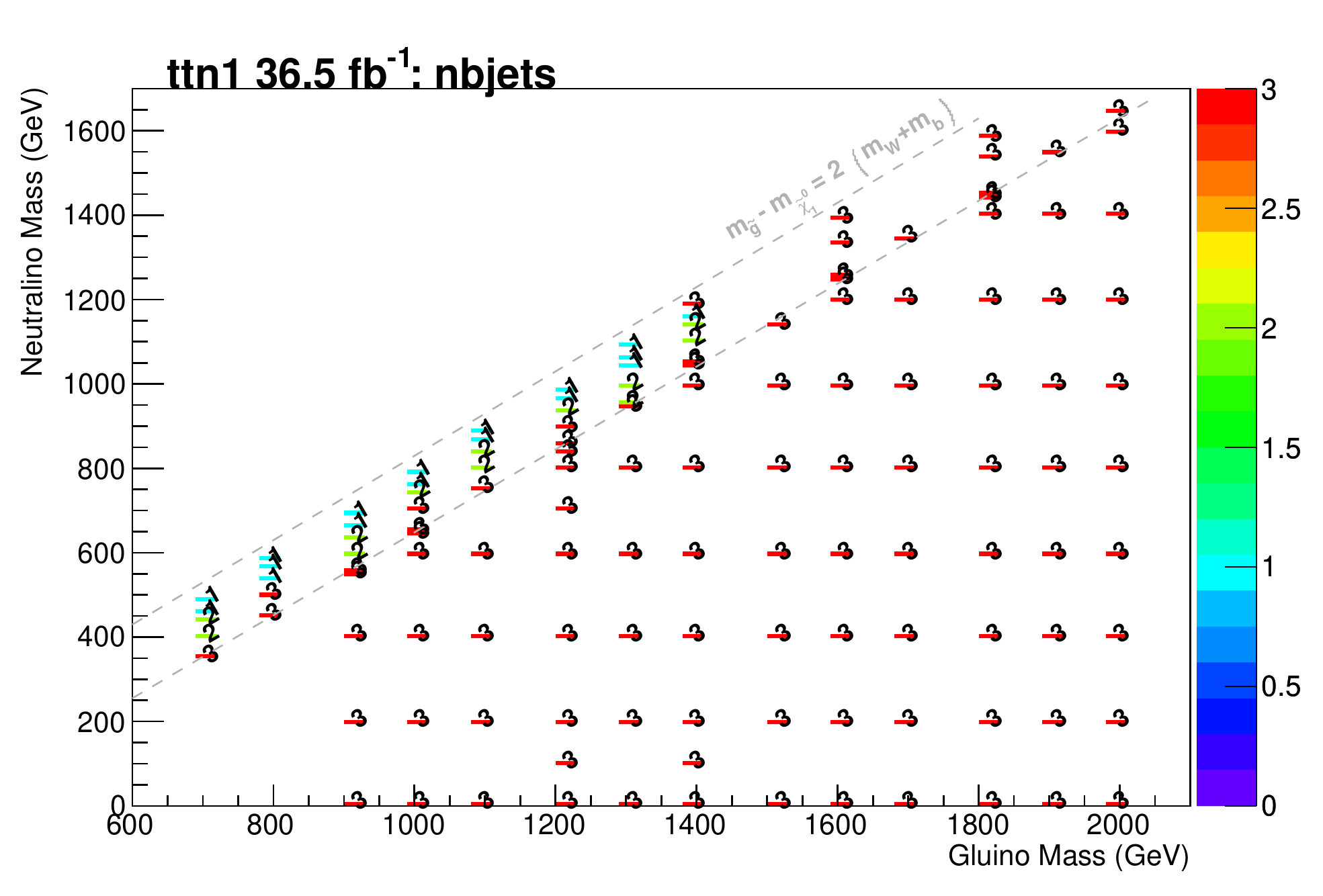}
\end{subfigure}
\begin{subfigure}[t]{0.48\textwidth}
\caption{Above the diagonal}
\includegraphics[width=\textwidth]{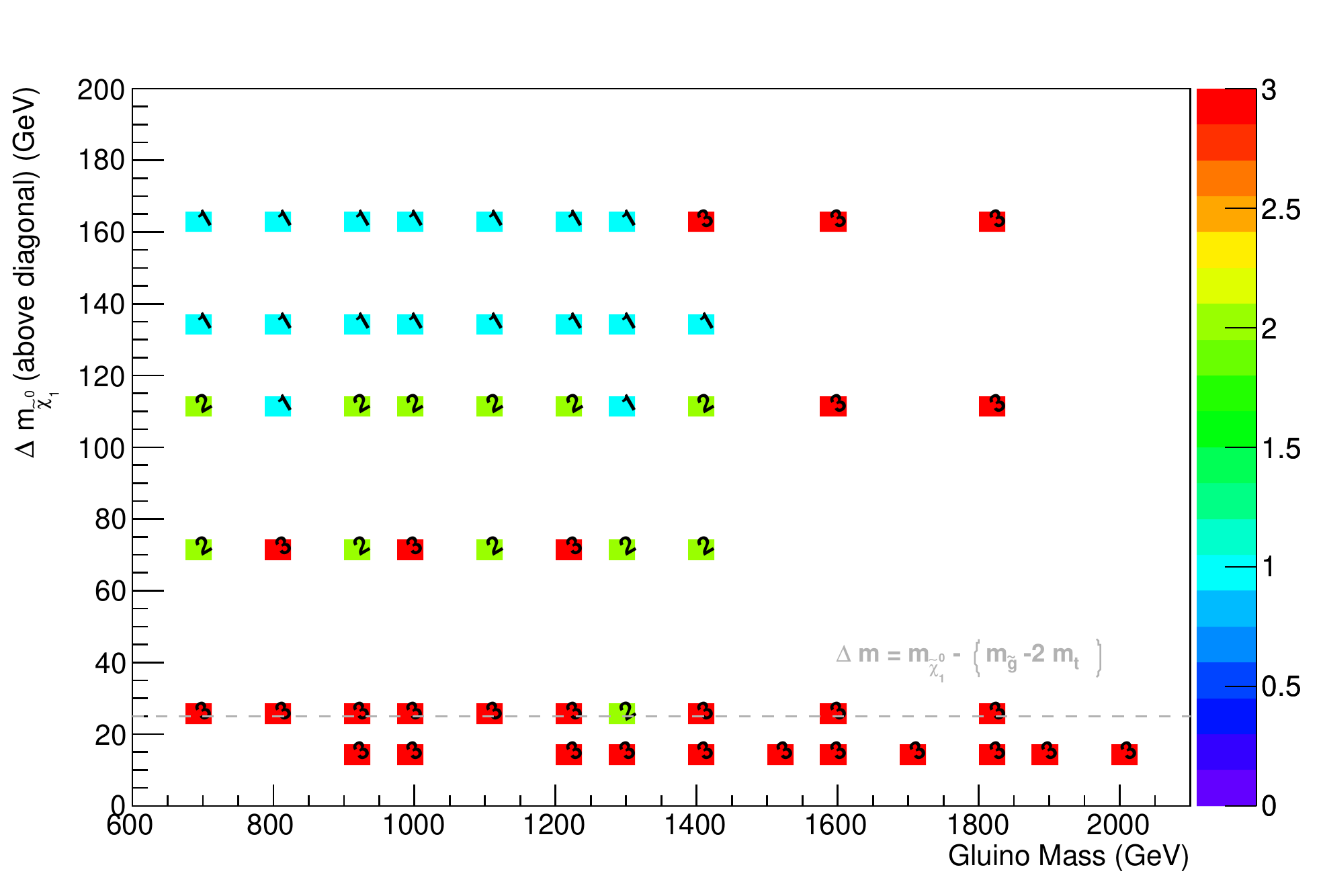}
\end{subfigure}
\caption{Optimal cut on the number of $b$-jets leading to the best discovery significance.
}
\label{fig:SR_Gtt_bjets}
\end{figure}



\begin{figure}
\centering
\begin{subfigure}[t]{0.49\textwidth}
\caption{Without $p_{\rm T}^{\ell_1}$ uppercut}
\includegraphics[width=\textwidth]{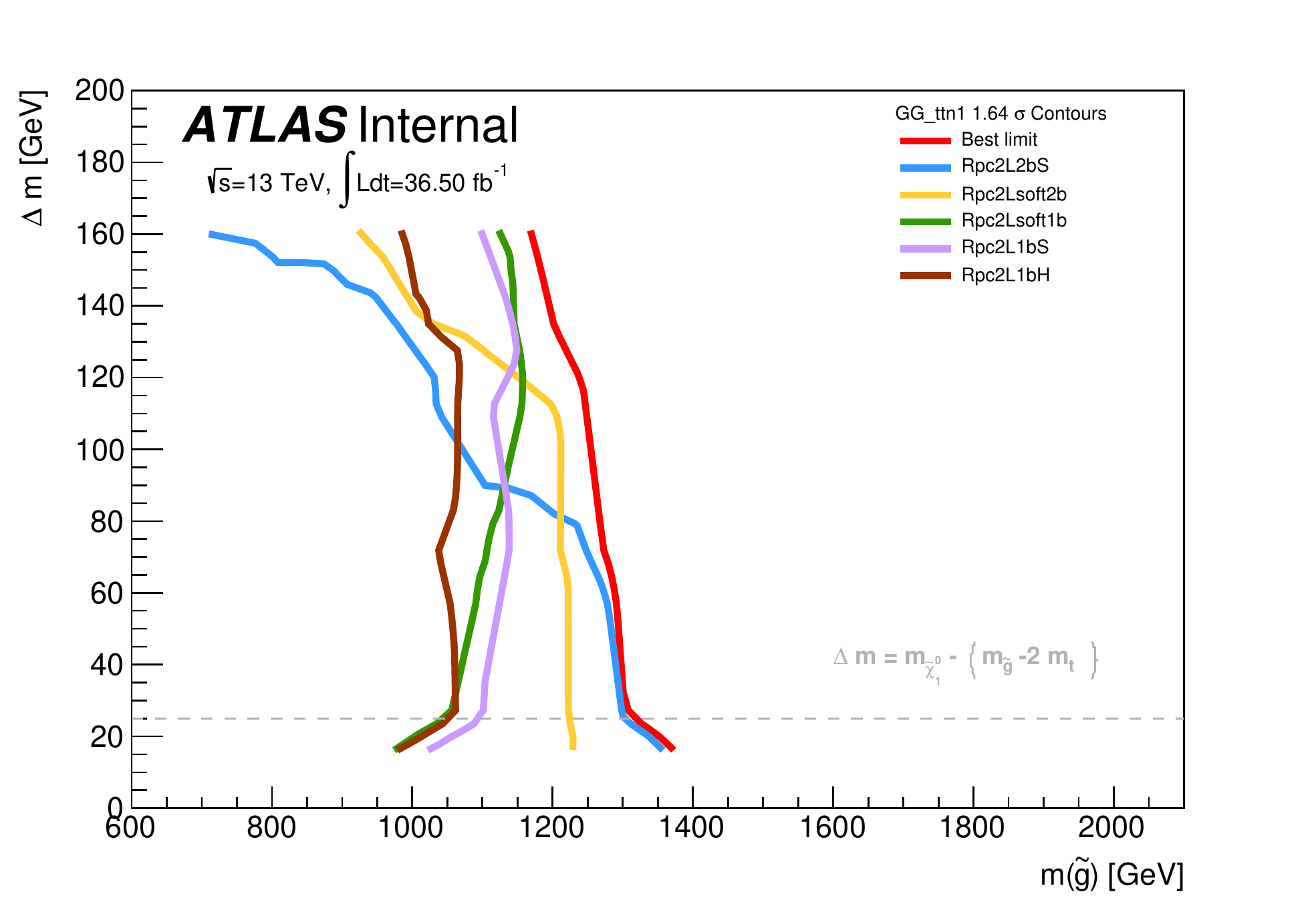}
\end{subfigure}
\begin{subfigure}[t]{0.49\textwidth}
\caption{With $p_{\rm T}^{\ell_1}$ uppercut}
\includegraphics[width=\textwidth]{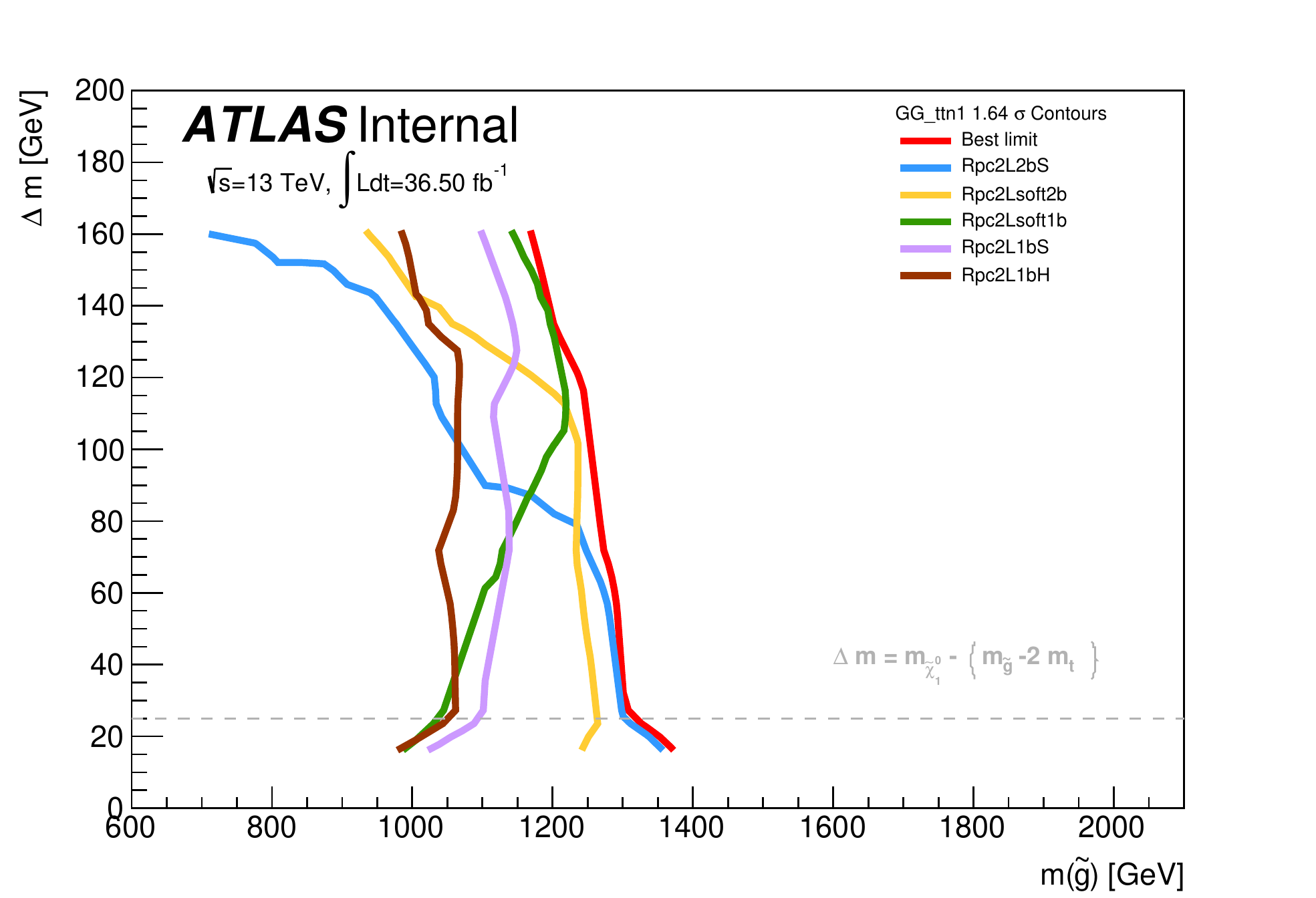}
\end{subfigure}
\caption{Comparison of significance contours at 1.64$\sigma$ for 36.5~\ifb~ between Rpc2Lsoft2b and Rpc2Lsoft1b and other 
signal regions in the off-diagonal region (left) without and (right) with an upper cut on the leading lepton \pt.}
\label{fig:SR_ptbound}
\end{figure}

In addition, the SS/3L analysis has the unique potential to explore the region of phase space at high LSP masses with a more compressed 
spectra. This scenario leads to softer decay products, in particular softer $b$-jets as seen in Figure~\ref{fig:SR_Gtt_bjets}, 
which makes the multi-$b$ analysis less sensitive. For this reason, two additional signal regions were introduced with at least 1 $b$-jet 
(Rpc2Lsoft1b) or 2 $b$-jets (Rpc2Lsoft2b) defined in Table~\ref{tab:SRdef3}. In addition, these signal regions are defined with an upper 
cut on the leading lepton \pt. The sensitivity is degraded if this upper cut is removed as shown in 
Figure~\ref{fig:SR_ptbound}.

Motivated by the $\stop$ production with $\stopone\to\neuttwo W$ model in Section~\ref{subsec:signals_3lss}, 
the signature of three leptons with the same electric charge (3LSS) is explored for the first time. As shown in Figure~\ref{fig:SR_3lss_incl}, 
after an inclusive 3LSS selection, the background is dominated by dibosons and $Z$+jets (with only one real lepton, and the two other leptons with either an electron
 with charge mis-identified or a fake lepton) both dominantly without $b$-jets. Once a $b$-jet requirement is applied in Figure~\ref{fig:SR_3lss}, the background is dominated by $\ttbar V$, 
with a clear peak at $m_{\ell\ell}\approx m_Z$ showing that a large fraction of these events are originated from charge mis-identification from events containing $Z\to ee$. 
After applying a $81<m_{e^\pm e^\pm}<101$~\GeV~veto, the background is reduced to only 1.7 events for 36.5~\ifb, almost removing the $Z$+jets and diboson backgrounds completely.
 The final background is dominated by $\ttbar+H,Z,W$, with $\sim$60\% originating from charge flips and $\sim$40\% from fakes and non-prompt leptons. 
With these very generic selections (Rpc3LSS1b in Table~\ref{tab:SRdef3}), a significance of 3.7$\sigma$ can be obtained for $m_{\stop}=550$~\GeV.
Figure~\ref{fig:SR_3lss_final} shows some lepton distributions, including the number of electrons, where most of the charge flip background populates the bins with 
2 or 3 electrons, although cutting away those bins would also have a large impact on the signal.
\begin{figure}[htb!] 
\centering
\includegraphics[width=0.85\textwidth]{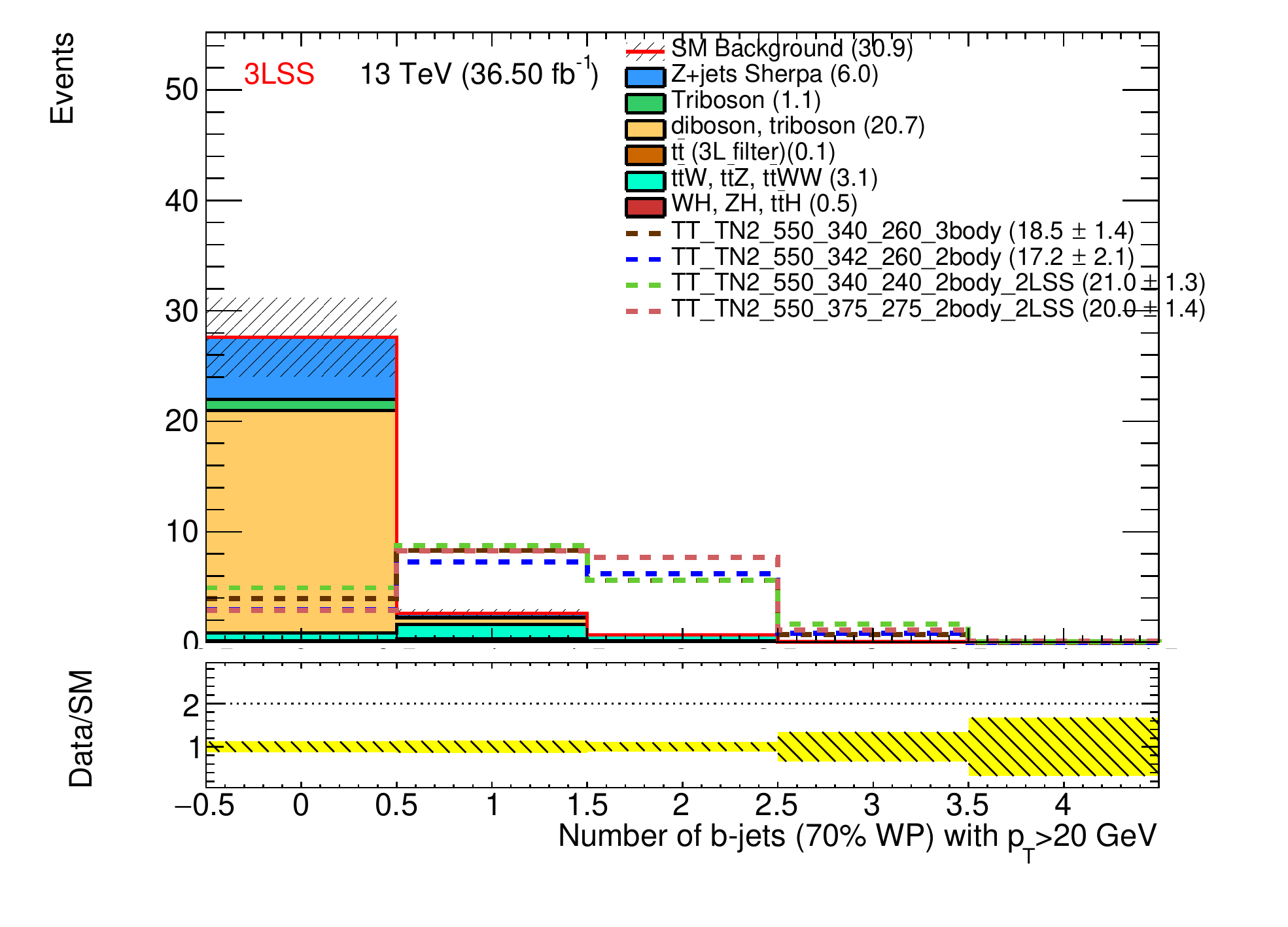}
\caption{$b$-jet multiplicity after a 3LSS selection. 
The background distributions are stacked, while the lines show the predictions for four signal points at $\stop$ mass of 550~\GeV.} 
\label{fig:SR_3lss_incl}
\end{figure}
\begin{figure}[htb!]
\centering
\includegraphics[width=0.49\textwidth]{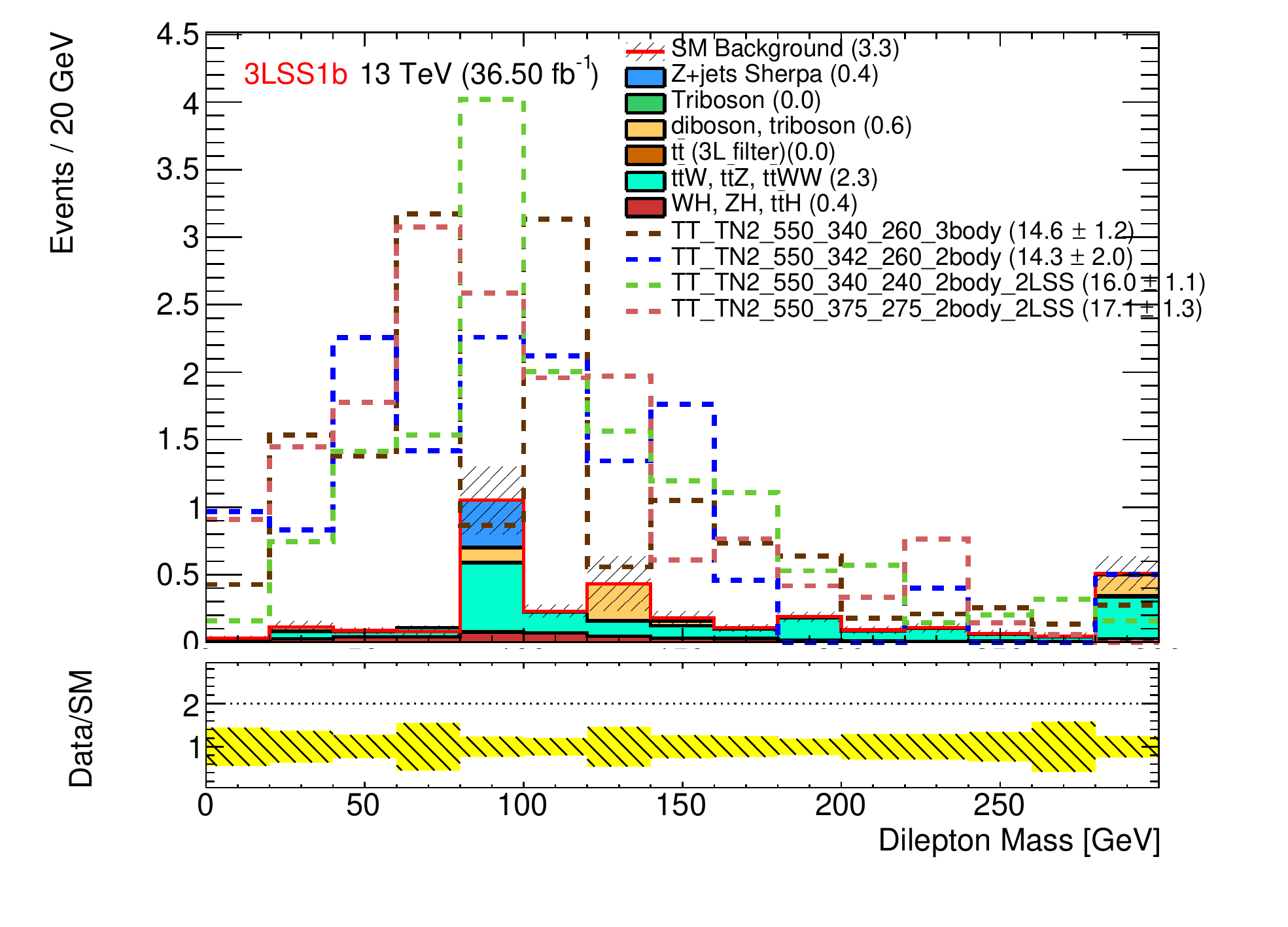} 
\includegraphics[width=0.49\textwidth]{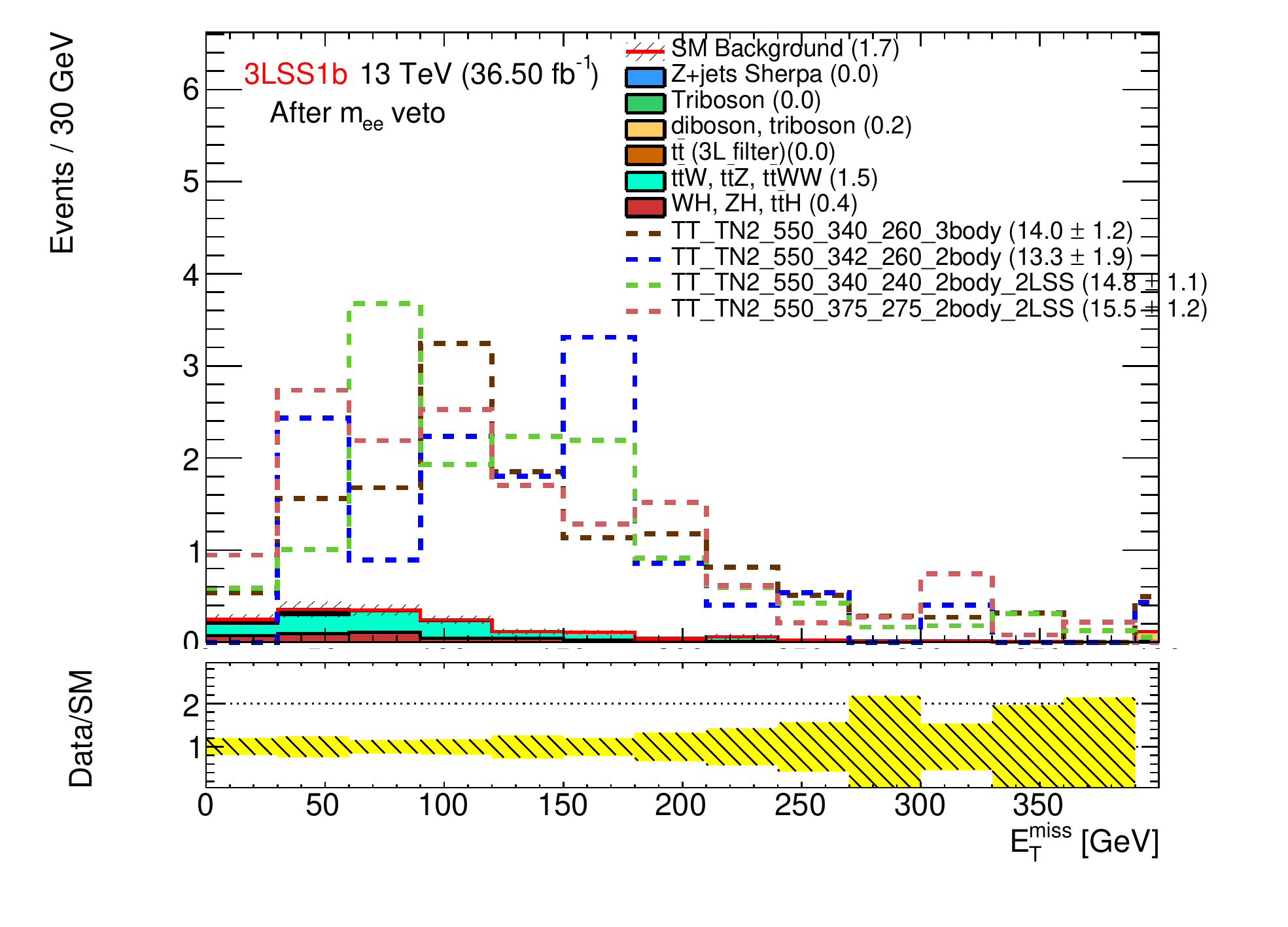}
\caption{Dilepton invariant mass distributions after a 3LSS plus $\geq$1 $b$-jet selection (right), and $\met$ distribution after a 3LSS, $\geq$1 $b$-jet and 
$81<m_{e^\pm e^\pm}<101$~\GeV~veto selection (bottom). 
}
\label{fig:SR_3lss}
\end{figure}
\begin{figure}[htb!]
\centering
\includegraphics[width=0.49\textwidth]{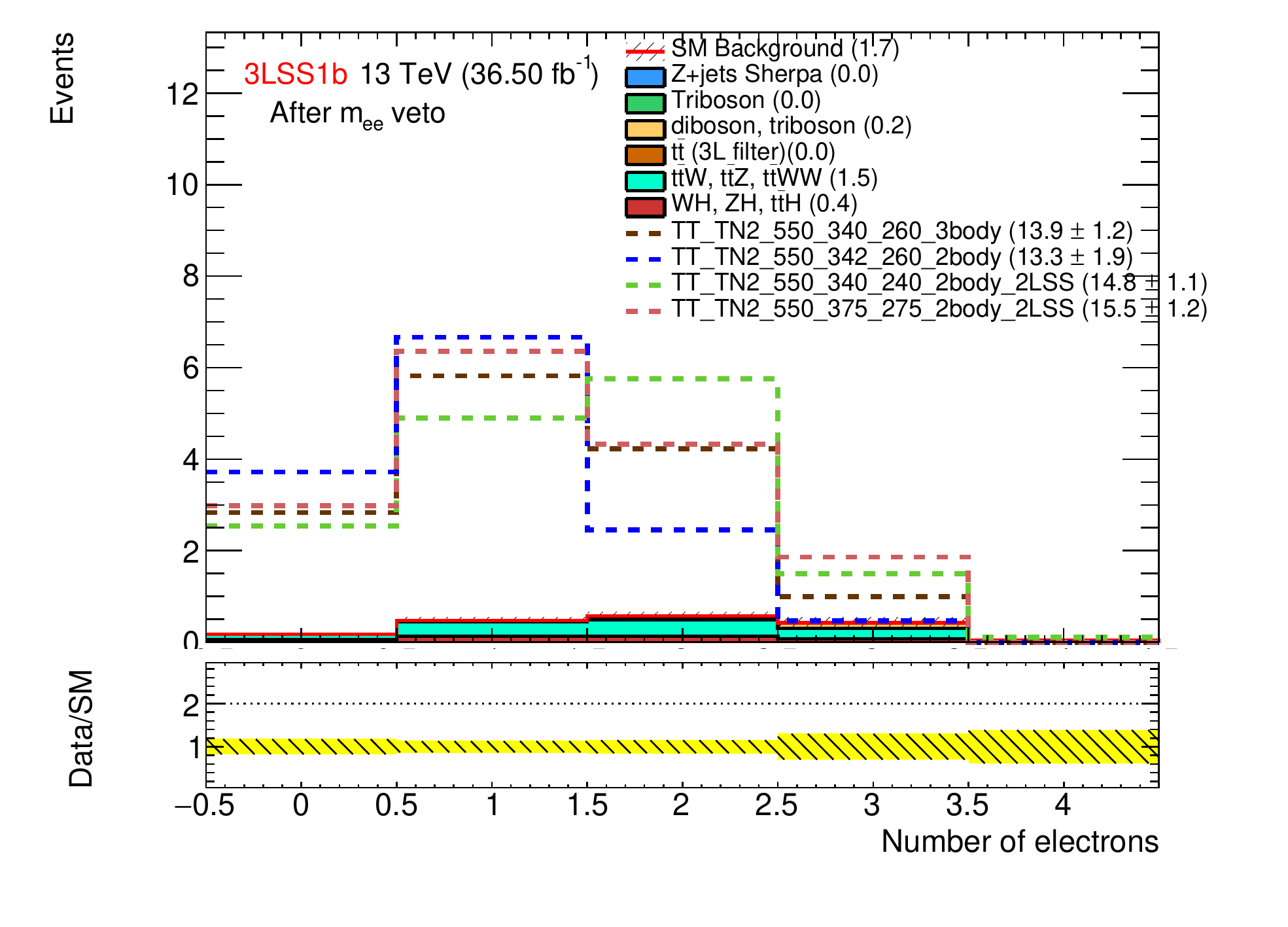} 
\includegraphics[width=0.49\textwidth]{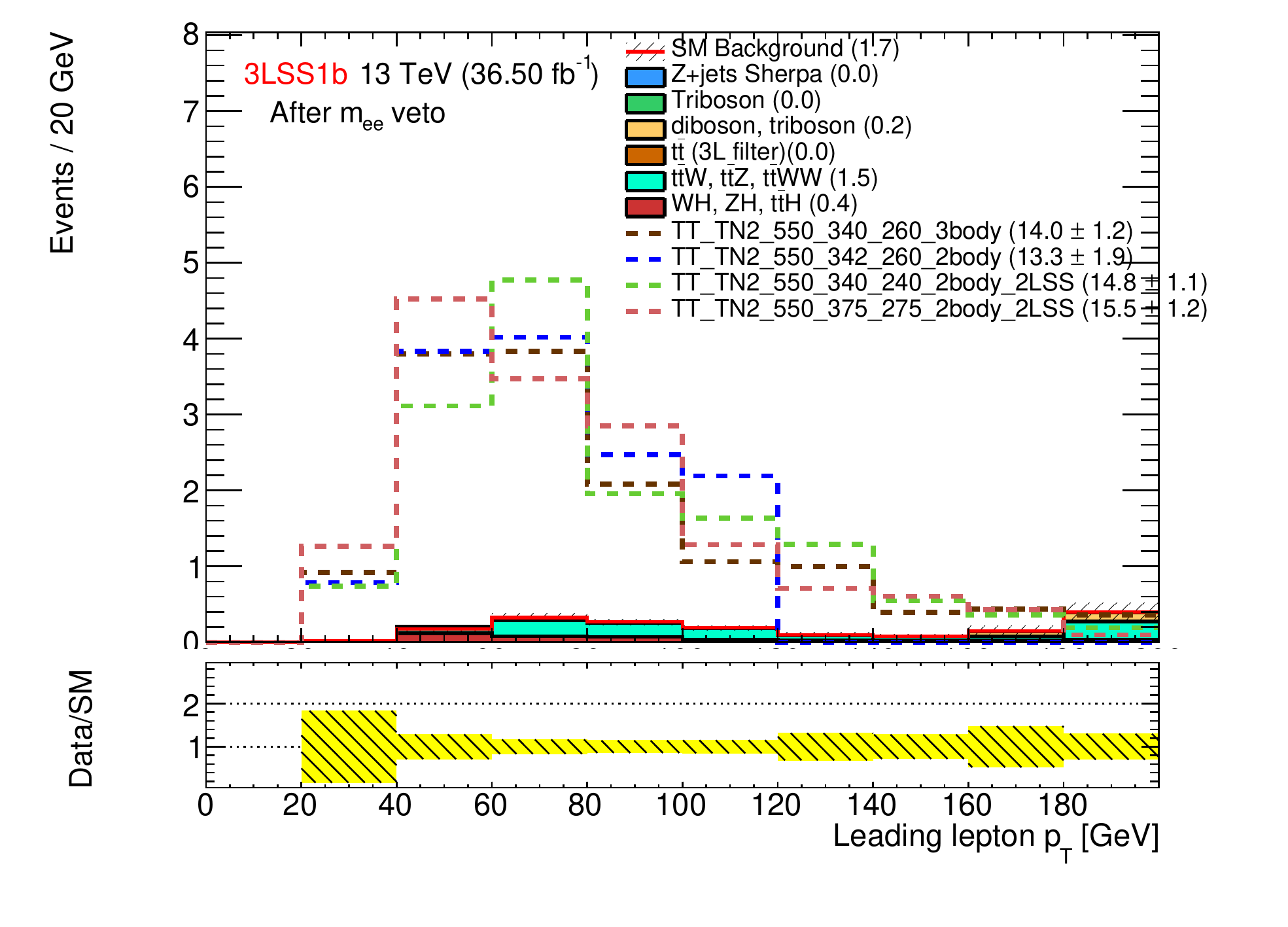} 
\includegraphics[width=0.49\textwidth]{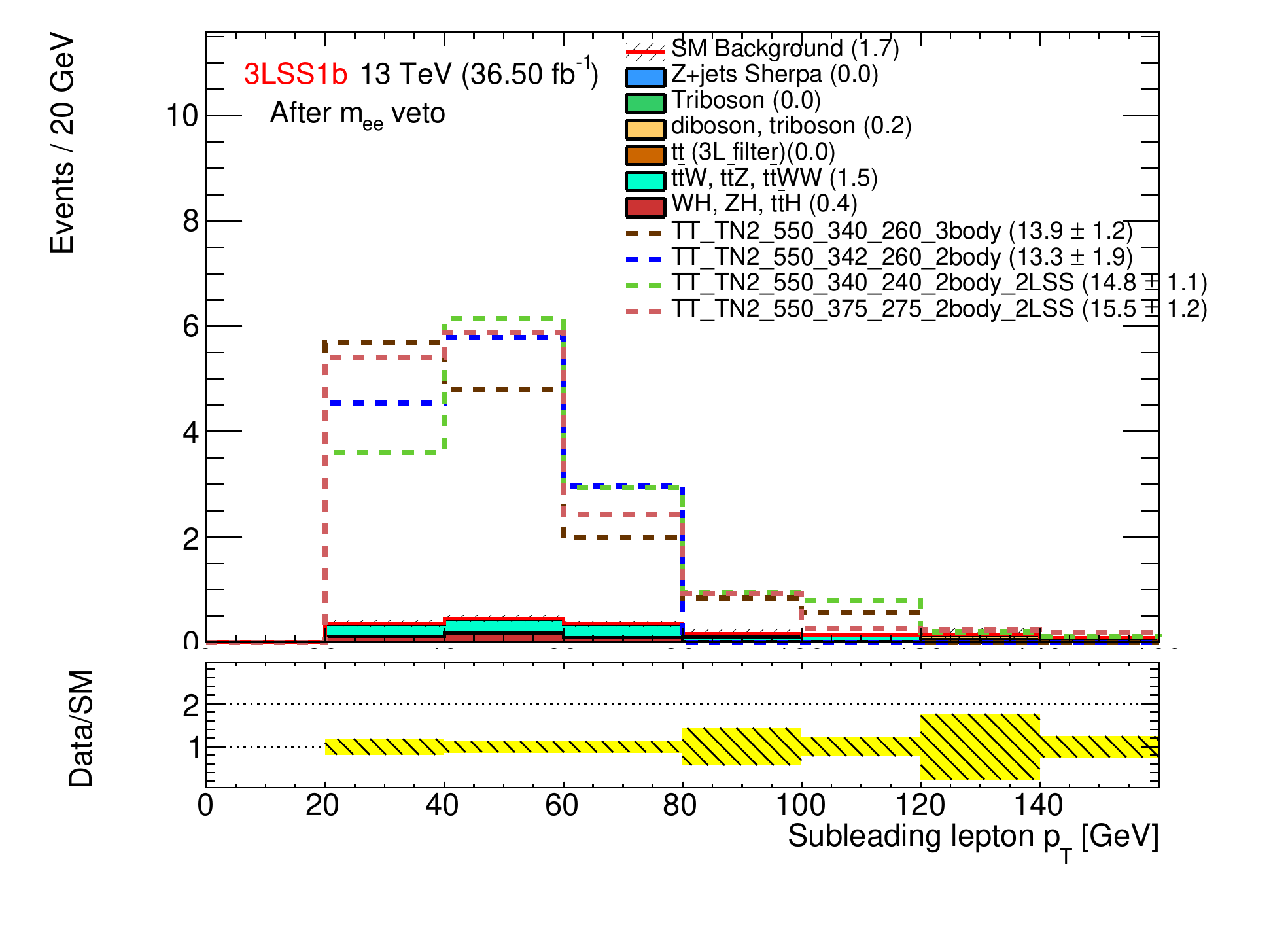} 
\includegraphics[width=0.49\textwidth]{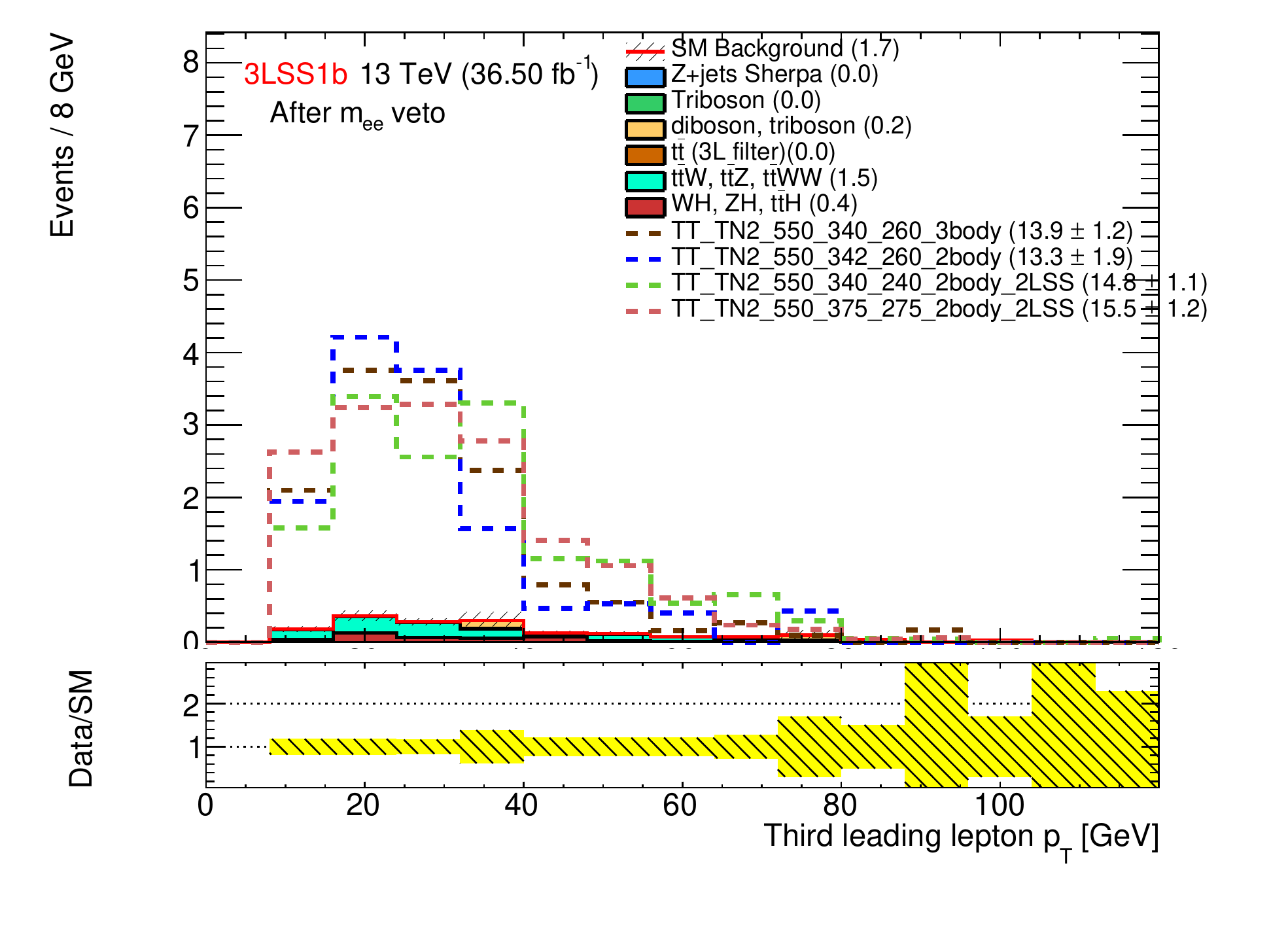} 
\caption{Number of electron (top left), and $\pT$ of the leading (top right), subleading (bottom left) and third leading lepton (bottom right) after a 3LSS, $\geq$1 $b$-jet and $81<m_{e^\pm e^\pm}<101$~\GeV~veto selection (bottom), all corresponding to 36.5~\ifb. The background distributions are stacked, while the lines show the predictions for four signal points at $\stop$ mass of 550~GeV.}
\label{fig:SR_3lss_final}
\end{figure}

Finally, since the SRs defined for the \glgl\ production with $\gl\to q\bar q\ell\bar\ell\neut$ feature a $b$-jet veto (Rpc3L0bS and Rpc3L0bH), 
and to avoid leaving uncovered the 3 lepton plus $b$-jets signature, SRs with the same kinematic cuts as Rpc3L0bS and Rpc3L0bH but with a $\geq$1 $b$-jet requirement are also proposed 
in Table~\ref{tab:SRdef3} as Rpc3L1bS and Rpc3L1bH. 

%% file: texfiles/sec.strategy.acc.tex
Based on the signal regions defined in Section~\ref{sec:strategy.sr}, 
it is useful to evaluate the signal acceptance of the analysis using 
parton level MC simulation (truth study). The acceptance encodes all the 
kinematic cuts applied on the signal as well as the branching ratios of 
all the decay particles. This information 
will help us understand the sensitivity reach of the analysis and also allow 
theorists to use this information when comparing the simplified model 
results to their models. 
The signal acceptance is shown in Figure~\ref{fig:strategy.accRpc2L0bH} with 
the rest of the 
signal regions shown in Appendix~\ref{app:aux.AccEff}.
Table~\ref{tab:strategy.cut} shows an example of a detailed cut-flow 
for weighted signal MC events illustrating the impact of cuts from the 
Rpc2L0bH signal region on a signal model. More tables are shown in 
Appendix~\ref{app:aux.SRcut}. 
\begin{figure}[htb!]
\centering
\includegraphics[width=\textwidth]{EffAcc/acceptance_2StepRpc2L0bH}
\caption{Signal acceptance for simplified models of $\gluino\gluino$ production with $\gluino\to q\bar q^{'}WZ\ninoone$ decays, 
in the signal regions Rpc2L0bH.}
\label{fig:strategy.accRpc2L0bH}
\end{figure}
\begin{table}[ht]\centering\def\arraystretch{1.2}\begin{tabular}{|l|c|}\hline
   \multicolumn{2}{|l|}{Rpc2L0bH,\quad$\gluino\gluino$ production,\quad$\gluino\to q\bar q^{'}WZ\ninoone$}\\
   \multicolumn{2}{|l|}{$m_{\gluino}=1.6 \TeV$, $(m_{\chargino} - 750) = (m_{\tilde\chi_2^0} - 375) = m_{\ninoone}=100 \GeV$}\\\hline
   MC events generated  & 20000 \\\hline
   Expected for 36.1 \ifb  & $2.9\times 10^2$ \\
   $\geq 2$ SS leptons ($\pt>20 \GeV$)  & $12.8 \pm 0.5$ \\
   Trigger  & $12.5 \pm 0.5$ \\
   no $b$-jet ($\pt>20 \GeV$)  & $8.5 \pm 0.4$ \\
   $\ge 6$ jets ($\pt>40 \GeV$)  & $7.12 \pm 0.35$ \\
   $\met>250 \GeV$  & $5.13 \pm 0.29$ \\
   $\meff>0.9 \TeV$  & $5.13 \pm 0.29$ \\
\hline\end{tabular}
\caption{Number of signal events at different stages of the Rpc2L0bH signal region selection. 
Only statistical uncertainties are shown.}
\label{tab:strategy.cut}\end{table}

Another quantity of interest, to experimentalists in particular, is the detector
 efficiency that entails the reconstruction and identification efficiencies 
of the different particles used in the analysis. The efficiency $\epsilon$ 
can be obtained from the relation 
\begin{equation}
S = L_\text{int}\cdot\sigma_{\text{prod}}\cdot A\cdot\epsilon, 
\end{equation}
where $S$ is the expected number of signal events, $\sigma_{\text{prod}}$ is the 
production cross section of the signal process, $L_\text{int}$ is the integrated 
luminosity, and $A$ is the acceptance. 
Figure~\ref{fig:strategy.effRpc2L0bH} shows the efficiency map for one of the signal models 
with the rest of the signal models shown in Appendix~\ref{app:aux.AccEff}.
\begin{figure}[htb!]
\centering
\includegraphics[width=\textwidth]{EffAcc/efficiency_2StepRpc2L0bH}
\caption{Signal reconstruction efficiency 
for simplified models of $\gluino\gluino$ production with $\gluino\to q\bar q^{'}WZ\ninoone$ decays, 
in the signal regions Rpc2L0bH.}
\label{fig:strategy.effRpc2L0bH}
\end{figure}

%% file: texfiles/sec.fake.prob.tex
The reconstructed objects (leptons, photons, $b$-jets, etc.) in a collision event are used to perform a wide range of SM measurements 
or searches for evidence of BSM physics. The assumption is that these objects are `real` representing the desired particles 
in the final state used in the analysis. 
In practice, the reconstructed objects might not be always `real'. In fact, they may be something completely different that
were mistakenly reconstructed as the desired objects, called `fake' objects.
While these occurrences are rare, they do affect some analyses more than others.
The analysis presented in this dissertation is highly affected by 
 `fake' leptons. 
To illustrate the problem, a hadronic jet may deposit more energy in the electromagnetic calorimeter than the hadronic calorimeter, 
or it may leave a narrow deposit of energy, leading the reconstruction algorithms to mistake this jet for an electron.
From the analysis point of view, the `fake' electron will pass all the selection criteria and will be indistinguishable from 
a `real' electron. 
It is important for the analysis that it requires a reconstructed electron to model the fake electron background to get a sound 
result. This example was given with electrons, but can be generalized to muons as well. 
In short, any analysis that uses leptons in the final state must account for the `fake' lepton background. 
This background can be more or less important depending on the detector, the analysis selection, and the number of leptons required. 
To estimate this background it is important to first understand what type of processes lead to fake leptons.

%% file: texfiles/sec.fake.proc.tex
The reconstruction of `fake` leptons can be an instrumental effect related to the inability to identify the object based on 
its measured properties by the detector. In this case, the reconstructed lepton is not a real lepton and the production process 
will be different for electrons and muons.

The reconstruction of electrons relies on the observation of well aligned particle hits in the layers of the ID that are consistent 
with an energy deposition in the EM calorimeter. Photons can mimic this signature since they deposit energy in the EM 
calorimeter that might happen to be aligned with a charged track. A jet for example containing charged and neutral pions can 
lead to such scenarios. It is possible for the jet to have one charged pion leaving a track similar to that of an electron.
The decay of $\pi^0$ mesons to photons in this jet can deposit energy in the EM calorimeter leading to the required signature.
Another mechanism that can lead to fake electrons is the emission of photons via Brehmstrahlung from high energy muons. 
The muon track can be mistaken for that of an electron and the photons interact with the EM calorimeter leading to a
signature similar to that of electrons. An additional process is that of photon conversions into a $e^+e^-$.

The reconstruction of muons relies on the observation of tracks from the ID matched to tracks from the muon spectrometer.
It is possible for charged hadrons with long lifetime to traverse the calorimeter layers and leave hits in the muon spectrometer.
These hits may coincide with other hits from the ID due to the random activity in the event. As a result, a muon can get
reconstructed. Another instance may occur when pions or kaons decay in-flight to muons in the muon spectrometer
and happen to align with the primary vertex.

The leptons that are used in the physics analyses must come from the hard scatter, generally referred to as prompt leptons.
Non-prompt leptons are really reconstructed leptons that did not originate from 
the hard interaction, and are considered to be fake leptons for the purpose 
of this analysis. Non-prompt leptons can be produced from heavy flavor meson decays with a low energy activity 
around the lepton which allows it to pass isolation requirements. A good example of this type of process is the 
semi-leptonic decay of top quark pairs which contribute to final states with two leptons. 

For the rest of the thesis, the fake leptons will be referred to as fake/non-prompt (FNP) leptons.
There are several methods used to perform the estimation of FNP lepton backgrounds. 
A method that the author developed will be described next along with a standard method for estimating this type of backgrounds.
The benefit of having two methods for estimating the FNP lepton background is to
have two independent methods relying on different assumptions to estimate the 
same quantity. The agreement between the two methods will give confidence 
that the final estimate is reliable. 
Moreover, the final estimate of the FNP lepton background is taken as a statistical combination of the estimates from the 
two methods leading to a reduction of the systematic uncertainties on the estimate.

%% file: texfiles/sec.fake.mct.tex
\subsection{Motivation}
\label{sec:fake.mct}
The processes leading to FNP leptons depend on the selection applied in the analysis. For instance, a selection with same-sign leptons 
will have contributions from top quark pair production (\ttbar) or the associated production of a vector boson and jets 
($W$+jets or $Z$+jets). These processes do not give two real leptons of the 
same electric charge but can contribute to this final state when 
there is a charge mis-measurement or a FNP lepton was produced.
It is possible to generate the processes that can contribute to a FNP lepton, such as \ttbar or 
$V$+jets, with Monte Carlo event generators processed through the Geant4 
detector simulation of the ATLAS detector.  
This approach will yield an estimate, however it might not be reliable. For instance, the detector simulation itself might not 
reproduce the true behavior of the interaction of the physics objects with the detector, particularly when looking at rare processes 
such as the production of FNP leptons. The second limitation is in the generation of enough MC events to probe the region of the 
phase space targeted by the analysis which affects the statistical uncertainties in the estimates.
The latter concern is addressed by ensuring that the simulations for the major backgrounds (\ttbar and $V$+jets) have much 
higher event count than the corresponding number of events observed in the data sample.
In fact, these backgrounds have a large number of simulated events because they are important for many analyses 
(including SM measurements and BSM searches).
The rest of the section will concentrate on addressing the former limitation. 





\subsection{Description of the method}

The MC template method relies on the correct modelling of FNP leptons kinematics in 
MC simulation to extrapolate background predictions from control regions to the signal regions.
The method assumes that the kinematic shapes for each source of FNP lepton are correctly modeled in the simulations, 
and the normalization for each source is extracted in a combined fit to data control regions.
The number of normalization factors depends on the number of identified origins of the FNP lepton in the signal regions
and the control regions are designed to constrain these factors in regions enriched with FNP leptons from the same origin.

\begin{figure}[t!]
\centering
\includegraphics[width=0.7\textwidth]{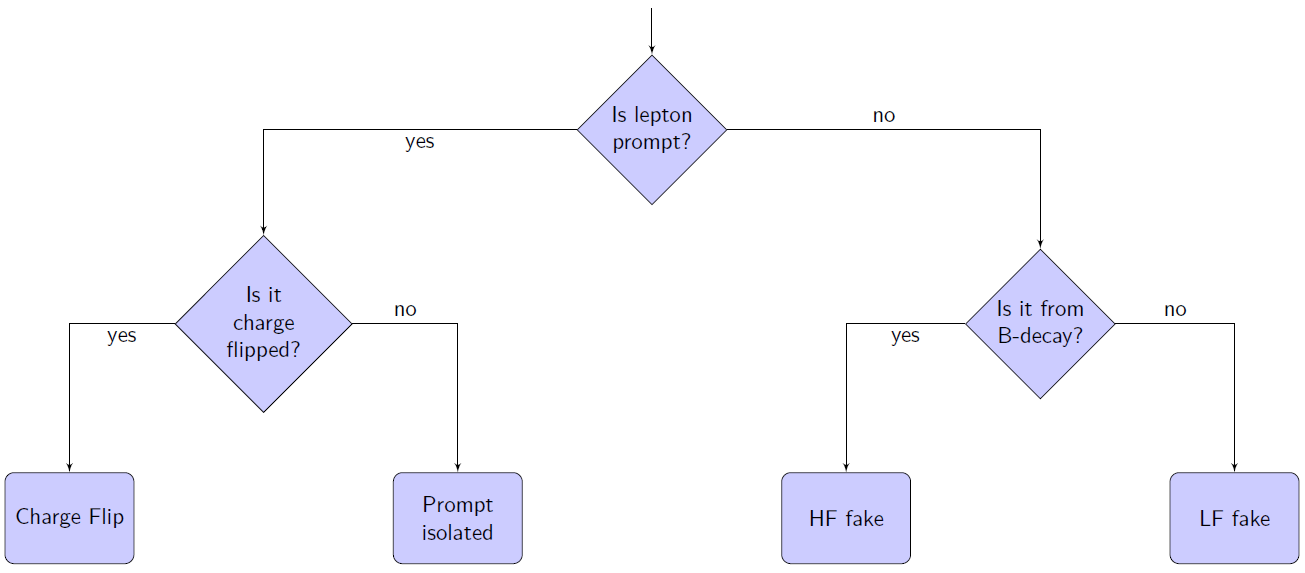}
\caption
{Lepton classification.
}
\label{Figurefakes_classification}
\end{figure}

To illustrate the approach, we describe the application of the method in 
the SS/3L analysis later described in this thesis.
The processes of interest that may lead to a FNP lepton or a charge flip are $\ttbar$ and $V$+jets. 
FNP leptons are classified using an algorithm that navigates the generator particle record to determine where the FNP lepton 
originated from. 
The lepton is classified as either an electron or a muon that is prompt from decays of on-shell $W$ and $Z$ bosons, 
non-prompt from a heavy flavor $b$ decay (HF), or fake from mis-identification of a light flavor jet or a photon (LF). 
In the case of an electron, we further classify the prompt electrons into electrons with the correct charge or with a 
charge mis-measurement, commonly named charge flip.
In total, five categories referred to as MC templates are constructed 
following the classification illustrated 
in Figure~\ref{Figurefakes_classification}.

\subsection{Correction factors}

The FNP estimate relies on kinematic extrapolation using processes expected to contribute via FNP leptons from control regions 
with low jet multiplicity and \met, to the signal regions that require high jet multiplicity and \met. 
The control regions are chosen to separate FNP leptons from HF origins and FNP leptons from LF origins.
For instance, a control sample characterized by the presence of a $b$-jet will be enriched in processes with one FNP lepton that is 
coming from a HF decay, while a sample characterized by the absence of a $b$-jet will have one FNP lepton from LF decay.
The presence of one FNP lepton in the control sample allows the correction of the production rate of these FNP leptons 
by performing a fit to data. 

For example, if a $Z\to\mu\mu$+LF jet event is reconstructed as a $\mu^+\mu^-e^+$ event, then the electron is fake.
Therefore, a correction of LF jet $\to e$ (Fr(LF$\to e$)) is applied to the rate of $\mu\mu e$ events. The correction Fr(LF$\to e$)
is constrained by a fit to data in control regions dominated by LF jet $\to e$ type fakes. 
Similarly, three other corrections are defined as LF jet $\to \mu$ (Fr(LF$\to\mu$)),  HF jet $\to e$ (Fr(HF$\to e$)),  
HF jet $\to\mu$ (Fr(HF$\to\mu$)). An additional correction is applied to correct the charge flip rate predicted by simulation.
For example, a $Z\to e^+e^-$ event is reconstructed as $e^+e^+$ or $e^-e^-$. The simulation takes into account the charge flip 
rate but the fraction of time it occurs may be wrong. The charge flip (Cf($e$)) correction, derived from a data fit, is expected to recover this mis-modeling.
The charge flip rate only concern electrons as the muon charge flip rate is negligible.

A likelihood fit is defined as the product of the Poisson probabilities describing the observed events in the binned 
distributions from the expected number of events rescaled by the five multipliers which are left free to float in the fit.  
These multipliers are applied to the MC predictions in the signal regions to obtain an estimation of the charge flip and FNP backgrounds.

\subsection{Control regions}

The corrections depend on the simulated sample,
the reconstructed final state, and the flavor of the leptons. As a result, care must be taken when designing the control regions 
used to perform the fit of the FNP leptons and electron charge flip templates. 
For instance, each template needs to be constrained in a selection that is representative of the processes leading to 
FNP leptons and charge flip electrons present in the kinematic region targeted by the search for BSM physics. 

In the SS/3L analysis discussed in this dissertation, the control regions are defined with at least two same-sign 
leptons, $\met>40$~GeV, and two or more jets. This pre-selection ensures that the FNP leptons are not from fakes originating from 
QCD like event topologies. 
They are further split in regions 
with or without $b$-jets to constrain the HF and LF leptons respectively. In addition, they are also split with different 
flavours of the same-sign lepton pair $ee$, $e\mu$, and $\mu\mu$, giving a total of six control regions. 
Any event entering the signal region is vetoed. The ee channel will constrain the charge flip correction factor, fake leptons 
 from LF decays in the selection without $b$-jets, and non-prompt decay from HF in the selection with $b$-jets. 
The $\mu\mu$ channel will constrain the muon fake rates in the LF and HF decays for the selection without or with $b$-jets, 
respectively. The $e\mu$ channel will constrain both the electron and muon fakes for events containing both lepton flavors. 

\begin{figure}[htb!]
   \includegraphics[width=.32\textwidth]{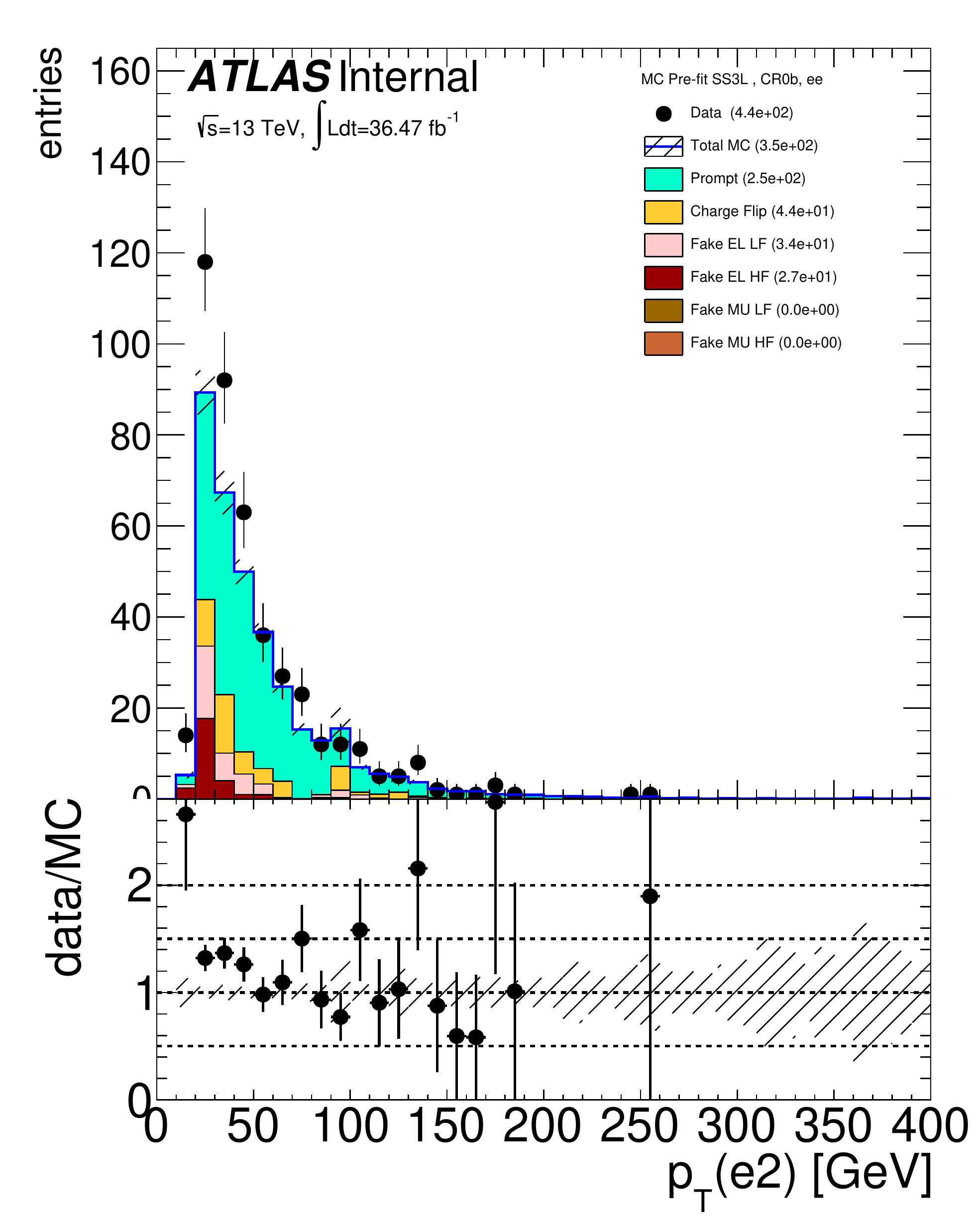}
   \includegraphics[width=.32\textwidth]{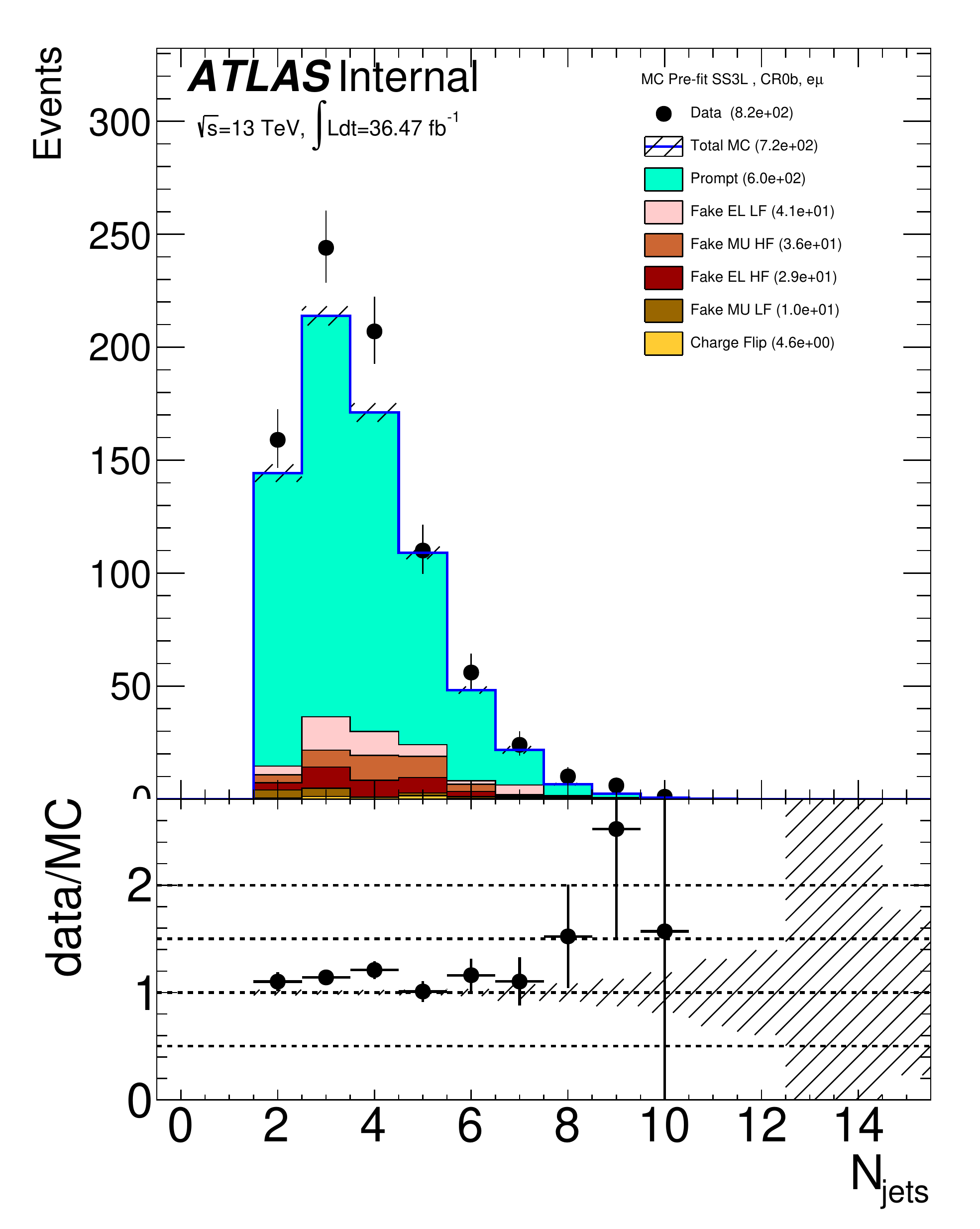}
   \includegraphics[width=.32\textwidth]{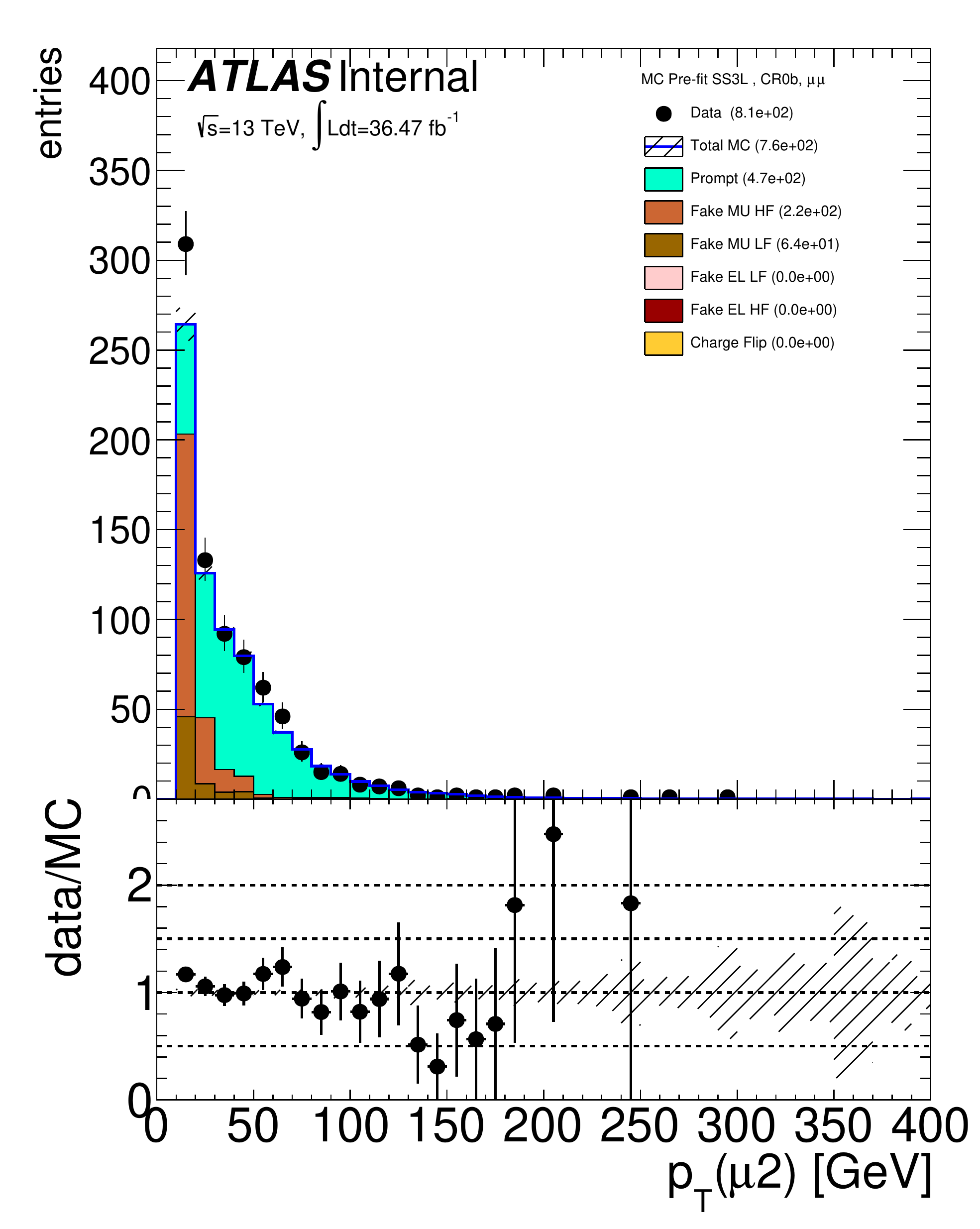}
 \caption{
 Pre-fit distributions for  $ee$ channel (left),  for  $e\mu$ channel (middle), and  for  $\mu\mu$ channel (right) from CR0b that were used in the fit to extract the FNP lepton and charge flip multipliers.
The generator used in these plots is  \POWHEGBOX+Pythia. The hashed band represents the sum of systematic uncertainties on the predictions.
 \label{f:prefit_CR0b}
 }
 \end{figure}

\begin{figure}[htb!]
  \includegraphics[width=.32\textwidth]{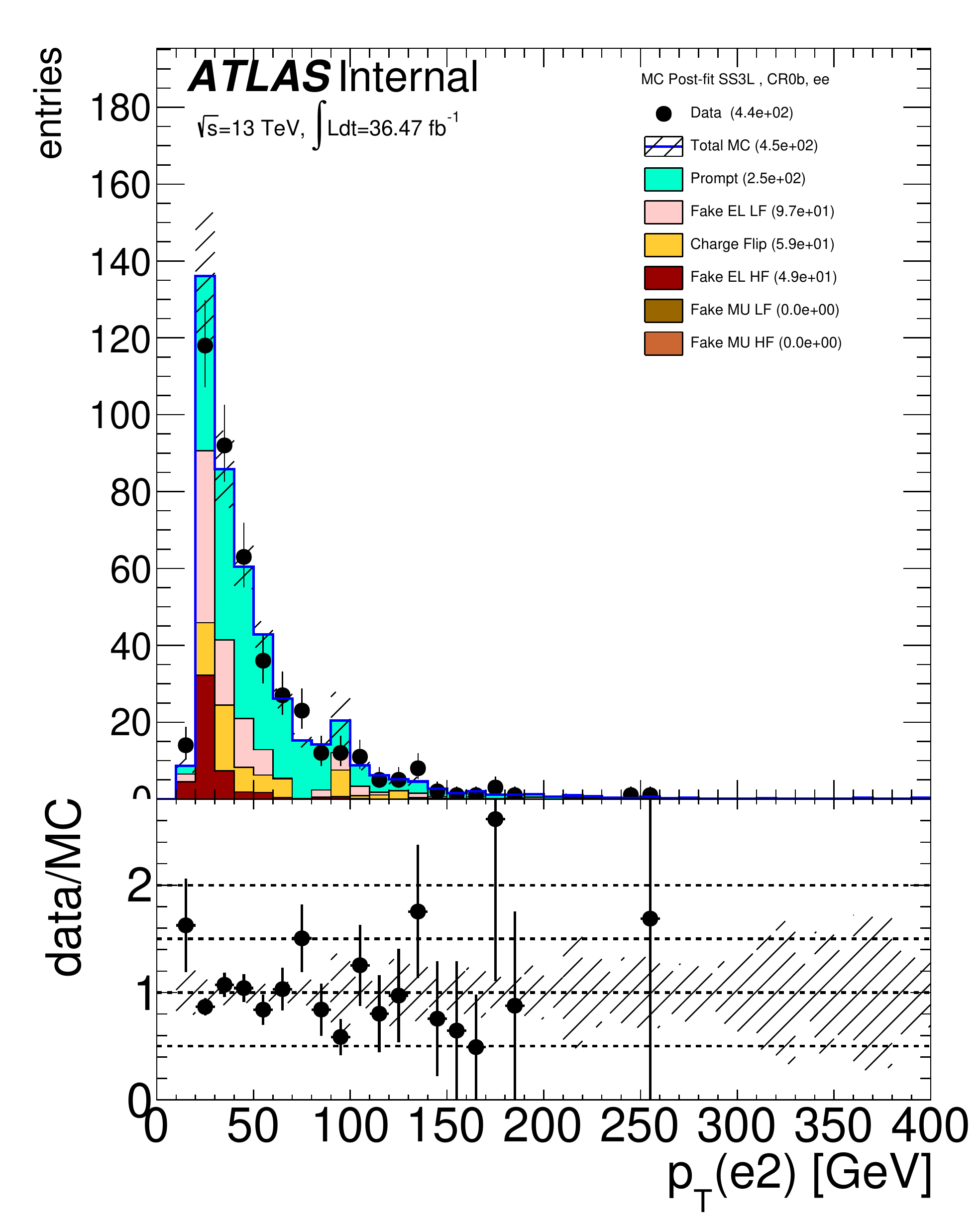}
  \includegraphics[width=.32\textwidth]{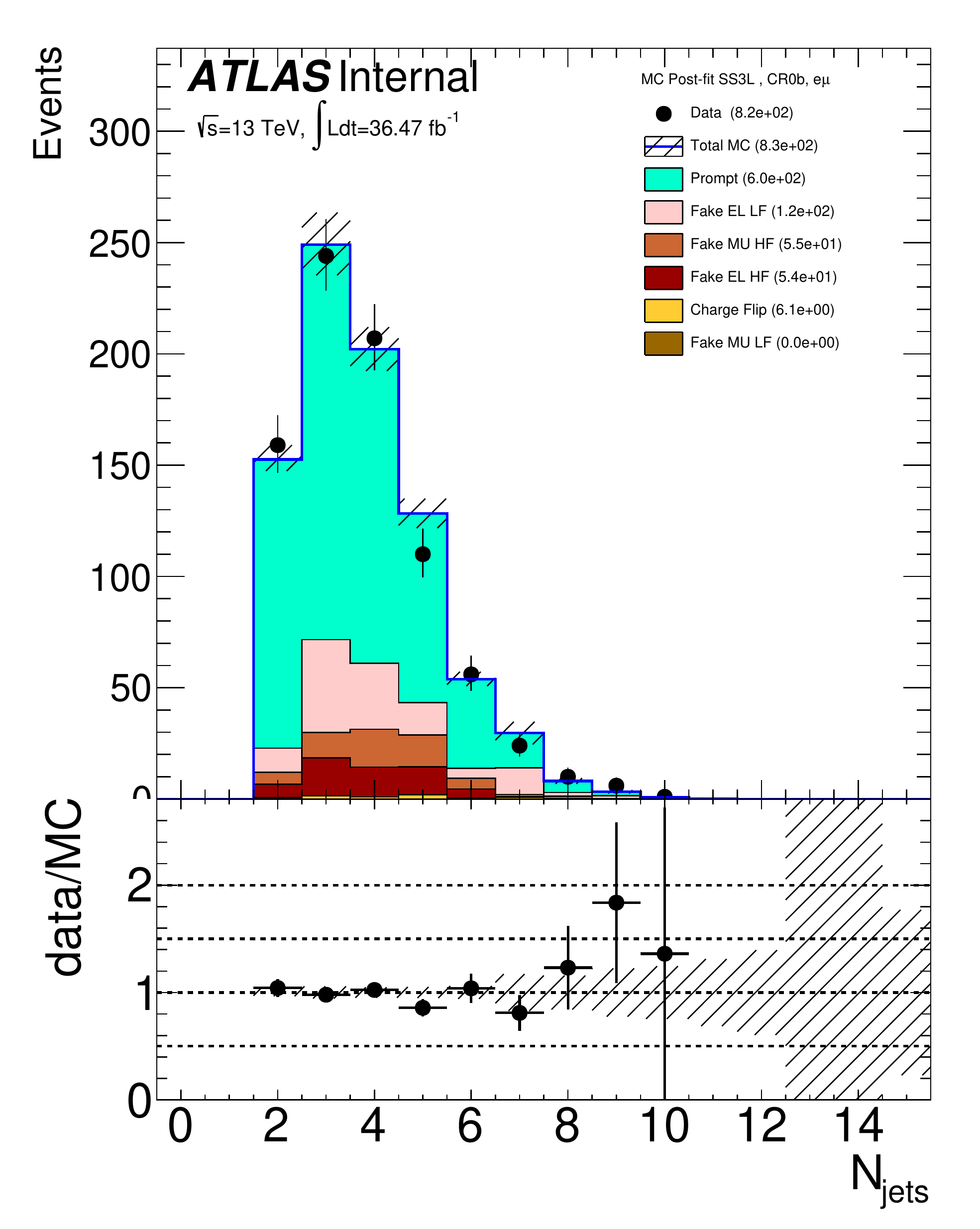}
  \includegraphics[width=.32\textwidth]{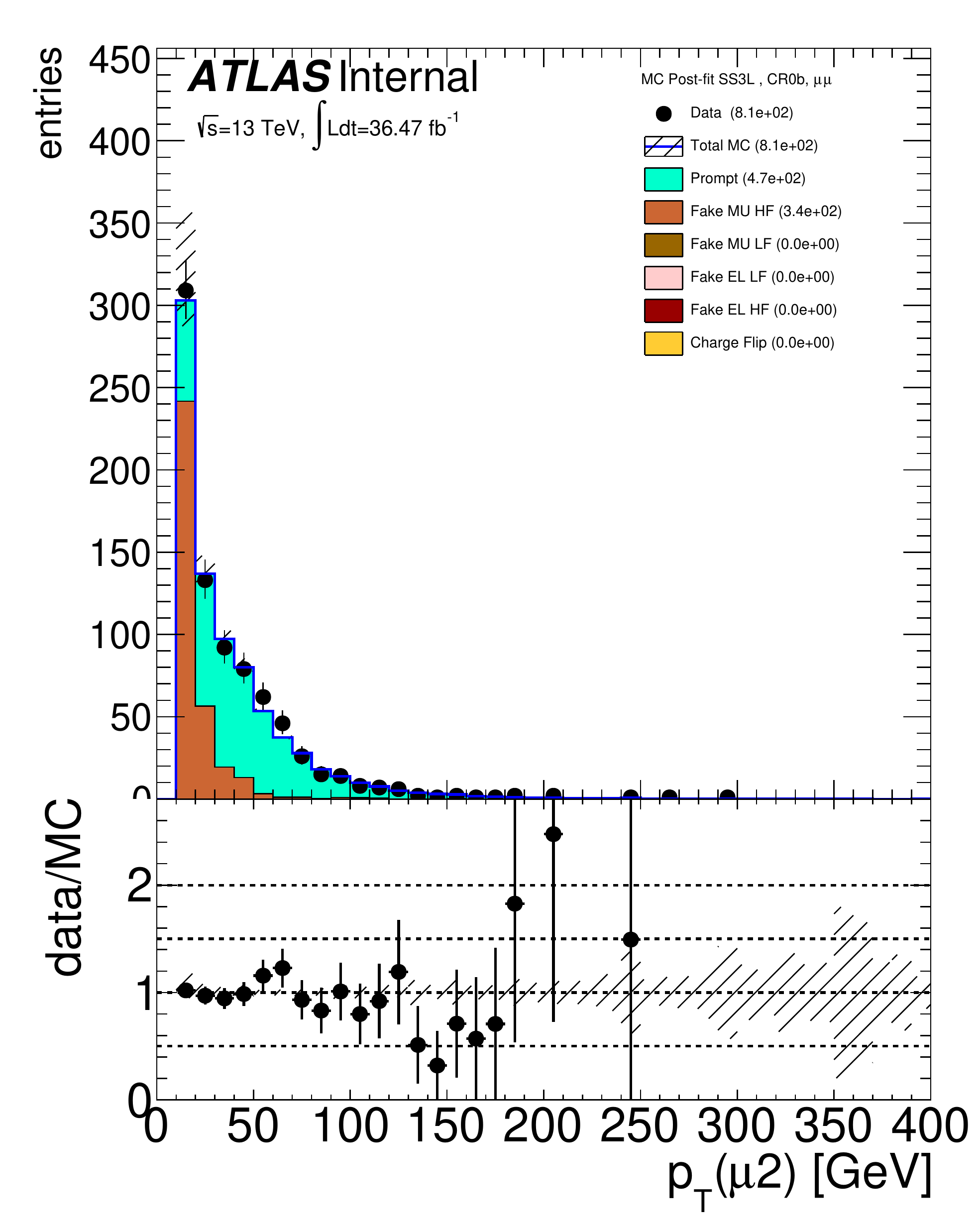}
\caption{
Post-fit distributions for  $ee$ channel (left),  for  $e\mu$ channel (middle), and  for  $\mu\mu$ channel (right) from CR0b that were used in the fit to extract the FNP lepton and charge flip multipliers.
The generator used in these plots is  \POWHEGBOX+Pythia . The hashed band represents the sum of systematic uncertainties on the predictions.
\label{f:postfit_CR0b}
}
\end{figure}

\begin{figure}[!htb]
   \includegraphics[width=.32\textwidth]{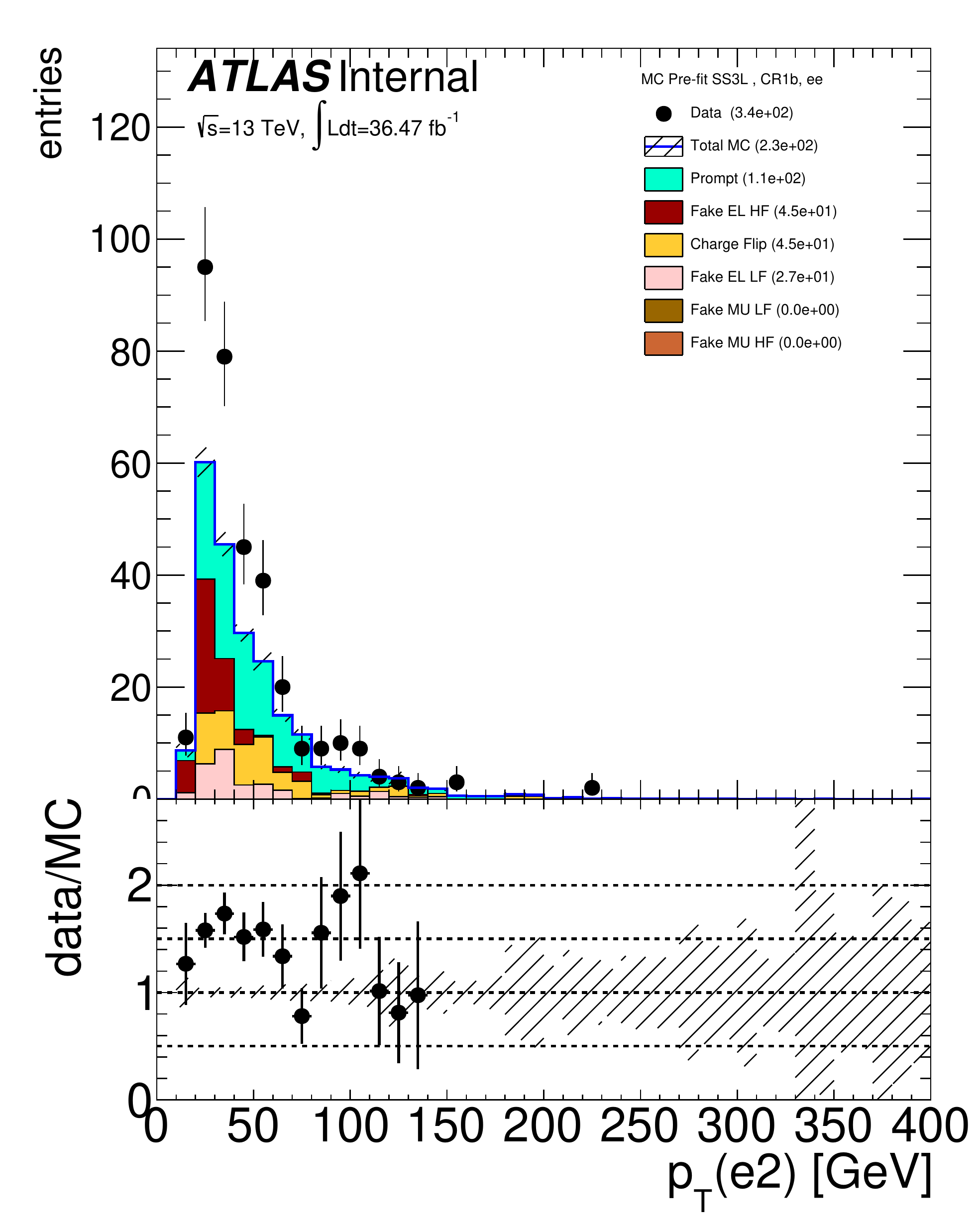}
   \includegraphics[width=.32\textwidth]{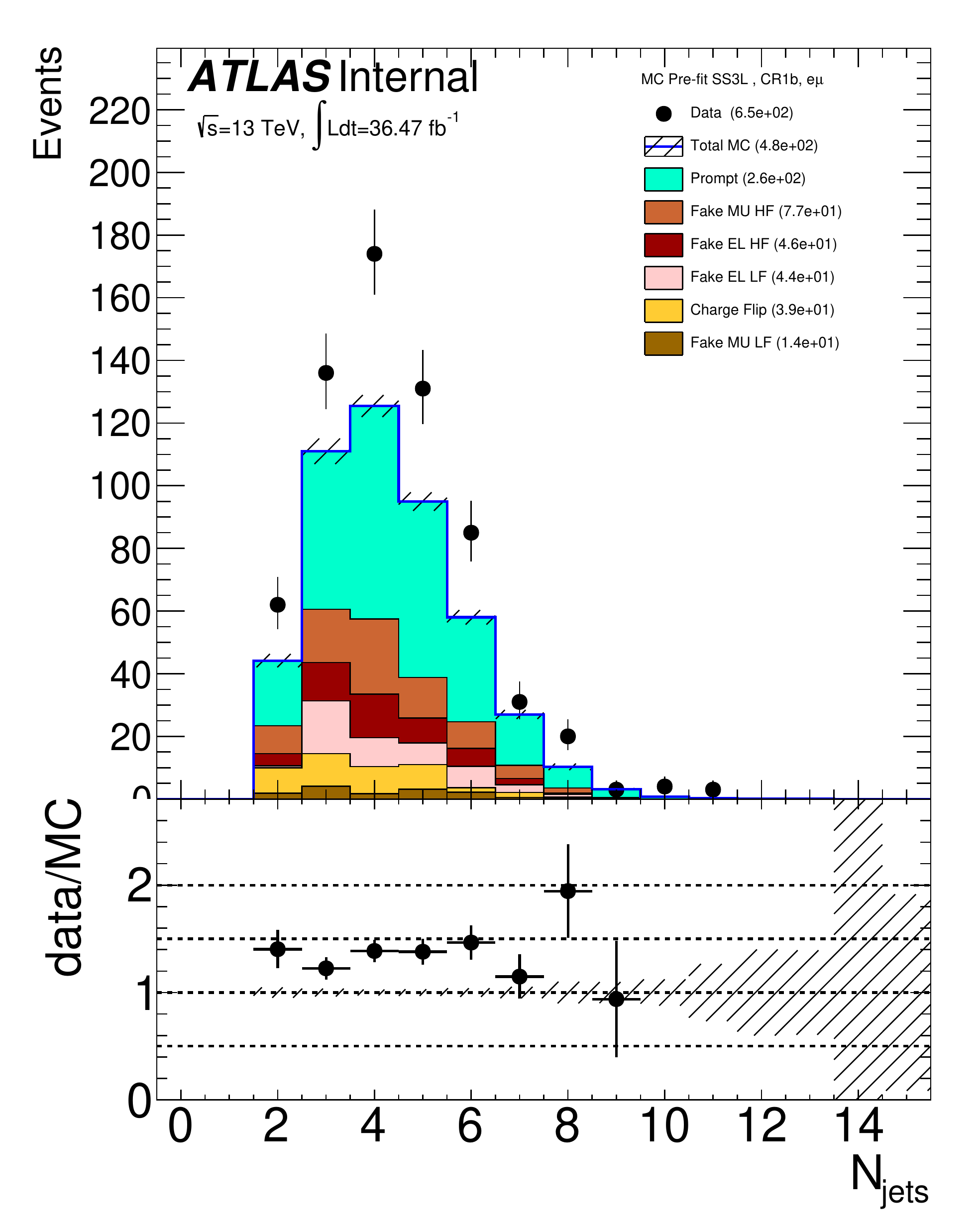}
   \includegraphics[width=.32\textwidth]{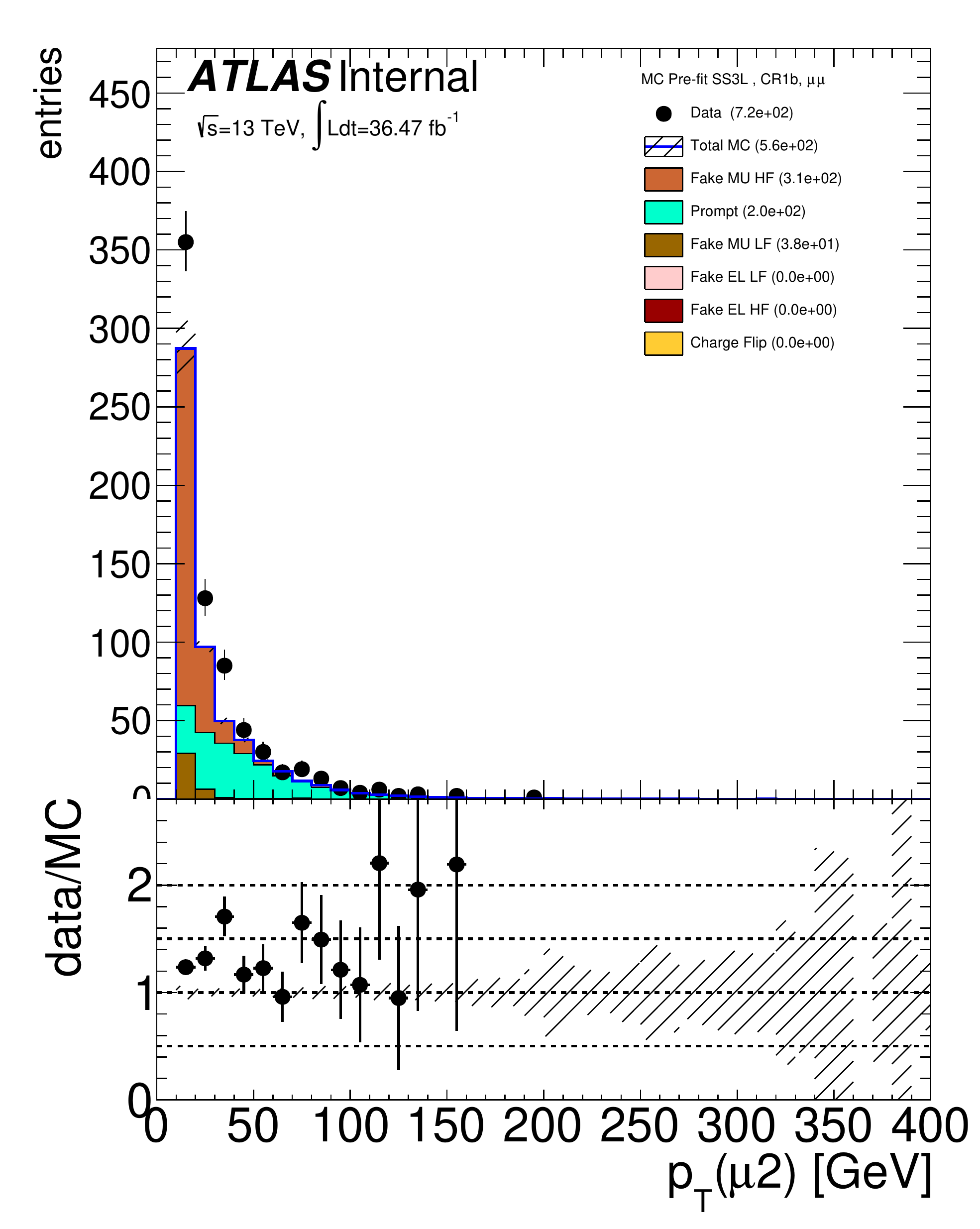}
 \caption{
 Pre-fit distributions for  $ee$ channel (left), for  $e\mu$ channel (middle), and  for  $\mu\mu$ channel (right) from CR1b that were used in the fit to extract the FNP lepton and charge flip multipliers.
The generator used in these plots is  \POWHEGBOX+Pythia. The hashed band represents the sum of systematic uncertainties on the predictions.
 \label{f:prefit_CR1b}
 }
 \end{figure}

\begin{figure}[!htb]
  \includegraphics[width=.32\textwidth]{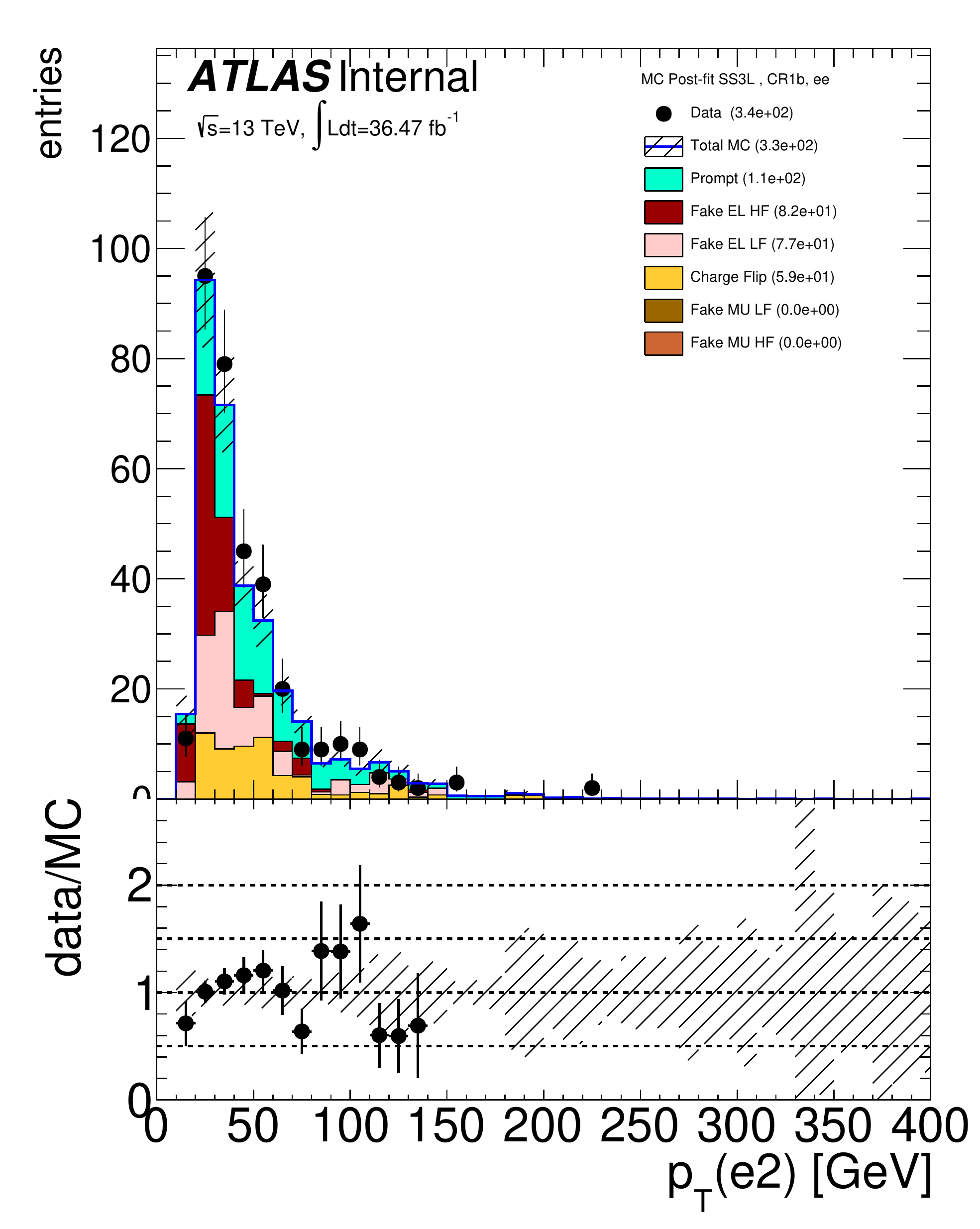}
  \includegraphics[width=.32\textwidth]{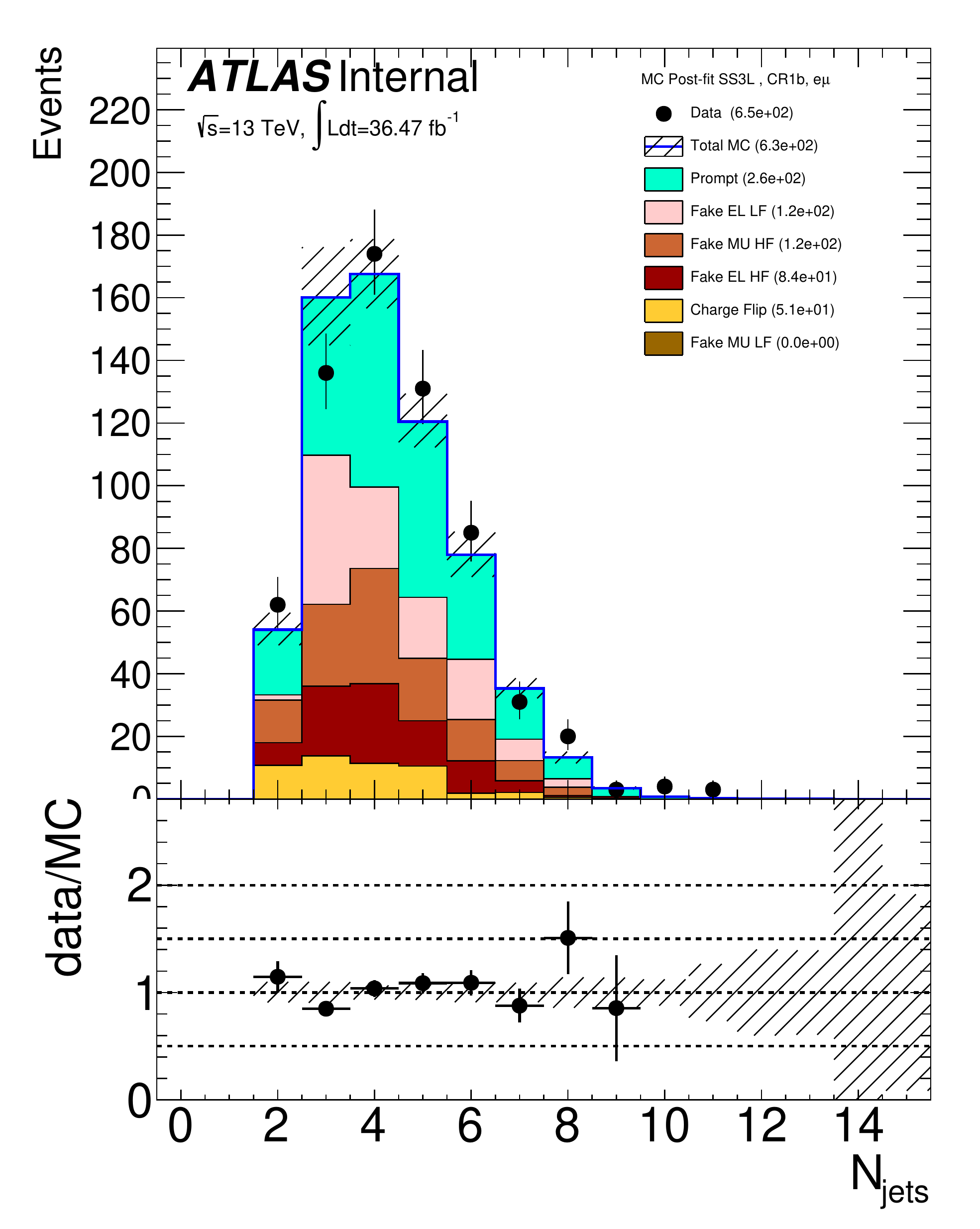}
  \includegraphics[width=.32\textwidth]{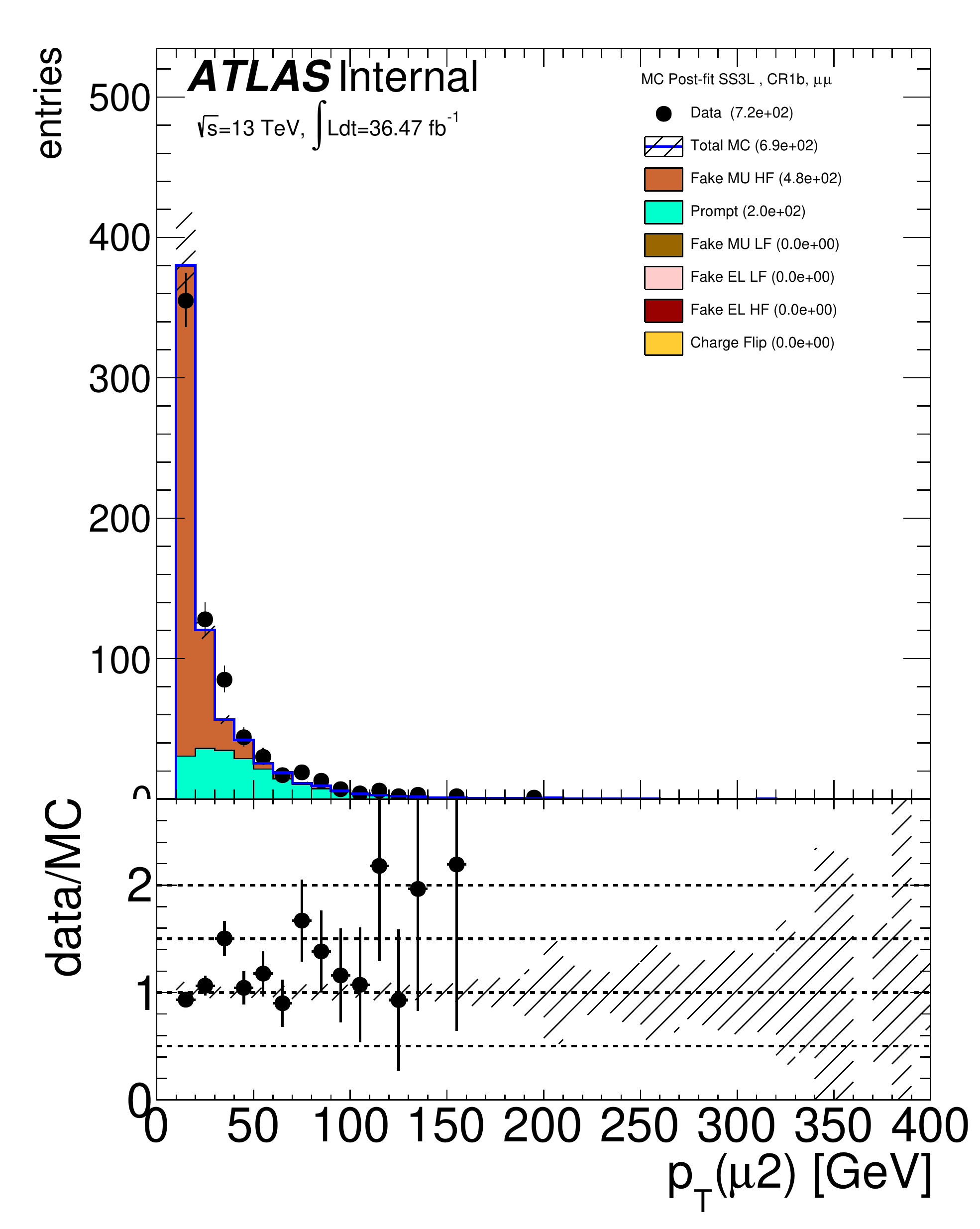}
\caption{
Post-fit distributions for  $ee$ channel (left), for  $e\mu$ channel (middle), and  for  $\mu\mu$ channel (right) from CR1b that were used in the fit to extract the FNP lepton and charge flip multipliers.
The generator used in these plots is  \POWHEGBOX+Pythia. The hashed band represents the sum of systematic uncertainties on the predictions.
\label{f:postfit_CR1b}
}
\end{figure}

The six distributions are chosen for variables that provide the best separation between processes with prompt leptons and processes with FNP leptons and charge flip and are shown 
before and after the fit in Figures \ref{f:prefit_CR0b}-\ref{f:prefit_CR1b} and Figures \ref{f:postfit_CR0b}-\ref{f:postfit_CR1b}, respectively. 

The minimization of the negative log likelihood using the \textsc{Minuit} package leads 
to the multipliers shown in Tables \ref{t:fake_factors_powheg} and \ref{t:fake_factors_sherpa}.
The tables represent the multipliers obtained from the fit upon using two different parton showers, \POWHEGBOX+Pythia and \SHERPA 
for the processes that lead to FNP leptons and charge flips.
The systematic uncertainty is obtained by varying the 
generator from \POWHEGBOX+Pythia to \SHERPA and evaluating the impact on the expected background from FNP and charge flip leptons. 
This is found to be the dominant contribution to the systematic uncertainty of the method (up to 80\%).
The uncertainties in the multipliers themselves correspond to how much the parameter needs to be varied for 
a one standard deviation change in the likelihood function. This uncertainty takes into account the limited number of simulated events and is included as a 
systematic uncertainty on the expected number of background events. 

\begin{table}[!htb]
  \caption{The FNP and charge flip multipliers obtained after minimizing the likelihood function using \POWHEGBOX+Pythia.
    The uncertainty in the multipliers takes into account the limited statistics of simulated events.
    \label{t:fake_factors_powheg}}
  \centering
   \begin{tabular}{|c|c|c|}
          \hline
          Category & Multiplier & Uncertainty  \\
          \hline
          chFlip & 1.49 & 0.58 \\ 
          HF EL & 2.80 & 0.98 \\
          LF EL & 2.89 & 0.88 \\
          HF MU & 1.59 & 0.31 \\
          LF MU & 1.00 & 1.34 \\
          \hline
        \end{tabular}
\end{table}

\begin{table}[!htb]
  \caption{The FNP and charge flip multipliers obtained after minimizing the likelihood function using \SHERPA.
    The uncertainty in the multipliers takes into account the limited statistics of simulated events.
    \label{t:fake_factors_sherpa}}
  \centering
  \begin{tabular}{|c|c|c|}
    \hline
    Category & Multiplier & Uncertainty  \\
    \hline
    chFlip & 1.34 & 0.58 \\ 
    HF EL & 2.40 & 0.85 \\
    LF EL & 1.83 & 1.04 \\
    HF MU & 1.17 & 0.16 \\
    LF MU & 2.40 & 0.81 \\
    \hline
  \end{tabular}                                                                                         
\end{table}

%% file: texfiles/sec.fake.mxm.tex
The FNP leptons do not often pass one of the 
lepton selection criteria but have non-zero impact parameter, and are often not 
well-isolated. These selection requirements are key ingredients to control the FNP leptons. 
The number of events with at least one FNP lepton is estimated using two classes of leptons: 
a real-enriched class of ``tight'' leptons corresponding to signal leptons and a fake-enriched class of ``loose'' leptons 
corresponding to candidate leptons with relaxed identification criteria\footnote{Signal leptons are leptons satisfying the signal lepton definition, while the candidate leptons are leptons satisfying some pre-selection cuts and usually passing the overlap removal requirements as discussed in the analysis 
Section~\ref{subsec:strategy.sel.obj}.}. 
In the next sections, a description of the simplest form of the matrix method will be given with events containing one object. 
Then a generalized treatment that can handle events with an arbitrary number of leptons in the final states will be discussed.

\subsection{Events with one object}

Given the probabilities $\varepsilon/\zeta$ for a real/FNP candidate lepton to satisfy the signal lepton criteria, 
one can relate the number of events with one candidate lepton passing/failing signal requirements ($n_\text{pass}/n_\text{fail}$) to the number of events with one real/FNP signal leptons ($n_\text{real}/n_\text{FNP}$):

\begin{align}
\begin{pmatrix}n_\text{pass}\\n_\text{fail}\end{pmatrix} 
= \begin{pmatrix}\varepsilon & \zeta\\ 1-\varepsilon & 1-\zeta\end{pmatrix}
\begin{pmatrix}n_\text{real}\\n_\text{FNP}\end{pmatrix}; 
\label{eqn:matrix_method}
\end{align}
allowing a determination of the unknown number of events $n_\text{FNP}$ from the observed $n_\text{pass}$ and $n_\text{fail}$ given measurements of the 
probabilities $\varepsilon/\zeta$. 

The predictive power of the matrix method comes from the fact that 
the real and FNP leptons have different composition in the two collections 
of tight and loose objects leading to $\varepsilon \neq \zeta$. In fact, 
the tight lepton collection will be dominated by real objects while the 
loose region will be dominated by fake objects. As a result, 
the inequality $\varepsilon >> \zeta$ will always hold true which 
guarantees that the matrix in Eq. \ref{eqn:matrix_method} is invertible 
and gives positive estimates. 

The next step is to invert the relation in Eq. \ref{eqn:matrix_method} to 
obtain

\begin{align}
\begin{pmatrix}n_\text{real}\\n_\text{FNP}\end{pmatrix} 
= \frac{1}{\varepsilon - \zeta} \begin{pmatrix}\bar\zeta & -\zeta\\ -\bar\varepsilon & \varepsilon\end{pmatrix}
\begin{pmatrix}n_\text{pass}\\n_\text{fail}\end{pmatrix}; 
\label{eqn:fake.inv_matrix_method}
\end{align}

where $\bar\varepsilon = 1 - \varepsilon$ and  $\bar\zeta = 1 - \zeta$. 
The FNP lepton component is: 

\begin{align}
n_\text{FNP} = \frac{1}{\varepsilon - \zeta}\left(\left(\varepsilon-1\right)n_\text{pass}+n_\text{fail}\right).
\label{eqn:fake.nfake}
\end{align}

However, the quantity of interest is the expected FNP lepton background that 
passes the tight selection criteria: 
$n_{\text{pass}~\bigcap~\text{FNP}} = \zeta n_\text{FNP}$.
 To obtain this quantity, 
the identity from Eq. \ref{eqn:matrix_method} is used to get:

\begin{align}
n_\text{FNP} = \frac{\zeta}{\varepsilon - \zeta}\left(\left(\varepsilon-1\right)n_\text{pass}+n_\text{fail}\right).
\label{eqn:fake.nFNPpass}
\end{align}

The linearity of Eq. \ref{eqn:fake.nFNPpass}  with respect to $n_\text{pass}$ 
and $n_\text{fail}$ allows the method to be applied on an event-by-event, 
effectively resulting in a weight being assigned to each event. 
By defining
\[
  n_\text{pass} = \sum_\text{all events} \mathbb{1}_\text{pass},~
  n_\text{fail} = \sum_\text{all events} \mathbb{1}_\text{fail},~ 
  \mathbb{1}_\text{fail} = 1 -  \mathbb{1}_\text{pass},
\]
where $\mathbb{1}_{\text{pass} \left(\text{fail}\right)} = 1$ if the object passes 
(fails) the tight selection requirement and $\mathbb{1}_{\text{pass}\left(\text{fail}\right)} = 0$ otherwise. Eq. \ref{eqn:fake.nFNPpass} can be written as
\[
n_\text{FNP} = \sum_\text{all events} \{
\frac{\zeta}{\varepsilon - \zeta}\left(\varepsilon - 
\mathbb{1}_\text{pass}\right)
\}
\\
=  \sum_\text{all events} \omega
\]
where 
\begin{align}
  \omega = \frac{\zeta}{\varepsilon - \zeta}\left(\varepsilon - 
\mathbb{1}_\text{pass}\right)
  \label{eqn:fake.nFNPpass.demo}
\end{align}
is the weight to be assigned to each event in the case of one FNP lepton 
in the event. 
The generalization of this formalism to higher dimensions 
with multiple objects will be covered next.

\subsection{Dynamic matrix method}

The one lepton case readily generalizes to events with more than one lepton
in a formalism that can handle an arbitrary number of leptons 
in the event. The method should be applied event-by-event, effectively 
resulting into a weight being assigned to each event. The predicted yield of 
events with FNP leptons is simply the sum of weights.
A general formula will be derived starting from the two objects case, 
then specific examples will be given to illustrate the application of the 
method.

If two objects are present in the event, the probabilities $\varepsilon/\zeta$
will depend on the kinematic properties of these objects. Typically 
the probability will vary as a function of \pt and $|\eta|$. For this reason,
the probabilities will be different and will have an index to 
identify the object under study: 
 $\varepsilon_i/\zeta_i$ where $i=1,2...$. 
An identity similar to Eq.  \ref{eqn:matrix_method} can be formed for 
two objects with a change in notation for simplicity:

\begin{align}
\left(\begin{array}{c}
N_{TT} \\  N_{TL} \\ N_{LT} \\ N_{LL}
\end{array}\right) = 
\Lambda \times 
\left(\begin{array}{c}
N_{RR} \\  N_{RF} \\ N_{FR} \\ N_{FF}
\end{array}\right), 
\label{eq:mxm_start}
\end{align}
where $(N_{RR},N_{RF},N_{FR},N_{FF})$ are the number of events with respectively two real, one real plus one FNP (two terms), and two FNP leptons before applying tight cuts, respectively, and $(N_{TT},N_{TL},N_{LT},N_{LL})$ are the observed number of events for which respectively both lepton pass the tight cut, only one of them (two terms), or both fail the tight cut, respectively. 

$\Lambda$ is given by:
\[
\Lambda=
\left(\begin{array}{cccc}
\varepsilon_1\varepsilon_2 & \varepsilon_1\zeta_2 & \zeta_1\varepsilon_2 & \zeta_1\zeta_2\\
\varepsilon_1(1-\varepsilon_2) & \varepsilon_1(1-\zeta_2) & \zeta_1(1-\varepsilon_2) & \zeta_1(1-\zeta_2)\\
(1-\varepsilon_1)\varepsilon_2 & (1-\varepsilon_1)\zeta_2 & (1-\zeta_1)\varepsilon_2 & (1-\zeta_1)\zeta_2\\
(1-\varepsilon_1)(1-\varepsilon_2) & (1-\varepsilon_1)(1-\zeta_2) & (1-\zeta_1)(1-\varepsilon_2) & (1-\zeta_1)(1-\zeta_2)
\end{array}\right) 
\]
which can also be written in terms of a Kronecker product in 
Eq. \ref{eq:mxm_start} to obtain:
\begin{align}
\left(\begin{array}{c}
N_{TT} \\  N_{TL} \\ N_{LT} \\ N_{LL}
\end{array}\right)
= \begin{pmatrix}\varepsilon_1 & \zeta_1\\ \bar\varepsilon_1 & \bar\zeta_1\end{pmatrix} \bigotimes \begin{pmatrix}\varepsilon_2 & \zeta_2\\ \bar\varepsilon_2 & \bar\zeta_2\end{pmatrix}
\left(\begin{array}{c}
N_{RR} \\  N_{RF} \\ N_{FR} \\ N_{FF}
\end{array}\right)
\label{eq:mxm_start_kroe}
\end{align}
To make the notation more compact, the set of 4 numbers $(N_{TT},N_{TL},N_{LT},N_{LL})$ can be represented by a rank 2 tensor $\mathcal{T}_{\alpha_1 \alpha_2}$ 
where $\alpha_i$ corresponds to one object that is either tight (T) or 
loose (L). 
Similarly the numbers $(N_{RR},N_{RF},N_{FR},N_{FF})$ can be represented 
by $\mathcal{R}_{\alpha_1 \alpha_2}$ where $\alpha_i$ corresponds to one object 
that is either real (R) or FNP (F). With this convention, the 
Kronecker product of Eq. \ref{eq:mxm_start_kroe} can be obtained by 
contracting each index $\alpha_i$ of the tensors $\mathcal{T}$ or $\mathcal{R}$
by the 2 $\times$ 2 matrix $\phi_i\tensor{\vphantom{\phi}}{_{\beta_i}^{\alpha_i}}$:

\begin{align}
\mathcal{T}_{\beta_1 \beta_2} = 
\phi_1\tensor{\vphantom{\phi}}{_{\beta_1}^{\alpha_1}} 
\phi_2\tensor{\vphantom{\phi}}{_{\beta_2}^{\alpha_2}}
\mathcal{R}_{\alpha_1 \alpha_2},~
\phi_i = 
 \begin{pmatrix}\varepsilon_i & \zeta_i\\ \bar\varepsilon_i & \bar\zeta_i\end{pmatrix}
\label{eq:fake.tensor_compact}
\end{align}

Following the same procedure as in the one object case, the matrix inversion 
of the 4$\times$4 $\Lambda$ matrix is simplified to a matrix inversion of 
the 2$\times$2 $\phi$ matrices. The quantity of interest is the FNP lepton 
background that passes the tight selection criteria as in 
Eq. \ref{eqn:fake.nFNPpass} which can be compactly written in the 
two objects case as: 

\begin{align}
\mathcal{T}_{\nu_1 \nu_2}^\text{FNP} = 
\phi\indices{_{\nu_1}^{\mu_1}} 
\phi\indices{_{\nu_2}^{\mu_2}}
\tensor*{\xi}{*^{\beta_1}_{\mu_1}^{\beta_2}_{\mu_2}}
\phi\indices{^{-1}_{\beta_1}^{\alpha_1}} 
\phi\indices{^{-1}_{\beta_2}^{\alpha_2}}
\mathcal{T}_{\alpha_1 \alpha_2}.
\label{eq:fake.FNP_compact}
\end{align}

The tensor $\xi$ encodes the component of tight and FNP lepton background. 
In the two objects case, $\xi$ needs to select the total background 
with at least one fake lepton $N_F = N_{RF}+N_{FR}+N_{FF}$ that are also 
passing the tight selection criteria corresponding to the region with 
signal leptons. As a result, $\xi$ takes the form: 

\[
\xi
=
\left(\begin{array}{cccc}
0 & 0 & 0 & 0\\
0 & 1 & 0 & 0\\
0 & 0 & 1 & 0\\
0 & 0 & 0 & 1\\
\end{array}\right) 
\]

To further illustrate, Eq. \ref{eq:fake.FNP_compact} can be written 
explicitly in the notation of Eq. \ref{eq:mxm_start} as:

\[
N_\text{FNP}^\text{signal} = 
\left(\begin{array}{cccc}
0 & \varepsilon_1\zeta_2 & \zeta_1\varepsilon_2 & \zeta_1\zeta_2\\
\end{array}\right) 
\Lambda^{-1}
\left(\begin{array}{c}
N_{TT} \\  N_{TL} \\ N_{LT} \\ N_{LL}
\end{array}\right)
\]

The generalization of Eq. \ref{eq:fake.FNP_compact} from the two objects case 
to $m$ number of objects in the final state is straightforward:

\begin{align}
\mathcal{T}_{\nu_1 \cdots \nu_m}^\text{FNP} = 
\phi\indices{_{\nu_1}^{\mu_1}}
\cdots 
\phi\indices{_{\nu_m}^{\mu_m}}
\tensor*{\xi}{*^{\beta_1}_{\mu_1}^{\cdots}_{\cdots}^{\beta_m}_{\mu_m}}
\phi\indices{^{-1}_{\beta_1}^{\alpha_1}} 
\cdots
\phi\indices{^{-1}_{\beta_m}^{\alpha_m}}
\mathcal{T}_{\alpha_1 \cdots \alpha_m}.
\label{eq:fake.FNP_compact_any}
\end{align}

The tensor $\xi$ is of the general form
\[
\tensor*{\xi}{*^{\beta_1}_{\mu_1}^{\cdots}_{\cdots}^{\beta_m}_{\mu_m}} = 
\tensor*{\delta}{*^{\beta_1}_{\mu_1}}
\cdots
\tensor*{\delta}{*^{\beta_m}_{\mu_m}}
h\left(\beta_1,\cdots,\beta_m,\nu_1,\cdots,\nu_m\right)
\]
where the function $h$ can take values 0 or 1 based on the tight or loose 
configuration being computed which is encoded in the dependence on the  
indices $\nu_i$. 


The application of the matrix method to multilepton final states comes with two important remarks. Firstly, contributions of events with charge-flip electrons would bias a straightforward matrix method estimate (in particular for a final state formed by two leptons with the same electric charge). This happens because the candidate-to-signal efficiency for such electrons is typically lower than for real electrons having a correctly-assigned charge. One therefore needs to subtract from $n_\text{pass}$ and $n_\text{fail}$ the estimated contributions from charge-flip. This can be performed by including events with pairs of opposite-sign candidate leptons in the matrix method estimate, but assigning them an extra weight corresponding to the charge-flip weight. Thanks (again) to the linearity of the matrix method with respect to $n_\text{pass}$ and $n_\text{fail}$, this weight-based procedure is completely equivalent (but more practical) to the aforementioned subtraction. 

Secondly, the analytic expression of the matrix method event weight depends on the lepton multiplicity of the final state. This concerns events with three or more candidate leptons: one such event takes part both in the evaluation of the FNP lepton background for a selection with two signal leptons or a selection with three signal leptons, but with different weights\footnote{This can appear for inclusive selections: for example an event with two signal leptons may or may not contain additional candidate leptons, in a transparent way}. Therefore, for a given event used as input to the matrix method, one should consider all possible leptons combinations, each with its own weight and its own set of kinematic variables. For example, a $e^+e^-\mu^+$ event is used in the background estimate both as an $e^+\mu^+$ event (with a weight $w_1$) and as an $e^+e^-\mu^+$ event (with a weight $w_2\neq w_1$).  

\subsection{Propagation of uncertainties}

The two parameters ($\varepsilon$ and $\zeta$ respectively) can be measured in data, and depend on the flavor and kinematics of the involved leptons.  Systematic uncertainties resulting from the measurement of these two parameters, and their extrapolation to the signal regions, can be propagated to uncertainties on the event weight through standard first-order approximations. The different sources of uncertainties should be tracked separately so that correlations of uncertainties across different events can be accounted for correctly. The resulting set of uncertainties on the cumulated event weights can be then added in quadrature to form the systematic uncertainty on the predicted FNP lepton background yield. The corresponding statistical uncertainty can be taken as the RMS of the event weights.

The methods described in this chapter will be employed to estimate the 
irreducible backgrounds in the search for supersymmetry presented in this 
dissertation.

%% file: texfiles/sec.bkg.overview.tex
In this analysis, two types of backgrounds can be distinguished.
 The first category is the irreducible background from events with two same-sign prompt 
leptons or at least three prompt leptons and is estimated using the MC simulation samples (Section~\ref{sec:bkg.irred}). 
Since diboson and $\ttbar V$ events are the main 
backgrounds in the signal regions, dedicated validation regions with an enhanced contribution from these processes, and small 
signal contamination, are defined to verify the background predictions from the simulation (Section~\ref{sec:bkg.red}). 
The second category is the reducible  
background, which includes events containing electrons with mis-measured charge, mainly from the production of top quark pairs, 
and events containing at least one fake or non-prompt (FNP) lepton.
The application of the data-driven methods of Chapter~\ref{chap:fake} 
is presented in Section~\ref{sec:bkg.irred}).

%% file: texfiles/sec.bkg.irred.tex
\subsection{Expected yields in the signal regions}
\label{sec:bkg.irred.prompt}

The predicted event yields in the signal regions are presented in Table~\ref{tab:prompt_sr_yields}, while the contributions of particular rare processes to the signal regions, relative to the summed contributions of all these processes, are shown in Table~\ref{tab:prompt_rare_contributions}.

\begin{table}[!htb]
\def\arraystretch{1.1}
\centering
\resizebox{0.8\textwidth}{!}{
\begin{tabular}{|c|c|c|c|c|c|}
\hline\hline
& $t\bar t V$ & $VV$ & $t\bar tH$ & $t\bar tt \bar t $ & rare  \\\hline\hline
 Rpc2L0bH  &     0.20 $\pm$ 0.05   &     1.14 $\pm$ 0.23  &     0.08 $\pm$ 0.04  &     0.02 $\pm$ 0.01    &     0.17 $\pm$ 0.04  \\
 Rpc2L0bS  &     0.82 $\pm$ 0.10   &     3.13 $\pm$ 0.21  &     0.26 $\pm$ 0.05  &     0.01 $\pm$ 0.00    &     0.20 $\pm$ 0.04  \\
 Rpc2L1bH  &     3.86 $\pm$ 0.20   &     0.61 $\pm$ 0.06  &     1.01 $\pm$ 0.10  &     0.53 $\pm$ 0.03    &     0.97 $\pm$ 0.12  \\
 Rpc2L1bS  &     3.94 $\pm$ 0.20   &     0.48 $\pm$ 0.05  &     1.28 $\pm$ 0.10  &     0.33 $\pm$ 0.03    &     0.87 $\pm$ 0.12  \\
 Rpc2L2bH  &     0.41 $\pm$ 0.05   &     0.04 $\pm$ 0.01  &     0.10 $\pm$ 0.03  &     0.17 $\pm$ 0.02    &     0.14 $\pm$ 0.04  \\
 Rpc2L2bS  &     1.57 $\pm$ 0.12   &     0.10 $\pm$ 0.03  &     0.44 $\pm$ 0.06  &     0.25 $\pm$ 0.02    &     0.32 $\pm$ 0.05  \\
 Rpc2Lsoft1b  &     1.24 $\pm$ 0.11   &     0.14 $\pm$ 0.02  &     0.44 $\pm$ 0.06  &     0.09 $\pm$ 0.01    &     0.18 $\pm$ 0.04  \\
 Rpc2Lsoft2b  &     1.15 $\pm$ 0.10   &     0.05 $\pm$ 0.02  &     0.37 $\pm$ 0.06  &     0.20 $\pm$ 0.02    &     0.17 $\pm$ 0.03  \\
 Rpc3L0bH  &     0.18 $\pm$ 0.04   &     2.64 $\pm$ 0.12  &     0.03 $\pm$ 0.02  &     0.01 $\pm$ 0.00    &     0.29 $\pm$ 0.04  \\
 Rpc3L0bS  &     0.99 $\pm$ 0.09   &     8.95 $\pm$ 0.21  &     0.12 $\pm$ 0.04  &     0.02 $\pm$ 0.01    &     0.75 $\pm$ 0.07  \\
 Rpc3L1bH  &     1.52 $\pm$ 0.11   &     0.48 $\pm$ 0.05  &     0.25 $\pm$ 0.06  &     0.28 $\pm$ 0.03    &     0.87 $\pm$ 0.12  \\
 Rpc3L1bS  &     7.02 $\pm$ 0.23   &     1.44 $\pm$ 0.10  &     1.36 $\pm$ 0.10  &     0.69 $\pm$ 0.04    &     2.51 $\pm$ 0.22  \\
 Rpc3LSS1b  &     0.00 $\pm$ 0.00   &     0.00 $\pm$ 0.00  &     0.21 $\pm$ 0.04  &     0.00 $\pm$ 0.00    &     0.09 $\pm$ 0.01  \\
\hline\hline
\end{tabular}}
\caption{Expected yields for background processes with prompt leptons, 
in the SRs proposed in Section~\ref{sec:strategy.sr}, for 36.1 \ifb. 
Quoted uncertainties include statistical sources only. 
Rare category includes $\ttbar WW$, $\ttbar WZ$, 3$t$, $tZ$, $tWZ$, $WH$, $ZH$ and $VVV$, and detailed contributions of these processes can be found in Table~\ref{tab:prompt_rare_contributions}. 
}
\label{tab:prompt_sr_yields}
\end{table}

\begin{table}[!htb]
\def\arraystretch{1.1}
\centering
\resizebox{0.7\textwidth}{!}{
\begin{tabular}{|c|c|c|c|c|c|c|c|}
\hline\hline
         & $VVV$ & $VH$ & 3$t$ & $tZ$ & $t \bar t WW$ & $t WZ$ & $t \bar t WZ$\\\hline\hline
 Rpc2L0bH  & 23\%  & 0\% & 2\% & 3\% & 25\% & 43\%  & 1\% \\
 Rpc2L0bS  & 50\%  & 0\% & 3\% & 15\% & 14\% & 16\%  & 0\% \\
 Rpc2L1bH  & 2\%  & 0\% & 7\% & 4\% & 41\% & 41\%  & 2\% \\
 Rpc2L1bS  & 2\%  & 0\% & 6\% & 3\% & 34\% & 50\%  & 2\% \\
 Rpc2L2bH  & 3\%  & 0\% & 15\% & 4\% & 47\% & 27\%  & 1\% \\
 Rpc2L2bS  & 2\%  & 0\% & 13\% & 2\% & 42\% & 36\%  & 2\% \\
 Rpc2Lsoft1b  & 3\%  & 0\% & 9\% & 0\% & 76\% & 7\%  & 2\% \\
 Rpc2Lsoft2b  & 2\%  & 0\% & 17\% & 4\% & 54\% & 19\%  & 2\% \\
 Rpc3L0bH  & 52\%  & 0\% & 0\% & 3\% & 1\% & 40\%  & 1\% \\
 Rpc3L0bS  & 50\%  & 0\% & 0\% & 4\% & 2\% & 39\%  & 1\% \\
 Rpc3L1bH  & 3\%  & 0\% & 3\% & 3\% & 17\% & 70\%  & 1\% \\
 Rpc3L1bS  & 2\%  & 0\% & 3\% & 7\% & 18\% & 64\%  & 2\% \\
 Rpc3LSS1b  & 25\%  & 0\% & 0\% & 0\% & 0\% & 0\%  & 74\% \\
\hline\hline
\end{tabular}}
\caption{Contributions of particular rare processes to the signal regions, relative to the summed contributions of all these processes.  
}
\label{tab:prompt_rare_contributions}
\end{table}

\begin{figure}[htb!]
\centering
\includegraphics[width=0.9\textwidth]{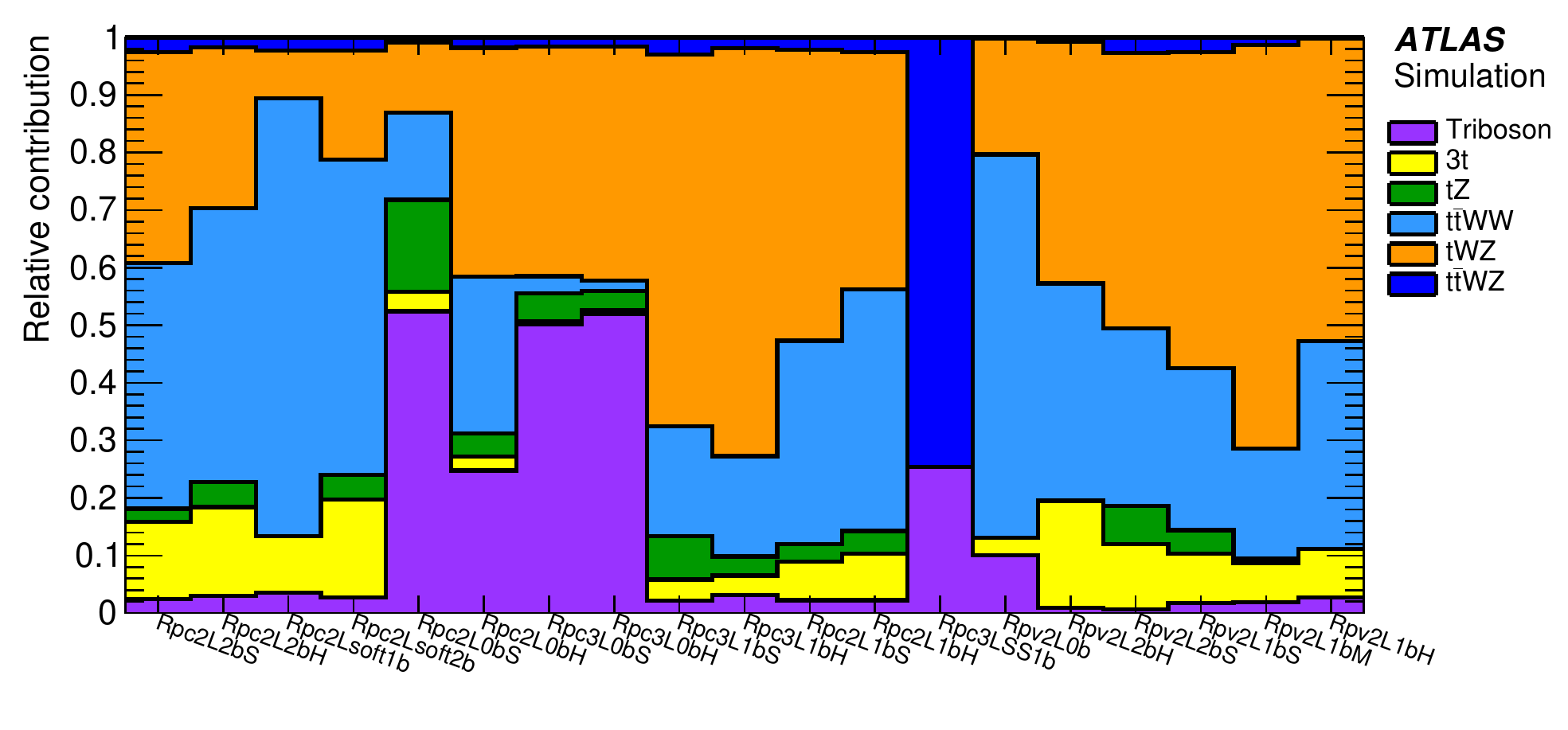}
\caption{Relative contribution in each signal region from the processes in the category labelled as rare in the paper ($\ttbar WW$, 
$\ttbar WZ$, $tZ$, $tWZ$, $t\ttbar$, $WH$, $ZH$ and triboson production). }
\label{fig:RareBreakdown} 
\end{figure} 

\subsection{Validation regions}
\label{sec:bkg.irred.def}

Dedicated validation regions (VR) are defined to verify the estimate
of the $W^\pm W^\pm jj$, the $WZjjjj(j)$, the $\ttbar W$ and $\ttbar Z$ 
processes in the signal regions. 
For a better validation of $WZ$ processes in association with a large jet 
multiplicity, two VRs are proposed : $WZ$4j and $WZ$5j, with 4 and
respectively 5 reconstructed jets in the event.
The corresponding selections are summarized in Table~\ref{tab:VRdef}.

\par
\begin{figure}[htb!]
\centering
\begin{subfigure}[t]{0.49\textwidth}
\includegraphics[width=\textwidth]{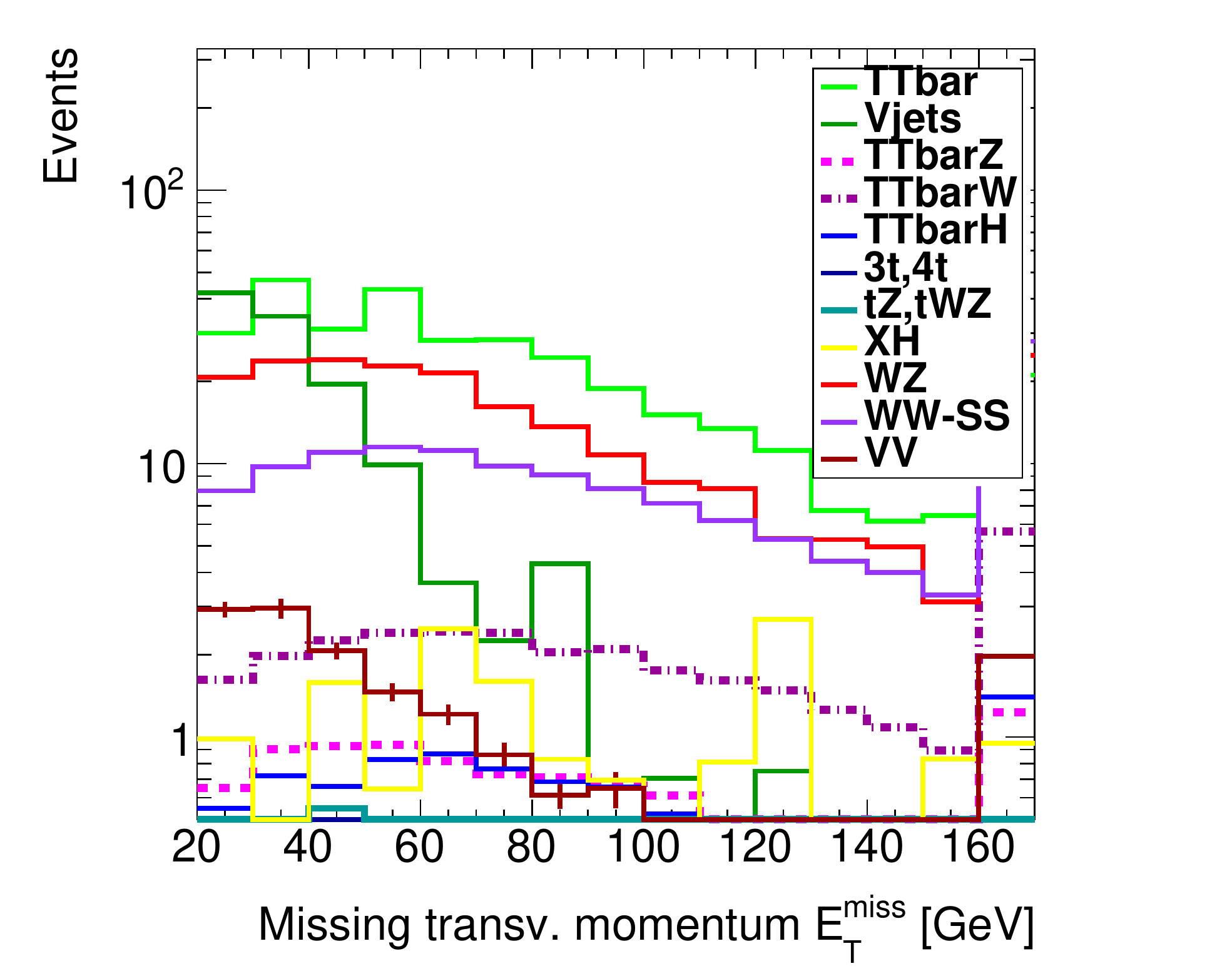}
\subcaption{}
\label{fig:wwa}
\end{subfigure}
\begin{subfigure}[t]{0.49\textwidth}
\includegraphics[width=\textwidth]{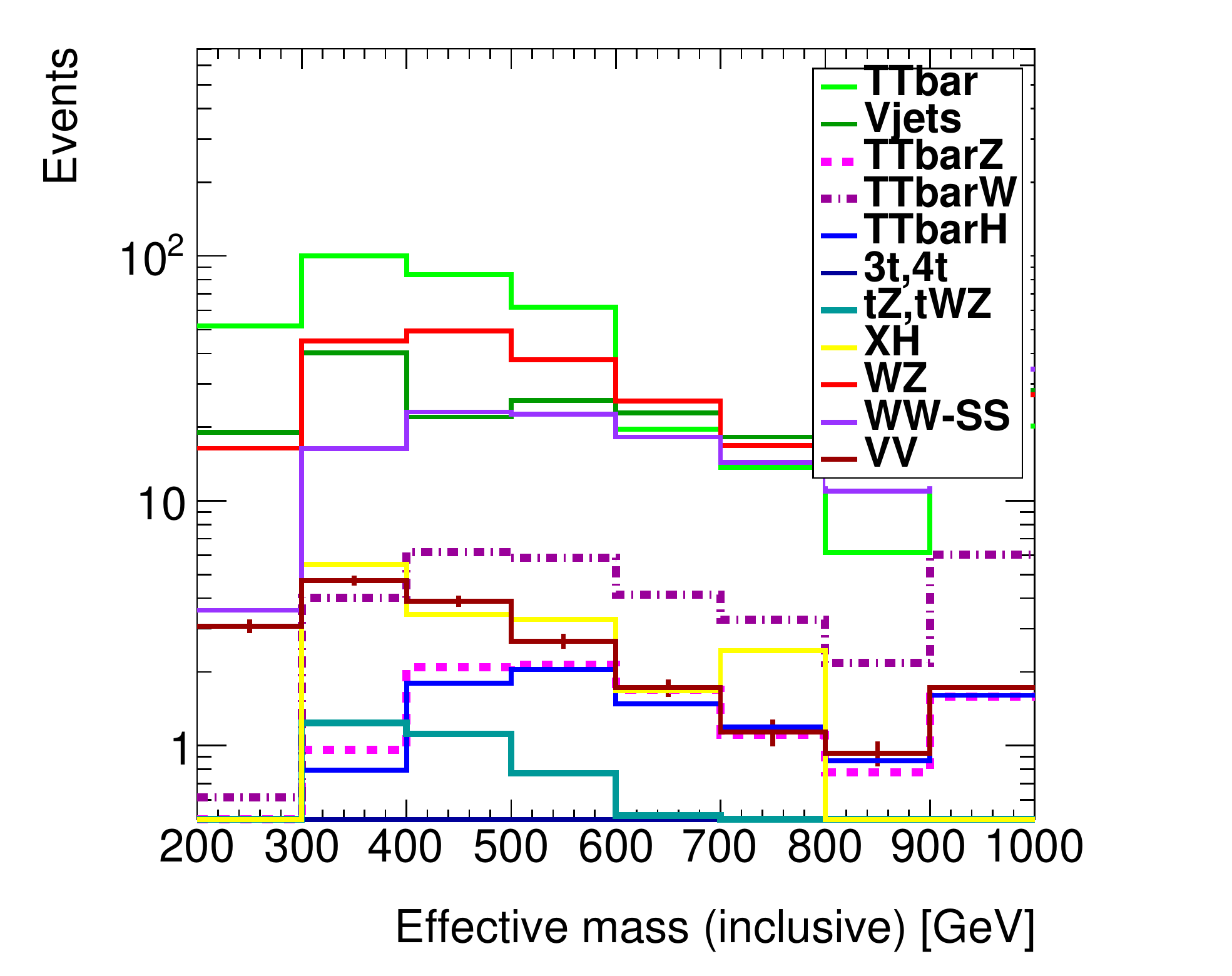}
\subcaption{}
\label{fig:wwb}
\end{subfigure}
\begin{subfigure}[t]{0.49\textwidth}
\includegraphics[width=\textwidth]{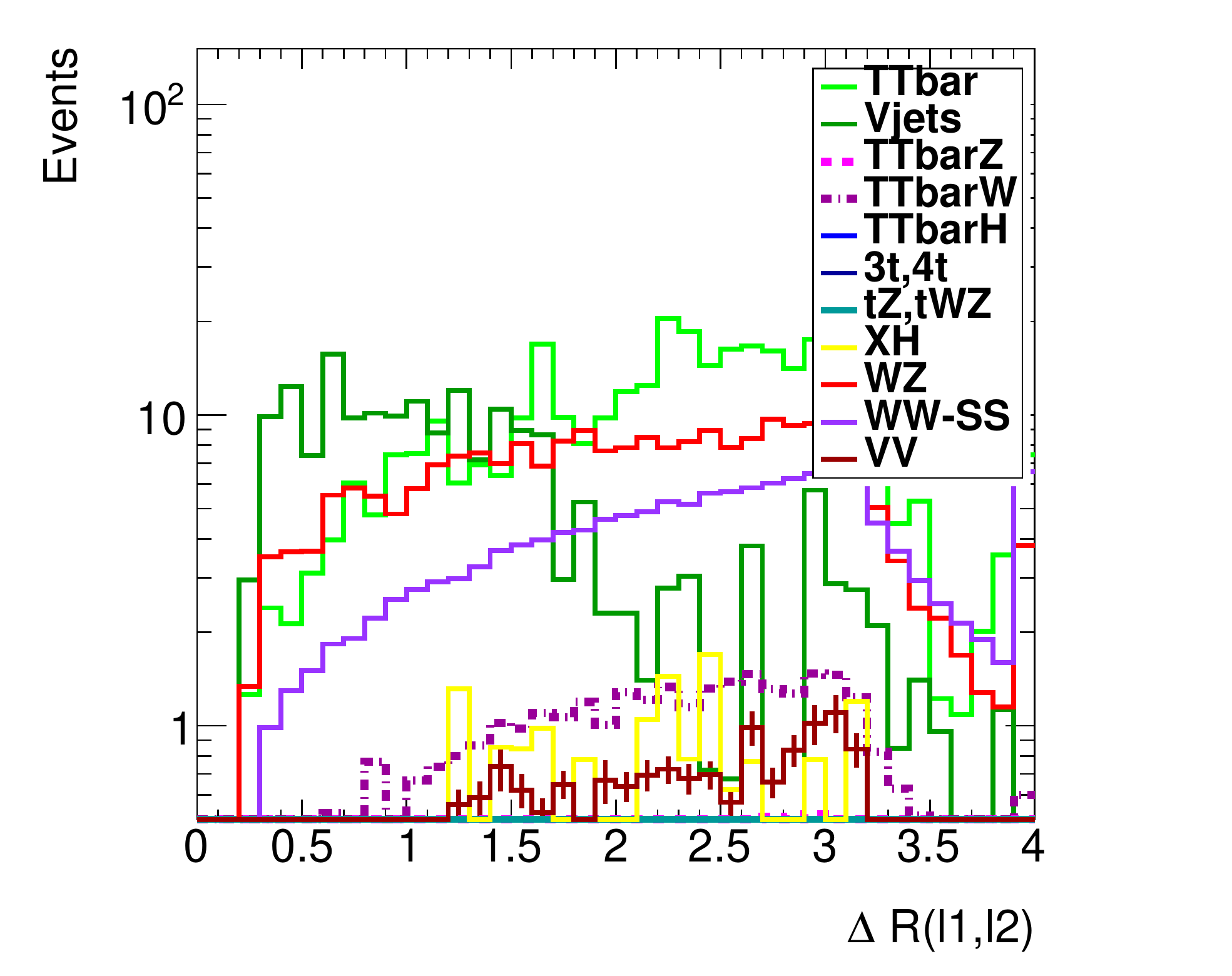}
\subcaption{}
\label{fig:wwc}
\end{subfigure}
\caption{(a) \met, (b) \meff, and (c) $\operatorname{min}\Delta R (\ell_{1}, \ell_2)$ after lepton and jet selection of the $W^\pm W^\pm$-VR (and no additional requirements). Signal regions are vetoed. All MC samples are normalized to a luminosity of 36.1 \ifb. The last bin includes overflow.
}
\label{fig:WW_VR_afterLepJetSel}
\end{figure}

{\bf $W^\pm W^\pm$+jets validation region\\}
The $W^{\pm}W^{\pm}$ + jets processes contribute mainly in the signal regions with no $b$-tagged jet requirement and two same-sign leptons. This validation region, $W^\pm W^{\pm}$-VR, has exactly one SS pair (and no additional baseline leptons), zero $b$-jets and at least two jets with \pt above 50 \GeV. Additional requirements on \met\ and \meff\ help to decrease the amount of detector background as shown in Figure~\ref{fig:WW_VR_afterLepJetSel} (a and b). To further improve the purity, the sub-leading lepton \pt is increased to 30 \GeV, and cuts on minimum angular separation between the leptons and jets, and between the two leptons (Figure~\ref{fig:WW_VR_afterLepJetSel}, c) are placed as detailed in Table~\ref{tab:VRdef}. The purity is around 34\% with this definition 
of the validation region. Signal contamination (highly reduced by applying a veto of all SRs) it is found to be at most 5\% when looking at $\gluino \rightarrow q\bar q (\tilde \ell \ell / \tilde \nu \nu)$ scenarios.

\par
\begin{figure}[htb!]
\centering
\begin{subfigure}[t]{0.49\textwidth}
\includegraphics[width=\textwidth]{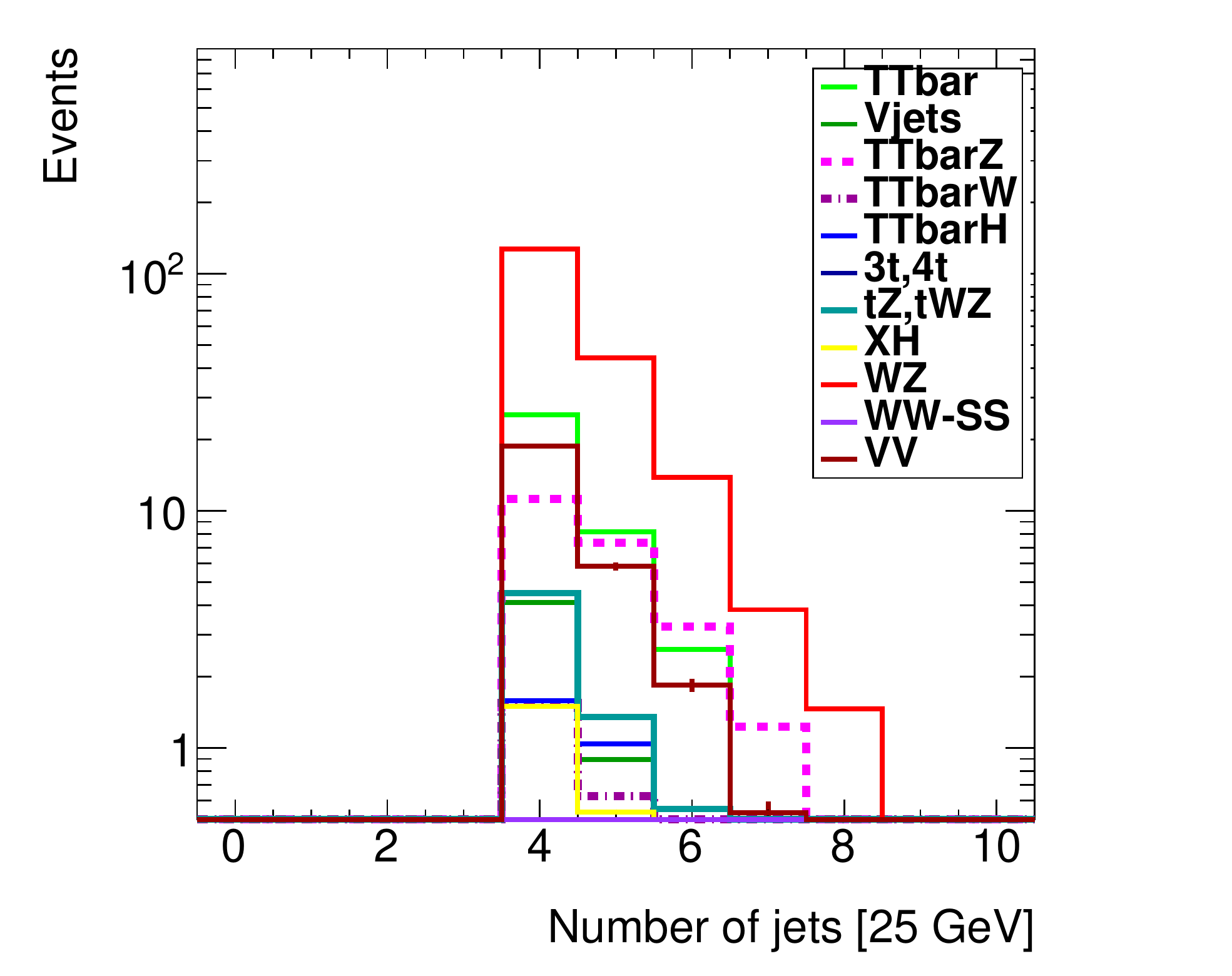}
\subcaption{}\label{fig:wza}\end{subfigure}
\begin{subfigure}[t]{0.49\textwidth}
\includegraphics[width=\textwidth]{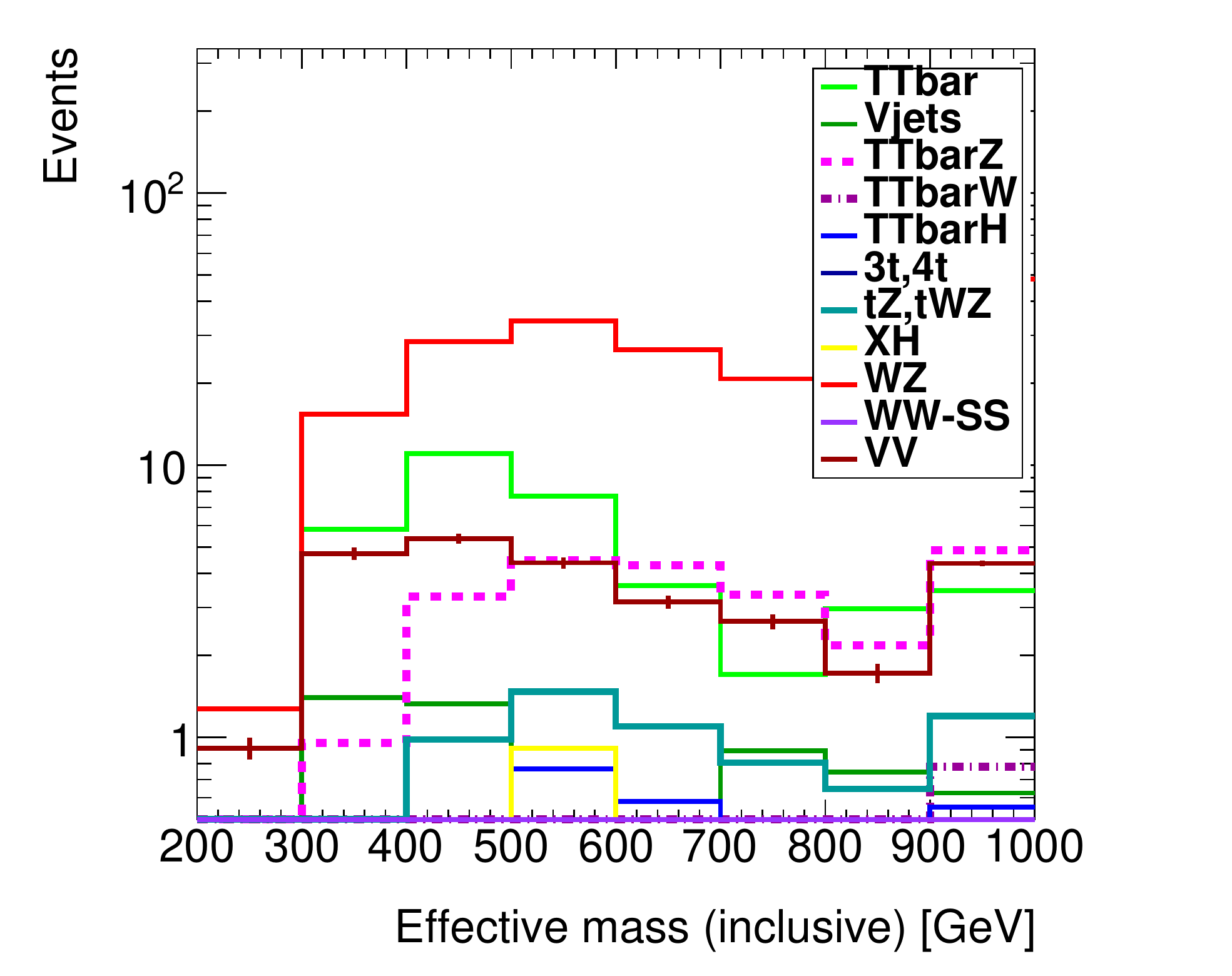}
\subcaption{}\label{fig:wzb}\end{subfigure}
\begin{subfigure}[t]{0.49\textwidth}
\includegraphics[width=\textwidth]{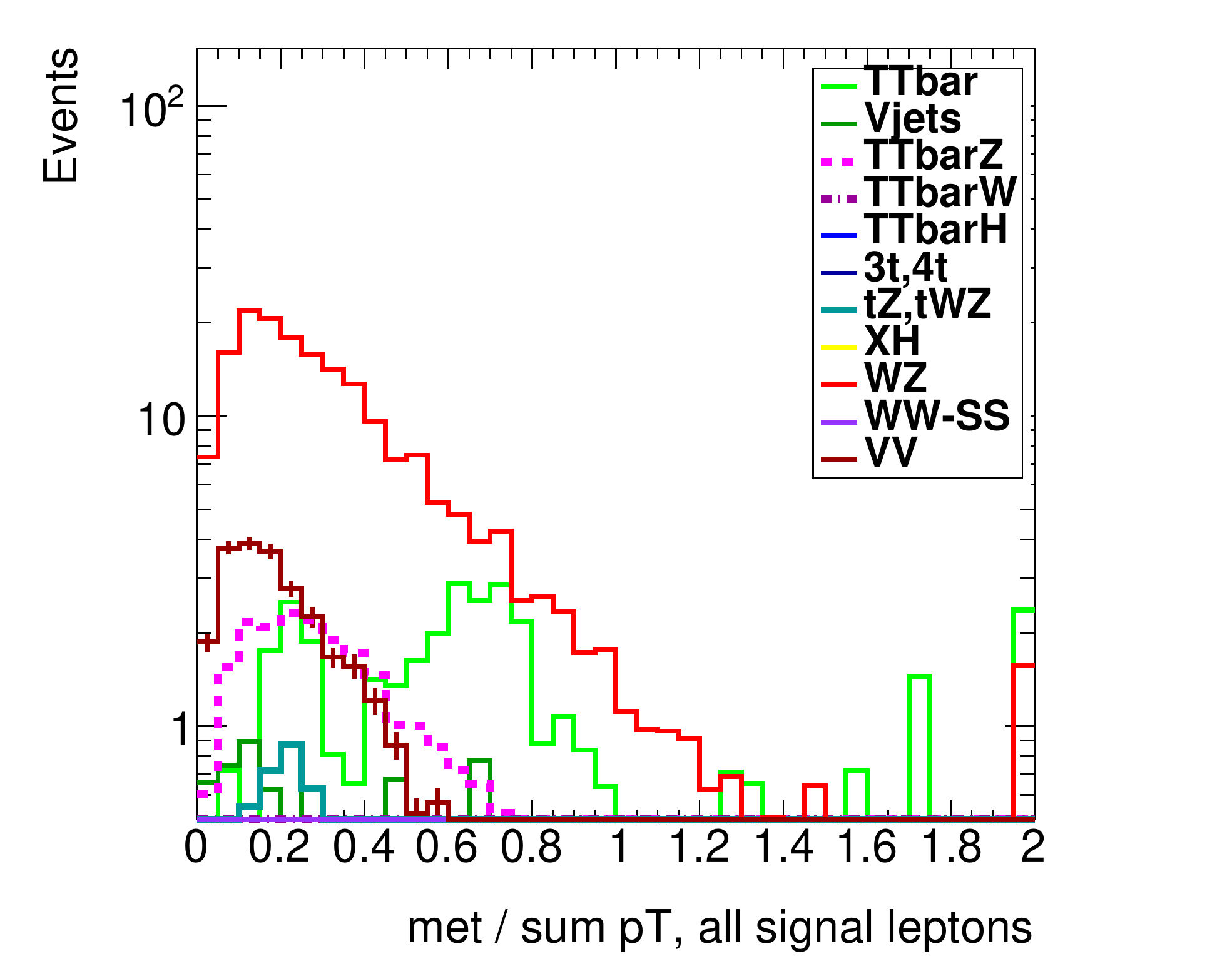}
\subcaption{}\label{fig:wzc}\end{subfigure}
\caption{(a) Number of jets with \pt $>$ 25 \GeV, (b) \meff, and (c) ratio between the \met\ in the event and the sum of all lepton \pt after lepton and four jet selection of the $WZ$-VR (and no additional requirements). Signal regions are vetoed as detailed in Table~\ref{tab:VRdef}. All MC samples are normalized to a luminosity of 36.1 \ifb. The last bin includes overflow.
}
\label{fig:WZ_VR_afterLepJetSel}
\end{figure}

{\bf $WZ$ + jets validation region\\}
Contributions from $WZ$+jets processes can be significant in regions vetoing the presence of $b$-jets and requiring three leptons. 
Given the large data sample collected, it is possible to design validation 
regions that require at least four and even at least five jets with \pt~$>$25 \GeV~in the event. Thus, two validation regions, $WZ$4j-VR and $WZ$5j-VR, are proposed to better probe the modelling of the jet multiplicity in $WZ$ processes. Both regions are defined with exactly three signal leptons and no fourth baseline lepton, to reduce the $ZZ$ background contamination. The \meff, and the upper cut on the ratio between the \met\ in the event and the sum of all lepton \pt are great discriminants against the reducible backgrounds. Some kinematic distributions after lepton and four jet selection are shown in Figure~\ref{fig:WZ_VR_afterLepJetSel}.

Purity in $WZ$4j-VR ($WZ$5j-VR) is around 67$\%$ (64\%). When looking at $\gluino \rightarrow q \bar q (\tilde \ell \ell / \tilde \nu \nu)$ scenarios, the signal contamination is below 5\% in most of the non-excluded phase space, except for small $\Delta M(\gluino,\neut)$ where it can go up to 30\% (15\%). Signal contamination is found to be much lower for $\gluino \rightarrow q \bar q WZ \neut$ scenarios.

\par
\begin{figure}[htb!]
\centering
\begin{subfigure}[t]{0.49\textwidth}
\includegraphics[width=\textwidth]{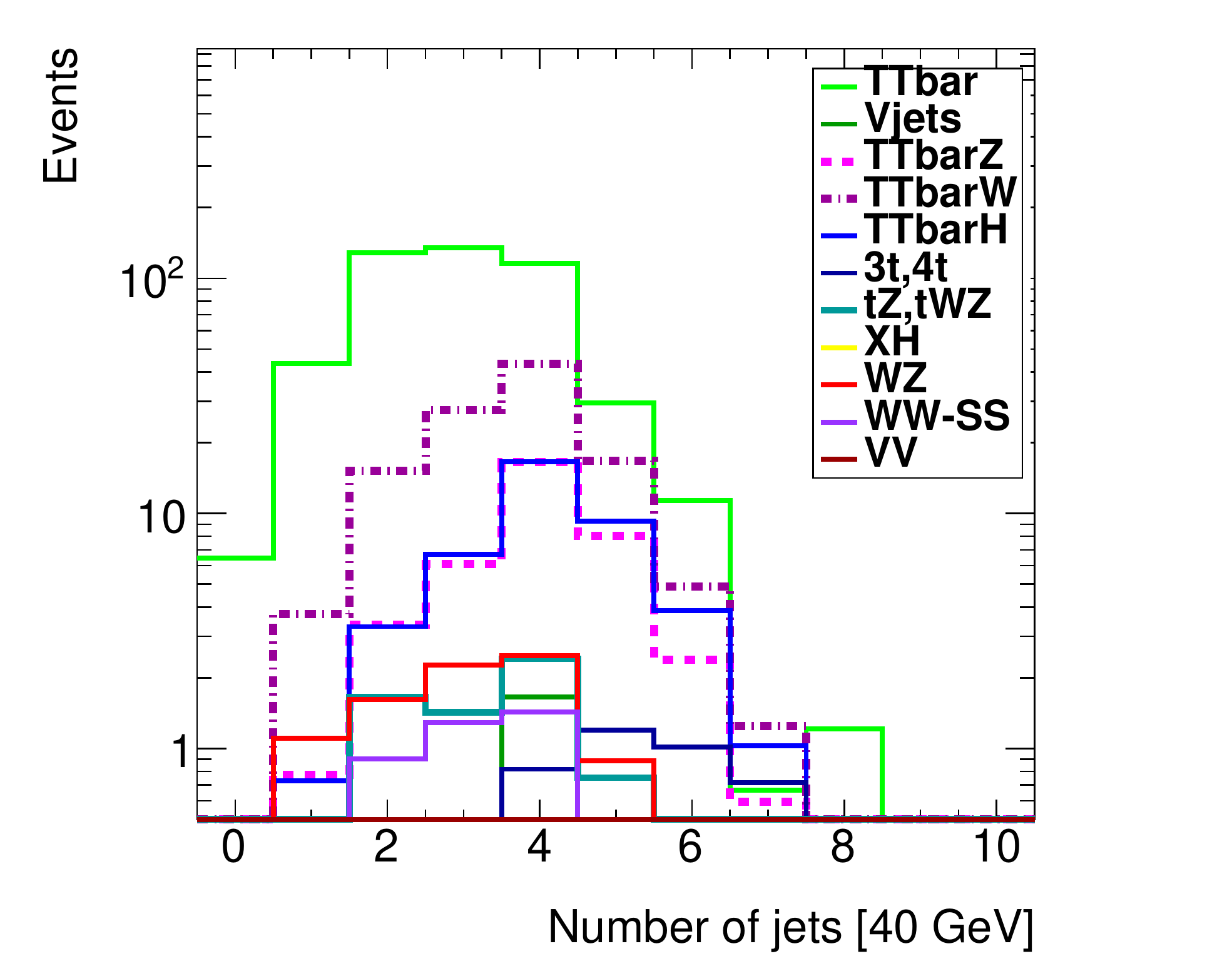}
\subcaption{}\label{fig:ttwa}\end{subfigure}
\begin{subfigure}[t]{0.49\textwidth}
\includegraphics[width=\textwidth]{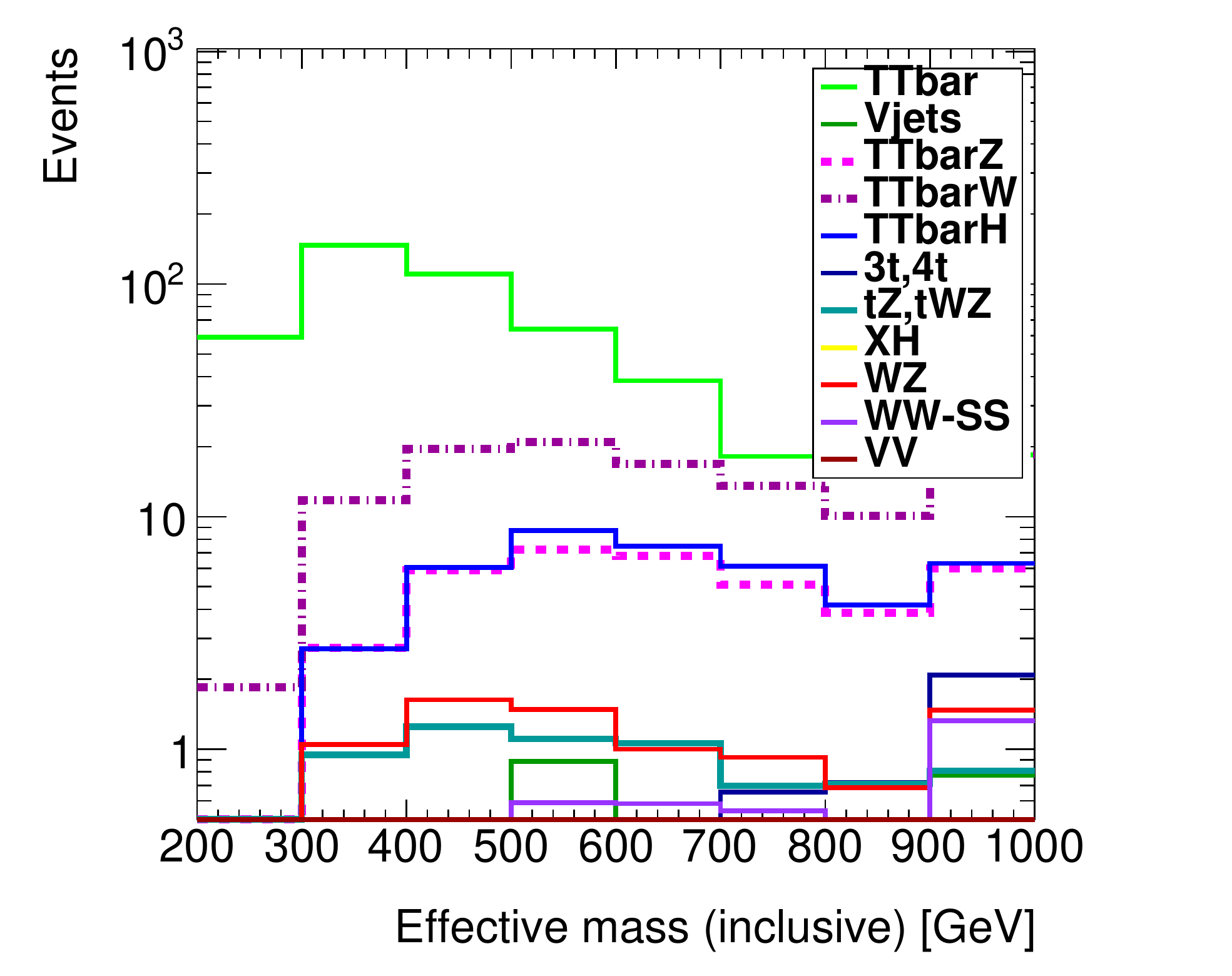}
\subcaption{}\label{fig:ttwb}\end{subfigure}
\begin{subfigure}[t]{0.49\textwidth}
\includegraphics[width=\textwidth]{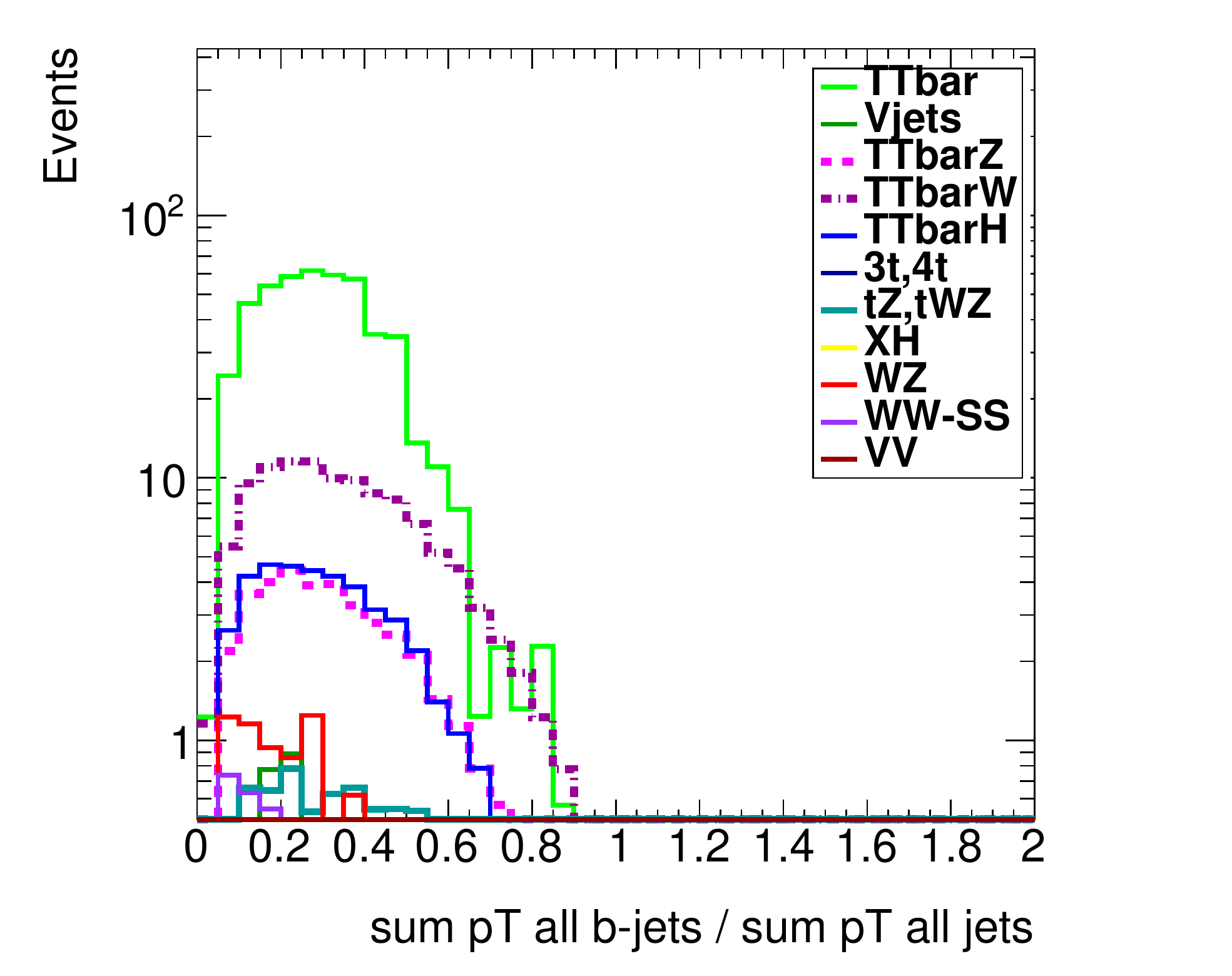}
\subcaption{}\label{fig:ttwc}\end{subfigure}
\caption{(a) Number of jets with \pt $>$ 40 \GeV~, (b) \meff, and (c) ratio between the sum of all $b$-jets \pt and the sum of all jets \pt after lepton and jet selection of the $\ttbar W$-VR (and no additional requirements). Signal regions are vetoed as detailed in Table~\ref{tab:VRdef}. All MC samples are normalized to a luminosity of 36.1 \ifb. The last bin includes overflow.
}
\label{fig:ttW_VR_afterLepJetSel}
\end{figure}

{\bf \ttbar\ + $W$ background validation\\}
A \ttbar\ + $W$ validation region ($\ttbar W$-VR) is defined with exactly one SS lepton pair and at least one $b$-jet. At least four jets are required in the $ee$ and $e\mu$ channels, while in the $\mu\mu$ channel the selection is relaxed to at least three jets  (less reducible background); also the jet \pt thresholds are different between these two cases (same motivation). As shown in Figure~\ref{fig:ttW_VR_afterLepJetSel}, the amount of \ttbar\ background after this pre-selection is still very large and additional requirements on \met, \meff\ and on the ratio between the sum of \pt of all $b$-jets and the sum of \pt of all jets are placed as mentioned in Table~\ref{tab:VRdef}. With this definition the achieved purity in $\ttbar W$-VR is 33\%. 
Signal contamination is around 20\% when looking at $\tilde b\tilde b$ SUSY 
models.

\par
\begin{figure}[htb!]
\centering
\begin{subfigure}[t]{0.49\textwidth}
\includegraphics[width=\textwidth]{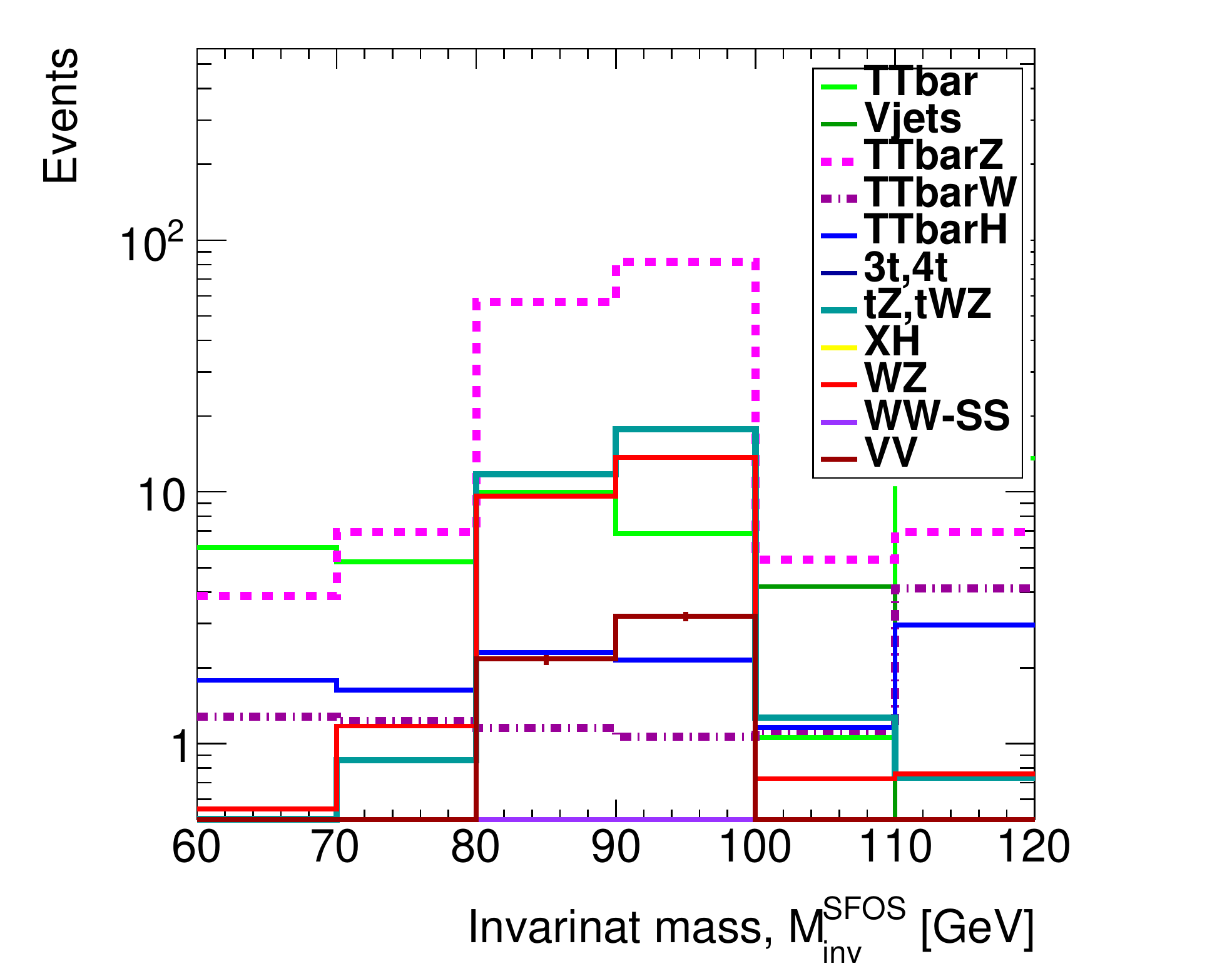}
\subcaption{}\label{fig:ttza}\end{subfigure}
\begin{subfigure}[t]{0.49\textwidth}
\includegraphics[width=\textwidth]{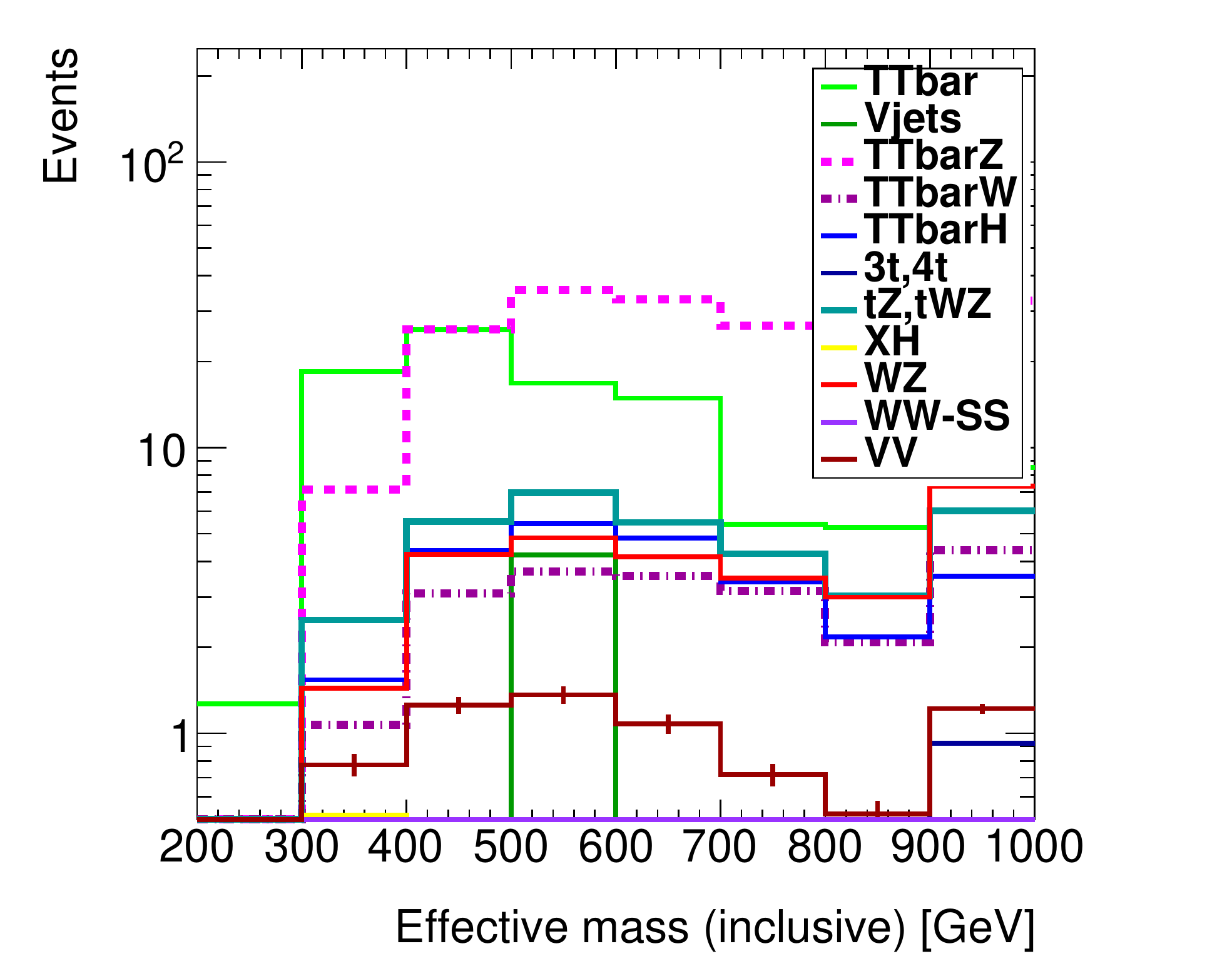}
\subcaption{}\label{fig:ttzb}\end{subfigure}
\begin{subfigure}[t]{0.49\textwidth}
\includegraphics[width=\textwidth]{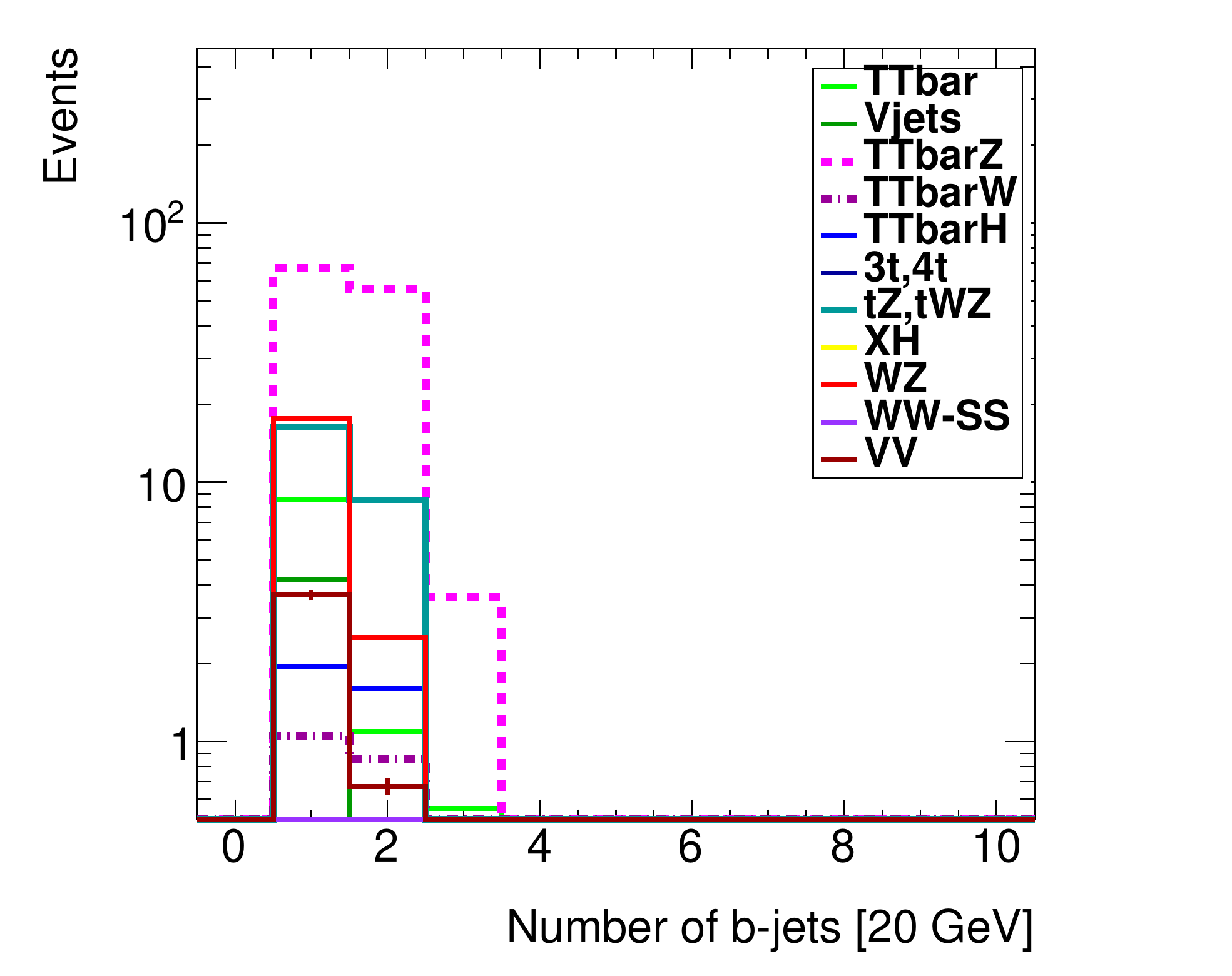}
\subcaption{}\label{fig:ttzc}\end{subfigure}
\caption{(a) Invariant mass of the SFOS lepton pair, (b) \meff\ after lepton and jet selection of the $\ttbar Z$-VR (and no additional requirements), and (c) number of $b$-jets with \pt $>$ 20 \GeV~after the $\ttbar Z$-VR selection. Signal regions are vetoed as detailed in Table~\ref{tab:VRdef}. All MC samples are normalized to a luminosity of 36.1 \ifb. The last bin includes overflow.
}
\label{fig:ttZ_VR_afterLepJetSel}
\end{figure}

{\bf \ttbar\ + $Z$ background validation\\}
A \ttbar\ + $Z$ enriched validation region ($ttZ$-VR) is defined with at least one same flavor opposite sign (SFOS) lepton pair and at least one $b$-jet. At least three jets are required in the event regardless of the lepton channel. Some kinematic distributions after this pre-selection are shown in Figure~\ref{fig:ttZ_VR_afterLepJetSel} (a and b). To increase the purity, the invariant mass of the SFOS lepton pair is selected to be $81 < m_{\ell\ell} < 101$ \GeV, where 
$m_{\ell\ell}$ is the invariant mass of the lepton pair,
and the \meff\ in the event should be greater than 450 \GeV,
where \meff is the sum of the transverse momenta of leptons and jets in the 
event in addition to the transverse missing momentum. 
With this selection, the purity is 58\%. One can increase it even further (by $\sim$10\%) if at least two $b$-jets are required in the event. However, with such a cut the statistics will be greatly reduced (up to a factor 2 lower as illustrated in Figure~\ref{fig:ttZ_VR_afterLepJetSel}, c), so it is not pursued. The signal contamination is found to be around 5\% for \sbsb\ pair production.

\begin{table}[htb!]
\hspace{0.5cm}
\def\arraystretch{1.1}
\centering
\resizebox{\textwidth}{!}
{\small
\begin{tabular}{|l|c|c|c|c|c|c|c|}
\hline    
Validation        &  $N_{\textrm{leptons}}^{\textrm{signal}}$    & $N_{b\textrm{-jets}}$  &  $N_{\textrm{jets}}$  & $\pt^{\textrm{jet}}$  & \met\ & \meff\  & Other \\
Region            &  &  &  & [GeV]  & [GeV] & [GeV]  & \\
\hline\hline
$\ttbar W$   	&$=2SS$     &$\geq 1$   & $\geq 4$ ($e^\pm e^\pm$, $e^\pm \mu^\pm$) & $>40$ & $> 45$  & $> 550$   & $\pt^{\ell_2}>40$~GeV\\
              	&           &       &  $\geq 3$ ($\mu^\pm \mu^\pm$)   &  $>25$ &      &          & $\sum \pt^{b\textrm{-jet}}/\sum \pt^{\textrm{jet}}>0.25$ \\ 
\hline
$\ttbar Z$    	&$\geq 3$  & $\geq 1$ & $\geq 3$ &  $>35$ &  --    & $> 450$  & $81<m_\text{SFOS}<101$~GeV~\\
                &$\geq 1$ SFOS pair&     &          &       &         &         &  \\
\hline
$WZ$4j            & $=3$      &  $=0$ & $\geq 4$ &  $>25$   & --    & $> 450$ & $\met/\sum \pt^{\ell} < 0.7$ \\
\hline
$WZ$5j            & $=3$      &  $=0$ & $\geq 5$ &  $>25$   & --    & $> 450$ & $\met/\sum \pt^{\ell} < 0.7$  \\ 
\hline
$W^{\pm} W^{\pm}jj$ & $=2SS$      &  $=0$ & $\geq 2$ &   $>50$ & $> 55$  & $> 650$ & veto $81<\mee<101$~GeV~ \\
               	  &               &	  &	     &         &   	&	  & $\pt^{\ell_2}>30$~GeV~\\
               	  &		  &	  &	     &         &   	&	  & $\Delta R_\eta (\ell_{1,2},j)>0.7$  \\
               	  &		  &	  &	     &         &   	&	  & $\Delta R_\eta (\ell_1, \ell_2)>1.3$ \\
\hline
All VRs           & \multicolumn{7}{c|}{Veto events belonging to any SR} \\
\hline
\end{tabular}
}
\caption{Summary of the event selection in the validation regions (VRs). 
Requirements are placed on the number of signal leptons ($N_{\textrm{leptons}}^{\textrm{signal}}$), 
the number of $b$-jets with $\pt>20 \GeV$ ($N_{b\textrm{-jets}}$) or the number of jets ($N_{\textrm{jets}}$) 
above a certain \pt threshold ($\pt^{\textrm{jet}}$). The two leading-\pt 
leptons are referred to as $\ell_{1,2}$ with decreasing \pt. Additional requirements are set 
on \met, \meff, the invariant mass of the two leading electrons \mee, the presence of SS 
leptons or a pair of same-flavour opposite-sign leptons (SFOS) and its invariant mass $m_\text{SFOS}$. 
A minimum angular separation between the leptons and the jets ($\Delta R_\eta (\ell_{1,2}, j)$) and between the two 
leptons ($\Delta R_\eta (\ell_{1}, \ell_2)$) is imposed in the $W^\pm W^\pm jj$ VR. 
For the two $WZ$ VRs the selection also relies on the ratio of the \met in the event to the sum of \pt of all signal leptons \pt (\met/$\sum{\pt^{\ell}}$). 
The ratio of the scalar sum of the \pt of all $b$-jets to that of all jets in the event 
($\sum \pt^{b{\textrm{-jet}}} / \sum{\pt^{\textrm{jet}}}$) is used in the $\ttbar W$ VR selection.}
\label{tab:VRdef}
\end{table}

\subsection{Validation of irreducible background estimates}
\label{sec:bkg.irred.res}

The observed yields, compared with the background predictions and uncertainties, 
can be seen in Table~\ref{tab:VR_yields}. There is good agreement between data and the estimated background in all
the validation regions. 

\begin{table}[htb!]
\hspace{0.5cm}
\def\arraystretch{1.1}
\centering
\resizebox{\textwidth}{!}{
\begin{tabular}{|l|c|c|c|c|c|}
\hline    
 Validation Region       & $\ttbar W$           & $\ttbar Z$           & $WZ$4j              & $WZ$5j                 & $W^\pm W^{\pm}jj$     \\
\hline\hline
$\ttbar Z/\gamma^*$      & $ 6.2 \pm 0.9 $      & $ 123 \pm 17\hpO $   & $ 17.8 \pm 3.5\hpO$ & $ 10.1\pm 2.3\hpO $    & $ 1.06 \pm 0.22 $     \\
$\ttbar W$               & $ 19.0 \pm 2.9\hpO $ & $ 1.71 \pm 0.27 $    & $ 1.30 \pm 0.32$    & $ 0.45 \pm 0.14 $      & $ 4.1 \pm 0.8 $     \\
$\ttbar H$               & $  5.8 \pm 1.2 $     & $ 3.6 \pm 1.8 $      & $ 1.8 \pm 0.6$      & $ 0.96 \pm 0.34 $      & $ 0.69 \pm 0.14 $     \\
4$t$                     & $ 1.02 \pm 0.22 $    & $ 0.27 \pm 0.14 $    & $ 0.04 \pm 0.02$    & $ 0.03 \pm 0.02 $      & $ 0.03 \pm 0.02 $     \\
$W^{\pm}W^{\pm}$         & $ 0.5 \pm 0.4 $      &   --                 &   --                &   --                   & $ 26 \pm 14$     \\
$WZ$                     & $ 1.4 \pm 0.8 $      & $ 29 \pm 17 $        & $ 200 \pm 110$      & $ 70 \pm 40 $          & $ 27 \pm 14 $    \\
$ZZ$                     & $ 0.04 \pm 0.03 $    & $  5.5 \pm  3.1$     & $ 22 \pm 12$        & $  9\pm 5 $            & $ 0.53\pm 0.30 $     \\
Rare                     & $ 2.2 \pm 0.5 $      & $ 26 \pm 13 $        & $ 7.3 \pm 2.1$      & $  3.0 \pm 1.0 $       & $ 1.8 \pm 0.5 $     \\
Fake/non-prompt leptons  & $ 18 \pm 16$         & $ 22 \pm 14$         & $ 49 \pm 31 $       & $  17 \pm 12 $         & $ 13 \pm 10 $    \\
Charge-flip              & $  3.4\pm 0.5 $      & --                   & --                  & --                     & $ 1.74 \pm 0.22 $     \\
\hline
Total SM background      & $ 57 \pm 16 $        & $212\pm 35\hpO$      & $300 \pm 130$       & $ 110 \pm 50\hpO $     & $ 77 \pm 31$     \\
\hline
Observed                 & $ 71 $               & $209$                & $257$               & $ 106 $                & $ 99  $            \\
\hline
\end{tabular}}
\caption{The numbers of observed data and expected background events in the validation regions. 
The rare category is defined in the text. Background categories with yields shown as ``--'' 
do not contribute to a given region (e.g. charge flips in three-lepton regions) or their estimates are below 0.01 events. 
The displayed yields include all statistical and systematic uncertainties.
}
\label{tab:VR_yields}
\end{table}

%% file: texfiles/sec.bkg.red.tex
The reducible backgrounds consist of the charge-flip background and 
the FNP lepton background.
The charge flip background is obtained by re-weighting opposite-sign 
lepton pairs data with the measured charge-flip probability and cross checked 
with the estimate obtained with the MC template method.
Two data-driven methods are used to estimate the FNP lepton background, 
the matrix method and the MC template method,
which are combined to obtain the final estimate.  
These methods were discussed in detail in Chapter~\ref{chap:fake}.
This section describes the application of these methods in the 
context of the SS/3L analysis and the validation of the reducible 
background estimates.

\subsection{Charge-flip Background}\label{sec:bkg.red.chflip}
\input{texfiles/subsec.bkg.chargeflip}

\subsection{Matrix Method}\label{sec:bkg.red.mxm}
\input{texfiles/subsec.bkg.mxm}

\subsection{MC Template Method}\label{sec:bkg.red.mct}
\input{texfiles/subsec.bkg.mct}

\subsection{Reducible Background Validation}\label{sec:bkg.red.val}
\input{texfiles/subsec.bkg.val}

%% file: texfiles/subsec.bkg.chargeflip.tex
The lepton charge mis-measurement commonly referred to as ``charge flip'', 
is an experimental background strongly associated to analyses relying on same-sign lepton final states. 
In those events, the electric charge of one of the two leptons forming an opposite-sign (OS) pair, 
coming from an abundant SM process ($pp\to Z$, \ttbar, $W^+W^-$\ldots), 
is mis-identified leading to a much rarer same-sign (SS) pair event.
In most cases, the source of such a mis-identification 
is the creation of additional close-by tracks $e^\pm\to e^\pm\gamma\to e^\pm e^\pm e^\mp$ 
via Brehmstrahlung of the original electron when interacting with the material of the inner tracker. 
If one of the secondary electron tracks is subsequently preferred to the original track in the reconstruction of the electron candidate, 
the charge assigned to the electron might be incorrect, leading to a charge-flip event. 
Errors on the track charge assignment itself may occur as well, but they are much rarer. 
In the case of muons, charge-flip is essentially negligible due to the much smaller interaction cross-section with matter, 
and the requirement of identical charges to be measured for the inner tracker and muon spectrometer tracks. 

A purely data-driven method is used to estimate event yields for the 
electron charge-flip background. 
Assuming that the electron charge flip rates $\xi(\eta,\pt)$ are known, 
a simple way to predict these yields is to select events with pairs of opposite-sign leptons in data and assign them a weight:

\begin{align}
w_\text{flip} = \xi_1(1-\xi_2) + (1-\xi_1)\xi_2
\label{eqn:chargeflip_weight}
\end{align}
where $\xi_{(i)}=0$ for muons.

The advantages of this method are a good statistical precision since the charge flip rate is quite small, 
and the absence of dependency to the simulation and related uncertainties. 
Obviously, it requires a precise measurement of the rates, which is described in this section. 
An inconvenience of this approach is that the reconstructed momentum for charge-flipped electrons  
tends to be negatively biased (too low by a few GeV), 
since such important Brehmstrahlung topologies represent only 
a very small fraction of the cases used to tune electron energy calibration. 
Simply re-weighting electrons from opposite-sign lepton pairs therefore does not predict correctly 
the charge-flip background shape for variables very sensitive to the electron momentum, for example the $m_{ee}$ line-shape. 
However, the kinematic range and variables used in the analysis are not
sensitive to this effect and can safely be neglected.

For the nominal (tight) estimate of the charge-flip background contributions, only events with exactly two OS signal electrons are considered. 
Corrections in the fake lepton estimate however require estimating as well charge-flip contributions for selections involving 
baseline electrons failing signal requirements; 
for that reason, the charge-flip (loose) rate is measured for these two categories of electrons. 

\subsection*{Methodology}
\label{subsec:chargeflip_method}

\begin{figure}[t!]
\centering
\begin{subfigure}[b]{0.45\textwidth}
	\includegraphics[width=\textwidth]{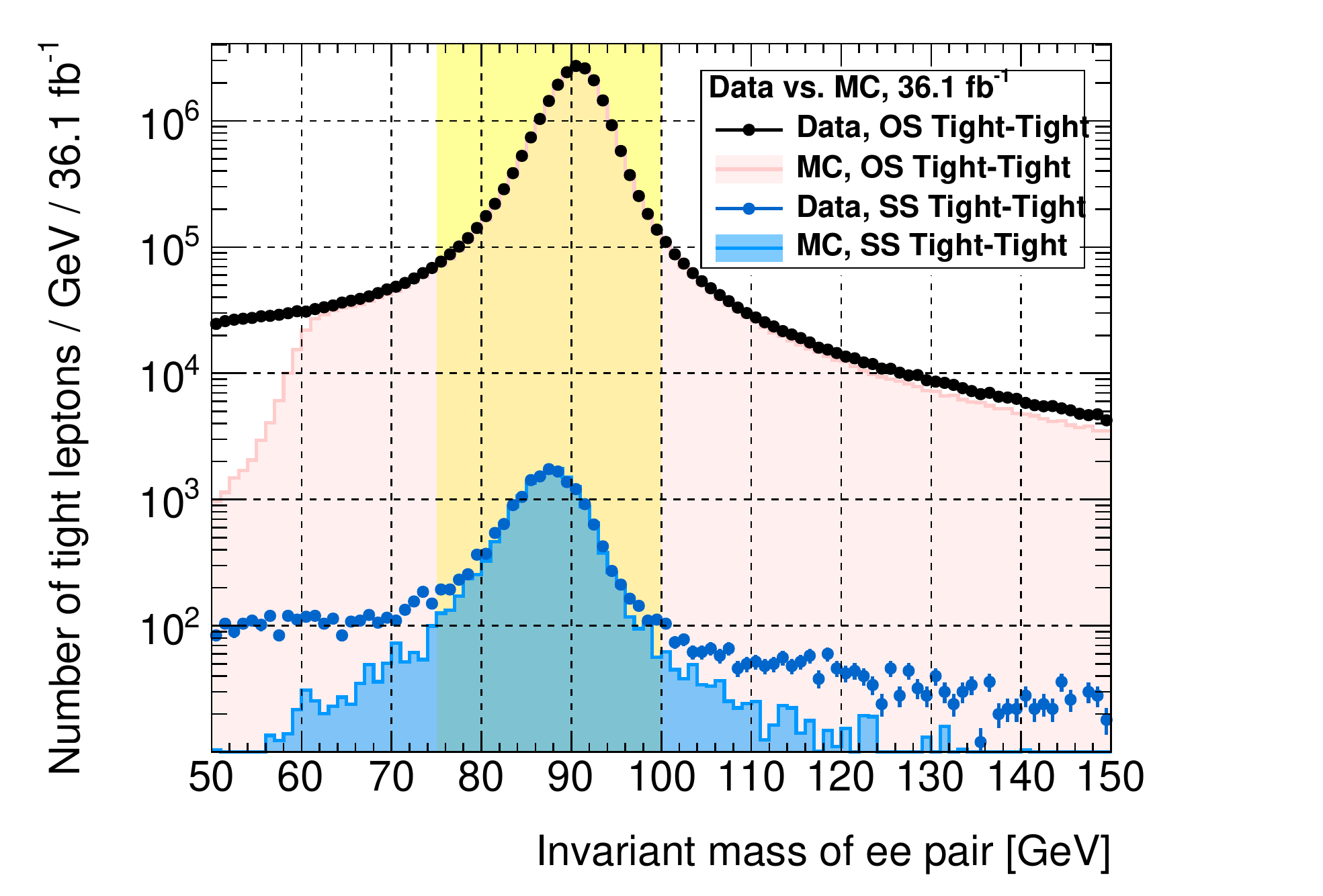}
\end{subfigure}
\begin{subfigure}[b]{0.45\textwidth}
	\includegraphics[width=\textwidth]{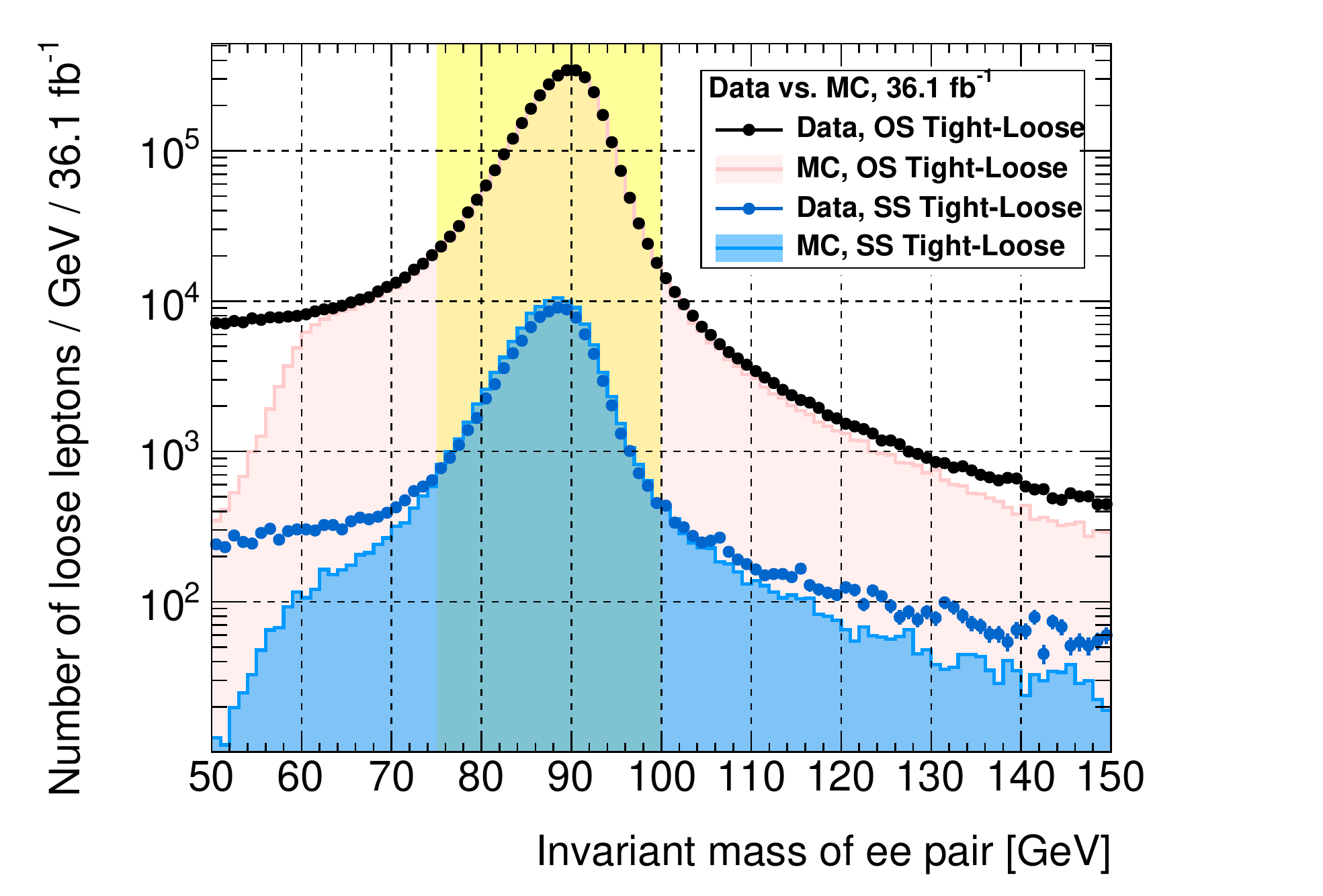}
\end{subfigure}
\caption{Invariant mass of opposite- and same-sign electron pairs, 
when both electrons satisfy signal requirements (left) or one of them fails them (right). Drell-Yan MC samples are not included, thus the drop in the MC distributions (light magenta filled area).}
\label{fig:chargeflip_mee}
\end{figure}

Charge-flip rates are measured in data relying on a clean $Z\to ee$ sample ($75<m_{ee}<100~\GeV$), 
in which the rates can be determined from the relative proportions of OS and SS electron pairs. 
Figure~\ref{fig:chargeflip_mee} illustrates this event selection. 
The rates are measured as function of $\eta$ and \pt, to follow their dependency to the distribution of material in the detector, 
the Brehmstrahlung emission rate, and the track curvature. 
Because of this binned measurements, and because the two electrons in a given pair generally have different kinematic properties, 
it has been found that the most efficient and least biased use of the available statistics 
is obtained by simultaneously extracting the rates in all bins via the maximization of the likelihood function describing the 
Poisson-expected yields of SS pairs: 

\begin{equation}
\begin{aligned}
{} & L(\{N^\text{SS,obs}_\varpi\}|\{\xi(\eta,\pt)\}) 
= \\
& \prod_{\varpi} \mathcal P\left(N^\text{SS,obs}_\varpi|w_\text{flip}(\xi(\eta_1,p_{\mathrm{T},1}),\xi(\eta_2,p_{\mathrm{T},2}))\times N^\text{OS+SS,obs}_\varpi\right)
\label{eqn:chargeflip_likelihood}
\end{aligned}
\end{equation}
with $\varpi=(\eta_1,p_{\mathrm{T},1},\eta_2,p_{\mathrm{T},2})$ indexing bins, where (arbitrarily) $p_{\mathrm{T},1}>p_{\mathrm{T},2}$; 
the expression of $w_\text{flip}$ is given by~(\ref{eqn:chargeflip_weight}). 
Statistical uncertainties on the extracted charge-flip rates are obtained (in a standard way) from the likelihood's numerically-computed Hessian matrix. 

In the nominal charge-flip measurement, the two electrons are required to satisfy signal requirements. 
To measure charge-flip rates for baseline electrons failing signal (noted $\bar\xi$ below), 
pairs with only one signal electron are used; 
this provides larger statistics than applying~(\ref{eqn:chargeflip_likelihood}) to electrons pairs where both fail the signal cuts. 
However, the expression of the likelihood has to be adapted due to the induced asymmetry between the two electrons forming the pair: 
\begin{equation}
\begin{aligned}
{} & L(\{N^\text{SS,obs}_\varpi\}|\{\xi(\eta_1,p_{\mathrm{T},1})\},\{\bar\xi(\eta_2,p_{\mathrm{T},2})\}) 
= \\
& \prod_{\varpi} \mathcal P\left(N^\text{SS,obs}_\varpi|w_\text{flip}(\xi(\eta_1,p_{\mathrm{T},1}),\bar\xi(\eta_2,p_{\mathrm{T},2}))\times N^\text{OS+SS,obs}_\varpi\right)
\label{eqn:chargeflip_likelihood_loose}
\end{aligned}
\end{equation}
where this time $(\eta_1,p_{\mathrm{T},1})$ corresponds to the signal electron. 
Using the same $\eta$ and \pt binning for both measurements, 
the number of free variables in the maximization of~(\ref{eqn:chargeflip_likelihood_loose}) 
-- as well as the number of terms in the product forming $L$ -- 
is twice as large as the nominal case~(\ref{eqn:chargeflip_likelihood}). 
In fact, a by-product of the maximization of~(\ref{eqn:chargeflip_likelihood_loose}) is another determination of the charge-flip rates for signal electrons, 
although with a more limited precision than obtained in the nominal measurement~(\ref{eqn:chargeflip_likelihood}).

Background subtraction is performed through a simple linear extrapolation of the invariant mass distribution sidebands; 
it matters mostly for low \pt in the nominal measurement, 
and for the additional measurement with baseline electrons failing signal requirements, where the level of background is larger. 

\subsection*{Measured rates}
\label{subsec:chargeflip_rates}

\begin{figure}[htb!]
\centering
\begin{subfigure}[b]{0.49\textwidth}
	\includegraphics[width=\textwidth]{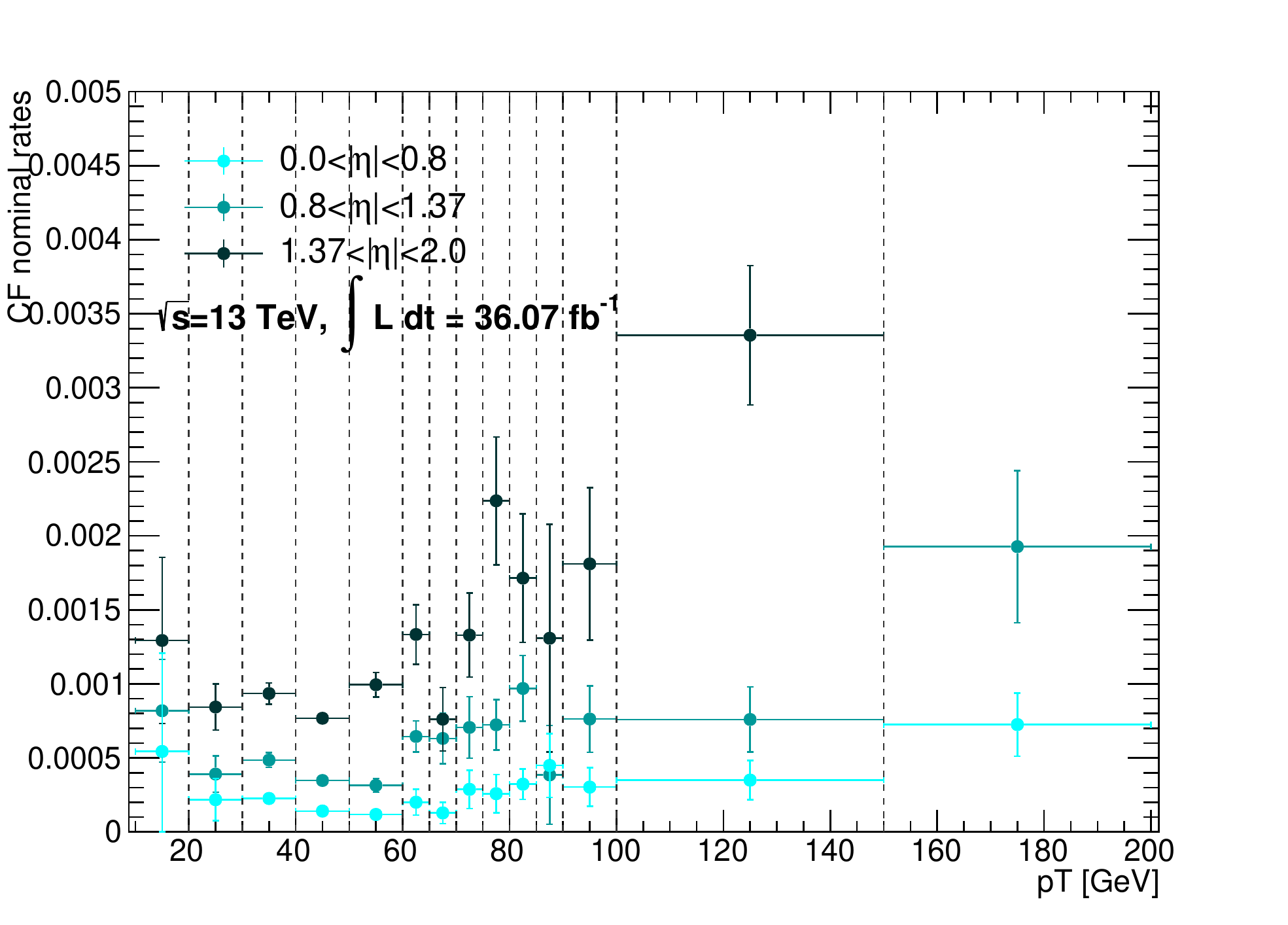}
	\caption{Data, signal electrons}\label{fig:Chflip_nominalData}
\end{subfigure}
\begin{subfigure}[b]{0.49\textwidth}
	\includegraphics[width=\textwidth]{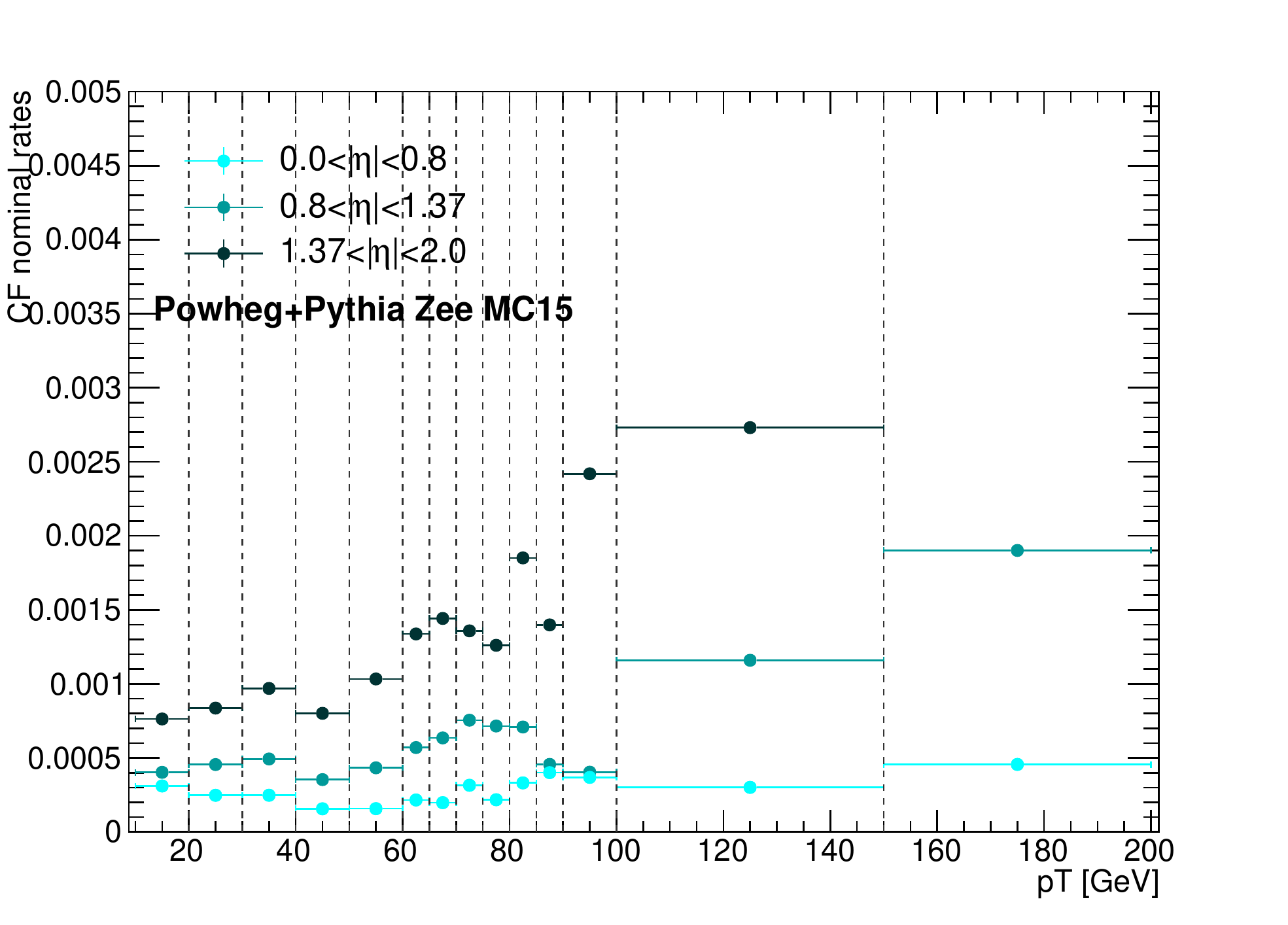}
	\caption{MC, signal electrons}\label{fig:Chflip_nominalMC}
\end{subfigure}
\begin{subfigure}[b]{0.49\textwidth}
	\includegraphics[width=\textwidth]{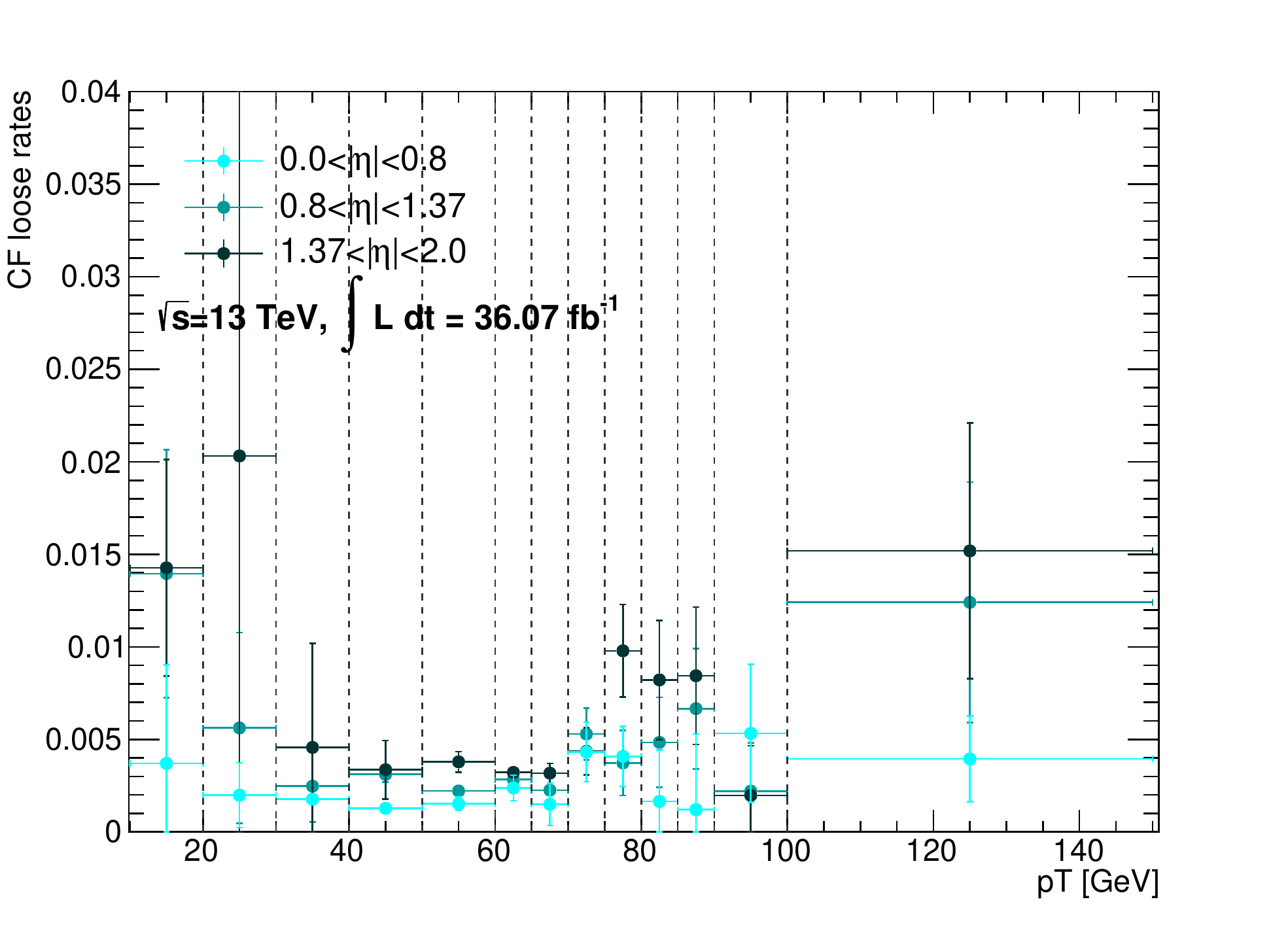}
	\caption{Data, baseline-failing-signal}\label{fig:Chflip_looseData}
\end{subfigure}
\begin{subfigure}[b]{0.49\textwidth}
	\includegraphics[width=\textwidth]{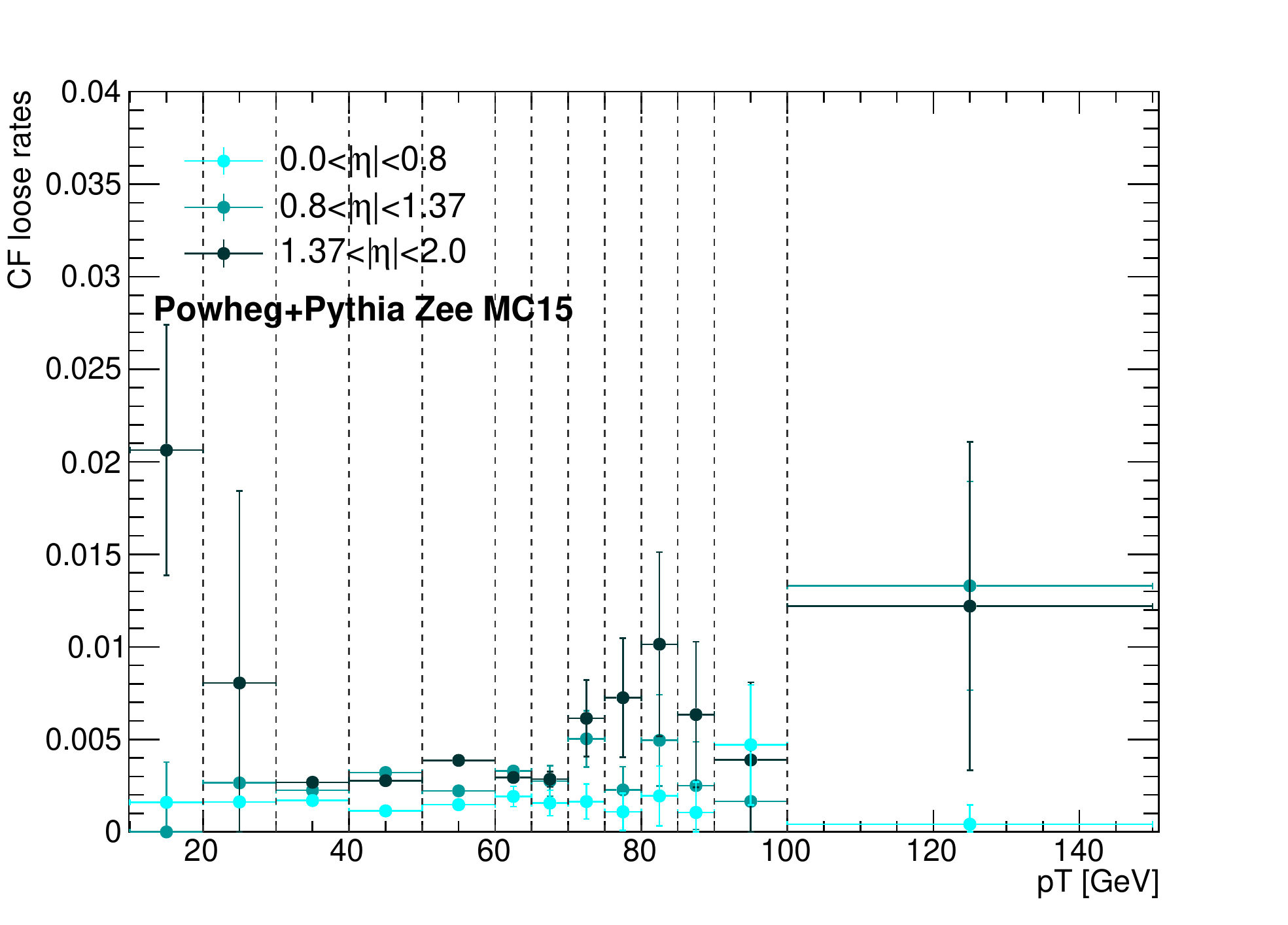}
	\caption{MC, baseline-failing-signal}\label{fig:Chflip_looseMC}
\end{subfigure}
\caption{Charge-flip rate as measured in data (left) and MC (right). 
Only the statistical uncertainty is displayed. The last \pt bin is inclusive.}
\label{fig:ChFlip_Rate}
\end{figure}

\begin{figure}[htb!]
\centering
\begin{subfigure}[t]{0.66\textwidth}\includegraphics[width=\textwidth]{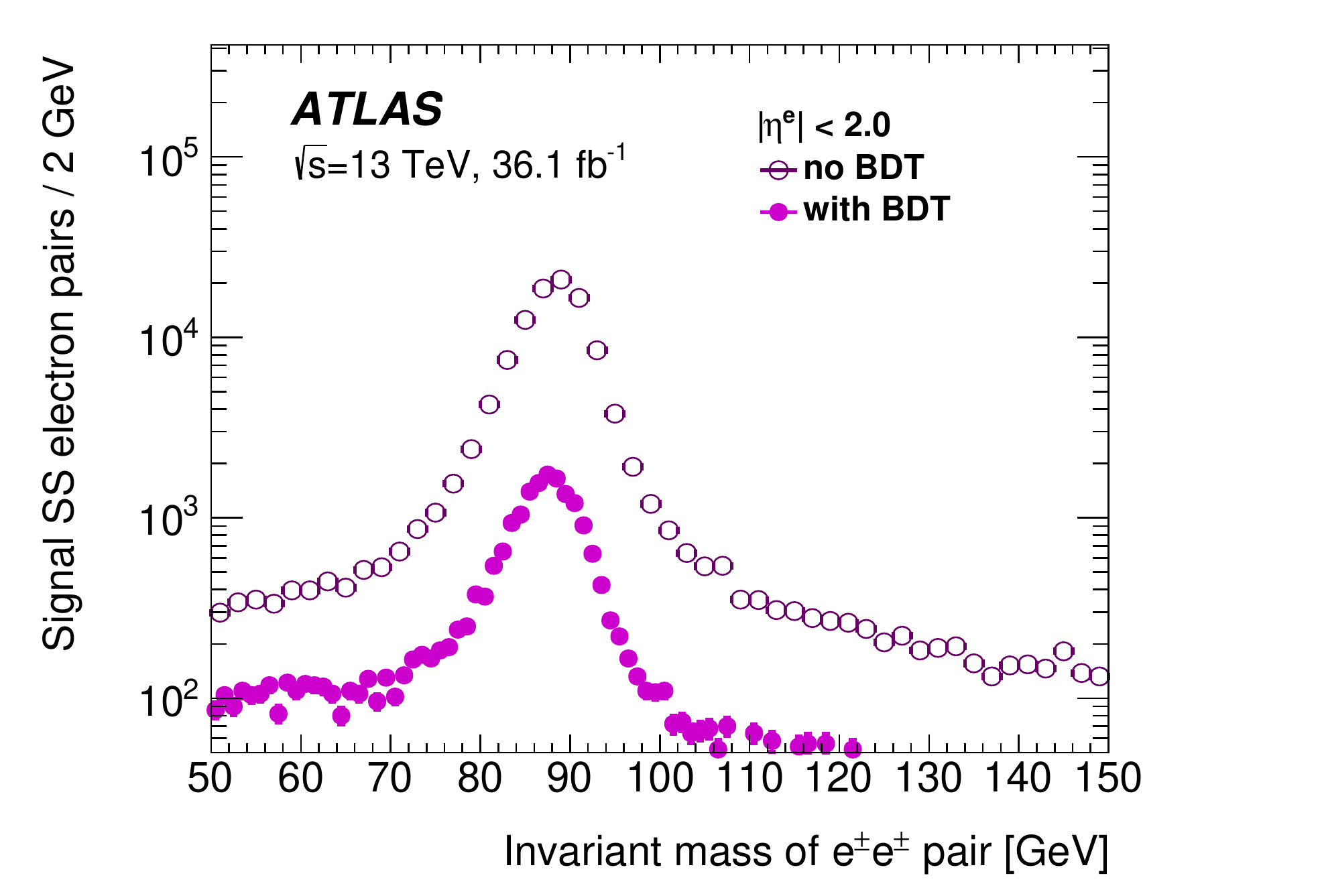}\end{subfigure}
\caption{Invariant mass of the signal $e^{\pm} e^{\pm}$ pair distribution with (full markers) and without (open markers) charge-flip electron BDT selection applied.
}
\label{fig:ETA_SS_BDTEL}
\end{figure}

The charge-flip rates measured in data and MC are shown on Figure~\ref{fig:ChFlip_Rate}. 
 In data, the nominal rates (Figure~\ref{fig:Chflip_nominalData}) go up to $\sim$0.1\% in the barrel region ($|\eta| < 1.37$), 
 while it increases up to $\sim$0.2\% in the end-cap region ($|\eta > 1.37|$). 
 For baseline electrons failing signal requirements (Figure~\ref{fig:Chflip_looseData}), 
 the rates are in general greater than the nominal ones in every bin, as expected. The charge-flip rates for these electrons go up to $\sim$0.5\% in the barrel region and up 1\% in the end-cap region. Compared to the rates used in the previous version of the analysis~\cite{ATLAS-CONF-2016-037}, the central values are much lower now. After suppressing the charge flip events with the charge-flip 
electron BDT classifier described in Section~\ref{subsec:strategy.sel.obj}, 
the charge flip rates are strongly reduced for both signal and baseline-failing-signal electrons (up to a factor 20 in some bins). Figure~\ref{fig:ETA_SS_BDTEL}
illustrates the charge flip background reduction in a loose selection.
Below 30~\GeV\, the statistics are very low for the loose measurement; however, these results are used only to measure the electron fake rate and, as illustrated in Figure~\ref{Figurefakes_preselection_electron}, in this \pt interval the charge flip background is negligible.

\begin{figure}[htb!]
\centering
{\includegraphics[width=0.49\textwidth]{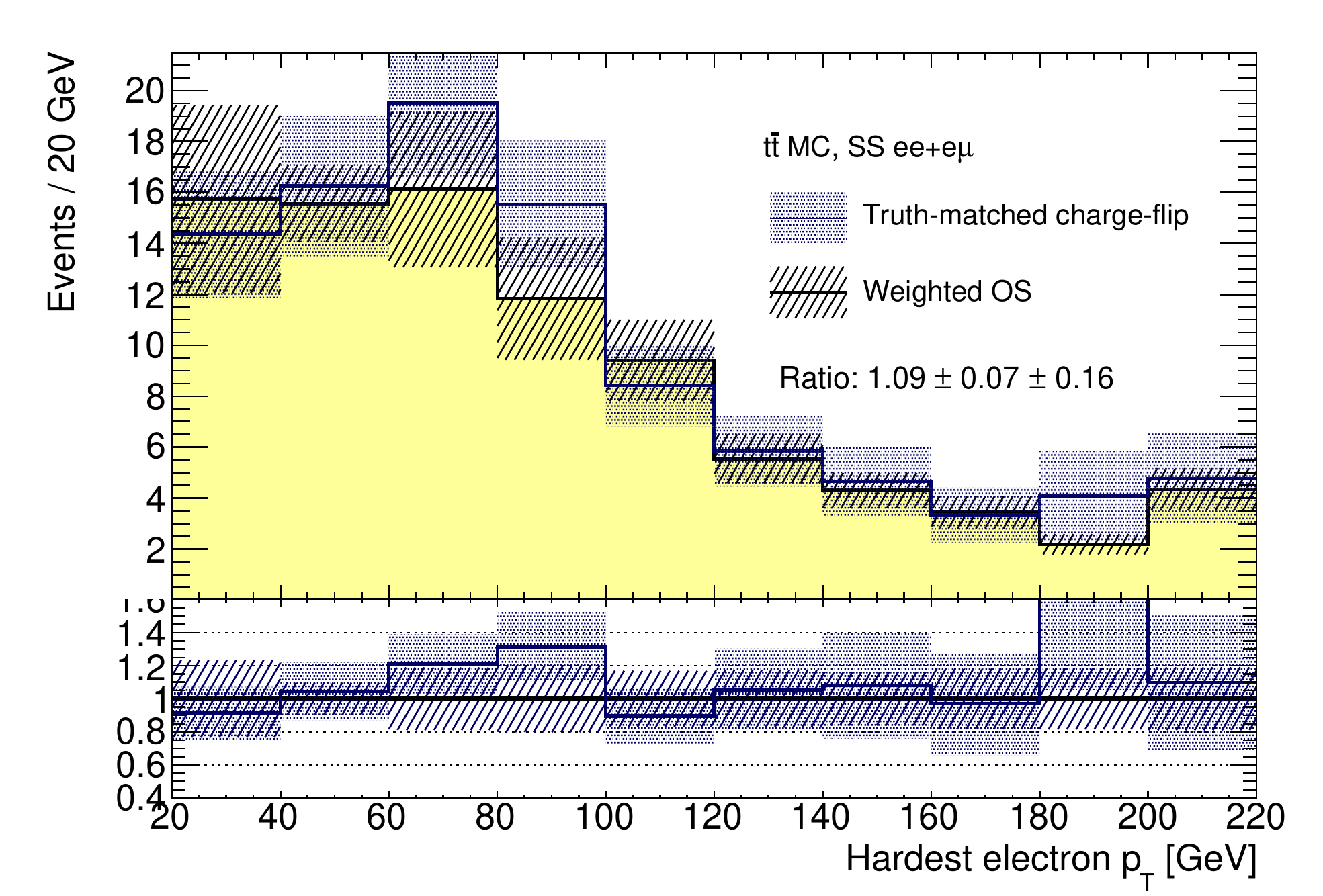}}
{\includegraphics[width=0.49\textwidth]{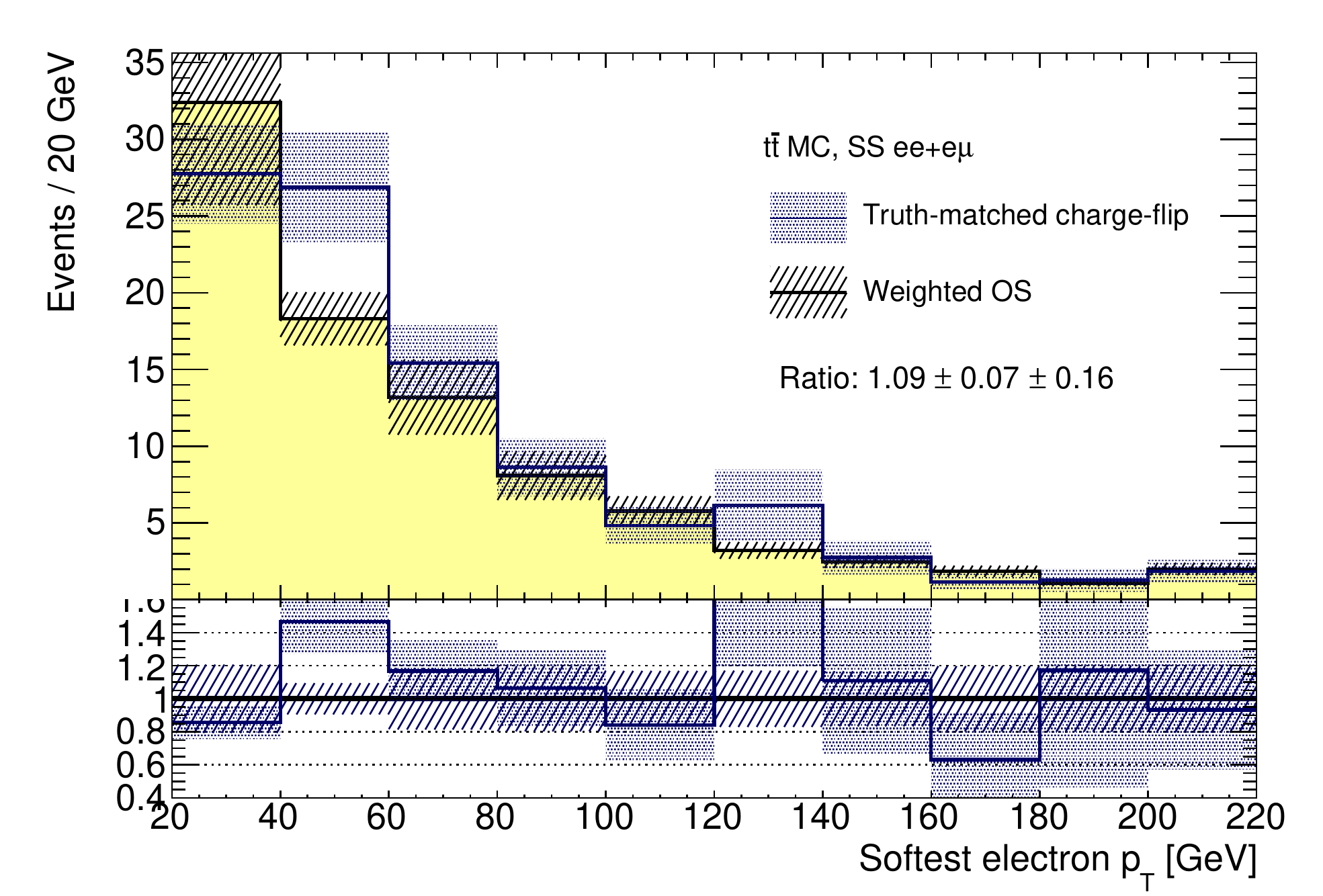}}
{\includegraphics[width=0.49\textwidth]{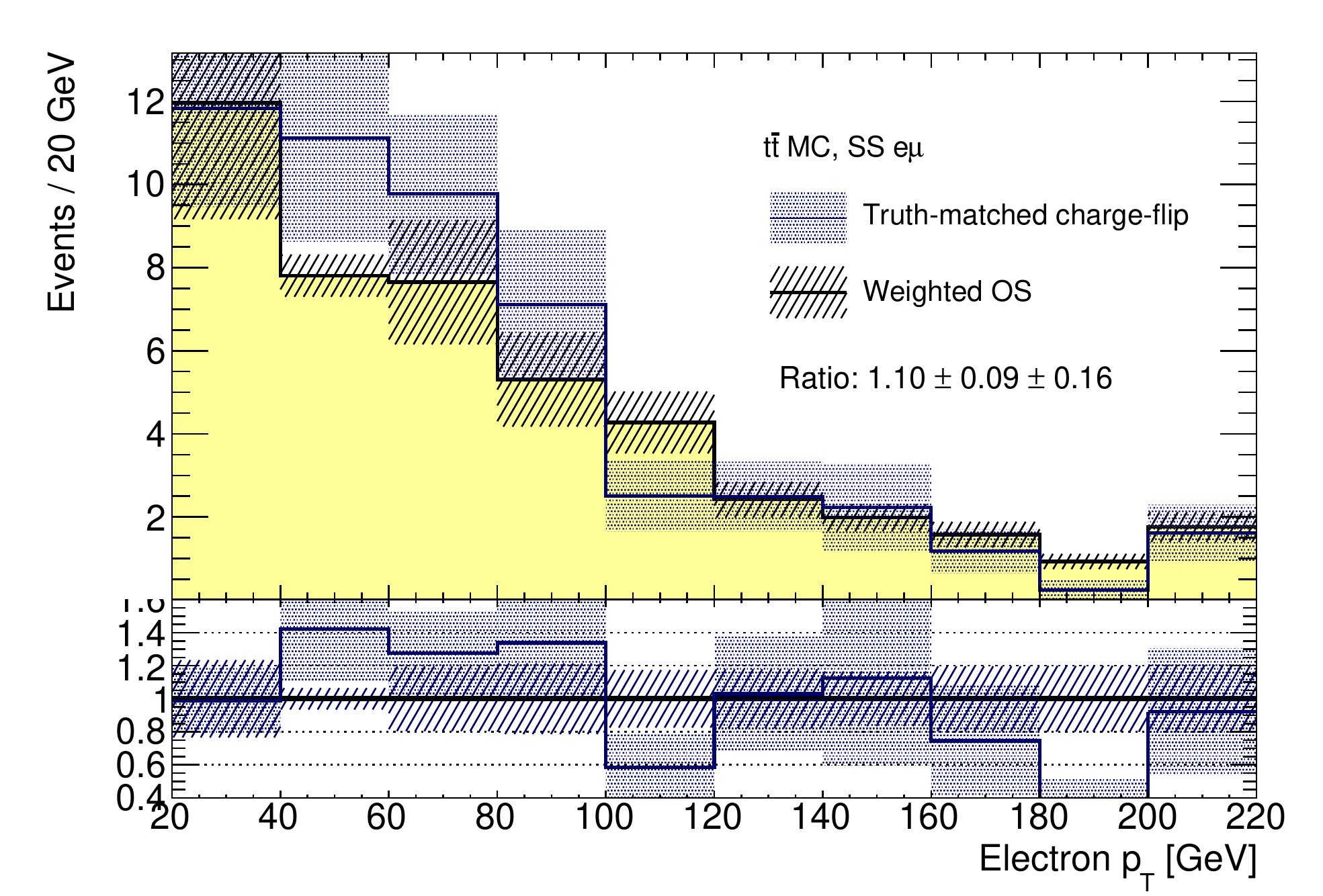}}
\caption{Closure test for the charge-flip background prediction, for simulated $t\bar t$ events 
using charge-flip rates measured in $Z\to ee$ MC 
(with the systematic uncertainties from the data measurements, though). 
Events are selected in the $e\mu$ and $ee$ channels, using signal leptons only, 
and charge-flipped electrons are identified by truth-matching. 
}
\label{fig:ChargeFlip_ClosureTest}
\end{figure}

The charge-flip rates in MC (Figure.~\ref{fig:Chflip_nominalMC},~\ref{fig:Chflip_looseMC}) 
are obtained by applying the same methodology as in data. 
Generally, the rates are not very far from data, validating the use of MC to predict charge-flip background 
in several of the optimization studies presented in this document. 
In addition, a closure test is performed on $t\bar t$ MC, 
checking that weighted OS events can reproduce the distribution of SS charge-flip events (identified by truth-matching). 
A good overall agreement is found, largely within the assigned uncertainties
as shown in Figure~\ref{fig:ChargeFlip_ClosureTest}. 

\subsection*{Systematic uncertainties}
\label{subsec:chargeflip_uncertainties}


The main uncertainties on the measured charge-flip rates come from the presence of background and the way it is estimated. To assess them, variations of the selection and background estimation are considered: 
\begin{itemize}
\item[1)] $75<m_{ee}<100~\GeV$, no background subtraction;
\item[2)] $75<m_{ee}<100~\GeV$, sidebands of 20~\GeV;
\item[3)] $75<m_{ee}<100~\GeV$, sidebands of 25~\GeV~(nominal measurement);
\item[4)] $75<m_{ee}<100~\GeV$, sidebands of 30~\GeV;
\item[5)] $80<m_{ee}<100~\GeV$, sidebands of 20~\GeV.
\end{itemize}

The effect of applying the background subtraction itself is evaluated by comparing configurations 1 and 3. 
The impact of the width of the $m_{ee}$ chosen for the measurement is  by comparing configurations 3 and 5, 
while the sideband width effects are evaluated by comparing configuration 3 and 2, or 3 and 4. 
The largest deviation in each bin is taken as the systematic uncertainty on the charge-flip rate.


For the signal electrons charge-flip rates the systematic uncertainties vary in general between 2\% and 20\% (increasing up to $>50\%$ in the region with $\pt < 10~\GeV$), whereas for baseline-failing-signal electrons they vary between 3\% and 30\% (increasing up to $>50\%$ in the region with $\pt < 10~\GeV$). Part of these large values, at low \pt and in the [80,90]~\GeV~\pt interval, can be explained by large statistical fluctuations between the different configurations.

%% file: texfiles/subsec.bkg.mxm.tex
The matrix method relates the number of events containing prompt or FNP leptons 
to the number of observed events with tight or loose-not-tight leptons 
using the probability for loose prompt or FNP leptons to satisfy the tight criteria.
The formalism for this method has been discussed in Section~\ref{sec:fake.mxm}.
The next sections will concentrate on the measurement of the 
two input variables needed for the matrix method:
the probability for loose FNP leptons to satisfy the tight selection
criteria ($\zeta$) and 
the probability for loose prompt leptons to satisfy the tight selection 
criteria ($\varepsilon$).

\subsection*{Baseline-to-signal efficiency for fake muons}

Baseline-to-signal efficiency for fake leptons (subsequently called ``fake rate'', ($\zeta$)) is measured 
in a sample enriched in fake leptons from \ttbar\ processes.
The MC simulations indicate that this background has the largest contribution to FNP lepton background in the signal regions, 
even those with $b$-jet vetoes, due to the requirements on jet multiplicity and \met. 
The events used for the measurements require exactly two same-sign muons (and no extra baseline lepton), 
at least one $b$-jet, and at least 3 jets that were acquired by di-muon triggers.
One of the muons in the event (referred to as ``tag'') is required to satisfy signal requirements, verify $\pt>25~\GeV$, 
and trigger the event recording. 
The measurement may then be performed on the other lepton (``probe''), likely to be the fake lepton of the pair. 

\begin{figure}[htb!]
\centering
\begin{tabular}{rr}
\begin{subfigure}[t]{0.5\textwidth}\includegraphics[width=\textwidth]{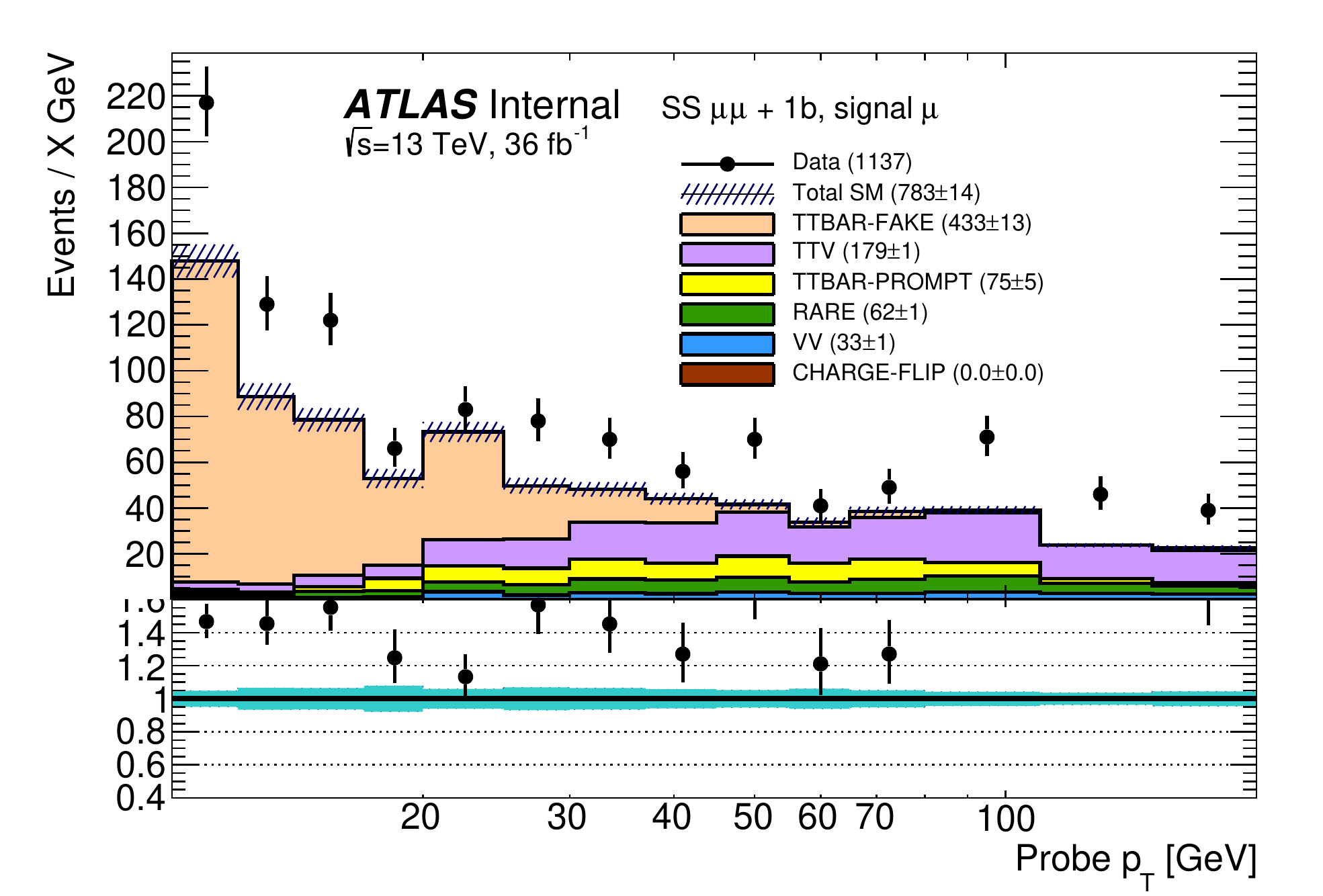}\caption{}\label{fig:bkg.mxm.INCLUSIVETAG_PROBE_PT_MUON_SIGNAL}\end{subfigure}&
\begin{subfigure}[t]{0.5\textwidth}\includegraphics[width=\textwidth]{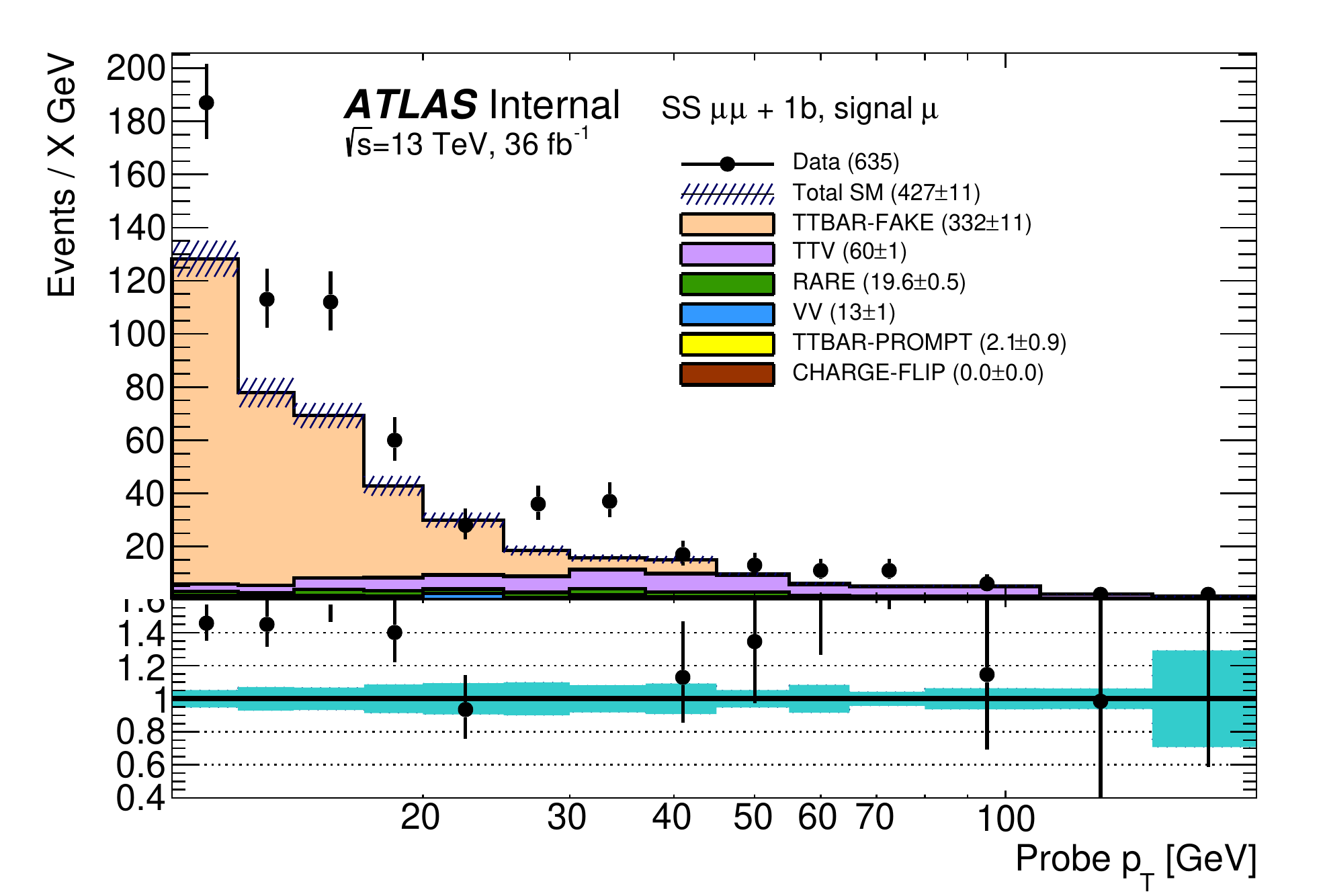}\caption{}\label{fig:bkg.mxm.IDEALTAG_PROBE_PT_MUON_SIGNAL}\end{subfigure} \\
\end{tabular}
\caption
{Signal probe muon \pt\ distribution in data and MC, after pre-selection (left) 
or further tightening of the tag muon requirements (right).
The yellow area indicates $t\bar t$ events in which the tag muon is fake and the probe real, 
leading to a measurement bias. 
}
\label{Figurefakes_preselection_muon}
\end{figure}

Figure~\ref{fig:bkg.mxm.INCLUSIVETAG_PROBE_PT_MUON_SIGNAL} shows the number of signal muon probes available after this pre-selection. 
It is clear that at this stage, measurements above 25 \GeV~would be very affected by the important fraction of events 
in which the tag muon is fake and the probe muon is real. 
To overcome this issue, two alternatives are considered: 
\begin{itemize}
\item tighten the \pt\ and isolation requirements of the tag muon beyond the ``signal'' requirements,
to reduce its probability of being a fake muon
\item use an identical selection for tag and probe muons, and require them to be in the same (\pt,$\eta$) bin for the measurement; 
after subtraction of estimated contributions from processes with two prompt muons, all events have one real and one fake muon, 
and the symmetry in the muon selection can be taken advantage of to obtain an unbiased measurement of the fake rate: 
$$
\zeta = \frac{\varepsilon n_2}{\varepsilon n_1+(2\varepsilon-1)n_2}
$$
with $n_1, n_2$ the number of events with 1 or 2 signal muons, 
and $\varepsilon$ the efficiency for prompt muons.

This method is limited to measurements in inclusive or wide bins. 
It also cannot be used at too low \pt, due to contributions from processes with two fake muons (e.g. from $B\bar B$ meson production). 
\end{itemize}
Comparisons made with $t\bar t$ MC indicated that when using a very tight isolation requirement on the tag muon 
($\operatorname{max}(E_\mathrm{T}^\text{topo, cone 40},p_\mathrm{T}^\text{cone 40})<0.02\times\pt$), 
the level of bias is always greatly inferior to the statistical uncertainty in the measurement, 
which itself is smaller than for the other two methods. 

Figure~\ref{fig:bkg.mxm.IDEALTAG_PROBE_PT_MUON_SIGNAL} shows the number of signal muon probes when applying those reinforced isolation criteria to the tag muon, 
as well as requiring $p_\mathrm{T}^\text{tag}>\operatorname{max}(40,p_\mathrm{T}^\text{probe}+10)$~\GeV. 
As expected, the number of pairs with a fake tag muons is down to a minor level, at least according to the simulation. 

\begin{figure}[htb!]
\centering
\begin{tabular}{rr}
\begin{subfigure}[t]{0.5\textwidth}\includegraphics[width=\textwidth]{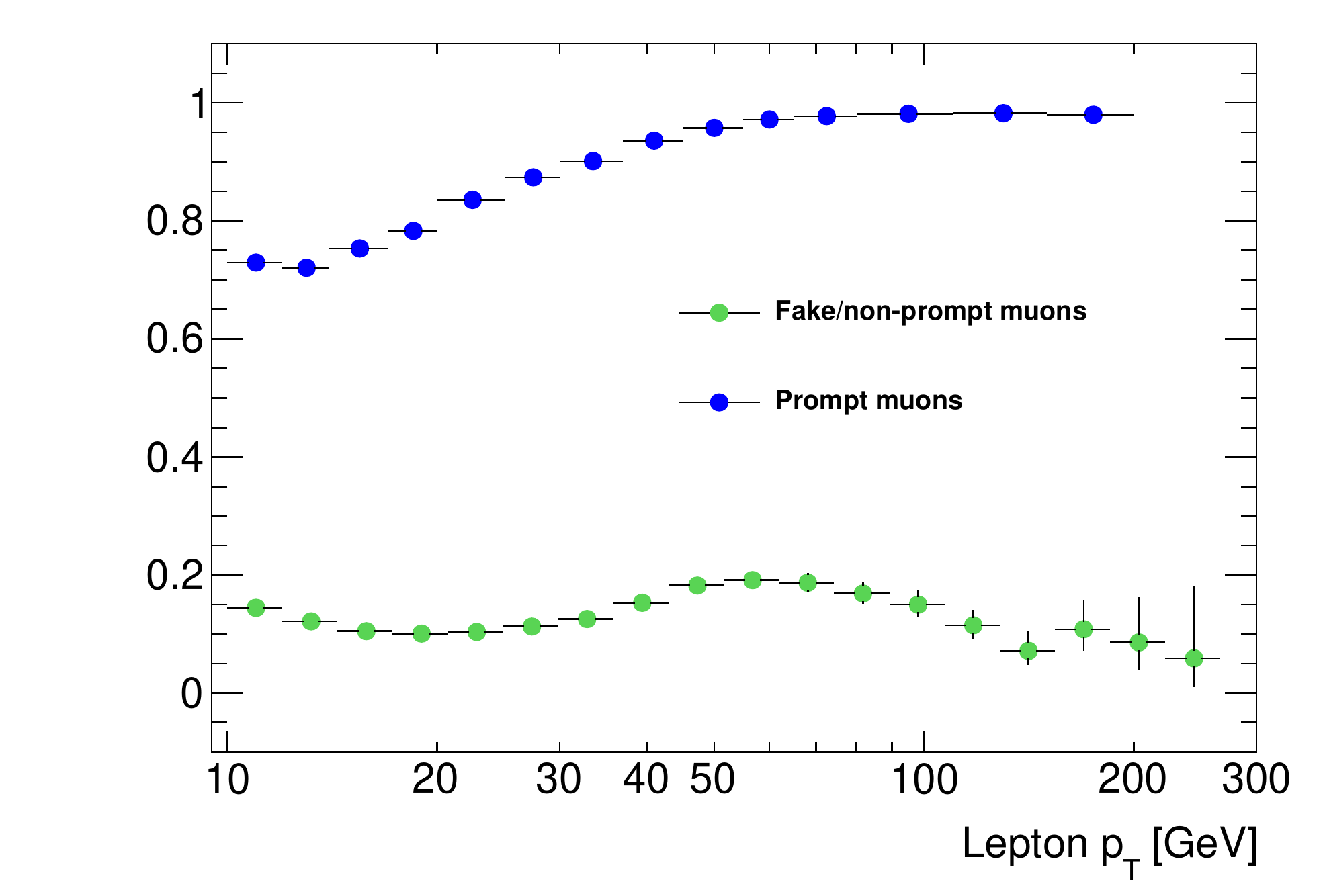}\caption{}\label{fig:TTBAR.Incl.FakeRate.Muon}\end{subfigure}&
\begin{subfigure}[t]{0.5\textwidth}\includegraphics[width=\textwidth]{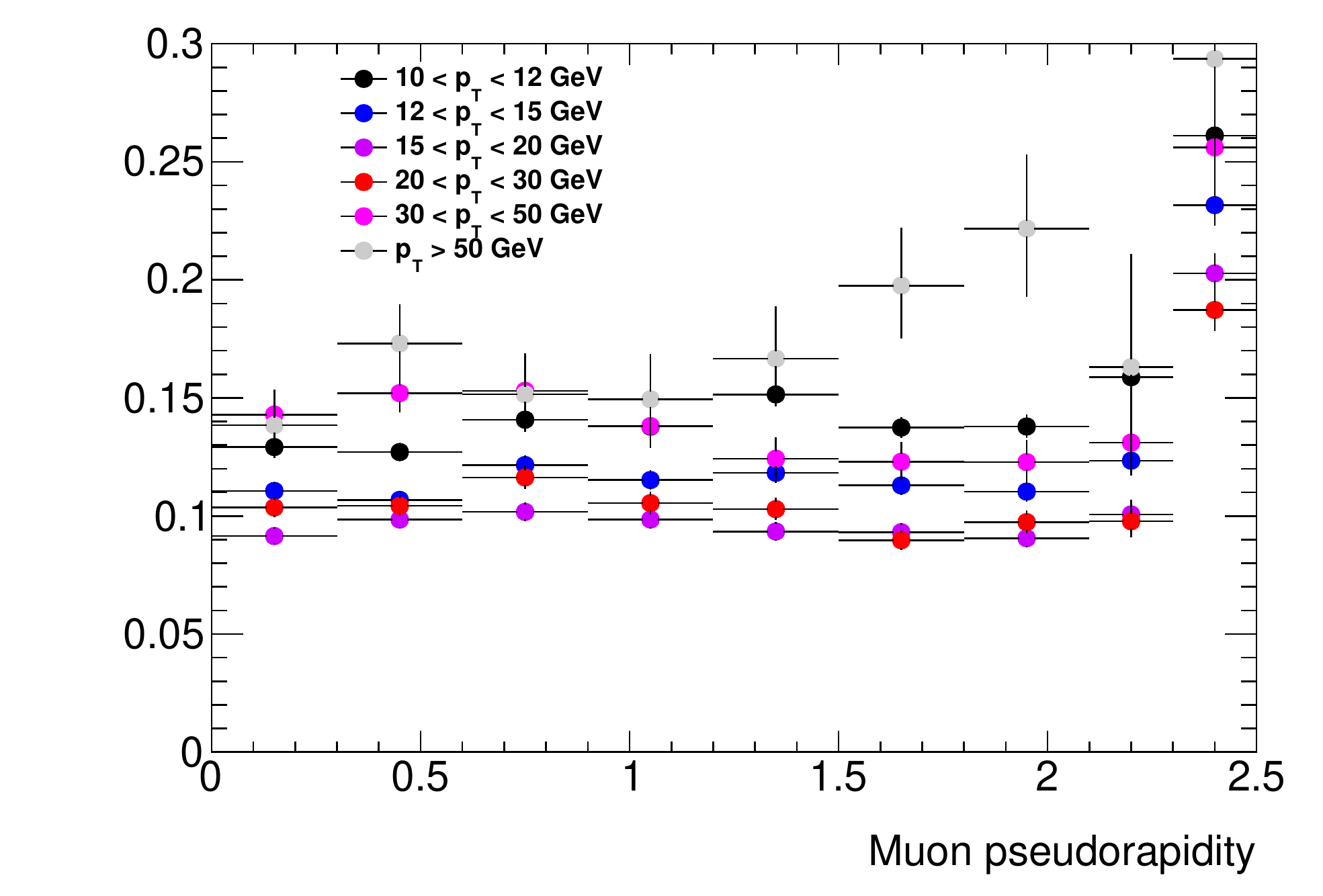}\caption{}\label{fig:TTBAR.Incl.FakeRateVsEta.Muon}\end{subfigure} \\
\end{tabular}
\caption
{
Muon fake rates in $t\bar t$ MC with an inclusive selection, 
as a function of \pt\ (left, green markers) or $|\eta|$ in different momentum ranges (right). 
}
\label{Figurefakes_MC_inclusive_rates_muon}
\end{figure}

Muon fake rates as predicted by the simulation ($t\bar t$, inclusive selection of leptons via truth-matching) 
are shown on Figure~\ref{Figurefakes_MC_inclusive_rates_muon} as function of \pt\ and $|\eta|$. 
One can expect a moderate dependency of the fake rates to the transverse momentum, with the strongest evolution at low \pt and a slight increase toward higher \pt. 
The fake rates are also essentially independent of the pseudorapidity, 
except at the edge ($|\eta|>2.3$) where there is a strongly pronounced increase of the rates. 
This motivates measurements in data as function of \pt\ in two $|\eta|$ bins. 

Observations in data seem to indicate that the rejection of fake tag muons by the reinforced isolation criteria 
is less important than in the simulation, or that the amount of fake muons at high \pt\ is larger than in the simulation, or both. 
This leads to an unknown level of bias in measurements performed with the straightforward tag-and-probe selection at high \pt. 
For that reason, the final rates measured in data are provided by the tag-and-probe method below 25~\GeV, 
and by the symmetric selection for $\pt>25~\GeV$. The former are obtained with 

\begin{align}
\zeta&=\frac{n_\text{signal}^\text{data} - n_\text{signal}^\text{MC}}{n_\text{baseline}^\text{data} - n_\text{baseline}^\text{MC}}\\
\text{with }&\Delta\zeta_\text{stat} = \frac{\sqrt{(1-2\zeta)n_\text{signal}^\text{data} + \zeta^2 n_\text{baseline}^\text{data}}}
{n_\text{baseline}^\text{data} - n_\text{baseline}^\text{MC}}\notag
\end{align}
while the latter are obtained with:
\begin{align}
\zeta &= \frac{\varepsilon (n_\text{both signal}^\text{data} - n_\text{both signal}^\text{MC})}
{\varepsilon (n_\text{only 1 signal}^\text{data} - n_\text{only 1 signal}^\text{MC})
+(2\varepsilon-1)(n_\text{both signal}^\text{data} - n_\text{both signal}^\text{MC})}\\
\text{with }&\Delta\zeta_\text{stat} 
= \frac{\zeta}{n_\text{both signal}^\text{data} - n_\text{both signal}^\text{MC}}
\sqrt{\zeta^2 n_\text{only 1 signal} + \left(1-\frac{2\varepsilon-1}{\varepsilon}\zeta\right)^2 n_\text{both signal}}.\notag
\end{align}
The efficiency for prompt muons $\varepsilon$ is assigned values compatible with section~\ref{subsubsec:fakes_matrix_real_efficiency}. 

The measured rates are presented in Table~\ref{table:fake_rates_muon}. 
The central values are shown together with the associated statistical uncertainty, 
as well as the propagation of the uncertainty on the subtracted backgrounds normalization, 
which is taken as a global $\Delta B/B=20\%$. 
The rates are of the order of $10\%$ up to 30 \GeV, beyond which they increase. 
Overall these values are not very different from those predicted by the simulation. 

\begin{table}[htb!]
\def\arraystretch{1.15}
\def\arraystretch{1.15}
\centering
\resizebox{\textwidth}{!}{
\begin{tabular}{|c|c|c|c|} \hline\hline
\multicolumn{2}{|c|}{$10<\pt<12~\GeV$}         & \multicolumn{2}{c|}{$12<\pt<14$}                  \\  
\hline 
$|\eta|<2.3$             & $|\eta|>2.3$             & $|\eta|<2.3$             & $|\eta|>2.3$            \\
\hline
$0.14 \pm 0.01 \pm 0.00$ & $0.22 \pm 0.05 \pm 0.00$ & $0.11 \pm 0.01 \pm 0.00$ & $0.24 \pm 0.06 \pm 0.00$ \\ 
\hline
\end{tabular}}

\resizebox{\textwidth}{!}{
\begin{tabular}{|c|c|c|c|} \hline
\multicolumn{2}{|c|}{$14<\pt<17$}                    & \multicolumn{2}{c|}{$17<\pt< 20~\GeV$}       \\       
\hline
$|\eta|<2.3$             & $|\eta|>2.3$             & $|\eta|<2.3$             & $|\eta|>2.3$            \\    
\hline
$0.12 \pm 0.01 \pm 0.00$ & $0.09 \pm 0.05 \pm 0.00$ & $0.09 \pm 0.01 \pm 0.00$ & $0.21 \pm 0.07 \pm 0.00$ \\
\hline 
\end{tabular}}

\resizebox{\textwidth}{!}{
\begin{tabular}{|c|c|c|c|c|} \hline
             $20<\pt<30$ &              $30<\pt<40$ &              $40<\pt<60$ &                $\pt>60$ \\
\hline
$0.07 \pm 0.02 \pm 0.00$ & $0.12 \pm 0.05 \pm 0.01$ & $0.16 \pm 0.09 \pm 0.04$ & $0.49 \pm 0.10 \pm 0.07$ \\
\hline \hline
\end{tabular}}
\caption{Muon fake rate measured in data and the associated statistical uncertainty. 
The systematic uncertainty originating from the subtraction of ``backgrounds'' with only prompt leptons is also displayed. }
\label{table:fake_rates_muon}
\end{table}

\begin{figure}[htb!]
\centering
\includegraphics[width=0.49\textwidth]{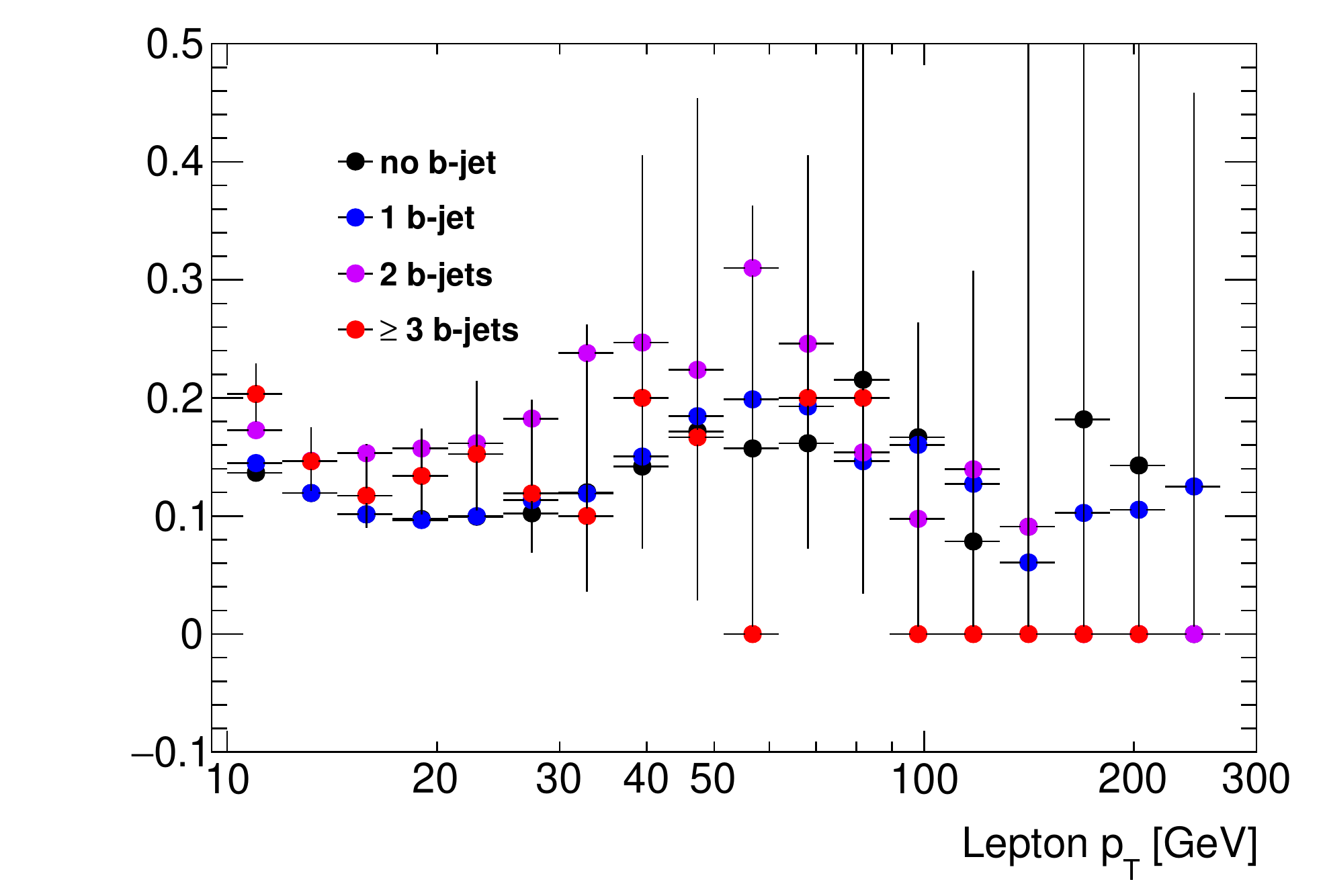}
\caption
{
Muon fake rates in $t\bar t$ MC with an inclusive selection, 
as function of \pt\ and split according to the number of $b$-tagged jets in the event. 
}
\label{fig:fakes_MC_vsBjets_muon}
\end{figure}

Some of the validation and signal regions require events with 2 or more $b$-tagged jets, 
which reduces the fraction of non-prompt muons coming from $B$ meson decays. 
Figure~\ref{fig:fakes_MC_vsBjets_muon} illustrates how this impacts 
the fake rates. 
Given the good agreement between data and simulation for the measured values, 
a correction is applied to the measured rates for events with $\ge 2$ $b$-jets, 
taken directly from simulated $t\bar t$ events. 
This correction factor varies between 1 and 2 with \pt, 
and the whole size of the correction is assigned as an additional systematic uncertainty (see Table~\ref{tab:fake_rates_muon_systematics}). 

\begin{table}[htb!]
\def\arraystretch{1.15}
\def\arraystretch{1.15}
\centering
\resizebox{\textwidth}{!}{
\begin{tabular}{|c|c|c|c|c|c|c|} \hline\hline
 \pt & $<14$ & $14-20$ & $20-30$ & $30-40$ & $40-60$ & $>60$\\\hline
$\Delta\zeta^\text{(syst)}$ & 30\% & 30\%  & 30\%     &  50\%   & \multicolumn{2}{c|}{
		\begin{tabular}{@{}c@{}}50\% for $H_\mathrm{T}<600$ \\ 70\% for $600<H_\mathrm{T}<1200$ \\85\% for $H_\mathrm{T}>1200$\end{tabular}} \\\hline
$\frac{\zeta_{\ge 2b}}{\zeta}$ & $1.2\pm 0.2$ & $1.5\pm 0.5$ & $1.7\pm 0.7$ & $2.0\pm 1.0$ & $1.5\pm 0.5$ & $-$\\\hline
\end{tabular}}
\caption{Additional systematic uncertainty on the muon fake rates, to address variations of the latter in different environments. 
The table also shows the correction factors and uncertainties applied to final states with $\ge 2$ $b$-tagged jets.}
\label{tab:fake_rates_muon_systematics}
\end{table}

\par{\bf Systematic uncertainties\\}
To cover potential differences in the fake rates between the measurement regions and the signal regions, 
that could be due to different origins or kinematic properties of the fake leptons, 
uncertainties are set based on the extent of those differences predicted by the simulation. 
The largest effect is the decrease of the fake rates with $H_\mathrm{T}$ (especially for high-\pt\ muons), 
which likely correlates to a harder jet at the origin of the non-prompt muon, hence a reduced likelihood to satisfy isolation requirements. 
Table~\ref{tab:fake_rates_muon_systematics} summarizes the additional systematic uncertainties applied to the muon fake rates. 
They vary from $30\%$ at low \pt, to up to 85\% for $\pt>40~\GeV$; in that range, the uncertainties are made $H_\mathrm{T}$-dependent. 

As already shown, Figure~\ref{fig:fakes_MC_vsBjets_muon} shows the variation of the fake rate in $t\bar t$ MC as a function of the number of $b$-tagged jets in the event. 
Unsurprisingly, the rates are very similar for $0b$ and $\ge 1b$ final states, 
justifying the use of the fake rates measured in this section (i.e. in a $\ge 1b$ region) to predict fake muon background in all signal regions. 

\subsection*{Baseline-to-signal efficiency for fake electrons}
\label{subsubsec:fakes_matrix_fake_rate_electrons}

\begin{figure}[htb!]
\centering
\includegraphics[width=0.49\textwidth]{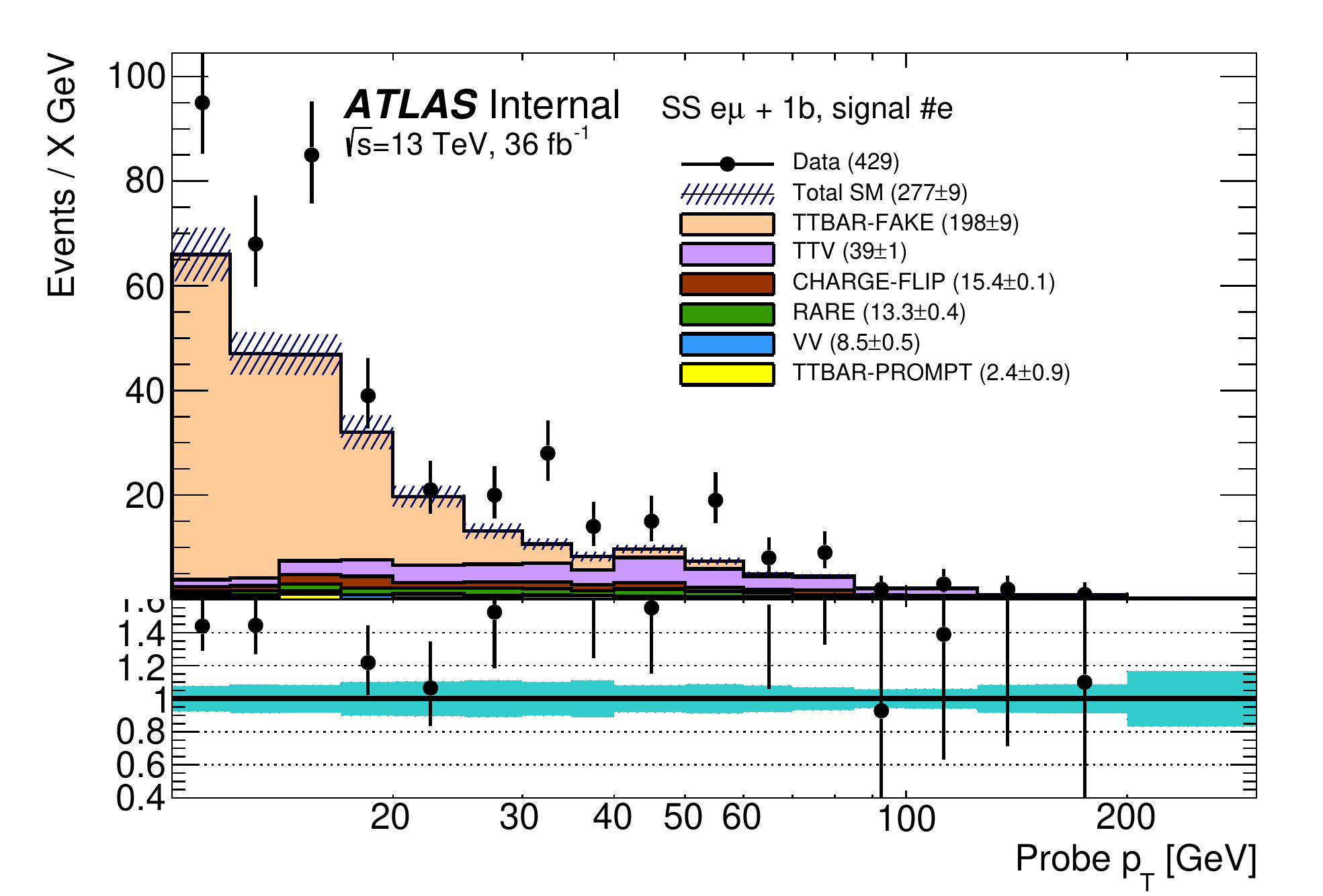}
\includegraphics[width=0.49\textwidth]{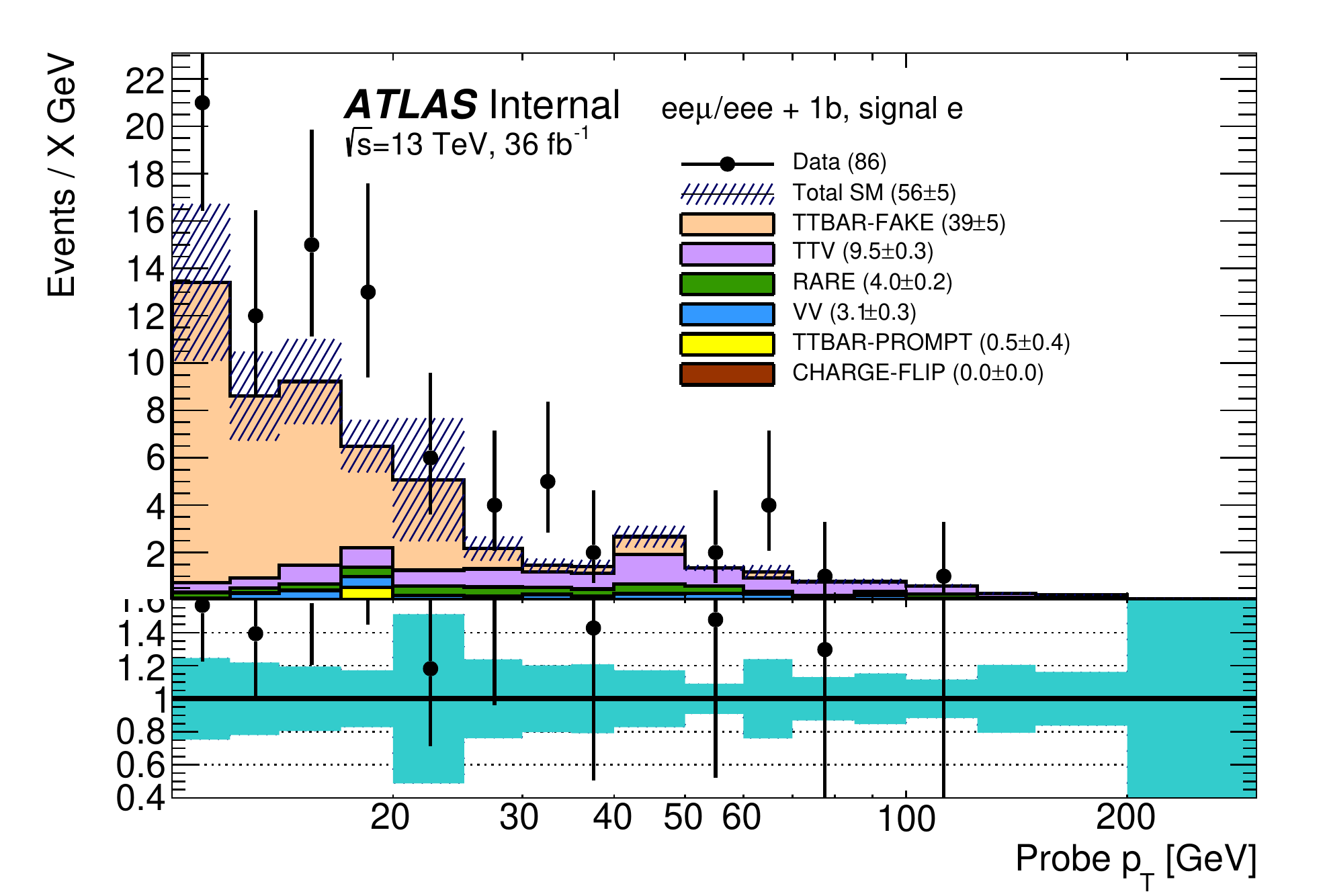}
\caption
{Signal probe electron \pt\ distribution in data and MC, for $e^\pm\mu^\pm$ pairs (left, with probe electrons satisfying  a reinforced tag muon selection) 
or $\ell^\pm e^\mp e^\mp$ pairs (right, with reinforced tag electron selection), 
as described in section~\ref{subsubsec:fakes_matrix_fake_rate_electrons}. 
The yellow area indicates $t\bar t$ events in which the tag lepton is fake and the probe electron real, 
leading to a measurement bias. 
}
\label{Figurefakes_preselection_electron}
\end{figure}

Electron fake rates are measured with a similar methodology, but the $e^\pm e^\pm$ channel is unusable due to the presence of a large charge-flip background. 
This is overcome by working with $e^\pm\mu^\pm$ pairs instead (with a tag muon), but mixing leptons of different flavours brings additional complications 
(for example, the unbiased measurement cannot be employed to measure muon fake rates at higher \pt, as there is no symmetry between the leptons). 
To improve confidence, measurements are performed in four different ways, which complement each other: 
\begin{itemize}
\item straightforward tag-and-probe with $e\mu$ pairs, with the same tag muon selection as in the previous section. 
\item same selection, but subtracting from the numerator the number of pairs with one fake tag muon and one prompt probe electron, 
itself estimated from the number of observed $e\mu$ events with a muon failing signal requirements, 
scaled by an efficiency correction factor $e\mu/\mu\mu$ taken from $t\bar t$ MC (only for pairs with one fake muon). 
This only works if the two muons satisfy the same kinematic requirements, therefore can be used only for measurements in wide or inclusive bins. 
\item selecting $\ell^\mp e^\pm e^\pm+\ge 1b$ events, with a $Z$ veto on SFOS pairs. 
This selection entirely suppresses contributions from charge-flip, or events with fake muons. 
One of the electron, with standard signal requirements, is required to satisfy the same reinforced \pt\ and isolation requirements as for the muon measurement,
and the measurement can be performed on the other electron. 
\item same selection, using the symmetry between the two same-sign electrons to measure the rates in an unbiased way, similarly to the muon case. 
\end{itemize}
Events are acquired with the combination of single-muon (as in previous section) and $e\mu$ triggers.

\begin{figure}[htb!]
\centering
\includegraphics[width=0.49\textwidth]{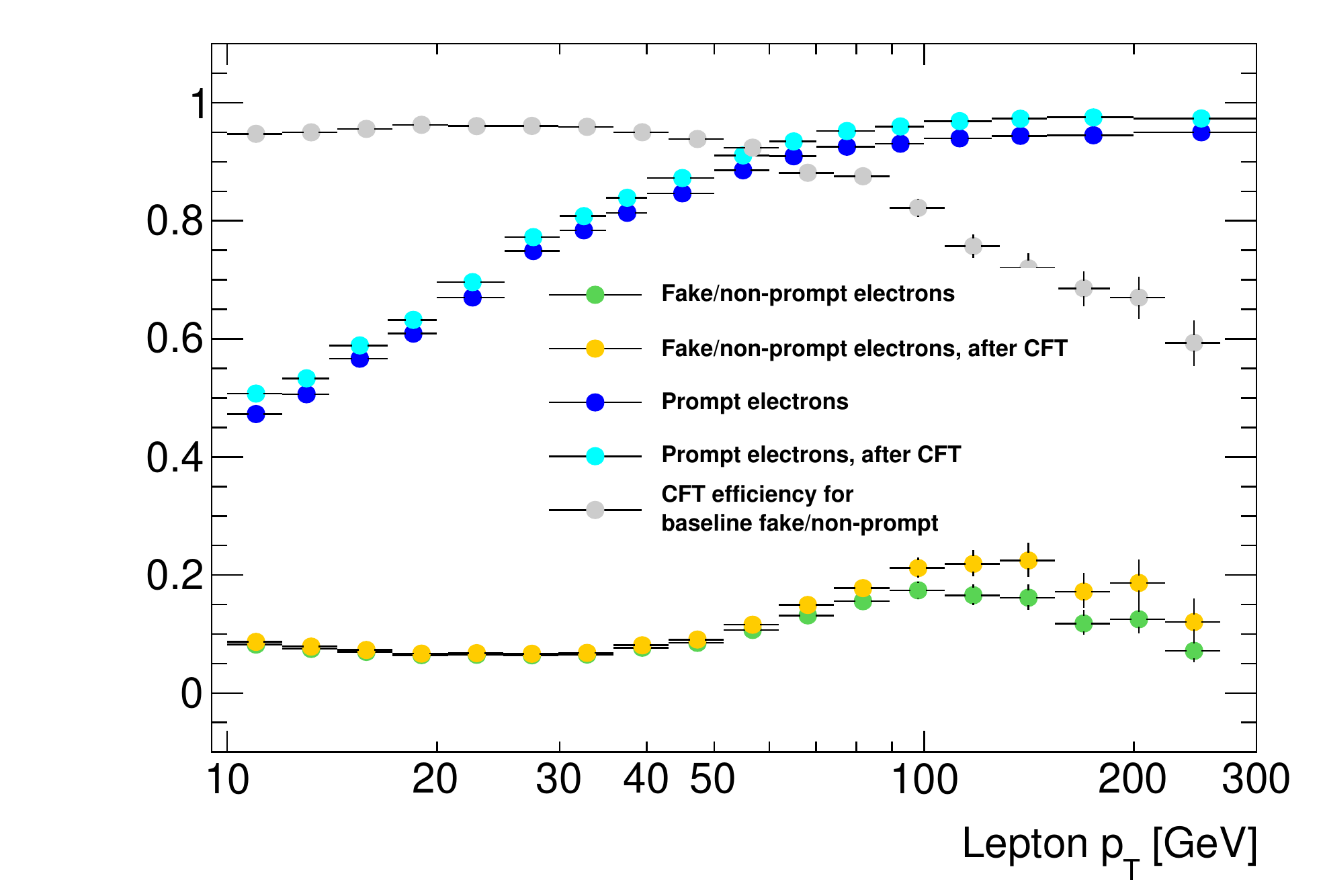}
\includegraphics[width=0.49\textwidth]{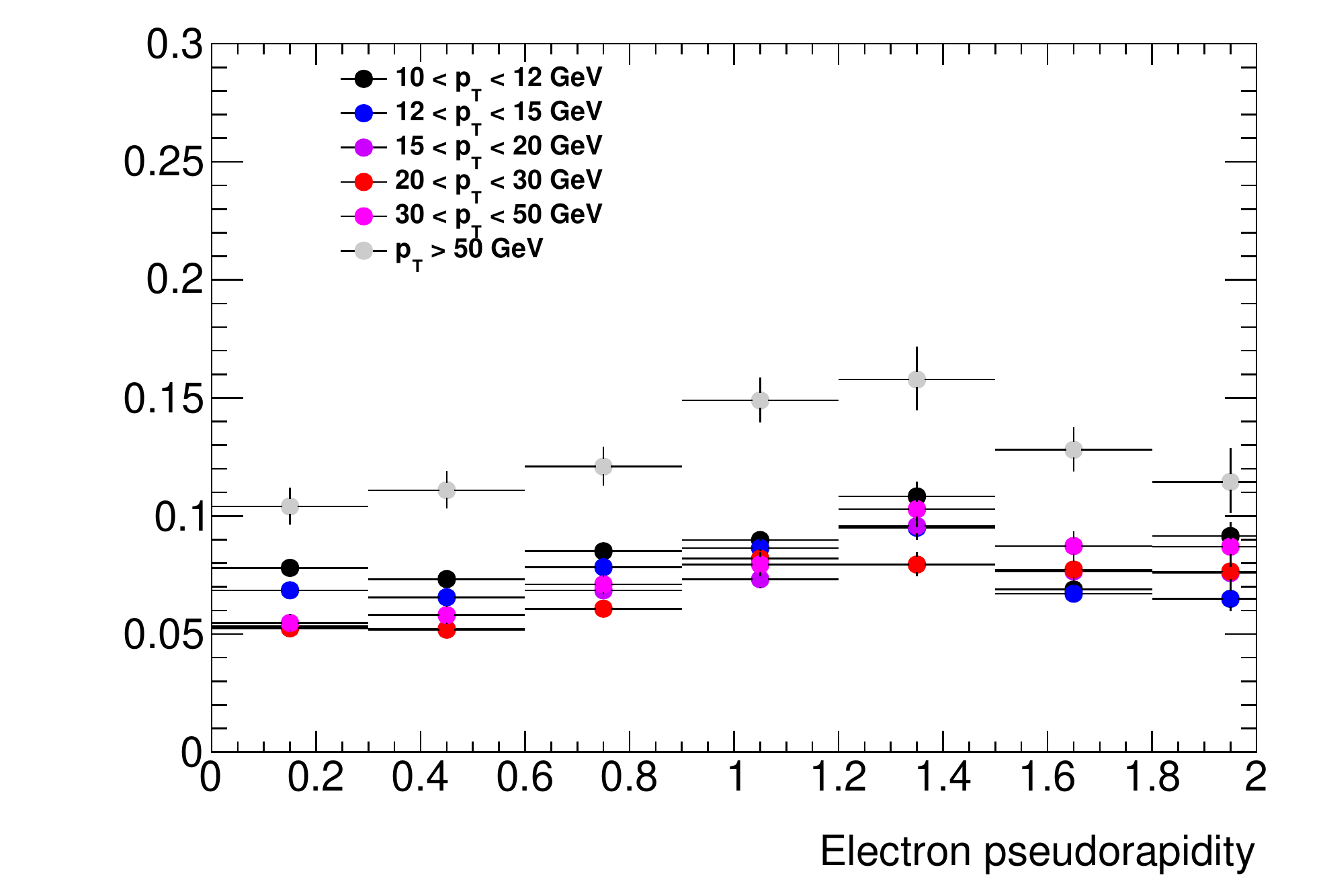}
\caption
{
Electron fake rates in $t\bar t$ MC with an inclusive selection, 
as function of \pt\ (left, yellow/green markers = with/without CFT cut applied) or $|\eta|$ in different momentum ranges (right). 
}
\label{Figurefakes_MC_inclusive_rates_electron}
\end{figure}

Figure~\ref{Figurefakes_preselection_electron} shows the number of signal probe electrons selected in the $e\mu$ and $\ell ee$ channels. 
There are significantly fewer events selected in the trilepton channel. 
Figure~\ref{Figurefakes_MC_inclusive_rates_electron} shows the electron fake rate as a function of \pt\ or $\eta$ in $t\bar t$ MC. 
The variations of the rates as function of the pseudorapidity are not very large, 
therefore measurements are only performed as a function of \pt. 
The low \pt\ range is dominated by non-prompt electrons from heavy flavour decays, while beyond 30~\GeV, 
electron fakes mostly come from conversions of photons produced inside jets, such as $\pi^0\to\gamma\gamma$ decays. 

Based on the estimated levels of bias, and achievable statistical precision of the different methods, 
the electron fake rate is measured with the tag-and-probe $e\mu$ selection up to 30~\GeV, 
and by combining ``unbiased'' evaluations in both $e\mu$ and $\ell ee$ channels beyond. 
The measured rates are presented in Table~\ref{table:fake_rates_electron}, 
together with the associated statistical and background-subtraction uncertainties. 
The rates are of the order of $10\%$ up to 30 \GeV, beyond which they increase 
up to 25\%. 

\begin{table}[htb!]
\def\arraystretch{1.15}
\def\arraystretch{1.15}
\centering
\resizebox{\textwidth}{!}{
\begin{tabular}{|c|c|c|c|} \hline\hline
             $10<\pt<12$ &              $12<\pt<14$ &              $14<\pt<17$ &             $17<\pt<20$ \\
$0.10 \pm 0.01 \pm 0.00$ & $0.10 \pm 0.01 \pm 0.01$ & $0.12 \pm 0.01 \pm 0.01$ & $0.08 \pm 0.02 \pm 0.00$ \\
\hline \hline
\end{tabular}}

\resizebox{\textwidth}{!}{
\begin{tabular}{|c|c|c|c|} \hline\hline
             $20<\pt<25$ &              $25<\pt<30$ &              $30<\pt<40$ &             $40>\pt$ \\
$0.07 \pm 0.02 \pm 0.01$ & $0.11 \pm 0.03 \pm 0.01$ & $0.20 \pm 0.07 \pm 0.03$ & $0.25 \pm 0.10 \pm 0.05$ \\
\hline \hline
\end{tabular}}
\caption{Electron fake rate measured in data and the associated statistical uncertainty. 
The systematic uncertainty originating from the subtraction of ``backgrounds'' with only prompt leptons is also displayed. }
\label{table:fake_rates_electron}
\end{table}

Unlike muons, MC-based correction factors are not applied for final states with $\ge 2$ $b$-tagged jets. 
This is because there is less good agreement between the measured rates and the simulation; 
in particular the former take larger values in the medium-\pt\ range. 

\par
\begin{figure}[htb!]
\centering
\includegraphics[width=0.49\textwidth]{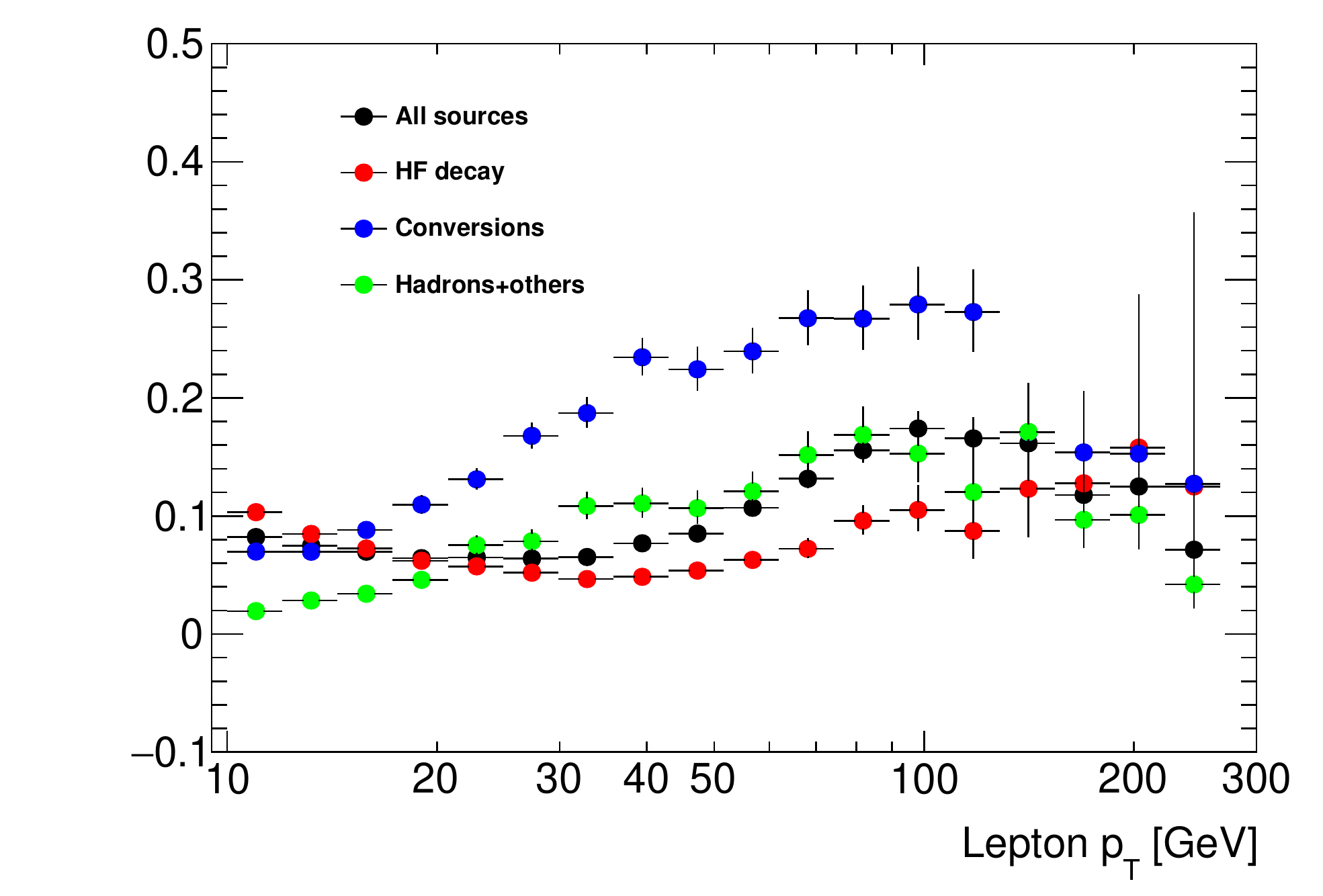}
\includegraphics[width=0.49\textwidth]{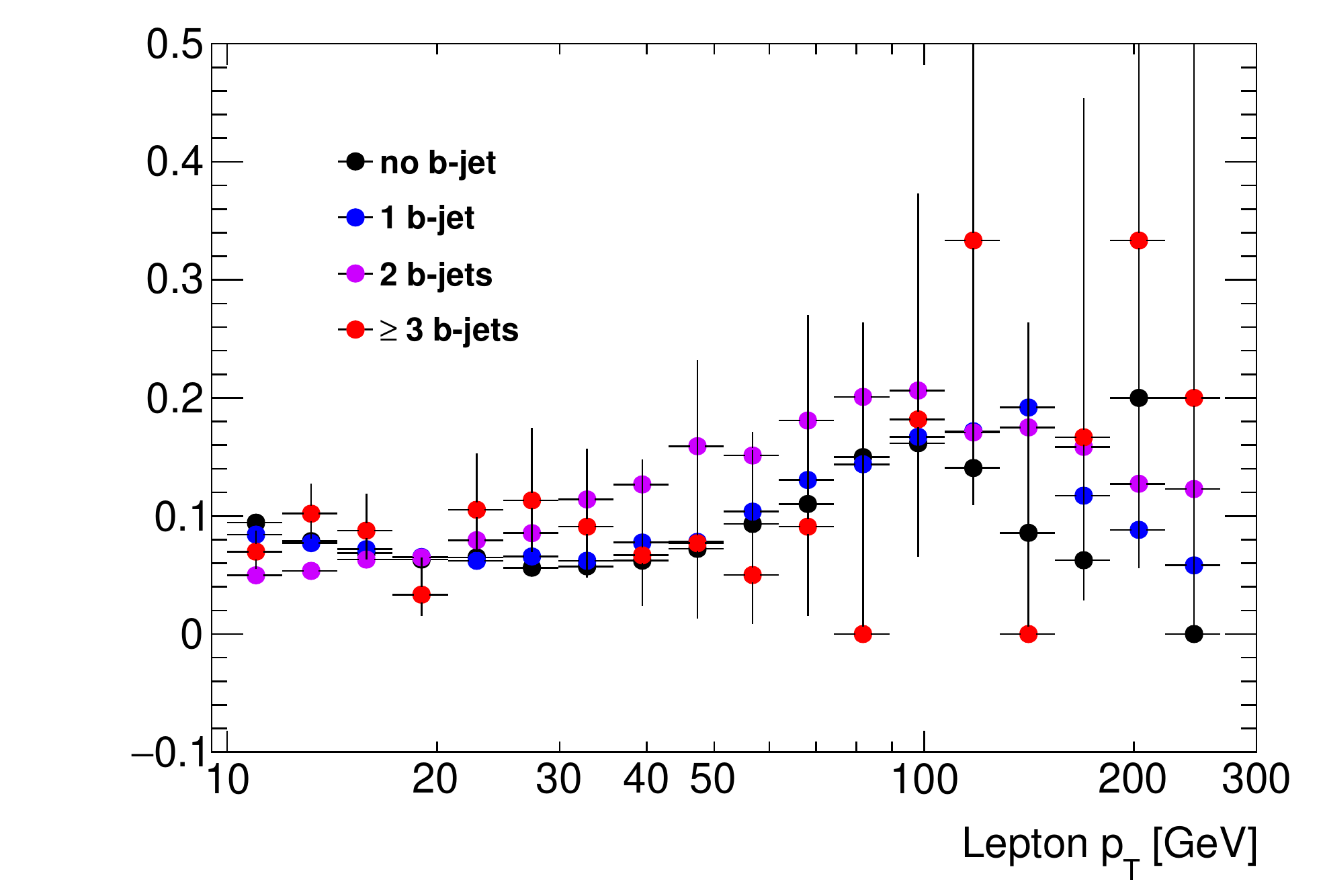}
\caption
{
Electron fake rates in $t\bar t$ MC with an inclusive selection, 
as a function of \pt\ and split according to the source of the fake electron (left). 
The relative contributions of each source (for signal electrons) are indicated on the right-hand-side. 
}
\label{fig:fakes_MC_perSource_electron}
\end{figure}

{\bf Systematic uncertainties\\}
Similarly to the muon case, systematic uncertainties are assigned to cover for difference in the rates in the measurement regions and in the signal regions
that would be due to different sources of fake leptons, or different kinematic properties of these sources. 
Unlike muons, there is much less of a dependency to $H_\mathrm{T}$. 
The dominant source of potential differences is therefore the origin of the fake electron (see Figure~\ref{fig:fakes_MC_perSource_electron}); 
for $\pt<20~\GeV$, non-prompt electrons from heavy--flavor hadron decays dominate,
which is confirmed by the good agreement between MC fake rates and those measured in data. 
In that range, an uncertainty of $30\%$ is assigned to the fake rates (inflated to $50\%$ for final states with $\ge 2b$-tagged jets). 
The rates measured in data are larger than those predicted by the simulation, 
and would for example be consistent with a larger amount of electrons from photon conversions than predicted. 
In that range, an uncertainty of $50\%$ is assigned to cover any arbitrary variation of the relative contributions of each source. 

Finally, Figure~\ref{fig:fakes_MC_perSource_electron} shows the variation of the fake rate in $t\bar t$ MC as function of the number of $b$-tagged jets in the event. 
As expected, the rates are very similar for $0b$ and $\ge 1b$ final states, 
justifying the use of the fake rates measured in this section (i.e. in a $\ge 1b$ region) to predict fake electron background in all signal regions. 

\subsection*{Baseline-to-signal efficiency for real leptons}
\label{subsubsec:fakes_matrix_real_efficiency}

\begin{figure}[htb!]
\centering
\begin{subfigure}[b]{0.45\textwidth}
	\includegraphics[width=\textwidth]{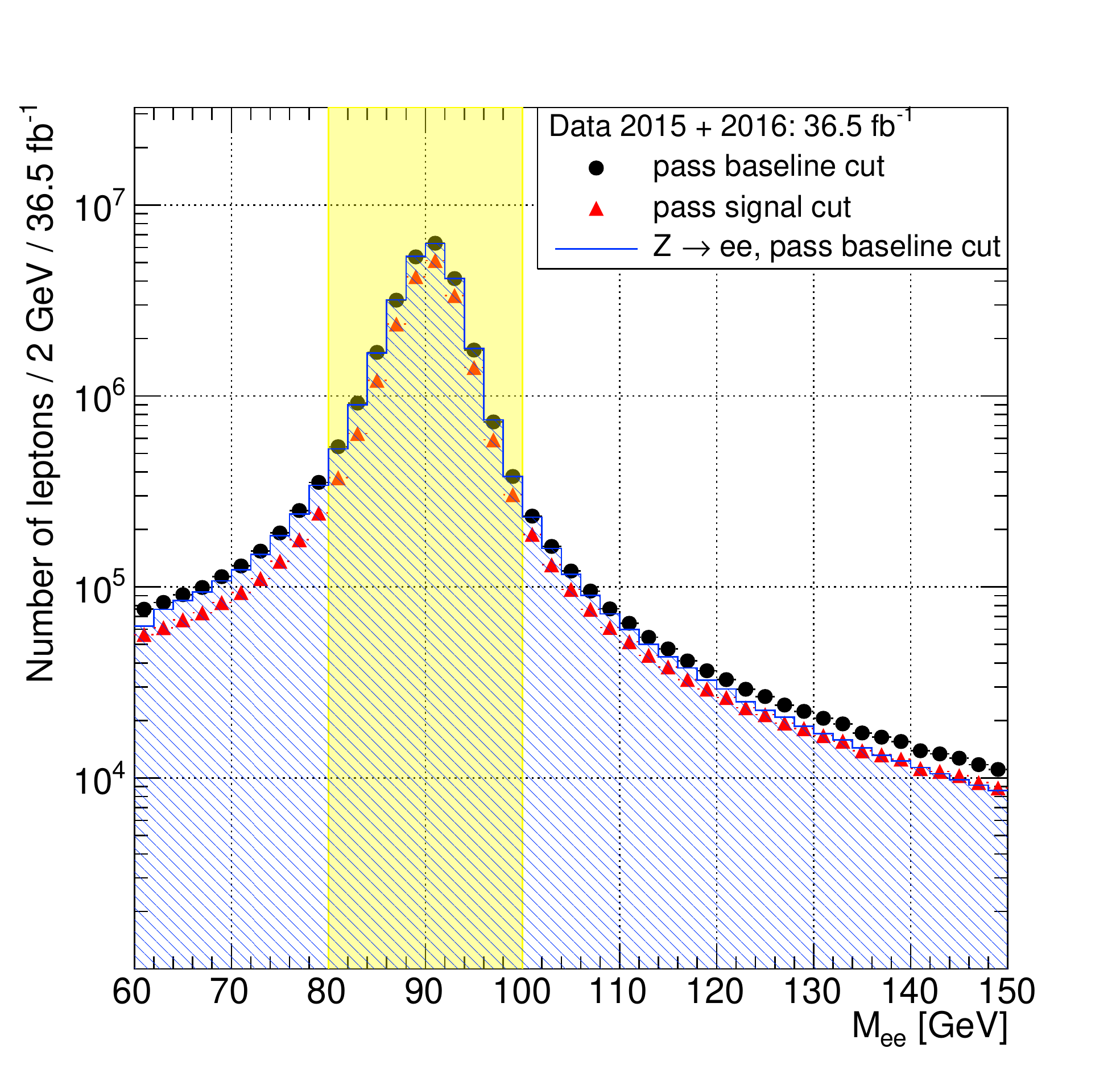}
\end{subfigure}
\begin{subfigure}[b]{0.45\textwidth}
	\includegraphics[width=\textwidth]{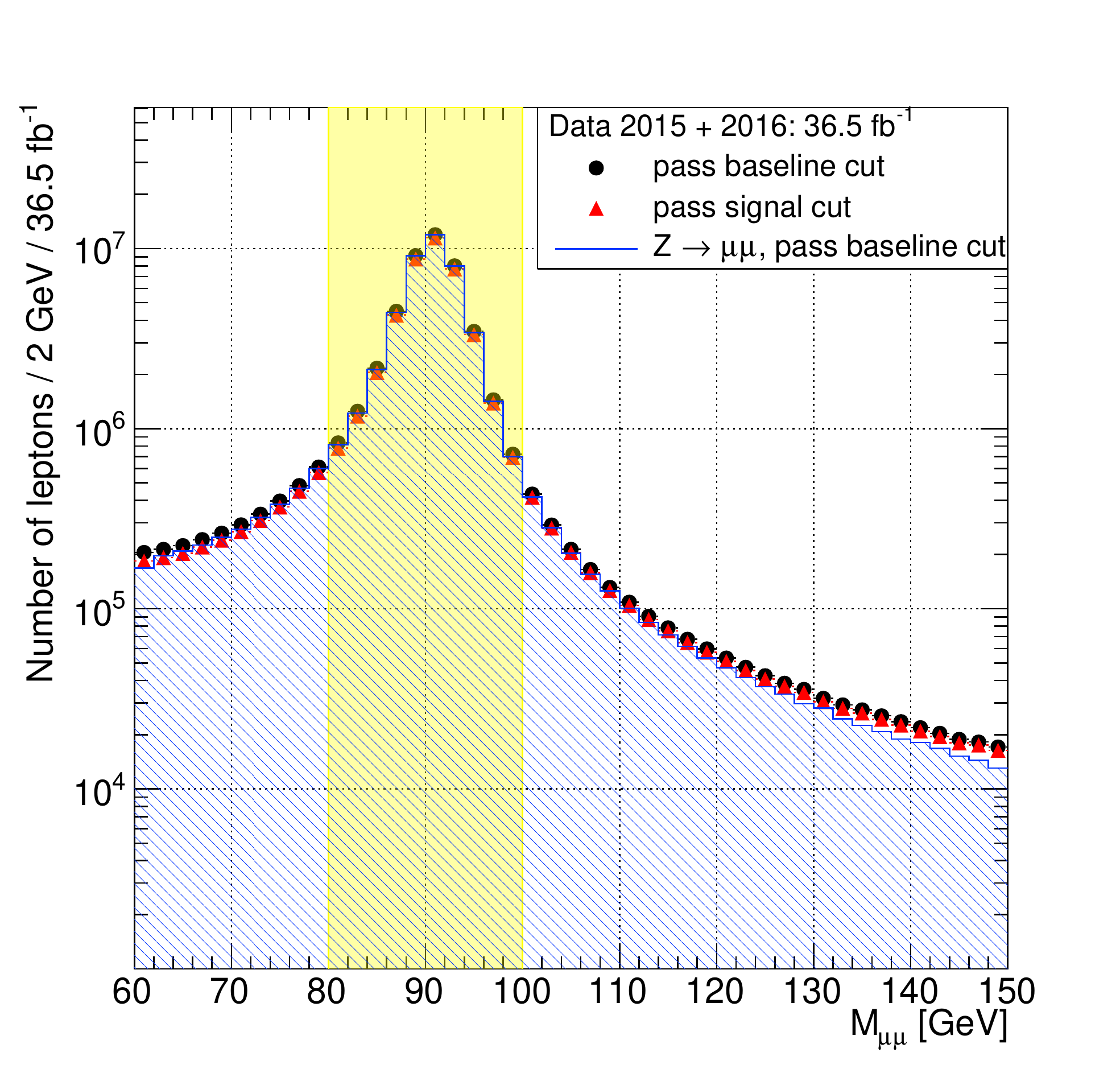}
\end{subfigure}
\caption{Invariant mass of opposite-sign same-flavor electrons (left) and muons (right), after the tag selection, 
where the probe satisfies the baseline requirements or the signal requirements.}
\label{fig:RLE_mll_distribution}
\end{figure}

Baseline-to-signal efficiency for real leptons is measured in a high purity data sample of opposite-sign same-flavor leptons with the standard $Z$ tag-and-probe method.
Events are selected by a single lepton trigger.
The tag lepton, required to have triggered the event recording, also satisfies signal requirements and verifies $\pt>25~\GeV$. 
The probe lepton used for the efficiency measurement satisfies baseline requirements. 
All possible tag-and-probe combinations are considered in an event (including permutation of the tag and probe leptons), 
as long as the invariant mass of the pair is comprised between 80 and 100~\GeV. 
Figure~\ref{fig:RLE_mll_distribution} illustrates this event selection.

A non-negligible background contamination in the electron channel affects measurements below $\pt=20~\GeV$. 
This contamination is taken into account in the measurement using a background template method inspired by the method used to measure reconstruction, identification, and 
isolation efficiencies documented in~\cite{ATLAS-CONF-2014-032}. 
This template is built from the tag-and-probe invariant mass distribution for baseline-level probe electrons that fail both tight identification
 and isolation requirements, smoothed by assuming an exponential shape whose parameters are determined by a fit in the interval $60<m_{ee}<120~\GeV$ excluding the $80<m_{ee}< 100~\GeV$ region. 
The background template is then normalized to the main tag-and-probe distribution in the background-dominated tail $120<m_{ee}<150~\GeV$. 
The estimated level of background goes up to $4\%$, reached for probe electrons with $\pt<15~\GeV$ and $|\eta|<0.8$. 

\begin{figure}[htb!]
\centering
\begin{subfigure}{0.49\textwidth}
\includegraphics[width=\textwidth]{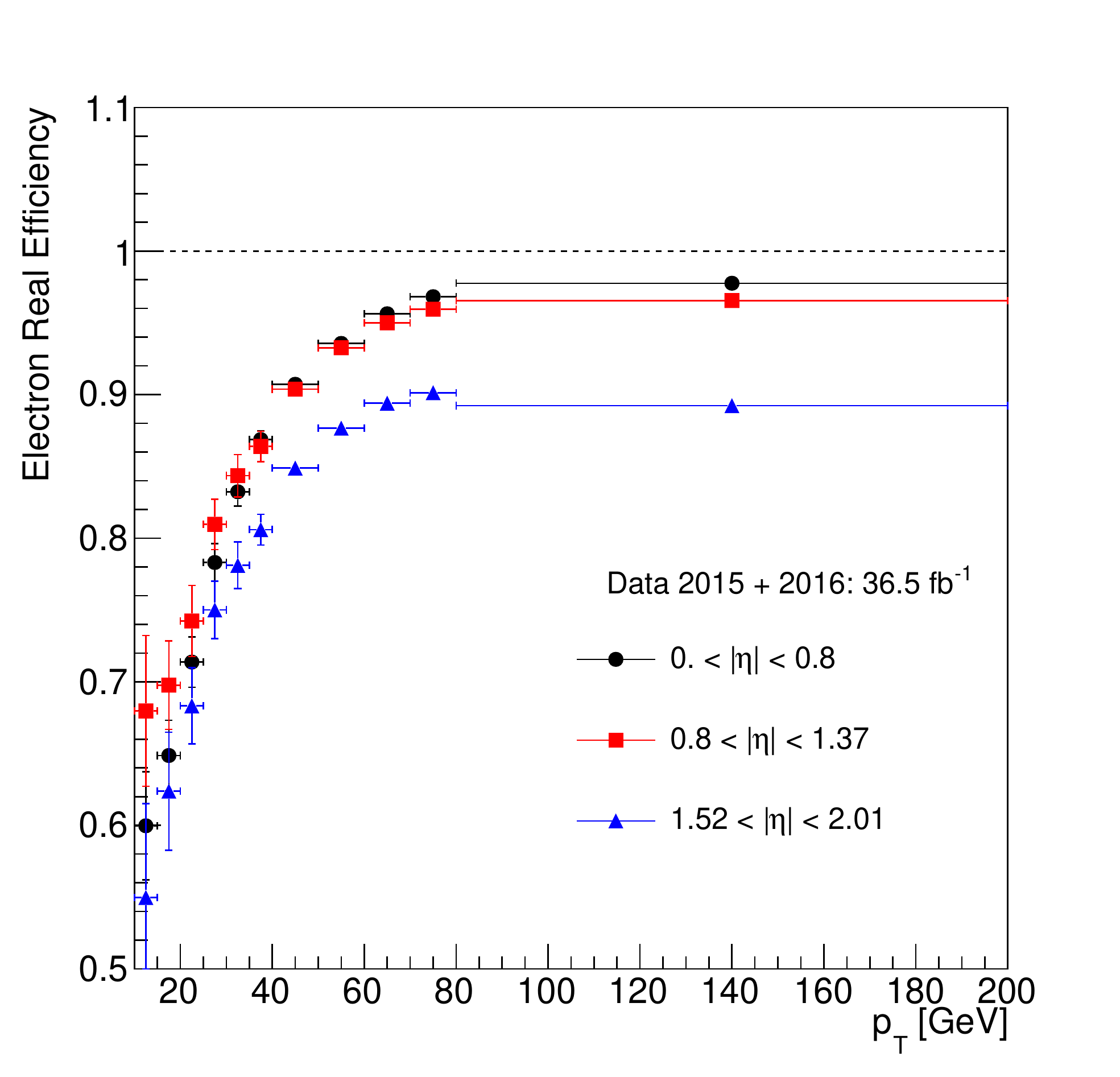}
\subcaption{Electrons}
\end{subfigure}
\begin{subfigure}{0.49\textwidth}
\includegraphics[width=\textwidth]{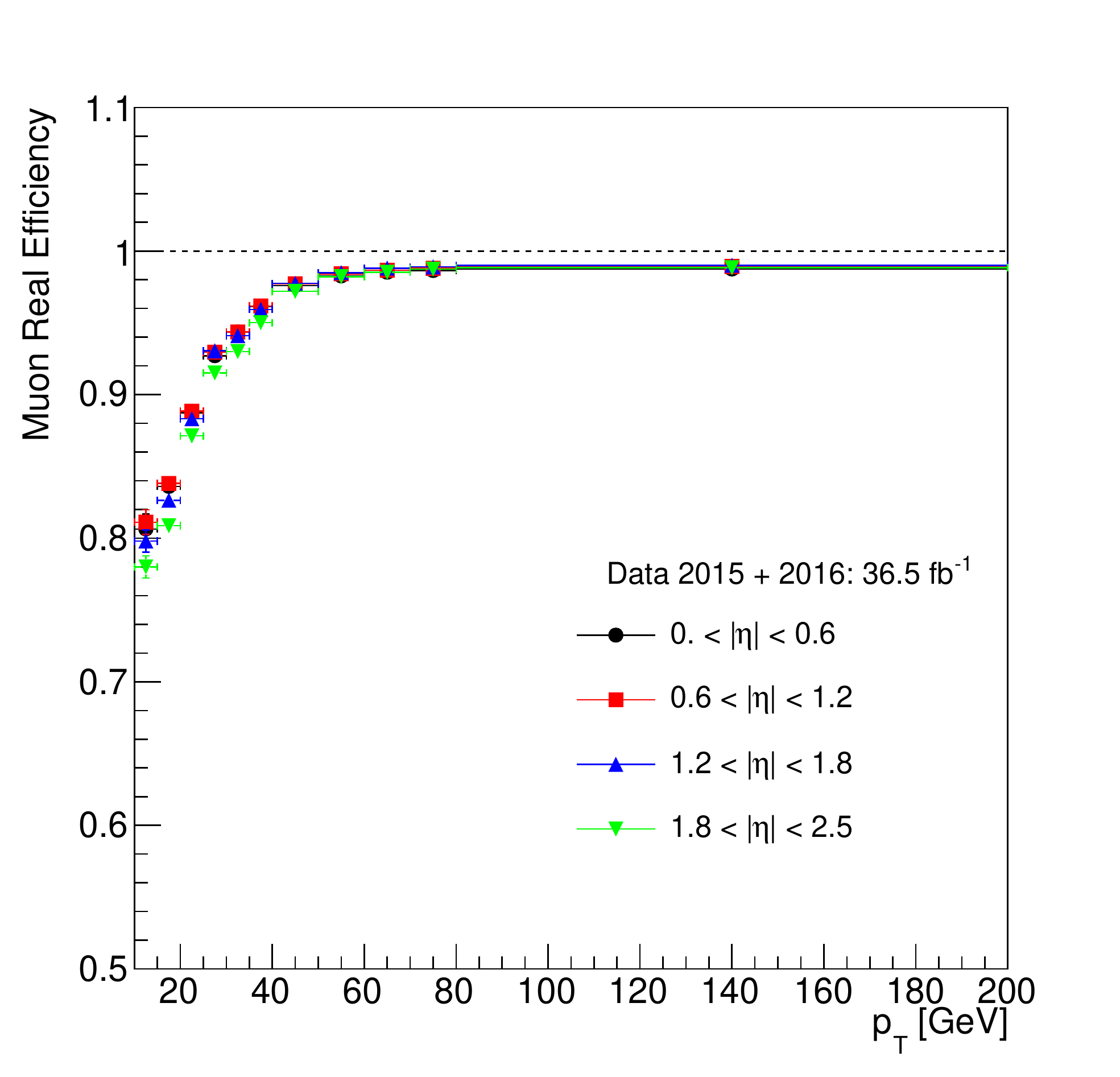}
\subcaption{Muons}
\end{subfigure}
\caption{Baseline-to-signal efficiencies as a function of $\pt$ and $|\eta|$ for real electrons (left) and muons (right), measured in 2015+2016 data.
The $|\eta|$ binning used in the electron case corresponds to the geometry of the electromagnetic calorimeter.
For muons a homogeneous $|\eta|$ binning is considered.
The error bars corresponds to the quadratic sum of the statistical and tag-and-probe measurement systematic uncertainties.}
\label{fig:prompt_leptons_eff}
\end{figure}

The efficiency is measured as a function of \pt\ and $\eta$, and the results are presented in Figure\ref{fig:prompt_leptons_eff} for electrons and muons. 
The background subtraction is applied on the electron channel only. 
The following systematic uncertainties are assigned to the measured efficiencies: 

\begin{itemize}
\item[$\bullet$] Background contamination: 27 variations of the tag-and-probe method are considered to assess the electron measurement systematics.
Three $m_{ee}$ windows and 9 variations of the background subtraction methods are considered.
The largest contribution to the systematics arises from the $m_{ee}$ window variation.
This is expected as the proportion of electrons affected by Brehmstrahlung depends on $m_{ee}$.
The resulting relative systematics vary from 6\% to 12\% in the $10<\pt<15~\GeV$ region, 3\% to 6\% in the $15<\pt<20~\GeV$ region, 1\% to 3\% in the $20<\pt<40~\GeV$ region, and less than 1\% for $\pt >$ 40 \GeV.
The systematic uncertainties associated to the muon efficiencies measurement vary from 1\% to 1.3\% in the $10<\pt<15~\GeV$ region and less than 1\% for $\pt >$ 15 \GeV. 
\item[$\bullet$] Trigger: a systematic uncertainty accounting for a potential bias at trigger level is considered and it varies between 0 and 4\%, depending on the \pt range.
\item[$\bullet$] Extrapolation to busy environments: efficiencies are typically lower in such environments due to the proximity of jets and leptons; 
an uncertainty is assigned by comparing efficiencies in simulated $Z\to\ell\ell$ and $\gluino\to\ttbar\neut$ events, for $\Delta m(\gluino,\neut)>1~\TeV$ which represents an extreme case of final states with highly boosted top quarks. 
The uncertainty, taken as the difference in efficiencies, is parametrized as a function of \pt\ and $\Delta R$ (the angular distance between the lepton and the closest jet). 
\end{itemize}
The resulting systematic uncertainties are summarized in Table~\ref{tab:RLE_systematics_all} and Table~\ref{tab:RLE_systematics_busy}.
\begin{center}
\begin{table}
\resizebox{\textwidth}{!}{%
\begin{tabular}{cccc|cccc}
\hline
\hline
& \multicolumn{3}{c}{Electrons} & \multicolumn{3}{c}{Muons}\\
& $0<|\eta|<0.8$ & $0.8<|\eta|<1.37$ & $1.52<|\eta|<2.01$ & $0<|\eta|<0.6$ & $0.6<|\eta|<1.2$ & $1.2<|\eta|<1.8$ & $1.8<|\eta|<2.5$\\
\hline
10~\GeV~$< p_{\text T} <$ 15~\GeV~& 0.047 & 0.063 & 0.089 & 0.014 & 0.010 & 0.008 & 0.011\\
15~\GeV~$< p_{\text T} <$ 20~\GeV~& 0.027 & 0.042 & 0.062 & 0.005 & 0.006 & 0.008 & 0.011\\
20~\GeV~$< p_{\text T} <$ 25~\GeV~& 0.018 & 0.031 & 0.041 & 0.003 & 0.006 & 0.010 & 0.010\\
25~\GeV~$< p_{\text T} <$ 30~\GeV~& 0.029 & 0.024 & 0.027 & 0.011 & 0.015 & 0.022 & 0.019\\
30~\GeV~$< p_{\text T} <$ 35~\GeV~& 0.023 & 0.021 & 0.023 & 0.007 & 0.009 & 0.014 & 0.011\\
35~\GeV~$< p_{\text T} <$ 40~\GeV~& 0.014 & 0.018 & 0.018 & 0.004 & 0.004 & 0.006 & 0.006\\
40~\GeV~$< p_{\text T} <$ 50~\GeV~& 0.007 & 0.010 & 0.010 & 0.002 & 0.001 & 0.002 & 0.001\\
50~\GeV~$< p_{\text T} <$ 60~\GeV~& 0.008 & 0.010 & 0.010 & 0.001 & 0.001 & 0.001 & 0.001\\
60~\GeV~$< p_{\text T} <$ 70~\GeV~& 0.007 & 0.010 & 0.010 & 0.001 & 0.001 & 0.001 & 0.002\\
70~\GeV~$< p_{\text T} <$ 80~\GeV~& 0.008 & 0.011 & 0.012 & 0.002 & 0.001 & 0.001 & 0.002\\
80~\GeV~$< p_{\text T} <$ 120~\GeV~& 0.010 & 0.010 & 0.011 & 0.004 & 0.002 & 0.002 & 0.002\\
120~\GeV~$< p_{\text T} <$ 150~\GeV~& 0.005 & 0.005 & 0.011 & 0.006 & 0.005 & 0.005 & 0.005\\
150~\GeV~$< p_{\text T} <$ 200~\GeV~& 0.005 & 0.002 & 0.019 & 0.005 & 0.005 & 0.005 & 0.006\\
\hline
\hline
\end{tabular}
}
\caption{
Systematic uncertainties on the measured real lepton efficiency, separating sources affecting the measurement itself (background subtraction, trigger bias, and different methods). 
}
\label{tab:RLE_systematics_all}
\end{table}
\end{center}

\begin{center}
\begin{table}
\resizebox{\textwidth}{!}{%
\begin{tabular}{ccccccccccc}
\hline
\hline
\multicolumn{9}{c}{electrons (busy environments)}\\
\hline
$\Delta R(e, jet)$ & [0, 0.1] & [0.1, 0.15] & [0.15, 0.2] & [0.2, 0.3] & [0.3, 0.35] & [0.35, 0.4] & [0.4, 0.6] & [0.6, 4]\\
\hline
10~\GeV~$< p_{\text T} <$ 20~\GeV~& - & - & - & - & - & - & 25.31\% & 6.5\%\\
20~\GeV~$< p_{\text T} <$ 30~\GeV~& - & - & - & - & - & 73.37\% & 10.21\% & 0.37\%\\
30~\GeV~$< p_{\text T} <$ 40~\GeV~& - & - & - & 97.71\% & 48.22\% & 15.54\% & 7.29\% & 0.58\%\\
40~\GeV~$< p_{\text T} <$ 50~\GeV~& - & - & - & 52.81\% & 22.80\% & 16.73\% & 7.68\% & 1.10\%\\
50~\GeV~$< p_{\text T} <$ 60~\GeV~& - & - & - & 29.96\% & 21.49\% & 20.23\% & 6.99\% & 2.78\%\\
60~\GeV~$< p_{\text T} <$ 80~\GeV~& - & - & 55.89\% & 24.31\% & 17.40\% & 24.77\% & 6.20\% & 2.87\%\\
80~\GeV~$< p_{\text T} <$ 150~\GeV~& - & 57.52\% & 30.24\% & 16.45\% & 12.73\% & 20.92\% & 4.44\% & 2.73\%\\
150~\GeV~$< p_{\text T} <$ 200~\GeV~& 88.54\% & 40.16\% & 19.34\% & 8.45\% & 14.66\% & 16.57\% & 2.57\% & 1.90\%\\
\hline
\hline
\multicolumn{9}{c}{muons (busy environments)}\\
\hline
$\Delta R(\mu, jet)$ & [0, 0.1] & [0.1, 0.15] & [0.15, 0.2] & [0.2, 0.3] & [0.3, 0.35] & [0.35, 0.4] & [0.4, 0.6] & [0.6, 4]\\
\hline
10~\GeV~$< p_{\text T} <$ 20~\GeV~& - & - & - & - & - & - & 33.59\% & 5.18\%\\
20~\GeV~$< p_{\text T} <$ 30~\GeV~& - & - & - & - & - & 82.34\% & 22.27\% & 3.39\%\\
30~\GeV~$< p_{\text T} <$ 40~\GeV~& - & - & -  & 98.54\% & 56.36\% & 31.89\% & 14.22\% & 2.24\%\\
40~\GeV~$< p_{\text T} <$ 50~\GeV~& - & - & - & 53.10\% & 21.33\% & 13.90\% & 6.81\% & 1.45\%\\
50~\GeV~$< p_{\text T} <$ 60~\GeV~& - & - & - & 24.98\% & 13.72\% & 9.62\% & 3.83\% & 0.79\%\\
60~\GeV~$< p_{\text T} <$ 80~\GeV~& - & - & 44.41\% & 13.75\% & 6.14\% & 4.76\% & 2.04\% & 0.15\%\\
80~\GeV~$< p_{\text T} <$ 150~\GeV~& - & 29.94\% & 7.14\% & 3.16\% & 1.30\% & 1.04\% & 0.07\% & 0.57\%\\
150~\GeV~$< p_{\text T} <$ 200~\GeV~& 82.26\% & 4.14\% & 1.02\% & 0.17\% & 0.29\% & 0.62\% & 1.02\% & 1.13\%\\
\hline
\hline
\end{tabular}
}
\caption{
Systematic uncertainties on the measured real lepton efficiency, due to the extrapolation to busy environments using $\gluino \to \ttbar \tilde{\chi_1^0}$ events. 
}
\label{tab:RLE_systematics_busy}
\end{table}
\end{center}

%% file: texfiles/subsec.bkg.mct.tex
As discussed in Section~\ref{sec:fake.mct}, the MC template method is
 a simulation-based method that provides an alternative estimate of the reducible backgrounds affecting the analysis.
It relies on the correct modelling of FNP leptons and charge-flipped electron kinematics in $\ttbar$ 
and $V$+jets.
FNP leptons are classified in four categories, namely electrons and muons coming 
from $b$ and light-quark jets. Normalisation factors for each of the five sources (including photon conversions) are computed to match the observed data 
in dedicated control regions. The fifth category is events with prompt electrons that have a charge mis-measurement 
(charge flip, electron heavy flavor (EL HF), 
muon heavy flavor (MU HF), 
electron light flavor (EL LF), 
muon light flavor (MU LF)).
Six non-overlapping control regions are defined by the presence of $b$-jets and by the flavors of the same sign lepton pair in the event:
\begin{itemize}
\item CR0b: events without $b$-jets in $ee$, $e\mu$, and $\mu\mu$ channels.
\item CR1b: events with at least one $b$-jet in $ee$, $e\mu$, and $\mu\mu$ channels.
\end{itemize}
All the selected events contain two or more same-sign signal leptons and \\$\met >40~\GeV$ and 2 or more jets. 
Events satisfying the signal regions requirements are excluded from the control regions. 
The purpose of the \met requirement is to remove multi-jet events that have two or more FNP leptons and tend to have low \met. 
The six distributions are chosen for variables that provide the best separation between processes with prompt leptons and processes with 
fake leptons and charge flip and are shown 
before and after the fit in Figures \ref{f:prefit_CR0b}-\ref{f:prefit_CR1b} and Figures \ref{f:postfit_CR0b}-\ref{f:postfit_CR1b}, 
respectively. 
The multipliers obtained after the minimization of the negative log likelihood were given 
in Tables \ref{t:fake_factors_powheg} and \ref{t:fake_factors_sherpa}.
The tables represent the correction factors obtained from the fit upon using two different parton showers, \POWHEGBOX+Pythia and \SHERPA
for the processes that lead to non-prompt leptons and charge flips.
The goal of varying the parton shower is to access the dependence of the fake and charge flip estimates on the choice of the 
parton shower. 

The MC template method is validated by looking at the agreement 
between observed data and prediction as shown in Figures~\ref{fig:bkg.val.met}.
In the MC template method, the systematic uncertainty is obtained by
changing the generator from \POWHEGBOX+Pythia to \SHERPA and propagating uncertainties from the control region fit to the global
normalization scale factors applied to the MC samples. 
The uncertainties in these scale factors are in the range 75--80\%,
depending on the SRs.
In practice, only \ttbar contributes to the SRs and the final yields with systematic uncertainties from 
fit uncertainty, theory uncertainties on \ttbar, and comparison of different showers (Pythia and Sherpa) are shown in Table \ref{tab:fakes_mcglobal}.
This table also shows a global correction factor derived by taking the ratio of the weighted \ttbar to raw MC \ttbar with
a global uncertainty that includes all systematic uncertainties used to obtain the final estimate. 

\begin{table}[!htb]
\centering
\resizebox{\textwidth}{!}{
\begin{tabular}{|c||c|c|c||}\hline
Region &              MC Template method  & Global correction & Shower systematic \\
\hline
Rpc2L0bH & $1.00 \pm 0.96 \pm 0.81$ & $2.80 \pm 2.10$ & 74\% \\
Rpc2L0bS & $1.68 \pm 1.02 \pm 1.26$ & $2.89 \pm 1.97$ & 65\% \\
Rpc2L1bH & $2.07 \pm 0.63 \pm 1.56$ & $1.22 \pm 1.14$ & 34\% \\
Rpc2L1bS & $2.33 \pm 1.17 \pm 2.10$ & $1.83 \pm 1.42$ & 81\% \\
Rpc2L2bH & $<0.5$ & $0 \pm 0$ & 0\% \\
Rpc2L2bS & $0.41 \pm 0.33 \pm 0.45$ & $1.47 \pm 1.12$ & 73\% \\
Rpc2Lsoft1b & $2.48 \pm 1.32 \pm 1.86$ & $1.59 \pm 1.31$ & 68\% \\
Rpc2Lsoft2b & $1.66 \pm 0.66 \pm 1.28$ & $1.72 \pm 1.29$ & 54\% \\
Rpc3L0bH & $<0.5$ & $0 \pm 0$ & 0\% \\
Rpc3L0bS & $0.21 \pm 0.15 \pm 0.16$ & $2.90 \pm 2.20$ & 71\% \\
Rpc3L1bH & $0.42 \pm 0.29 \pm 0.32$ & $1.59 \pm 1.25$ & 59\% \\
Rpc3L1bS & $3.55 \pm 1.80 \pm 2.76$ & $1.76 \pm 1.32$ & 67\% \\
Rpc3LSS1b & $0.90 \pm 0.14 \pm 0.69$ & $2.34 \pm 1.44$ & 56\% \\
\hline
\hline
\end{tabular}
}
\caption{Expected yields for background processes with fake leptons,
in the signal regions with a global correction factor that represents the ratio of weighted \ttbar to raw MC \ttbar with
a global uncertainty that includes: fit uncertainty, theory uncertainties on \ttbar, comparison of different showers. 
The fraction of the systematic uncertainty from the comparison between two showers (Pythia and Sherpa) is also shown.
}
\label{tab:fakes_mcglobal}
\end{table}

%% file: texfiles/subsec.bkg.val.tex
The reducible backgrounds estimated with the methods described in the previous 
sections are validated by comparing observed data to 
predicted background after various kinematic requirements. 
The next sections will validate 
the MC template method used to estimate the charge flip and the FNP lepton 
background, and the matrix method used to estimate the FNP lepton background.

\subsection*{Data/MC comparisons}

\begin{figure}[htb!]
\centering
\includegraphics[width=0.8\textwidth]{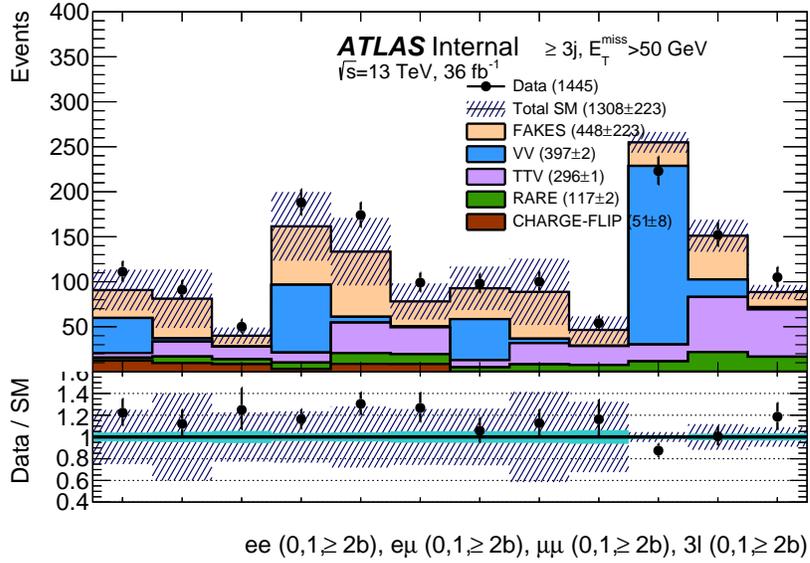}
\caption{Summarized level of agreement between observed data (2015+2016, 36.5 \ifb) and expected SM+detector backgrounds 
for events with $\ge 2$ same-sign leptons ($\pt>20 \GeV$), $\met>50 \GeV$ and $\ge 3$ jets ($\pt>40 \GeV$), 
split as function of the lepton flavours and the number of $b$-tagged jets. 
Uncertainties include statistical sources, as well as systematic uncertainties for the data-driven backgrounds; 
for illustration, statistical uncertainties alone are shown in the light-colored error bands in the ratio plots. 
Events belonging to any of the signal regions are rejected, both in data and MC. 
}
\label{fig:distributions_summary}
\end{figure}

\begin{figure}[htb!]
\centering
\begin{subfigure}[t]{0.49\textwidth}\includegraphics[width=\textwidth]{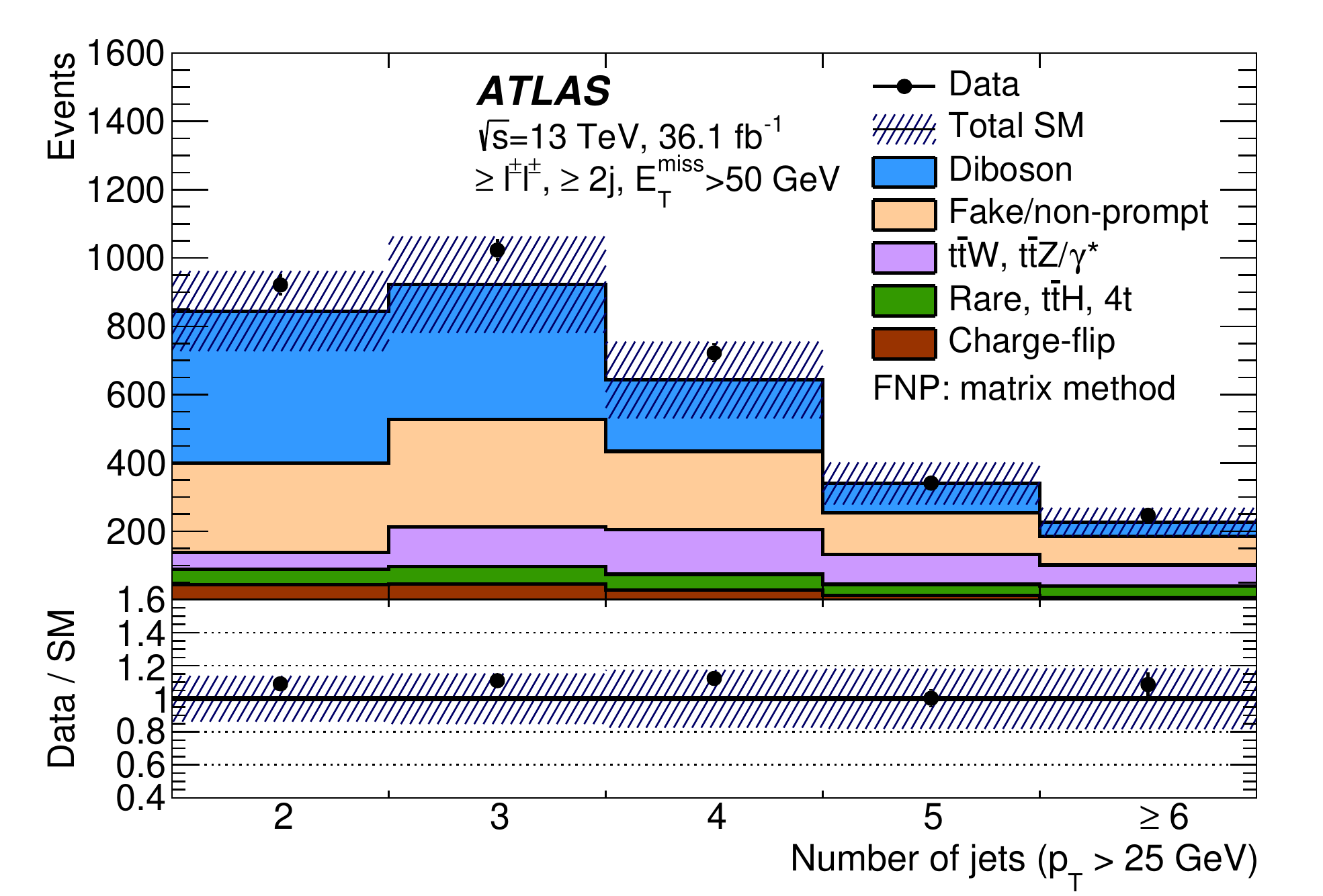}\caption{}\label{fig:VRnj}\end{subfigure}
\begin{subfigure}[t]{0.49\textwidth}\includegraphics[width=\textwidth]{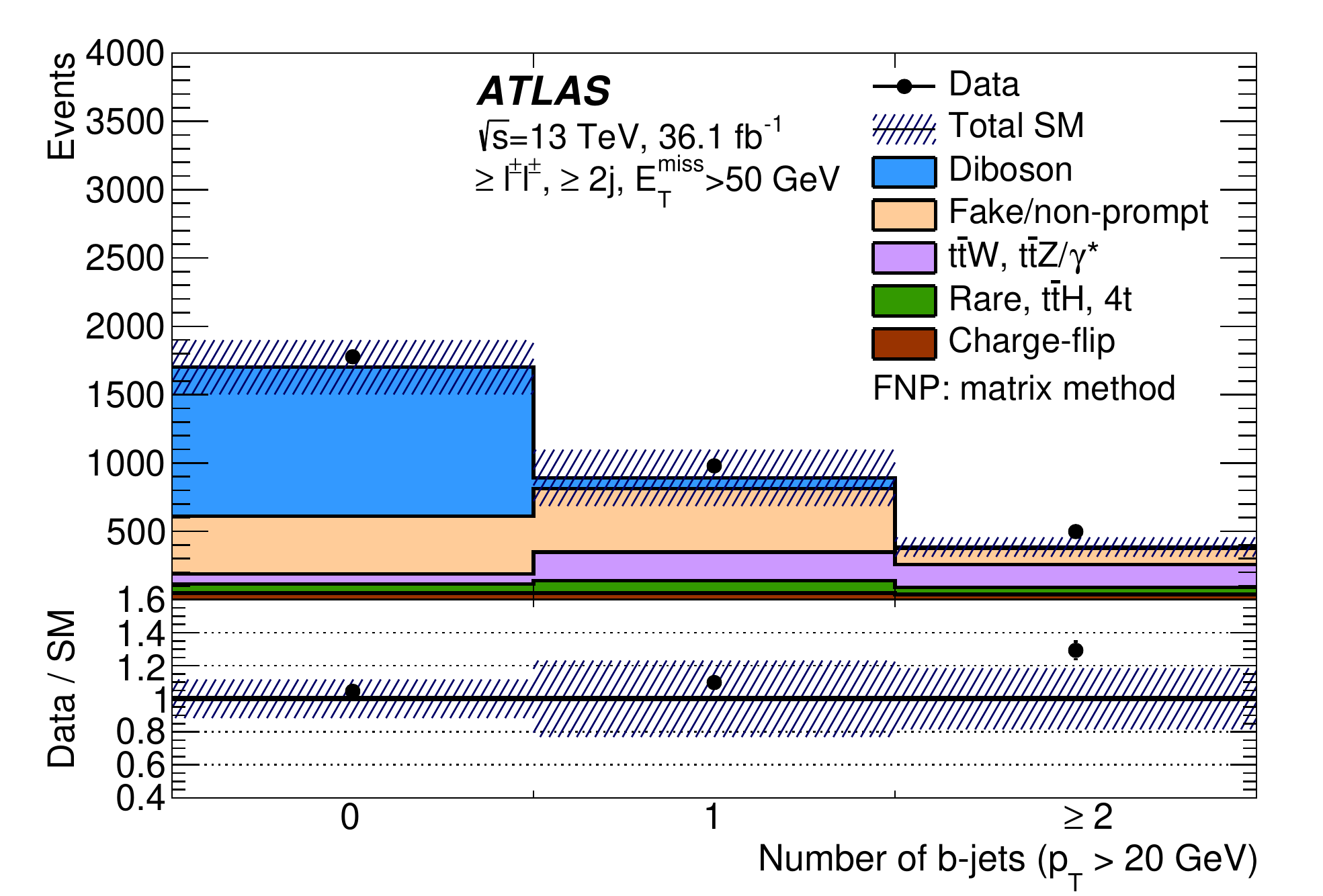}\caption{}\label{fig:VRnb}\end{subfigure}
\begin{subfigure}[t]{0.49\textwidth}\includegraphics[width=\textwidth]{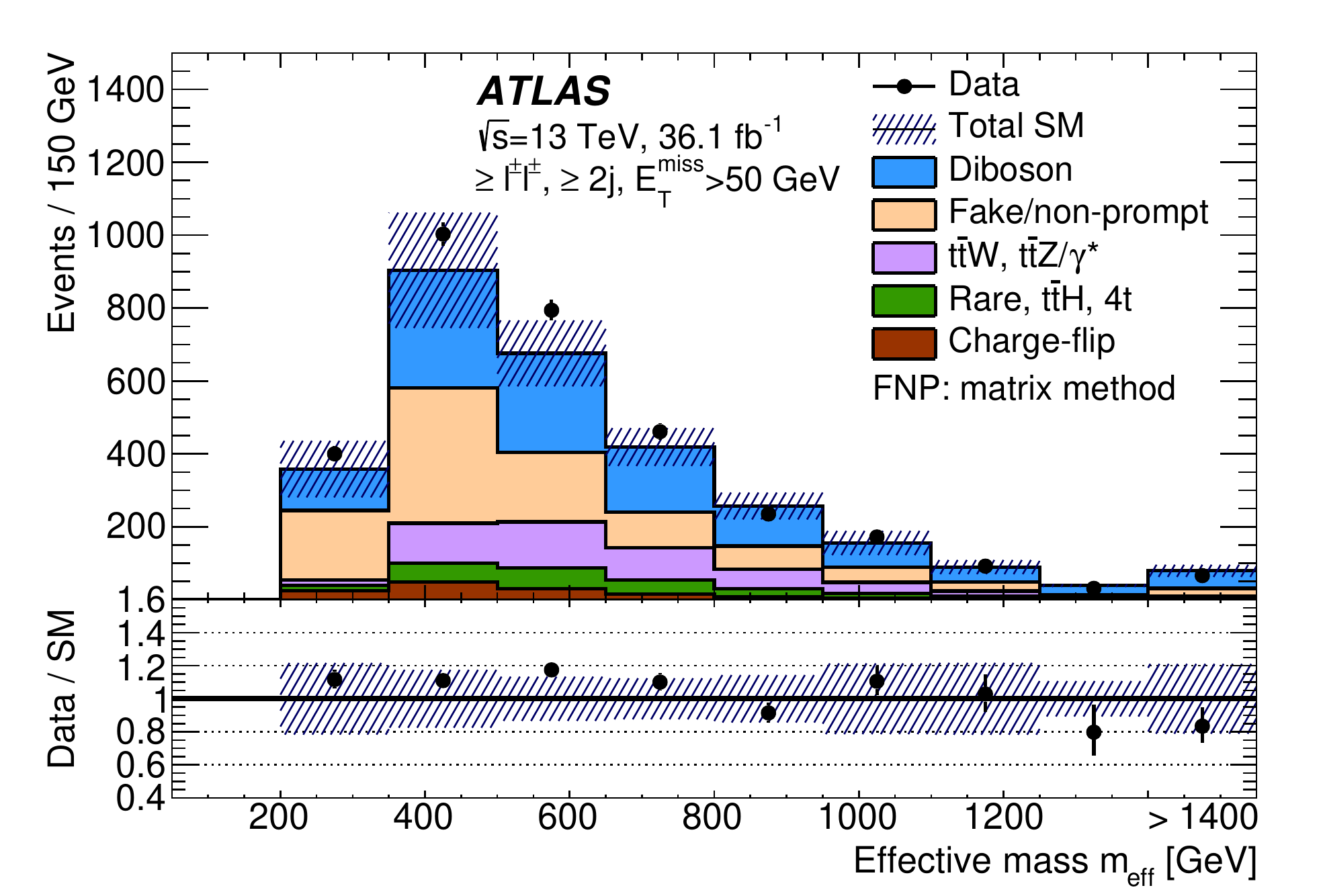}\caption{}\label{fig:VRmeff1}\end{subfigure}
\begin{subfigure}[t]{0.49\textwidth}\includegraphics[width=\textwidth]{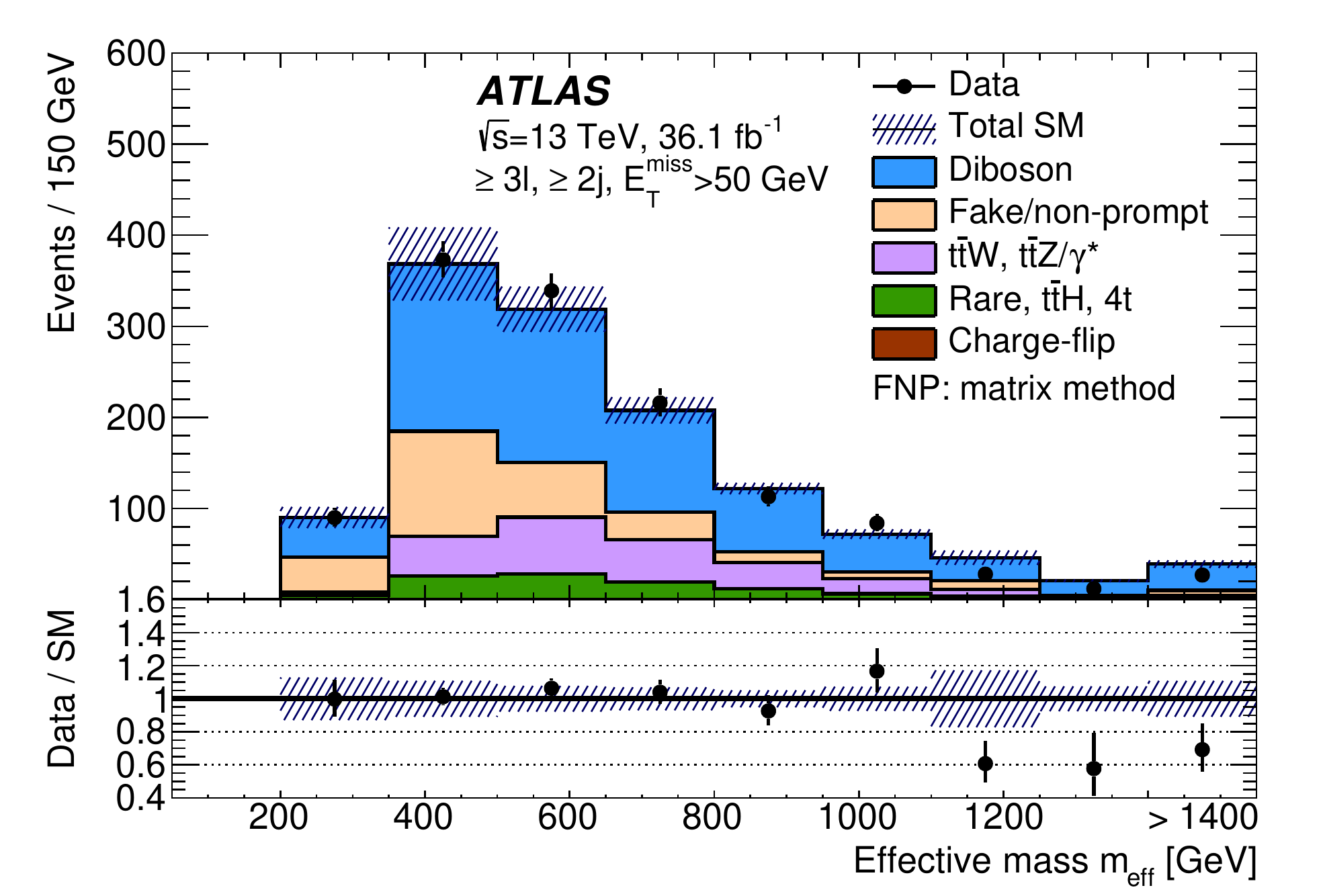}\caption{}\label{fig:VRmeff2}\end{subfigure}
\caption{
Distributions of (a) the number of jets, (b) the number of $b$-tagged jets and (c), (d) the effective mass. The distributions are made 
after requiring at least two jets ($\pT>40$ \GeV) and $\met>50$ \GeV, as well as at least two same-sign leptons ((a), (b), (c)) 
or three leptons (d). The uncertainty bands include the statistical uncertainties for the background prediction as well as the 
systematic uncertainties for fake- or non-prompt-lepton backgrounds (using the matrix method) and charge-flip electrons. Not included
are theoretical uncertainties in the irreducible background contributions.
The rare category is defined in the text.}
\label{fig:Bkg_distribs} 
\end{figure}

The overall level of agreement obtained by the matrix method
can be seen in different channels in Figure\ref{fig:distributions_summary}. 
The distributions of several variables are shown on Figure\ref{fig:Bkg_distribs} for an inclusive same-sign lepton selection
after various requirements on the number of jets and $b$-jets.
These distributions illustrate that the data are described by the prediction 
within uncertainties. 
The apparent disagreement 
for \meff\ above 1 TeV~in Figure~\ref{fig:VRmeff2} is covered by the large 
theory uncertainty for the diboson background, which is not shown 
but amounts to about 30\% for \meff\ above 1 \TeV. 
To avoid the presence of signal in the tails of these distributions, 
events belonging to any of the signal regions are vetoed 
both in data and for the predicted backgrounds. 

\begin{figure}[htb!]
\begin{subfigure}[t]{0.49\textwidth}\includegraphics[width=\textwidth]{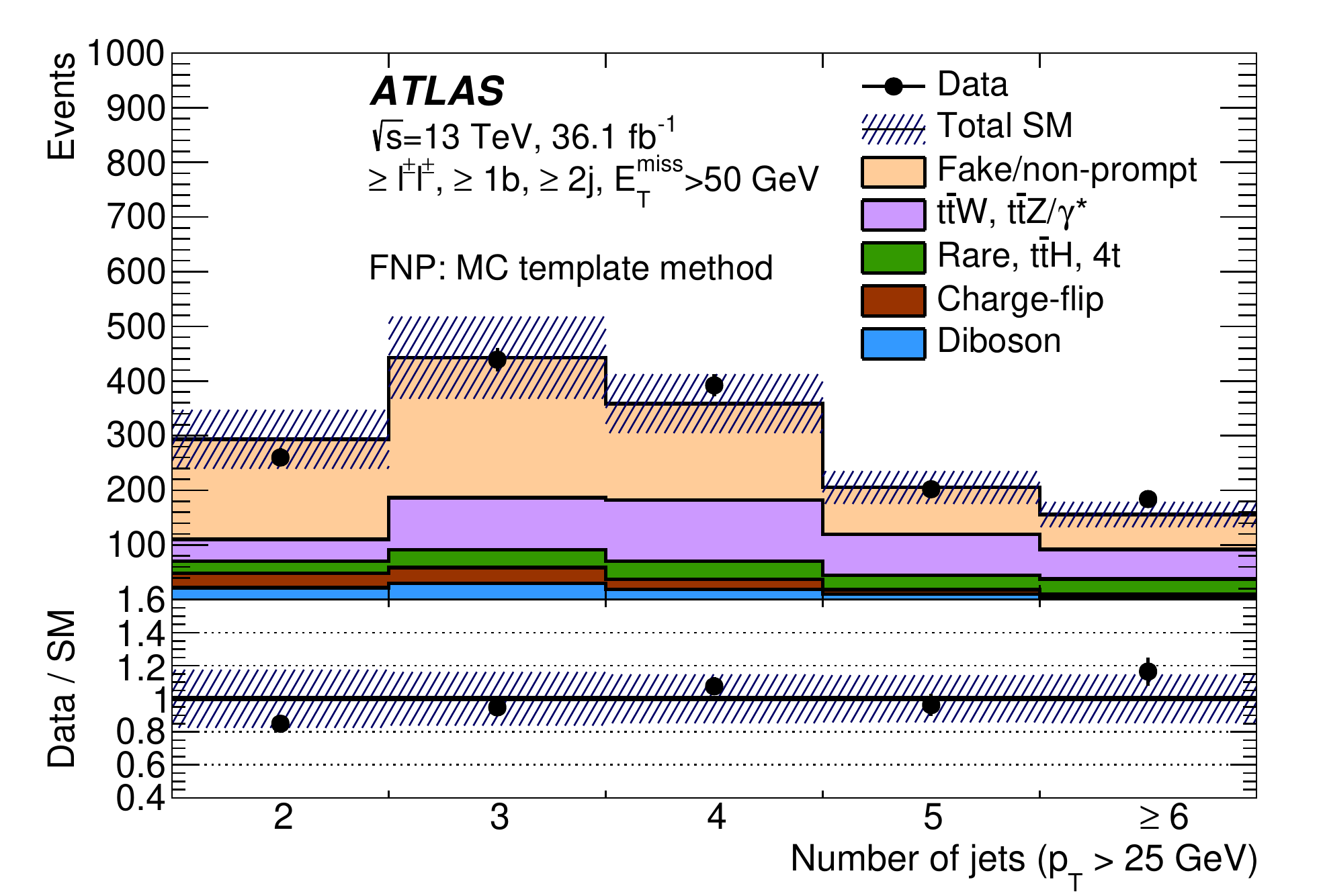}\caption{}\label{fig:VR1b2j_MxM}\end{subfigure}
\begin{subfigure}[t]{0.49\textwidth}\includegraphics[width=\textwidth]{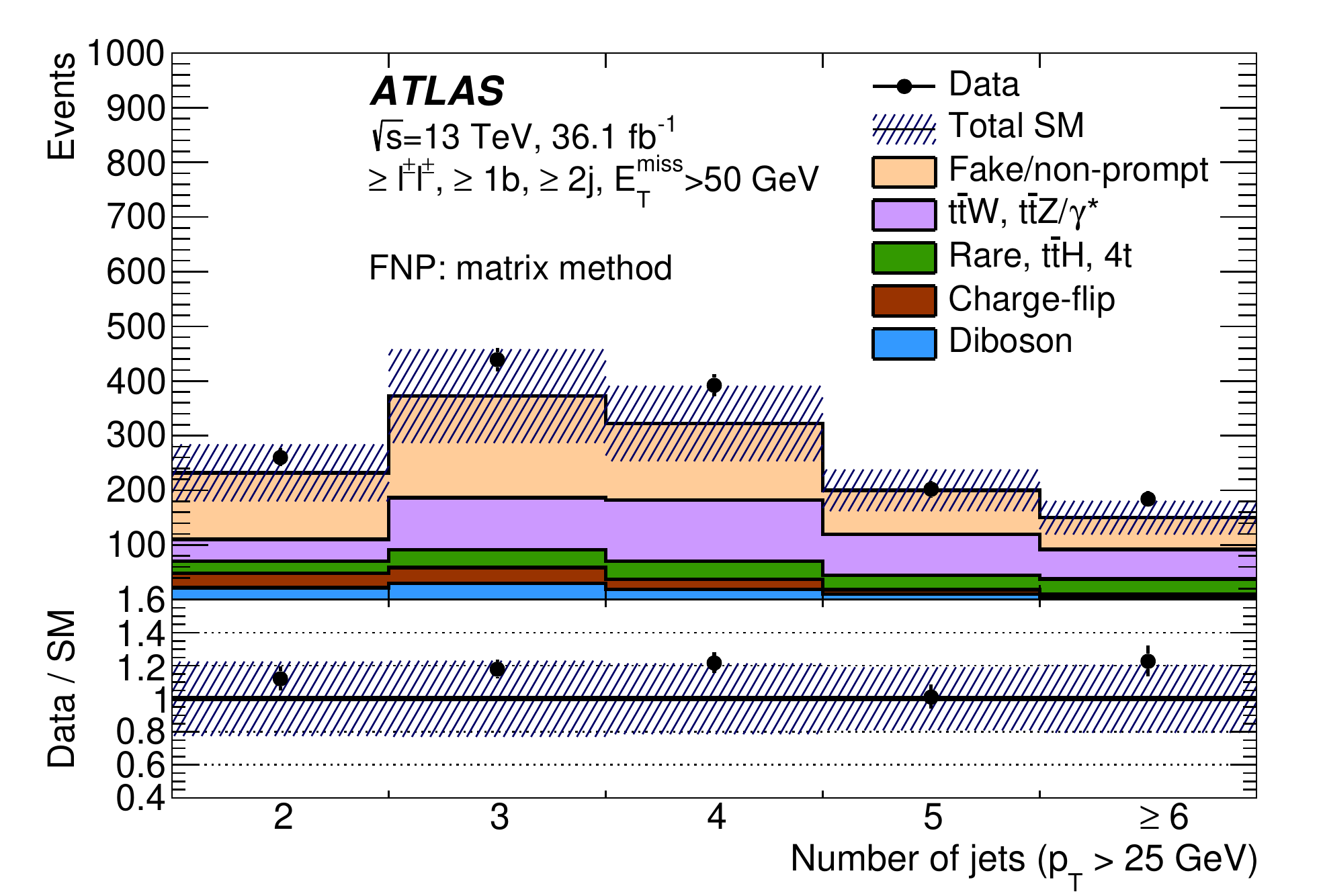}\caption{}\label{fig:VR1b2j_MCT}\end{subfigure}
\caption{
Distributions of the number of jets after requiring at least two jets ($\pT> 40 \GeV$) and $\met> 50 \GeV$, 
as well as at least two same-sign leptons. 
The fake or non-prompt leptons backgrounds are estimated alternatively with the MC template method (\ref{fig:VR1b2j_MCT}) or the matrix method (\ref{fig:VR1b2j_MxM}). 
The uncertainty band includes the statistical uncertainties for the background prediction as well as the
full systematic uncertainties for fake or non-prompt leptons backgrounds or charge-flip electrons. 
The rare category is defined in the text. In both figures, the last bin contains the overflow.
}
\label{fig:bkg.val.mctVSmxm}
\end{figure}

\begin{figure}[htb!]
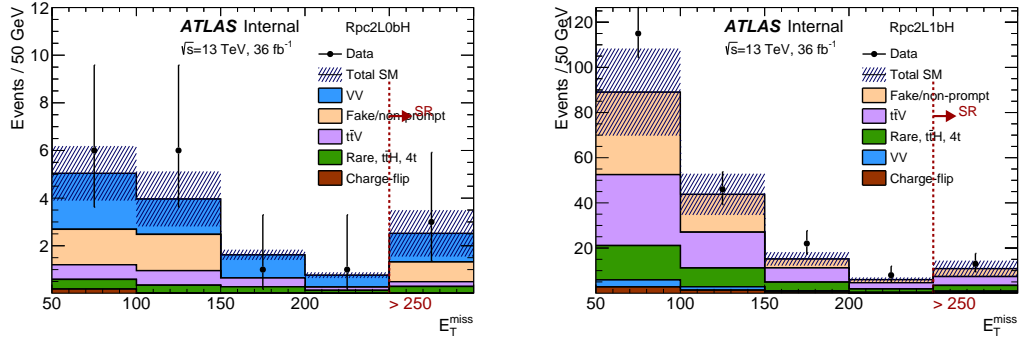
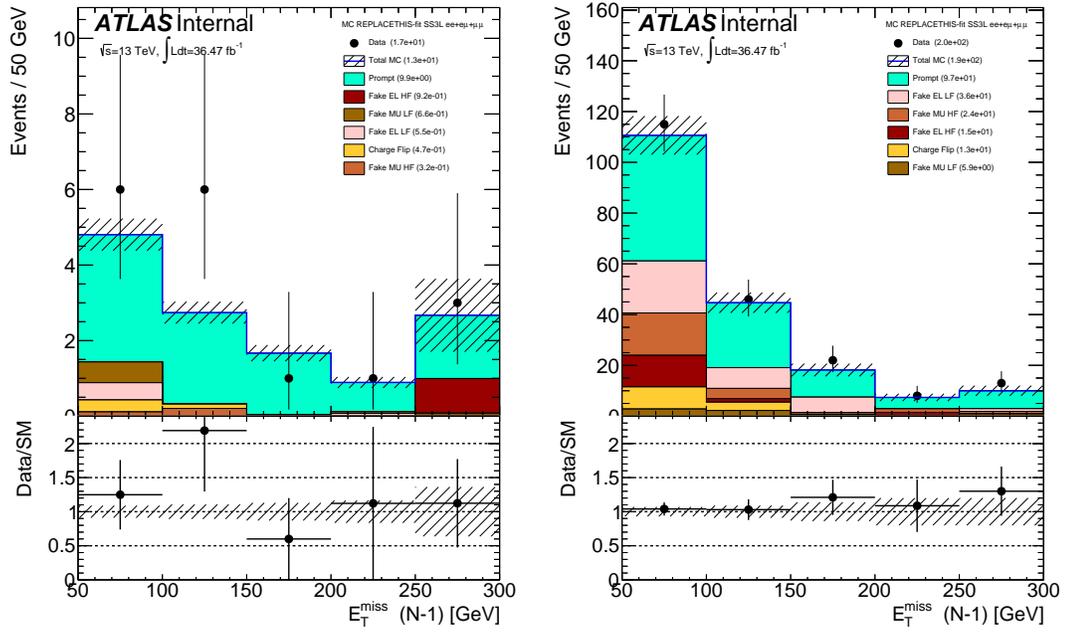

\begin{subfigure}[t]{0.49\textwidth}\includegraphics[width=\textwidth]{LooseRpc2L0bH.pdf}\caption{}\label{fig:bkg.val.Rpc2L0bH.MxM}\end{subfigure}
\begin{subfigure}[t]{0.49\textwidth}\includegraphics[width=\textwidth]{LooseRpc2L1bH.pdf}\caption{}\label{fig:bkg.val.Rpc2L1bH.MxM}\end{subfigure} \\
\begin{subfigure}[t]{0.49\textwidth}\includegraphics[width=\textwidth]{MET_comb_SR0b2mMET_SS3L.pdf}\caption{}\label{fig:bkg.val.Rpc2L0bH.MCT}\end{subfigure}
\begin{subfigure}[t]{0.49\textwidth}\includegraphics[width=\textwidth]{MET_comb_SR1b2mMET_SS3L.pdf}\caption{}\label{fig:bkg.val.Rpc2L1bH.MCT}\end{subfigure} \\

\caption{
Missing transverse momentum distributions after 
(a--c)
Rpc2L0bH 
and 
(b--d)
Rpc2L1bH selection, except the \met~requirement. 
Estimates with the matrix method are in the upper plots 
(a--b)
while estimates with the MC template method are in the lower plots 
(c--d)
.
The results in the signal regions are shown in the last (inclusive) bin of each plot. 
The statistical uncertainties and the full systematic uncertainties for backgrounds with fake or non-prompt leptons, or charge-flip
are included.
}
\label{fig:bkg.val.met}
\end{figure}

Figure~\ref{fig:bkg.val.mctVSmxm} shows a comparison between the estimates of the 
MC template method and the matrix method in a loose selection.
Other \met distributions with events satisfying the signal region 
requirements except the \met\ cut are shown in 
Figure~\ref{fig:bkg.val.met} comparing the two methods. 

\subsection*{Reducible background estimates}

The expected yield for processes with FNP leptons and charge-flip electrons, 
estimated with the matrix method, likelihood method (charge-flip), and the MC template method 
are presented in Table~\ref{tab:fakes_sr_yields} and Table~\ref{tab:chflips_sr_yields} for the signal regions and Table~\ref{tab:VR_Comparison} for the validation regions. 
Since the predictions from the MC template and matrix methods in the signal and validation regions are consistent 
with each other, 
the final numbers retained for the FNP lepton background estimate (also shown in the tables) 
are taken as the weighted-average of the predictions from the matrix method and the MC template; 
the weights are based on the statistical component, and the systematic uncertainties are propagated 
assuming conservatively a full correlation between the two methods (although they are in fact largely independent!). 
The central value and statistical/systematic uncertainties are therefore: 

\begin{equation}
\begin{aligned}
\left(w\zeta_1 + (1-w)\zeta_2\right) 
& \pm \sqrt{w^2\left(\Delta\zeta_1^\text{(stat)}\right)^2 + (1-w)^2\left(\Delta\zeta_2^\text{(stat)}\right)^2} \\
& \pm \left(w\Delta\zeta_1^\text{(syst)} + (1-w)\Delta\zeta_2^\text{(syst)}\right)
\end{aligned}
\end{equation}

$$\qquad\text{ with }w=\frac{\left(\Delta\zeta_2^\text{(stat)}\right)^2}{\left(\Delta\zeta_1^\text{(stat)}\right)^2+\left(\Delta\zeta_2^\text{(stat)}\right)^2}\notag$$

When the estimated value is too small(below 0.15), the expected yield is set to $0.15\pm 0.15$, 
to cover for possibilities of an under-fluctuation of the number of baseline-not-signal leptons 
when applying the matrix method, as well as lack of statistics in the MC samples for the other method. 

The charge flip background is not combined between the MC template method and the likelihood method due to the very large uncertainty in the MC template estimate.
The OS data has a much larger number of events which makes a precise prediction of this background. The MC template result is used as a cross check.

\begin{table}[!htb]
\centering
\resizebox{\textwidth}{!}{
\begin{tabular}{|c||c|c|c||c|}\hline
      Region &              Matrix method   &   Template method   &     Retained estimate  \\\hline
    Rpc2L0bH & $ 0.83 \pm  0.56 \pm  0.74$  &  $1.00 \pm 0.96 \pm 0.81$   &  $ 0.87 \pm  0.48 \pm  0.76$  \\
    Rpc2L0bS & $ 1.51 \pm  0.60 \pm  0.66$  &  $1.68 \pm 1.02 \pm 1.26$  &  $ 1.55 \pm  0.52 \pm  0.81$  \\
    Rpc2L1bH & $ 3.54 \pm  1.62 \pm  3.12$  &  $2.07 \pm 0.63 \pm 1.56$   &  $ 2.26 \pm  0.59 \pm  1.76$  \\
    Rpc2L1bS & $ 2.65 \pm  1.21 \pm  1.89$  &  $02.33 \pm 01.17 \pm 02.10$   &  $ 2.48 \pm  0.84 \pm  2.00$  \\ 
    Rpc2L2bH & $-0.11 \pm  0.11 \pm  0.18$  &  $<0.5$  &  $ 0.15 \pm 0.15 \pm  0.00$  \\
    Rpc2L2bS & $ 1.31 \pm  1.07 \pm  1.65$  &  $0.41 \pm 0.33 \pm 0.45$   &  $ 0.49 \pm  0.32 \pm  0.55$  \\
    Rpc2Lsoft1b & $ 4.75 \pm  1.42 \pm  2.64$  &  $2.48 \pm 1.32 \pm 1.86$  &  $ 3.53 \pm  0.97 \pm  2.22$  \\
    Rpc2Lsoft2b & $ 1.91 \pm  1.18 \pm  1.63$  &  $1.66 \pm 0.66 \pm 1.28$  &  $ 1.72 \pm  0.58 \pm  1.36$  \\
    Rpc3L0bH & $-0.01 \pm  0.11 \pm  0.10$  &  $<0.5$  &  $ 0.15 \pm  0.15 \pm  0.00$  \\
    Rpc3L0bS & $ 2.31 \pm  1.50 \pm  2.63$  &  $0.21 \pm 0.15 \pm 0.16$  &  $ 0.23 \pm  0.15 \pm  0.18$  \\
    Rpc3L1bH & $ 0.57 \pm  0.43 \pm  0.50$  &  $0.42 \pm 0.29 \pm 0.32$  &  $ 0.47 \pm  0.24 \pm  0.38$  \\
    Rpc3L1bS & $ 4.94 \pm  1.83 \pm  2.96$  &  $3.55 \pm 1.80 \pm 2.76$  &  $ 4.23 \pm  1.28 \pm  2.86$  \\
    Rpc3LSS1b & $-0.18 \pm  1.24 \pm  2.85$  &  $0.90 \pm 0.14 \pm 0.69$  &  $ 0.89 \pm  0.14 \pm  0.72$  \\
\hline
\hline
\end{tabular}
}
\caption{Expected yields for background processes with fake leptons,
in the signal regions  shown for 36 \ifb. 
Estimates from the matrix method and the MC template method are shown along with the retained estimates.
Uncertainties include all statistical and systematic sources for the nominal estimate.
}
\label{tab:fakes_sr_yields}
\end{table}

\begin{table}[!htb]
\centering
\begin{tabular}{|c||c|c|}\hline
 Region      &   Weighted OS data          &     Template method \\\hline
    Rpc2L0bH & $ 0.01 \pm  0.00 \pm  0.00$ & $<0.4$ \\
    Rpc2L0bS & $ 0.05 \pm  0.01 \pm  0.01$ & $ 00.02 \pm 00.02 \pm 00.00 $ \\
    Rpc2L1bH & $ 0.25 \pm  0.03 \pm  0.04$ & $ 00.21 \pm 00.32 \pm 00.16 $ \\
    Rpc2L1bS & $ 0.25 \pm  0.02 \pm  0.04$ & $ 00.35 \pm 00.37 \pm 00.26 $ \\
    Rpc2L2bH & $ 0.02 \pm  0.01 \pm  0.00$ & $<0.4$ \\
    Rpc2L2bS & $ 0.10 \pm  0.01 \pm  0.02$ & $<0.4$ \\
 Rpc2Lsoft1b & $ 0.08 \pm  0.01 \pm  0.02$ & $<0.4$ \\
 Rpc2Lsoft2b & $ 0.08 \pm  0.01 \pm  0.02$ & $<0.4$ \\
   Rpc3LSS1b & $ 0.39 \pm  0.03 \pm  0.07$ & $ 00.81 \pm 00.53 \pm 00.34 $ \\
\hline
\hline
\end{tabular}
\caption{Expected yields for background processes with charge-flipped electrons,
in the signal regions shown for 36 \ifb. 
Estimates from the likelihood method and the MC template method are shown.
Uncertainties include all statistical and systematic sources. 
Charge-flip processes do not contribute to signal regions which require $\ge 3$ leptons. 
}
\label{tab:chflips_sr_yields}
\end{table}

\begin{table}[!htb]
\def\arraystretch{1.1}
\centering
\resizebox{0.8\textwidth}{!}{
\begin{tabular}{|c|c|c|c|c|c|}
\hline 
               & VR-$t\bar t W$ & VR-$t\bar t Z$ & VR-$WZ$4j & VR-$WZ$5j & VR-$W^\pm W^{\pm}$  \\\hline	   
Fakes DD       & $23 \pm 5 \pm 24$      & $30 \pm 4 \pm 14$   & $53 \pm 6 \pm 27$   & $21 \pm 4 \pm 10$  & $14 \pm 3 \pm 10$ \\
Fakes MC       & $14 \pm 4 \pm 10$      & $18 \pm 3 \pm 13 $  & $46 \pm 5  \pm 34$  & $16 \pm 2 \pm 12$  & $13 \pm 2 \pm 10$ \\\hline
Combined       & $18 \pm 3 \pm 15$      & $22 \pm 2 \pm 13$   & $49 \pm 4 \pm 30$   & $17 \pm 2 \pm 12$  & $13 \pm 2 \pm 10$ \\\hline
Charge-flip DD & $3.4 \pm 0.1 \pm 0.5$  & $-$                 & $-$                 & $-$                & $1.7 \pm 0.1\pm 0.2$ \\
Charge-flip MC & $3.8  \pm 1.0 \pm 1.9$ & $-$                 & $-$                 & $-$                & $1.0 \pm 0.3 \pm 0.2$\\
\hline
\end{tabular}}
\caption{Comparison of expected yields for background processes with fake leptons,
in the validation regions, shown for 36 \ifb~between the data driven (DD) estimates and the MC template method (MC) estimates. 
}
\label{tab:VR_Comparison}
\end{table}

%% file: texfiles/sec.syst.overview.tex
The systematic uncertainties related to the estimated background from same-sign prompt leptons arise from the experimental uncertainties 
as well as theoretical modelling and theoretical cross-section uncertainties.
The statistical uncertainty of the simulated event samples is also taken into account.

%% file: texfiles/sec.syst.theory.tex
The cross-sections used to normalize the MC samples are varied according to the uncertainty in the 
cross-section calculation, which is 13\% for $\ttbar W$, 12\% for $\ttbar Z$ production~\cite{YR4}, 6\% for diboson
production~\cite{ATL-PHYS-PUB-2016-002}, 8\% for $\ttbar H$~\cite{YR4} and 30\% for 4$t$~\cite{Alwall:2014hca}. 
Additional uncertainties are assigned to some of these backgrounds to account for the theoretical modelling of the kinematic 
distributions in the MC simulation.

\subsection*{Associate $t\bar t+W/Z$ production}

The theoretical uncertainties on the $ttW$ and $ttZ^{(*)}$ processes are evaluated by several variations added in quadrature:

\begin{itemize}
\item Normalization and factorization scales varied independently up and down by a factor of two from the central scale $\mu_0 = H_{\rm T}/2$ as detailed in Ref.~\cite{ATL-PHYS-PUB-2016-005}. 
The largest deviation with respect to the nominal is used as the symmetric uncertainty.
\item Variation of the PDF used.
The standard deviation of the yields obtained using different PDF sets was used as the absolute uncertainty due to PDF. 
The relative uncertainty is then computed by dividing the standard deviation by the mean yield.
\item Comparison of the nominal \AMCATNLO MC samples to alternative \textsc{Sherpa} (v2.2) samples produced at leading-order with one extra parton in the matrix element for $ttW$ and 2 extra partons 
for $ttZ$~\cite{ATL-PHYS-PUB-2016-005}. 
The yield comparison for all SRs is shown in Table~\ref{tab:ttVGenComp}, with negligible differences in some SRs and up to 28\% in the worst case.
\end{itemize}
As a result of these studies, the total theory uncertainty for these processes 
is at the level of 15-35\% in the signal and validation regions used in the 
analysis. 
\begin{table}[!htb]
\caption{Comparison of the event yields for the $\ttbar V$ background processes between \AMCATNLO (default generator) and \textsc{Sherpa} in the SRs, as well as their relative difference.
}
\label{tab:ttVGenComp}
\def\arraystretch{1.1}
\centering
\begin{tabular}{|c|c|c|c|}
\hline\hline
   SR    & \textsc{Sherpa} & aMCATNLO & Relative diff.\\ \hline
Rpc2L0bH   &   0.25 $\pm$ 0.03   &   0.20 $\pm$ 0.05   &   25\% \\
Rpc2L0bS   &   0.60 $\pm$ 0.06   &   0.82 $\pm$ 0.10   &   -26\% \\
Rpc2L1bH   &   3.84 $\pm$ 0.14   &   3.86 $\pm$ 0.20   &   $<$1\% \\
Rpc2L1bS   &   3.55 $\pm$ 0.13   &   3.94 $\pm$ 0.20   &   -9\% \\
Rpc2L2bH   &   0.35 $\pm$ 0.04   &   0.41 $\pm$ 0.05   &   -14\% \\
Rpc2L2bS   &   1.57 $\pm$ 0.08   &   1.57 $\pm$ 0.12   &   $<$1\% \\
Rpc2Lsoft1b   &   1.01 $\pm$ 0.07   &   1.24 $\pm$ 0.11   &   -18\% \\
Rpc2Lsoft2b   &   1.13 $\pm$ 0.07   &   1.15 $\pm$ 0.10   &   -1\% \\
Rpc3L0bH   &   0.23 $\pm$ 0.02   &   0.18 $\pm$ 0.04   &   27\% \\
Rpc3L0bS   &   0.90 $\pm$ 0.05   &   0.99 $\pm$ 0.09   &   -9\% \\
Rpc3L1bH   &   1.54 $\pm$ 0.08   &   1.52 $\pm$ 0.11   &   1\% \\
Rpc3L1bS   &   6.95 $\pm$ 0.16   &   7.02 $\pm$ 0.23   &   $<$1\% \\
Rpc3LSS1b   &   0.00 $\pm$ 0.00   &   0.00 $\pm$ 0.00   &   - \\
\hline\hline
\end{tabular}
\end{table}

\subsection*{Diboson $WZ, ZZ, W^\pm W^\pm$ production}

The theoretical uncertainties on the  $WZ$ and $ZZ$ processes are evaluated by several variations added in quadrature:
\begin{itemize}
\item Normalization and factorization scales varied independently up and down by a factor of two from the central scale choice. The largest deviation with respect to the nominal is used as the 
symmetric uncertainty.
\item The standard deviation of the yields obtained using different PDF sets was used as the absolute uncertainty due to PDF. 
The relative uncertainty is then computed by dividing the standard deviation by the mean yield.
\item Re-summation scale varied up and down by a factor of two from the nominal value.
\item The scale for calculating the overlap between jets from the matrix element and the parton shower is varied from the nominal value of 20 GeV~down to 15 GeV~
 and up to 30 GeV. The largest deviation with respect to the nominal is used as the symmetric uncertainty due to matrix element matching.
\item An alternative recoil scheme is considered to estimate the uncertainty associated with mis-modeling of jet multiplicities larger than three.
\end{itemize}

Based on these studies and the cross-section uncertainties, the total theory uncertainty for these processes is at the level of 25-40\% in the signal and validation regions used in the analysis. 

No theoretical uncertainties have been evaluated specifically for the $W^\pm W^\pm jj$ process, 
to which we assign the same uncertainties as for $WZ$, by lack of a better choice. 
But it should be noted that contributions from this process are minor in the SRs and typically  
smaller than those from $WZ$ and $ZZ$.

\subsection*{Other rare processes}
A conservative 50\% uncertainty is assigned on the summed contributions 
of all these processes ($\ttbar H$, $tZ$, $tWZ$, $\ttbar\ttbar$, $\ttbar WW$, $\ttbar WZ$, $WH$, $ZH$, $VVV$), 
which is generally quite larger than the uncertainties on their inclusive production cross-sections, 
and assumes a similar level of mis-modelling as for diboson or $t\bar t V$ processes. 

%% file: texfiles/sec.syst.exp.tex
Uncertainties associated with the measurement and reconstruction of the 
physics objects used in the analysis (leptons, jets, etc.) must be accounted
 for when interpreting the results.
The systematic uncertainties from the data-driven method have already been 
discussed in Section~\ref{sec:bkg.red}. In fact, these data-driven backgrounds 
are affected by the same systematic uncertainty as in data to which they 
are being compared to. As a result, only systematic uncertainties on 
backgrounds estimated with MC simulation and detector simulation needs to 
be considered. The uncertainties considered for the analysis and 
recommended by the ATLAS SUSY group are:\\
\textbf{Jet energy scale (JES)}  \\  
In order to account for inefficiencies in the calorimeter cells
and the varying response to charged and neutral particles passing through 
them, the energies of the jets used in this analysis were corrected. 
The calibration procedure uses a combination of simulation and test beam 
and in situ data ~\cite{Aaboud:2017jcu} with an uncertainty correlated 
between all events.
As a result, all distributions used in the final result are produced 
with the nominal calibration as well as an up and down variation of the 
the jet energy scale (in a fully correlated way) by the 
$\pm 1\sigma$ uncertainty of each nuisance parameter.
A combined version of several independent sources contributing to the 
calibration was used in the analysis 
to reduce the number of nuisance variables in the fitting procedure.\\
\textbf{Jet energy resolution (JER)} \\ 
An extra $\pt$ smearing is added to the jets based on their $\pt$ and $\eta$ 
to account for a possible underestimate of the jet energy resolution 
in the MC simulation. A systematic
uncertainty is considered to account for this defect on the final result. 
The JER in data has previously been estimated by ATLAS in dijet events. \\ 
\textbf{Jet vertex tagger}
The uncertainties account for the residual contamination from pile-up jets 
after pile-up suppression and the MC generator choice
~\cite{ATLAS-CONF-2014-018}.\\
\textbf{Flavor tagging}
The MC simulation does not reproduce correctly the $b$-tagging, 
charm identification, and light jet reject efficiencies of the detector. 
A \ttbar MC simulation and di--jet measurements are used to derive 
correction factors to be applied to MC simulation
~\cite{ATL-PHYS-PUB-2015-022,ATL-PHYS-PUB-2016-012}.
These correction factors are then varied within
their uncertainties to produce up and down variations.\\
\textbf{Lepton energy scale, resolution, and Identification efficiencies}
Similar to the case of jets, electrons and muons also have corresponding 
energy scale and resolution systematic uncertainties. Corrections are 
also applied to take into account any variations in the identification 
efficiency in the detector and its simulation~\cite{ATLAS-CONF-2016-024,Aad:2016jkr,ATLAS-CONF-2016-024}.\\
\textbf{\met\ soft term uncertainties}
The main effect come from the hard object uncertainties (most notably JES and 
JER) that are propagated to the $\met$.\\
\textbf{Pileup re-weighting}
This uncertainty is obtained by re-scaling the $\mu$ value in data by 1.00 and 1/1.18, 
covering the full difference between applying and not-applying the nominal $\mu$ correction of 1/1.09, 
as well as effects resulting from uncertainties on the luminosity measurements, which are expected to dominate.\\
\textbf{Luminosity}
The integrated luminosities in data corresponds to 3.2 \ifb and 32.9 \ifb 
for 2015 and 2016 respectively. The combined luminosity error for 2015 and 2016 is 3.2\%, assuming partially correlated uncertainties in 2015 and 2016.\\
\textbf{Trigger}
To account for any differences between the trigger efficiency in simulation 
and data, corrections factors are derived to correct for them. 
Uncertainties on the correction factors as well as inefficiencies 
related to the plateau of the trigger are propagated to the final result.

The uncertainty on the beam energy is neglected. 
All the experimental uncertainties are applied also on the signal samples when computing exclusion limits on SUSY scenarios.

All of these uncertainties are fed into the fitting and limit setting 
machinery by treating them as uncorrelated uncertainties, and thus 
treated independently. 

%% file: texfiles/sec.stat.overview.tex
The goal of the analysis is to maximize the information that can be 
extracted from comparing the observed data to the background 
prediction in the signal regions designed to search for new physics 
topologies. Statistical tools are essential to tell in the most 
powerful way and to the best of our knowledge if there is a new physics 
signal beyond what is already known in the observed data. 
At the same time, it is important to properly treat the systematic 
uncertainties associated with the complexity of the experimental 
apparatus (the ATLAS detector) and the background predictions when 
presenting an interpretation of the results. 
This chapter describes the statistical methodology employed to 
test the compatibility between data and prediction while taking into 
account the systematic uncertainties. 
The analysis' possible outcomes are represented by a likelihood function 
that combines observations, predictions, and associated uncertainties. 
At this point the 
hypothesis testing is performed with the corresponding one-sided profile 
likelihood ratio~\cite{Cowan:2010js}, 
and upper limits are provided as one-sided $95\%$ confidence level intervals in the CL$_\text{s}$ formalism~\cite{Read:2002}. 
The statistical tool used to perform the quantification of the significance 
of hypothetical excesses seen in data
or upper limits setting on new physics contributions as implemented in 
this analysis will be described in this chapter.

%% file: texfiles/sec.stat.like.tex
The likelihood for a set of parameters $\left(\mu,  \vect{\theta}\right)$
given all the data that might have been observed  $\vect{X}$ is the probability 
of observing the data given the parameters

\begin{equation}
\mathcal{L} \left( \mu, \vect{\theta} \rvert \vect{X} \right) = 
\Pr\left(\vect{X} \rvert \mu, \vect{\theta} \right).
\end{equation}

The data $\vect{X}$ includes observation in the signal regions as well as 
other auxiliary experiments such as control regions used to constrain 
backgrounds. 
In this analysis, hypothesis tests are performed on one signal region 
at the time, single-binned, and without control regions. 
As a result, the observed data $\vect{X}$ has a one-dimensional component 
with value $X$ representing the count of events in the signal region.
The first parameter of interest represents the `strength' of 
the signal process $\mu > 0$ that will increase the number of 
expected events in the signal region given that the signal of the new physics 
model tested is present. In practice, the signal strength $\mu$ is used to 
scale the nominal expected cross section for the signal process, or the
number of expected signal events $s$. 
Thus, the predicted background will be of the form $\mu s + \sum_i b_i$ where 
$b_i$ represents the Standard Model background processes expected to 
contribute to the signal region. 
The parameter $\vect{\theta}$ refers to 
the nuisance parameters used to parametrize the systematic uncertainties
(luminosity, JES, JER, etc.)\footnote{The parameters represented by 
$\vect{\theta}$ are called nuisance 
parameters since the aim is not to set a limit on them.}.
Thus, the likelihood is built as the product of a Poisson probability density 
function describing the observed number of events in the signal 
region and Gaussian distributions for each of the sources of systematic 
uncertainties.
The likelihood takes the simple form
\begin{equation}
\Pr\left(X,  \vect{\theta^0} \rvert \mu, \vect{\theta} \right) = 
\mathcal{P} \left(X~\rvert~\mu s\left(\vect{\theta}\right)
+ \sum_i b_i\left(\vect{\theta}\right)
\right)
\times  
\prod_{j}  \mathcal{G} \left(\theta^0 \rvert \theta_{j}\right)
\label{eq:stat.like.Pr}
\end{equation}
where $\mathcal{P} \left(X~\rvert \nu\right) = e^{-\nu}\nu^X/X!$ 
and $\mathcal{G} \left(\theta\right)  = \frac{1}{\sqrt{2\pi\sigma_{\theta}^2}}
e^{-\frac{\left(\theta-\theta^0\right)^2}{2\sigma_\theta^2}}$
 are the Poisson and Gaussian probability density functions. 
In addition to the observed data $X$, there is also auxiliary measurements
$ \vect{\theta^0}$ that constrain the nuisance parameters.
Both signal and backgrounds depend on the nuisance parameter
$\vect{\theta}$ which controls all independent sources of uncertainty 
and will be \textit{profiled} (or constrained) in the $CL_s$ procedure
described next.
Correlations of a given nuisance parameter between the different sources of backgrounds 
and the signal are taken into account when relevant. 
To give an example, the luminosity uncertainty will have a mean 
as the luminosity central value $ \theta_{\text{lumi}}^{\left(0\right)}$ and the 
width as an experimentally determined 
uncertainty $\sigma_{\theta_{\text{lumi}}}$. When evaluating the effect of the 
luminosity uncertainty on the likelihood function, all terms involving the 
nuisance parameter $\theta_{\text{lumi}}$ will be scaled in the same way 
since luminosity is correlated across all backgrounds and signal.

The likelihood function described in this section is used in a fit to data by 
the maximum likelihood method that aims at finding the value of the 
signal strength $\mu$ 
that makes the likelihood a maximum. The procedure relies on an iterative 
minimization algorithm implemented in \textsc{Minuit} \cite{minuit} and accessed by \textit{RooFit} \cite{Moneta:2010pm} within \textsc{ROOT}, a high energy physics data analysis  framework
\cite{Antcheva:2009zz}.
The final uncertainties on the nuisance parameters in $\vect{\theta}$, constrained 
by the fit, are obtained from the covariance matrix of these parameters.

%% file: texfiles/sec.stat.limit.tex
\subsection{Profile Likelihood Ratio}

The procedure for setting exclusion limits using the likelihood function 
(Eq. \ref{eq:stat.like.Pr})
relies on a profile-likelihood-ratio test~\cite{Cowan:2010js}.
The null hypothesis considered is that of background only with $\mu=0$ and the alternate hypothesis 
is the presence of a signal with strength $\mu>0$.
According to the Neyman-Pearson Lemma \cite{Neyman289}, 
the most powerful test when performing a hypothesis test between two simple hypotheses, $\mu=0$ and $\mu>0$,
is the profile-likelihood-ratio test, which rejects $\mu=0$ in favor of $\mu>0$.
The profile-likelihood-ratio test $q_{\mu}$ is defined to be
\begin{equation}
q_{\mu}  \left( \vect{X} \right)= 
\begin{cases}
  -2\ln\left(
  \frac
      {\mathcal{L} \left( \mu, \hat{\hat{\vect{\theta}}}\left(\mu\right) \rvert \vect{X} \right)}
      {\mathcal{L} \left( \hat{\mu}, \hat{\vect{\theta}} \rvert \vect{X} \right)}
 \right), & \text{if}\ \mu > \hat{\mu} \\
      0, & \text{otherwise}
    \end{cases}
\label{eq:stat.limit.qmu}
\end{equation}
For a given observation $\vect{X}$, the parameters $\hat{\mu}$ and $\vect{\theta}$ are obtained from maximizing the likelihood. 
If the value of $\mu$ is also fixed, the set of values $\hat{\hat{\vect{\theta}}}$ corresponds to the set of values that 
maximize the likelihood for that particular value of $\mu$. The variable $q_{\mu}$ is always positive ($q_{\mu}\geq0$).
Qualitatively, small values of $q_{\mu}$ correspond to a better compatibility between the observed data and the 
tested hypothesis with a signal strength $\mu$, while large values indicate a very improbable signal hypothesis.
The condition on the sign of $ \mu - \hat{\mu}$ is necessary in order not to interpret an upward fluctuation of data 
($X > \mu s + b$) as incompatible with the tested hypothesis. 
The form of $q_{\mu}$ is motivated by Wald and Wilk's \cite{Wilks:1938dza,Wald:ams1943} approximation formulas in the large sample limit
where the distributions of $q_{\mu}$ also become independent of nuisance parameters as it will be discussed next.
The test statistic for discovery will try to reject the background only hypothesis ($\mu=0$). Thus Eq. \ref{eq:stat.limit.qmu} becomes
\begin{equation}
q_{0}  \left( \vect{X} \right)= 
\begin{cases}
  -2\ln\left(
  \frac
      {\mathcal{L} \left( 0, \hat{\hat{\vect{\theta}}}\left(0\right) \rvert \vect{X} \right)}
      {\mathcal{L} \left( \hat{\mu}, \hat{\vect{\theta}} \rvert \vect{X} \right)}
 \right), & \text{if}\ \hat{\mu} < 0 \\
      0, & \text{otherwise}
    \end{cases}
\end{equation}

\subsection{The $p$-value}

\begin{figure}[t]
\centering
\begin{subfigure}[t]{0.45\textwidth}\includegraphics[width=\textwidth]{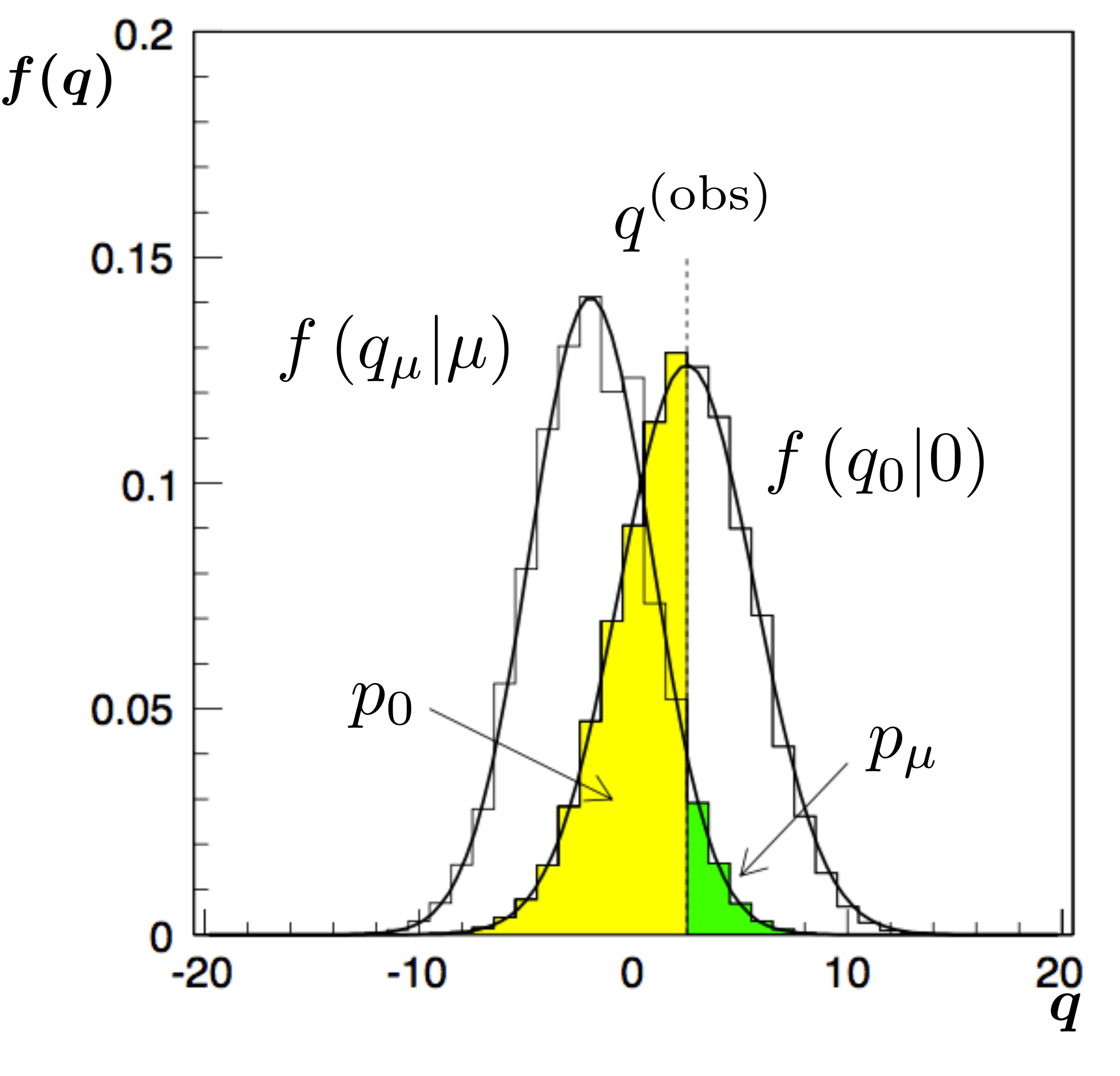}\caption{}\label{fig:stat.limit.qmunormal}\end{subfigure}
\begin{subfigure}[t]{0.467\textwidth}\includegraphics[width=\textwidth]{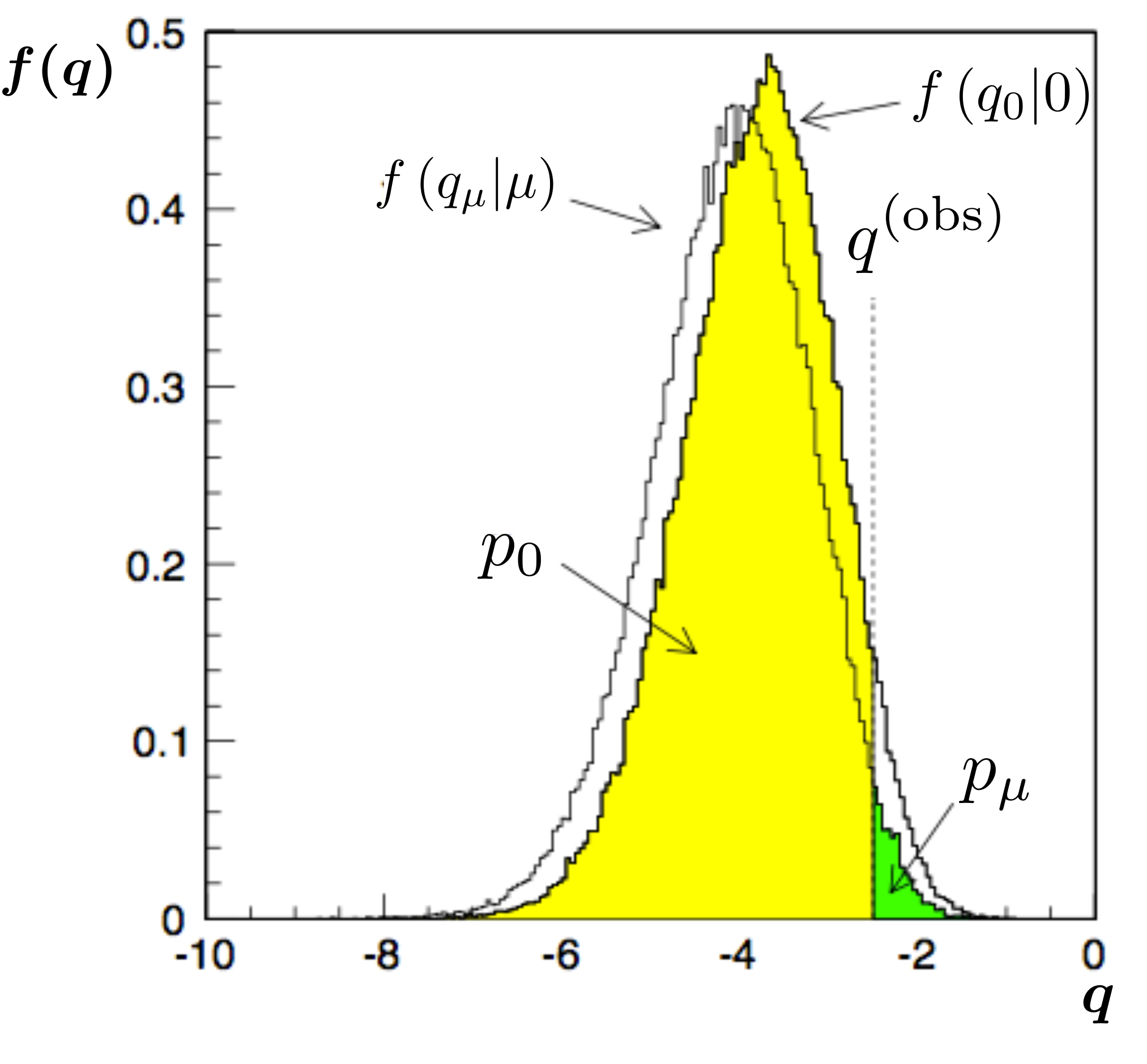}\caption{}\label{fig:stat.limit.qmunosens}\end{subfigure}
\caption{Distributions of the test variable $q_{\mu}$ under the $\mu>0$ and $\mu=0$ hypotheses: (a) typical case, 
  (b) case where there is very little sensitivity to the signal model}
\label{fig:stat.limit.qmu}
\end{figure}

At this stage, a test statistic $q_{\mu}$ has been constructed to distinguish between the hypothesis 
that the data contains signal and background $\mu>0$ and that of background only $\mu=0$.
To illustrate the limit setting procedure, we consider distributions of the test statistic under each 
hypothesis: $f\left(q_{\mu} \rvert \mu\right)$ for $\mu>0$ and $f\left(q_{0} \rvert 0\right)$ for $\mu=0$.
These distributions are shown in Figure~\ref{fig:stat.limit.qmunormal} and details on how to obtain their functional forms
 will be discussed later. 

Given that the actual observed data leads to a test variable $q_{\mu}^{\left(\text{obs}\right)}$, it is possible to 
quantify the level of discrepancy between the observed data and the tested hypothesis ($\mu>0$ or $\mu=0$) using 
a $p$-value. 
The $p$-value of the signal hypothesis ($\mu>0$) is then defined as the probability, under assumption of the signal hypothesis, 
to find a value of $q_{\mu}$ with equal or lesser compatibility with the signal model considered relative to what is 
found with $q_{\mu}^{\left(\text{obs}\right)}$. In other words,  higher values of $q_{\mu}$ indicate an increasing disagreement
between data and the signal model. The mathematical expression of the $p$-value with $\mu>0$ is taken as the probability 
to find $q_{\mu}$ greater than or equal $q_{\mu}^{\left(\text{obs}\right)}$, under the signal hypothesis\footnote{Note that 
 the background-only distribution $f\left(q_{0} \rvert 0\right)$ in the example given in 
Figure~\ref{fig:stat.limit.qmunormal} is shifted to the right.} is given by
\begin{equation}
p_{\mu} = \Pr\left(q_{\mu} \geq q_{\mu}^{\left(\text{obs}\right)} \rvert \mu \right) = \int_{q_{\mu}^{\left(\text{obs}\right)}}^{\infty} f\left(q_{\mu} \rvert \mu\right) \diff q_{\mu}
\end{equation}
where $q_{\mu}^{\left(\text{obs}\right)}$ is the value of the statistic test observed in data and the function $f$
denotes the probability distribution function of $q_{\mu}$ under the signal hypothesis. 
Similarly, the $p$-value of the background-only hypothesis with $\mu=0$ takes the form of 
\begin{equation}
p_{0} = \Pr\left(q_{0} \geq q_{0}^{\left(\text{obs}\right)} \rvert 0 \right) = \int^{q_{0}^{\left(\text{obs}\right)}}_{-\infty} f\left(q_{0} \rvert 0\right) \diff q_{0}
\end{equation}
and can be interpreted as the probability of the observation to be consistent with the background only hypothesis:
the smaller the $p_{0}$ value is, the less compatible data is with the background only hypothesis. 
In order to claim a discovery in particle physics, 
the background-only hypothesis must be rejected using the $p_0$-value. 
However, there is an ambiguity related to how small the $p_0$-value needs to 
be, before declaring a discovery. 
The problem can also be formulated in terms of significance.
Assuming a Gaussian distributed variable, how many $Z$ standard deviations 
$\sigma$ above the mean are required to cover an upper-tail probability equal 
to $p$ expressed as 
\begin{equation}
Z = \Phi^{-1}\left(1-p\right)
\end{equation}
 where $\Phi$ is the Gaussian cumulative distribution.
The particle physics community considers a discovery if $Z=5$, commonly 
used ``five sigma excess'' statement, which corresponds to 
$p_0 = 2.87\times 10^{-7}$ or one in 3.5 million probability for the background
 to be as extreme as the observation. In order to announce evidence for a 
new particle, a significance of $Z=3$ or $p_0 = 1.35\times 10^{-3}$ is required.

\subsection{The $CL_\mathrm{s}$ Prescription}
The next step is to define a confidence interval ($CI$) that includes the parameter $\mu$ at a specified confidence level ($CL$). 
In other words, instead of estimating the parameter $\mu$ by a single value, an interval $CI$ likely to include the parameter$\mu$ is given.
The $CL$ provides a quantitative statement on how likely the interval $CI$ is to contain the parameter $\mu$. 
Given the measurement of a parameter $\mu_{meas}$, we deduce that there is
a 95\% $CI$ $[\mu_1,\mu_2]$. The statement means
that in an ensemble of experiments 95\% of the obtained $CI$s will contain 
the true value of $\mu$\footnote{The statement \textbf{does not} mean that
there is a 95\% probability that the interval $[\mu_1,\mu_2]$ contains the 
true value of the parameter $\mu$.}.
The upper limit is simply the case where the 95\% $CI$ is $[0,\mu_{up}]$:
In an ensemble of experiments 95\% of the obtained $CI$s will contain the true value of $\mu$, including $\mu=0$. 
The conclusion is that $\mu < \mu_{up}$ at the 95\% $CL$ or that $ \mu_{up}$ is an upper limit. 

Applying this concept to the problem at hand where we want to place an upper limit on the expected number of 
signal events $S$ ($S=\mu s$, $s$ being the expected number of signal events for $\mu=1$) in one or more signal regions.
The $CL$ is obtained from a standard statistical test of the signal model ($\mu>0$) which can establish the exclusion of the signal 
model at confidence level $1-\alpha = 95 \%$ if 
\begin{equation}
  CL_\mathrm{s+b}  = p_{\mu} < \alpha
\end{equation}
where $\alpha=0.05$. 
Since the result section will present $CL$ values in terms of the expected number of signal events from beyond the Standard Model 
processes, we continue this discussion of estimating an upper limit on $S$ rather than $\mu$.
Thus, a $CI$ at confidence level $CL=1-\alpha$ for the expected number of signal events $S$ can be constructed from those values of
$S$ (or $\mu$) that are not excluded, and the upper limit $S_{up}^{1-\alpha}$ is the largest value of $S$ not excluded.
By construction, the interval $[0,S_{up}^{95}]$ will cover the expected number of signal events $S$ with a probability of at least 
95\%, regardless of the value of $S$.

An anomaly arises with the $CL_\mathrm{s+b}$ prescription when the number of expected signal events is much less than that of 
the background and the data observation had a downward fluctuation below the expected background.
The procedure will lead to excluding, with probability close to $\alpha$, hypotheses to which the experiment has no sensitivity.
For $\alpha=5\%$, it means that one out of twenty tests for different signal models where one has no sensitivity will result in exclusion.
In fact, the desired behavior of the exclusion probability in this case is to approach zero rather than $\alpha$.
This scenario is illustrated in Figure~\ref{fig:stat.limit.qmunosens} where the distribution of $q_{\mu}$
under both the signal and background-only hypotheses are almost similar. 
To remedy this problem, a different procedure is used where a model is regarded as excluded if 
\begin{equation}
  CL_\mathrm{s}  = \frac
{p_{\mu}}
{1-p_{0}}
< \alpha.
\end{equation}
In this form, the $p$-value is penalized by dividing by $1-p_{0}$. 
If the distribution of $q_{\mu}$ under a signal or background-only hypotheses 
are widely separated, then 
the quantity $1-p_{0}$ is close to unity which recovers the $CL_\mathrm{s+b}$ value. However, if
the distributions of both hypotheses are similar, due to the lack of sensitivity, $1-p_{0}$ becomes smaller
and the $CL_\mathrm{s+b}$ value is increased more leading to a weaker upper limit. 
Similar to the case of $CL_\mathrm{s+b}$, the upper limit on $S_{up}^{95}$ is taken as the largest value of the parameter
$S$ not excluded.

\subsection{Approximate Sampling Distributions}

The remaining task is to determine the sampling distributions
$f\left(q_{0} \rvert 0\right)$ and $f\left(q_{\mu} \rvert \mu\right)$
needed to compute the $p$-values
used in the case of discovery and setting upper limits, respectively.
These distributions do not have an analytic form but can be obtained from 
pseudo-experiments or asymptotic approximations.
The pseudo-experiments are more accurate than the asymptotic 
approximations since a large number of 
datasets are generated which are drawn from a distribution that is consistent 
with those observed.
However, for a complex likelihood function where the procedure of generating 
pseudo-datasets needs to be repeated many times for each parameter point
of each model being considered, the pseudo-data set method is computing 
intensive and not practical. For this reason, the current analysis 
uses asymptotic formulae to approximate $q_{\mu}$ and shown to be valid in the 
large sample size limit \cite{Cowan:2010js}. The approximation is based on an important result 
by Wilks \cite{Wilks:1938dza} and Wald \cite{Wald:ams1943} who showed that for a single parameter of interest,
\begin{equation}
  q_{\mu} =
  \begin{cases}
    \frac{\left(\mu - \hat{\mu}\right)^2}{\sigma^2} + \mathcal{O}\left(\frac{1}{\sqrt{N}}\right)
    , & \text{if}\ \mu > \hat{\mu} \\
    0, & \text{otherwise}.
  \end{cases}
\label{eq:stat.model.asymp}
\end{equation}
where $\mu$ is a Gaussian distribution with a mean $\bar{\mu}$ and 
standard deviation $\sigma$, and a sample size $N$. In the case where 
$\mu = \hat{\mu}$, the test statistic $q_{\mu}$ follows a $\chi^2$ distribution
with one degree of freedom. 
The variance $\sigma^2$ is obtained from an artificial data set called the 
``Asimov data set''
\footnote{Inspired from the short story \textit{Franchise}\cite{asimov} by Isaac Asimov that entails using a single voter that represents the entire electorate 
population in an election. Similarly, an ensemble of pseudo-experiments 
can be replaced by a single representative data set.}
 that verifies $X_A = \bar{\mu} s + b$. 
From Eq. \ref{eq:stat.model.asymp}, the variance is then 
$\sigma^2 = \left( \mu - \bar{\mu} \right)^2/q_{\mu,A}$, where $q_{\mu,A}$ 
is evaluated from the exact expression of $q_{\mu}$ using the Asimov data set.

The results obtained from the asymptotic approximations have been compared to 
exclusion limits obtained with a limited number of pseudo-experiments.
A reasonable agreement has been observed which validated the use of 
the asymptotic formalism to obtain the exclusion limits on the different models.

%% file: texfiles/sec.stat.impl.tex
The statistical interpretations of the observations 
are performed with the \texttt{HistFitter} framework~\cite{Baak:2014wma}, commonly in use within the SUSY ATLAS working group. 
It consists of a user-friendly abstraction layer to the \texttt{HistFactory} and \texttt{RooFit} frameworks~\cite{Cranmer:1456844,Verkerke:2003ir}, 
to which it delegates the tasks of building a probabilistic representation of the analysis, and performing hypothesis tests. 
\texttt{HistFitter} also provides several related utilities, 
such as the creation of summary tables of yields and uncertainties, or the creation of exclusion plots such as those presented in Chapter~\ref{chap:res}.

The likelihood function implemented in \texttt{HistFitter} is built as the 
product of a Poisson probability density function describing the observed 
number of events in the signal region 
and, to constrain the nuisance parameters associated with the systematic 
uncertainties, 
Gaussian distributions whose widths correspond to the sizes of these 
uncertainties.
Poisson distributions are used instead for MC simulation statistical 
uncertainties.
Correlations of a given nuisance parameter between the backgrounds and the 
signal are taken into account when relevant. 
The hypothesis tests are performed for each of the signal regions 
independently. 

%% file: texfiles/chap.res.tex
\section{Predictions and Observations}

Observed data (36.1 \ifb) and predicted SM background event yields 
are compared in Figures~\ref{fig:results_datamc_rpc1}-\ref{fig:results_datamc_rpc2} for signal regions addressing $R$-parity-conserving signal scenarios, 
for events satisfying the SR requirements except the \met\ cut. Adjustments to the selections are made to allow more events close to the signal regions (SR) as follows: 
\begin{itemize}
\item SRs involving an effective mass cut: it is relaxed 
together with the \met\ cut to avoid indirectly tightening requirements 
on the visible leptonic and hadronic contributions to \meff\ for low values of \met; 
the \meff\ requirement is therefore changed to:
$$
\meff > \left(m_\mathrm{eff}^\text{SR cut}\right) - \operatorname{max}\left[\left(E_\mathrm{T}^\text{miss, SR cut}\right)-\met,0\right]
$$
\item SRs involving a cut on the $\met/\meff$ ratio: it is relaxed together with the \met\ cut 
to allow populating the low \met\ tail of the distributions: 
$$
\frac{\met}{\meff} > \left(\frac{\met}{\meff}\right)^\text{SR cut}\times \frac{\met}{\left(E_\mathrm{T}^\text{miss, SR cut}\right)^\text{SR cut}}
$$
in addition, an upper cut on \meff\ is added. 
\end{itemize}

\begin{figure}[htb!]
\centering
 \begin{subfigure}{0.42\textwidth}
 \includegraphics[width=\textwidth]{LooseRpc2L0bH.pdf}
 \subcaption{Rpc2L0bH before the \met\ requirement}
 \end{subfigure}
 \begin{subfigure}{0.42\textwidth}
 \includegraphics[width=\textwidth]{LooseRpc2L0bS.pdf}
 \subcaption{Rpc2L0bS before the \met\ requirement}
 \end{subfigure}
  \begin{subfigure}{0.42\textwidth}
 \includegraphics[width=\textwidth]{LooseRpc2L1bH.pdf}
 \subcaption{Rpc2L1bH before the \met\ requirement}
 \end{subfigure}
  \begin{subfigure}{0.42\textwidth}
 \includegraphics[width=\textwidth]{LooseRpc2L2bH.pdf}
 \subcaption{Rpc2L2bH before the \meff\ requirement}
 \end{subfigure}
   \caption{
Missing transverse momentum distributions for observed data and predicted backgrounds 
after the signal regions selection, beside the \met requirement.
The effective mass and/or $\met/\meff$ ratio cuts are also relaxed for \met\ values below the SR threshold (see text for details). 
The signal regions correspond to the last (inclusive) bins of the figures. 
The shaded area represents uncertainties on the total SM background estimate, 
which include all sources of statistical uncertainties, 
as well as the systematic uncertainties for fake lepton and charge-flip backgrounds. 
}
\label{fig:results_datamc_rpc1}
\end{figure}

\begin{figure}[htb!]
\centering
   \begin{subfigure}{0.42\textwidth}
 \includegraphics[width=\textwidth]{LooseRpc2L2bS.pdf}
 \subcaption{Rpc2L2bS before the \met\ requirement}
 \end{subfigure}
   \begin{subfigure}{0.42\textwidth}
 \includegraphics[width=\textwidth]{LooseRpc2Lsoft2b.pdf}
 \subcaption{Rpc2Lsoft2b before the \met\ requirement}
 \end{subfigure}
   \begin{subfigure}{0.42\textwidth}
 \includegraphics[width=\textwidth]{LooseRpc3L0bS.pdf}
 \subcaption{Rpc3L0bS before the \met\ requirement}
 \end{subfigure}
    \begin{subfigure}{0.42\textwidth}
 \includegraphics[width=\textwidth]{LooseRpc3L1bS.pdf}
 \subcaption{Rpc3L1bS before the \met\ requirement}
 \end{subfigure}
   \caption{
Missing transverse momentum distributions for observed data and predicted backgrounds 
after the signal regions selection, beside the \met requirement.
The effective mass and/or $\met/\meff$ ratio cuts are also relaxed for \met\ values below the SR threshold (see text for details). 
The signal regions correspond to the last (inclusive) bins of the figures. 
The shaded area represents uncertainties on the total SM background estimate, 
which include all sources of statistical uncertainties, 
as well as the systematic uncertainties for fake lepton and charge-flip backgrounds. 
}
\label{fig:results_datamc_rpc2}
\end{figure}

Figure~\ref{fig:PlotSR} shows the event yields for data and the expected background contributions 
in all signal regions. Detailed information about the yields can be found in Table~\ref{tab:SR_yields}.
The 95\% confidence level (CL) upper limits are shown on the observed and expected numbers of BSM events, $S_{\textrm{obs}}^{95}$ and $S_{\textrm{exp}}^{95}$ 
(as well as the $\pm 1\sigma$ excursions from the expected limit)
, respectively. The 95\% CL upper limits on the visible cross-section 
($\sigma_{\textrm{vis}}$) are also given. Finally the $p$-values ($p_{0}$) give the probabilities of the observations being consistent 
with the estimated backgrounds. The number of equivalent Gaussian standard deviations ($Z$) is also shown when $p_{0}<0.5$. 
In all 13 SRs the number of observed data events is consistent with the expected background within the uncertainties. 
The contributions listed in the rare category are dominated by triboson, $tWZ$ and $\ttbar WW$ production.
The triboson processes generally dominate in the SRs with no $b$-jets, while $tWZ$ and $\ttbar WW$
dominate in the SRs with one and two $b$-jets, respectively. Contributions from $WH$, $ZH$, $tZ$ and $\ttbar t$ production 
never represent more than 20\% of the rare background.

\begin{figure}[htb!]
\begin{center}
\begin{subfigure}[t]{0.98\textwidth}\includegraphics[width=\textwidth]{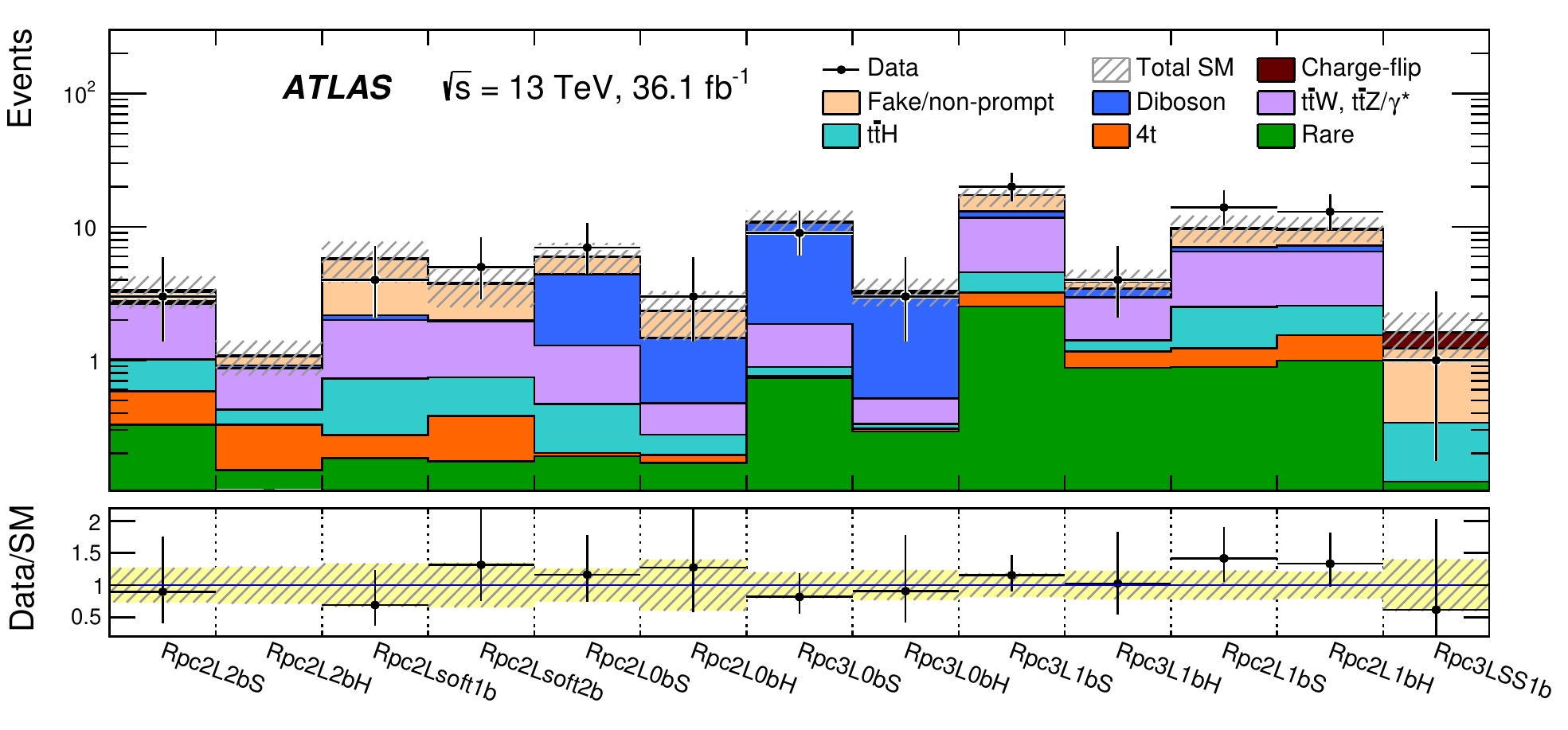}\caption{}\label{fig:Results_SRSum}\end{subfigure}
\begin{subfigure}[t]{1.08\textwidth}\includegraphics[width=\textwidth]{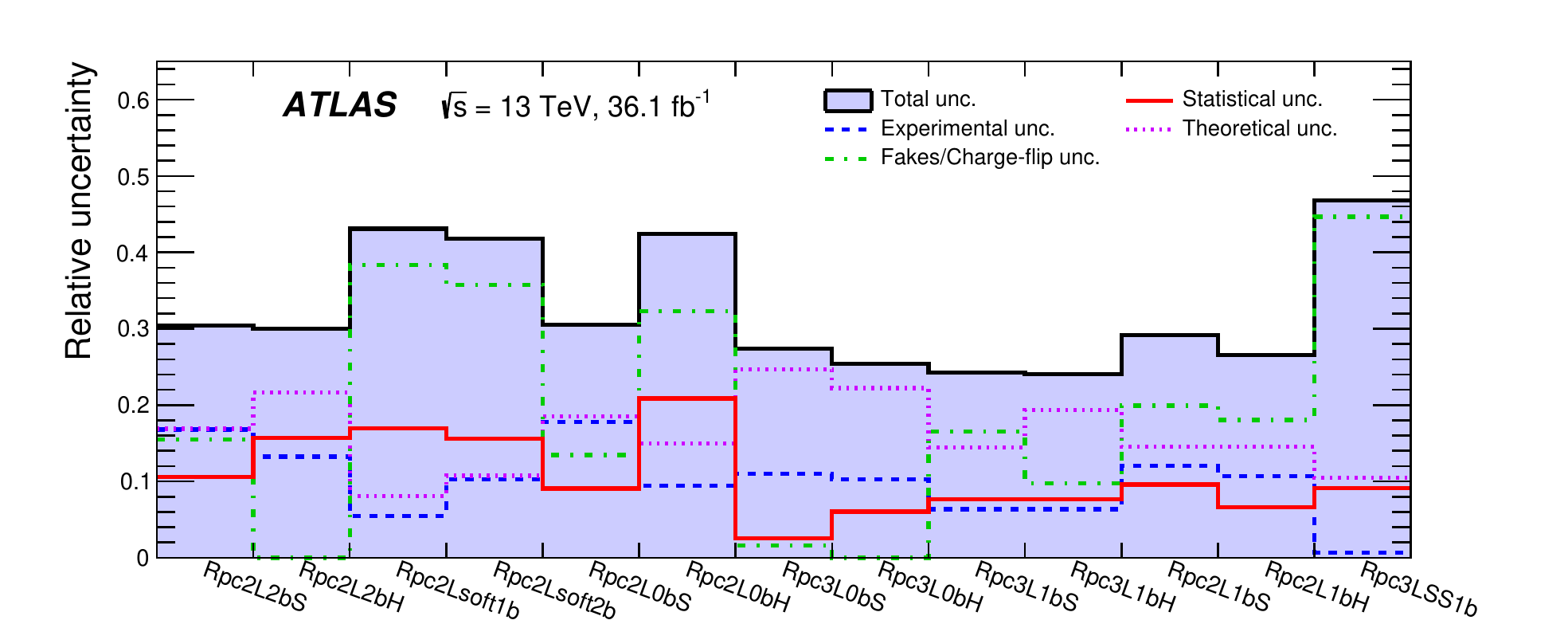}\caption{}\label{fig:Results_SystSum}\end{subfigure}
\end{center}
\caption{Comparison of (a) the observed and expected event yields in each signal region and (b) the relative uncertainties in the total 
background yield estimate. For the latter, ``statistical uncertainty'' corresponds to reducible and irreducible background 
statistical uncertainties. The background predictions correspond to those presented in Table~\ref{tab:SR_yields} and the 
rare category is explained in the text. } 
\label{fig:PlotSR}
\end{figure}

\begin{table}[htb!]
\scriptsize
\begin{center}
\vspace*{-0.035\textwidth}
\resizebox{1.\textwidth}{!}{
\begin{tabular}{|l|c|c|c|c|c|c|}
\hline
Signal Region                  & \textbf{Rpc2L2bS} & \textbf{Rpc2L2bH}    & \textbf{Rpc2Lsoft1b} &\textbf{Rpc2Lsoft2b}& \textbf{Rpc2L0bS} & \textbf{Rpc2L0bH}\\
\hline
\hline
$\ttbar W$, $\ttbar Z\gamma^*$ & $1.6\pm0.4$       & $0.44\pm0.14$     & $1.3\pm0.4$       & $1.21\pm0.33$       & $0.82\pm0.31$       & $0.20\pm0.10$  \\
$\ttbar H$                     & $0.43\pm0.25$     & $0.10\pm0.06$     & $0.45\pm0.24$     & $0.36\pm0.21$       & $0.27\pm0.15$       & $0.08\pm0.07$   \\
4$t$                           & $0.26\pm0.13$     & $0.18\pm0.09$     & $0.09\pm0.05$     & $0.21\pm0.11$       & $0.01\pm0.01$       & $0.02\pm0.02$   \\
Diboson                        & $0.10\pm0.10$     & $0.04\pm0.02$     & $0.17\pm0.09$     & $0.05\pm0.03$       & $3.1\pm1.4$         & $1.0\pm0.5$   \\
Rare                           & $0.33\pm0.18$     & $0.15\pm0.09$     & $0.18\pm0.10$     & $0.17\pm0.10$       & $0.19\pm0.11$       & $0.17\pm0.10$   \\
Fake/non-prompt leptons        & $0.5\pm0.6$       & $0.15\pm0.15$     & $3.5\pm2.4$       & $1.7\pm1.5$         & $1.6\pm1.0$         & $0.9\pm0.9$   \\
Charge-flip                    & $0.10\pm0.01$     & $0.02\pm0.01$     & $0.08\pm0.02$     & $0.08\pm0.02$       & $0.05\pm0.01$       & $0.01\pm0.01$   \\ 
\hline
Total Background               & $3.3\pm1.0$       & $1.08\pm0.32$     & $5.8\pm2.5$       & $3.8 \pm1.6$        & $6.0\pm1.8$         & $2.4\pm1.0$   \\
\hline
Observed                       & $3$               & $0$               & $4$               & $5$                 & $7$                 & $3$   \\
\hline\hline
$S_{\textrm{obs}}^{95}$        & \ral{$5.5$}               & \ral{$3.6$}               & \ral{$6.3$}               & \ral{$7.7$}               & \ral{$8.3$}               & \ral{$6.1$}  \\
$S_{\textrm{exp}}^{95}$        & \ral{$5.6_{-1.5}^{+2.2}$} & \ral{$3.9_{-0.4}^{+1.4}$} & \ral{$7.1_{-1.5}^{+2.5}$} & \ral{$6.2_{-1.5}^{+2.6}$} & \ral{$7.5_{-1.8}^{+2.6}$} & \ral{$5.3_{-1.3}^{+2.1}$} \\
$\sigma_{\textrm{vis}}$ [fb]   & \ral{$0.15$}              & \ral{$0.10$}              & \ral{$0.17$}              & \ral{$0.21$}              & \ral{$0.23$}              & \ral{$0.17$}  \\
$p_{0}$ ($\textrm{Z}$)         & \ral{$0.71$ (--)}         & \ral{$0.91$ (--)}         & \ral{$0.69$ (--)}         & \ral{$0.30\ (0.5\sigma)$} & \ral{$0.36\ (0.4\sigma)$} & \ral{$0.35\ (0.4\sigma)$}  \\
\hline 
\end{tabular}}

\vspace*{0.5cm}
\resizebox{1.\textwidth}{!}{
\begin{tabular}{|l|c|c|c|c|c|c|c|}
\hline
Signal Region 		& \textbf{Rpc3L0bS } 	& \textbf{Rpc3L0bH } 	& \textbf{Rpc3L1bS } 	& \textbf{Rpc3L1bH } 	& \textbf{Rpc2L1bS } 	& \textbf{Rpc2L1bH } 	& \textbf{Rpc3LSS1b }\\
\hline
\hline
$\ttbar W$, $\ttbar Z\gamma^*$   & $0.98\pm0.25$       	& $0.18\pm0.08$       	& $7.1\pm1.1$       	& $1.54\pm0.28$       	& $4.0\pm1.0$ 		& $4.0\pm0.9$	  	&  --     		\\
$\ttbar H$              & $0.12\pm0.08$       	& $0.03\pm0.02$       	& $1.4\pm0.7$       	& $0.25\pm0.14$       	& $1.3\pm0.7$ 		& $1.0\pm0.6$	  	& $0.22\pm0.12$    	\\
4$t$	   		& $0.02\pm0.01$	  & $0.01\pm0.01$	  & $0.7\pm0.4$ 	  & $0.28\pm0.15$	  & $0.34\pm0.17$	  & $0.54\pm0.28$	  &  -- 		  \\
Diboson                  & $8.9\pm2.9$       	& $2.6\pm0.8$       	& $1.4\pm0.5$       	& $0.48\pm0.17$       	& $0.5\pm0.3$ 		& $0.7\pm0.3$ 		&  --     		\\
Rare                     & $0.7\pm0.4$       	& $0.29\pm0.16$       	& $2.5\pm1.3$       	& $0.9\pm0.5$       	& $0.9\pm0.5$		& $1.0\pm0.6$		& $0.12\pm0.07$    	\\
Fake/non-prompt leptons  & $0.23\pm0.23$       	& $0.15\pm0.15$       	& $4.2\pm3.1$       	& $0.5\pm0.5$       	& $2.5\pm2.2$ 		& $2.3\pm1.9$  		& $0.9\pm0.7$    	\\
Charge-flip              &  --   		&  --    		&  --    		&  --     		& $0.25\pm0.04$ 	& $0.25\pm0.05$  	& $0.39\pm0.08$		\\	
\hline
Total Background         & $11.0\pm3.0\hpO$	       & $3.3\pm0.8$       	& $17\pm4\hpO$       	& $3.9\pm0.9$       	& $9.8\pm2.9$ 		& $9.8\pm2.6$  		& $1.6\pm0.8$	   	\\
\hline
Observed                 & $9$       		& $3$       		& $20$		       	& $4$       		& $14$  		&  $13$    		&  $1$			  \\
\hline\hline
$S_{\textrm{obs}}^{95}$       & \ral{$8.3$}	  	& $5.4$	   		& \ral{$14.7$}	    	& \ral{$6.1$}	     		& \ral{$13.7$}  		& \ral{$12.4$}   		& \ral{$3.9$}     		\\
$S_{\textrm{exp}}^{95}$       & \ral{$9.3_{-2.3}^{+3.1}$}	& \ral{$5.5_{-1.5}^{+2.2}$}	& \ral{$12.6_{-3.4}^{+5.1}$} 	& \ral{$5.9_{-1.8}^{+2.2}$}	& \ral{$10.0_{-2.6}^{+3.7}$}	& \ral{$9.7_{-2.6}^{+3.4}$}   & \ral{$4.0_{-0.3}^{+1.8}$}    \\
$\sigma_{\textrm{vis}}$ [fb] & \ral{$0.23$}		& \ral{$0.15$}  		& \ral{$0.41$}      		& \ral{$0.17$}      		& \ral{$0.38$}  		& \ral{$0.34$}   		& \ral{$0.11$}     		\\
$p_{0}$ ($\textrm{Z}$)        & \ral{$0.72$ (--)}  	& \ral{$0.85$ (--)}  		& \ral{$0.32\ (0.5\sigma)$}  & \ral{$0.46\ (0.1\sigma)$}  	& \ral{$0.17\ (1.0\sigma)$}  	& \ral{$0.21\ (0.8\sigma)$}	& \ral{$0.56$ (--)}	\\
\hline 
\end{tabular}}

\vspace*{0.5cm}

\vspace*{-0.01\textheight}\caption{Numbers of events observed in the signal regions compared with the expected backgrounds. 
The table shows the 95\% confidence level (CL) upper limits on the observed and expected numbers of BSM events
$S_{\textrm{obs}}^{95}$ and $S_{\textrm{exp}}^{95}$,
the 95\% CL upper limits on the visible cross-section ($\sigma_{\textrm{vis}}$), the  $p$-values ($p_{0}$), 
and the number of equivalent Gaussian standard deviations ($Z$).
Background categories with yields shown as a ``--'' 
do not contribute to a given region (e.g. charge flips in three-lepton regions).
}
\label{tab:SR_yields}
\end{center}
\end{table}

Figure~\ref{fig:Results_SystSum} summarizes the contributions from the different sources of systematic uncertainty 
to the total SM background predictions in the signal regions. The uncertainties amount to 25--45\% of the 
total background depending on the signal region, dominated by systematic uncertainties coming from the reducible background or the theory. 
The breakdown of the systematic uncertainties for each signal region is given in Table~\ref{tab:res.sys.break}

\begin{table}[htb!]
\scriptsize
\begin{center}
\vspace*{-0.035\textwidth}
\resizebox{1.\textwidth}{!}{
\begin{tabular}{|l|c|c|c|c|c|c|}
\hline
Signal Region                  & \textbf{Rpc2L2bS} & \textbf{Rpc2L2bH}    & \textbf{Rpc2Lsoft1b} &\textbf{Rpc2Lsoft2b}& \textbf{Rpc2L0bS} & \textbf{Rpc2L0bH}\\
\hline
\hline
Total background expectation    &    $3.35 $    &   $1.08 $    &    $5.78 $    &    $3.80 $    &    $6.02 $    &    $2.35 $    \\
\hline
Total statistical    &    $10.56 \%$    &   $15.67 \%$    &    $16.93 \%$    &    $15.61 \%$    &    $9.08 \%$    &    $20.87 \%$    \\
Total background systematic    &    $30.41 \%$    &   $29.97 \%$    &    $43.10 \%$    &    $41.79 \%$    &    $30.51 \%$    &    $42.39 \%$    \\
\hline\hline
Fake/non-prompt    &    $15.46 \%$    &   $0.00 \%$    &    $38.39 \%$    &    $35.75 \%$    &    $13.46 \%$    &    $32.31 \%$    \\
Charge-flip    &    $0.06 \%$    &   $0.00 \%$    &    $0.35 \%$    &    $0.53 \%$    &    $0.17 \%$    &    $0.00 \%$    \\
\hline
Jet Energy Scale    &    $15.19 \%$    &   $11.37 \%$    &    $5.27 \%$    &    $9.28 \%$    &    $17.28 \%$    &    $8.11 \%$    \\
Other Jet Unc.    &    $2.09 \%$    &   $2.71 \%$    &    $0.80 \%$    &    $0.99 \%$    &    $2.31 \%$    &    $3.42 \%$    \\
Flavor Tagging    &    $6.27 \%$    &   $5.55 \%$    &    $0.81 \%$    &    $3.96 \%$    &    $3.33 \%$    &    $3.27 \%$    \\
Electrons    &    $1.20 \%$    &   $1.72 \%$    &    $0.51 \%$    &    $0.51 \%$    &    $0.76 \%$    &    $0.74 \%$    \\
Muons    &    $0.90 \%$    &   $1.39 \%$    &    $0.35 \%$    &    $0.51 \%$    &    $0.83 \%$    &    $0.93 \%$    \\
Missing transverse momentum    &    $2.24 \%$    &   $1.68 \%$    &    $0.85 \%$    &    $1.50 \%$    &    $0.65 \%$    &    $0.54 \%$    \\
\hline
Diboson Th. Unc.    &    $1.07 \%$    &   $1.39 \%$    &    $1.07 \%$    &    $0.50 \%$    &    $17.68 \%$    &    $13.54 \%$    \\
ttV Th. Unc.    &    $7.33 \%$    &   $8.86 \%$    &    $5.01 \%$    &    $4.48 \%$    &    $4.06 \%$    &    $2.44 \%$    \\
Rare Th. Unc.    &    $15.18 \%$    &   $19.67 \%$    &    $6.28 \%$    &    $9.75 \%$    &    $3.89 \%$    &    $5.87 \%$    \\
PDF    &    $0.00 \%$    &   $0.00 \%$    &    $0.00 \%$    &    $0.00 \%$    &    $0.00 \%$    &    $0.00 \%$    \\
\hline 
\end{tabular}}

\vspace*{0.5cm}
\resizebox{1.\textwidth}{!}{
\begin{tabular}{|l|c|c|c|c|c|c|c|}
\hline
Signal Region 		& \textbf{Rpc3L0bS } 	& \textbf{Rpc3L0bH } 	& \textbf{Rpc3L1bS } 	& \textbf{Rpc3L1bH } 	& \textbf{Rpc2L1bS } 	& \textbf{Rpc2L1bH } 	& \textbf{Rpc3LSS1b }\\
\hline
\hline
       Total background expectation    &    $11.02 $    &   $3.31 $    &    $17.33 $    &    $3.90 $    &    $9.88 $    &    $9.75 $       &    $1.62 $     \\
        \hline
        Total statistical    &    $2.57 \%$    &   $6.05 \%$    &    $7.66 \%$    &    $7.70 \%$    &    $9.59 \%$    &    $6.65 \%$       &    $9.15 \%$     \\
        Total background systematic    &    $27.37 \%$    &   $25.40 \%$    &    $24.22 \%$    &    $24.02 \%$    &    $29.19 \%$    &    $26.52 \%$       &    $46.79 \%$     \\
        \hline\hline
        Fake/non-prompt    &    $1.63 \%$    &   $0.00 \%$    &    $16.50 \%$    &    $9.73 \%$    &    $19.93 \%$    &    $18.05 \%$       &    $44.45 \%$     \\
        Charge-flip    &    $0.00 \%$    &   $0.00 \%$    &    $0.00 \%$    &    $0.00 \%$    &    $0.40 \%$    &    $0.41 \%$       &    $4.32 \%$     \\
        \hline
        Jet Energy Scale    &    $9.78 \%$    &   $8.98 \%$    &    $5.54 \%$    &    $4.20 \%$    &    $11.71 \%$    &    $10.40 \%$       &    $0.02 \%$     \\
        Other Jet Unc.    &    $3.41 \%$    &   $2.55 \%$    &    $0.70 \%$    &    $2.30 \%$    &    $1.42 \%$    &    $1.46 \%$       &    $0.20 \%$     \\
        Flavor Tagging    &    $2.79 \%$    &   $2.93 \%$    &    $2.22 \%$    &    $2.82 \%$    &    $1.32 \%$    &    $1.38 \%$       &    $0.32 \%$     \\
        Electrons    &    $1.78 \%$    &   $2.16 \%$    &    $1.66 \%$    &    $2.47 \%$    &    $0.67 \%$    &    $0.89 \%$       &    $0.41 \%$     \\
        Muons    &    $1.73 \%$    &   $2.12 \%$    &    $1.25 \%$    &    $1.79 \%$    &    $0.80 \%$    &    $0.92 \%$       &    $0.41 \%$     \\
        Missing transverse momentum    &    $0.78 \%$    &   $0.53 \%$    &    $0.38 \%$    &    $0.59 \%$    &    $1.70 \%$    &    $1.06 \%$       &    $0.00 \%$     \\
        \hline
        Diboson Th. Unc.    &    $24.28 \%$    &   $21.58 \%$    &    $2.57 \%$    &    $3.78 \%$    &    $1.87 \%$    &    $2.50 \%$       &    $0.00 \%$     \\
        ttV Th. Unc.    &    $1.49 \%$    &   $1.76 \%$    &    $5.34 \%$    &    $5.56 \%$    &    $6.96 \%$    &    $5.72 \%$       &    $0.00 \%$     \\
        Rare Th. Unc.    &    $4.02 \%$    &   $5.02 \%$    &    $13.19 \%$    &    $18.11 \%$    &    $12.68 \%$    &    $13.16 \%$       &    $10.49 \%$     \\
        PDF    &    $0.00 \%$    &   $0.00 \%$    &    $0.00 \%$    &    $0.00 \%$    &    $0.00 \%$    &    $0.00 \%$       &    $0.00 \%$     \\
\hline 
\end{tabular}}

\vspace*{0.5cm}

\vspace*{-0.01\textheight}\caption{Breakdown of the dominant systematic uncertainties on background estimates in the various signal regions.
Note that the individual uncertainties can be correlated, and do not necessarily add up quadratically to 
the total background uncertainty. The percentages show the size of the uncertainty relative to the total expected background.}
\label{tab:res.sys.break}
\end{center}
\end{table}

\section{Statistical Interpretation}

In the absence of any significant deviation from the Standard Model predictions,
the results will be interpreted to 
 establish 95\% confidence intervals using the CL$_\mathrm{s}$ prescription~\cite{Read:2002hq} 
based on a profile-likelihood-ratio test~\cite{Cowan:2010js}
 described in Chapter~\ref{chap:stat}.
The interpretation will include upper limits on possible beyond the Standard Model contributions to the signal 
regions in a model independent way,
as well as exclusion limits on the masses of 
SUSY particles in the benchmark scenarios of Figure~\ref{fig:strategy.pheno.feynman}. 

\subsection{Model independent discovery and upper limits}

Table~\ref{tab:SR_yields} presents 95\% confidence level (CL) observed (expected) model-independent upper limits 
on the number of BSM events, $S_{\textrm{obs}}^{95}$ ($S_{\textrm{exp}}^{95}$), that may contribute to the signal regions. 
Normalizing these by the integrated luminosity $L$ of the data sample, they can be interpreted as upper limits on the visible 
BSM cross-section ($\sigma_{\textrm{vis}}$), defined as $\sigma_{\textrm{vis}}=\sigma_{\textrm{prod}}\times A \times\epsilon=S_{\textrm{obs}}^{95}/L$, where 
$\sigma_{\textrm{prod}}$ is the production cross-section, $A$ the acceptance and $\epsilon$ the reconstruction efficiency. The largest 
deviation of the data from the background prediction corresponds to an excess of 1.0 standard deviation in the Rpc2L1bS SR.

\subsection{Model dependent exclusion limits}

Exclusion limits at 95\% CL are set on the masses of the superpartners involved in the SUSY benchmark scenarios considered. 
Apart from the NUHM2 model, simplified models are used, corresponding to a single production mode and with 100\% branching ratio to a specific decay chain, 
with the masses of the SUSY particles not involved in the process set to very high values. 

In order to determine which signal region is used to set an exclusion limit 
on a particular model, the expected $CL_s$ value is computed for each signal 
region at a given point in the signal parameter space. The signal 
region with the smallest expected  $CL_s$ value (more disagreement with data 
under the signal hypothesis) is used to set an exclusion limit on the model.
An example on what signal region is performing best at a given model using the 
decay is shown in Figure~\ref{fig:res.best_gtt}.

\begin{figure}[htb!]
\centering
\includegraphics[width=.75\textwidth]{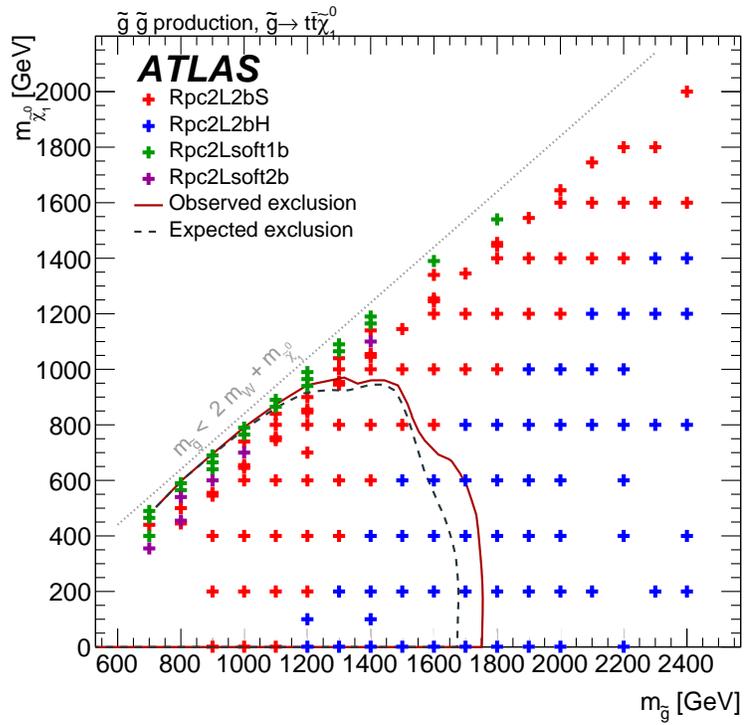}
\caption{Illustration of the best expected signal region per signal grid point for the 
$\gluino \to t\bar t\neut$ (Figure~\ref{fig:strategy.pheno.feynman_gtt}) model. 
This mapping is used for the final combined exclusion limits.}
\label{fig:res.best_gtt}
\end{figure}

Figures~\ref{fig:Results_Limits_RPC1} -- \ref{fig:Results_Limits_NUHM2} show the exclusion limits in all 
the models considered in Figure~\ref{fig:strategy.pheno.feynman} and the NUHM2 model. The assumptions about the decay chain considered for the different SUSY particles are 
stated above each figure. For each region of the signal parameter space, the SR with the best expected sensitivity is chosen.

Each one of the Figures~\ref{fig:Results_Limits_RPC1} -- \ref{fig:Results_Limits_RPC2} contain:
\begin{itemize}
\item Observed limit (thick solid red line): all uncertainties are included in the fit as nuisance 
parameters, with the exception of the theoretical signal uncertainties (PDF, scales) on the 
inclusive cross section.
\item Expected limit (less thick long-dashed black line): all uncertainties are included in the fit 
as nuisance parameters, with the exception of the theoretical signal uncertainties (PDF, scales) 
on the inclusive cross section. 
\item $\pm 1\sigma$ lines around the observed limit (thin dark-red dotted): re-run limit calculation
 while increasing or decreasing the signal cross section by the theoretical signal uncertainties 
  (PDF, scales).
\item $\pm 1\sigma$  band around expected limit (yellow band): represents the  $\pm 1\sigma$ 
uncertainty from the fit.
\end{itemize}

\begin{figure}[htb!]
\centering
\begin{subfigure}[t]{0.49\textwidth}\includegraphics[width=\textwidth]{Gtt_SRbest}\caption{Rpc2L2bS/H, Rpc2Lsoft1b/2b}\label{fig:limits_feynman_gtt}\end{subfigure}
\begin{subfigure}[t]{0.49\textwidth}\includegraphics[width=\textwidth]{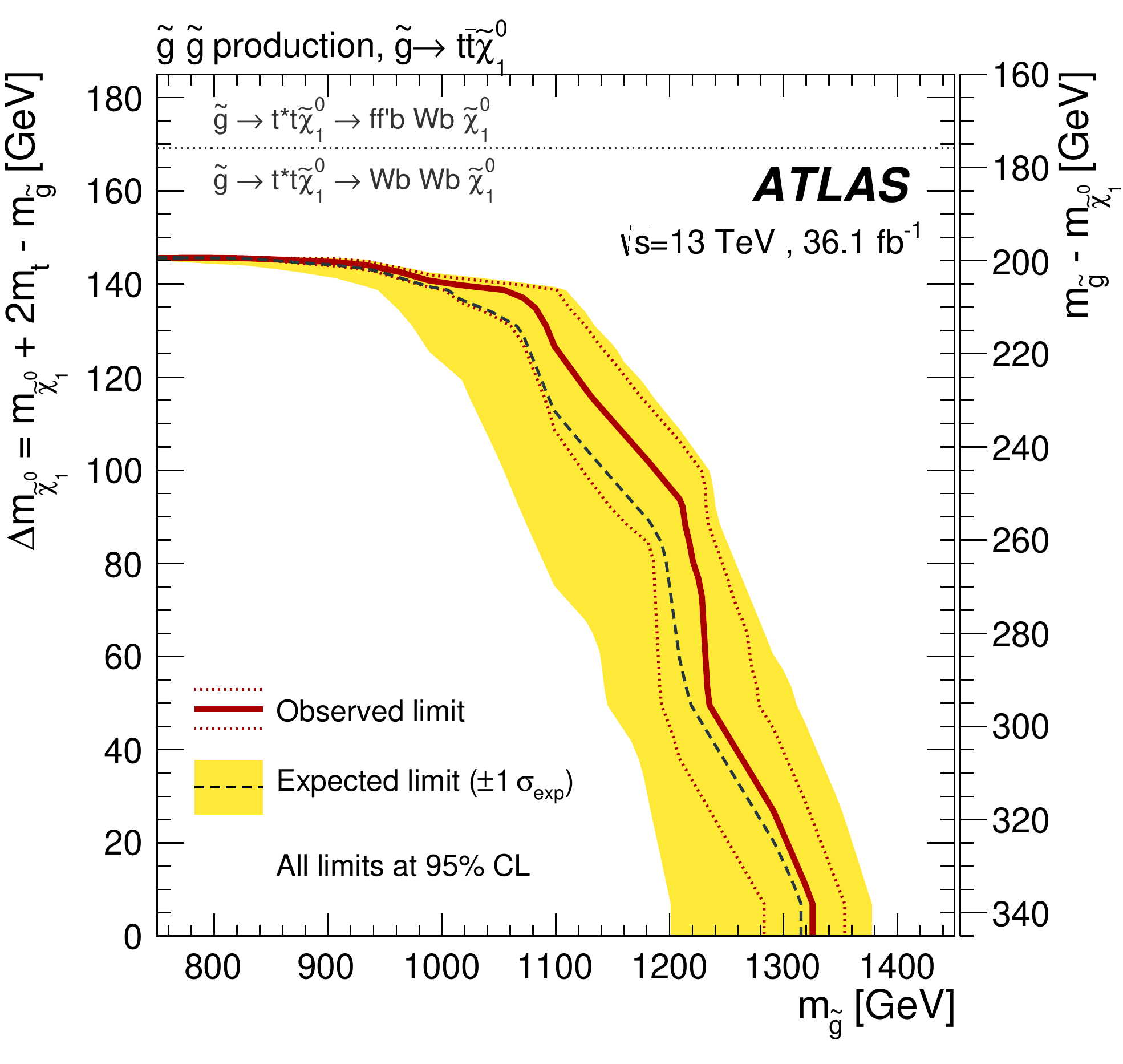}\caption{Rpc2Lsoft1b, Rpc2Lsoft2b}\label{fig:limits_feynman_gttOffshell}\end{subfigure}
\caption{Observed and expected exclusion limits on the $\tilde{g}$ and \ninoone masses for (a) the model in Figure~\ref{fig:strategy.pheno.feynman_gtt} 
and (b) the model in Figure~\ref{fig:strategy.pheno.feynman_gttOffshell}. 
Figure (b) is a zoomed version of Figure (a) in the mass-parameter space where there is at least one top-quark off-shell decay.
All limits are computed at 95\% CL. The dotted lines around the observed
limit illustrate the change in the observed limit as the nominal signal cross-section is scaled up and down
by the theoretical uncertainty. The contours of the band around the expected 
limit are the $\pm$1$\sigma$ results, 
including all uncertainties except the theoretical ones in the signal cross-section. 
}
\label{fig:Results_Limits_RPC1} 
\end{figure}

\begin{figure}[htb!]
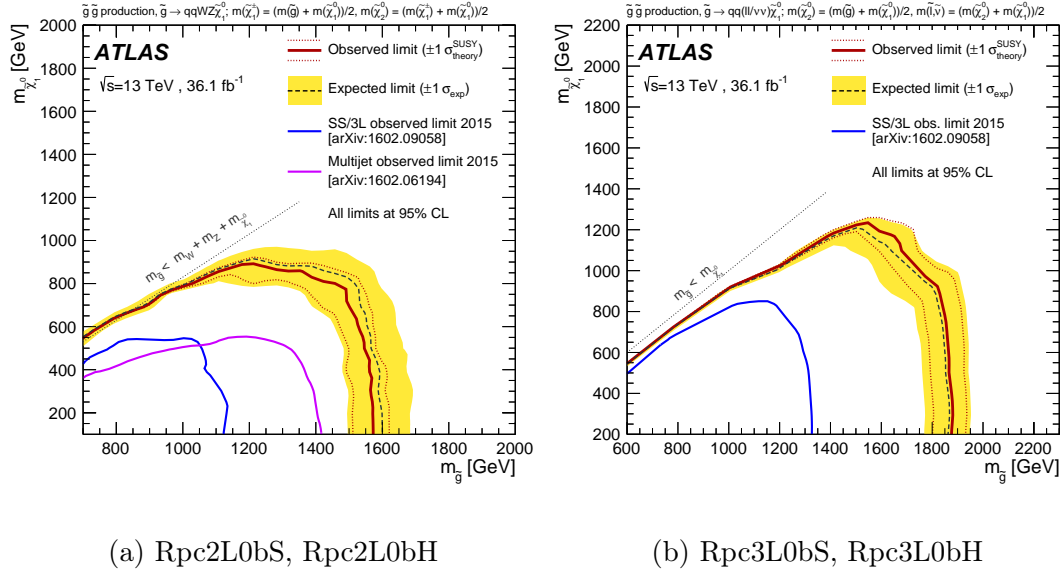

\centering
\begin{subfigure}[t]{0.49\textwidth}\includegraphics[width=\textwidth]{2stepWZ_SRbest}\caption{Rpc2L0bS, Rpc2L0bH}\label{fig:limits_feynman_gg2WZ}\end{subfigure}
\begin{subfigure}[t]{0.49\textwidth}\includegraphics[width=\textwidth]{GSL_SRbest.pdf}\caption{Rpc3L0bS, Rpc3L0bH}\label{fig:limits_feynman_gg2sl}\end{subfigure}
\caption{Observed and expected exclusion limits on the $\tilde{g}$ and \ninoone masses 
for (a) the model in Figure~\ref{fig:strategy.pheno.feynman_gg2WZ} 
and (b) the model in Figure~\ref{fig:strategy.pheno.feynman_gg2sl}. 
All limits are computed at 95\% CL. The dotted lines around the observed
limit illustrate the change in the observed limit as the nominal signal cross-section is scaled up and down
by the theoretical uncertainty. The contours of the band around the expected 
limit are the $\pm$1$\sigma$ results,
including all uncertainties except the theoretical ones in the signal cross-section.}
\label{fig:Results_Limits_RPC2} 
\end{figure} 

\begin{figure}[htb!]
\centering
\begin{subfigure}[t]{0.49\textwidth}\includegraphics[width=\textwidth]{Btt_SRbest.pdf}\caption{Rpc2L1bS, Rpc2L1bH}\label{fig:limits_feynman_b1b1}\end{subfigure}
\begin{subfigure}[t]{0.49\textwidth}\includegraphics[width=\textwidth]{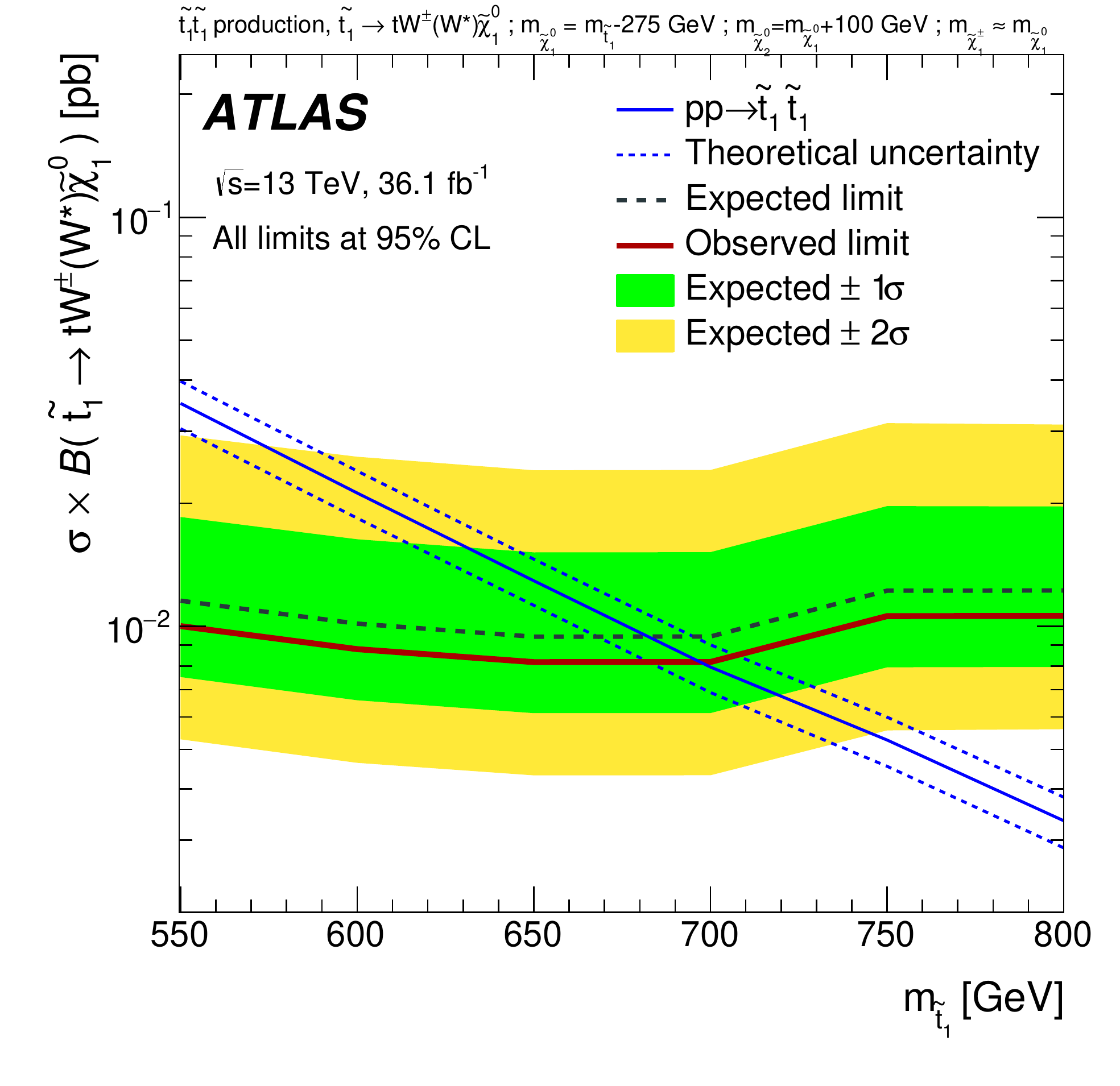}\caption{Rpc3LSS1b}\label{fig:limits_feynman_t1t1}\end{subfigure}
\caption{
Observed and expected exclusion limits on the  \sbottomone, \stopone, and \ninoone masses 
for (a) the model in Figure~\ref{fig:strategy.pheno.feynman_b1b1}
and (b) the model in Figure~\ref{fig:strategy.pheno.feynman_t1t1}. 
The two models are complementary where the one-dimensional change in stop masses (b) is along the grayed diagonal in the sbottom plot (b).
All limits are computed at 95\% CL. The dotted lines around the observed
limit illustrate the change in the observed limit as the nominal signal cross-section is scaled up and down
by the theoretical uncertainty. The contours of the band around the expected 
limit are the $\pm$1$\sigma$ results  ($\pm$2$\sigma$ is also considered in (b)),
including all uncertainties except the theoretical ones in the signal cross-section.
}
\label{fig:Results_Limits_NUHM2} 
\end{figure} 

\begin{figure}[htb!]
\centering
\includegraphics[width=0.7\textwidth]{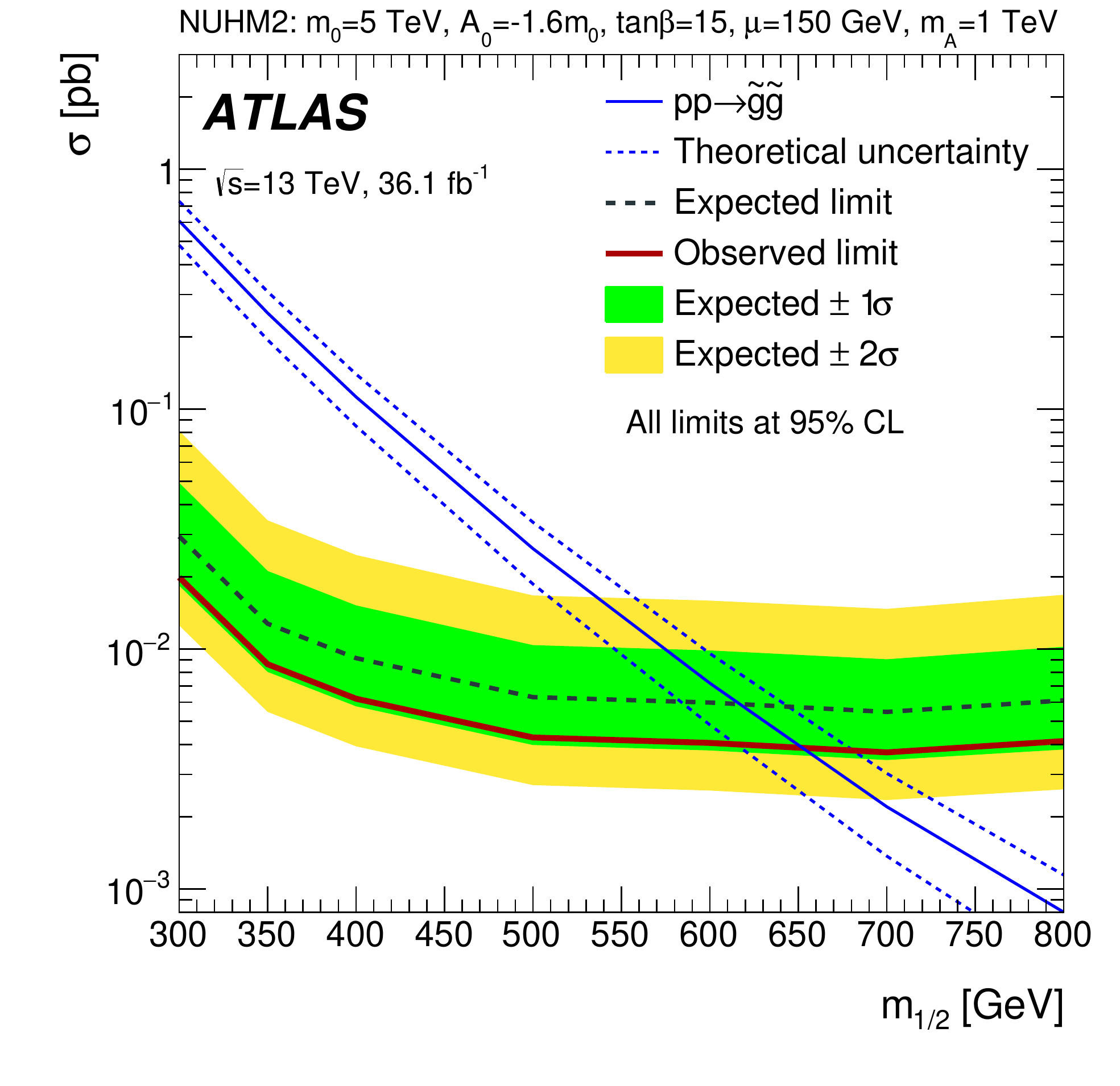}\label{fig:limits_feynman_nuhm2}
\caption{Observed and expected exclusion limits as a function of $m_{1/2}$ in the NUHM2 model~\cite{Ellis:2002iu,Ellis:2002wv}.
The signal region Rpc2L2bH is used to obtain the limits. 
The contours of the green (yellow) band around the expected limit are the $\pm$1$\sigma$ ($\pm$2$\sigma$) results, including all uncertainties. The limits are computed at 95\% CL.}
\label{fig:Results_Limits_NUHM2} 
\end{figure} 

The limits set are compared with the existing limits set by other ATLAS SUSY 
searches~\cite{paperSS3L,Aad:2016jxj}. For the models shown in Figure~\ref{fig:Results_Limits_RPC1} -- \ref{fig:Results_Limits_RPC2}, 
the mass limits on gluinos and bottom squarks are up to 400 \GeV~higher than the previous limits, reflecting the improvements 
in the signal region definitions as well as the increase in integrated luminosity. Gluinos with masses up to 1.75 \TeV~
are excluded in scenarios with a light $\ninoone$ in Figure~\ref{fig:limits_feynman_gtt}. This limit is extended to 1.87 \TeV~ when 
$\ninotwo$ and slepton masses are in between the gluino and the $\ninoone$ masses (Figure~\ref{fig:limits_feynman_gg2sl}). More generally, gluino masses 
below 1.57 \TeV~and bottom squarks with masses below 700 \GeV
are excluded in models with a massless LSP. The ``compressed'' regions, where SUSY particle masses are close to each other, are also better covered 
and LSP masses up to 1200 and 250 \GeV~are excluded in the gluino and bottom squark pair-production models, respectively. Of particular
interest is the observed exclusion of models producing gluino pairs with an off-shell top quark in the decay (Figure~\ref{fig:strategy.pheno.feynman_gttOffshell}), 
see Figure~\ref{fig:limits_feynman_gtt}. In this case, models are excluded for mass differences between the gluino and neutralino of 205 GeV~(only ~35 GeV
larger than the minimum mass difference for decays into two on-shell $W$ bosons and two $b$-quarks) for a gluino mass below 0.9
TeV. The Rpc3LSS1b SR allows the exclusion of top squarks with masses below 700 \GeV~when the top squark decays to a top quark and a cascade of electroweakinos 
$\ninotwo \to \chinoonepm W^{\mp} \to W^{*} W^{\mp} \ninoone$ (see Figure~\ref{fig:limits_feynman_t1t1} for the conditions 
on the sparticle masses).

Finally, in the NUHM2 model with low fine-tuning, values of the parameter $m_{1/2}$ below 615 \GeV~are excluded, 
corresponding to gluino masses below 1500 \GeV~(Figure~\ref{fig:Results_Limits_NUHM2}).

%% file: texfiles/concl.tex
A search for supersymmetry in events with two same-sign leptons or at least three leptons, multiple jets, 
$b$-jets and large $\met$ and/or large $\meff$ was presented in this dissertation. 
The analysis is performed with proton--proton collision data at $\sqrt{s}=13$ \TeV 
collected in 2015 and 2016 with the ATLAS detector at the Large Hadron Collider 
corresponding to an integrated luminosity of 36.1 fb$^{-1}$. 
With no significant excess over the Standard Model prediction observed,
results are interpreted in the framework of simplified models featuring gluino 
and squark production in $R$-parity-conserving (RPC) scenarios. Lower limits on particle 
masses are derived at 95\% confidence level. 
In the $\gluino\gluino$ simplified RPC models considered, gluinos with masses up to 1.87 \TeV~
are excluded in scenarios with a light $\ninoone$. RPC models with bottom squark masses below 700~GeV
are also excluded in a $\sbottomone\sbottomonebar$ simplified model with $\sbottomone\to tW^-\ninoone$ and a light $\ninoone$. 
All models with gluino masses below 1.3 \TeV~are excluded, greatly extending the previous exclusion limits.
Model-independent limits on the cross-section of a possible signal contribution to the signal regions are set.

%% file: texfiles/app.aux.tex
\section{Signal region with best exclusion}
\label{app:aux.bestSR}

\begin{figure}[htb!]
\centering
\begin{subfigure}[t]{0.49\textwidth}\includegraphics[width=\textwidth]{best_Rpc2L2b}\caption{}\label{fig:best_Rpc2L2b}\end{subfigure}
\begin{subfigure}[t]{0.49\textwidth}\includegraphics[width=\textwidth]{best_Rpc2L0b}\caption{}\label{fig:best_Rpc2L0b}\end{subfigure}
\caption{Illustration of the best expected signal region per signal grid point for the (a)  
$\gluino\to q\bar q (\ell\ell/\nu\nu)\tilde\chi^0_1$ and (b) $\gluino\to q\bar q' WZ\tilde\chi^0_1$ models. 
This mapping is used for the final combined exclusion limits.}
\label{fig:best_SR1}
\end{figure}

\begin{figure}[htb!]
\centering
\begin{subfigure}[t]{0.49\textwidth}\includegraphics[width=\textwidth]{best_Rpc3L0b}\caption{}\label{fig:best_Rpc3L0b}\end{subfigure}
\begin{subfigure}[t]{0.49\textwidth}\includegraphics[width=\textwidth]{best_Rpc2L1b}\caption{}\label{fig:best_Rpc2L1b}\end{subfigure}
\caption{Illustration of the best expected signal region per signal grid point for the 
(a) $\gluino\to q\bar q' \ell/\nu \ell/\nu \tilde\chi^0_1$ and (b) $\sbottom\to t W \tilde\chi^0_1$ models. 
This mapping is used for the final combined exclusion limits.}
\label{fig:best_SR2}
\end{figure}

\newpage

\section{Upper limit on cross section}
\label{app:aux.ULcs}

\begin{figure}[htb!]
\centering
\begin{subfigure}[t]{0.49\textwidth}\includegraphics[width=\textwidth]{GreyNumbers/Gtt_SRbest}\caption{Rpc2L2bS/H, Rpc2Lsoft1b/2b}\label{fig:GN_limits_feynman_gtt}\end{subfigure}
\begin{subfigure}[t]{0.49\textwidth}\includegraphics[width=\textwidth]{GreyNumbers/2stepWZ_SRbest}\caption{Rpc2L0bS, Rpc2L0bH}\label{fig:GN_limits_feynman_gg2WZ}\end{subfigure}

\caption{Observed and expected exclusion limits on the $\tilde{g}$ and \ninoone masses 
in the context of RPC SUSY scenarios with simplified mass spectra. The signal regions used to obtain the limits are specified in the subtitle of each scenario. All limits are computed at 95\% CL. 
The grey numbers show 95\% CL upper limits on production cross-sections (in fb) obtained using the signal efficiency and acceptance specific to each model.}
\label{fig:Results_Limits_RPC_GN} 
\end{figure} 

\begin{figure}[htb!]
\centering
\begin{subfigure}[t]{0.49\textwidth}\includegraphics[width=\textwidth]{GreyNumbers/GSL_SRbest.pdf}\caption{Rpc3L0bS, Rpc3L0bH}\label{fig:GN_limits_feynman_gg2sl}\end{subfigure}
\begin{subfigure}[t]{0.49\textwidth}\includegraphics[width=\textwidth]{GreyNumbers/Btt_SRbest.pdf}\caption{Rpc2L1bS, Rpc2L1bH}\label{fig:GN_limits_feynman_b1b1}\end{subfigure}

\caption{Observed and expected exclusion limits on the $\tilde{g}$, \sbottomone, and \ninoone masses 
in the context of RPC SUSY scenarios with simplified mass spectra. The signal regions used to obtain the limits are specified in the subtitle of each scenario. All limits are computed at 95\% CL. 
The grey numbers show 95\% CL upper limits on production cross-sections (in fb) obtained using the signal efficiency and acceptance specific to each model.}
\label{fig:Results_Limits_RPC_GN} 
\end{figure}

\section{Signal region cutflow}
\label{app:aux.SRcut}

\begin{table}[htb!]\centering\def\arraystretch{1.2}\begin{tabular}{|l|c|}\hline
   \multicolumn{2}{|l|}{Rpc2L2bS,\quad$\gluino\gluino$ production,\quad$\gluino\to \ttbar\ninoone$}\\
   \multicolumn{2}{|l|}{$m_{\gluino}= 1.5 \TeV$, $m_{\ninoone}= 800 \GeV$}\\\hline
   MC events generated  & 98000 \\\hline
   Expected for 36.1 \ifb  & $5.1\times 10^2$ \\
   $\geq 2$ SS leptons ($\pt>20 \GeV$)  & $19.96 \pm 0.35$ \\
   Trigger  & $19.17 \pm 0.35$ \\
   $\ge 2$ $b$-jets ($\pt>20 \GeV$)  & $16.10 \pm 0.32$ \\
   $\ge 6$ jets ($\pt>25 \GeV$)  & $13.11 \pm 0.28$ \\
   $\met>200 \GeV$  & $10.17 \pm 0.26$ \\
   $\meff>0.6 \TeV$  & $10.17 \pm 0.26$ \\
   $\met>0.25\times\meff$  & $5.94 \pm 0.20$ \\
\hline\end{tabular}
\caption{Number of signal events at different stages of the Rpc2L2bS signal region selection. 
Only statistical uncertainties are shown.}
\end{table}

\begin{table}[htb!]\centering\def\arraystretch{1.2}\begin{tabular}{|l|c|}\hline
   \multicolumn{2}{|l|}{Rpc2L2bH,\quad$\gluino\gluino$ production,\quad$\gluino\to \ttbar\ninoone$}\\
   \multicolumn{2}{|l|}{$m_{\gluino}=1.7 \TeV$, $m_{\ninoone}=200 \GeV$}\\\hline
   MC events generated  & 98000 \\\hline
   Expected for 36.1 \ifb  & $1.7\times 10^2$ \\
   $\geq 2$ SS leptons ($\pt>20 \GeV$)  & $7.32 \pm 0.13$ \\
   Trigger  & $7.19 \pm 0.13$ \\
   $\ge 2$ $b$-jets ($\pt>20 \GeV$)  & $5.81 \pm 0.11$ \\
   $\ge 6$ jets ($\pt>40 \GeV$)  & $4.92 \pm 0.11$ \\
   $\meff>1.8 \TeV$  & $3.93 \pm 0.09$ \\
   $\met>0.15\times\meff$  & $3.12 \pm 0.08$ \\
\hline\end{tabular}
\caption{Number of signal events at different stages of the Rpc2L2bH signal region selection. 
Only statistical uncertainties are shown.}\end{table}

\begin{table}[htb!]\centering\def\arraystretch{1.2}\begin{tabular}{|l|c|}\hline
   \multicolumn{2}{|l|}{Rpc2Lsoft1b,\quad$\gluino\gluino$ production,\quad$\gluino\to tWb\ninoone$}\\
   \multicolumn{2}{|l|}{$m_{\gluino}=1.2 \TeV$, $m_{\ninoone}=940 \GeV$}\\\hline
   MC events generated  & 50000 \\\hline
   Expected for 36.1 \ifb  & $3.1\times 10^3$ \\
   $\geq 2$ SS leptons ($100>\pt>20,10$~GeV)  & $101.9 \pm 2.7$ \\
   Trigger  & $89.3 \pm 2.5$ \\
   $\ge 1$ $b$-jet ($\pt>20 \GeV$)  & $75.1 \pm 2.3$ \\
   $\ge 6$ jets ($\pt>25 \GeV$)  & $31.5 \pm 1.5$ \\
   $\met>100 \GeV$  & $23.0 \pm 1.3$ \\
   $\met>0.3\times\meff$  & $6.5 \pm 0.7$ \\
\hline\end{tabular}
\caption{Number of signal events at different stages of the Rpc2Lsoft1b signal region selection. 
Only statistical uncertainties are shown.}\end{table}

\begin{table}[htb!]\centering\def\arraystretch{1.2}\begin{tabular}{|l|c|}\hline
   \multicolumn{2}{|l|}{Rpc2Lsoft2b,\quad$\gluino\gluino$ production,\quad$\gluino\to tWb\ninoone$}\\
   \multicolumn{2}{|l|}{$m_{\gluino}=1.2 \TeV$, $m_{\ninoone}=900 \GeV$}\\\hline
   MC events generated  & 50000 \\\hline
   Expected for 36.1 \ifb  & $3.1\times 10^3$ \\
   $\geq 2$ SS leptons ($100>\pt>20,10$~GeV)  & $91.8 \pm 2.6$ \\
   Trigger  & $79.7 \pm 2.4$ \\
   $\ge 2$ $b$-jets ($\pt>20 \GeV$)  & $41.3 \pm 1.7$ \\
   $\ge 6$ jets ($\pt>25 \GeV$)  & $21.4 \pm 1.2$ \\
   $\met>200 \GeV$  & $8.7 \pm 0.7$ \\
   $\meff>0.6 \TeV$  & $8.7 \pm 0.7$ \\
   $\met>0.25\times\meff$  & $6.7 \pm 0.6$ \\
\hline\end{tabular}
\caption{Number of signal events at different stages of the Rpc2Lsoft2b signal region selection. 
Only statistical uncertainties are shown.}\end{table}

\begin{table}[htb!]\centering\def\arraystretch{1.2}\begin{tabular}{|l|c|}\hline
   \multicolumn{2}{|l|}{Rpc2L0bS,\quad$\gluino\gluino$ production,\quad$\gluino\to q\bar q^{'}WZ\ninoone$}\\
   \multicolumn{2}{|l|}{$m_{\gluino}=1.2 \TeV$, $(m_{\chargino} - 150) = (m_{\tilde\chi_2^0} - 75) = m_{\ninoone}=900 \GeV$}\\\hline
   MC events generated  & 19000 \\\hline
   Expected for 36.1 \ifb  & $3.1\times 10^3$ \\
   $\geq 2$ SS leptons ($\pt>20 \GeV$)  & $64 \pm 4$ \\
   Trigger  & $58.6 \pm 3.3$ \\
   no $b$-jet ($\pt>20 \GeV$)  & $46.3 \pm 3.0$ \\
   $\ge 6$ jets ($\pt>25 \GeV$)  & $26.6 \pm 2.4$ \\
   $\met>150 \GeV$  & $16.3 \pm 2.0$ \\
   $\met>0.25\times\meff$  & $9.0 \pm 1.3$ \\
\hline\end{tabular}
\caption{Number of signal events at different stages of the Rpc2L0bS signal region selection. 
Only statistical uncertainties are shown.}\end{table}

\begin{table}[htb!]\centering\def\arraystretch{1.2}\begin{tabular}{|l|c|}\hline
   \multicolumn{2}{|l|}{Rpc2L0bH,\quad$\gluino\gluino$ production,\quad$\gluino\to q\bar q^{'}WZ\ninoone$}\\
   \multicolumn{2}{|l|}{$m_{\gluino}=1.6 \TeV$, $(m_{\chargino} - 750) = (m_{\tilde\chi_2^0} - 375) = m_{\ninoone}=100 \GeV$}\\\hline
   MC events generated  & 20000 \\\hline
   Expected for 36.1 \ifb  & $2.9\times 10^2$ \\
   $\geq 2$ SS leptons ($\pt>20 \GeV$)  & $12.8 \pm 0.5$ \\
   Trigger  & $12.5 \pm 0.5$ \\
   no $b$-jet ($\pt>20 \GeV$)  & $8.5 \pm 0.4$ \\
   $\ge 6$ jets ($\pt>40 \GeV$)  & $7.12 \pm 0.35$ \\
   $\met>250 \GeV$  & $5.13 \pm 0.29$ \\
   $\meff>0.9 \TeV$  & $5.13 \pm 0.29$ \\
\hline\end{tabular}
\caption{Number of signal events at different stages of the Rpc2L0bH signal region selection. 
Only statistical uncertainties are shown.}\end{table}

\begin{table}[htb!]\centering\def\arraystretch{1.2}\begin{tabular}{|l|c|}\hline
   \multicolumn{2}{|l|}{Rpc3L0bS,\quad$\gluino\gluino$ production,\quad$\gluino\to q\bar q(\tilde\ell\ell/\tilde\nu\nu)$}\\
   \multicolumn{2}{|l|}{$m_{\gluino}=1.4 \TeV$, $(m_{\tilde\chi_2^0} - 150) = (m_{\tilde\ell, \tilde\nu} - 75) = m_{\ninoone}=1100 \GeV$}\\\hline
   MC events generated  & 20000 \\\hline
   Expected for 36.1 \ifb  & $9.1\times 10^2$ \\
   $\geq 3$ leptons ($\pt>20,20,10$~GeV)  & $76.9 \pm 2.1$ \\
   Trigger  & $76.0 \pm 2.0$ \\
   no $b$-jet ($\pt>20 \GeV$)  & $67.5 \pm 1.9$ \\
   $\ge 4$ jets ($\pt>40 \GeV$)  & $31.6 \pm 1.3$ \\
   $\met>200 \GeV$  & $17.1 \pm 1.0$ \\
   $\meff>0.6 \TeV$  & $17.1 \pm 1.0$ \\
\hline\end{tabular}
\caption{Number of signal events at different stages of the Rpc3L0bS signal region selection. 
Only statistical uncertainties are shown.}\end{table}

\begin{table}[htb!]\centering\def\arraystretch{1.2}\begin{tabular}{|l|c|}\hline
   \multicolumn{2}{|l|}{Rpc3L0bH,\quad$\gluino\gluino$ production,\quad$\gluino\to q\bar q(\tilde\ell\ell/\tilde\nu\nu)$}\\
   \multicolumn{2}{|l|}{$m_{\gluino}=1.8 \TeV$, $(m_{\tilde\chi_2^0} - 850) = (m_{\tilde\ell, \tilde\nu} - 375) = m_{\ninoone}=100 \GeV$}\\\hline
   MC events generated  & 20000 \\\hline
   Expected for 36.1 \ifb  & $1.0\times 10^{2}$ \\
   $\geq 3$ leptons ($\pt>20,20,10$~GeV)  & $9.98 \pm 0.25$ \\
   Trigger  & $9.94 \pm 0.25$ \\
   no $b$-jet ($\pt>20 \GeV$)  & $8.44 \pm 0.23$ \\
   $\ge 4$ jets ($\pt>40 \GeV$)  & $7.79 \pm 0.22$ \\
   $\met>200 \GeV$  & $6.58 \pm 0.21$ \\
   $\meff>1.6 \TeV$  & $6.56 \pm 0.21$ \\
\hline\end{tabular}
\caption{Number of signal events at different stages of the Rpc3L0bH signal region selection. 
Only statistical uncertainties are shown.}\end{table}

\begin{table}[htb!]\centering\def\arraystretch{1.2}\begin{tabular}{|l|c|}\hline
   \multicolumn{2}{|l|}{Rpc2L1bS,\quad$\sbottomone\sbottomonebar$ production,\quad$\sbottomone\to t\tilde\chi_1^{-}\to t W^-
\ninoone$}\\
   \multicolumn{2}{|l|}{$m_{\sbottomone}=600 \GeV$, $m_{\chargino} = 350 \GeV$, $m_{\ninoone}=250 \GeV$}\\\hline
   MC events generated  & 10000 \\\hline
   Expected for 36.1 \ifb  & $6.3\times 10^3$ \\
   $\geq 2$ SS leptons ($\pt>20 \GeV$)  & $221 \pm 4$ \\
   Trigger  & $201 \pm 4$ \\
   $\ge 1$ $b$-jet ($\pt>20 \GeV$)  & $173 \pm 4$ \\
   $\ge 6$ jets ($\pt>25 \GeV$)  & $66.3 \pm 2.2$ \\
   $\met>150 \GeV$  & $36.5 \pm 1.7$ \\
   $\meff>0.6 \TeV$  & $36.1 \pm 1.7$ \\
   $\met>0.25\times\meff$  & $15.1 \pm 1.1$ \\
\hline\end{tabular}
\caption{Number of signal events at different stages of the Rpc2L1bS signal region selection. 
Only statistical uncertainties are shown.}\end{table}

\begin{table}[htb!]\centering\def\arraystretch{1.2}\begin{tabular}{|l|c|}\hline
   \multicolumn{2}{|l|}{Rpc2L1bH,\quad$\sbottomone\sbottomonebar$ production,\quad$\sbottomone\to t\tilde\chi_1^{-}$}\\
   \multicolumn{2}{|l|}{$m_{\sbottomone}=750 \GeV$, $m_{\chargino} = 200 \GeV$, $m_{\ninoone}=100 \GeV$}\\\hline
   MC events generated  & 10000 \\\hline
   Expected for 36.1 \ifb  & $1.6\times 10^3$ \\
   $\geq 2$ SS leptons ($\pt>20 \GeV$)  & $71.1 \pm 1.2$ \\
   Trigger  & $66.4 \pm 1.2$ \\
   $\ge 1$ $b$-jet ($\pt>20 \GeV$)  & $56.6 \pm 1.1$ \\
   $\ge 6$ jets ($\pt>25 \GeV$)  & $27.7 \pm 0.7$ \\
   $\met>250 \GeV$  & $12.5 \pm 0.5$ \\
   $\met>0.2\times\meff$  & $9.5 \pm 0.4$ \\
\hline\end{tabular}
\caption{Number of signal events at different stages of the Rpc2L1bH signal region selection. 
Only statistical uncertainties are shown.}\end{table}

\begin{table}[htb!]\centering\def\arraystretch{1.2}\begin{tabular}{|l|c|}\hline
   \multicolumn{2}{|l|}{Rpc3LSS1b,\quad$\stopone\stoponebar$ production,\quad$\stopone\to t\ninotwo\to \tilde t W^\pm \chi_1^\mp$}\\
   \multicolumn{2}{|l|}{$m_{\stopone}=700 \GeV$, $m_{\ninotwo}=525 \GeV$, $m_{\chargino}\approx m_{\ninoone}=425 \GeV$}\\\hline
   MC events generated  & 5000 \\\hline
   Expected for 36.1 \ifb  & $2.4\times 10^3$ \\
   $\geq 3$ SS leptons ($\pt>20,20,10$~GeV), $Z\to e^\pm e^\pm$ veto  & $4.6 \pm 0.5$ \\
   Trigger  & $4.5 \pm 0.5$ \\
   $\ge 1$ $b$-jet ($\pt>20 \GeV$)  & $3.6 \pm 0.4$ \\
\hline\end{tabular}
\caption{Number of signal events at different stages of the Rpc3LSS1b signal region selection. 
Only statistical uncertainties are shown.}\end{table}

\section{Acceptance and Efficiency}
\label{app:aux.AccEff}
\begin{figure}[htb!]
\centering
\begin{subfigure}[t]{0.49\textwidth}\includegraphics[width=\textwidth]{EffAcc/acceptance_2StepRpc2L0bH}\caption{Rpc2L0bH acceptance}\end{subfigure}
\begin{subfigure}[t]{0.49\textwidth}\includegraphics[width=\textwidth]{EffAcc/efficiency_2StepRpc2L0bH}\caption{Rpc2L0bH efficiency}\end{subfigure}
\begin{subfigure}[t]{0.49\textwidth}\includegraphics[width=\textwidth]{EffAcc/acceptance_2StepRpc2L0bS}\caption{Rpc2L0bS acceptance}\end{subfigure}
\begin{subfigure}[t]{0.49\textwidth}\includegraphics[width=\textwidth]{EffAcc/efficiency_2StepRpc2L0bS}\caption{Rpc2L0bS efficiency}\end{subfigure}
\caption{Signal acceptance (a,c) and reconstruction efficiency (b,d) 
for simplified models of $\gluino\gluino$ production with $\gluino\to q\bar q^{'}WZ\ninoone$ decays, 
in the signal regions Rpc2L0bH (a,b) and Rpc2L0bS (c,d).}
\end{figure}

\begin{figure}[htb!]
\centering
\begin{subfigure}[t]{0.49\textwidth}\includegraphics[width=\textwidth]{EffAcc/acceptance_GSLRpc3L0bH}\caption{Rpc3L0bH acceptance}\end{subfigure}
\begin{subfigure}[t]{0.49\textwidth}\includegraphics[width=\textwidth]{EffAcc/efficiency_GSLRpc3L0bH}\caption{Rpc3L0bH efficiency}\end{subfigure}
\begin{subfigure}[t]{0.49\textwidth}\includegraphics[width=\textwidth]{EffAcc/acceptance_GSLRpc3L0bS}\caption{Rpc3L0bS acceptance}\end{subfigure}
\begin{subfigure}[t]{0.49\textwidth}\includegraphics[width=\textwidth]{EffAcc/efficiency_GSLRpc3L0bS}\caption{Rpc3L0bS efficiency}\end{subfigure}
\caption{Signal acceptance (a,c) and reconstruction efficiency (b,d) 
for simplified models of $\gluino\gluino$ production with $\gluino\to q\bar q(\ell\ell/\nu\nu)\ninoone$ decays, 
in the signal regions Rpc3L0bH (a,b) and Rpc3L0bS (c,d).}
\end{figure}

\begin{figure}[htb!]
\centering
\begin{subfigure}[t]{0.49\textwidth}\includegraphics[width=\textwidth]{EffAcc/acceptance_GttRpc2L2bH}\caption{Rpc2L2bH acceptance}\end{subfigure}
\begin{subfigure}[t]{0.49\textwidth}\includegraphics[width=\textwidth]{EffAcc/efficiency_GttRpc2L2bH}\caption{Rpc2L2bH efficiency}\end{subfigure}
\begin{subfigure}[t]{0.49\textwidth}\includegraphics[width=\textwidth]{EffAcc/acceptance_GttRpc2L2bS}\caption{Rpc2L2bS acceptance}\end{subfigure}
\begin{subfigure}[t]{0.49\textwidth}\includegraphics[width=\textwidth]{EffAcc/efficiency_GttRpc2L2bS}\caption{Rpc2L2bS efficiency}\end{subfigure}
\caption{Signal acceptance (a,c) and reconstruction efficiency (b,d) 
for simplified models of $\gluino\gluino$ production with $\gluino\to t\bar t\ninoone$ decays, 
in the signal regions Rpc2L2bH (a,b) and Rpc2L2bS (c,d).}
\end{figure}

\begin{figure}[htb!]
\centering
\begin{subfigure}[t]{0.49\textwidth}\includegraphics[width=\textwidth]{EffAcc/acceptance_GttRpc2Lsoft1b}\caption{Rpc2Lsoft1b acceptance}\end{subfigure}
\begin{subfigure}[t]{0.49\textwidth}\includegraphics[width=\textwidth]{EffAcc/efficiency_GttRpc2Lsoft1b}\caption{Rpc2Lsoft1b efficiency}\end{subfigure}
\begin{subfigure}[t]{0.49\textwidth}\includegraphics[width=\textwidth]{EffAcc/acceptance_GttRpc2Lsoft2b}\caption{Rpc2Lsoft2b acceptance}\end{subfigure}
\begin{subfigure}[t]{0.49\textwidth}\includegraphics[width=\textwidth]{EffAcc/efficiency_GttRpc2Lsoft2b}\caption{Rpc2Lsoft2b efficiency}\end{subfigure}
\caption{Signal acceptance (a,c) and reconstruction efficiency (b,d) 
for simplified models of $\gluino\gluino$ production with $\gluino\to t\bar t\ninoone$ decays, 
in the signal regions Rpc2Lsoft1b (a,b) and Rpc2Lsoft2b (c,d).}
\end{figure}

\begin{figure}[htb!]
\centering
\begin{subfigure}[t]{0.49\textwidth}\includegraphics[width=\textwidth]{EffAcc/acceptance_ComprGttRpc2L2bH}\caption{Rpc2L2bH acceptance}\end{subfigure}
\begin{subfigure}[t]{0.49\textwidth}\includegraphics[width=\textwidth]{EffAcc/efficiency_ComprGttRpc2L2bH}\caption{Rpc2L2bH efficiency}\end{subfigure}
\begin{subfigure}[t]{0.49\textwidth}\includegraphics[width=\textwidth]{EffAcc/acceptance_ComprGttRpc2L2bS}\caption{Rpc2L2bS acceptance}\end{subfigure}
\begin{subfigure}[t]{0.49\textwidth}\includegraphics[width=\textwidth]{EffAcc/efficiency_ComprGttRpc2L2bS}\caption{Rpc2L2bS efficiency}\end{subfigure}
\caption{Signal acceptance (a,c) and reconstruction efficiency (b,d) 
for simplified models of $\gluino\gluino$ production with $\gluino\to tWb\ninoone$ decays (region with $\Delta m(\gluino,\ninoone)<2m_t$), 
in the signal regions Rpc2L2bH (a,b) and Rpc2L2bS (c,d).}
\end{figure}

\begin{figure}[htb!]
\centering
\begin{subfigure}[t]{0.49\textwidth}\includegraphics[width=\textwidth]{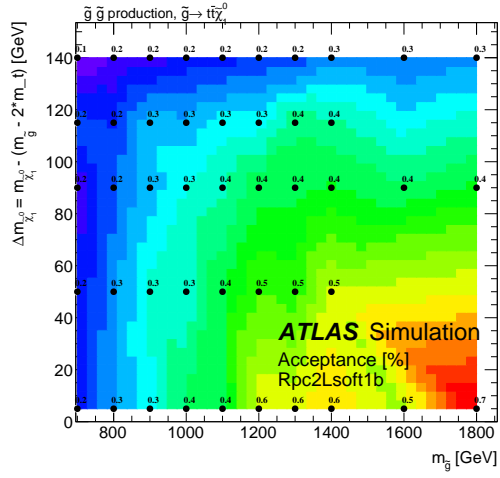}\caption{Rpc2Lsoft1b acceptance}\end{subfigure}
\begin{subfigure}[t]{0.49\textwidth}\includegraphics[width=\textwidth]{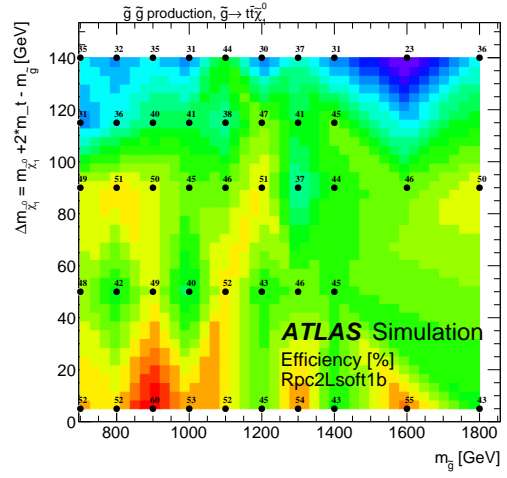}\caption{Rpc2Lsoft1b efficiency}\end{subfigure}
\begin{subfigure}[t]{0.49\textwidth}\includegraphics[width=\textwidth]{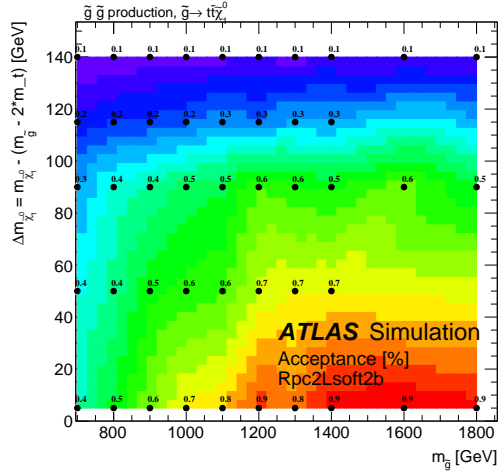}\caption{Rpc2Lsoft2b acceptance}\end{subfigure}
\begin{subfigure}[t]{0.49\textwidth}\includegraphics[width=\textwidth]{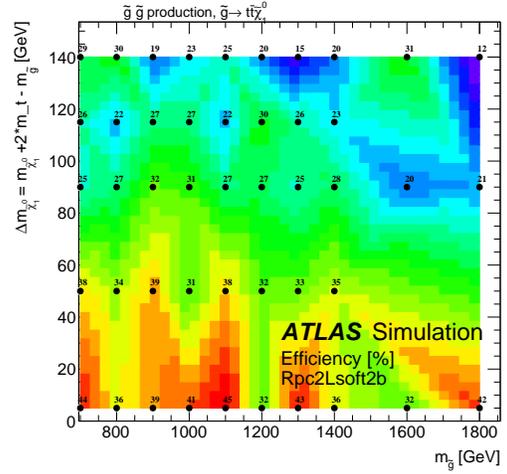}\caption{Rpc2Lsoft2b efficiency}\end{subfigure}
\caption{Signal acceptance (a,c) and reconstruction efficiency (b,d) 
for simplified models of $\gluino\gluino$ production with $\gluino\to tWb\ninoone$ decays (region with $\Delta m(\gluino,\ninoone)<2m_t$), 
in the signal regions Rpc2Lsoft1b (a,b) and Rpc2Lsoft2b (c,d).}
\end{figure}

\begin{figure}[htb!]
\centering
\begin{subfigure}[t]{0.49\textwidth}\includegraphics[width=\textwidth]{EffAcc/acceptance_BttRpc2L1bH}\caption{Rpc2L1bH acceptance}\end{subfigure}
\begin{subfigure}[t]{0.49\textwidth}\includegraphics[width=\textwidth]{EffAcc/efficiency_BttRpc2L1bH}\caption{Rpc2L1bH efficiency}\end{subfigure}
\begin{subfigure}[t]{0.49\textwidth}\includegraphics[width=\textwidth]{EffAcc/acceptance_BttRpc2L1bS}\caption{Rpc2L1bS acceptance}\end{subfigure}
\begin{subfigure}[t]{0.49\textwidth}\includegraphics[width=\textwidth]{EffAcc/efficiency_BttRpc2L1bS}\caption{Rpc2L1bS efficiency}\end{subfigure}
\caption{Signal acceptance (a,c) and reconstruction efficiency (b,d) 
for simplified models of $\sbottomone\sbottomonebar$ production with $\sbottomone\to tW^{-}\ninoone$ decays, 
in the signal regions Rpc2L1bH (a,b) and Rpc2L1bS (c,d).}
\end{figure}

\begin{table}[htb!]\centering\begin{tabular}{|l|c|c|c|c|c|c|c|}\hline
\multicolumn{8}{|l|}{Rpc2L2bH,\quad $\gluino\gluino$ production in the NUHM2 model}\\\hline
$m_{\gluino}$ [GeV] & 300 & 350 & 400 & 500 & 600 & 700 & 800 \\\hline
Acceptance & 0.8\% & 1.6\% & 2.1\% & 3.2\% & 3.5\% & 4.4\% & 4.0\% \\
Efficiency & 43\% & 49\% & 50\% & 49\% & 48\% & 43\% & 49\%\\\hline
\end{tabular}
\caption{Rpc2L2bH signal region acceptance and reconstruction efficiency for $\gluino\gluino$ production in the NUHM2 model.}
\end{table}

\begin{table}[htb!]\centering\begin{tabular}{|l|c|c|c|c|c|c|}\hline
\multicolumn{7}{|l|}{Rpc3LSS1b,\quad $\stopone\stoponebar$ production,\quad $\stopone\to t\ninotwo$, $\ninotwo\to W^\mp\chargino$}\\\hline
$m_{\stopone}$ [GeV] & 550 & 600 & 650 & 700 & 750 & 800 \\\hline
Acceptance & 0.3\% & 0.3\% & 0.3\% & 0.4\% & 0.4\% & 0.4\%\\
Efficiency & 36\% & 42\% & 44\% & 37\% & 33\% & 30\%\\\hline
\end{tabular}
\caption{Rpc3LSS1b signal region acceptance and reconstruction efficiency for $\stopone\stoponebar$ production with $\stopone\to t\ninotwo$ ($\ninotwo\to W\chargino$) decays.}
\end{table}